\documentclass[11pt,a4paper]{article}

\usepackage[utf8]{inputenc}
\usepackage[T1]{fontenc}
\usepackage{amsmath, amssymb}
\usepackage{commath}
\usepackage{bm}
\usepackage[pdftex]{graphicx}
\usepackage{graphicx}
\usepackage{multirow}

\usepackage{float}
\usepackage{placeins}

\usepackage{subfiles}
\usepackage[pdftex,linkcolor=black,pdfborder={0 0 0}]{hyperref}
\usepackage{blindtext}
\usepackage{geometry}
\usepackage[acronym,nonumberlist]{glossaries}
\makeglossaries

\usepackage{xcolor}
\usepackage{siunitx}
\usepackage{calc}
\usepackage{enumitem}
\usepackage{comment}
\usepackage{caption}
\usepackage[all]{nowidow}
\usepackage[protrusion=true,expansion=true]{microtype}
\usepackage{algorithm}
\usepackage{algpseudocode}

\usepackage{wrapfig}
\usepackage{lipsum} 

\newenvironment{biography}[2]{
    \begin{minipage}{\textwidth}
    \begin{wrapfigure}[7]{l}{2.54cm} 
        \centering
        #1
    \end{wrapfigure}
    \noindent\textbf{#2}\\
}{
    \end{minipage}
    \vspace{0.2cm}
}

\newcommand{\mycomment}[1]{}

\newacronym{qam}{QAM}{Quadrature Amplitude Modulation}
\newacronym{psk}{PSK}{Phase Shift Keying}
\newacronym{ask}{ASK}{Amplitude Shift Keying}
\newacronym{2-ask}{2-ASK}{Binary Amplitude Shift Keying}
\newacronym{4-ask}{4-ASK}{Quaternary Amplitude Shift Keying}
\newacronym{8-ask}{8-ASK}{Octal Amplitude Shift Keying}
\newacronym{bpsk}{BPSK}{Binary Phase Shift Keying}
\newacronym{qpsk}{QPSK}{Quadrature Phase Shift Keying}
\newacronym{qask}{QASK}{Quadrature Amplitude Shift Keying}

\newacronym{ftn}{FTN}{Faster-Than-Nyquist}
\newacronym{sftn}{SFTN}{Single-Carrier FTN}
\newacronym{mftn}{MFTN}{Multi-Carrier FTN}

\newacronym{sc}{SC}{Single-Carrier}
\newacronym{mc}{MC}{Multi-Carrier}

\newacronym{fir}{FIR}{Finite Impulse Response}
\newacronym{iir}{IIR}{Infinite Impulse Response}

\newacronym{isi}{ISI}{Inter-Symbol interference}
\newacronym{sisi}{SISI}{Self-Inter-Symbol Interference}
\newacronym{ici}{ICI}{Inter-Carrier Interference}
\newacronym{ber}{BER}{Bit Error Rate}
\newacronym{bep}{BEP}{Bit Error Probability}
\newacronym{sep}{SEP}{Symbol Error Probability}
\newacronym{blep}{BLEP}{BLock Error Probability}
\newacronym{bler}{BLER}{BLock Error Rate}

\newacronym{awgn}{AWGN}{Additive White Gaussian Noise}
\newacronym{snr}{SNR}{Signal-to-Noise Ratio}
\newacronym{prs}{PRS}{Partial Response Signaling}
\newacronym{ms-prs}{MS-PRS}{Multi-Stream Partial Response Signaling}

\newacronym{ldpc}{LDPC}{Low-Density Parity Check}
\newacronym{pccc}{PCCC}{Parallel Concatenated Convolutional Code}

\newacronym{ldgm}{LDGM}{Low-Density Generating Matrix}
\newacronym{aldgm}{ALDGM}{Analog Low-Density Generating Matrix}
\newacronym{msed}{MSED}{Minimum Squared Euclidean Distance}
\newacronym{med}{MED}{Minimum Euclidean Distance}
\newacronym{sed}{SED}{Squared Euclidean Distance}
\newacronym{sen}{SEN}{Squared Euclidean Norm}
\newacronym{mmse}{MMSE}{Minimum Mean Square Error}

\newacronym{nsm}{NSM}{Nyquist Signaling Modulation}

\newacronym{1d}{1D}{One-Dimensional}
\newacronym{2d}{2D}{Two-Dimensional}
\newacronym{3d}{3D}{Three-Dimensional}
\newacronym{4d}{4D}{Four-Dimensional}

\newacronym{md}{MD}{Main Diagonal}
\newacronym{sd}{SD}{Secondary Diagonal}

\newacronym{dft}{DFT}{Discrete Fourier Transform}
\newacronym{idft}{IDFT}{Inverse Discrete Fourier Transform}

\newacronym{tf}{TF}{Transfer Function}
\newacronym{rtf}{RTF}{Reduced Transfer Function}
\newacronym{atf}{ATF}{Augmented Transfer Function}

\newacronym{noma}{NOMA}{Non-Orthogonal Multiple Access}
\newacronym{pd-noma}{PD-NOMA}{Power-Domain NOMA}
\newacronym{cd-noma}{CD-NOMA}{Code-Domain NOMA}
\newacronym{musa}{MUSA}{Multiuser Shared Access}
\newacronym{scma}{SCMA}{Sparse Code Multiple Access}
\newacronym{lds}{LDS}{Low-Density Spreading}
\newacronym{lds-cdma}{LDS-CDMA}{Low-Density Signature CDMA}
\newacronym{pdma}{PDMA}{Pattern Division Multiple Access}
\newacronym{idma}{IDMA}{Interleave Division Multiple Access}
\newacronym{ssc}{SSC}{Sparse Superposition Coding}

\newacronym{amp}{AMP}{Approximate Message Passing}

\newacronym{rsc}{RSC}{Recursive Systematic Convolutional}

\newacronym{otfs}{OTFS}{Orthogonal Time Frequency Space}
\newacronym{ofdm}{OFDM}{Orthogonal Frequency Division Multiplexing}

\newacronym{papr}{PAPR}{Peak-to-Average Power Ratio}
\newacronym{psd}{PSD}{Power Spectral Density}

\newacronym{mimo}{MIMO}{Multiple-Input Multiple-Output}

\newacronym{e2e}{E2E}{Edge-to-Edge}
\newacronym{f2f}{F2F}{Face-to-Face}
\newacronym{v2v}{V2V}{Vertex-to-Vertex}

\newacronym{qc}{QC}{Quasi-Cyclic}

\newacronym{dct}{DCT}{Discrete Cosine Transform}
\newacronym{klt}{KLT}{Karhunen–Loève Transform}
\newacronym{pca}{PCA}{Principal Component Analysis}

\newacronym{pso}{PSO}{Particle Swarm Optimization}
\newacronym{gwo}{GWO}{Grey Wolf Optimizer}
\newacronym{cs}{CS}{Cuckoo Search}
\newacronym{sso}{SSO}{Social Spider Optimization}
\newacronym{kh}{KH}{Krill Herd}
\newacronym{woa}{WOA}{Whale Optimization Algorithm}

\newacronym{rc}{RC}{Raised Cosine}
\newacronym{rrc}{RRC}{Root Raised Cosine}
\newacronym{pswf}{PSWF}{Prolate Spheroidal Wave Functions}

\newacronym{ml}{ML}{Maximum-Likelihood}
\newacronym{map}{MAP}{Maximum \emph{a posteriori}}
\newacronym{bcjr}{BCJR}{Bahl, Cocke, Jelinek, and Raviv}
\newacronym{llr}{LLR}{Log-Likelihood Ratio}

\newacronym{adsl}{ADSL}{Asymmetric Digital Subscriber Line}

\newacronym{tcm}{TCM}{Trellis-Coded Modulation}
\newacronym{ccdm}{CCDM}{Constant Composition Distribution Matching}
\newacronym{cdma}{CDMA}{Code Division Multiple Access}
\newacronym{thp}{THP}{Tomlinson-Harashima Precoding}

\newacronym{gcd}{gcd}{greatest common divisor}

\newacronym{ar}{AR}{AutoRegressive}
\newacronym{ma-mov}{MA}{Moving Average}
\newacronym{arma}{ARMA}{AutoRegressive Moving Average}

\newacronym{ma-mac}{MA}{Multiple Access}

\newacronym{epr4}{EPR4}{Extended Partial-Response Class 4}

\newacronym{jscc}{JSCC}{Joint Source–Channel Coding}

\newacronym{clt}{CLT}{Central Limit Theorem}

\newacronym{mn}{MN}{MacKay–Neal}

\newacronym{cpm}{CPM}{Continuous Phase Modulation}

\hypersetup{ 	
pdfsubject = {},
pdftitle = {},
pdfauthor = {}
}

\begin{document}

\pagenumbering{roman}   

\begin{titlepage}
    \centering
    \vspace*{2cm} 

    {\LARGE \textbf{Nyquist Signaling Modulation (NSM): An FTN-Inspired Paradigm Shift in Modulation Design for 6G and Beyond}\par}
    \vspace{2cm}    
    {\large
    Mohamed Siala\textsuperscript{1}\\
    Abdullah Al-Nafisah\textsuperscript{2}\\
    Tareq Al-Naffouri\textsuperscript{2}\\[0.5cm] 
    
    \textsuperscript{1}Higher School of Communications of Tunis (SUP'COM), Carthage University, Tunis, Tunisia\\
    \textsuperscript{2}Computer, Electrical and Mathematical Sciences and Engineering (CEMSE), King Abdullah University of Science and Technology (KAUST), Thuwal, Saudi Arabia\\
    
    \texttt{mohamed.siala@supcom.tn}\\
    \texttt{\{abdullah.alnafisah, tareq.alnaffouri\}@kaust.edu.sa}
    \par}
    
    \vfill
    {\large \today\par}
\end{titlepage}
\thispagestyle{empty} 
\newpage

\setcounter{page}{2}

\begin{abstract}
Nyquist Signaling Modulations (NSMs) are a novel signaling paradigm inspired by faster-than-Nyquist principles but following a fundamentally different approach that enables controlled inter-symbol interference through carefully designed finite-impulse-response filters. NSMs can be realized in one-, two-, and three-dimensional configurations using rational- or real-tapped filters, providing increased degrees of freedom for filter design, energy balancing, and preservation of minimum squared Euclidean distances (MSED). This structured approach offers a flexible and analytically tractable framework that can enhance robustness, spectral efficiency, and performance relative to classical modulation schemes.

NSMs are systematically optimized through rational- or real-valued finite-impulse-response filters. Rational-tap filters allow direct derivation of closed-form filter expressions and support tractable exhaustive searches using equivalence-class reductions, even for relatively long filter lengths. Real-tap filters, in contrast, provide a larger space of candidate NSMs, which can improve achievable performance, although their optimization is more intricate and rapidly becomes intractable as filter length increases. For these filters, closed-form expressions for the filter structures can be derived indirectly via symbolic calculus, but only for relatively short filter lengths within the tractable optimization range. Multidimensional extensions of NSMs further increase design flexibility, enabling the preservation of the $2$-ASK minimum squared Euclidean distance across substreams. The performance of NSMs is analyzed using a transfer-function framework, including full, reduced, and truncated representations as well as iterative computational methods, which accurately capture dominant error events and yield BEP bounds in very tight agreement with simulated BER across extended SNR ranges. These transfer functions are structurally similar to those of convolutional codes but exhibit radically distinct error behaviors due to the exponential decay of error-event significance with the Hamming weight of their input differences.

Beyond one-dimensional structures, NSMs generalize and extend the multistream FTN (MFTN) concept to two-, three-, and mixed-dimensional configurations, offering additional degrees of freedom for arranging substreams and controlling inter-substream interference while preserving minimum squared Euclidean distances. Detection complexity naturally increases with dimensionality. To address this, we propose a two-stage maximum-likelihood detector that substantially reduces computational effort. Nevertheless, its complexity remains non-negligible, motivating further research into reduced-complexity and iterative detection schemes such as sphere decoding.

A unified analytical framework supports the optimization and evaluation of NSMs, particularly for rational-tapped filters, where equivalence relations and MSED-preserving transformations drastically reduce the search space and enable systematic exploration of high-dimensional configurations. Reduced and truncated transfer-function methodologies, validated through extensive simulations, provide tight analytical bounds on bit-error probability and confirm the predictive accuracy of the theoretical models across wide SNR ranges.

NSMs are further generalized within an analog Low-Density Generator Matrix (LDGM) framework, which relaxes the rigid structure of conventional NSMs and introduces controlled sparsity and randomness across generating matrices. This added flexibility enables better control of energy distribution and structural regularity, helping to mitigate low-distance error events and enhance robustness. The analog LDGM formulation also provides a coherent foundation for the seamless and harmonious integration of channel coding, or jointly channel and source coding, with the modulation process, forming a unified and highly adaptable signal representation paradigm. Within this framework, energy balancing, interleaving, and message-passing strategies play a central role in mitigating low-distance and low-Hamming-weight error events, thereby reinforcing the structural and performance robustness of NSMs in both coded and uncoded scenarios.

Collectively, these developments establish a coherent foundation for the next generation of physical-layer architectures. By naturally combining modulation with source and/or channel coding within a single, continuous design space, the NSM–analog LDGM paradigm introduces a flexible, scalable, and performance-driven alternative to classical modulations. Its structural compatibility with modern LDPC-based coding schemes and its ability to harmonize controlled interference, sparsity, and joint optimization make it a strong candidate for future 6G and beyond communication standards.
\end{abstract}

\textbf{Keywords:} Nyquist Signaling Modulations (NSMs), Faster-than-Nyquist (FTN), Multistream Faster-than-Nyquist (MFTN), Minimum Squared Euclidean Distance (MSED), Analog LDGM NSMs, Multidimensional NSMs, Multi-Stream Partial-Response Signaling (MS-PRS), Energy balancing, Tightness, Reduced transfer function, Truncated transfer function, Closed-form filter derivation, Digital–analog packing, Joint source-channel-modulation design, 6G communications, Next-generation wireless systems, Future physical-layer design
\newpage

\tableofcontents
\newpage
\listoffigures
\newpage
\listoftables
\newpage
\listofalgorithms
\newpage

\glsaddall
\printglossary[type=\acronymtype, title={List of Acronyms}, toctitle={List of Acronyms}]\newpage

\clearpage                 
\pagenumbering{arabic}     
\setcounter{page}{1}       


\section{Introduction}

The exponential growth of wireless connectivity and high-speed multimedia services has placed unprecedented pressure on spectral resources. Bandwidth efficiency, defined as the average data rate per unit bandwidth, has therefore become a central design metric for next-generation mobile systems. Conventional transmission architectures follow the Nyquist criterion. This ensures time-orthogonality and eliminates \gls{isi} at a symbol spacing tied to the signal bandwidth.

\gls{ftn} signaling departs from this orthogonality constraint by intentionally increasing the symbol packing density in time. The technique accepts controlled \gls{isi} in exchange for higher spectral efficiency. Mazo \cite{Mazo75} demonstrated that, for ideal “sinc” pulses in continuous time, the symbol rate may be increased by up to 25\% without reducing the \gls{msed} between distinct transmitted sequences. This establishes an information-theoretic and geometrical basis for \gls{ftn}’s potential to improve throughput without immediate penalty to error performance at high \gls{snr}.

While Mazo’s original treatment was cast in continuous time, practical realizations require discrete-time models and finite-length processing. McGuire and Sima \cite{McGuire10} introduced a block-based discrete-time reformulation of \gls{ftn}. In their approach, rectangular frequency-domain shaping over finite blocks yields aliased, “sinc”-like discrete pulses and limits the \gls{isi} span to a finite number of samples. This block formulation bounds receiver complexity and enables parallelizable detection algorithms. It makes \gls{ftn} detection tractable for real systems while retaining the distance-preserving intuition of the continuous-time theory.  

A conceptually related but differently motivated line of work is represented by the Signal Codes of Shalvi \cite{Shalvi08}. In their original interpretation, Signal Codes modify the signal geometry in Euclidean space with the goal of maximizing the \gls{msed} between constellation points for a given data rate. The approach emphasizes coding gain rather than bandwidth gain. A complementary interpretation, which is of particular relevance here, is to view the same mechanism as a means of preserving the \gls{msed} when the data rate is increased. In this dual perspective, Signal Codes conceptually achieve the same overarching goal as \gls{ftn}: enabling data rate growth without sacrificing distance properties. Although practical implementations of Signal Codes are hindered by the need for additional shaping and power-control procedures to ensure spectral efficiency, their structural resemblance to filter-based \gls{ftn} frameworks makes them an instructive analogue for the schemes proposed in this work.

\gls{ftn} signaling has also been extended to \gls{mc} systems, giving rise to \gls{mftn}, also referred to as time-frequency packing. The first study of \gls{mftn} was reported by Rusek and Anderson \cite{Rusek05}. In \gls{mftn}, non-orthogonal symbol packing occurs along two axes, creating a multi-dimensional signaling structure. This generalization allows higher spectral efficiency while retaining manageable interference through advanced receiver processing. \gls{mftn} serves as an inspiration for considering multi-dimensional modulation frameworks. In the proposed framework, these additional dimensions are abstract and not tied to time or frequency interpretations, and the structure can, in principle, be generalized beyond two dimensions.  

Beyond \gls{mc} systems, \gls{ftn} principles have also been applied to \gls{mimo} and higher-order modulation frameworks \cite{Dasalukunte14b}. In addition, \gls{ftn} shares a strong conceptual analogy with \gls{prs}, and several studies treat \gls{prs} and \gls{ftn} in a unified framework or employ \gls{prs}-inspired equalization techniques within \gls{ftn} systems \cite{Anderson06, Rusek09, Barbieri09, Anderson13b, Dasalukunte14a, Rusek06c}. These extensions illustrate the flexibility of \gls{ftn} in enhancing spectral and energy efficiency and highlight the wide applicability of bandwidth-efficient signaling techniques.

Over the past two decades, research on \gls{ftn} has evolved from foundational theoretical studies to extensive investigations into receiver architectures, coding strategies, and advanced system designs. The sustained academic and industrial attention it has received underscores \gls{ftn}’s relevance as both a conceptual and practical paradigm for achieving higher spectral efficiency.

\begin{figure}[!htbp]
\centering
\includegraphics[width=1.0\linewidth]{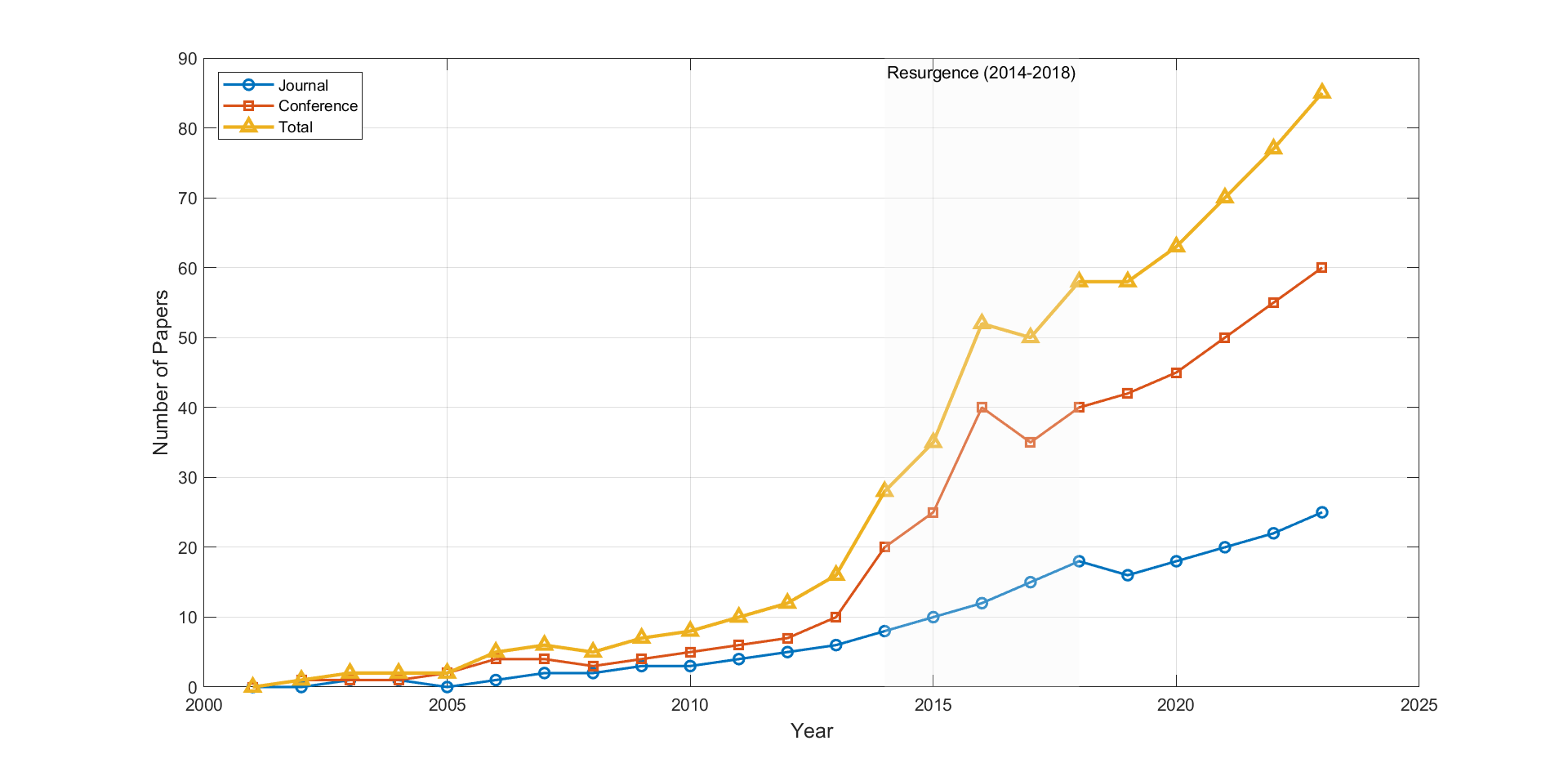}
\caption{Estimated number of FTN-related publications per year (2001--2023), showing journal and conference contributions. The statistics confirm the resurgence of interest around 2014--2018 and stable, continuing activity through 2023.}
\label{fig:FTN_statistics}
\end{figure}

Figure~\ref{fig:FTN_statistics} provides an overview of the evolution of \gls{ftn}-related research activity between 2001 and 2023, illustrating the estimated number of publications per year. The trend clearly indicates a notable resurgence of interest during the 2014–2018 period, coinciding with the introduction of low-complexity detection algorithms, iterative receivers, and practical precoding techniques that made \gls{ftn} increasingly feasible for implementation. Following this peak, the figure shows continued and stable publication activity through 2023, suggesting that \gls{ftn} has matured into a stable and active area of research rather than a transient topic of interest. This sustained output reflects the community’s ongoing efforts to extend \gls{ftn} principles to broader system paradigms, including multi-dimensional modulation, coded systems, and machine-learning-assisted detection.

Collectively, these studies outline the principal algorithmic tools underpinning modern \gls{ftn} implementations, including trellis-based detection, sphere decoding, frequency-domain equalization, and precoding \cite{Anderson06, Rusek09, Barbieri09, Anderson13, Dasalukunte14a, Rusek06b}. The continuous exploration of these tools reinforces \gls{ftn}’s position as a cornerstone in the pursuit of spectrally efficient signaling for next-generation communication systems.

The present work adopts a strictly discrete-time perspective inspired by the continuous-time \gls{ftn} intuition of Mazo \cite{Mazo75} and the discrete block formulation of McGuire and Sima \cite{McGuire10}. Mazo’s original contribution established that introducing intentional \gls{isi} through \gls{ftn} signaling can enable higher data rates without degrading performance. McGuire later demonstrated that this principle can be reformulated in a discrete-time framework, showing that \gls{ftn} signaling can be implemented in practice using block-based methods. In parallel, the work of Shalvi on Signal Codes \cite{Shalvi08} provides a structural analogue in which filter-based signal transformations preserve distance properties and inspire our modulation and shaping strategies. Together, these three perspectives form the conceptual foundation of our approach and naturally lead to the problem formulation, research objectives, and methodological contributions that are developed in the remainder of this introduction.  

Building on the seminal work of Mazo \cite{Mazo75} on \gls{ftn} signaling, we propose a new paradigm and architecture for our main modulation scheme, which we refer to as \gls{nsm}. While Mazo used continuous-time “sinc” waveforms with an oversampling of the modulated signal at the same rate as the data symbols, our approach reinterprets the continuous-time modulated signal as a discrete-time signal sampled at exactly the Nyquist rate, in accordance with the Shannon sampling theorem. This ensures that no information about the transmitted symbols is lost.  

\begin{figure}[!htbp]
	\centering
	\includegraphics[width=\linewidth]{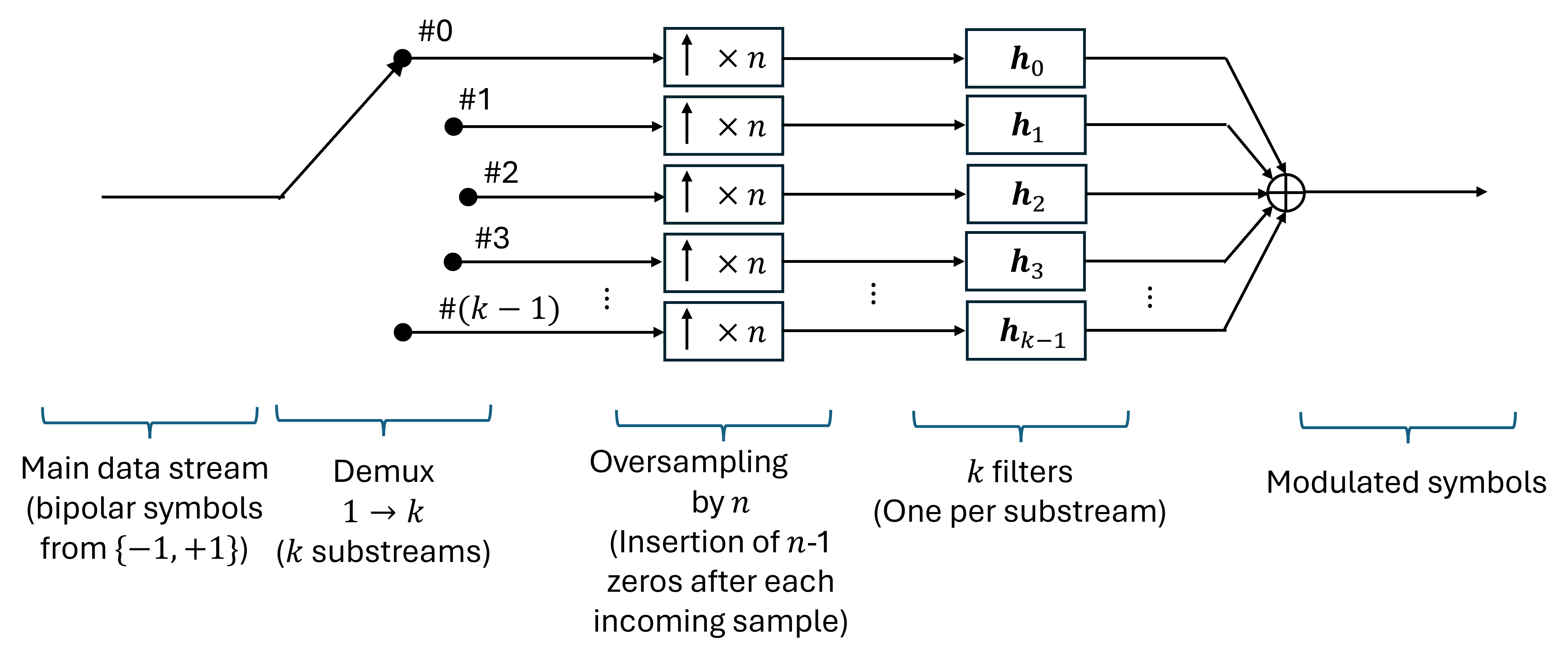}
	\caption{Proposed \gls{ms-prs} structure with rate $\rho_m = k/n$. Each data stream is over-sampled, filtered, and combined to produce the modulated signal.}
	\label{fig:mpr_modulations}
\end{figure}

In our framework, the Mazo-modulated signal can be viewed as a \emph{multi-stream polyphase structure}, where each stream is filtered with an appropriately phased version of the original “sinc” waveform. This interpretation naturally leads to a general family of \gls{nsm} schemes, referred to as \gls{ms-prs} modulations. In these modulations, the data stream from the encoder is first de-multiplexed into $k$ parallel sub-streams. Each sub-stream is then over-sampled by a factor of $n$, inserting $n-1$ zeros between successive samples, and filtered by an independently designed digital filter. The outputs of all filtered sub-streams are summed to form the composite modulated signal. Consequently, the overall modulation rate is scaled by a factor of $\rho_m = k/n$, while the total system rate becomes $\rho = \rho_c \rho_m$ when combined with an error-correction code of rate $\rho_c$. This architecture, illustrated in Figure~\ref{fig:mpr_modulations}, highlights the structural simplicity and flexibility of the proposed \gls{ms-prs} modulation.

A distinctive property of the \gls{ms-prs} framework is that the filters applied to each sub-stream can be chosen freely and independently from any underlying continuous-time waveform. These filters can be of \gls{fir} type, allowing simplified signal generation at the transmitter, reduced detection complexity at the receiver, and elimination of truncation effects inherent to classical \gls{ftn} implementations, where oversampled “sinc” filters typically have \glspl{iir}. This decoupling of the digital sub-stream filters from the analog shaping filter provides a high degree of design flexibility and enables direct optimization of pulse characteristics according to desired system objectives, such as maximizing the \gls{msed} or minimizing the \gls{sep}, with or without channel coding.

From a broader perspective, the \gls{nsm} family provides a unified framework encompassing these multi-stream modulations. The proposed \gls{ms-prs} scheme represents its fundamental one-dimensional realization, conceptually aligned with the classical \gls{sc} framework: a single encoded data sequence is distributed across multiple parallel filtered paths that jointly form the transmitted modulated signal. This structure can be interpreted as a form of analog-domain error correction that complements digital coding, offering a more flexible and efficient alternative to conventional \gls{ftn} signaling while retaining compatibility with existing transceiver architectures.

Inspired by \gls{mftn} \cite{Rusek09}, we extend the \gls{nsm} framework to \gls{2d} and \gls{3d} systems. In the \gls{2d} case, the main data stream is demultiplexed into multiple sub-streams organized on a \gls{2d} grid, filtered with freely chosen \gls{2d} \gls{fir} filters, and transmitted via single- or multi-carrier systems, optionally using \gls{mimo} systems. Similarly, \gls{3d} \glspl{nsm} demultiplex data streams along a \gls{3d} grid and apply freely chosen \gls{3d} \gls{fir} filters. These multi-dimensional \glspl{nsm} provide additional design degrees of freedom and facilitate achieving the \gls{msed} of $2$-ASK more easily than the \gls{1d}, \gls{sc} case.

We consider specific families of \glspl{nsm} based on their transmission rates. We study rate-$2$ \glspl{nsm}, which have been the main focus of historical \gls{ftn} research \cite{Anderson06}, corresponding in terms of spectral efficiency to \gls{1d} $4$-ASK or \gls{2d} $16$-QAM modulations. We also study rate-$3$ \glspl{nsm}, equivalent to \gls{1d} $8$-ASK or \gls{2d} $64$-QAM modulations in spectral efficiency. Additionally, we consider \glspl{nsm} with rates below $2$, expressed as $(Q+1)/Q$ for an integer $Q \ge 2$. These lower-rate \glspl{nsm} provide a tradeoff between transmission rate and performance, and are inspired by the original Mazo \gls{ftn} work, which focused on rates around $5/4$ (the Mazo’s maximum compression factor of $0.802$, which preserves the \gls{msed} of $2$-ASK, yields an effective rate close to $5/4$).

Within the \gls{1d} \gls{nsm} framework, we propose two families of \gls{ms-prs} modulations using \gls{fir} filters. The first family employs filters with real-valued taps, while the second uses rational or integer-valued taps after renormalization. Filter lengths are fixed to maintain manageable processing complexity at both transmitter and receiver. For rate-$2$ \glspl{nsm} with real-valued filters, two categories are considered: unconstrained, where the sub-stream filters can share the overall average modulated symbol energy freely, and constrained, where this average energy is distributed equally among the sub-stream filters. Unconstrained \glspl{nsm} achieve higher \glspl{msed} and provide a better bootstrap for iterative turbo-equalization at low \gls{snr}, whereas constrained \glspl{nsm} offer balanced energy distribution that benefits performance at later iterations.

For \gls{1d} rate-$2$ \glspl{nsm} with real-valued taps, the optimization of filters begins with a random search procedure that approximates the identification of filter sets maximizing the \gls{msed}, up to the limits of numerical precision. This initial step is refined through a three-stage methodology. First, promising candidate filters are retained in numerical form. Second, the error events corresponding to the \gls{msed} are determined exactly. Finally, exploiting the fact that these error events all yield the same \gls{msed}, we derive a system of equations whose symbolic solution provides closed-form expressions for both the optimal filters and the associated \glspl{msed}, yielding exact analytical characterizations for many of the considered configurations.

A notable structural observation concerns rate-$2$ \glspl{nsm} where one filter has length 1 and the other has variable length. In this case, achieving an \gls{msed} equal to that of $2$-ASK requires the longer filter to have length at least $10$. For shorter lengths, the attainable distance is strictly smaller. In this context, we also introduce the notion of tightness, which designates configurations in which the minimum distance is attained with high multiplicity of error events. While the implications of this property will be addressed later, the definition itself provides an additional criterion for analyzing and classifying \glspl{nsm} beyond minimum distance alone.

For rate-$3$ \glspl{nsm} with real-valued taps, no attempt is made to derive closed-form symbolic characterizations. Two complementary optimization approaches are considered. In the first, filter optimization is carried out entirely through numerical search procedures, yielding approximate identification of filters that maximize the \gls{msed}. In the second, rate-$3$ \glspl{nsm} are constructed by inheriting optimized rate-$2$ filters and appending a third sub-stream filtered by a properly scaled impulse response of length one. Depending on the configuration, these two approaches either converge to the same filters or yield distinct solutions. The reasons underlying these similarities and differences are analyzed in detail, providing additional insight into the structure of rate-$3$ \glspl{nsm}.

A second family of \glspl{nsm} is based on rational-tap filters (or equivalently, integer taps after renormalization). In the \gls{1d} rate-$2$ case, we first focus on the special scenario where the non-null taps are bipolar. This allows a rigorous algebraic characterization of performance, employing tools such as $m$-sequences and their correlation properties to analyze precisely how the structure of these integer-tap filters determines the \gls{msed}. This algebraic study provides exact insights into the performance of this constrained class of \glspl{nsm}. Subsequently, we consider the more general \gls{1d} rate-$2$ \glspl{nsm} with arbitrary rational-tap filters. While these more general filters are not analyzed algebraically, they are studied thoroughly through numerical design and optimization. The additional freedom in selecting tap values enables improved \gls{nsm} performance by expanding the design space beyond the constraints of the bipolar-tap case.

Multi-dimensional rate-$2$ \glspl{nsm}, as well as \gls{1d} rate-$(Q+1)/Q$ \glspl{nsm} with $Q \geq 2$, make use of rational-tap filters (equivalently, integer-tap filters after renormalization) to simplify the optimization process and keep the search space tractable. In the \gls{2d} and \gls{3d} cases, full characterization of the \gls{msed} is not feasible due to the absence of a trellis structure. Therefore, a partial optimization is performed by extracting \gls{1d} \glspl{nsm} along key axes—such as horizontal, vertical, and diagonal axes in two dimensions, or horizontal, vertical, depth, and \gls{e2e} axes in three dimensions—and maximizing the minimum of the \glspl{msed} of these extracted \gls{1d} systems. While this approach provides a tractable design methodology, it does not guarantee that the overall multi-dimensional \gls{nsm} will achieve the same \gls{msed}, since error events that are not captured by the extracted \gls{1d} \glspl{nsm} may exist and could lead to lower \glspl{sed} in the full \gls{2d} or \gls{3d} system.

Our framework enables the use of very short filters for each sub-stream, avoiding the truncation issues encountered in classical \gls{ftn} and \gls{mftn}. This significantly simplifies detection while maintaining performance. One-dimensional \glspl{nsm} naturally correspond to \gls{sc} systems. In contrast, \gls{2d} and \gls{3d} \glspl{nsm} demultiplex the data streams along \gls{2d} or \gls{3d} grids, respectively, and apply freely chosen \gls{2d} or \gls{3d} \gls{fir} filters. These multi-dimensional \glspl{nsm} can be transmitted over single- or multi-carrier channels, optionally using \gls{mimo}. With two- and three-dimensional \glspl{nsm}, the \gls{msed} of $2$-ASK is achieved more easily and naturally, allowing most of the time intuitive selection of the appropriate multi-dimensional filters.

To evaluate the performance of the proposed \glspl{nsm}, simulation results are obtained in terms of raw \gls{ber}. To complement these simulations and provide analytical insight, \gls{bep} bounds are derived from state diagrams that characterize all possible error events (including both dominant and non-dominant ones). These diagrams provide a comprehensive view of the \gls{nsm} error structure and form the foundation for subsequent \gls{tf} analyses.

In order to analyze the error events of \glspl{nsm}, we build on the classical framework of convolutional codes, where \glspl{tf} are widely used to enumerate and characterize error sequences. Similarly, the \gls{tf} of an \gls{nsm} depends on two symbolic parameters: $N$, which accounts for differences in input sequences, and $D$, which characterizes the corresponding differences in output sequences. By depending on these two parameters, the \gls{tf} provides a complete representation of all possible error events. A simplified tool, the \gls{rtf}, is introduced to ease evaluation. It depends only on the parameter $D$, preserving the meaning of $D$ from the full \gls{tf} while discarding the explicit tracking of input differences. This simplification makes the \gls{rtf} far more tractable for both symbolic and numerical analysis.

Truncated versions of either the \gls{tf} or the \gls{rtf} are considered only when direct, one-shot symbolic computation of the full function is either impossible or prohibitively time-consuming. These truncated versions allow focusing on the most relevant error events, particularly those associated with the lowest \glspl{sed}, without losing the ability to characterize the dominant contributors to \gls{nsm} performance. For non-truncated \glspl{tf} or \glspl{rtf}, exact symbolic determination is performed directly whenever feasible. When working with truncated versions, a finite iterative procedure is applied to compute them. Convergence of the truncated \gls{rtf} is guaranteed after a finite number of iterations in all cases. In contrast, convergence of the truncated full \gls{tf} is ensured only for non-degenerate \glspl{nsm}, where degeneracy is defined as the presence of an infinite number of error events that all lead to the same \gls{msed}, arising from an infinite set of possible input-sequence differences. Degenerate \glspl{nsm} require special attention, as this phenomenon can prevent guaranteed convergence of the truncated full \gls{tf}, whereas the \gls{rtf} effectively mitigates this issue and remains a reliable tool for performance evaluation.

These \gls{tf} analyses are also applied to optimize multi-dimensional \glspl{nsm}. By extracting appropriate \gls{1d} components embedded within the higher-dimensional structures—along horizontal, vertical, and diagonal axes in \gls{2d} grids, or  \gls{f2f} (horizontal, vertical, and depth) and \gls{e2e} axes in \gls{3d} grids—the \glspl{msed} of these components can be evaluated. Maximizing the minimum among these extracted \gls{1d} \glspl{nsm} provides guidance for designing the multi-dimensional \gls{nsm} filters. It should be noted, however, that this does not strictly guarantee the overall \gls{msed} of the higher-dimensional \gls{nsm}, as error events may exist outside the extracted components. Nevertheless, this approach enables a systematic and computationally manageable method for characterizing and optimizing \glspl{nsm}, linking simulation results, analytical bounds, and filter design in a coherent framework.

Beyond the conception and evaluation of the proposed \glspl{nsm}, several structural and theoretical aspects are worth highlighting, which reflect both the philosophy behind \gls{nsm} design and their broader implications for communication systems:
\begin{itemize}

    \item Classical ASK and QAM modulations, whether Gray- or non-Gray-coded, are in fact particular instances of \glspl{nsm}. This underscores that the \gls{nsm} framework generalizes conventional modulation schemes while providing additional degrees of freedom for optimization.
    
    \item All \glspl{nsm} presented are primarily designed for the Gaussian channel. Nevertheless, the framework allows for potential redesigns to exploit diversity in selective or fading channels, illustrating the adaptability of the \gls{nsm} architecture to different propagation environments.
    
    \item Due to filtering, the \gls{psd} of \glspl{nsm} is generally non-white. Scrambling techniques can be applied to whiten the spectrum without impairing the performance metrics, such as the \gls{msed}, thus preserving the essential characteristics of the modulation.
    
    \item Performance enhancement through increased filter lengths inevitably leads to more significant overlap between filtered symbol contributions, which increases the \gls{papr} in \gls{sc} systems. To mitigate this, real-valued filters may be replaced with complex-valued filters, providing extra degrees of freedom for reducing \gls{papr} while maintaining the desired performance and keeping memory and complexity requirements manageable.
    
    \item The structural similarity between \glspl{ms-prs} and convolutional codes is striking. While convolutional codes operate over discrete alphabets, typically Galois fields, \glspl{nsm} operate over real or rational sets, which motivates the designation \emph{analog convolutional codes}. Classical convolutional codes may thus be viewed as digital counterparts to \glspl{ms-prs}.
    
    \item Techniques inspired by convolutional codes, such as tail-biting, can be applied to \gls{1d} \glspl{nsm} to increase spectral efficiency without significant performance loss, by effectively reusing the \gls{nsm} structure across multiple symbol blocks. This principle naturally extends to \gls{2d} and higher-dimensional \glspl{nsm}, where the cyclic reuse of the \gls{nsm} structure can be applied across the entire \gls{2d} or \gls{3d} grid, allowing the same efficiency benefits in multi-dimensional signal arrangements.
    
    \item A deep structural duality exists between \glspl{nsm} and source coding using overcomplete frames and sigma-delta quantization. Source coding transforms analog signals into bipolar symbols, while \glspl{nsm} map these bipolar symbols back to analog modulated signals. Error correction coding, such as \gls{ldpc}, operates on the bipolar symbols and naturally links source coding and \glspl{nsm}. This duality enables a unified framework where source coding, channel coding, and \glspl{nsm} are efficiently integrated, leveraging the analog and digital domains in a complementary manner for end-to-end system design.
    
    \item Analog packing provided by \glspl{nsm} and digital packing offered by error correction codes are complementary. What analog packing can achieve efficiently in terms of packing density would require extremely high complexity if implemented solely with digital codes, while what digital packing can accomplish naturally and with low complexity may be very difficult or even prohibitive for analog packing alone. By assigning tasks according to the strengths of each domain—using analog packing where it is more efficient and digital packing where it is more natural—both approaches support each other, leveraging the best of each. This synergy demonstrates that relying solely on digital packing, as in conventional modulations such as 16-QAM or 64-QAM, may lead to suboptimal performance and unnecessary complexity.
    
    \item By summing the contributions of overlapping filtered sub-streams, \glspl{nsm} naturally generate modulated symbols whose distributions increasingly resemble Gaussian distributions as filter lengths grow. According to information theory, achieving the capacity of the Gaussian channel requires that the channel input itself follow a Gaussian probability distribution. The \gls{nsm} framework aligns asymptotically with this fundamental optimality condition on the channel inputs, demonstrating that as filter lengths increase, the transmitted symbols approach the Gaussian distribution needed for capacity-approaching transmission.
    
    \item Higher-dimensional generalizations of \glspl{nsm} can be constructed using analog \glspl{ldgm}, which structurally mirror \gls{ldpc} codes. This correspondence allows \glspl{nsm} to retain flexibility in filter design while benefiting from the structured sparsity of \gls{ldgm} architectures. The resulting systems facilitate a natural joint design of source coding, \glspl{nsm}, and error correction coding, producing a structured and efficient synergy between the analog and digital domains.

\end{itemize}

Taken together, these aspects demonstrate that \glspl{nsm} form a versatile and high-performance family of modulation schemes, capable of flexible implementation across \gls{1d} and multi-dimensional systems. By reinterpreting Mazo’s modulation and combining it with multi-stream polyphase structures and freely chosen \gls{fir} filters, the \gls{nsm} framework not only achieves strong performance but also provides a unifying structure that bridges modulation, coding, and source representation in a coherent and theoretically grounded manner. This extended framework naturally supports joint source coding, channel coding, and modulation designs, highlighting the broad applicability and foundational significance of the proposed \glspl{nsm}.

Building on the framework developed above, the principal contributions of this report are summarized below. They cover the design of new \glspl{nsm} and multi-stream partial-response schemes, theoretical and algebraic analyses (including closed-form filter characterizations and transfer-function methods), practical optimization procedures for one- and multi-dimensional systems, and a unified architecture that tightly integrates source coding, channel coding and modulation via analog \gls{ldgm} and \gls{ldpc} structures. The main contributions are as follows:

\begin{itemize}

    \item A variety of novel modulation schemes inspired by \gls{ftn} signaling, establishing an innovative, reunifying paradigm that enables flexible design of simple yet highly efficient modulations.
    
    \item Proposal of new \gls{ms-prs} schemes, in which a main data stream is demultiplexed and processed through multiple parallel partial response filters. Unlike classical \gls{ftn} systems, the shaping filter used to generate the analog modulated signal can be chosen freely and independently of the modulation scheme. Parallel filtering operations can also be freely optimized, with full control over impulse response lengths, providing a direct handle on processing complexity and performance.
    
    \item Introduction of \glspl{nsm} that are not limited to \gls{1d} or \gls{2d} structures. \glspl{nsm} can naturally extend to higher dimensions, providing additional design flexibility and favorable spatial characteristics unavailable at lower dimensions. Some \glspl{nsm} exploit fully \gls{3d} frameworks, while others combine one-, two-, and four-dimensional architectures. In terms of performance and spectral efficiency, these \glspl{nsm} can guarantee the asymptotic performance of $2$-ASK (equivalently BPSK, or equivalently QPSK in two dimensions), while offering much higher spectral efficiencies, equivalent to those of $4$-ASK (equivalently $16$-QAM in two dimensions) and $8$-ASK (equivalently $64$-QAM in two dimensions).
    
    \item Design and analysis of \gls{1d} rate-$2$ \glspl{nsm} (equivalent in spectral efficiency to $4$-ASK or $16$-QAM) with real filter taps. Filters are optimized using two criteria: unconstrained average symbol energy allocation (allowing maximum \gls{msed}) and balanced energy allocation (equal filter energies). Unconstrained designs improve performance at the start of iterative detection (turbo-equalization), while balanced designs enhance performance at later iterations.
    
    \item Derivation of closed-form expressions for the optimal filters and the corresponding \glspl{msed} in \gls{1d} rate-$2$ \glspl{nsm} with real taps. The process involves three stages: (i) numerical optimization to approximate the best filter taps, (ii) exact determination of error events corresponding to the \gls{msed}, and (iii) symbolic calculus exploiting the fact that these error events lead to the same \gls{msed}, enabling closed-form determination of the filters and distances.
    
    \item Identification of minimum filter lengths required for rate-$2$ \glspl{nsm} with real filter taps to perfectly achieve the \gls{msed} of $2$-ASK. In particular, when one of the two filters has length $1$, the other filter must have length $10$ or greater in order for the scheme to reach the asymptotic bit error probability performance of $2$-ASK.
    
    \item Proposal and optimization of one-, two-, and three-dimensional rate-$2$ \glspl{nsm} using rational (or equivalently, integer) filter taps to achieve the $2$-ASK \gls{msed}.
    
    \item Design of \gls{1d} rate-$2$ \glspl{nsm} with bipolar non-null filter taps, where error events with \gls{sed} lower than that of $2$-ASK cannot be entirely prevented, but their multiplicity decreases extremely rapidly towards zero as the length of one filter increases while the other filter is fixed to length $1$. This vanishing multiplicity property enables these \glspl{nsm} to asymptotically approach the bit error probability performance of $2$-ASK. They are thoroughly characterized algebraically using binary Galois field tools, including primitive polynomials and m-sequences.
    
    \item Development of \glspl{tf} analogous to convolutional codes. For \glspl{nsm} with real taps, \glspl{tf} are represented using lists due to non-integer exponents of $D$, while \glspl{nsm} with rational (or equivalently, integer) taps are represented symbolically using two parameters: $N$ (bipolar input difference Hamming weight) and $D$ (real output difference \gls{sed}). These \glspl{tf} enable tight \gls{bep} bounds, with $N$ instantiated as $1/2$, reflecting the exponentially vanishing occurrence probability of error events as $(1/2)^{w}$, where $w$ is the Hamming weight of the input sequence difference of the error event. This suppression of higher Hamming-weight input differences is a surprising and innovative property, reminiscent of the mechanism behind \glspl{pccc}, although here the exponential suppression applies specifically to input sequence differences rather than output differences as in turbo codes.
    
    \item Introduction of the \gls{rtf}, a compact symbolic representation depending solely on $D$. It is obtained from the full \gls{tf} by taking the partial derivative with respect to $N$, multiplying the result by $N$, and then instantiating $N=1/2$. This construction simplifies analysis, particularly for \glspl{nsm} with large filter lengths or degenerate \glspl{nsm} (having an infinite number of error events with identical \gls{msed}).
    
    \item Use of sophisticated iterative algorithms applied specifically to truncated versions of the \gls{tf} or to the \gls{rtf} to provide tight \gls{bep} bounds, validating simulated \gls{ber} results. These algorithms operate on numerical lists for real-tap \glspl{nsm} and symbolically for rational-tap \glspl{nsm}. For degenerate \glspl{nsm}, truncated \glspl{rtf} enable exact iterative computation, providing reliable and tight upper bounds where standard truncated \glspl{tf} may fail.
    
    \item Partial and indirect optimization of multi-dimensional \glspl{nsm} by optimizing embedded \gls{1d} \glspl{nsm} extracted along key axes of the \gls{2d} or \gls{3d} grid.
    
    \item Extension of the \gls{nsm} framework to analog \gls{ldgm} modulations, generalizing the classic filtered \gls{nsm} approach. Low-density integer-valued matrices are favored for implementation simplicity, with the resulting modulation referred to as analog \gls{ldgm} \gls{nsm}.
    
    \item Proposal of a new architecture combining analog \gls{ldgm} \gls{nsm} modulation with \gls{ldpc} error correction coding. Analog \gls{ldgm} \glspl{nsm} possess Tanner-like bipartite graphs in the real domain, structurally matching the Tanner graph of the associated \gls{ldpc} code in the binary domain, enabling natural and efficient turbo-detection (turbo-equalization) at the receiver. Standard non-Gray-coded modulations (e.g., $4$-ASK, $16$-QAM, $8$-ASK, $64$-QAM) appear as particular cases within this framework.
    
    \item Establishment of a duality between source coding using overcomplete frames with sigma-delta quantization and the proposed \gls{ms-prs} modulations, which share a similar structural framework. When \gls{ms-prs} modulations are generalized to analog \gls{ldgm} \glspl{nsm}, the overcomplete-frame-based source coding can likewise be extended to a more general source coding paradigm using an analog \gls{ldgm} structure. This leads to a unified framework encompassing source coding, channel coding, and modulation, where source coding employs an analog \gls{ldgm} for quantization, modulation uses another analog \gls{ldgm} for signal generation, and \gls{ldpc} coding links the operations. This framework enhances both transmitter source coding and receiver demodulation performance without requiring Gray mappings and is expected to provide groundbreaking efficiency and performance improvements in future 6G communication systems.

\end{itemize}

The report is organized as follows:

\begin{itemize}

    \item \textbf{Section~\ref{Background}} provides background on the various proposed modulation schemes. It begins with a comparison between \gls{ftn} and Nyquist signaling modulations, demonstrating that \gls{ftn} with the “sinc” shaping filter, as first proposed by J. Mazo, is equivalent to \gls{ms-prs} operating at the Nyquist signaling rate. The section then presents two illustrative examples of simple modulation schemes that follow the \gls{ms-prs} paradigm, offering either increased data rate or enhanced performance via an improved \gls{msed}.
    
    \item \textbf{Section~\ref{sec:Rate-2 NSMs}} focuses on rate-$2$ \glspl{nsm}. The first part addresses \gls{1d} rate-$2$ \glspl{nsm} with real-valued filter taps, considering both unconstrained cases with unbalanced filter energies and constrained cases with balanced filter energies. The second part is dedicated to one-, two-, and three-dimensional rate-$2$ \glspl{nsm} with rational (or equivalently, integer) filter taps.
    
    \item \textbf{Section~\ref{sec:Rate-3 NSMs}} is dedicated to rate-$3$ \glspl{nsm}. The first part examines \gls{1d} rate-$3$ \glspl{nsm} with real-valued filter taps, comparing \glspl{nsm} optimized independently with \glspl{nsm} inherited from rate-$2$ designs. The second part discusses how combining one-, two-, and four-dimensional filters within the same rate-$3$ \gls{nsm} can preserve the \gls{msed} of $2$-ASK.
    
    \item \textbf{Sections~\ref{Rate-3/2 Approaching NSMs}, \ref{Rate-4/3 Approaching NSMs}, and~\ref{Rate-5/4 Approaching NSMs}} address \glspl{nsm} with rational (or equivalently, integer) filter coefficients and fractional rates of $3/2$, $4/3$, and $5/4$, respectively.
    
    \item \textbf{Section~\ref{Discussion and conceptual parallels}} provides a broad discussion on topics directly or indirectly related to the proposed \glspl{nsm}. Among the topics, conventional modulations—whether Gray- or non-Gray-coded—are shown to be special cases of the \gls{nsm} framework. The section also establishes an analogy between the structure of convolutional codes and the proposed \gls{ms-prs} modulations. A key discussion highlights the structural similarity and duality between source coding using overcomplete frames with sigma-delta quantization and \gls{ms-prs} modulations. This similarity underpins innovative joint designs for error correction coding and modulation, as well as unified schemes for source coding, error correction coding, and modulation, which could significantly impact the standardization of the 6G physical layer.

\end{itemize}



\section{Background} \label{Background}

\subsection{Potential Ways to Increase Spectral Efficiency}

In conventional communication systems, Nyquist signaling is systematically used to guarantee zero interference between transmitted symbols. For any single- or a \gls{mc} system, without guard interval insertion, this means that the use of a frequency band of width $\mathcal{W}$ during a period of time $\mathcal{T}$ offers $2\mathcal{W} \!\!\! \mathcal{T}$ interference-free transmission opportunities, for real modulated symbols. As such, the only degree of freedom left for increasing spectral efficiency is to increase the modulation order, going for example from $2$-ASK (or BPSK) to $4$-ASK, for real modulated symbols, and from $4$-PSK (or QPSK) to $16$-QAM, for complex modulated symbols.

For the sake presentation simplicity, we next exclusively focus on real modulated symbols. Complex modulations of interest, such as those of famous systems like WiFi and 4G/5G, can be seen as two parallel real modulations transmitted separately on inphase and quadrature-phase components. 

When modulation order is increased in Nyquist signaling systems, in order to reach a targeted spectral efficiency, the energy cost per transmitted bit should be significantly increased, to guarantee the same targeted performance. Performance could be measured either theoretically, in terms of \gls{bep} or \gls{blep}, or experimentally, in terms of \gls{ber} or \gls{bler}.

For example, when we go from a $2$-ASK, with symbol alphabet $\{\pm 1\}$, to a $4$-ASK, with symbol alphabet $\{\pm 1, \pm 3\}$, the normalized average energy per symbol goes from $1$ to $5$. 

This $5$-fold increase in average symbol energy is due to the fact that the additional amplitudes, $\pm 3$, added to double the offered bit rate, cost as much as $9$ times the basic amplitudes, $\pm 1$, in terms of energy. It goes without saying that this dramatic increase in energy can in no way be compensated by the two-fold increase in binary rate offered by $4$-ASK, with respect to $2$-ASK. Indeed, the new energy cost per bit is now $5/2$ times that of $2$-ASK, leading to a degradation of $10 \log_{10}(5/2) \approx 3.98$ dB. This dramatic cost increase phenomenon remains also valid when moving from any ASK modulation to the next ASK modulation, with twice the alphabet size. 

To keep the energy cost increase to a minimum, when the offered bit rate is required to be increased, another way to proceed is to keep the modulation order as low as possible while using \gls{ftn} signaling. For \gls{sc} communications, \gls{ftn} signaling leads to a compression in time by a factor $\alpha$, $0 < \alpha < 1$, meaning that the effective working symbol period is decreased to $\alpha T$ with respect to $T$, the symbol period used for Nyquist signaling. For \gls{mc} communications, besides time compression, \gls{ftn} signaling brings an additional compression in frequency by a factor $\beta$, $0 < \beta < 1$, meaning that the working sub-carrier spacing is decreased to $\beta F$, $F$ being the sub-carrier spacing used in Nyquist signaling. The overall compression factor experienced by \gls{mc} systems is hence equal to $\alpha \beta$, offering $2\mathcal{W} \!\! \! \mathcal{T} / \alpha \beta$ real modulated symbol transmission opportunities, within a frequency bandwidth $\mathcal{W}$ and a time duration $\mathcal{T}$. As such, the spectral efficiency gets increased by a factor of $1/ \alpha \beta$, every thing else being equal.

The benefits of \gls{ftn} signaling, in terms of increased spectral efficiency, come always at a price. First, \gls{sc} communications experience \gls{isi}, while \gls{mc} communications experience both \gls{isi} and \gls{ici}. Secondly, the generated modulated signals can be subject to a reduction in \gls{med}, leading systematically to an unacceptable degradation in performance.

In 1975, Mazo~\cite{Mazo75} investigated the \gls{med} of \gls{sc} binary \gls{ftn} signaling and proved that the ideal “sinc” pulse preserves modulation \gls{med}, as long as the compression factor, $\alpha$, remains above $0.802$. This important result means that the offered binary rate can be boosted by $25 \%$, at no performance loss, and that the only penalty is in the higher receiver complexity resulting from interference processing and suppression.

In 2009, Rusek and Anderson~\cite{Rusek09} investigated \gls{mftn} systems that use \gls{qpsk} modulation and preserve the \gls{med}, with an aggregate compression factor $\alpha \beta$ approaching $0.5$. Nevertheless, the considered low compression factors lead to strong \gls{isi} and \gls{ici}, originating from the nearest as well as the farthest neighbors in time-frequency, and making detection a challenging problem at the receiver, that requires the additional help of error correction coding.

\subsection{Common-Framework Comparison of Nyquist Signaling and FTN Signaling}

\begin{figure}[!htbp]
    \centering
    \includegraphics[width=0.8\textwidth]{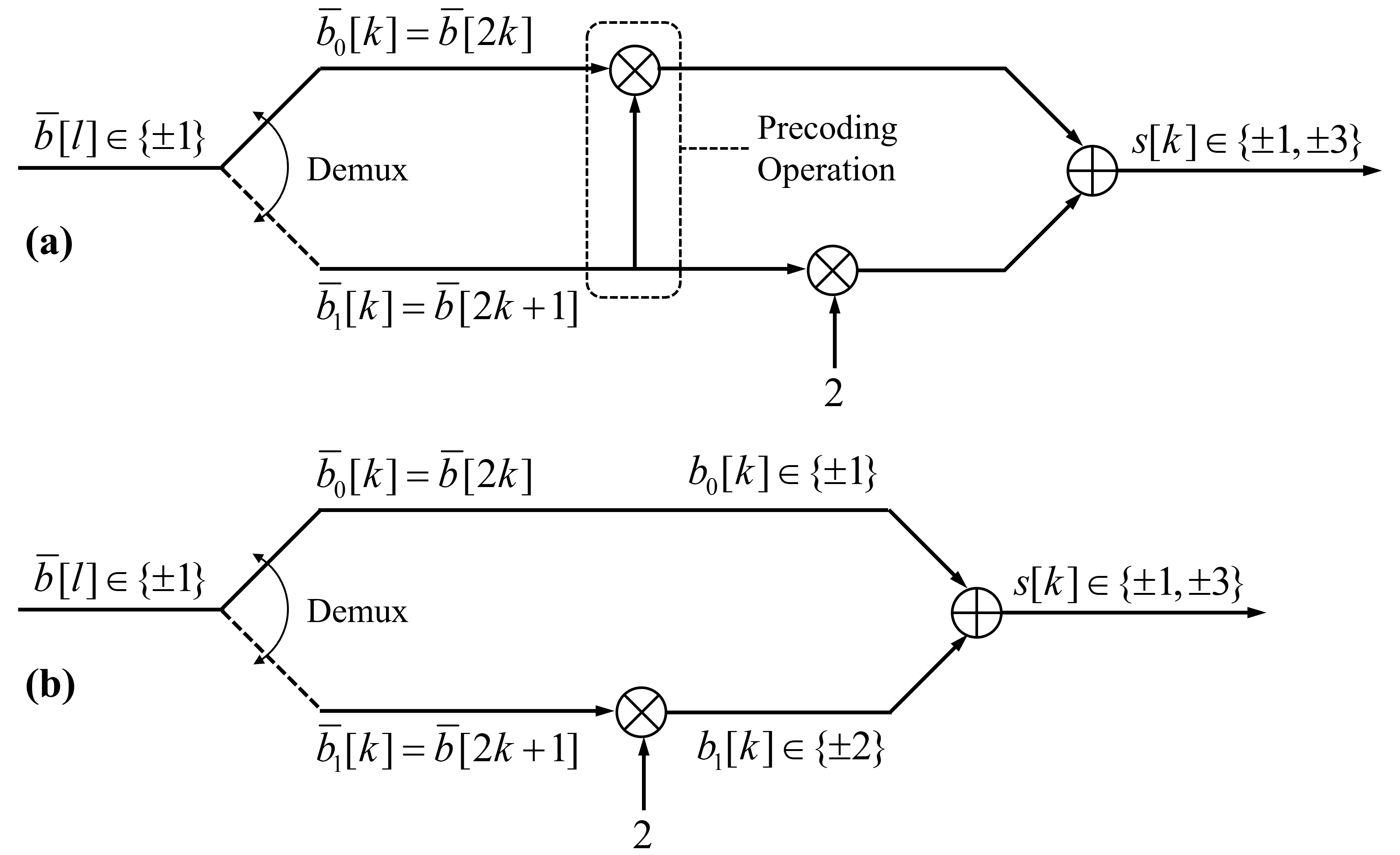}
    \caption{Block diagram of the Nyquist signaling modulator of $4$-ASK: (a) Gray-coded $4$-ASK, (b) Non Gray-coded $4$-ASK.}
    \label{fig:Block Diagram 4-ASK}
\end{figure}

To get a deep understanding of the mechanism underpinning the superiority of \gls{ftn} signaling, when compared to Nyquist Signaling, a comparison on a common background of both is required. From the Nyquist signaling side, we consider $4$-ASK, the simplest amplitude modulation, immediately following $2$-ASK (or \gls{bpsk}) in spectral efficiency. To enable comparison with \gls{ftn} signaling, we divert from Gray-coded $4$-ASK, the modulator of which is shown in Figure~\ref{fig:Block Diagram 4-ASK}(a), to non Gray-coded $4$-ASK, the modulator of which is shown in Figure~\ref{fig:Block Diagram 4-ASK} (b). Gray coding, which is acquired by an elementary precoding operation, as illustrated in Figure~\ref{fig:Block Diagram 4-ASK}(a), leads to a slight enhancement in performance, with respect to non Gray coding. Although it is used in practical systems, it goes against the sought common comparison framework between Nyquist signaling and \gls{ftn} signaling.

Based on Figure~\ref{fig:Block Diagram 4-ASK}(b), we can write the modulated sequence, $s[k] = \bar{b}_0[k] + 2 \bar{b}_1[k]$, at time $kT$, for $4$-ASK, as
\begin{equation} \label{eq:Mod Seq 4-ASK}
    s[k] = \sum_{l} \bar{b}_0[l] h_0[k-l] + \sum_{l} \bar{b}_1[l] h_1[k-l] = \sum_{l} b_0[l] \bar{h}_0[k-l] + \sum_{l} b_1[l] \bar{h}_1[k-l],
\end{equation}
where $\bar{b}_m[k] = \bar{b}[2k+m],$ $m \in \{0,1\}$, $\bar{b}[k] \in \{\pm 1\}$ is the input sequence, in its bipolar form, $h_m[k] = 2^m \delta[k],$ $m \in \{0,1\}$ is the discrete-time filter having $\bar{b}_m[k]$ as input, $\delta[k]$ is the Kronecker delta function, $\bar{h}_m[k]=h_m[k]/\|h_m\| = \delta[k],$ $m \in \{0,1\}$ is the normalized version of $h_m[k]$, $b_m[k]=\|h_m\| \bar{b}_m[k],$ $m \in \{0,1\}$, and $\|h_m\|, m \in \{0,1\},$ is the Euclidean norm of discrete-time filter $h_m[k]$. As such, $\bar{b}_m[k]=b_m[k]/|b_m[k]|$ can be seen as the normalized version of $b_m[k]$. Each of the two written forms of the modulated sequence, $s[k]$, in (\ref{eq:Mod Seq 4-ASK}), will better serve one or another of the facets of the discussions and comparisons to follow.

From the \gls{ftn} signaling side, we consider, for illustration purpose, the “sinc” function $h(t)=\operatorname{sinc}(t/T)$, first used in~\cite{Mazo75} to illustrate the concept and merits of \gls{ftn} signaling, and a compression factor $\alpha=0.8$. As portrayed in Figure~1 of~\cite{Anderson13}, the concept of \gls{ftn} signaling is typically explained using a sampling rate equal to the \gls{ftn} rate, $1/\alpha T$, corresponding to the compressed sampling period $\alpha T$. For the compression factor $\alpha=0.8$ at hand, this means that the transmission of $5$ consecutive elements from the input sequence $\bar{b}[k]$ requires exactly a time duration of $4T$. Regrettably, sampling at the \gls{ftn} rate, $1/\alpha T$, prevents any possible comparison with the Nyquist signaling side which, obviously, operates at the Nyquist sampling rate. Fortunately, owing to the frequency occupancy of the rectangular function, Fourier transform of the carefully chosen “sinc” function, the continuous-time \gls{ftn} modulated signal, $s(t)$, spans exactly the interval $[-1/2T, 1/2T]$. As a consequence, no modulated signal spectrum aliasing occurs in frequency, and a reversible sampling of the modulated signal at the Nyquist rate, $1/T$, can be accomplished, although \gls{ftn} signaling is used.

\begin{figure}[!htbp]
    \centering
    \includegraphics[width=1.0\textwidth]{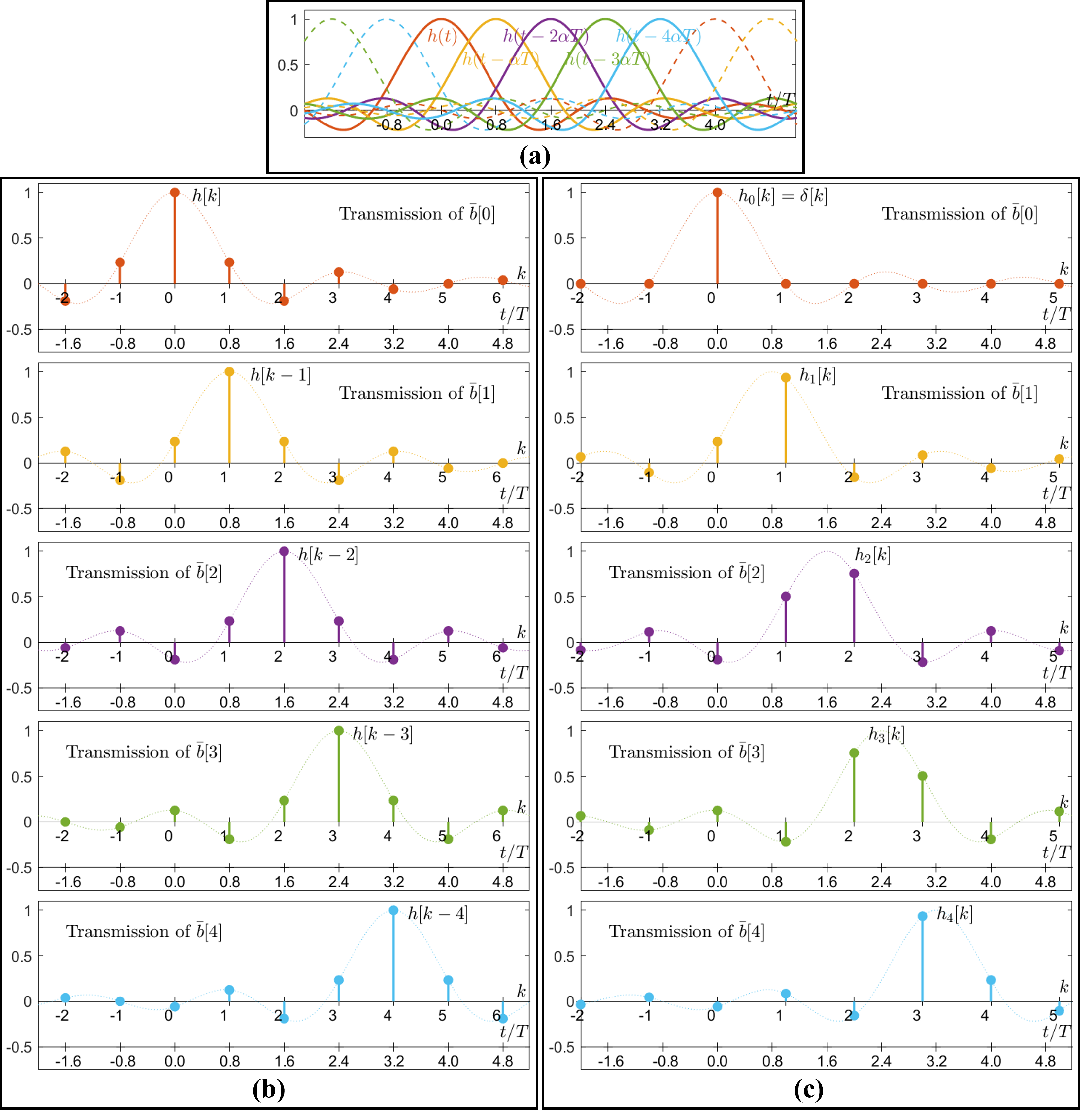}
    \caption{FTN time shifted versions of the “sinc” function, with compression factor $\alpha=0.8$: (a) Continuous time, (b) FTN sampled, discrete time, (c) Nyquist sampled, discrete time.}
    \label{fig:Filters Sinc Function}
\end{figure}

In Figure~\ref{fig:Filters Sinc Function}(a), we show the succession, in continuous time, of the \gls{ftn} time shifted versions of the “sinc” function $h(t)$, with time shifts $k\alpha T$ corresponding to the transmissions of the bipolar inputs $\bar{b}[k]$. To nuance the illustrative \gls{ftn} sampling, adopted in~\cite{Anderson13}, from the Nyquist sampling, adopted here for the sake of a common comparison background, we show in Figure~\ref{fig:Filters Sinc Function}(b) the discrete-time filters involved in the construction of the \gls{ftn} sampled version of the modulated signal. Clearly, a unique discrete-time filter, namely $h[k] = h(k\alpha T)$, is brought into play in the generation of the \gls{ftn} sampled version, $s[k] = \bar{b}[k] \circledast h[k]$, of the modulated signal $s(t)$, $\circledast$ being the discrete-time convolution operator.

To complete the whole picture, we provide in Figure~\ref{fig:Filters Sinc Function}(c) an insightful illustration of how \gls{ftn} signaling works when observed from a different standpoint, where Nyquist sampling is used instead of \gls{ftn} sampling. We notice here, for $\alpha=0.8$, that every consecutive $5$ bipolar elements in the input sequence are transmitted in the same way using $5$ discrete-time filters, which can be interpreted as $5$ polyphase filters, emanating from sampling the “sinc” function $h(t)$ at $5$ times the Nyquist rate $1/T$. More precisely, the sampled version of the modulated signal, $s(t)$, at the Nyquist rate, can be expressed as
\begin{equation}  \label{eq:Mod Seq FTN}
    s[k] = \sum_{m=0}^4 \sum_l \bar{b}_m[l] h_m[k-4l],
\end{equation}
where $\bar{b}_m[k]=\bar{b}[5k+m],$ $0 \le m \le 4$ and $h_m[k] = h(kT-m \alpha T)$, $\alpha$ being obviously equal to $0.8$. Notice here that, owing to the specific choice of the “sinc” function and the absence of frequency aliasing, all involved filters $h_m[k],$ $0 \le m \le 4$ are identical to their normalized counterparts, $\bar{h}_m[k]$, and therefore $\bar{b}_m[k]=b_m[k]$. Notice equally that the first filter, which is a \gls{fir} filter, given by $\bar{h}_0[k]=\delta[k]$, is identical to the normalized filters used in (\ref{eq:Mod Seq 4-ASK}) for Nyquist signaling. More importantly, notice that, apart from the first filter $\bar{h}_0[k]$, all other involved filters, namely $\bar{h}_1[k], \bar{h}_2[k], \bar{h}_3[k]$ and $\bar{h}_4[k]$, are \gls{iir} filters.

The first outstanding lesson to be learned from (\ref{eq:Mod Seq FTN}) is that \gls{ftn} signaling, as proposed in~\cite{Mazo75}, can also be seen, from a different angle, as a regular Nyquist signaling, with more transmission opportunities than modulated signal samples. By way of comparison, conventional Nyquist signaling with $4$-ASK, as portrayed by (\ref{eq:Mod Seq 4-ASK}), offers two transmission opportunities per sample, while \gls{ftn} signaling with a compression factor $\alpha=0.8$, as illustrated in (\ref{eq:Mod Seq FTN}), offers 5 transmission opportunities for every 4 samples. 

Before proceeding to other lessons and insights, it is important to underline that rewriting \gls{ftn} signals as a conventional Nyquist signals was also implicitly accomplished in the past, in~\cite{McGuire10}, without any explicit mention to that in the paper. More specifically, \gls{ftn} signaling has been achieved in a vector discrete time framework, with $M$ transmission opportunities offered over $N$ samples, with $N<M$. The so-called transfer matrix, translating every $M$ data inputs into $N$ consecutive samples of the modulated signal, sampled at the Nyquist rate, is a succession of a \gls{dft}, a filtering in the frequency domain, using a rectangular window, and an \gls{idft}. As such, the $N$ modulated signal samples, corresponding to every $M$ data inputs, can be written in the same way as in (\ref{eq:Mod Seq FTN}), with the nuance that the polyphase filters $h_m[k]$ are replaced by their time-aliased counterparts.

The previous interpretation of \gls{ftn} signaling as a classical Nyquist signaling, when it comes to \gls{sc} systems employing the “sinc” function, can be extended to \gls{mftn}. To see this, consider an \gls{mftn} system, with compression factors $\alpha$ in time and $\beta$ in frequency, offering $2\mathcal{W} \!\! \! \mathcal{T} / \alpha \beta$ bipolar transmission opportunities, within a frequency bandwidth $\mathcal{W}$ and a time duration $\mathcal{T}$. The underlying signal space on which bank all the previous transmission opportunities has $2\mathcal{W} \!\! \! \mathcal{T}$ as real dimension, meaning that there are, in average $1/\alpha \beta$ transmission opportunities par dimension. As such, each transmission opportunity is subtended by a vector, which stems from a representation of the underlying time-frequency shifted version of the \gls{mftn} base pulse, in an orthonormal base of the signal space. A simple analogy between \gls{sftn} and \gls{mftn} show that each dimension (respectively, subtending vector) in \gls{mftn} plays the same role as a sample (respectively, polyphase filter) in \gls{sftn}.  

\begin{figure}[!htbp]
    \centering
    \includegraphics[width=0.8\textwidth]{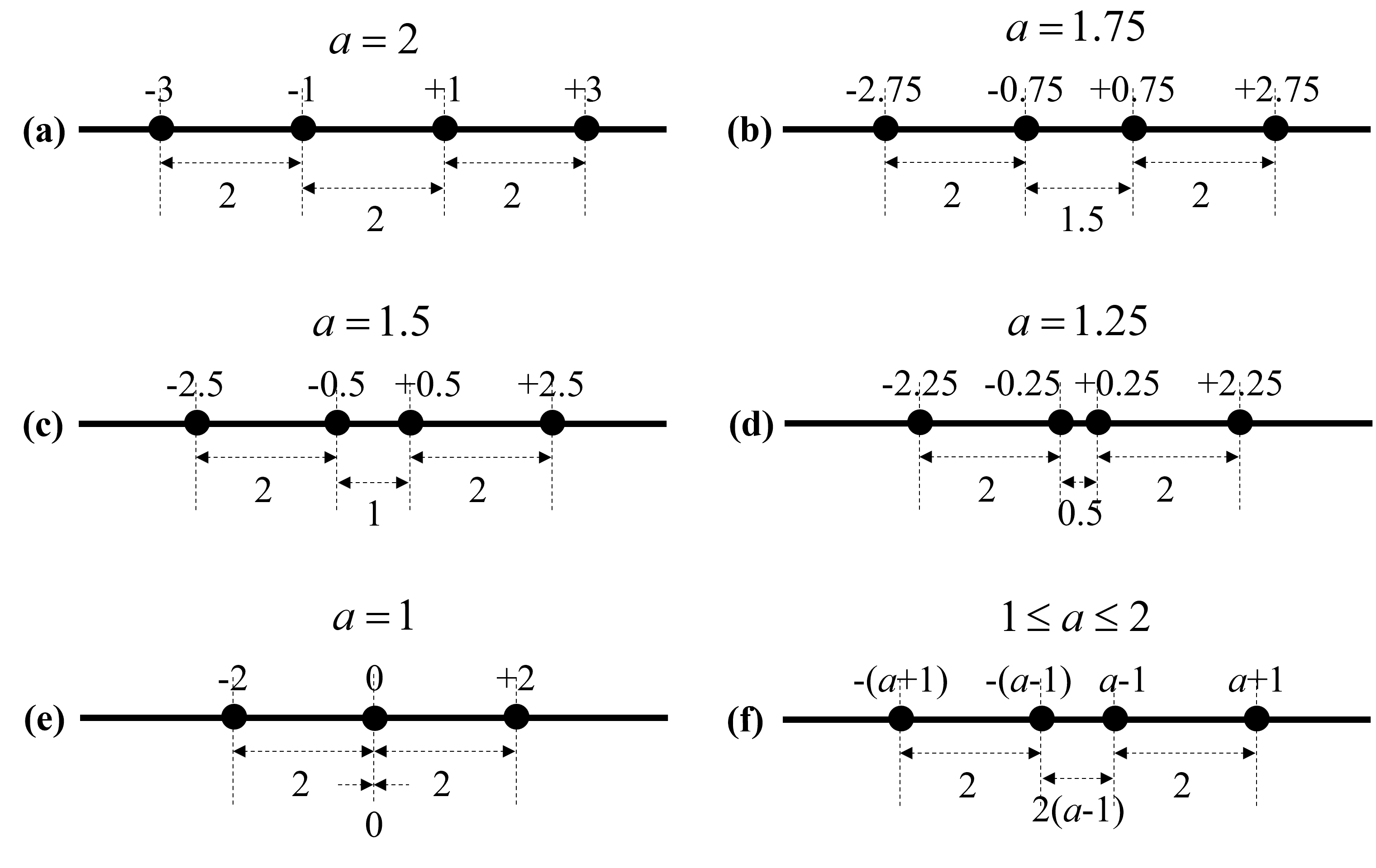}
    \caption{Constellations of $4$-ASK for values of $a$ going from 2 down to 1: (a) $a=2$, (b) $a=1.75$, (c) $a=1.5$, (d) $a=1.25$, (e) $a=1$, (f) $1 \le a \le 2.$}
    \label{fig:Different Configurations 4-ASK}
\end{figure}

The second lesson to be learned from a comparison of (\ref{eq:Mod Seq 4-ASK}), for Nyquist signaling, and (\ref{eq:Mod Seq FTN}), for \gls{ftn} signaling, is that, while both approaches guarantee the same (or almost the same) \gls{med} as $2$-ASK, the second transmitted bipolar input sequence, $b_1[l]$, in $4$-ASK, belongs to the bipolar alphabet $\{\pm 2\}$, and, as such, costs four times the energy of first transmitted bipolar sequence, $b_0[l]$, which has $\{\pm 1\}$ as bipolar alphabet. Unfortunately, any attempt to reduce the corresponding contribution to energy for $4$-ASK, by using $\{\pm a\}$, $1 \le a < 2$, as alternative bipolar alphabet for $b_1[l]$, is accompanied by an even more accentuated reduction in the \gls{med}. Figure~\ref{fig:Different Configurations 4-ASK} shows different configurations of the $4$-ASK constellation, when the parameter $a$ goes from $2$ down to $1$, with a step of $0.25$, in addition to a generic configuration for an arbitrary value of $a,$ $1 \le a \le 2$, to help assess the strong decrease in \gls{med}, as a function of $a$, and compare it to the less pronounced decrease in average $4$-ASK symbol energy. The first $5$ sub-figures of Figure~\ref{fig:Different Configurations 4-ASK} clearly show the pronounced decrease in \gls{med} when $a$ decreases from $2$ to $1$, which outright vanishes to $0$, when the bipolar input $b_1[l]$ reaches the same energy cost as the bipolar input $b_0[l],$ in \gls{sftn} and \gls{mftn}. The last sub-figure of Figure~\ref{fig:Different Configurations 4-ASK} shows that the \gls{med} of $4$-ASK, with $a,$ $1 \le a \le 2$ is equal to $2(a-1)$. Given that the average energy per $4$-ASK symbol is now equal to $1+a^2$, this leads to $4(a-1)^2/(1+a^2)$ as normalized \gls{sed}. This ratio is the right measure $4$-ASK modulation at high signal-to-noise ratios. It decreases from $4/5$, for conventional $4$-ASK, with $a=2$, to $0$, for $a=1$, value that precisely leads to the same cost in energy of bipolar input $b_1[l]$ as other bipolar inputs in \gls{sftn} and \gls{mftn}.

An important question arises now. What really makes the \gls{med} stay at the level of $2$-ASK, when the $5$ bipolar inputs of \gls{ftn} signaling, in (\ref{eq:Mod Seq FTN}), have the same minimum energy cost, but vanishes to 0 when the $2$ bipolar inputs of Nyquist signaling, in (\ref{eq:Mod Seq 4-ASK}) are similarly set to the same minimum energy cost? And the response, which will be corroborated in the next subsection, is particularly shocking. A quick comparison of the forms of the filters, originating from Nyquist and \gls{ftn} signaling, suggests that the interference exhibited in \gls{ftn} signaling by the \gls{iir} polyphase filters, other than $h_0[k]$, is the true mechanism that prevents the \gls{med} from collapsing to zero, although all bipolar inputs are of common minimum energy. This observation, which promotes interference as a technique to preserve distance, is in complete contradiction with the common sens and the quasi unanimous attitude reported towards interference, in most if not all past research on \gls{ftn} signaling. Actually, researchers on \gls{ftn} signaling see interference as a bitter bill that must be payed in return for any increase in throughput through \gls{ftn} signaling. Additionally, they recognize the difficulty of suppressing the interference resulting from \gls{iir} filters, such as those in (\ref{eq:Mod Seq FTN}), with a reasonable complexity and without sacrificing detection performance. Indeed, when a “sinc” or \gls{rrc} filter is used, these polyphase \gls{iir} filters generally demonstrate a slow decay in time, towards $\pm \infty$. As a consequence, their truncation in time hardly achieves a good and satisfactory compromise between complexity and performance.

Another no less important question arises from the above. Is \gls{ftn} signaling the best way to create interference from performance and complexity perspectives? And the response is no.
On the one hand, for \gls{sc} systems, the discrete time filter $h[k]$ of \gls{ftn} signaling (or its equivalent Nyquist signaling polyphase filters, $h_m[k]$) is uniquely and indirectly determined by the compression factor $\alpha$, once the transmission filter, $h(t)$ has been chosen. This limits the number of degrees of freedom in choosing $h[k]$, leading typically to a discrete-time \gls{iir} filter, the truncation of which fails to achieve a good compromise between performance and complexity. On the other hand, for \gls{mc} systems, while we have the same kind of concerns as \gls{sc} systems, the resulting \gls{ftn} signaling filter, $h[k,l],$ includes a frequency dimension, played by the parameter $l,$ in addition to the time dimension , played by the parameter $k.$ This two-dimensional filter, which accounts for \gls{ici}, in addition to \gls{isi}, prevents from using any simple trellis-based detection scheme for interference processing and suppression.

Now, using the Nyquist signaling representation framework, exemplified in (\ref{eq:Mod Seq FTN}), can we, with maximum freedom, directly and arbitrarily choose polyphase filters to implement \gls{ftn}-like systems? Additionally, can we optimize these filters, for a targeted complexity, determined by the polyphase filters lengths? And the common response is yes. A foretaste of this response will be made clear in the following sub-sections.

The contradictory interpretation of the interference underlying \gls{ftn} signaling, raised above, is not the first case of its kind in the field of wireless communications. Before the advent of MIMO wireless communications~\cite{Biglieri07}, the multi-path Rayleigh fading channel was considered as an unwanted and wicked channel, as compared to the most desirable and most pleasant Gaussian channel. As a consequence, a plethora of research work was carried on new modulation and error correction coding schemes capable of enhancing the performance of wireless communications over Rayleigh fading channels~\cite{Boutros96, Belfiore05, Giraud97, Boulle92}. This displeasing multi-path Rayleigh fading has become the cornerstone of spatial multiplexing~\cite{Bolcskei02, Foschini96, Telatar99}, one of the categories of MIMO wireless communications, which is now an essential element of modern wireless communication standards. Spatial multiplexing is a very powerful technique for increasing channel capacity at higher signal-to-noise ratios, that performs well only for MIMO systems with Rayleigh fading channels. 

Before moving to the experimental analysis, which will support interference as the mechanism at the origin of enhanced modulation performance, it is very instructive, next, to make some thoughts on \gls{ftn} signaling, in relation to Nyquist signaling. Firstly, we notice that the polyphase filters, $h_m[k]$, appearing in (\ref{eq:Mod Seq FTN}), offer, for the same truncation precision, a rough reduction by $\alpha$ of the number of taps used by the unique \gls{ftn} discrete filter $h[k]=h(k\alpha T)$. Secondly, Nyquist signaling, offers a reduction by $\alpha$ of the the processing rate, compared to \gls{ftn} signaling, since the former is operated at the Nyquist rate, $1/T,$ while the latter is operated at rate $1/\alpha T.$ Combining the two previous observations, we can conclude that the Nyquist signaling interpretation, of any \gls{ftn} signaling modulation, can offer a substantial reduction in computational complexity. 

\subsection{Uncovering the Benefits of Interference through Experimentation} \label{ssec:Merits of Interference}

In what follows, we confirm, through experimental results, the importance of interference as a mechanism to protect data inputs, when transmitting, at the same spectral efficiency as \gls{ftn} signaling, using Nyquist signaling. For simplicity sake, we focus on alternative \glspl{nsm} with the same spectral efficiency and normalized bipolar data inputs as $4$-ASK. We keep the same symbolic expression for modulated symbols, as for $4$-ASK in (\ref{eq:Mod Seq 4-ASK}), with the exception that the two filters at stake, $h_m[k],$ $m = 0, 1,$ have now a common length, $L \ge 1$, and can be written as $h_m[k] = \sum_{l=0}^{L-1} h_m[l] \delta[k-l]$. Obviously, once increased, the length, $L,$ must give rise to extended degrees of freedom, and therefore to an improvement in performance, at the cost of an increase in reception complexity.

To have a common comparison framework with conventional $4$-ASK, we freeze the average energy per modulated symbol to the common value of $5$. If we denote by $\eta_m$ the average energy contribution of data inputs $\bar{b}_m[k],$ $m = 0, 1,$ to each modulated symbol, then $\eta_0 + \eta_1 = 5$. To simplify the specification of the distribution of average symbol energy between the two types of data inputs, $\bar{b}_m[k],$ $m = 0, 1$, we introduce of the generic parameter $1 \le \eta \le 4$, with $\eta_0 = \eta$ and $\eta_1 = 5 - \eta$. In this setup, filters $h_m[k],$ $m=0, 1,$ play similar and symmetric roles, with the extreme values, $\eta = 1$ and $\eta = 4$, being also those of conventional $4$-ASK.

To emphasize the role played by the generic parameter $\eta$ in determining the performance of the corresponding modulation, we write the discrete-time filters in terms of their normalized counterparts, namely $h_0[k]=\sqrt{\eta} \bar{h}_0[k]$ and $h_1[k]=\sqrt{5-\eta} \bar{h}_1[k]$. Now, given that the average energy per modulated symbol is frozen, we look to the evolution of the achievable \gls{med} of the modulation, resulting from a bunch of random choice of the normalized filters, $\bar{h}_m[k],$ $m = 0, 1,$ when the parameter $\eta$ is varied from $1$ to $4$. For this, we make use of Algorithm~\ref{alg:d_min^2 Rate-2 NSM L_0>1, L_1>1}, introduced in Subsection~\ref{sapp:d_min^2 Rate-2 NSM L_0>1, L_1>1}, which is capable of determining the \gls{med} of the alternative modulations at hand.

The \gls{med}, $d_{\text{min}},$ plays a simple, yet fundamental, role in assessing the behavior of error probability of modulations, at high signal-to-noise ratios. For illustration purpose, consider all the modulated signals, generated according to (\ref{eq:Mod Seq 4-ASK}). Now, let $s^i[k]$ and $s^j[k],$ $i \ne j$, be two such sequences, with respective pairs of normalized input bipolar sequences $(\bar{b}_0^i[k], \bar{b}_1^i[k])$ and $(\bar{b}_0^j[k], \bar{b}_1^j[k])$. Moreover, assume that these pairs of input data differ in a finite number of components. Then, $d_{\text{min}}^2,$ is the least \gls{sed}, $\|s^j-s^i\|^2$, between any such pairs of modulated sequences.

\begin{figure}[!htbp]
    \centering
    \includegraphics[width=0.65\textwidth]{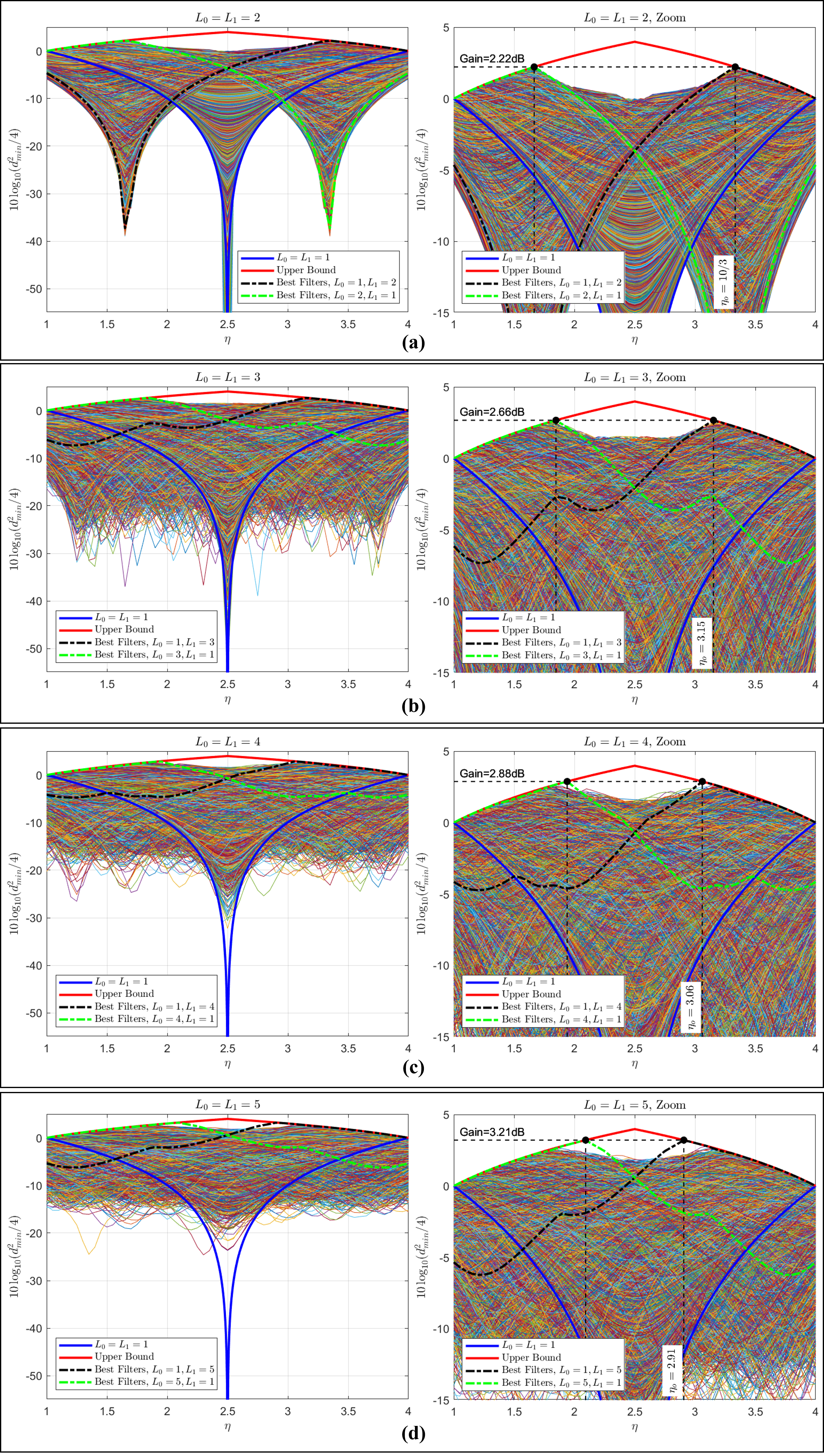}
    \caption{Achievable gain in terms of $d_{\text{min}}^2$ for $4$-ASK as a function of $\eta$, $1\le \eta \le 4$, for $10^6$ random uniform draws of $\bar{h}_m,$ $m=0,1,$ in the unit sphere of dimension $L$: (a) $L_0=L_1=L=2$, (b) $L_0=L_1=L=3$, (c) $L_0=L_1=L=4$, (d) $L_0=L_1=L=5$. Shown also the gain evolution for conventional $4$-ASK, with $L_0=L_1=1$, an upper-bound of achievable gain, for $L_0=L_1=L$, and the performance of optimum filters with $L_0=1, L_1=L$ and $L_0=L, L_1=1.$}
    \label{fig:Achievable Gain d_min2 Random Draws}
\end{figure}

In Figures~\ref{fig:Achievable Gain d_min2 Random Draws} (a)-(d), we show the evolution of the gain, in terms of $d_{\text{min}}^2$, with respect to conventional $4$-ASK, as a function of $\eta,$ for a myriad of random draws of normalized discrete-time filters $\bar{h}_m,$ $m=0,1,$ when the common filter length, $L=L_0=L_1,$ varies respectively from $2$ to $5$. The determination of $d_{\text{min}}^2,$ for each random draw of $\bar{h}_m,$ $m=0,1,$ and each associated value of $\eta,$ is obtained thanks to Algorithm~\ref{alg:d_min^2 Rate-2 NSM L_0>1, L_1>1}, which stems from the discussions and considerations in Subsection~\ref{sapp:d_min^2 Rate-2 NSM L_0>1, L_1>1}.

To be able to find one's bearings and, consequently, draw some instructive insights, we also draw two benchmark curves in each of Figures~\ref{fig:Achievable Gain d_min2 Random Draws}(a)--\ref{fig:Achievable Gain d_min2 Random Draws}(d). Also, for completeness, we show the gain evolution for the best filters, which will be proposed and characterized in Section~\ref{Rate-2 guaranteeing, minimum Euclidean distance approaching NSMs with real filters' coefficients}, with one of them having length $1$ and the other having length $L$. 

\begin{figure}[!htbp]
    \centering
    \includegraphics[width=0.85\textwidth]{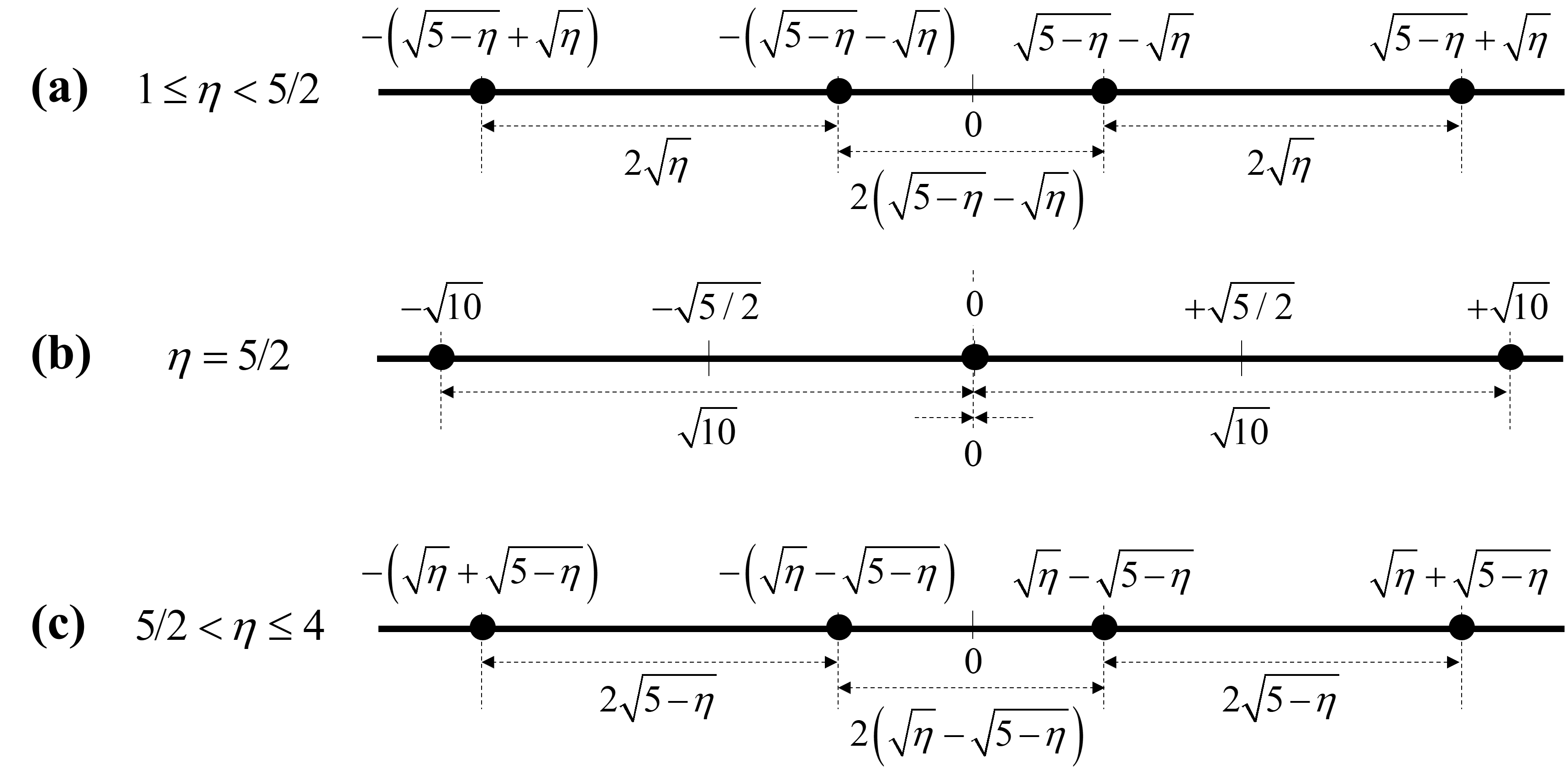}
    \caption{Constellations of $4$-ASK for different values of $\eta$: (a) $1 \le \eta < 5/2$, (b) $\eta=5/2$, (c) $5/2 < \eta \le 4.$}
    \label{fig:Different Energy Distributions4-ASK}
\end{figure}

The first benchmark specifies the \gls{msed}, $d_{\text{min}}^2,$ of interference-free modulations, corresponding to $L=1$ and including conventional $4$-ASK as a special case. Figures~\ref{fig:Different Energy Distributions4-ASK}(a)--\ref{fig:Different Energy Distributions4-ASK}(c) show, respectively, the resulting modulation constellations, when $\eta \in [1, 4]$ falls strictly below, is equals, or exceeds $5/2$ ($5/2$ corresponds to an even distribution of symbol energy between the input data streams $b_0[k]$ and $b_1[k]$). Clearly, the \gls{msed}, for $\eta \in [1, 4]$, is given by $d_{\text{min}}^2 = 4(\sqrt{5-\eta} - \sqrt{\eta})^2$. As a first particular case, conventional $4$-ASK, which corresponds to either $\eta=1$ or $\eta=4$, lead to the expected value of $4$ for $d_{\text{min}}^2$. As a second particular case, with $\eta=5/2$, the input data streams, $b_0[k]$ and $b_1[k]$, use the same transmitted average energy, leading to a null \gls{med}. To the latter case corresponds a disastrous situation in \gls{pd-noma}~\cite{Islam17}, when $b_0[k]$ and $b_1[k]$ are understood as the superimposed signals of two users in the radio mobile network.

The second benchmark provides an upper-bound for the \gls{msed}, $d_{\text{min}}^2,$ as a function of $\eta$. It corresponds to the minimum achievable Euclidean distance when a pair of input sequences, $(b_0^n[k],b_1^n[k]),$ $n=0,1,$ differ in only one input bipolar symbol. On the one hand, if the difference in input sequences occurs in the first entry, $b_0^n[k],$ $n=0,1,$ of the input sequence, meaning that $\Delta \bar{b}_1[k] \triangleq \bar{b}_1^1[k]-\bar{b}_1^0[k] = 0$ and that $\Delta \bar{b}_0[k] \triangleq \bar{b}_0^1[k]-\bar{b}_0^0[k] = \pm 2 \delta(k-l)$, for an arbitrary value of $l,$ then the achievable distance is equal to $4 \eta$. On the other hand, if the difference in input sequences occurs in the second entry, $b_1^n[k],$ $n=0,1,$ of the input sequence, meaning that $\Delta \bar{b}_0[k] = 0$ and that $\Delta \bar{b}_1[k] = \pm 2 \delta(k-l)$, for an arbitrary value of $l,$ then the achievable distance is equal to $4 (5-\eta)$. As a consequence, the upper-bound in \gls{msed} is given by
\begin{equation} \label{Upper-Bound Minimum Squared Euclidean Distance}
d_{\text{min}}^2 = 4 \min (\eta, 5-\eta).
\end{equation}

A first lesson to learn from Figures~\ref{fig:Achievable Gain d_min2 Random Draws}(a)--\ref{fig:Achievable Gain d_min2 Random Draws}(d) is that artificially created interference on top of input data sequences can, for some draws, provide some positive gain in \gls{med}, with respect to conventional $4$-ASK. A second lesson is that the maximum achievable gain increases with the common filters length, $L=L_0=L_1$, when $L$ goes from $2$ to $5$. Hence, as expected intuitively, any increase in filters length provides additional protection to the input data sequences. A third lesson is that the optimum value, $\eta_\text{o},$ of $\eta$, corresponding to the maximum achievable gain, should steer towards the medium value, $5/2,$ of $\eta,$ corresponding to input data sequences, $b_m[k],$ $m=0,1,$ with the same allocated energy ($\eta = 5-\eta$). This medium value of $\eta,$ if reached by $\eta_\text{o}$ (this is indeed the case, as will be shown in Section~\ref{Rate-2 guaranteeing, minimum Euclidean distance approaching NSMs with real filters' coefficients}), through an increase of $L,$ provides the same performance as $2$-ASK, from a \gls{med} point of view. A fourth lesson to draw is that this ultimate value, $5/2,$ of $\eta$, while being the most promising one, is equally the most critical for some random draws of the filters, with gains going towards $-\infty.$ However, as expected, and showcased by a comparison of Figures~\ref{fig:Achievable Gain d_min2 Random Draws}(a)-\ref{fig:Achievable Gain d_min2 Random Draws}(d), the criticality of this promising value of $\eta$ is alleviated by increasing filters length, emphasizing again the effectiveness of interference as a mechanism for protecting input data. Another lesson to take is that conventional $4$-ASK remains optimal for the extreme values, $1$ and $4,$ meaning that filtering cannot be of any help. Finally, a last fact to notice is that, due to an increase of $L=L_0=L_1,$ an exploration of the unit sphere of dimension $L,$ from which $\bar{h}_m,$ $m=0,1,$ are uniformly drawn become more and more difficult. This facts explains why, in Figure~\ref{fig:Achievable Gain d_min2 Random Draws}(d), the $10^6$ carried filters draws, for $L=L_0=L_1=5,$ are far from achieving the performance of the best filters for $(L_0=1, L_1=5)$ and $(L_0=5, L_1=1),$ which are located in the same sphere.

To complete and close the previous discussion, it should be noted that the critical value, $5/2,$ of $\eta,$ which was deemed to be catastrophic for \gls{pd-noma}, raises no concern for \gls{cd-noma}, such as \gls{musa}~\cite{Yuan15}, \gls{scma}~\cite{Nikopour13}, and \gls{lds}~\cite{Hoshyar08}. This can be explained by the fact that the codes used in \gls{cd-noma} introduce interference between users signals and, as such, play similar roles to the filters used in \glspl{nsm}.

Now that experimental results clearly showcase the effectiveness of interference as a mechanism for protecting input data, we propose in the next subsection two simple, yet efficient, modulation schemes, banking on the concept of interference protection.

\subsection{Two Simple yet Efficient NSMs Advocating Interference}

Our perception of interference as the right enabler for efficient modulation schemes operating at the Nyquist signaling rate will next be supported by two simple modulation schemes. The first modulation scheme, to be described next, will provide a foretaste of a large family of \glspl{nsm} with guaranteed \gls{med}, equal to that of $2$-ASK. The second modulation scheme will provide a taster for another family of guaranteed spectral efficiency. On the one hand, the first family of modulation schemes will, in the sequel, be dubbed interchangeably as “minimum distance guaranteeing modulations” or “asymptotic spectral efficiency achieving modulations.” On the other hand, the second family will be referred to as “spectral efficiency guaranteeing modulations” or “asymptotic minimum distance achieving modulations.” As will be made clear in Section~\ref{Rate-2 guaranteeing, minimum Euclidean distance approaching NSMs with real filters' coefficients}, the intersection of these families is not empty, meaning that there will be some \gls{nsm} schemes that are simultaneously minimum-distance and spectral efficiency achieving.

\subsubsection{Minimum Euclidean Distance Guaranteeing Modulation of Rate 5/4} \label{ssec:Minimum Euclidean Distance Guaranteeing 5/4-NSM}

The \gls{nsm} proposed next, of rate $5/4,$ offers $5$ transmission opportunities of bipolar input data, for every $4$ symbol periods. As such, it offers the same spectral efficiency as the \gls{ftn} signaling modulation with compression factor $\alpha=4/5=0.8.$ This value of the compression factor has the merit of being both rational and near the Mazo limit, $0.802$, guaranteeing no loss in \gls{med} with respect to $2$-ASK.

The sampled signal of the proposed modulation obeys the same symbolic expression as in \ref{eq:Mod Seq FTN}, for \gls{ftn} signaling. However, the involved filters are now of \gls{fir} nature and are explicitly given by $\bar{h}_m[k] = \delta[k-m],$ for $m=0, 1, 2$ and $3$, and $\bar{h}_4[k] = \tfrac{1}{2}(\delta[k]+\delta[k-1]+\delta[k-2]+\delta[k-3]).$ These filters have the clear advantage of guaranteeing no overlap between consecutive blocks of $4$ modulated signal samples. Indeed, every $l$-th modulated block, $\bm{s}[l] \triangleq (s[4l], s[4l+1], s[4l+2], s[4l+3]),$ of $4$ consecutive signal samples, exclusively depends on the $l$-th input block, $\bm{b}[l] \triangleq (b_0[l], b_1[l], b_2[l], b_3[l], b_4[l]),$ of $5$ bipolar input data symbols. As such, it can be interpreted as an analog block modulation, with $\bm{s}[l]$ and $\boldsymbol{b}[l]$ related by $\bm{s}[l] = \bm{b}[l] \bm{G},$ where
\begin{equation} \label{eq:Generating Matrix NSM Rate 5/4}
    \bm{G} =
    \begin{pmatrix}
    1 & 0 & 0 & 0 \\
    0 & 1 & 0 & 0 \\
    0 & 0 & 1 & 0 \\
    0 & 0 & 0 & 1 \\
    \tfrac{1}{2} & \tfrac{1}{2} & \tfrac{1}{2} & \tfrac{1}{2}
    \end{pmatrix}
\end{equation}
plays the role of an “analog” generating matrix.

To alleviate the notations, we omit, in what follows, the indexing parameter $l$ in the expressions of $\bm{s}[l]$ and $\bm{b}[l].$ Next, we prove that the \gls{med} of $2$-ASK is preserved by the proposed \gls{nsm} scheme. For that, we need to show that the squared Euclidean norm of the modulated block difference, $\Delta \bm{s} \triangleq \bm{s}^1-\bm{s}^0,$ corresponding to the non-null input block difference, $\Delta \bm{b} \triangleq \bm{b}^1-\bm{b}^0,$ of two different input blocks, $\bm{b}^0$ and $\bm{b}^1,$ is always greater than or equal to $4.$

Firstly, we consider the case where the last component of $\Delta \bm{b}$ is null. This case corresponds exactly to that of $2$-ASK, and at least one component of $\Delta \bm{s}$ is in $\{ \pm 2 \}$, because at least one of the $4$ first components, $\Delta \bm{b}$ is non-null, and therefore equal to $\pm 2.$ Hence, the squared Euclidean norm of $\Delta \bm{s}$ is at least equal to $4.$

Secondly, we consider the case where the last component of $\Delta \bm{b}$ is non-null, and therefore equal to $\pm 2.$ In this case, since the entries corresponding to the last raw of the generating matrix $\bm{G}$ are all equal to $1 \! / \! 2,$ the contributions of the first $4$ components of $\Delta \bm{b}$ to $\Delta \bm{s},$ which are in $\{ 0, \pm 2 \},$ are shifted by either $+1$ or $-1.$ Therefore, the modulated block difference, $\Delta \bm{s},$ has its components in the set $\{ \pm 1, \pm 3 \}.$ It follows that all components of $\Delta \bm{s}$ have a modulus greater than or equal to $1.$ Since $\Delta \bm{s}$ has $4$ such components, it follows that the squared Euclidean norm of $\Delta \bm{s}$ is also at least equal to $4$ in this case. This establishes the fact that the proposed modulation scheme preserves the \gls{med} of $4$-ASK.

The previous considerations, which led to the determination of the \gls{msed} of the modulation at hand, remain unchanged, if any of the non-null entries of generating matrix $\bm{G}$ changes sign. Hence, we can freely change the sign of any of the non-null entries of $\bm{G},$ without altering the modulation \gls{med}. 

\begin{figure}[!htbp]
    \centering
    \includegraphics[width=1.0\textwidth]{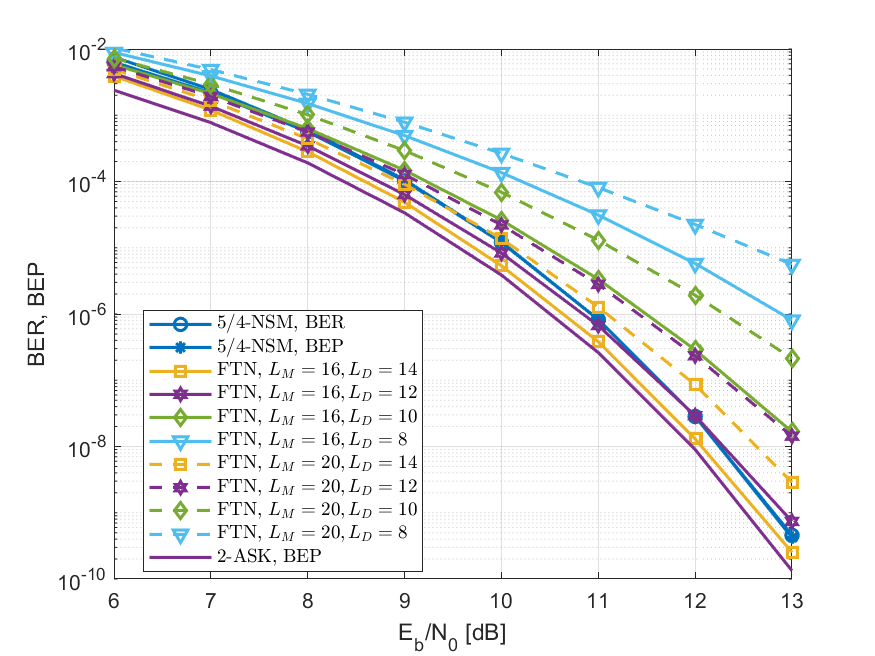}
    \caption{BER and BEP bound of the NSM of rate $5/4$. Several \gls{ftn} signaling modulations, using the “sinc” filter, with compression factor $\alpha=0.8$, are used as benchmarks.}
    \label{fig:BER-BEP-NSM-5_4}
\end{figure}

As depicted in Figure~\ref{fig:BER-BEP-NSM-5_4}, we, next, assess the performance of the proposed \gls{nsm}, through an evaluation of the achieved \gls{ber}. For comparison purpose, we also evaluate the \gls{ber} of the Mazo “sinc” based \gls{ftn} signaling, with a compression factor, $\alpha,$ of $0.8$ and filter truncation lengths, $L_M$ and $L_D,$ at the modulator and the demodulator, respectively. On the one hand, the modulator's truncated filter length, $L_M,$ must be large enough to meet power spectrum mask requirements, while remaining small enough to maintain reasonable filtering complexity at the transmitter. On the other hand, the demodulator's truncated filter length, $L_D,$ must be large enough to capture most of the useful energy conveyed by each transmitted symbol and, as a result, reduce residual \gls{isi}, while remaining small enough to maintain affordable equalization complexity at the receiver.

Notice that the modulator's filtering complexity is always proportional to the corresponding truncated filter length $L_M.$ Also take note that the demodulation complexity can be kept proportional to $L_D,$ if a sub-optimal equalizer, such as a linear or decision-feedback equalizer, is used at the demodulator. However, when optimum \gls{ml} detection is applied using the Viterbi algorithm~\cite{Forney72, Forney73}, as in the simulations shown in Figure~\ref{fig:BER-BEP-NSM-5_4}, equalization complexity at the demodulator grows exponentially as a function of $L_D.$ Before delving into the analysis of Figure~\ref{fig:BER-BEP-NSM-5_4}, keep in mind that the preceding real-world considerations remain fictitious for the “sinc” function, which is never used in practice, as it produces a closed eye diagram.

Additionally, for a validation sake, we complete the assessment with a tight estimate of the proposed modulation achievable \gls{bep}, at high \gls{snr}, using an approximate version of the union bounding technique. For the derivation of this \gls{bep} estimate, we get past all input block differences, $\Delta \bm{b},$ of error events leading to modulated block differences, $\Delta \bm{s},$ of \gls{med}. The detailed considerations relative to \gls{bep} estimate derivation are provided in Appendix~\ref{app:Tight Estimate BEP Rate 5/4}, and lead to the \gls{bep} estimate expression
\begin{equation} \label{eq:BEP Estimate Rate 5/4}
    \text{BEP} \approx \frac{253}{160} \operatorname{erfc} \left( \sqrt{\frac{E_b}{N_0}} \right),
\end{equation}
where $E_b/N_0$ is the energy per bit to noise ratio.

Figure~\ref{fig:BER-BEP-NSM-5_4} reveals a perfect match, at high \gls{snr}, between the \gls{ber} and the \gls{bep} estimate, given in (\ref{eq:BEP Estimate Rate 5/4}), for the proposed \gls{nsm} of rate $5/4.$ It also demonstrates that \gls{ftn} performs well, and even better than or comparable to the proposed \gls{nsm}, as long as $L_D$ is kept close to $L_M$ (cases where $L_M=16$ and $L_D=14$ or $12$). However, when $L_M$ is raised from $16$ to $20,$ to respond to the need of meeting a target spectral mask, or when $L_D$ is reduced from $14$ to $8,$ to express the need of reducing demodulation complexity, the obtained \gls{ber} is worse than that of the proposed modulation. The degradation in \gls{ber} performance relative to the proposed \gls{nsm} becomes more pronounced as the filter truncation length, $L_D,$ drops below $12.$ This is true even though the proposed equalization has a much higher equalization complexity than the proposed modulation, as will be made clear at the end of this subsection.

Having said that, does the fact that \gls{ftn} modulations with $L_D$ kept close to $L_M$ outperform the proposed \gls{nsm}, as demonstrated in Figure~\ref{fig:BER-BEP-NSM-5_4}, imply that Nyquist signaling has performance limitations? The response is no. Indeed, as will be demonstrated in Section~\ref{Rate-5/4 Approaching NSMs}, increasing the length of the fifth filter, $\bar{h}_4[k],$ from $4$ to $8$, results in a new rate $5/4$ \gls{nsm} modulation that outperforms the \gls{ftn} “sinc” modulation, while having an incredibly low demodulation complexity, due to a simple two-state demodulation trellis.

Before moving to the second illustrative \gls{nsm} scheme, we investigate a simple demodulation technique for the proposed $5/4$-rate modulation. If $\bm{r}[l] \triangleq (r[4l], r[4l+1], r[4l+2], r[4l+3])$ denotes the block, received on a Gaussian channel, corresponding to input block $\bm{b}[l]$ and modulated block $\bm{s}[l],$ then the \gls{ml} decision, $\hat{\bm{b}}[l],$ on $\bm{b}[l],$ is given by
\begin{equation} \label{eq:ML Detection Rate-5/4 NSM}
    \hat{\bm{b}}[l] = \arg \min_{\bm{b}[l]} \| \bm{r}[l] - \bm{b}[l] \bm{G} \|^2.
\end{equation}

To simplify the \gls{ml} detection process, we proceed into two steps. We start by determining the \gls{ml} decision on the fifth component, $b_4[l],$ then proceed to the determination of the \gls{ml} decisions on the other $4$ remaining components of $\bm{b}[l]$. The \gls{ml} decision on $b_4[l]$ is expressed as
\begin{equation} \label{eq:First Step Detection Algorithm Rate-5/4 NSM}
    \hat{b}_4[l] = \arg \min_{b_4[l]} \left( \min_{b_m[l], \; 0 \le m \le 3}\| \bm{r}[l] - \bm{b}[l] \bm{G} \|^2 \right).
\end{equation}

For $b_4[l]=a,$ $a \in \{ \pm 1 \},$ the values of $b_m[l],$ $0 \le m \le 3$ that minimize the squared norm, $\| \bm{r}[l] - \bm{b}[l] \bm{G} \|^2,$ are given by $b_m[l] = \operatorname{sgn} (r[4l+m]-a/2),$ where $\operatorname{sgn}(\cdot)$ is the signum function. The minimum of the squared norm over $b_m[l],$ $0 \le m \le 3,$ is therefore given by
\begin{equation} \label{eq:Closed Form Expression Metric Rate-5/4 NSM}
    \min_{b_m[l], \; 0 \le m \le 3}\| \bm{r}[l] - \bm{b}[l] \bm{G} \|^2 = \sum_{m=0}^3 (|r[4l+m]-a/2|-1)^2.
\end{equation}
Taking, now, into account the fact that $a \in \{ \pm 1 \},$ we can write
\begin{equation} \label{eq:Second Step Detection Algorithm Rate-5/4 NSM}
    \hat{b}_4[l] = \operatorname{sgn} \left( \sum_{m=0}^3 \zeta(r[4l+m]) \right),
\end{equation}
where $\zeta(x)$ is an odd function, expressed as $-x,$ for $0 \le x < 1/2,$ and as $x-1,$ for $x \ge 1/2.$

Once the decision on ${b}_4[l]$ has been made, we can inject it into the expressions of the decisions on the other components, leading to
\begin{equation} \label{eq:Last Step ML Decision Rate-5/4 NSM}
    \hat{b}_m[l] = \operatorname{sgn} \left( r[4l+m]-\hat{b}_4[l]/2 \right), \; 0 \le m \le 3.
\end{equation}

\subsubsection{Spectral Efficiency Guaranteeing Modulation of Rate 2} \label{ssec:Modulation of Rate 2}

Before moving on to the description and characterization of the illustrative \gls{nsm} of rate $2,$ it is considered to be highly helpful to have a brief discussion on some relevant aspects of \glspl{nsm}.

In the first place, it is important to keep in mind that any \gls{nsm} only requires the specification of the filters, up to an arbitrary non-null multiplicative factor. So, if, for a particular multiplicative factor, $\varepsilon,$ the scaled filters, $\mathring{h}_m[k] = \varepsilon \bar{h}_m[k],$ have simpler expressions than the normalized filters, $\bar{h}_m[k],$ then they can be extremely insightful when designing and characterizing \glspl{nsm}.

Second, for an infinite train of input bipolar data, Nyquist signaling systems with different filters can be perfectly equivalent in terms of characteristics and performance. The opposite, $-h_m[k],$ of any of the filters, $h_m[k],$ preserves Nyquist signaling system characteristics and performance. Furthermore, shifting in time any of the filters by an arbitrary time advance or time delay has no effect on performance. Besides, time reversal, in which all filters, $h_m[k]=\sum_{l=0}^{L-1} h_m[l] \delta[k-l],$ are replaced by their time reversal counterparts, $h_m^-[k]=\sum_{l=0}^{L-1} h_m[L-1-l] \delta[k-l],$ is one of the filter operations that preserves characteristics and performance. Also, alternately changing the signs of the coefficients of an arbitrary filter, $h_m[l],$ yields the equivalent filter $\tilde{h}_m[l] = \sum_{l=0}^{L-1} (-1)^l h_m[l] \delta[k-l].$ To see this, consider the fact that if an input sequence difference, $\Delta b[k],$ leads to the modulated sequence difference $\Delta s[k],$ when filtered by $h_m[l],$ then the alternate input sequence difference, $\Delta \tilde{b}[k] \triangleq (-1)^k \Delta b[k],$ leads to the alternate modulated sequence difference, $\Delta \tilde{s}[k] = (-1)^k \Delta s[k],$ which has precisely the same norm. To finish, it is crucial to remember that the equivalence ceases for several of the previous filter operations, when input data sequences of finite lengths are taken into account.

In light of the foregoing, we will now explore the intuitive derivation of one of the most straightforward rate-$2$ \glspl{nsm}. This derivation can be made by just modifying one of the filters, $h_m[k] = 2^m \delta[k],$ $m=0,1,$ that are used in traditional $4$-ASK. For this, notice the important discrepancy in conveyed modulated signal energy between $h_1[k]$ and $h_0[k],$ with the former bearing four times as much energy as the latter. The proposed Nyquist signaling rate-$2$ modulation addresses this disparity in conveyed energy by keeping the most energetic filter, $h_1[k],$ as it is and replacing the least energetic filter, $h_0[k]=\delta[k],$ with a two-fold more energetic filter. More specifically, in their simplest scaled forms, the proposed filters are $\mathring{h}_0[k]=\delta[k]+\delta[k-1]$ and $\mathring{h}_1[k]=2\delta[k],$ resulting in a reduction by a factor of two of the discrepancy.

Prior to examining and describing the suggested modulation, through its most basic scaled filters, $\mathring{h}_m[k],$ $m=0,1,$ it is worthwhile to specify it in terms of the ASK-aligned filters, $h_m[k]$ and $\bar{h}_m[k],$ $m=0,1,$ and parameters, $\eta_m,$ $m=0,1,$ and $\eta,$ introduced in Subsection~\ref{ssec:Merits of Interference}. Notice that the average energy per symbol conveyed by the scaled filters $\mathring{h}_m[k],$ $m=0,1,$ is equal to $6,$ while that conveyed by conventional $4$-ASK is $5$. Hence, $h_m[k] = \sqrt{\tfrac{5}{6}} \mathring{h}_m[k],$ $m=0,1,$ leading to the explicit expressions $h_0[k] = \sqrt{5/6} (\delta[k]+\delta[k-1])$ and $h_1[k] = \sqrt{10/3} \delta[k].$ Equivalently, we have $\bar{h}_0[k] = \tfrac{1}{\sqrt{2}} (\delta[k]+\delta[k-1]),$ $\bar{h}_1[k] = \delta[k],$ $\eta_0=\eta=5/3$ and $\eta_1=5-\eta=10/3.$ These values of $\eta_0$ and $\eta_1$ are precisely the values corresponding to the maximum achievable gain, shown in the right-hand side of Figure~\ref{fig:Achievable Gain d_min2 Random Draws}(a), for $L_0=2$ and $L_1=1.$

\begin{figure}[!htbp]
    \centering
    \includegraphics[width=0.9\textwidth]{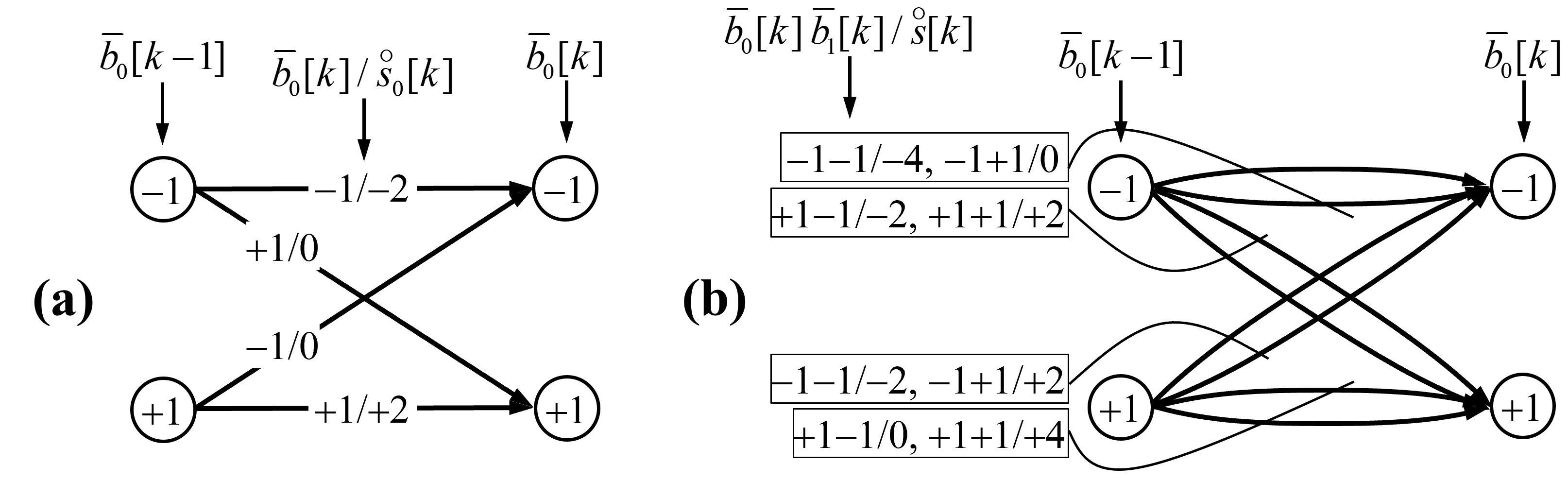}
    \caption{Trellises of the “duobinary” channel and the rate-$2$ NSM: (a) “Duobinary” channel, (b) Rate-$2$ modulation.}
    \label{fig:Trellis Duobinary Channel Rate-2 Modulation}
\end{figure}

The proposed rate-$2$ modulation can be characterized using its trellis. To accomplish this, we will take two steps. First, in Figure~\ref{fig:Trellis Duobinary Channel Rate-2 Modulation}(a), we depict one section of the trellis that characterizes the filtering performed by the first filter, $\mathring{h}_0[k]=\delta[k]+\delta[k-1],$ which is referred to in the literature as the “duobinary” channel. Then, in Figure~\ref{fig:Trellis Duobinary Channel Rate-2 Modulation}(b), we augment the resulting trellis section with the contributions of the second memoryless filter, $\mathring{h}_1[k] = 2 \delta[k],$ to obtain the final trellis section of the \gls{nsm} at hand.

The states delimiting the $k$-th section for both trellises are the admissible values of $\bar{b}_0[k-1],$ to the left, and $\bar{b}_0[k],$ to the right. Each branch in the $k$-th section of the first trellis, depicted in Figure~\ref{fig:Trellis Duobinary Channel Rate-2 Modulation}(a), connecting state $\bar{b}_0[k-1]$ to state $\bar{b}_0[k],$ has $\bar{b}_0[k]$ as input label and $\mathring{s}_0[k] \triangleq \sum_l \bar{b}_0[l] \mathring{h}_0[k-l] = \bar{b}_0[k] + \bar{b}_0[k-1]$ as output label. Each branch in the second trellis, depicted in Figure~\ref{fig:Trellis Duobinary Channel Rate-2 Modulation}(b), has an input label of $\bar{b}_0[k] \, \bar{b}_1[k]$ and an output label of $\mathring{s}[k] \triangleq \sum_l \bar{b}_0[l] \mathring{h}_0[k-l] + \sum_l \bar{b}_1[l] \mathring{h}_1[k-l] = \bar{b}_0[k] + \bar{b}_0[k-1] + 2 \bar{b}_1[k].$ As a result, the second lattice is formed by duplicating each branch of the first trellis, leading to two “parallel” branches, one with the label $\bar{b}_0[k] + \bar{b}_0[k-1] -2,$ corresponding to input $\bar{b}_1[k]=-1,$ and the other with the label $\bar{b}_0[k] + \bar{b}_0[k-1] +2,$ corresponding to input $\bar{b}_1[k]=+1.$

It is common knowledge that the “duobinary” channel, $\mathring{h}_0[k]=\delta[k]+\delta[k-1],$ preserves the \gls{msed} of $8,$ when the input data sequence, $\bar{b}_0[k],$ takes its values in the bipolar alphabet $\{ \pm 1\}.$ In contrast, it is known that this channel exhibits some degenerate behavior that, in some ways, recalls the behavior of catastrophic convolutional codes. The proposed rate-$2$ modulation obviously inherits this degeneracy. Fortunately, unlike convolution codes, where additions and multiplications are carried in the Galois field, $\text{GF}_2 = \{0,1\},$ here they are carried in the field of real numbers, $\mathbb{R}$. This has the advantage of greatly decreasing the adverse effects of degeneracy, as a result of the exponential decrease in the occurrence probability of error events of \gls{med}, as a function of their length.

To see the degenerate nature of the “duobinary” channel, consider two arbitrary input bipolar data sequence, $\bar{b}_0^0[k]$ and $\bar{b}_0^1[k],$ that are identical, except for $K \ge 1$ consecutive indices $0 \le k < K,$ where $\bar{b}_0^0[k] = -\bar{b}_0^1[k] = (-1)^k.$ The corresponding “duobinary” channel outputs, $\mathring{s}[k],$ which have the particularity of being both perfectly null for $0<k<K$, are identical everywhere, except for $k = 0$ and $K,$ where they take values in $\{ \pm 2\}.$ These pairs of input sequences achieve the \gls{msed}, $(\pm 2)^2 + (\pm 2)^2 = 8,$ of the “duobinary” channel, $\mathring{h}_0[k] = \delta[k]+\delta[k-1],$ which, when normalized by $\sqrt{2}$ leads to the value of $4,$ the same value as the \gls{msed} of $2$-ASK. They are infinite in number and lead to $K$ bipolar errors, for each value of $K,$ $K \ge 1$. Fortunately, for the same value of $K,$ their occurrence probability, which is equal to $1/2^{K-1},$ vanishes exponentially as a function of $K,$ and out-weights the underlying $K$ bipolar errors. Using an approximate union bounding technique, where only minimum-distance neighbor output sequences are kept in the error probability upper-bound, a tight estimate of the binary (or bipolar) error probability of the “duobinary” channel is given by
\begin{equation} \label{eq:BEP Estimate Duobinary Channel}
    \text{BEP} \approx \sum_{K \ge 1} \frac{K}{2^{K-1}} \frac{1}{2} \operatorname{erfc} \left( \sqrt{\frac{E_b}{N_0}} \right) = 2 \operatorname{erfc} \left( \sqrt{\frac{E_b}{N_0}} \right).
\end{equation}

\begin{figure}[!htbp]
    \centering
    \includegraphics[width=1\textwidth]{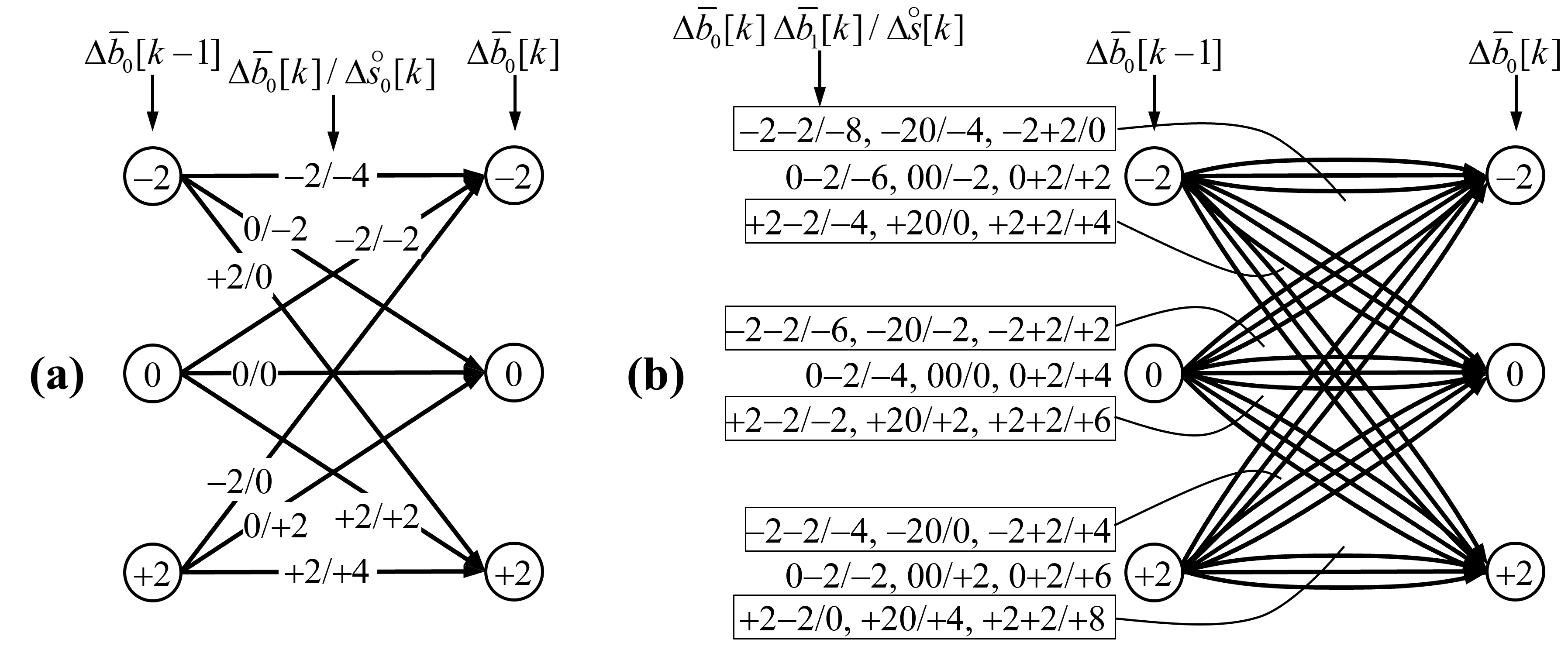}
    \caption{Trellises of the input/output sequences differences of the “duobinary” channel and the rate-$2$ NSM: (a) “Duobinary” channel, (b) Rate-$2$ modulation.}
    \label{fig:Trellis Input Difference Duobinary Channel Rate-2 Modulation}
\end{figure}

To further investigate the degeneracy of the “duobinary” channel and the proposed modulation, and prepare the ground for the computation of a tight estimate of the \gls{bep} of the rate-$2$ \gls{nsm}, we show in Figures~\ref{fig:Trellis Input Difference Duobinary Channel Rate-2 Modulation}(a) and \ref{fig:Trellis Input Difference Duobinary Channel Rate-2 Modulation}(b) the trellises corresponding to the differences in input and output sequences, for the “duobinary” channel and the proposed modulation, respectively.

For the “duobinary” channel, the input sequences differences, $\Delta \bar{b}_0[k],$ take their values in the ternary alphabet $\{0, \pm 2 \}.$ Therefore, the $k$-th section of corresponding trellis, depicted in Figure~\ref{fig:Trellis Input Difference Duobinary Channel Rate-2 Modulation}(a), is delimited by three states to the left, corresponding to the admissible values of $\Delta \bar{b}_0[k-1],$ and by three other states to the right, corresponding to the admissible values of $\Delta \bar{b}_0[k].$ The branch in the $k$-th section of this trellis, delimited by state $\Delta \bar{b}_0[k-1],$ in the left, and by state $\Delta \bar{b}_0[k],$ in the right, has $\Delta \bar{b}_0[k],$ as input difference label and $\Delta \mathring{s}_0[k] = \sum_l \Delta \bar{b}_0[l] \mathring{h}_0[k-l] = \Delta \bar{b}_0[k] + \Delta \bar{b}_0[k-1],$ as output difference label.

In direct connection with the previous discussion on input bipolar sequences achieving the \gls{med} of the “duobinary” channel, we see, from Figure~\ref{fig:Trellis Input Difference Duobinary Channel Rate-2 Modulation}(a), that error events of length $K,$ $K \ge 1,$ with input sequences differences of the forms $\Delta \bar{b}_0[k] = \sum_{l=0}^{K-1} (-1)^l 2 \delta[k-l]$ and $\Delta \bar{b}_0[k] = -\sum_{l=0}^{K-1} (-1)^l 2 \delta[k-l],$ lead to output sequences differences of the form $\Delta \mathring{s}_0[k] = 2 (\delta[k] + (-1)^{K-1}\delta[k-K])$ and $\Delta \mathring{s}_0[k] = -2 (\delta[k] + (-1)^{K-1}\delta[k-K]),$ respectively. Again, these output sequences differences, which are infinite in number, lead to the \gls{msed} of $8,$ for the “duobinary” channel $\mathring{h}_0[k].$ Fortunately, and in accordance with the discussion at the beginning of Appendix~\ref{app:Tight Estimate BEP Rate 5/4}, the probability of occurrence of each of these error events is governed by the number, $K,$ of non-null input sequence differences, and vanishes exponentially as $(\tfrac{1}{2})^K.$

To obtain a section of the input sequences differences trellis of the rate-$2$ \gls{nsm}, shown in Figure~\ref{fig:Trellis Input Difference Duobinary Channel Rate-2 Modulation}(b), we triplicate each branch of the “duobinary” channel trellis section shown in Figure~\ref{fig:Trellis Input Difference Duobinary Channel Rate-2 Modulation}(a). Each branch of the “duobinary” channel trellis, with $\Delta \bar{b}_0[k]/\Delta \mathring{s}_0[k]$ as input/output difference label, leads to three branches, with $\Delta \bar{b}_0[k] \Delta \bar{b}_1[k]/\Delta \mathring{s}[k]$ as input/output difference label, corresponding to the three values taken by $\Delta \bar{b}_1[k]$ in $\{0, \pm 2 \}.$ The output label is the output sequence difference $\Delta \mathring{s}[k] = \sum_l \Delta \bar{b}_0[l] \mathring{h}_0[k-l] + \sum_l \Delta 
 \bar{b}_1[l] \mathring{h}_1[k-l] = \Delta \bar{b}_0[k] + \Delta \bar{b}_0[k-1] + 2 \Delta \bar{b}_1[k]$ in this case.

The degeneracy of the proposed \gls{nsm} is more pronounced than that of the “duobinary” channel, because of the addition of the term $2 \Delta \bar{b}_1[k]$ to the labels of the “duobinary” channel to obtain the labels of the proposed modulation. On the one hand, when we move from state $\Delta \bar{b}_0[k-1] = -2,$ (respectively, $+2$) to state $\Delta \bar{b}_0[k] = -2,$ (respectively, $+2$), with $\Delta \bar{b}_1[k] = +2,$ (respectively, $-2$), the output sequence difference, $\Delta \mathring{s}[k],$ is null. On the other hand, when we move from state $\Delta \bar{b}_0[k-1] = -2,$ (respectively, $+2$) to state $\Delta \bar{b}_0[k] = +2,$ (respectively, $-2$), with $\Delta \bar{b}_1[k] = 0,$ the output sequence difference, $\Delta \mathring{s}[k],$ is also null. This former case is new while the last case is directly inherited from the degeneracy of the “duobinary” channel.

Fortunately, the proposed modulation's \gls{msed} remains the same as that of the “duobinary” channel. One way to see this, is to notice that the output sequences differences of the “duobinary” channel, corresponding to error events, have at least two non-null components in the set $\{\pm 2\}.$ To get the corresponding output sequences differences components of the proposed modulation, these components undergo the addition of components from the input sequences differences, $\Delta \bar{b}_1[k],$ which take their values in $\{0, \pm 4 \}.$ Hence, these two components take their values in the set $\{\pm 2, \pm 6 \},$ and therefore contribute by at least $(\pm 2)^2+(\pm 2)^2=8$ to the \gls{msed} of the propose modulation.

Another way to be convinced that the proposed modulation preserves the minimum distance of the “duobinary” channel is to examine the trellis in Figure~\ref{fig:Trellis Input Difference Duobinary Channel Rate-2 Modulation}(b). On the one hand, notice that each error event starts by leaving state $0$ to one of states $-2$ and $+2$ of the modulation trellis, leading to an output sequence difference in the sets $\{-6, \pm 2\}$ and $\{\pm 2,+6\},$ respectively. This contributes to the \gls{msed} by at least $(\pm 2)^2 = 4.$ On the other hand, notice that each error event ends by leaving one of states $-2$ and $+2$ to state $0$ of the modulation trellis, leading to an output sequence difference in the sets $\{-6, \pm 2\}$ and $\{\pm 2,+6\},$ respectively. Similarly, this contributes to the \gls{msed} by at least $(\pm 2)^2 = 4.$ Adding both contributions, we can ensure that the \gls{msed} of the proposed modulation is equal to $8.$

Now, to evaluate the gain in performance, with respect to conventional $4$-ASK, we should use the filters $h_m[k]=\sqrt{5/6} \mathring{h}_m[k],$ $m=0,1,$ instead of their scaled versions, used so far for the characterization of the proposed modulation. These filters enable the alignment of the the average symbol energy of the proposed modulation with the average symbol energy of $4$-ASK. As a consequence, the effective \gls{msed} of the proposed modulation is $(5/6)8=20/3.$ Comparing this squared distance with $4,$ the \gls{sed} of conventional $4$-ASK, we get a gain of $10 \log_{10}(5/3) = 10 \log_{10}(\eta_0) \approx 2.22$ dB. This maximum gain is explicitly shown in the right-hand side of Figure~\ref{fig:Achievable Gain d_min2 Random Draws}(a), for $(L_0=2, L_1=1)$ but also for the symmetrical case of $(L_0=1, L_1=2).$

\begin{figure}[!htbp]
    \centering
    \includegraphics[width=1.0\textwidth]{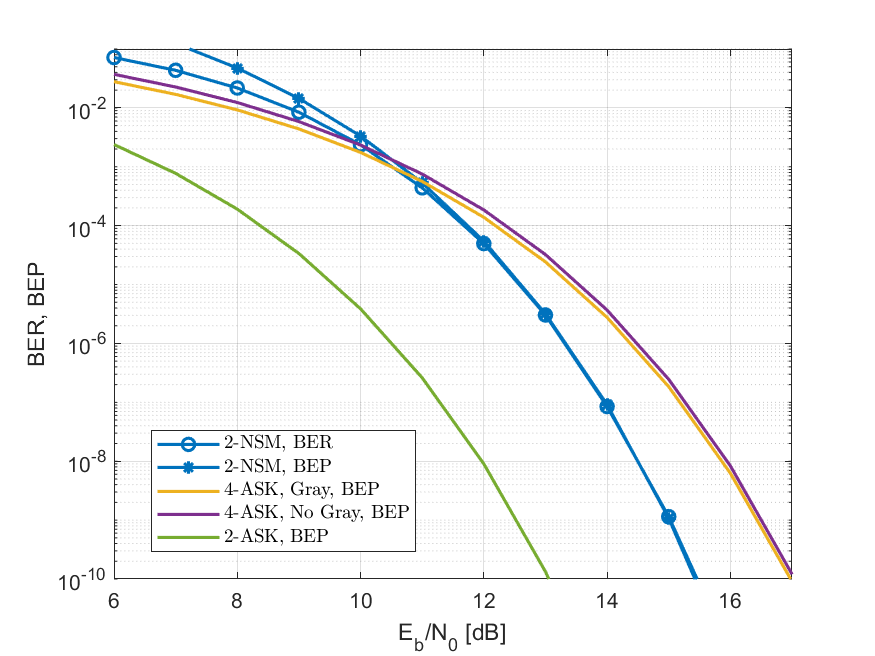}
    \caption{BER and BEP approximation of the NSM of rate $2,$ with $L_0 = 2$ and $L_1 = 1$. For reference, $2$-ASK and Gray and non-Gray precoded $4$-ASK conventional modulations are used.}
    \label{fig:BER-BEP-NSM-2}
\end{figure}

To properly complete the previous evaluation of the effectiveness of the suggested modulation, which relies on minimum squared distance evaluation, several simulation and theoretical results, regarding \gls{ber} and \gls{bep}, are provided in Figure~\ref{fig:BER-BEP-NSM-2}. It is important to note that the \gls{med} criterion is only precise at high useless $E_b/N_0$ values, where it provides an exact asymptotic gain, with respect to conventional $4$-ASK modulation. However, it is not always accurate and insightful at moderate and practical $E_b/N_0$ values, where the multiplicity of nearest modulated sequences is a key factor in determining \gls{bep} performance. This inaccuracy is even exacerbated by the proposed modulation's extreme degeneracy, with most of the gain promised by the increase in \gls{med} lost at practical values of $E_b/N_0.$ To sort this out and get a clearer appreciation the misleading effect of extreme degeneracy, it suffice to examine the expression of the approximate estimate,
\begin{equation} \label{eq:Tight Estimate BEP Rate 2}
        \text{BEP} \approx \frac{51}{2} \frac{1}{2} \operatorname{erfc} \left( \sqrt{\frac{8}{12} \frac{E_b}{N_0}} \right) = \frac{51}{4} \operatorname{erfc} \left( \sqrt{\frac{2}{3} \frac{E_b}{N_0}} \right),
\end{equation}
of the \gls{bep} for the proposed modulation of rate $2.$  This expression is the outcome of an in-depth analysis presented in Appendix~\ref{app:Tight Estimate BEP Rate 2}. On the one hand, it reveals a $51/2$ multiplicity as a result, which seriously hinders the achievement of the asymptotic gain promised by the increase in \gls{med}, at moderate \glspl{snr}. On the other hand, it discloses that, in contrast to conventional $4$-ASK, where the multiplicative factor inside the square-root within the $\operatorname{erfc}(\cdot)$ function is $2/5$, it achieves the expected asymptotic gain factor of $5/3$, stated above, using a different argumentation.

Figure~\ref{fig:BER-BEP-NSM-2} backs up the preceding declarations and reflections. First, at moderate to high \glspl{snr}, it shows a perfect match between the \gls{ber} curve, obtained by simulation for the proposed \gls{nsm} of rate $2,$ and the approximate \gls{bep} curve, expressed in (\ref{eq:Tight Estimate BEP Rate 2}). As a result, at high \gls{snr}, we can rely on this \gls{bep} estimate and confidently confirm the achievement of an asymptotic gain of $10 \log_{10}(5/3) \approx 2.22$ dB. Second, comparing the \gls{ber} curve of the proposed \gls{nsm} with the \gls{bep} curve of conventional $4$-ASK, without Gray precoding, reveals a gain of only $1.61$ dB, at the extremely low and unworkable \gls{ber}/\gls{bep} target of $10^{-10}.$ It is crucial to note that practical communication systems employ error correction codes, so that, depending on the strength of the coding scheme, a target coded \gls{ber}/\gls{bep}, in the practical range $10^{-9}\!-\!10^{-6},$ at the decoder output, typically translates to a target row \gls{ber}/\gls{bep}, in the range of $10^{-4}\!-\!10^{-2},$ at the demodulator output. The huge multiplicity factor of $51/2$ is clearly the reason why the proposed \gls{nsm} is either useless or only produces a marginal gain in the latter range of row \glspl{ber}/\glspl{bep}.

To complete the characterization of the proposed \gls{nsm}, it is instructive to investigate aspects relating to \gls{papr}, for the proposed modulation. Notice, from Figure~\ref{fig:Trellis Duobinary Channel Rate-2 Modulation}(b), that the alphabet used by the said modulation is $\{ \pm 4, \pm 2, 0\}.$ Given that the scaled filters $\mathring{h}_m[k],$ $m=0,1,$ which are at the origin of this alphabet, lead to an average symbol energy of $6,$ we may conclude that the \gls{papr} is equal to $4^2/6=8/3.$ Dividing by the gain in \gls{sed} achieved by the proposed modulation, we get the normalized \gls{papr} $(8/3)/(5/3)=8/5.$ The good, but unexpected, news is that this value is lower than $3^2/5=9/5,$ the \gls{papr} of conventional $4$-ASK, even with its reduced amplitude alphabet $\{\pm 1, \pm 3 \}.$

Before moving on, it is worth noting that, thanks to the relationships established at the start of the subsection between the filters of equivalent \glspl{nsm}, the scaled filter $\mathring{h}_0[k]=\delta[k]+\delta[k-1],$ associated with the “duobinary” channel, can be replaced by the scaled filter $\mathring{h}_0[k]=\delta[k]-\delta[k-1],$ known as the “dicode” channel. The “dicode” channel has been extensively studied in the field of digital magnetic recording~\cite{Siala95,Siegel91, Karabed91}.

It is important to note that both variants of $\mathring{h}_0[k],$ corresponding to the “duobinary” and “dicode” channels, result in the identical \gls{nsm} characteristics and performance, except when the tail-biting approach~\cite{Ma86} is applied to recover the small rate loss due to trellis termination. When tail-biting is enabled, the two instances of filter $\mathring{h}_0[k]$ can act in unexpected ways. On the one hand, when using the “dicode” channel filter, $\mathring{h}_0[k]=\delta[k]-\delta[k-1],$ with modulated sequences of finite length, $K,$ and input sequences differences $\Delta \bar{b}_0[k] = \sum_{l=0}^{K-1} 2 \delta[k-l]$ and $\Delta \bar{b}_1[k] = 0,$ the tail-bitten output sequence difference, $\Delta \mathring{s}[k],$ is absolutely null. On the other hand, if the “duobinary” channel filter, $\mathring{h}_0[k]=\delta[k]+\delta[k-1],$ is applied, when the modulated sequence length, $K,$ is odd, the tail-bitten output sequence difference, $\Delta \mathring{s}[k],$ can never be null. However, when $K$ is even, the input sequences differences $\Delta \bar{b}_0[k] = \sum_{l=0}^{K-1} (-1)^l 2 \delta[k-l]$ and $\Delta \bar{b}_1[k] = 0,$ result in a fully null tail-bitten output sequence difference, $\Delta \mathring{s}[k].$ Fortunately, in all cases, the likelihood of such a catastrophic scenario occurring reduces exponentially, as $(\tfrac{1}{2})^K,$ as a function of sequence length, $K.$



\section{Rate-2 NSMs} \label{sec:Rate-2 NSMs}

In Subsection~\ref{ssec:Modulation of Rate 2}, we introduced a rate-$2$ guaranteeing \gls{nsm} based on short filter lengths $L_0 = 2$ and $L_1 = 1$, illustrating how constructive \gls{isi} can be exploited to achieve significant performance gains, while maintaining a structure simpler than that of \gls{ftn} signaling.

The design of optimal rate-$2$ \glspl{nsm}, however, hinges on the ability to determine the \gls{msed} they can offer for a given choice of filters. This evaluation is addressed in Appendix~\ref{app:d_min^2 Rate-2 NSM}, which provides the framework and algorithms necessary to compute this distance for any candidate \gls{nsm}. This foundational tool is used extensively in both Subsection~\ref{Rate-2 guaranteeing, minimum Euclidean distance approaching NSMs with real filters' coefficients} and Subsection~\ref{Rate 2 Approaching NSM Simple Rational Coefficients} to guide the optimization of the filters. In addition, Appendix~\ref{app:Iterative Determination Transfer Function Rate-2 NSM} supports the derivation of \glspl{rtf} used in estimating upper bounds on the \gls{bep}, specifically for \glspl{nsm} with real-valued filter coefficients. This makes it relevant mainly to the studies presented in Subsection~\ref{Rate-2 guaranteeing, minimum Euclidean distance approaching NSMs with real filters' coefficients}.

Subsection~\ref{Rate-2 guaranteeing, minimum Euclidean distance approaching NSMs with real filters' coefficients} focuses on rate-$2$ \glspl{nsm} characterized by two filters, $h_0[k]$ and $h_1[k]$, with lengths $L_0$ and $L_1$ respectively. The framework builds on the experimental motivation introduced earlier in Subsection~\ref{ssec:Merits of Interference}, where we explored how increasing the filter length $L$ improves performance by leveraging carefully designed interference. The optimization process begins with a numerical search—based on a simulated-annealing-inspired algorithm—to identify high-performing \glspl{nsm} that maximize the \gls{msed}, as determined using the techniques from Appendix~\ref{app:d_min^2 Rate-2 NSM}. Once promising filter candidates are obtained, we analyze the associated error events that lead to the minimum Euclidean distances. This paves the way for a symbolic resolution step, enabling the derivation of closed-form expressions for optimal filters whenever possible.

In this process, two distinct optimization frameworks are used for designing \glspl{nsm} with real filter coefficients, as detailed in Subsection~\ref{Rate-2 guaranteeing, minimum Euclidean distance approaching NSMs with real filters' coefficients}: the \emph{unconstrained} and \emph{constrained} approaches. In both cases, the total average symbol energy is normalized to match that of $4$-ASK, i.e., $\eta_0 + \eta_1 = 5$, where $\eta_0 = \|h_0[k]\|^2$ and $\eta_1 = \|h_1[k]\|^2$ denote the respective energies of the shaping filters applied to the two input bipolar symbol streams. This normalization ensures fair performance comparison under equal spectral efficiency. The distinction between the two frameworks lies in how this energy is distributed across the symbol streams.

In the \emph{unconstrained} framework, the filter energies $\eta_0$ and $\eta_1$ are allowed to differ, with $\eta_0 = \eta$ and $\eta_1 = 5 - \eta$. This flexibility expands the design space and often leads to \glspl{nsm} with higher \glspl{msed}, improving uncoded performance. However, the resulting modulation is unbalanced: half of the bipolar symbols carry energy $\eta_0$, while the other half carry energy $\eta_1$. This imbalance impacts behavior under turbo-equalization. Specifically, unconstrained \glspl{nsm} tend to perform better during early decoding iterations when interference dominates, but their unequal symbol energies limit convergence toward the performance of a balanced, coded $2$-ASK system in later iterations.

In contrast, the \emph{constrained} framework enforces an equal energy condition, $\eta_0 = \eta_1 = 2.5$, guaranteeing uniform energy across both symbol streams. While this restriction can slightly reduce the achievable \gls{msed} for given filter lengths $L_0$ and $L_1$, it proves advantageous in coded systems using iterative detection. As iterations progress, the receiver becomes increasingly capable of isolating the contributions from each input stream, causing the \gls{nsm} to behave like two parallel $2$-ASK modulations. In this context, energy symmetry ensures that both streams contribute equally, allowing the overall system to approach the performance of an ideal, coded $2$-ASK scheme in the final stages of turbo decoding.

The resulting closed-form filters and minimum distance expressions for both design cases are presented in Appendices~\ref{app:Closed-Form Expressions Optimum Filters Non-Degenerate Rate-2 NSMs} and~\ref{app:Closed-Form Expressions Optimum Filters Rate-2 NSMs Constrained Optimization}, respectively. The upcoming results in Subsection~\ref{Rate-2 guaranteeing, minimum Euclidean distance approaching NSMs with real filters' coefficients} focus primarily on the unconstrained optimization case, for which \gls{ber} performance is systematically evaluated. To support this analysis, Appendix~\ref{app:Iterative Determination Transfer Function Rate-2 NSM} details how truncated versions of three distinct \glspl{tf} associated with a given \gls{nsm} are computed iteratively. These \glspl{tf} form the basis for deriving tight upper bounds on the \gls{bep}. The purpose of this appendix is to enable a rigorous cross-validation between the simulated \gls{ber} curves and their theoretical \gls{bep} approximations, thereby indirectly validating the accuracy of the \glspl{rtf} and confirming the overall consistency and reliability of the analytical framework used to assess \gls{nsm} performance under both optimization strategies.

Subsection~\ref{Rate 2 Approaching NSM Simple Rational Coefficients} investigates a class of rate-$2$ \glspl{nsm} whose filter taps are constrained to be \emph{rational-valued} (which is equivalent to assuming integer-valued taps, given the guaranteed equivalence through suitable rescaling of the involved filters). This restriction simplifies hardware implementation—particularly under fixed-point arithmetic—and reduces the complexity of the optimization process. Instead of searching over the continuous space of real-valued coefficients, the design is guided by a discrete \emph{pattern vector} with integer components, which defines the structure and energy of the primary filter $h_0$. The secondary filter $h_1$ is reduced to a single non-zero tap, whose value is chosen to match the squared norm of the pattern vector, thereby ensuring balanced energy contributions between the two filters. These structural constraints yield low-complexity, hardware-friendly \glspl{nsm} that aim to closely match the performance of their more general real-valued counterparts.

The study in Subsection~\ref{Rate 2 Approaching NSM Simple Rational Coefficients} spans \emph{one-, two-, and three-dimensional} filter configurations. For \gls{1d} \glspl{nsm}, the evaluation relies on exact computation of the \gls{msed} and the shortest associated error events, as detailed in Appendix~\ref{app:d_min^2 Rate-2 NSM}. For \gls{2d} and \gls{3d} \glspl{nsm}, direct distance analysis becomes intractable, so the assessment is instead performed through projections onto multiple \gls{1d} directions—horizontal, vertical, and diagonal in two dimensions, and both \gls{f2f} and \gls{e2e} in three dimensions. Each projected \gls{1d} \gls{nsm} is analyzed using the \gls{rtf} approach, which reveals the complete set of low-distance error events along with their multiplicities. This richer information enables a filtering process that retains only those higher-dimensional configurations whose \gls{1d} projections all exhibit strong distance properties and low multiplicities.

While the \glspl{rtf} provide valuable insight into the quality of each projected \gls{1d} \gls{nsm}, they cannot fully characterize the performance of complete two- or three-dimensional \glspl{nsm} with rational-valued filter taps. This framework applies only to the \gls{1d} projections and thus cannot directly capture higher-dimensional interference patterns or spatial error-event structures. Consequently, the final validation of these \gls{2d} and \gls{3d} \glspl{nsm} relies on full \gls{ber} simulations. These simulations aim to confirm that the best-performing rational-valued configurations—selected based on the strength and consistency of their projected \gls{1d} behaviors—closely match the \gls{ber} performance of $2$-ASK. This outcome validates the effectiveness of the projection-based selection strategy, despite its indirect nature, for identifying near-optimal rational-tap \glspl{nsm} across dimensions.

\subsection{NSMs with Real Filter Taps}
\label{Rate-2 guaranteeing, minimum Euclidean distance approaching NSMs with real filters' coefficients}

This subsection focuses on rate-$2$ \glspl{nsm} constructed from real-valued shaping filters, developed under both the unconstrained and constrained design paradigms introduced in the preceding section. Each \gls{nsm} is defined by two finite-length real-valued filters: $h_0[k]$, of length $L_0$, and $h_1[k]$, of length $L_1$, subject to the constraint $1 \leq L_1 \leq L_0$. During early stages of the optimization process, filter lengths $L_1 > 1$—specifically $L_1 = 2$, $3$, and $4$—were actively investigated. However, numerical results consistently indicated that the optimal configurations, under equivalent detection complexities, always resulted in $h_1[k]$ having a single non-zero tap. Therefore, in the remainder of this work, we have chosen to restrict our attention to the case $L_1 = 1$, which is adopted for all subsequent designs and analyses. This choice, while potentially limiting for broader scenarios involving larger values of $L_0$ and $L_1$, has the merit of drastically simplifying the \gls{nsm} optimization process, which remains the primary focus of this study.

Building on the optimized filters obtained through numerical search, our analysis aims to identify the specific error events responsible for the \gls{msed} $d_{\min}^2$. These dominant error patterns are extracted directly from the best-performing filter configurations and serve as a foundation for the symbolic resolution phase. By leveraging the structure of these events, we derive closed-form expressions for both the filters and their associated distance metrics, whenever feasible.

This symbolic derivation is carried out for both the unconstrained and constrained design cases. In the unconstrained case, the unequal energy allocation between the filters introduces additional degrees of freedom that can enhance $d_{\min}^2$, albeit at the cost of energy imbalance between the two bipolar symbol streams. In contrast, the constrained case enforces equal energy sharing, $\eta_0 = \eta_1 = 5/2$, which reduces the dimensionality of the search space and streamlines the identification of symbolic solutions.

These structural differences between the two design frameworks translate into distinct behaviors under iterative detection schemes such as turbo-equalization. Unconstrained \glspl{nsm}, due to their typically larger $d_{\min}^2$, offer stronger performance in the early iterations, where interference is most critical. However, as the iterations progress and the receiver becomes capable of distinguishing the two symbol streams $\bar{b}_0[k]$ and $\bar{b}_1[k]$, each stream behaves like an independent $2$-ASK modulation. In the unconstrained case, the unequal energies $\eta_0$ and $\eta_1$ cause the system to mimic a coded $2$-ASK with two distinct energy levels, which ultimately limits convergence to ideal performance. In contrast, the energy symmetry of constrained \glspl{nsm} ensures that both streams contribute equally, allowing the system to approach the performance of a balanced coded $2$-ASK scheme in the final decoding stages.

This dual-track approach underscores how energy allocation strategies fundamentally shape both the structure of the optimal filters and the achievable performance gains. In both design frameworks, symbolic characterization not only confirms the numerical findings but also facilitates structured comparisons with conventional $4$-ASK and $2$-ASK modulations, illustrating how intentional interference and shaping filter design jointly contribute to performance improvements.

Both unconstrained and constrained optimization problems are addressed using a simulated annealing-like metaheuristic framework. In the unconstrained case, however, the process is more computationally demanding due to the variable energy split between the two filters. Specifically, the energy allocation parameter $\eta$ defines the share of the total normalized symbol energy (equal to $5$) between $h_0[k]$ and $h_1[k]$, with $\eta_0 = \eta$ and $\eta_1 = 5 - \eta$. For each value of $L_0 \in \{3, \dots, 10\}$, multiple values of $\eta$ must be explored, and for each $\eta$, a separate optimization over the filter coefficients is performed to identify the configuration that maximizes $d_{\min}^2$. This nested optimization introduces a significant computational overhead. Numerical results reveal that the optimal value of $\eta$ increases with $L_0$, reaching the balanced value $\eta = 5/2$ for the first time at $L_0 = 10$ (with no smaller $L_0$ exhibiting this balance). This monotonic trend enables a progressive narrowing of the $\eta$ search interval as $L_0$ increases, thereby slightly reducing the computational load in higher filter-length regimes.

In contrast, the constrained optimization case eliminates the need to search over $\eta$ by design, as it fixes $\eta_0 = \eta_1 = 5/2$. This simplification reduces the dimensionality of the optimization space and significantly lowers the computational burden. Since the search is conducted in a more compact and uniform space, the optimization procedure tends to be more consistent across multiple runs, avoiding local irregularities caused by varying energy splits. This makes the constrained case especially attractive in practice when dealing with longer filters, where the expanded parameter space of the unconstrained case becomes increasingly costly to explore exhaustively.

Two special configurations depart from the main optimization flow when $L_0 = 2$, both of which are rooted in a reinterpretation of the conventional $4$-ASK modulation as a baseline rate-$2$ \gls{nsm}. In this view, the $4$-ASK signal can be modeled using two shaping filters: $\bm{h}_0 = (1)$ and $\bm{h}_1 = (2)$, resulting in an energy imbalance between the two bipolar input streams in the ratio $1\!:\!4$. To improve performance, particularly in terms of distance properties, one strategy is to \emph{reinforce the energy} of the weaker stream by extending the filter $\bm{h}_0$. This leads to a two-tap configuration, such as $\bm{h}_0 = (1, 1)$ (up to sign changes), which doubles the energy allocated to the first stream and reduces the energy imbalance to $2\!:\!4$. This memory-introducing configuration—interpreted as a form of \emph{energy consolidation}—is explored in the \emph{unconstrained} case with $L_1 = 1$, as discussed in Subsection~\ref{ssec:Modulation of Rate 2}. It was proposed early in the study based on guided reasoning rather than formal optimization, but still illustrates how modest filter extensions can yield improved distance properties. 

In the \emph{constrained} case, a different but similarly motivated approach is examined, in which both filters are extended to length $2$ and designed to have equal energy: for example, $\bm{h}_0 = (2,1)$ and $\bm{h}_1 = (1,2)$, up to sign changes and normalization. This configuration, detailed in Subsection~\ref{app:Closed-form expressions for L0 = L1 = 2} of Appendix~\ref{app:Closed-Form Expressions Optimum Filters Rate-2 NSMs Constrained Optimization}, achieves exact energy balancing between the two input streams and preserves both the average symbol energy and the \gls{msed} of $4$-ASK, after normalization by $\sqrt{2}$. While this balancing requires memory and hence trellis-based detection, it becomes especially valuable when error correction coding and iterative decoding methods—such as turbo-equalization—are applied. Equal energy distribution between the streams facilitates convergence and enables the overall system to closely match the bit error performance of coded $2$-ASK. These two conceptually-driven configurations, though not derived via formal optimization, offer valuable insight into \gls{nsm} behavior and demonstrate how thoughtful filter design can effectively trade complexity for performance gain.

These special configurations not only provide design intuition but also serve as a foundation for the symbolic analysis that follows. Closed-form expressions for the optimal filters and their associated \glspl{msed} are derived, enabling rigorous performance characterization. To estimate the \gls{bep}, upper bounds are constructed using \glspl{rtf} computed from modified state diagrams that capture the memory effects introduced by the filters. These modified state diagrams are analogous to those used in convolutional code analysis and allow for efficient calculation of truncated \glspl{tf}, as detailed in Appendix~\ref{app:Iterative Determination Transfer Function Rate-2 NSM}. In the unconstrained case, the analytical \gls{bep} upper bounds obtained from these \glspl{tf} are compared against \gls{ber} simulations for \glspl{nsm} with $L_0 \in \{3, \dots, 8\}$ and $L_1 = 1$, for which closed-form filter expressions exist. The close agreement between the theoretical bounds and simulation results confirms the validity of both the symbolic framework and the numerical optimization techniques.

\subsubsection*{Numerical Optimization of NSM Filters}

The optimization of filter coefficients for non-degenerate rate-$2$ \glspl{nsm}, whether under unconstrained or constrained settings, proceeds in three main stages:

\begin{enumerate}

    \item \textbf{Random initialization and parameter sampling.} For each pair of filter lengths $(L_0, L_1)$ satisfying $1 \leq L_1 \leq L_0$, multiple random initializations of normalized filter coefficients $\bar{h}_m[k]$, $m=0,1$, are generated. Each candidate configuration is associated with a trial value of the energy allocation parameter $\eta$, governing the partition of the total symbol energy $\eta_0 + \eta_1 = 5$ between $h_0$ and $h_1$.

    \begin{itemize}
    
        \item In the \textit{unconstrained} case, $\eta$ is sampled in the interval $[1, 5/2]$, ensuring that $\eta_0 = \eta \leq \eta_1 = 5 - \eta$ to satisfy $L_0 \geq L_1$ and reflect typical \gls{nsm} asymmetry. To reduce the search complexity and leverage monotonic trends observed in prior optimizations, the range of $\eta$ for a given $L_0$ is restricted to lie between the previously optimized value for $L_0 - 1$ and $5/2$. This is motivated by the theoretical upper bound on $d_{\min}^2$ (cf. Equation~\ref{Upper-Bound Minimum Squared Euclidean Distance}) and the empirical observation that this upper bound is attained for each $L_0$ at a unique value of $\eta$ increasing toward $5/2$ as $L_0$ grows.

        \item In the \textit{constrained} case, energy balancing is imposed by setting $\eta_0 = \eta_1 = 5/2$ and keeping $\eta$ fixed. Only the normalized filter coefficients $\bar{h}_0[k]$ and $\bar{h}_1[k]$ are randomly drawn and optimized.
    
    \end{itemize}
    For each generated candidate, the \gls{msed} $d_{\min}^2$ is estimated using Algorithm~\ref{alg:d_min^2 Rate-2 NSM L_0>1, L_1>1} or Algorithm~\ref{alg:d_min^2 Rate-2 NSM L_0>1, L_1=1}, both presented in Appendix~\ref{app:d_min^2 Rate-2 NSM}, depending on the value of $L_1$.

    \item \textbf{Local refinement of top candidates.} The ten best candidate configurations—ranked by their estimated $d_{\min}^2$—undergo local optimization to further enhance distance properties. In the unconstrained case, this refinement involves both the normalized filter coefficients $\bar{h}_m[k]$ and the energy allocation parameter $\eta$, yielding updated values $\hat{\eta}$ and $\bar{\hat{h}}_m[k]$. In the constrained case, $\eta$ remains fixed at $5/2$, and only the filter coefficients are fine-tuned.
 
    \item \textbf{Error event analysis and \gls{tf} construction.} For each optimized configuration with $L_0 > 2$, the truncated \gls{atf} $\hat{T}(J, N, D; P)$ is computed using Algorithm~\ref{alg:Truncated Augmented Transfer Function T(N,D;P) Rate-2 NSM}, as described in Appendix~\ref{app:Iterative Determination Transfer Function Rate-2 NSM}. The truncation parameter $P$ is chosen to ensure that the \gls{tf} contains multiple terms of the form $J^k N^\ell D^m$, especially around the region corresponding to the estimated \gls{msed} $\hat{d}_{\min}^2$. As the filter length $L_0$ increases, the number of dominant error events near the minimum distance typically grows, which in turn requires larger values of $P$—often on the order of tens—to capture all relevant terms in the truncated \gls{tf}. From this \gls{tf}, the maximum error event length $K_{\max}$ is defined as the largest exponent of $J$ among all terms $J^k N^\ell D^m$ for which $m$ takes values near $\hat{d}_{\min}^2$. This quantity reflects the largest number of modulated symbol intervals spanned by any error event that contributes to the \gls{msed}. Alternatively, $K_{\max}$ can be determined directly by modifying Algorithms~\ref{alg:d_min^2 Rate-2 NSM L_0>1, L_1>1} and~\ref{alg:d_min^2 Rate-2 NSM L_0>1, L_1=1} so that, in addition to computing the \gls{msed}, they also track the largest modulated symbol interval over which any error event achieving $d_{\min}^2$ extends. This dual approach provides a consistent and reliable means of determining the temporal extent of the dominant error patterns, which is critical for the symbolic derivations discussed in the next subsection.

\end{enumerate}

Table~\ref{table:Best Filters Numerical Form Rate-2 NSM} lists, in numerical form, the characteristics of the best fine-tuned rate-$2$ \glspl{nsm} under unconstrained design, for the case $L_1 = 1$ and $L_0$ ranging from $2$ to $10$. The table includes the normalized filter coefficients $\bar{\hat{h}}_0[k]$ of the first filter $\hat{h}_0[k]$, the optimized energy allocation parameter $\hat{\eta}$, and the corresponding \gls{msed} $\hat{d}_{\min}^2$. Performance metrics are also reported, specifically: the gain in dB with respect to conventional $4$-ASK and the gap in dB with respect to $2$-ASK. It is worth noting that the gain and gap in dB for each \gls{nsm} configuration add up to the fixed gap between $2$-ASK and $4$-ASK, given by $10 \log_{10}(5/2) \approx 3.9794$~dB.

\begin{table}[H]
\caption{Numerical characteristics of fine-tuned Rate-$2$ NSMs, with $L_0$ from $2$ to $10$ ($L_1=1$ and $\bar{\hat{h}}_1[k]=\delta[k]$), under Unconstrained Optimization.}
\label{table:Best Filters Numerical Form Rate-2 NSM}
\centering
\begin{tabular}{|c||c|c|c|} 
\hline
$L_0$ & $2$ & $3$ & $4$ \\ \hline
$\bar{\hat{h}}_0[0]$ & $0.707106781186547$ & $0.653281482438188$ & $0.627963030199554$ \\
$\bar{\hat{h}}_0[1]$ & $0.707106781186547$ & $0.382683432365090$ & $0.398112608509063$ \\
$\bar{\hat{h}}_0[2]$ & - & $0.653281482438188$ & $0.229850421690492$ \\
$\bar{\hat{h}}_0[3]$ & - & - & $0.627963030199554$ \\ \hline
$\hat{\eta}$ & $1.666666666666667$ & $1.846990312590646$ & $1.939976905650510$ \\ \hline
$\hat{d}_{\text{min}}^2$ & $6.666666666666667
$ & $7.387961250362585$ & $7.759907622602041$ \\ \hline
Gain [dB] & $2.218487496163564$ & $2.664646175845513$ & $2.877965599259757$ \\ \hline
Gap [dB] & 1.760912590556812 & $1.314753910874863$ & $1.101434487460618$ \\ \hline
$K_{\text{max}}$ & $\infty$ & $4$ & $8$ \\ [0.5ex] 
 \hline\hline
$L_0$ & $5$ & $6$ & $7$ \\ \hline
$\bar{\hat{h}}_0[0]$ & $0.589216898252358$ & $0.554700196225229$ & $0.542805974620653$ \\
$\bar{\hat{h}}_0[1]$ & $0.498472935623522$ & $0$ & $0.542805974620653$ \\
$\bar{\hat{h}}_0[2]$ & $0$ & $0.277350098112615$ & $0$ \\
$\bar{\hat{h}}_0[3]$ & $0.239105888841342$ & $0.554700196225229$ & $0.284912667360959$ \\
$\bar{\hat{h}}_0[4]$ & $0.589216898252358$ & $0$ & $0$ \\
$\bar{\hat{h}}_0[5]$ & - & $0.554700196225229$ & $0.186841627389698$ \\
$\bar{\hat{h}}_0[6]$ & - & - & $0.542805974620653$ \\ \hline
$\hat{\eta}$ & $2.093183319624400$ & $2.241379310344827$ & $2.295101060896494$ \\ \hline
$\hat{d}_{\text{min}}^2$ & $8.372733278497600$ & $8.965517241379310$ & $9.180404243585976$ \\ \hline
Gain [dB] & $3.208072652306588$ & $3.505153587438994$ & $3.608018137178788$ \\ \hline
Gap [dB] & $0.771327434413788$ & $0.474246499281382$ & $0.371381949541588$ \\ \hline
$K_{\text{max}}$ & $11$ & $12$ & $14$ \\ [0.5ex]
\hline\hline
 
$L_0$ & $8$ & $9$ & $10$ \\ \hline
$\bar{\hat{h}}_0[0]$ & $0.516339056518458$ & $0.504766583589045$ & $0.498560784768823$ \\
$\bar{\hat{h}}_0[1]$ & $0.269675949396294$ & $0.112573847375572$ & $0.081935655036829$ \\
$\bar{\hat{h}}_0[2]$ & $0.516339056518458$ & $-0.250553050832785$ & $-0.255744209387574$ \\
$\bar{\hat{h}}_0[3]$ & $0$ & $-0.287654680029553$ & $0.248296837518889$ \\
$\bar{\hat{h}}_0[4]$ & $0$ & $0.475333192855289$ & $0.477569733208884$ \\
$\bar{\hat{h}}_0[5]$ & $0$ & $0.223474387842203$ & $-0.265066957307874$ \\
$\bar{\hat{h}}_0[6]$ & $0.357010950053487$ & $-0.230910865105507$ & $-0.096128146262105$ \\
$\bar{\hat{h}}_0[7]$ & $0.516339056518458$ & $-0.040831369188616$ & $0.157001189616759$ \\
$\bar{\hat{h}}_0[8]$ & - & $-0.504866577978142$ & $-0.504766583589045$ \\
$\bar{\hat{h}}_0[9]$ & - & - & $0.497101127210722$ \\ \hline
$\hat{\eta}$ & $2.419638851454818$ & $2.476280678537178$ & $2.5$ \\ \hline
$\hat{d}_{\text{min}}^2$ & $9.678555405819273
$ & $9.905122714148712$ & $10$ \\ \hline
Gain [dB] & $3.837505492346264$ & $3.937998690360034$ & $\bm{3.979400086720376}$ \\ \hline
Gap [dB] & $0.141894594374111$ & $0.041401396360342$ & $\bm{0}$ \\ \hline
$K_{\text{max}}$ & $17$ & $27$ & - \\ [1ex] 
\hline
\end{tabular}
\end{table}

The table also allows reconstruction of the weighted filter versions using the relationships $\hat{h}_0[k] = \sqrt{\hat{\eta}}\, \bar{\hat{h}}_0[k]$ and $\hat{h}_1[k] = \sqrt{5 - \hat{\eta}}\, \bar{\hat{h}}_1[k]$. In this specific case $L_1 = 1$, and the second filter reduces to a scaled unit impulse, so that $\bar{\hat{h}}_1[k] = \delta[k]$ and hence $\hat{h}_1[k] = \sqrt{5 - \hat{\eta}}\, \delta[k]$.

Before continuing, it is important to highlight a key asymptotic observation. As $L_0$ increases, the gap in dB with respect to $2$-ASK consistently decreases and eventually vanishes for the first time at $L_0 = 10$. In the optimization of unconstrained rate-$2$ \glspl{nsm}, the filter energies are generally allowed to be unbalanced, providing greater degrees of freedom. For $L_0 < 10$, the optimized filters always exhibit some degree of energy imbalance, and the \gls{msed} of $2$-ASK is never achieved. By contrast, at $L_0 = 10$, energy balance is attained for the first time, and simultaneously the \gls{msed} of $2$-ASK is achieved. These observations align perfectly with the theoretical upper bound on the \gls{msed} discussed in \ref{Upper-Bound Minimum Squared Euclidean Distance}, which explicitly depends on the energy imbalance. This bound establishes that achieving the \gls{msed} of $2$-ASK requires perfect energy balancing between the filters—a condition that is both necessary and sufficient for the bound to coincide exactly with the \gls{msed} of $2$-ASK.

We have also examined the constrained rate-$2$ \glspl{nsm}, in which the filter energies are explicitly enforced to be balanced. This constraint naturally limits the solution space and reduces the flexibility available during optimization. In the unconstrained case, we observed that the optimal solution achieving the \gls{msed} of $2$-ASK is already energy-balanced at $L_0 = 10$. This finding immediately implies that the constrained (energy-balanced) formulation can also reach the same asymptotic performance at this length. But notice that, for any value of $L_0$, the constrained case cannot outperform the unconstrained one, since the additional energy-balancing requirement restricts the feasible filter set. Consequently, the balanced configuration also attains the asymptotic performance of $2$-ASK for the first time at $L_0 = 10$, in the same way as the unconstrained case.

This is a significant result: it shows that a single-stream \gls{nsm} with carefully optimized real-valued filters can asymptotically match the distance performance of a standard $2$-ASK modulation, without resorting to the complexity of advanced signaling techniques such as those described in~\cite{Rusek09}. In particular, \gls{ftn} signaling, as studied in~\cite{Rusek09}, achieves a similar asymptotic goal but does so using multistream architectures and iterative detection techniques such as turbo-equalization, often combined with forward error correction coding.

\subsubsection*{Identification of Minimum-Distance Error Events} \label{Identification of Minimum-Distance Error Events}

We now proceed to identify the input sequence differences that correspond to the \gls{msed} between transmitted modulated symbol sequences. This analysis is central to understanding the dominant error events that impact system performance. In the unconstrained case, the filters and the parameter $\hat{\eta}$ have been jointly optimized in the previous steps to maximize this distance. In contrast, in the constrained case, $\hat{\eta}$ was fixed to the value $5/2$, and the filter design was carried out under this constraint. For both cases, the maximum error event length $K_{\max}$, which corresponds to the longest error events achieving the \gls{msed}, has been assessed for each of the best optimized \glspl{nsm}.

Although all optimized \glspl{nsm} of interest ultimately correspond to the case $L_1 = 1$, we begin by considering the general setting where $L_0 > 1$ and $L_1 \ge 1$, in order to present a complete and flexible approach. We then specialize the discussion to the practically relevant case $L_1 = 1$, which admits significant simplifications and encompasses all optimized configurations retained in this study.

To begin, we define the $K_{\max} \times (K_{\max} - L_m + 1)$ Sylvester matrices associated with the weighted filters:
\begin{equation}
\hat{\bm{H}}_m \triangleq
    \begin{pmatrix}
    \hat{h}_m[0] & 0 & \cdots & 0 & 0 \\
    \hat{h}_m[1] & \hat{h}_m[0] & \ddots & \vdots & \vdots \\
    \vdots & \hat{h}_m[1] & \ddots & 0 & 0 \\
    \hat{h}_m[L_m-2] & \vdots & \ddots & \hat{h}_m[0] & 0 \\
    \hat{h}_m[L_m-1] & \hat{h}_m[L_m-2] & \vdots & \hat{h}_m[1] & \hat{h}_m[0] \\
    0 & \hat{h}_m[L_m-1] & \ddots & \vdots & \hat{h}_m[1] \\
    0 & 0 & \ddots & \hat{h}_m[L_m-2] & \vdots \\
    \vdots & \vdots & \ddots & \hat{h}_m[L_m-1] & \hat{h}_m[L_m-2] \\
    0 & 0 & \cdots & 0 & \hat{h}_m[L_m-1]
    \end{pmatrix},
\end{equation}
for $m = 0, 1$. As in earlier sections, the weighted filters are defined by $\hat{h}_0[k] = \sqrt{\hat{\eta}}\, \bar{\hat{h}}_0[k]$ and $\hat{h}_1[k] = \sqrt{5 - \hat{\eta}}\, \bar{\hat{h}}_1[k]$.

Next, we define the $(K_{\max} - L_m + 1) \times 1$ input difference vectors:
\begin{equation}
\Delta \bar{\bm{b}}_m \triangleq (\Delta \bar{b}_m[0], \Delta \bar{b}_m[1], \ldots, \Delta \bar{b}_m[K_{\max} - L_m])^T,
\end{equation}
for $m = 0, 1$, where $(\cdot)^T$ denotes the transpose operator. The components $\Delta \bar{b}_m[n]$ belong to the set $\{0, \pm 2\}$ and represent symbol-wise differences between two distinct pairs of bipolar input sequences used to evaluate the \gls{sed} between their corresponding output modulated signals.

We begin with the general case where $L_0 > 1$ and $L_1 \ge 1$, for which the estimated output difference vector, based on the modulation expression in~(\ref{eq:Mod Seq 4-ASK}), is given by
\begin{equation} \label{eq:Output Differences Vector L_0 > 1 L_1 >= 1}
\Delta \hat{\bm{s}} = \hat{\bm{H}}_0 \Delta \bar{\bm{b}}_0 + \hat{\bm{H}}_1 \Delta \bar{\bm{b}}_1.
\end{equation}

For the determination of error events with minimum Euclidean distance, we consider all pairs of input difference vectors $(\Delta \bar{\bm{b}}_0, \Delta \bar{\bm{b}}_1)$ for which at least $\Delta \bar{b}_0[0]$ or $\Delta \bar{b}_1[0]$ is not null. This constraint avoids unnecessary expansion of the Sylvester matrices when both vectors begin with leading zeros.

To reduce the number of pairs under consideration, we observe that the pairs $(\Delta \bar{\bm{b}}_0, \Delta \bar{\bm{b}}_1)$ and their opposites $(-\Delta \bar{\bm{b}}_0, -\Delta \bar{\bm{b}}_1)$ yield the same Euclidean distance. Therefore, it suffices to assume that at least one of the first components, $\Delta \bar{b}_0[0]$ or $\Delta \bar{b}_1[0]$, is equal to $+2$.

In addition to their opposites, all discovered pairs that result in an estimated Euclidean distance $\| \Delta \hat{\bm{s}} \|$ which is within a tolerance $\varepsilon$ of the estimated minimum Euclidean distance $\hat{d}_{\min}^2$ are considered to correspond to error events of the true minimum Euclidean distance $d_{\min}^2$. The accuracy parameter $\varepsilon$ is selected in a way that separates true minimum-distance events from those associated with larger Euclidean distances.

We now turn to the special case $L_1 = 1$, for which the analysis becomes significantly more tractable. In this setting, the filter $\hat{h}_1[k]$ reduces to a scaled unit impulse, that is, $\hat{h}_1[k] = \hat{h}_1[0] \delta[k]$, and the output difference vector simplifies to
\begin{equation} \label{eq:simplified_output_diff}
\Delta \hat{\bm{s}} = \hat{\bm{H}}_0 \Delta \bar{\bm{b}}_0 + \hat{h}_1[0] \Delta \bar{\bm{b}}_1,
\end{equation}
where $\Delta \bar{\bm{b}}_1$ is now a $K_{\max} \times 1$ vector with entries in $\{0, \pm 2\}$.

A key observation concerns error events for which $\Delta \bar{\bm{b}}_0$ is the null vector. In this case, the output difference simplifies to $\Delta \hat{\bm{s}} = \hat{h}_1[0]\, \Delta \bar{\bm{b}}_1$. The smallest possible \gls{sed} under this condition arises when all components of $\Delta \bar{\bm{b}}_1$ are zero except the first, which must be equal to $2$ in order to satisfy the requirement that not both first components of $\Delta \bar{\bm{b}}_0$ and $\Delta \bar{\bm{b}}_1$ are zero. This yields $\| \Delta \hat{\bm{s}} \|^2 = 4(\hat{h}_1[0])^2 = 4(5 - \hat{\eta})$. On the other hand, the \gls{sed} of any error event is upper-bounded by $4\hat{\eta}$, as established in~(\ref{Upper-Bound Minimum Squared Euclidean Distance}). When $\hat{\eta} < 5/2$---which occurs in the unconstrained case for $L_0 < 10$---we have $4\hat{\eta} < 4(5 - \hat{\eta})$. Since $\hat{\eta} < 5/2$, the bound $4\hat{\eta}$ is always strictly less than $4(5 - \hat{\eta})$, confirming that the \gls{msed} cannot be attained unless $\Delta \bar{\bm{b}}_0$ is non-null. Therefore, all minimum-distance error events in this regime necessarily involve nonzero contributions from $\Delta \bar{\bm{b}}_0$.

In contrast, when $\hat{\eta} = 5/2$---as is the case in the constrained setting (for all values of $L_0$) and in the unconstrained setting when $L_0 \ge 10$---the inequality becomes an equality. In this situation, the configuration where $\Delta \bar{\bm{b}}_0$ is null and $\Delta \bar{\bm{b}}_1$ is a vector whose first component is $\pm 2$ and the rest are zero does attain the \gls{msed}. However, this particular configuration corresponds to a trivial input structure and is not instrumental in the symbolic determination of the closed-form expressions of the filters. Consequently, even when $\hat{\eta} = 5/2$, only the minimum-distance error events for which $\Delta \bar{\bm{b}}_0$ is non-null are useful for the analysis. This leads to the same conclusion as in the case $\hat{\eta} < 5/2$: for the purpose of deriving closed-form filter expressions, it suffices to consider only those error events for which $\Delta \bar{\bm{b}}_0$ is a nonzero vector. Accordingly, we focus exclusively on this subset of error events, which also happens to be the more involved case from a computational standpoint.

We begin by drawing a vector $\Delta \bar{\bm{b}}_0$ with components in $\{0, \pm 2\}$ such that $\Delta \bar{b}_0[0] = +2$, and compute its partial contribution to the output difference vector:
\begin{equation} \label{eq:Partial Output Differences Vector L_0 > 1 L_1 = 1}
\Delta \hat{\bm{s}}_0 \triangleq \hat{\bm{H}}_0 \Delta \bar{\bm{b}}_0.
\end{equation}
We then search for the vector $\Delta \bar{\bm{b}}_1$ that minimizes the total \gls{sed} $\| \Delta \hat{\bm{s}} \|^2$, where the full output difference vector is given by
\begin{equation} \label{eq:Output Differences Vector L_0 > 1 L_1 = 1}
\Delta \hat{\bm{s}} = \Delta \hat{\bm{s}}_0 + \hat{h}_1[0] \Delta \bar{\bm{b}}_1.
\end{equation}

Since the Sylvester matrix $\hat{\bm{H}}_1$ reduces to a scalar $\hat{h}_1[0]$, this minimization simplifies significantly and decouples across the components of the vector. It becomes a straightforward aggregated, component-wise minimization:
\begin{equation} \label{eq:Partial Minimization}
m(\Delta \bar{\bm{b}}_0) \triangleq \min_{\Delta \bar{\bm{b}}_1 \in \{0, \pm 2\}^{K_{\text{max}}}} \| \Delta \hat{\bm{s}} \|^2 = \sum_{k=0}^{K_{\text{max}} - 1} \min_{\Delta \bar{b}_1[k] \in \{0, \pm 2\}} \left( \Delta \hat{s}_0[k] + \hat{h}_1[0] \Delta \bar{b}_1[k] \right)^2,
\end{equation}
where $\Delta \hat{s}_0[k]$ is the $k$-th component of the vector $\Delta \hat{\bm{s}}_0$.

This formulation offers a major computational benefit: the partial minimization in \ref{eq:Partial Minimization} has a linear complexity in $K_{\text{max}}$. Moreover, this complexity can be further reduced by limiting the summation to the nonzero components of $\Delta \hat{\bm{s}}_0$.

The metric $m(\Delta \bar{\bm{b}}_0)$ is evaluated for all candidate vectors $\Delta \bar{\bm{b}}_0 \in \{0, \pm 2\}^{K_{\text{max}} - L_0 + 1}$ with the first component fixed to $\Delta \bar{b}_0[0] = +2$. Vectors for which $\Delta \bar{b}_0[0] = -2$ do not need to be considered, since their opposites, $-\Delta \bar{\bm{b}}_0$, yield the same \gls{sed}.

All vectors $\Delta \bar{\bm{b}}_0$ that produce a \gls{sed} within a margin $\varepsilon$ of the estimated \gls{msed} are included in the set of first input difference vectors corresponding to error events with minimum Euclidean distance. If $\Delta \bar{\hat{\bm{b}}}_0$ is one such vector, then the associated second input difference vector $\Delta \bar{\hat{\bm{b}}}_1$ is determined such that each of its components $\Delta \bar{\hat{b}}_1[k]$ satisfies
\begin{equation}
(\Delta \hat{s}_0[k] + \hat{h}_1[0] \Delta \bar{\hat{b}}_1[k])^2 \le \min_{\Delta \bar{b}_1[k] \in \{0, \pm 2 \}} (\Delta \hat{s}_0[k] + \hat{h}_1[0] \Delta \bar{b}_1[k])^2 + \varepsilon.
\end{equation}
As a result, some components of $\Delta \bar{\hat{\bm{b}}}_1$ may admit more than one value in $\{0, \pm 2\}$. In practice, when this ambiguity occurs, the values are typically either $\{0, +2\}$ or $\{0, -2\}$.

Tables~\ref{table:Input Differences Vectors L_0=3 L_1=1}--\ref{table:Input Differences Vectors L_0=9 L_1=1} list the minimum-distance error events identified for the \textit{unconstrained} optimization setting, with $L_0$ ranging from $3$ to $9$. Each table provides the index $n$ of each event and the corresponding vectors $\Delta \bar{\hat{\bm{b}}}_0^n$ and $\Delta \bar{\hat{\bm{b}}}_1^n$, which represent the symbol-wise input differences giving rise to the smallest estimated \glspl{sed}. A similar collection of results is presented in Tables~\ref{table:Input Differences Vectors L_0=3 L_1=1 Constrained Optimization}--\ref{table:Input Differences Vectors L_0=6 L_1=1 Constrained Optimization}, which detail the most significant error events obtained under \textit{constrained} optimization for $L_0 = 3$ through $6$. A reduced list is included in Table~\ref{table:Input Differences Vectors L_0=7 L_1=1 Constrained Optimization} for $L_0 = 7$, as the corresponding optimal filter $h_0[k]$ in this case admits a straightforward symbolic representation.

\begin{table}[!htbp]
\caption{Input differences vectors of error events, with minimum Euclidean distance, for $L_0=3$ and $L_1=1,$ under unconstrained optimization ($K_{\text{max}} = 4$).}
\label{table:Input Differences Vectors L_0=3 L_1=1}
\centering
\begin{tabular}{|c|c|c|} 
 \hline
 $n$ & $\Delta \bar{\hat{\bm{b}}}_0^n$ & $\Delta \bar{\hat{\bm{b}}}_1^n$ \\ [0.5ex] 
 \hline\hline
 $0-7$ & $\pm (2,-2)$ & $\pm (-2\lor0,0,0,2\lor0)$ \\ \hline
 $8-15$ & $\pm (2)$ & $\pm (-2\lor0,0,-2\lor0)$ \\ \hline
$16-23$ & $\pm (2,2)$ & $\pm (-2\lor0,-2,-2,-2\lor0)$  \\ [1ex] 
 \hline
\end{tabular}
\end{table}

\begin{table}[!htbp]
\caption{Input differences vectors of error events, with minimum Euclidean distance, for $L_0=4$ and $L_1=1,$ under unconstrained optimization ($K_{\text{max}} = 8$).}
\label{table:Input Differences Vectors L_0=4 L_1=1}
\centering
\begin{tabular}{|c|c|c|} 
 \hline
 $n$ & $\Delta \bar{\hat{\bm{b}}}_0^n$ & $\Delta \bar{\hat{\bm{b}}}_1^n$ \\ [0.5ex] 
 \hline\hline
 $0-7$ & $\pm (2)$ & $\pm (-2\lor0,0,0,-2\lor0)$ \\ \hline
 $8-15$ & $\pm (2,2,2,0,-2)$ & $\pm (-2\lor0,-2,-2,-2,0,0,0,2\lor0)$ \\ \hline
$16-23$ & $\pm (2,2,2)$ & $\pm (-2\lor0,-2,-2,-2,-2,-2\lor0)$  \\
\hline
$24-31$ & $\pm (2,2,2,0,2)$ & $\pm (-2\lor0,-2,-2,-2,-2,-2,0,-2\lor0)$  \\ [1ex] 
 \hline
\end{tabular}
\end{table}

\begin{table}[!htbp]
\caption{Input differences vectors of error events, with minimum Euclidean distance, for $L_0=5$ and $L_1=1,$ under unconstrained optimization ($K_{\text{max}} = 11$).}
\label{table:Input Differences Vectors L_0=5 L_1=1}
\centering
\begin{tabular}{|c|c|c|} 
 \hline
 $n$ & $\Delta \bar{\hat{\bm{b}}}_0^n$ & $\Delta \bar{\hat{\bm{b}}}_1^n$ \\ [0.5ex] 
 \hline\hline
 $0-7$ & $\pm (2,-2,2,-2,0,0,-2)$ & $\pm (-2\lor0,0,0,0,0,0,0,2,0,0,2\lor0)$ \\ \hline
 $8-15$ & $\pm (2,-2,2,-2)$ & $\pm (-2\lor0,0,0,0,0,0,0,2\lor0)$ \\ \hline
$16-23$ & $\pm (2,-2,2,-2,0,0,2)$ & $\pm (-2\lor0,0,0,0,0,0,-2,0,0,0,-2\lor0)$  \\
\hline
$24-31$ & $\pm (2)$ & $\pm (-2\lor0,0,0,0,-2\lor0)$  \\ \hline
$32-39$ & $\pm (2,2,2,2,0,0,-2)$ & $\pm (-2\lor0,-2,-2,-2,-2,-2,0,0,0,0,2\lor0)$  \\ \hline
$40-47$ & $\pm (2,2,2,2)$ & $\pm (-2\lor0,-2,-2,-2,-2,-2,-2,-2\lor0)$  \\ \hline
$48-55$ & $\pm (2,2,2,2,0,0,2)$ & $\pm (-2\lor0,-2,-2,-2,-2,-2,-2,-2,0,0,-2\lor0)$  \\ [1ex] 
 \hline
\end{tabular}
\end{table}

\begin{table}[!htbp]
\caption{Input differences vectors of error events, with minimum Euclidean distance, for $L_0=6$ and $L_1=1,$ under unconstrained optimization ($K_{\text{max}} = 12$).}
\label{table:Input Differences Vectors L_0=6 L_1=1}
\centering
\begin{tabular}{|c|c|c|} 
 \hline
 $n$ & $\Delta \bar{\hat{\bm{b}}}_0^n$ & $\Delta \bar{\hat{\bm{b}}}_1^n$ \\ [0.5ex] 
 \hline\hline
$0-7$ & $\pm (2,0,-2,-2,-2,0,-2)$ & $\pm (-2\lor0,0,0,0,2,0,2,2,2,2,0,2\lor0)$ \\ \hline
 $8-15$ & $\pm (2,0,-2,-2,-2,0,2)$ & $\pm (-2\lor0,0,0,0,2,0,0,2,0,0,0,-2\lor0)$ \\ \hline
$16-23$ & $\pm (2,0,-2,-2,2,0,-2)$ & $\pm (-2\lor0,0,0,0,0,0,2,0,2,0,0,2\lor0)$  \\
\hline
 $24-31$ & $\pm (2,0,-2,-2,2,0,2)$ & $\pm (-2\lor0,0,0,0,0,0,0,0,0,-2,0,-2\lor0)$  \\ \hline
$32-39$ & $\pm (2,0,-2,2,-2,0,-2)$ & $\pm (-2\lor0,0,0,-2,2,0,0,2,0,2,0,2\lor0)$  \\ \hline
 $40-47$ & $\pm (2,0,-2,2,-2,0,2)$ & $\pm (-2\lor0,0,0,-2,2,0,-2,2,-2,0,0,-2\lor0)$  \\ \hline
$48-55$ & $\pm (2,0,-2,2,2,0,-2)$ & $\pm (-2\lor0,0,0,-2,0,0,0,0,0,0,0,2\lor0)$  \\ \hline
 $56-63$ & $\pm (2,0,-2,2,2,0,2)$ & $\pm (-2\lor0,0,0,-2,0,0,-2,0,-2,-2,0,-2\lor0)$  \\ \hline
$64-79$ & $\pm (2)$ & $\pm (-2\lor0,0,0,-2\lor0,0,-2\lor0)$  \\ \hline
 $80-87$ & $\pm (2,0,2,-2,-2,0,-2)$ & $\pm (-2\lor0,0,-2,0,0,-2,2,0,2,2,0,2\lor0)$  \\ \hline
$88-95$ & $\pm (2,0,2,-2,-2,0,2)$ & $\pm (-2\lor0,0,-2,0,0,-2,0,0,0,0,0,-2\lor0)$  \\ \hline
 $96-103$ & $\pm (2,0,2,-2,2,0,-2)$ & $\pm (-2\lor0,0,-2,0,-2,-2,2,-2,2,0,0,2\lor0)$  \\ \hline
$104-111$ & $\pm (2,0,2,-2,2,0,2)$ & $\pm (-2\lor0,0,-2,0,-2,-2,0,-2,0,-2,0,-2\lor0)$  \\ \hline
 $112-119$ & $\pm (2,0,2,2,-2,0,-2)$ & $\pm (-2\lor0,0,-2,-2,0,-2,0,0,0,2,0,2\lor0)$  \\ \hline
$120-127$ & $\pm (2,0,2,2,-2,0,2)$ & $\pm (-2\lor0,0,-2,-2,0,-2,-2,0,-2,0,0,-2\lor0)$  \\ \hline
 $128-135$ & $\pm (2,0,2,2,2,0,-2)$ & $\pm (-2\lor0,0,-2,-2,-2,-2,0,-2,0,0,0,2\lor0)$  \\ \hline
$136-143$ & $\pm (2,0,2,2,2,0,2)$ & $\pm (-2\lor0,0,-2,-2,-2,-2,-2,-2,-2,-2,0,-2\lor0)$  \\ [1ex] 
 \hline
\end{tabular}
\end{table}

\begin{table}[!htbp]%
\caption{Input differences vectors of error events, with minimum Euclidean distance, for $L_0=7$ and $L_1=1,$ under unconstrained optimization ($K_{\text{max}} = 14$).}
\label{table:Input Differences Vectors L_0=7 L_1=1}
\centering
\begin{tabular}{|c|c|c|} 
 \hline
 $n$ & $\Delta \bar{\hat{\bm{b}}}_0^n$ & $\Delta \bar{\hat{\bm{b}}}_1^n$ \\ [0.5ex] 
 \hline\hline
$0-7$ & $\pm (2,-2,-2,0,-2,-2)$ & $\pm (-2\lor0,0,2,0,2,2,0,2,2,0,2,2\lor0)$ \\ \hline
 $8-15$ & $\pm (2,-2,-2,0,-2,2)$ & $\pm (-2\lor0,0,2,0,2,0,-2,2,0,0,0,-2\lor0)$ \\ \hline
$16-23$ & $\pm (2,-2,-2)$ & $\pm (-2\lor0,0,2,0,0,0,0,2,2\lor0)$  \\ \hline
 $24-31$ & $\pm (2,-2,2,0,-2,0,0,-2)$ & $\pm (-2\lor0,0,0,-2,2,0,0,2,0,0,2,0,0,2\lor0)$  \\ \hline
$32-39$ & $\pm (2,-2,2,0,-2,0,0,2)$ & $\pm (-2\lor0,0,0,-2,2,0,0,0,-2,0,0,0,0,-2\lor0)$  \\ \hline
 $40-55$ & $\pm (2)$ & $\pm (-2\lor0,-2\lor0,0,0,0,0,-2\lor0)$  \\ \hline
$56-63$ & $\pm (2,2,-2,0,-2,-2)$ & $\pm (-2\lor0,-2,0,0,0,2,0,0,2,0,2,2\lor0)$  \\ \hline
 $64-71$ & $\pm (2,2,-2,0,-2,2)$ & $\pm (-2\lor0,-2,0,0,0,0,-2,0,0,0,0,-2\lor0)$  \\ \hline
$72-79$ & $\pm (2,2,-2)$ & $\pm (-2\lor0,-2,0,0,0,0,-2,0,2\lor0)$  \\ \hline
 $80-87$ & $\pm (2,2,2,0,-2,0,0,-2)$ & $\pm (-2\lor0,-2,-2,-2,0,0,-2,0,0,0,2,0,0,2\lor0)$  \\ \hline
$88-95$ & $\pm (2,2,2,0,-2,0,0,2)$ & $\pm (-2\lor0,-2,-2,-2,0,0,-2,-2,-2,0,0,0,0,-2\lor0)$  \\ [1ex] 
\hline
\end{tabular}
\end{table}

\begin{table}[!htbp]%
\caption{Input differences vectors of error events, with minimum Euclidean distance, for $L_0=8$ and $L_1=1,$ under unconstrained optimization ($K_{\text{max}} = 17$).}
\label{table:Input Differences Vectors L_0=8 L_1=1}
\small
\centering
\begin{tabular}{|c|c|c|}
 \hline 
$n$ & $\Delta \bar{\hat{\bm{b}}}_0^n$ & $\Delta \bar{\hat{\bm{b}}}_1^n$ \\ [0.5ex] 
 \hline\hline
$0-7$ & $\pm (2,-2,-2,-2,-2)$ & $\pm (-2\lor0,0,0,2,2,2,0,0,2,2,2,2\lor0)$ \\ \hline
$8-15$ & $\pm (2,-2,-2,-2,2)$ & $\pm (-2\lor0,0,0,2,0,0,-2,0,2,2,0,-2\lor0)$ \\ \hline
$16-23$ & $\pm (2,-2,-2,2,-2)$ & $\pm (-2\lor0,0,0,0,2,0,0,0,2,0,0,2\lor0)$  \\ \hline
$24-31$ & $\pm (2,-2,-2,2,2)$ & $\pm (-2\lor0,0,0,0,0,-2,-2,0,2,0,-2,-2\lor0)$  \\ \hline
$32-39$ & $\pm (2,-2,0,-2,-2,0,0,0,-2,2)$ & $\pm (-2\lor0,0,0,2,2,2,0,0,2,0,2,0,0,0,0,0,-2\lor0)$  \\ \hline
$40-47$ & $\pm (2,-2,0,-2,2,0,0,0,-2,-2)$ & $\pm (-2\lor0,0,0,2,0,0,-2,0,2,2,2,0,0,0,0,2,2\lor0)$  \\ \hline
$48-55$ & $\pm (2,-2,0,2,-2,0,0,0,2,2)$ & $\pm (-2\lor0,0,0,0,0,0,0,0,0,-2,-2,0,0,0,0,-2,-2\lor0)$  \\ \hline
$56-63$ & $\pm (2,-2,0,2,2,0,0,0,2,-2)$ & $\pm (-2\lor0,0,0,0,-2,-2,-2,0,0,0,-2,0,0,0,0,0,2\lor0)$  \\ \hline
$64-71$ & $\pm (2,-2,2,-2,-2)$ & $\pm (-2\lor0,0,-2,2,0,2,0,0,0,0,2,2\lor0)$  \\ \hline
$72-79$ & $\pm (2,-2,2,-2,2)$ & $\pm (-2\lor0,0,-2,2,-2,0,-2,0,0,0,0,-2\lor0)$  \\ \hline
$80-87$ & $\pm (2,-2,2,2,-2)$ & $\pm (-2\lor0,0,-2,0,0,0,0,0,0,-2,0,2\lor0)$  \\ \hline
$88-95$ & $\pm (2,-2,2,2,2)$ & $\pm (-2\lor0,0,-2,0,-2,-2,-2,0,0,-2,-2,-2\lor0)$  \\ \hline
$96-111$ & $\pm (2)$ & $\pm (-2\lor0,0,-2\lor0,0,0,0,0,-2\lor0)$  \\ \hline
$112-119$ & $\pm (2,2,-2,-2,-2)$ & $\pm (-2\lor0,-2,0,0,2,2,0,-2,0,2,2,2\lor0)$  \\ \hline
$120-127$ & $\pm (2,2,-2,-2,2)$ & $\pm (-2\lor0,-2,0,0,0,0,-2,-2,0,2,0,-2\lor0)$  \\ \hline
$128-135$ & $\pm (2,2,-2,2,-2)$ & $\pm (-2\lor0,-2,0,-2,2,0,0,-2,0,0,0,2\lor0)$  \\ \hline
$136-143$ & $\pm (2,2,-2,2,2)$ & $\pm (-2\lor0,-2,0,-2,0,-2,-2,-2,0,0,-2,-2\lor0)$  \\ \hline
$144-151$ & $\pm (2,2,0,-2,-2,0,0,0,-2,2)$ & $\pm (-2\lor0,-2,-2,0,2,2,0,-2,0,0,2,0,0,0,0,0,-2\lor0)$  \\ \hline
$152-159$ & $\pm (2,2,0,-2,2,0,0,0,-2,-2)$ & $\pm (-2\lor0,-2,-2,0,0,0,-2,-2,0,2,2,0,0,0,0,2,2\lor0)$  \\ \hline
 $160-167$ & $\pm (2,2,0,2,-2,0,0,0,2,2)$ & $\pm (-2\lor0,-2,-2,-2,0,0,0,-2,-2,-2,-2,0,0,0,0,-2,-2\lor0)$  \\ \hline
 $168-175$ & $\pm (2,2,0,2,2,0,0,0,2,-2)$ & $\pm (-2\lor0,-2,-2,-2,-2,-2,-2,-2,-2,0,-2,0,0,0,0,0,2\lor0)$  \\ \hline
 $176-183$ & $\pm (2,2,2,-2,-2)$ & $\pm (-2\lor0,-2,-2,0,0,2,0,-2,-2,0,2,2\lor0)$  \\ \hline
 $184-191$ & $\pm (2,2,2,-2,2)$ & $\pm (-2\lor0,-2,-2,0,-2,0,-2,-2,-2,0,0,-2\lor0)$  \\ \hline
 $192-199$ & $\pm (2,2,2,2,-2)$ & $\pm (-2\lor0,-2,-2,-2,0,0,0,-2,-2,-2,0,2\lor0)$  \\ \hline
$200-207$ & $\pm (2,2,2,2,2)$ & $\pm (-2\lor0,-2,-2,-2,-2,-2,-2,-2,-2,-2,-2,-2\lor0)$  \\ [1ex] 
\hline
\end{tabular}
\end{table}

\begin{table}[!htbp]
\caption{Input differences vectors of error events, with minimum Euclidean distance, for $L_0=9$ and $L_1=1,$ under unconstrained optimization ($K_{\text{max}} = 27$).}
\label{table:Input Differences Vectors L_0=9 L_1=1}
\small
\centering
\begin{tabular}{|c|c|c|}
 \hline 
 $n$ & $\Delta \bar{\hat{\bm{b}}}_0^n$ & $\Delta \bar{\hat{\bm{b}}}_1^n$ \\ [0.5ex] 
 \hline\hline
$0-7$ & $\pm (2)$ & $\pm (-2 \lor 0,0,0,0,0,0,0,0,2 \lor 0)$ \\ \hline
$8-15$ & $\pm (2,0,2)$ & $\pm (-2 \lor 0,0,0,0,0,0,0,0,2,0,2 \lor 0)$ \\ \hline
$16-23$ & $\pm (2,0,0,0,-2)$ & $\pm (-2\lor0,0,0,0,0,0,0,0,2,0,0,0,-2\lor0)$ \\ \hline
$24-31$ & $\pm (2,0,0,2,2,-2,0,0,2)$ & $\pm (-2 \lor 0,0,0,-2,-2,2,0,-2,0,2,0,0,0,0,0,0,2 \lor 0)$ \\ \hline
$32-39$ & $\pm (2,0,-2,0,-2,2,-2,-2,2,0,2)$ & $\pm (-2\lor0,0,2,0,0,0,2,2,0,-2,0,0,-2,0,-2,0,2,0,2\lor0)$ \\ \hline
$40-47$ & $\pm (2,0,2,2,-2,-2,-2,0,2,-2,0,$ & $\pm (-2 \lor 0,0,0,-2,0,2,0,0,2,2,2,0,-2,0,0,$ \\
 & $-2)$ & $0,0,-2,0,-2 \lor 0)$ \\ \hline
$48-55$ & $\pm (2,-2,-2,-2,2,-2,0,2,-2,-2,$ & $\pm (-2 \lor 0,0,2,0,-2,2,2,-2,0,2,-2,-2,2,0,0,0,-2,0,$ \\
 & $0,0,2)$ & $0,0,2\lor0)$ \\ \hline

$56-63$ & $\pm (2,0,2, -2,0,-2, 0,2,-2, 2,-2,2,$ & $\pm (-2 \lor 0,0,0,0,0,0,0,0,2,0,0,-2,0,0,0,-2,0,2,-2,0,$ \\
 & $0,-2,2, 2,0,0, -2)$ & $0,0,2,0,0,0,-2 \lor 0)$ \\ [1ex]
\hline
\end{tabular}
\end{table}

\begin{table}[!htbp]
\caption{Input differences vectors of error events, with minimum Euclidean distance, for $L_0=3$ and $L_1=1,$ under constrained optimization.}
\label{table:Input Differences Vectors L_0=3 L_1=1 Constrained Optimization}
\centering
\begin{tabular}{|c|c|c|} 
 \hline
 $n$ & $\Delta \bar{\hat{\bm{b}}}_0^n$ & $\Delta \bar{\hat{\bm{b}}}_1^n$ \\ [0.5ex] 
 \hline\hline
 $0-7$ & $\pm (2,-2)$ & $\pm (-2\lor0,0,0,2\lor0)$ \\ \hline
 $8-15$ & $\pm (2)$ & $\pm (-2\lor0,-2,-2\lor0)$ \\ \hline
$16-23$ & $\pm (2,2)$ & $\pm (-2\lor0,-2,-2,-2\lor0)$  \\ [1ex] 
 \hline
\end{tabular}
\end{table}

\begin{table}[!htbp]
\caption{Input differences vectors of error events, with minimum Euclidean distance, for $L_0=4$ and $L_1=1,$ under constrained optimization.}
\label{table:Input Differences Vectors L_0=4 L_1=1 Constrained Optimization}
\centering
\begin{tabular}{|c|c|c|} 
 \hline
 $n$ & $\Delta \bar{\hat{\bm{b}}}_0^n$ & $\Delta \bar{\hat{\bm{b}}}_1^n$ \\ [0.5ex] 
 \hline\hline
 $0-7$ & $\pm (2)$ & $\pm (-2\lor0, -2, 0, 2\lor0)$ \\ \hline
 $8-15$ & $\pm (2,2,2,0,-2)$ & $\pm (-2\lor0,-2,-2,0,2,2,0,-2\lor0)$ \\ \hline
 $16-23$ & $\pm (2,2,2)$ & $\pm (-2\lor0,-2,-2,0,2,2\lor0)$ \\ \hline
 $24-31$ & $\pm (2,2,2,0,2)$ & $\pm (-2\lor0,-2,-2,0,0,0,0,2\lor0)$  \\ [1ex] 
 \hline
\end{tabular}
\end{table}

\begin{table}[!htbp]
\caption{Input differences vectors of error events, with minimum Euclidean distance, for $L_0=5$ and $L_1=1,$ under constrained optimization.}
\label{table:Input Differences Vectors L_0=5 L_1=1 Constrained Optimization}
\centering
\begin{tabular}{|c|c|c|} 
 \hline
 $n$ & $\Delta \bar{\hat{\bm{b}}}_0^n$ & $\Delta \bar{\hat{\bm{b}}}_1^n$ \\ [0.5ex] 
 \hline\hline
 $0-7$ & $\pm (2,-2,-2,-2)$ & $\pm (-2\lor0, 0, 2, 0, 0, 2, 2, 2\lor0)$ \\ \hline
 $8-15$ & $\pm (2,-2,2)$ & $\pm (-2\lor0,0,0,-2,0,0,-2\lor0)$ \\ \hline
 $16-23$ & $\pm (2)$ & $\pm (-2\lor0, 0, 0, -2, -2\lor0)$ \\ \hline
 $24-31$ & $\pm (2,2,0,2,-2)$ & $\pm (-2\lor0,-2,0,-2,-2,0,-2,0,2\lor0)$  \\ [1ex] 
 \hline
\end{tabular}
\end{table}

\begin{table}[!htbp]
\caption{Input differences vectors of error events, with minimum Euclidean distance, for $L_0=6$ and $L_1=1,$ under constrained optimization.}
\label{table:Input Differences Vectors L_0=6 L_1=1 Constrained Optimization}
\centering
\begin{tabular}{|c|c|c|} 
 \hline
 $n$ & $\Delta \bar{\hat{\bm{b}}}_0^n$ & $\Delta \bar{\hat{\bm{b}}}_1^n$ \\ [0.5ex] 
 \hline\hline
 $0-7$ & $\pm (2,-2)$ & $\pm (-2\lor0, 0, 2, 0, 0, -2, 2\lor0)$ \\ \hline
 $8-15$ & $\pm (2,0,-2)$ & $\pm (-2\lor0,0,2,2,0,-2,0,2\lor0)$ \\ \hline
 $16-23$ & $\pm (2,0,-2,2,-2,-2)$ & $\pm (-2\lor0, 0, 2, 0, 0, 0, 0, 0, -2, 0, 2\lor0)$ \\ \hline
 $24-31$ & $\pm (2,0,2,0,-2)$ & $\pm (-2\lor0,0,0,0,2,0,0,-2,0,2\lor0)$ \\ \hline
 $32-39$ & $\pm (2,0,2)$ & $\pm (-2\lor0,0,0,0,2,0,0,-2\lor0)$ \\ [1ex]
 \hline
\end{tabular}
\end{table}

\begin{table}[!htbp]
\caption{Input differences vectors of a reduced set of error events, with minimum Euclidean distance, for $L_0=7$ and $L_1=1,$ under constrained optimization.}
\label{table:Input Differences Vectors L_0=7 L_1=1 Constrained Optimization}
\centering
\begin{tabular}{|c|c|c|} 
 \hline
 $n$ & $\Delta \bar{\hat{\bm{b}}}_0^n$ & $\Delta \bar{\hat{\bm{b}}}_1^n$ \\ [0.5ex] 
 \hline\hline
 $0-7$ & $\pm (2,0,0,2,2,2,0,0,2,2)$ & $\pm (-2\lor0,0,0,0,-2,-2,2,0,-2,-2,0,2,0,-2,0,2\lor0)$ \\ [1ex]
 \hline
\end{tabular}
\end{table}

The entries in these tables are fundamental to the symbolic determination of the optimal filters. Once the dominant minimum-distance error events are known, the normalized filters $\bar{h}_0[k]$ and $\bar{h}_1[k]$ can be derived in closed form. In the \textit{unconstrained} case, the energy allocation parameter $\eta$—which governs the distribution of the total symbol energy between the two filters—is also derived in closed form. The modulated filters are then given by $h_0[k] = \sqrt{\eta} \, \bar{h}_0[k]$ and $h_1[k] = \sqrt{5 - \eta} \, \bar{h}_1[k]$, with the total symbol energy normalized to 5. In the \textit{constrained} case, by contrast, the parameter $\eta$ is fixed to $\eta = 5/2$, such that both filters share the symbol energy equally. The second filter is defined as $h_1[k] = \sqrt{5/2} \, \delta[k]$ (or equivalently $-\sqrt{5/2} \, \delta[k]$), since $L_1 = 1$ and the energy constraint $\| h_1[k] \|^2 = 5/2$ holds. In both scenarios, the closed-form expressions of the normalized filters and the energy allocation parameter fully determine the modulated filters and enable the \gls{msed} $d_{\min}^2$ to be computed analytically.

In all tables, the notation $a \lor b$ indicates that a given component may take either value. For example, an entry of the form $(-2 \lor 0, 0, 0, 2 \lor 0)$ implies four possible vectors: $(0, 0, 0, 0)$, $(-2, 0, 0, 0)$, $(0, 0, 0, 2)$, or $(-2, 0, 0, 2)$. Consequently, a single $\Delta \bar{\hat{\bm{b}}}_0$ vector may correspond to multiple distinct $\Delta \bar{\hat{\bm{b}}}_1$ vectors (and vice versa), depending on how the combinations of ambiguous components resolve. Each pairing should be understood to generate one or more valid minimum-distance events, including their sign-inverted versions.

As can be observed from Tables~\ref{table:Input Differences Vectors L_0=3 L_1=1}--\ref{table:Input Differences Vectors L_0=9 L_1=1} for the unconstrained case, and from Tables~\ref{table:Input Differences Vectors L_0=3 L_1=1 Constrained Optimization}--\ref{table:Input Differences Vectors L_0=6 L_1=1 Constrained Optimization} for the constrained case, the minimum-distance error events exhibit consistent structural patterns for $L_0 < 10$ in the unconstrained case and for $L_0 \le 7$ in the constrained case. These patterns stem from a key phenomenon described earlier: for a given input-difference vector $\Delta \bar{\hat{\bm{b}}}_0$, the associated second input-difference vector $\Delta \bar{\hat{\bm{b}}}_1$ is constructed such that each of its components minimizes a local quadratic expression within an $\varepsilon$-neighborhood. When this local minimum is not uniquely attained, the component $\Delta \bar{\hat{b}}_1[k]$ may admit more than one value---typically $0$ and either $+2$ or $-2$. As noted in the preceding paragraph, this leads to multiple valid combinations of $(\Delta \bar{\hat{\bm{b}}}_0, \Delta \bar{\hat{\bm{b}}}_1)$ that differ only in one such component, yet all correspond to the same \gls{msed}. It is precisely these families of closely related error events that give rise to what we refer to as \textit{tightness conditions}: structural dependencies that impose strong relationships among the filter coefficients. In the majority of cases, these conditions constrain only the first and last coefficients of the first filter $h_0[k]$, namely $h_0[0]$ and $h_0[L_0 - 1]$, along with the scalar coefficient $h_1[0]$ defining the second filter $h_1[k]$. In some exceptional configurations, however, an additional coefficient of $h_0[k]$ may also be involved. These constraints reflect the precise balance required to preserve ambiguity in certain components of $\Delta \bar{\hat{\bm{b}}}_1$ while still achieving the minimum Euclidean distance.

The scenario $L_0 = 10$ in the \textit{unconstrained} setting constitutes a remarkable and singular case. Here, the \gls{msed} exactly matches that of a conventional $2$-ASK transmission—the theoretical maximum achievable by any rate-$2$ \gls{nsm}. Since this value is attained while preserving the spectral efficiency of $4$-ASK, this configuration achieves the maximum possible asymptotic \gls{snr} gain. Compared to the $L_0 = 9$ case—where the performance closely approaches this limit but yields a unique optimal solution (up to equivalence)—the $L_0 = 10$ case introduces additional structure and flexibility, as detailed next.

Crucially, for $L_0 = 10$, the only input-difference vectors that lead to the \gls{msed} are the trivial ones of the form $\Delta \bar{\hat{\bm{b}}}_0 = \pm(2)$ with $\Delta \bar{\hat{\bm{b}}}_1 = \bm{0}$, or vice versa. All other, nontrivial error events yield strictly higher \glspl{sed}. This behavior reflects the fact that the extended span and increased design freedom offered by $L_0 = 10$ make it possible to construct filters that effectively avoid all harmful interference patterns associated with nontrivial input differences. As a result, there is a performance margin: the smallest \gls{sed} over nontrivial error events is strictly greater than that of the trivial events, imparting robustness to the design. Consequently, small perturbations in the filter coefficients are not expected to reduce the \gls{msed} from its maximum value, which equals that of $2$-ASK.

This structural property helps explain the experimental observation that multiple distinct filter pairs $(h_0[k], h_1[k])$ achieve the same optimal performance. We conjecture the existence of an infinite family of such optimal solutions, in contrast to the unique (up to equivalence transformations) solutions found for $L_0 < 10$. Among these, Table~\ref{table:Best Filters Numerical Form Rate-2 NSM} reports a representative filter pair for $L_0 = 10$. The existence of a rich solution space also opens the door to further performance optimization: by selecting, among the optimal filters, those that maximize the second \gls{msed}, it is possible to enhance system behavior at low to moderate \glspl{snr}, tightening the bit error probability gap with $2$-ASK in this regime.

In contrast to the flexibility and abundance of optimal solutions observed in the \textit{unconstrained} setting with $L_0 = 10$, the \textit{constrained} case with $L_0 = 7$ presents a more rigid structure. Here, a full enumeration of minimum-distance error events is not necessary: the optimal filters exhibit a clear numerical pattern, supported by tightness conditions that establish simple relationships between the first and last non-null taps of $h_0[k]$ and the unique non-null tap of $h_1[k]$. This structure enables a symbolic closed-form solution to be directly obtained from a representative subset of the dominant error events, which is reported in the corresponding table.

\subsubsection*{Determination of the Projected Distance Spectra}

Given the closed-form expressions of the real-tapped filters of the top-performing rate-$2$ \glspl{nsm} for $3 \le L_0 \le 8$, and leveraging the steps previously followed to identify the corresponding error events of \gls{msed}, we can gain further insight into these \glspl{nsm} through their projected distance spectra.  

The computation of the projected distance spectrum can be viewed as a sequence-based (scalar) counterpart of the matrix-form derivation presented in Subsection~\ref{Identification of Minimum-Distance Error Events}, where the sizes of the input vectors and matrices were fixed from the outset. In contrast, the sequence-based approach considered here does not impose such limitations, allowing us to handle longer input sequences and a wider variety of error events. Specifically, the projected spectrum captures all \glspl{sed} larger than or equal to the minimum distance, up to a chosen threshold beyond which the distance spectrum is no longer of practical interest.

The computation proceeds through the following steps:
\begin{itemize}

    \item For each finite-length first input sequence difference $\Delta \bar{b}_0[k]$, the corresponding partial modulated sequence is computed as
    $\Delta \hat{s}_0[k] = \Delta \bar{b}_0[k] \circledast \hat{h}_0[k]$, which is the sequence-based counterpart of the matrix-form expression in (\ref{eq:Partial Output Differences Vector L_0 > 1 L_1 = 1}).

    \item The \gls{sed} is obtained by minimizing the \gls{sen} of 
    $\Delta \hat{s}[k] = \Delta \hat{s}_0[k] + \Delta \hat{s}_1[k]$ over all possible second input sequence differences $\Delta \bar{b}_1[k]$, with
    $\Delta \hat{s}_1[k] = \Delta \bar{b}_1[k] \circledast \hat{h}_1[k] = \hat{h}_1[0] \Delta \bar{b}_1[k]$. This is the sequence-based counterpart of the matrix-form expression in (\ref{eq:Output Differences Vector L_0 > 1 L_1 = 1}).

    \item The projected \gls{sed} for each $\Delta \bar{b}_0[k]$ is then
    $\sum_k \min_{\Delta \bar{b}_1[k] \in \{0, \pm 2\}} (\Delta \hat{s}_0[k] + \hat{h}_1[0] \Delta \bar{b}_1[k])^2$, which corresponds to the sequence-based version of the matrix-form expression in (\ref{eq:Partial Minimization}).

\end{itemize}

\begin{figure}[!htbp]
    \centering
    \includegraphics[width=0.8\textwidth]{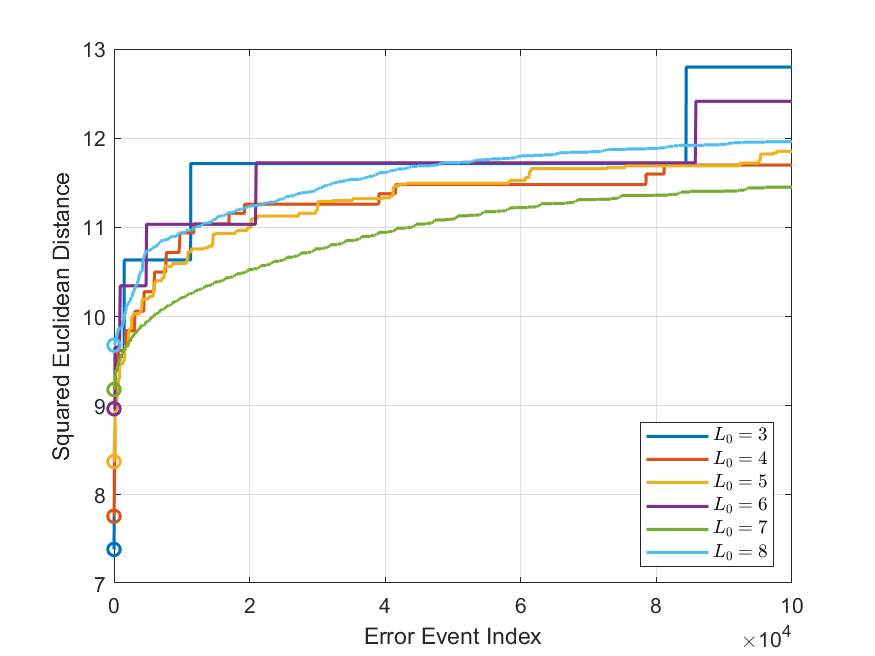}
    \caption{Projected SED spectrum for the NSM of rate $2$, for $3 \le L_0 \le 8$ and $L_1 = 1$, showing the first $10^5$ error events.}
    \label{fig:Distance-Spectrum-NSM-2}
\end{figure}

\begin{figure}[!htbp]
    \centering
    \includegraphics[width=0.8\textwidth]{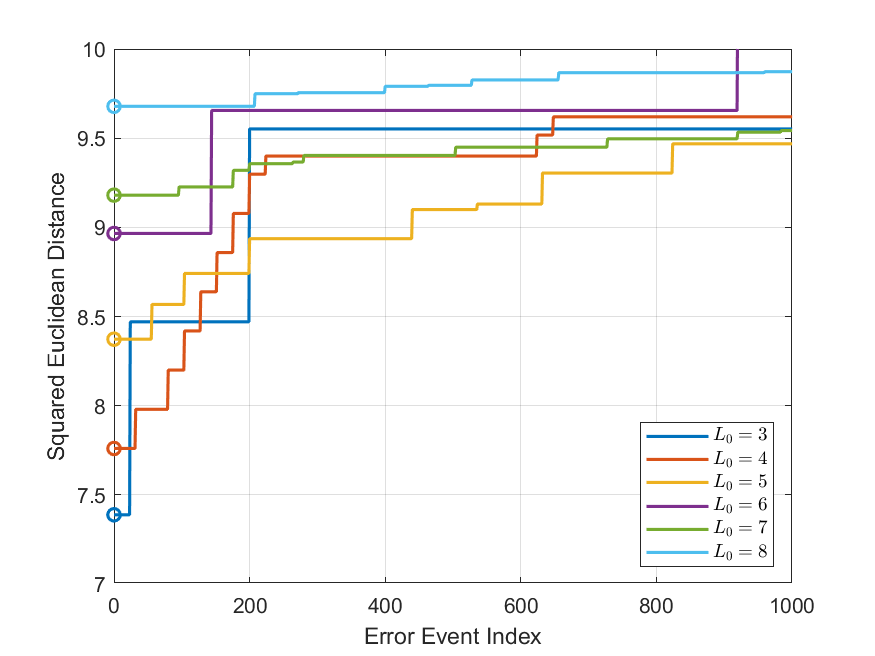}
    \caption{Zoomed view of the projected SED spectrum for the NSM of rate $2$, for $3 \le L_0 \le 8$ and $L_1 = 1$, showing the first $1000$ error events.}
    \label{fig:Distance-Spectrum-NSM-2-Zoom}
\end{figure}

Figures~\ref{fig:Distance-Spectrum-NSM-2} and~\ref{fig:Distance-Spectrum-NSM-2-Zoom} depict the projected \gls{sed} spectra for the top-performing rate-$2$ \glspl{nsm} with real-tapped filters, for $3 \le L_0 \le 8$ and $L_1 = 1$, for which closed-form filter expressions and \glspl{msed} have been determined. Figure~\ref{fig:Distance-Spectrum-NSM-2} shows the first $10^5$ error events, selected for having the lowest \glspl{sed} and arranged in order of increasing squared distance, with each event corresponding to a distinct input sequence $\bar{b}_0[k]$. Figure~\ref{fig:Distance-Spectrum-NSM-2-Zoom} provides a zoomed view focusing on the first $1000$ error events to better highlight the behavior of the lowest-distance events.

Both figures show a monotonic increase starting from the \gls{msed} for each optimum \gls{nsm}. These initial values mark the beginning of each \gls{nsm}’s projected distance spectrum curve and correspond to the error events achieving the respective minimum distances.

From both figures, we further observe that the squared distance curves, which are generally increasing, exhibit some crossings. Notably, the \gls{nsm} with $L_0 = 3$, which has the worst \gls{msed} among the considered cases, quickly attains better squared distances than the \glspl{nsm} with $L_0 = 4$ and $5$ after a few error events, even though the latter configurations achieve better \glspl{msed}. This illustrates that the relative ordering of \glspl{nsm} in terms of squared distances can vary across error events, and that a smaller \gls{msed} does not necessarily imply uniformly better performance across the entire projected spectrum.

\subsubsection*{Comparison Between Simulated BERs and BEP Upper Bounds}

\begin{figure}[!htbp]
    \centering
    \includegraphics[width=1.0\textwidth]{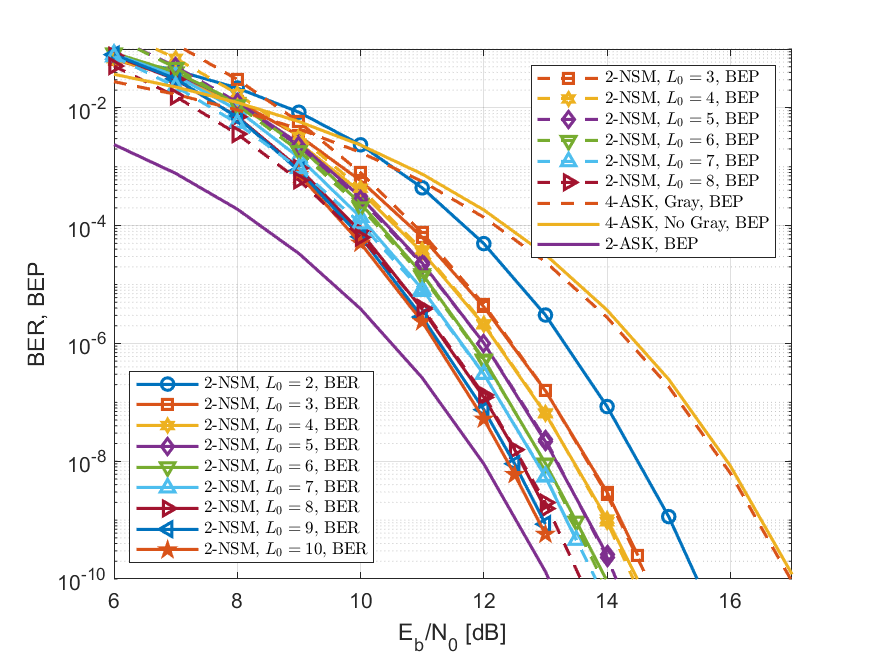}
    \caption{BER and BEP approximation of the NSM of rate $2,$ for $2 \le L_0 \le 10$ and $L_1 = 1$. For reference, the Bit Error Probabilities (BEPs) of $2$-ASK and Gray and non-Gray precoded $4$-ASK conventional modulations are used.}
    \label{fig:BER-BEP-NSM-2-Full}
\end{figure}

Figure~\ref{fig:BER-BEP-NSM-2-Full} presents the simulated \gls{ber} curves of unconstrained rate-$2$ \glspl{nsm} with real-valued filters, for filter lengths $L_0$ ranging from $2$ to $10$, and fixed $L_1 = 1$. For each configuration with $L_0 \geq 3$, the filters were obtained through numerical optimization, as detailed in Table~\ref{table:Best Filters Numerical Form Rate-2 NSM}, which provides the normalized coefficients, energy allocation parameter $\hat{\eta}$, and the corresponding \gls{msed} $\hat{d}_{\min}^2$ for each optimized \gls{nsm}. The special case $L_0 = 2$ was treated in detail in Section~\ref{ssec:Modulation of Rate 2} and is included here for completeness through its \gls{ber} curve.

For the cases $L_0 = 3$ through $8$, the theoretical upper bounds on the \gls{bep} are also plotted in the figure, as functions of $E_b/N_0$. These bounds are derived from the truncated \gls{rtf} $\dot{T}(D;P)$, computed iteratively using Algorithm~\ref{alg:Truncated Reduced Transfer Function T(D;P) Rate-2 NSM} in Appendix~\ref{app:Iterative Determination Transfer Function Rate-2 NSM}, specifically in Section~\ref{ssec:Determination Truncated Reduced Transfer Function}. We recall that, in the preceding sections, the design process began with the numerical optimization of rate-$2$ \gls{nsm} configurations, which yielded the best energy allocation parameter $\eta$ and corresponding filter coefficients $h_0[k]$ and $h_1[k]$ for each value of $L_0$. We also recall that these optimized filters were used to identify the dominant error events responsible for the \gls{msed}. Based on this error event structure, symbolic derivations were then carried out in Appendix~\ref{app:Closed-Form Expressions Optimum Filters Non-Degenerate Rate-2 NSMs} to obtain closed-form expressions for the optimal filters. These symbolic expressions, together with the exact energy partition, served as input to compute the \glspl{rtf} and the associated \gls{bep} upper bounds. The results are explicitly reported in Tables~\ref{table:Transfer Functions L_0=3 L_1=1} through~\ref{table:Transfer Functions L_0=8 L_1=1}, providing analytical confirmation of the high-$E_b/N_0$ behavior observed in simulation.

A striking outcome, evident in Figure~\ref{fig:BER-BEP-NSM-2-Full}, is the perfect match between the simulated \gls{ber} curves and the \gls{bep} upper bounds for all \glspl{nsm} with $L_0 = 3$ to $8$ over the moderate-to-high $E_b/N_0$ range (i.e., above 10 dB). This tight agreement confirms the correctness of both our simulation procedures and our \gls{tf}-based upper bound derivation, thus providing strong cross-validation of the entire framework.

In addition to \gls{nsm} performance, the figure includes benchmark \gls{bep} curves for conventional $2$-ASK and $4$-ASK modulations, with and without Gray coding. These benchmarks are critical for contextualizing the performance of \glspl{nsm}: the $4$-ASK modulation offers the same spectral efficiency (rate-$2$) as the \glspl{nsm} but inferior error performance; conversely, the $2$-ASK modulation achieves the best possible uncoded performance but with half the spectral efficiency (rate-$1$). The goal of the \gls{nsm} design is to bridge this gap—achieving $2$-ASK-like reliability while maintaining $4$-ASK-like spectral efficiency.

As seen in the figure, the optimized \glspl{nsm} exhibit steady performance improvement as $L_0$ increases. Notably, the \gls{nsm} with $L_0 = 10$ achieves a \gls{ber} within less than $0.5$~dB of that of $2$-ASK at \gls{ber} levels below $10^{-8}$. This observation is consistent with the results in Table~\ref{table:Best Filters Numerical Form Rate-2 NSM}, which shows that the optimized \gls{nsm} for $L_0 = 10$ is the first configuration to attain the \textbf{same \gls{msed}} as $2$-ASK. In contrast, for $L_0 = 9$, although the distance gap is very small (less than $0.05$~dB), this nonzero gap prevents the exact asymptotic matching of $2$-ASK performance in terms of \gls{bep}. Nevertheless, the performance for $L_0 = 9$ closely approaches that of $2$-ASK at high $E_b/N_0$. These results highlight that only at $L_0 = 10$ is it possible to perfectly replicate the \gls{msed} and thus the asymptotic \gls{bep} of $2$-ASK, establishing $L_0 = 10$ as the minimal filter length required for exact distance-optimality in the rate-$2$ \gls{nsm} design.

\gls{bep} upper bounds were not shown in Figure~\ref{fig:BER-BEP-NSM-2-Full} for the cases $L_0 = 2$, $9$, and $10$. The case $L_0 = 2$ was already treated in detail in Section~\ref{ssec:Modulation of Rate 2}, where both the simulated \gls{ber} and the \gls{bep} bound were provided in Figure~\ref{fig:BER-BEP-NSM-2}. In this degenerate configuration, the \gls{nsm} produces an infinite multiplicity of minimum-distance error events, making the truncated \gls{rtf} $\dot{T}(D; P)$ inadequate for bounding the \gls{bep}. For $L_0 = 9$ and $10$, no closed-form expressions for the optimized filters were available. Consequently, symbolic computation of the \glspl{rtf} based on these numerically obtained filters was rendered unreliable due to floating-point imprecision, which introduces ambiguity in the labeling of distance terms. These limitations prevented a trustworthy derivation of \gls{bep} upper bounds for these configurations in Figure~\ref{fig:BER-BEP-NSM-2-Full}.

Despite their excellent performance at high $E_b/N_0$, all \glspl{nsm} exhibit relatively poor behavior at low $E_b/N_0$. This degradation is primarily caused by a large number of error events with low \glspl{sed}—that is, a high multiplicity of near-minimum-distance events. For \glspl{nsm} with $L_0 < 10$, this issue is further linked to the tightness conditions observed in the optimized filters. These tightness conditions are not imposed during the filter optimization process but rather arise as an inherent outcome of it. While such tightness simplifies the derivation of closed-form expressions, it unfortunately increases the multiplicity of minimum-distance error events, adversely impacting the \gls{bep} performance across all $E_b/N_0$ ranges. The value $L_0 = 10$ is significant as it is the first configuration that eliminates these tightness conditions. Extending $L_0$ beyond $10$ not only reduces the multiplicity of critical error events but also avoids the tightness conditions, thereby offering the potential to improve performance in the low-$E_b/N_0$ regime.

In summary, the joint examination of simulated \gls{ber} and \gls{bep} upper bounds confirms that optimized unconstrained rate-$2$ \glspl{nsm} can closely approach, and eventually match, the performance of $2$-ASK modulation. This is achieved with simpler transmitter architectures and without requiring iterative detection techniques or forward error correction. These findings validate the potential of \glspl{nsm} as efficient modulation schemes for high-performance communication systems.

\begin{table}[!htbp]
\caption{Partially determined TFs, $T(D,N)$ and $\dot{T}(D),$ and \gls{bep} approximation for $L_0=3$ and $L_1=1$ (First $100.000$ error events used for the determination of the first terms of $T(D,N)$ and $\dot{T}(D)$).}
\label{table:Transfer Functions L_0=3 L_1=1}
\small
\centering
\begin{tabular}{|c||c|} 
 \hline
\multirow{11}{*}{$T(D,N)$} & $(2N+6N^2+6N^3+4N^4+4N^5+2N^6) D^{j_0}$ \\ 
 & $+(2N^2+10N^3+26N^4+36N^5+34N^6+30N^7+20N^8+10N^9+6N^{10}+2N^{11}) D^{j_1}$ \\
 & $+(2N^3+16N^4+58N^5+138N^6+206N^7+222N^8+216N^9+180N^{10}$\\
& $+124N^{11}+80N^{12}+42N^{13}+18N^{14}+8N^{15}+2N^{16}) D^{j_2}$ \\
& $+(2N^4+22N^5+116N^6+352N^7+756N^8+1166N^9+1404N^{10}+1498N^{11}$ \\
& $+1396N^{12}+1120N^{13}+826N^{14}+542N^{15}+312N^{16}+166N^{17}+74N^{18}$ \\
& $+28N^{19}+10N^{20}+2N^{21}) D^{j_3}$ \\ 
& $+ (2N^5+30N^6+194N^7+782N^8+2098N^9+4252N^{10}+6736N^{11}+8736N^{12}$ \\
& $+9958N^{13}+10054N^{14}+8994N^{15}+7370N^{16}+5488N^{17}+3690N^{18}+2294N^{19}$ \\
& $+1288N^{20}+650N^{21}+300N^{22}+118N^{23}+40N^{24}+12N^{25}+2N^{26}) D^{j_4}$ \\
& $(2N+2N^2+4N^3+2N^4) D^{j_5} + \cdots$ \\
\hline
$\dot{T}(D)$ & $\frac{129}{16}D^{j_0} + \frac{3544}{157}D^{j_1} + \frac{4583}{89}D^{j_2} + \frac{12717}{119}D^{j_3} + \frac{4824}{23}D^{j_4} + 4D^{j_5} + \cdots$ \\ [0.5ex]
 \hline 
\multirow{2}{*}{\gls{bep} approximation} & $\frac{129}{64} \operatorname{erfc}\left( \sqrt{\frac{j_0}{10} \frac{E_b}{N_0}} \right)
+ \frac{886}{157} \operatorname{erfc}\left( \sqrt{\frac{j_1}{10} \frac{E_b}{N_0}} \right)
+ \frac{4583}{356} \operatorname{erfc}\left( \sqrt{\frac{j_2}{10} \frac{E_b}{N_0}} \right)$ \\
& $+ \frac{12717}{476} \operatorname{erfc}\left( \sqrt{\frac{j_3}{10} \frac{E_b}{N_0}} \right)
+ \frac{603}{23} \operatorname{erfc}\left( \sqrt{\frac{j_4}{10} \frac{E_b}{N_0}} \right)
+ \operatorname{erfc}\left( \sqrt{\frac{j_5}{10} \frac{E_b}{N_0}} \right)$ \\ [1ex]
 \hline
 \multirow{2}{*}{Parameters} & $j_m = d_{\text{min}}^2 + 4 m h_0^2[1],$ $0 \le m \le 4,$ $j_5 = 4 h_1^2[0],$ $d_{\text{min}}^2 = 20(4-\sqrt{2})/7,$ \\
 & $h_0^2[1] = 5(5-3\sqrt{2})/14$ and $h_1^2[0]=5(3+\sqrt{2})/7$ \\ \hline
\multicolumn{2}{|l|}{\# of terms used in the \gls{bep} approximation $ = 6$} \\ \hline
\multicolumn{2}{|l|}{\# of iterations to stabilize the transfer functions $ = 18$} \\ [1ex] 
 \hline
\end{tabular}
\end{table}

\begin{table}[!htbp]
\caption{Partially determined TFs, $T(D,N)$ and $\dot{T}(D),$ and BEP approximation for $L_0=4$ and $L_1=1$ (First $100.000$ error events used for the determination of the first terms of $T(D,N)$ and $\dot{T}(D)$).}
\label{table:Transfer Functions L_0=4 L_1=1}
\small
\centering
\begin{tabular}{|c||c|} 
\hline
\multirow{11}{*}{$T(D,N)$} & $(2N+4N^2+2N^3+4N^7+8N^8+6N^9+4N^{10}+2N^{11}) D^{j_0}$ \\ 
& $+(2N^2+4N^3+6N^4+8N^5+8N^6+8N^7+6N^8+4N^9+2N^{10}) D^{j_1}$ \\
& $+(4N^9+8N^{10}+6N^{11}+4N^{12}+2N^{13})$ \\
& $\times (D^{j_2} + ND^{j_3} + N^2 D^{j_4} + N^3 D^{j_5} + N^4 D^{j_6} + N^5 D^{j_7})$ \\
& $+(4N^2+10N^3+12N^4+14N^5+16N^6+14N^7+18N^8+30N^9+34N^{10}+28N^{11}$ \\
& $+22N^{12}+16N^{13}+20N^{14}+36N^{15}+40N^{16}+36N^{17}+28N^{18}+14N^{19}$ \\
& $+6N^{20}+2N^{21}) D^{j_8}$ \\
& $+(4N^9+8N^{10}+6N^{11}+4N^{12}+2N^{13})N^6 D^{j_9} + \cdots$ \\ 
& $+(4N^{15}+10N^{16}+10N^{17}+40N^{18}+86N^{19}+114N^{20}   136N^{21}+170N^{22}+196N^{23}$ \\ 
& $+268N^{24}+404N^{25}+416N^{26}+352N^{27}+264N^{28}+130N^{29}+54N^{30}$ \\ 
& $+18N^{31}) D^{j_{32}} + \cdots$ \\ 
\hline
\multirow{2}{*}{$\dot{T}(D)$} & $\frac{1684}{385}D^{j_0} + \frac{1719}{256}D^{j_1} + \frac{801}{4096}D^{j_2} + \frac{441}{4096}D^{j_3} + \frac{74}{1259}D^{j_4} + \frac{261}{8192}D^{j_5}$ \\
& $+ \frac{59}{3437}D^{j_6} + \frac{60}{6521}D^{j_7} + \frac{1417}{95}D^{j_8} + \frac{51}{10388}D^{j_9} + \cdots + \frac{115}{6628}D^{j_{32}} + \cdots$ \\ [0.5ex]
\hline 
\multirow{4}{*}{\gls{bep} approximation} &
$\frac{421}{385} \operatorname{erfc}\left( \sqrt{\frac{j_0}{10} \frac{E_b}{N_0}} \right)
+ \frac{1719}{1024} \operatorname{erfc}\left( \sqrt{\frac{j_1}{10} \frac{E_b}{N_0}} \right)
+ \frac{801}{16384} \operatorname{erfc}\left( \sqrt{\frac{j_2}{10} \frac{E_b}{N_0}} \right)$ \\
& $+ \frac{441}{16384} \operatorname{erfc}\left( \sqrt{\frac{j_3}{10} \frac{E_b}{N_0}} \right)
+ \frac{37}{2518} \operatorname{erfc}\left( \sqrt{\frac{j_4}{10} \frac{E_b}{N_0}} \right)
+ \frac{261}{32768} \operatorname{erfc}\left( \sqrt{\frac{j_5}{10} \frac{E_b}{N_0}} \right)$ \\
& $+ \frac{59}{13748} \operatorname{erfc}\left( \sqrt{\frac{j_6}{10} \frac{E_b}{N_0}} \right)
+ \frac{15}{6521} \operatorname{erfc}\left( \sqrt{\frac{j_7}{10} \frac{E_b}{N_0}} \right)
+ \frac{1417}{380} \operatorname{erfc}\left( \sqrt{\frac{j_8}{10} \frac{E_b}{N_0}} \right)$ \\
& $+ \frac{51}{41552} \operatorname{erfc}\left( \sqrt{\frac{j_9}{10} \frac{E_b}{N_0}} \right) + \cdots
+ \frac{115}{26512} \operatorname{erfc}\left( \sqrt{\frac{j_{32}}{10} \frac{E_b}{N_0}} \right)$ \\ [1ex]
\hline
\multirow{2}{*}{Parameters} & $j_m = d_{\text{min}}^2 + m \Delta,$ $0 \le m \le 7,$ $d_{\text{min}}^2 = 20(6-\sqrt{3})/11,$ $\Delta = 5(23-13\sqrt{3})/11,$ \\
 & $j_8 = 10(19-5\sqrt{3})/11,$ $j_9 = d_{\text{min}}^2 + 8 \Delta, \ldots$ and $j_{32} \approx 11.596766791071444$ \\ \hline
\multicolumn{2}{|l|}{\# of terms used in the \gls{bep} approximation $ = 33$} \\ \hline
\multicolumn{2}{|l|}{\# of iterations to stabilize the transfer functions $ = 30$} \\ [1ex] 
\hline
\end{tabular}
\end{table}

\begin{table}[!htbp]
\caption{Partially determined TFs, $T(D,N)$ and $\dot{T}(D),$ and BEP approximation for $L_0=5$ and $L_1=1$ (First $100.000$ error events used for the determination of the first terms of $T(D,N)$ and $\dot{T}(D)$).}
\label{table:Transfer Functions L_0=5 L_1=1}
\small
\centering
\begin{tabular}{|c||c|}
\hline
\multirow{8}{*}{$T(D,N)$} & $(2N+4N^2+2N^3+2N^4+4N^5+6N^6+8N^7+4N^8+4N^{10}+8N^{11}+6N^{12}$ \\ 
& $+4N^{13}+2N^{14}) D^{j_0}$ \\
& $+(6N^5+12N^6+12N^7+12N^8+6N^9) D^{j_1}$ \\
& $+(2N^9+4N^{10}+6^{11}+12^{12}+12^{13}+8^{14}+8^{15}+8^{16}+12N^{17}+12N^{18}+6N^{19}$ \\
& $+4N^{20}+2N^{21}) D^{j_2}$ \\
& $+(2N^{10}+6N^{11}+6N^{12}+2N^{13}+4N^{14}+12N^{15}+22N^{16}+28N^{17}+28N^{18}$ \\
& $+36N^{19}+44N^{20}+44N^{21}+42N^{22}+46N^{23}+52N^{24}+42N^{25}+28N^{26}+24N^{27}$ \\ 
& $+28N^{28}+24N^{29}+12N^{30}+8N^{31}+4N^{32}) D^{j_{62}} + \cdots$ \\ 
\hline
$\dot{T}(D)$ & $\frac{3744}{613}D^{j_0} 
+ \frac{819}{256}D^{j_1} + \frac{562}{3197}D^{j_2}
+ \frac{1135}{677}D^{j_3} + \cdots + \frac{523}{5482}D^{j_{62}} + \cdots$ \\ [0.5ex]
\hline 
\multirow{2}{*}{\gls{bep} approximation} & $\frac{936}{613} \operatorname{erfc}\left( \sqrt{\frac{j_0}{10} \frac{E_b}{N_0}} \right) + \frac{819}{1024} \operatorname{erfc}\left( \sqrt{\frac{j_1}{10} \frac{E_b}{N_0}} \right) + \frac{281}{6394} \operatorname{erfc}\left( \sqrt{\frac{j_2}{10} \frac{E_b}{N_0}} \right)$ \\
& $+ \frac{1135}{2708} \operatorname{erfc}\left( \sqrt{\frac{j_3}{10} \frac{E_b}{N_0}} \right)
+ \cdots + \frac{523}{21928} \operatorname{erfc}\left( \sqrt{\frac{j_{62}}{10} \frac{E_b}{N_0}} \right)
$ \\ [1ex]
\hline
\multirow{2}{*}{Parameters} & $j_0 = d_{\text{min}}^2 \approx 8.372733278497531,$ $j_1 \approx 8.567388799829605,$ $j_2 \approx 8.741322115953427$\\
 & $j_3 \approx 8.935977637285481, \ldots$ and $j_{62} \approx 11.832433215057055$ \\ \hline
\multicolumn{2}{|l|}{\# of terms used in the \gls{bep} approximation $ = 63$} \\ \hline
\multicolumn{2}{|l|}{\# of iterations to stabilize the transfer functions $ = 36$} \\ [1ex] 
\hline
\end{tabular}
\end{table}

\begin{table}[!htbp]
\caption{Partially determined TFs, $T(D,N)$ and $\dot{T}(D),$ and BEP approximation for $L_0=6$ and $L_1=1$ (First $100.000$ error events used for the determination of the first terms of $T(D,N)$ and $\dot{T}(D)$).}
\label{table:Transfer Functions L_0=6 L_1=1}
\small
\centering
\begin{tabular}{|c||c|} 
 \hline
\multirow{13}{*}{$T(D,N)$} & $(
2N+6N^2+6N^3+2N^4+4N^6+14N^7+16N^8+12N^9+24N^{10}+32N^{11}$ \\ 
& $+16N^{12}+4N^{13}+4N^{14}+2N^{15}) D^{j_0}$ \\
& $+(4N^2+12N^3+16N^4+16N^5+14N^6+12N^7+26N^8+74N^9+122N^{10}$ \\
& $+130N^{11}+134N^{12}+124N^{13}+64N^{14}+16N^{15}+8N^{16}+4N^{17}) D^{j_1}$ \\
& $+(4N^3+18N^4+38N^5+48N^6+40N^7+46N^8+112N^9+23236N^{10}+41036N^{11}$ \\
& $+63836N^{12}+74836N^{13}+64436N^{14}+46636N^{15}+28636N^{16}+11236N^{17}$ \\ 
& $+2636N^{18}+1436N^{19}+636N^{20}) D^{j_2}$ \\

& $(2N+10N^4+42N^5+86N^6+146N^7+210N^8+224N^9+288N^{10}+628N^{11}$ \\
& $+1266N^{12}+1988N^{13}+2604N^{14}+2878N^{15}+2504N^{16}+1668N^{17}+914N^{18}$ \\
& $+440N^{19}+176N^{20}+58N^{21}+20N^{22}+8N^{23}+2N^{24}) D^{j_3}$ \\
& $+(8N^5+66N^6+192N^7+354N^8+548N^9+726N^{10}+930N^{11}+1556N^{12}$ \\
& $+3148N^{13}+5802N^{14}+8578N^{15}+10292N^{16}+10612N^{17}+9250N^{18} +6338N^{19}$ \\
& $+3480N^{20}+1770N^{21}+818N^{22}+272N^{23}+74N^{24}+32N^{25}+10N^{26}) D^{j_4} + \cdots$ \\
\hline
$\dot{T}(D)$ & $\frac{6518}{719}D^{j_0} 
+ \frac{4769}{243}D^{j_1} + \frac{5473}{180}D^{j_2}
+ \frac{10660}{197}D^{j_3} + \frac{187441}{2533}D^{j_4} + \cdots$ \\ [0.5ex]
\hline 
\multirow{2}{*}{\gls{bep} approximation} & $\frac{3259}{1438} \operatorname{erfc}\left( \sqrt{\frac{j_0}{10} \frac{E_b}{N_0}} \right) + \frac{4769}{972} \operatorname{erfc}\left( \sqrt{\frac{j_1}{10} \frac{E_b}{N_0}} \right) + \frac{5473}{720} \operatorname{erfc}\left( \sqrt{\frac{j_2}{10} \frac{E_b}{N_0}} \right)$ \\
& $+ \frac{2665}{197} \operatorname{erfc}\left( \sqrt{\frac{j_3}{10} \frac{E_b}{N_0}} \right)
+  \frac{187441}{10132} \operatorname{erfc}\left( \sqrt{\frac{j_4}{10} \frac{E_b}{N_0}} \right)
$ \\ [1ex]
\hline
Parameters & $j_m = d_{\text{min}}^2 + m \Delta,$ $0 \le m \le 4,$ $d_{\text{min}}^2 = 260/29,$ $\Delta = 20/29$\\
\hline
\multicolumn{2}{|l|}{\# of terms used in the \gls{bep} approximation $ = 5$} \\ \hline
\multicolumn{2}{|l|}{\# of iterations to stabilize the transfer functions $ = 36$} \\ [1ex] 
\hline
\end{tabular}
\end{table}

\begin{table}[!htbp]
\caption{Partially determined TFs, $T(D,N)$ and $\dot{T}(D),$ and BEP approximation for $L_0=7$ and $L_1=1$ (First $100.000$ error events used for the determination of the first terms of $T(D,N)$ and $\dot{T}(D)$).}
\label{table:Transfer Functions L_0=7 L_1=1}
\small
\centering
\begin{tabular}{|c||c|}
\hline
\multirow{11}{*}{$T(D,N)$} & 
$(2N+6N^2+6N^3+2N^4+4N^5+8N^6+6N^7+6N^8+12N^9+16N^{10}+14N^{11}$ \\ 
& $+10N^{12}+4N^{13}) D^{j_0}$ \\
& $+(2N^5+4N^6+4N^7+8N^8+12N^9+12N^{10}+12N^{11}+10N^{12}+6N^{13}+4N^{14}$ \\
& $+4N^{15}+2N^{16}) D^{j_1}$ \\

& $+(2N^7+4N^8+2N^9+2N^{11}+4N^{12}+4N^{13}+4N^{14}     2N^{15}) D^{j_2}$ \\
& $+(2N^7+6N^8+8N^9+8N^{10}+8N^{11}+6N^{12}+6N^{13}+8N^{14}+6N^{15}+4N^{16}$ \\
& $+2N^{17}) D^{j_3}$ \\
& $+(2N^8+10N^9+20N^{10}+32N^{11}+56N^{12}+94N^{13}+136N^{14}+174N^{15}$ \\
& $+214N^{16}+276N^{17}+336N^{18}+344N^{19}+300N^{20}+220N^{21}+130N^{22}+64N^{23}$ \\
& $+36N^{24}+22N^{25}+6N^{26}  +2N^{57}+4N^{58}+2N^{59}+4N^{144}+8N^{145}+8N^{146}$ \\
& $+8N^{147}+4N^{148}) D^{j_{212}} + \cdots$ \\ 
\hline
$\dot{T}(D)$ & $\frac{538}{59}D^{j_0} 
+ \frac{819}{514}D^{j_1} + \frac{668}{2207}D^{j_2}
+ \frac{513}{860}D^{j_3} + \cdots + \frac{934}{749}D^{j_{212}} + \cdots$ \\ [0.5ex]
\hline 
\multirow{2}{*}{\gls{bep} approximation} & $\frac{269}{118} \operatorname{erfc}\left( \sqrt{\frac{j_0}{10} \frac{E_b}{N_0}} \right) + \frac{819}{2056} \operatorname{erfc}\left( \sqrt{\frac{j_1}{10} \frac{E_b}{N_0}} \right) + \frac{167}{2207} \operatorname{erfc}\left( \sqrt{\frac{j_2}{10} \frac{E_b}{N_0}} \right)$ \\
& $+ \frac{513}{3440} \operatorname{erfc}\left( \sqrt{\frac{j_3}{10} \frac{E_b}{N_0}} \right)
+ \cdots + \frac{467}{1498} \operatorname{erfc}\left( \sqrt{\frac{j_{212}}{10} \frac{E_b}{N_0}} \right)
$ \\ [1ex]
\hline
\multirow{2}{*}{Parameters} & $j_0 = d_{\text{min}}^2 \approx 9.180404243585986,$ $j_1 \approx 9.226750056674900,$ $j_2 \approx 9.319441682852730$\\
 & $j_3 \approx 9.356997193814088, \ldots$ and $j_{212} \approx 11.442066233276130$ \\ \hline
\multicolumn{2}{|l|}{\# of terms used in the \gls{bep} approximation $ = 213$} \\ \hline
\multicolumn{2}{|l|}{\# of iterations to stabilize the transfer functions $ = 154$} \\ [1ex] 
\hline
\end{tabular}
\end{table}

\begin{table}[!htbp]
\caption{Partially determined TFs, $T(D,N)$ and $\dot{T}(D),$ and BEP approximation for $L_0=8$ and $L_1=1$ (First $100.000$ error events used for the determination of the first terms of $T(D,N)$ and $\dot{T}(D)$).}
\label{table:Transfer Functions L_0=8 L_1=1}
\small
\centering
\begin{tabular}{|c||c|}
\hline
\multirow{10}{*}{$T(D,N)$} & 
$(2N+6N^2+6N^3+2N^4+4N^7+8N^8+18N^9+30N^{10}+34N^{11}+38N^{12}+26N^{13}$ \\ 
& $10N^{14}+10N^{15}+10N^{16}+4N^{17}) D^{j_0}$ \\
& $+(2N^7+4N^8+6N^9+10N^{10}+10N^{11}+10N^{12}+10N^{13}+6N^{14}+4N^{15}$ \\
& $+2N^{16}) D^{j_1}$ \\
& $+(2N^{11}+10N^{12}+20N^{13}+24N^{14}+24N^{15}+20N^{16}+12N^{17}+8N^{18}+6N^{19}$ \\
& $+2N^{20}) D^{j_2}$ \\
& $+(2N^7+10N^8+18N^9+14N^{10}+6N^{11}+6N^{12}+6N^{13}+2N^{14}) D^{j_3}$ \\
& $+(2N^{12}+4N^{13}+8N^{14}+34N^{15}+86N^{16}+140N^{17}+174N^{18}+164N^{19}+122N^{20}$ \\
& $+104N^{21}+108N^{22}+126N^{23}+202N^{24}+290N^{25}+328N^{26}+364N^{27}+364N^{28}$ \\
& $+298N^{29}+226N^{30}+128N^{31}+40N^{32}+12N^{33}+4N^{34}) D^{j_{240}} + \cdots$ \\ 
\hline
$\dot{T}(D)$ & $\frac{2781}{340}D^{j_0} 
+ \frac{788}{1449}D^{j_1} + \frac{433}{3918}D^{j_2}
+ \frac{1917}{2048}D^{j_3} + \cdots + \frac{337}{3485}D^{j_{240}} + \cdots$ \\ [0.5ex]
\hline 
\multirow{2}{*}{\gls{bep} approximation} & $\frac{2781}{1360} \operatorname{erfc}\left( \sqrt{\frac{j_0}{10} \frac{E_b}{N_0}} \right) + \frac{197}{1449} \operatorname{erfc}\left( \sqrt{\frac{j_1}{10} \frac{E_b}{N_0}} \right) + \frac{433}{15672} \operatorname{erfc}\left( \sqrt{\frac{j_2}{10} \frac{E_b}{N_0}} \right)$ \\
& $+ \frac{1917}{8192} \operatorname{erfc}\left( \sqrt{\frac{j_3}{10} \frac{E_b}{N_0}} \right)
+ \cdots + \frac{337}{13940} \operatorname{erfc}\left( \sqrt{\frac{j_{240}}{10} \frac{E_b}{N_0}} \right)
$ \\ [1ex]
\hline
\multirow{2}{*}{Parameters} & $j_0 = d_{\text{min}}^2 \approx 9.678555405819292,$ $j_1 \approx 9.749530044767258,$ $j_2 \approx 9.755225239167640$\\
 & $j_3 \approx 9.790712558641621, \ldots$ and $j_{240} \approx 11.955346421670228$ \\ \hline
\multicolumn{2}{|l|}{\# of terms used in the \gls{bep} approximation $ = 241$} \\ \hline
\multicolumn{2}{|l|}{\# of iterations to stabilize the transfer functions $ = 50$} \\ [1ex] 
\hline
\end{tabular}
\end{table}

\subsection{NSMs with Rational Filter Taps}
\label{Rate 2 Approaching NSM Simple Rational Coefficients}

This subsection investigates \glspl{nsm} where the shaping filters $h_0$ and $h_1$ have \emph{rational-valued taps}. Such a restriction facilitates practical implementation, especially in fixed-point arithmetic systems. The filter $h_0$ is indirectly specified via a \emph{pattern} $\bm{\pi}_0$ whose entries are positive integers (with some pathological \gls{3d} cases allowing zeros), while $h_1$ has a single non-zero tap. The filters are scaled to have equal squared Euclidean norms, a balanced energy allocation that is fundamental for enabling the \gls{nsm} to asymptotically approach the performance of coded $2$-ASK when combined with appropriate error-correcting codes and iterative decoding techniques such as turbo-equalization.

In the \gls{1d} setting, a first family of rational-tap \glspl{nsm} extends the ideas from the rate-$5/4$ \gls{nsm} of Subsection~\ref{ssec:Minimum Euclidean Distance Guaranteeing 5/4-NSM}, which is based on the pattern $\bm{\pi}_0 = (1,1,1,1)$. Because this pattern has squared norm $2$, the corresponding filter $h_1[k]$ consists of a single non-zero tap equal to $2$ at $k=0$. Consequently, the filters $h_0[k]$ for this family have non-zero taps drawn from the bipolar set $\{-1,+1\}$. Empirical optimization using an intuitive, straightforward algorithm inspired by metaheuristic approaches such as simulated annealing, performed in Subsection~\ref{Rate-2 guaranteeing, minimum Euclidean distance approaching NSMs with real filters' coefficients} on filters with real-valued taps, consistently shows that the best-performing \glspl{nsm} adopt this structure with $h_1[k]$ containing exactly one non-null coefficient. This configuration arises naturally from performance considerations and, additionally, greatly facilitates analytical treatment.

These \gls{1d} \glspl{nsm}, with pattern $\bm{\pi}_0 = (1,1,1,1)$, consistently exhibit an \gls{msed} that is \emph{exactly half} that of $2$-ASK. The second-lowest \gls{sed}, however, matches that of $2$-ASK. Coupled with the fact that the multiplicity of minimum-distance error events decreases exponentially as a function of their length, which itself grows exponentially with the filter length $L_0$, this behavior allows such \glspl{nsm} to achieve performance comparable to $2$-ASK for practical \gls{bep} ranges (\gls{bep} values greater than $10^{-10}$) when $L_0 \geq 8$.

To overcome the inherent distance limitation associated with the pattern $\bm{\pi}_0 = (1,1,1,1)$, additional \gls{1d} \glspl{nsm} based on larger patterns will be investigated. These include, for example, $\bm{\pi}_0 = (1,1,1,1,1,1,1,1,1)$, $\bm{\pi}_0 = (2,2,2,2,2,2,1)$, and $\bm{\pi}_0 = (3,3,3,2,2,1)$. These patterns define longer filters $h_0$, and all non-equivalent candidates (under performance-preserving transformations) will be systematically assessed in terms of their achievable \gls{msed}. Some of these candidates may be unable to reach the $2$-ASK \gls{msed} regardless of $L_0$. Others may show increasing minimum distances with $L_0$, while still remaining below that of $2$-ASK. Only a select few are expected to achieve the \gls{msed} of $2$-ASK for sufficiently large $L_0$, though such conclusions can only be drawn based on the analysis of error events up to a finite maximum length (e.g., $10^4$). Therefore, the possibility remains that longer error events could exhibit smaller distances.

Moving to \gls{2d} \glspl{nsm}, the filter $h_0[k,l]$ adopts a \gls{2d} footprint, initially based on the same pattern $\bm{\pi}_0 = (1,1,1,1)$, now interpreted over a $2 \times 2$ support. The filter $h_1[k,l]$ remains a single tap at $(0,0)$, with value exactly $2$, ensuring balanced energy with $h_0[k,l]$ due to the norm of the pattern being $2$. Unlike in \gls{1d}, where a unit shift causes an overlap of $L_0 - 1$ taps, a unit shift in \gls{2d} results in reduced overlaps thanks to the spatial layout of the footprint. Specifically, for the $2 \times 2$ support, a horizontal or vertical unit shift causes an overlap of $2$ taps, while a diagonal unit shift results in an overlap of only $1$ tap. These reduced overlaps significantly mitigate \gls{isi} and contribute to lower multiplicities of error events, distinguishing \gls{2d} \glspl{nsm} from their \gls{1d} counterparts.

To mitigate the higher multiplicities inherent in small-footprint designs, such as those based on the $2 \times 2$ support and the pattern $\bm{\pi}_0 = (1,1,1,1)$, the study will extend to \glspl{nsm} built upon $3 \times 3$ footprints for the filter $h_0[k,l]$. These configurations are guided by integer-valued patterns of size $9$, and the non-zero taps of $h_0[k,l]$ are obtained through sign changes and interleaving of the components of the chosen pattern. The accompanying filter $h_1[k,l]$ consists of a single non-zero tap at $(0,0)$, whose value is set such that $h_0[k,l]$ and $h_1[k,l]$ have equal squared norms. These larger $3 \times 3$ footprints are considered promising for their potential to reduce the multiplicity of low-distance error events, albeit at the cost of reduced rate for finite spatial supports. Nonetheless, as the \gls{2d} spatial extent increases, the achievable rate asymptotically approaches $2$, matching the rate of $4$-ASK modulation.

The best-performing \gls{2d} \glspl{nsm} are identified by projecting the \gls{2d} configurations onto \gls{1d} \glspl{nsm} along four directions: horizontal, vertical, and the two diagonals. These projections are then analyzed using the \gls{rtf} introduced in Appendix~\ref{app:Tight Estimate BEP Rate 2}, specifically Equation~\eqref{eq:Multiplicity and Distance Rate-2 Modulation}, and computed via the simplified Algorithm~\ref{alg:Simplified One-Shot Two-Step Reduced Transfer Function T(D) Rate-(Q+1)/Q NSM}. Only those \gls{2d} configurations whose directional projections all achieve the \gls{msed} of $2$-ASK \emph{and} exhibit the lowest associated multiplicities for that minimum distance are retained for further analysis. This ensures that the full \gls{2d} \gls{nsm} inherits the favorable performance characteristics of its \gls{1d} projections.

\begin{algorithm}
\caption{Simplified one-shot, \emph{two-step}, symbolic determination of the RTF, $\dot{T}(D),$ of a rate-$(Q+1)/Q$ \gls{nsm} and its truncated version $\dot{T}(D;P),$ \emph{with} prior computation of TF $T(N,D)$}\label{alg:Simplified One-Shot Two-Step Reduced Transfer Function T(D) Rate-(Q+1)/Q NSM}
\begin{algorithmic}[1]
\Require $\mathring{\bm{h}}_0 = (\mathring{h}_0[0], \mathring{h}_0[1], \ldots, \mathring{h}_0[L_0-1]),$ $\mathring{\bm{h}}_1 = (\mathring{h}_1[0]),$ $Q \ge 1$ and $Q \mid L_0,$ $P$
\Ensure $\dot{T}(D),$ $\dot{T}(D;P)$

\State $k \gets 0$
\State $M \gets L_0/Q$ 
\State $\Sigma \gets \{ \bm{\sigma} = (\Delta \bar{b}_0[k-(M-2)], \ldots, \Delta \bar{b}_0[k-1], \Delta \bar{b}_0[k]), \Delta \bar{b}_0[k-m]) \in \{0, \pm 2\}, 0 \le m < M-1 \}$ 
\State $\Sigma^* \gets \{ \bm{\sigma} \in \Sigma | \bm{\sigma} \ne \bm{0}\}$ 
\State $\Sigma^{+*} \gets \{ \bm{\sigma} = (\sigma_0, \sigma_1, \ldots, \sigma_{M-2}) \in \Sigma^*, \sigma_m > 0 \text{ where } m = \min\{k | \sigma_k \ne 0\} \}$ 

\For{$\Delta \bar{b}_0[k-m] \in \{0, \pm 2\}, m=0,1,\ldots,M-1,$}

\State $\bm{\sigma}[k-1] \gets (\Delta \bar{b}_0[k-(M-1)], \ldots, \Delta \bar{b}_0[k-2], \Delta \bar{b}_0[k-1])$
\State $\bm{\sigma}[k] \gets (\Delta \bar{b}_0[k-(M-2)], \ldots, \Delta \bar{b}_0[k-1], \Delta \bar{b}_0[k])$
\State $i_0 \gets \tfrac{1}{2} |\Delta \bar{b}_0[k]|$
\State $\Delta \mathring{\bm{s}}_0 = (\Delta \mathring{s}_0[kQ], \Delta \mathring{s}_0[kQ+1], \ldots, \Delta \mathring{s}_0[kQ+Q-1]) \gets \sum_{m=0}^{M-1} \Delta \bar{b}_0[k-m] (\mathring{h}_0[mQ], \mathring{h}_0[mQ+1], \ldots, \mathring{h}_0[mQ+Q-1])$
\State Let $ N, D $ be two symbolic variables
\State $L_{\bm{\sigma}[k-1]\bm{\sigma}[k]}(N,D) \gets 1$

\For{$l=0,1,\ldots,Q-1,$} \Comment{We work subsection-by-subsection in a section of the state diagram}
\State $\xi(N,D) \gets 0$
\For{$\Delta \bar{b}_1[kQ+l] \in \{0, \pm 2\}$}
 \State $\Delta \mathring{\bm{s}}[kQ+l] \gets \Delta \mathring{s}_0[kQ+l] + \mathring{h}_1[0] \Delta \bar{b}_1[kQ+l]$
 \State $i_1 \gets \tfrac{1}{2} |\Delta \bar{b}_1[kQ+l]|$
 \State $j \gets |\Delta \mathring{\bm{s}}[kQ+l]|^2$
 \State $\xi(N,D) \gets \xi(N,D) + N^{i_1} D^j$
\EndFor
\State $L_{\bm{\sigma}[k-1]\bm{\sigma}[k]}(N,D) \gets \xi(N,D) L_{\bm{\sigma}[k-1]\bm{\sigma}[k]}(N,D) $
\EndFor

\State $L_{\bm{\sigma}[k-1]\bm{\sigma}[k]}(N,D) \gets N^{i_0} L_{\bm{\sigma}[k-1]\bm{\sigma}[k]}(N,D)$
\EndFor

\State $L_{\bm{0}\bm{0}}(N,D) \gets L_{\bm{0}\bm{0}}(N,D)-1$ \label{step:last simplified two-step algorithm common step}

\State $L_{\bm{\sigma}^\prime \bm{\sigma}}^+(N,D) \gets L_{\bm{\sigma}^\prime \bm{\sigma}}(N,D)+L_{-\bm{\sigma}^\prime \bm{\sigma}}(N,D), \bm{\sigma}^\prime \in \Sigma^+,  \bm{\sigma} \in \Sigma^+ \cup \{ \bm{0} \}$
\State $L_{\bm{\sigma}^\prime \bm{\sigma}}^+(N,D) \gets L_{\bm{\sigma}^\prime \bm{\sigma}}(N,D), \bm{\sigma}^\prime = \bm{0}, \bm{\sigma} \in \Sigma^+ \cup \{ \bm{0} \}$


\State $\bm{L}^+(N,D) \gets (L_{\bm{\sigma}^\prime \bm{\sigma}}^+(N,D))_{(\bm{\sigma}^\prime, \bm{\sigma}) \in (\Sigma^+ \cup \{ \bm{0} \}) \times (\Sigma^+ \cup \{ \bm{0} \})}$


\State Let $\bm{T}_+^e(N,D) = (T_{\bm{\sigma}}^e(N,D))_{\bm{\sigma} \in \Sigma^+ \cup \{ \bm{0} \}}$ be a symbolic $1 \times (|\Sigma^+|+1)$ row vector

\State $\bm{T}_+^s(N,D) \gets \bm{T}_+^e(N,D)$
\State $T^s_{\bm{0}}(N,D) \gets 1$

\State $\bm{T}_+^e(N,D) \gets  \text{Solve} \left( \bm{T}_+^e(N,D) =  \bm{T}_+^s(N,D) \bm{L}^+(N,D)  \right)$

\State $T(N,D) \gets T^e_{\bm{0}}(N,D)$

\State $\dot{T}(D) \gets \left. N \tfrac{\partial }{\partial N} T(N,D)\right|_{N=1/2}$
\State $\dot{T}(D;P) \gets \text{Taylor} \left(\dot{T}(D), P\right)$

\end{algorithmic}
\end{algorithm}

The approach developed for evaluating \gls{2d} \glspl{nsm} extends naturally to the \gls{3d} case. In this setting, the filter $h_0[k,l,m]$ is supported over a $2 \times 2 \times 2$ footprint defined by a pattern of size $8$ with non-negative integer components, chosen to ensure that both $h_0[k,l,m]$ and $h_1[k,l,m]$ have rational-valued taps and balanced squared Euclidean norms. The filter $h_1[k,l,m]$ remains a single non-zero tap at position $(0,0,0)$. Performance is evaluated through projections of the \gls{3d} \gls{nsm} onto multiple \gls{1d} directions: \gls{f2f}, including horizontal, vertical, and in-depth directions, as well as \gls{e2e} projections, which capture diagonal traversals along the cube’s edges. As in the \gls{2d} case, each \gls{1d} projection is analyzed using the \gls{rtf} framework, allowing consistent assessment of achievable minimum distances and multiplicities across dimensions.

Not all \gls{2d} or \gls{3d} \glspl{nsm} are expected to satisfy the critical performance criteria required to match $2$-ASK modulation. From a logical standpoint, some candidate configurations may fail to achieve the \gls{msed} of $2$-ASK in one or more of their \gls{1d} projections, which would immediately disqualify them. Others might achieve the required minimum distance across all directions, yet still exhibit a higher multiplicity of minimum-distance error events in at least one projection. In such cases, the dominant term in the high \gls{snr} \gls{bep} upper bound would exceed that of $2$-ASK, rendering the design suboptimal. Only those configurations that are logically expected to meet both conditions across all projections—correct minimum squared distance and matching multiplicity—are considered promising and will be investigated further. This discussion is meant to clarify the types of outcomes and filtering criteria that guide the forthcoming analysis.

Performance-preserving transformations play an essential role in all cases (\gls{1d}, \gls{2d}, and \gls{3d}). In \gls{1d}, these include time-reversal and sign inversion. In \gls{2d}, richer symmetry operations such as rotations, reflections, and permutations reduce the number of configurations to explore. Three dimensions offer an even richer set of transformations, crucial for managing computational complexity. These equivalence relations help identify non-equivalent configurations without sacrificing analytical rigor.

While increasing the spatial extent improves the achievable rate—enabling it to asymptotically converge toward $2$, the rate of $4$-ASK modulation—the \gls{msed} remains constant, and the multiplicities of low-distance error events are \emph{expected} to increase with the dimensionality of the signal space. Moreover, achieving comparable rates below $2$ in higher dimensions generally requires larger spatial extents: \gls{3d} configurations demand greater extents than \gls{2d} ones, which in turn require larger extents than their \gls{1d} counterparts. Therefore, designing effective filters in higher dimensions involves a trade-off: one must judiciously choose the footprint size, the pattern used to define filter taps, and the spatial extent, so that the resulting \gls{nsm} maintains \gls{bep} performance comparable to that of $2$-ASK while approaching the higher rate of $4$-ASK.

\subsubsection{One-Dimensional NSMs}
\label{sssec:One-dimensional designed NSMs}

\paragraph*{\textbf{Foundational NSMs inspired by rate-5/4 designs}}

As pointed out above, in its normalized form, the first filter, $h_0[k],$ of length $L_0,$ is of the form $\bar{h}_0[k] = \tfrac{1}{2} (\delta[k] \pm \delta[k-k_0] \pm \delta[k-k_1] \pm \delta[k-L_0+1]),$ where $0 < k_0 < k_1 < L_0-1.$ The characteristics of the resulting \gls{nsm} are fully determined by the choice of the indices, $k_0$ and $k_1,$ of the intermediate filter taps and the signs of the coefficients of all taps, except the first one. If $L_0$ is even, then, without loss of generality, the sign of the last tap, $\delta[k-L_0+1]$, could be taken positive, due to the alternate sign change equivalence stated at the beginning of Subsection~\ref{ssec:Modulation of Rate 2}.

The objective of the optimization is to guarantee the same asymptotic performance as $2$-ASK, at high \glspl{snr}. For this the average energy per symbol, which is inherited from $4$-ASK with symbol alphabet $\{ \pm 1, \pm 3 \},$ should be split equally between normalized data inputs $\bar{b}_m,$ $m=0,1.$ Hence, as stated earlier, the generic parameter $\eta$ should be taken equal to $5/2,$ implying that $\|h_0\|^2 = \|h_1\|^2 = 5/2.$

We consider a pair of input sequences, $(b_0^l[k],b_1^l[k]),$ $l=0,1,$ and their modulated sequences $s^l[k],$ $l=0,1,$ for a characterization of the minimum achievable Euclidean distance of the generated \glspl{nsm}. As in Subsection~\ref{ssec:Modulation of Rate 2}, we introduce the associated normalized input sequences differences $\Delta \bar{b}_0[k] \triangleq \bar{b}_0^1[k]-\bar{b}_0^0[k]$ and $\Delta \bar{b}_1[k] \triangleq \bar{b}_1^1[k]-\bar{b}_1^0[k],$ which take their values in the ternary alphabet $\{0, \pm 2\},$ as well as the modulated sequence difference $\Delta s[k] \triangleq s^1[k]-s^0[k].$

Based on (\ref{eq:Mod Seq 4-ASK}), which was initially developed for $4$-ASK but is still applicable to all rate-$2$ \glspl{nsm} considered thus far, we may write
\begin{equation}
    \Delta s[k] = \sum_l \Delta \bar{b}_0[l] h_0[k-l] + \sum_l \Delta \bar{b}_1[l] h_1[k-l] =
    \Delta \bar{b}_0[k] \circledast h_0[k] + \Delta \bar{b}_1[k] \circledast h_1[k].
\end{equation}
To make the investigation easier, we consider the non-scaled version, $\Delta \bar{s}[k],$ of $\Delta s[k]$, specified as $\Delta \bar{s}[k] \triangleq \Delta s[k]/\sqrt{\eta}.$ We also introduce the scaled filters $\mathring{h}_m[k] \triangleq 2 \bar{h}_m[k],$ $m=0,1,$ as well as the scaled input sequences differences $\Delta \mathring{b}_m[k] \triangleq \Delta \bar{b}_m[k]/2,$ $m=0,1.$ Then, the non-scaled version of $\Delta s[k]$ can be appropriately recast as
\begin{equation} \label{eq:Scaled Modulated Sequence Difference One-Dimensional NSM}
\Delta \bar{s}[k] = \sum_l \Delta \mathring{b}_0[l] \mathring{h}_0[k-l] + 2 \Delta \mathring{b}_1[k] = \Delta \mathring{b}_0[k]  \circledast \mathring{h}_0[k] + 2 \Delta \mathring{b}_1[k],
\end{equation}
where we have leveraged the fact that $\mathring{h}_1[k] = 2 \delta[k].$ Both $\mathring{h}_0[k] = \delta[k] \pm \delta[k-k_0] \pm \delta[k-k_1] \pm \delta[k-L_0+1]$ and $\Delta \mathring{b}_m[k],$ $m=0,1,$ now have the desirable attribute of having their non-null taps in the bipolar set $\{ \pm 1\}.$

When an error event has a single non-null coefficient in either $\Delta \mathring{b}_0[k]$ or $\Delta \mathring{b}_1[k],$ the scaled modulated sequence difference is of the form $\Delta \bar{s}[k] = \pm \mathring{h}_0[k-l]$ or $\Delta \bar{s}[k] = \pm 2 \delta[k-l],$ for some $l \in  \mathbb{Z}.$ As a result, and in accordance with (\ref{Upper-Bound Minimum Squared Euclidean Distance}), the \gls{msed}, $d_{\text{min}}^2,$ of any resulting \gls{nsm}, is always upper bounded by $10,$ when down-scaled by $\eta = 5/2,$ because $\| \Delta \bar{s} \|^2 = 4$ in this instance.

Besides, it is useful to notice that the down-scaled \gls{sed} of any error event is always integral and even. To demonstrate this, first notice that $\Delta \bar{s}[k]$ takes values in the integer range $[-6, 6].$ Second, notice that when the second term, $2 \Delta \mathring{b}_1[k],$ in (\ref{eq:Scaled Modulated Sequence Difference One-Dimensional NSM}), is added to the first term, $\sum_l \Delta \mathring{b}_0[l] \mathring{h}_0[k-l],$ it does not modify the parity of $\Delta \bar{s}[k].$ Last but not least, notice that the partial difference sequence $\sum_l \Delta \mathring{b}_0[l] \mathring{h}_0[k-l],$ whose coefficients are in the integer range $[-4,4],$ has always an even number of odd coefficients (indeed notice that for any finite-length partial difference sequence, the sum of the components, $\sum_k (\sum_l \Delta \mathring{b}_0[l] \mathring{h}_0[k-l]) = \sum_l \Delta \mathring{b}_0[l] (\sum_k \mathring{h}_0[k-l])$ is always even, given that $\sum_k \mathring{h}_0[k-l] = \sum_k \mathring{h}_0[k]$ is even).

In summary, the down-scaled \gls{msed}, $d_{\text{min}}^2/\eta,$ may be either $2$ or $4,$ (corresponding to an effective \gls{msed} of either $5$ or $10$) with the latter value being desirable as it guarantees the asymptotic performance of $2$-ASK. To check whether it is possible to guarantee the best scaled \gls{msed} of $4$ or not, Algorithm~\ref{alg:d_min^2 Rate-2 NSM L_0>1, L_1=1} in Subsection~\ref{ssec:Determination Minimum Squared Euclidean Distance L_0>1 L_1=1} has been used for \glspl{nsm} with first filter lengths, $L_0,$ between $5$ and $15.$ The primary findings, summarized in Tables~\ref{table:Simple Filters Coefficients L_0=5 L_1=1}--\ref{table:Simple Filters Coefficients L_0=15 L_1=1}, provide, for each consider value of $L_0,$ the total number of simple normalized $h_0[k]$ filters, $\bar{h}_0[k]$, as well as the total number of non-equivalent filters. We recall that two $h_0[k]$ filters are said to be equivalent if they are identical up to a global sign change, an alternating sign change, a time shift or a time reversal. As expected, as the filter length, $L_0,$ increases, so does the total number of candidate filters $h_0[k].$
However, with regard to non-equivalent filters, whether being among the best or not, we do not see a steady increase in the number of candidates when $L_0$ increases. For example, when $L_0$ goes from $13$ to $14,$ the number of non-equivalent filters goes down from $140$ to $138.$ Also, when $L_0$ goes from $9$ to $10,$ the number of non-equivalent filters goes down from $16$ to $8.$ Tables~\ref{table:Simple Filters Coefficients L_0=5 L_1=1}--\ref{table:Simple Filters Coefficients L_0=15 L_1=1} also reveal that all investigated \glspl{nsm} have a common scaled \gls{msed} of $2,$ implying that $d_{\text{min}}^2$ is equal to $5,$ resulting in a poor asymptotic gain with respect to $4$-ASK of only $10 \log_{10}(5/4) \approx 0.9691$ dB.

Despite the fact that all of the \glspl{nsm} under investigation have subpar asymptotic gains with respect to $4$-ASK, there are substantial differences among them in terms of the shortest lengths of error events with the minimum Euclidean distance. Tables~\ref{table:Simple Filters Coefficients L_0=5 L_1=1}--\ref{table:Simple Filters Coefficients L_0=15 L_1=1} list only the equivalent representatives of those filter $h_0[k]$ candidates, whose minimum Euclidean distance shortest error events have the largest lengths, in addition to listing their number. They reveal that for all studied values of $L_0,$ the maximum possible length, $K_{\text{min}},$ of the shortest error event with minimum Euclidean distance, is equal to $2^{L_0-2},$ except for $L_0=14$, where it is equal to $2^{L_0-2}-1.$

Before providing a preliminary algebraic explanation for the reported maximum lengths of the shortest error events with minimum Euclidean distance, it is necessary to explain why $K_{\text{min}},$ the length of the shortest error event with minimum Euclidean distance, was selected as the appropriate criterion to maximize. As previously stated, every error event has an even integral value for the down-scaled \gls{sed}. As a result, for Nyquist signaling modulated data packets comprising fewer than $K_{\text{min}}$ modulated symbols, all potential error events can only have lengths that are strictly smaller than $K_{\text{min}}.$ Since an error event must have at least $K_{\text{min}}$ as length of to achieve the down-scaled \gls{msed} of $2,$ all error events within the packet must have down-scaled \glspl{sed} of at least $4.$ As a result, whenever the data packet size is taken below $K_{\text{min}},$ the optimal minimum Euclidean distance achieves asymptotically the performance of $2$-ASK at high \gls{snr}. As a direct consequence, maximizing $K_{\text{min}}$ for each filter length $L_0$ allows the adoption of the least complex modulation detection technique (with least value of $L_0$ and therefore the least number of states in the detection trellis) for each targeted data packet size.

This being said, we will now address in detail the case where $K_{\text{min}} = 2^{L_0-2},$ followed by the case when $K_{\text{min}} = 2^{L_0-2}-1.$ In this endeavor for $K_{\text{min}} = 2^{L_0-2},$ we proceed in two steps. In the first step, we establish a lower bound on the length of the shortest error event with \gls{msed}, if the latter happens to be equal to $2,$ and demonstrate that this lower bound is precisely equal to $K_{\text{min}}.$ All that is left to do in the second step is to provide evidence that at least an error event with a \gls{sed} of $2$ actually exists.

Assume that there are two finite-length input sequence differences, $\Delta \mathring{b}_0[k]$ and $\Delta \mathring{b}_1[k],$ that result in an error event with a minimum Euclidean distance of $2.$ This indicates that the squared norm, $\| \Delta \bar{s}[k] \|^2,$ of the related modulated sequence difference $\Delta \bar{s}[k],$ whose expression is provided in (\ref{eq:Scaled Modulated Sequence Difference One-Dimensional NSM}), is equal to $2.$ As a result, because $\Delta \bar{s}[k]$ takes integer values in the range $[-6,6],$ it should take the form $\pm \delta[k] \pm \delta[k-(K-1)],$ where $K$ is the length of the corresponding error event. Without loss of generality, we can assume that $\Delta \bar{s}[k]$ is of the form
$\delta[k] \pm \delta[k-(K-1)],$ at the risk of changing the signs of $\Delta \mathring{b}_0[k]$ and $\Delta \mathring{b}_1[k].$ Taking advantage of the fact that all variables in (\ref{eq:Scaled Modulated Sequence Difference One-Dimensional NSM}) take integer values, we propose to work modulo $2$ and carry out algebraic operations in the Galois Field, $\text{GF}(2)=\{0, 1\}.$

Let $\Delta \ddot{s}[k],$ $\Delta \ddot{b}_0[k],$ $\Delta \ddot{b}_1[k]$ and $\ddot{h}_0[k]$ denote respectively the binary counterparts of $\Delta \bar{s}[k],$ $\Delta \mathring{b}_0[k],$ $\Delta \mathring{b}_1[k]$ and $\mathring{h}_0[k],$ which are obtained via a modulo $2$ operation. Taking into account the above-mentioned special form of $\Delta \bar{s}[k],$ (\ref{eq:Scaled Modulated Sequence Difference One-Dimensional NSM}) transforms into
\begin{equation} \label{eq:Galois Field One-Dimensional NSM}
\Delta \ddot{s}[k] = \sum_l \Delta \ddot{b}_0[l] \ddot{h}_0[k-l] = \Delta \ddot{b}_0[k] \circledast \ddot{h}_0[k] = \delta[k] + \delta[k-(K-1)].
\end{equation}
When written in polynomial form, this equality becomes
\begin{equation} \label{eq:Galois Field One-Dimensional NSM Polynomial Form}
\Delta \ddot{s}(x) = \Delta \ddot{b}_0(x) \ddot{h}_0(x) = x^{K-1} + 1,
\end{equation}
where $\Delta \ddot{s}(x) \triangleq \sum_k \Delta \ddot{s}[k] x^k,$ $\Delta \ddot{b}_0(x) \triangleq \sum_k \Delta \ddot{b}_0[k] x^k$ and $\ddot{h}_0(x) \triangleq \sum_{k=0}^{L_0-1} \ddot{h}_0[k] x^k.$ But, with the exception of Table~\ref{table:Simple Filters Coefficients L_0=14 L_1=1}, Tables~\ref{table:Simple Filters Coefficients L_0=9 L_1=1}--\ref{table:Simple Filters Coefficients L_0=15 L_1=1} indicate that $q^n(x) \triangleq \ddot{h}_0(x)$ is always of the form $(x+1)p^n(x),$ where $p^n(x)$ is a primitive polynomial in $\text{GF}(2)[x]$ of degree $L_0-2.$ As a result, $x+1$ and $p^n(x)$ must each divide $x^{K-1} + 1.$ On the one hand, $x+1,$ which is also a primitive polynomial of degree $1,$ trivially divides $x^{K-1} + 1.$ On the other hand, algebraic theory \cite{Berlekamp68, Peterson72} shows that, because $p^n(x)$ is a primitive polynomial of degree $l \triangleq L_0-2,$ the lowest positive integer $m$ such that $p^n(x)$ divides $x^m + 1$ is $m=2^l-1.$ As a result, we can argue that $K-1 \ge m = 2^l-1 = 2^{L_0-2}-1,$ and hence that $2^{L_0-2}$ is precisely the desired lower bound on the length, $K_{\text{min}},$ of the shortest error event with $2$ as potential \gls{msed}.

We now demonstrate that there exist input differences sequences $\Delta \mathring{b}_0[k]$ and $\Delta \mathring{b}_1[k]$ that ensure $\Delta \bar{s}[k] = \delta[k] \pm \delta[k-(K-1)],$ for $K = 2^{L_0-2},$ and so prove that the lower bound, $2^{L_0-2},$ is met for $K_{\text{min}}.$ But before that, it is worth noting that if $\Delta \mathring{b}_0[k]$ and $\Delta \mathring{b}_1[k],$ with components in $\{0, \pm 1\},$ are arbitrarily chosen so that the binary sequence $\Delta \ddot{b}_0[k]$ achieves the equality in (\ref{eq:Galois Field One-Dimensional NSM}), with $K = 2^{L_0-2},$ then nothing prevents the values of $\Delta \bar{s}[k],$ $0 < k < K-1,$ which are null modulo $2,$ from taking non-null values in the set $\{\pm 2, \pm 4, \pm 6 \}.$ Obviously, this set may be reduced to the rigorous minimum, $\{\pm 2 \},$ because, from (\ref{eq:Galois Field One-Dimensional NSM}), $\sum_l \Delta \mathring{b}_0[l] \mathring{h}_0[k-l]$ must take values in $\{\pm 2, \pm 4 \}$ (remember that $\mathring{h}_0[k]$ has exactly $4$ non-null components in the set $\{\pm 1\}$) and $2 \Delta \mathring{b}_1[k],$ which takes values in $\{\pm 2 \},$ can precisely be taken opposite in sign to this sum. Nevertheless, this fact still does not guarantee that, for some carefully selected input difference sequences, $\Delta \bar{s}[k]$ is perfectly equal to $\delta[k] \pm \delta[k-(K-1)],$ with $K = 2^{L_0-2}.$

We proceed step by step, to construct a set of input differences sequences that guarantee that $\Delta \bar{s}[k] = \delta[k] \pm \delta[k-(K-1)],$ for $K = 2^{L_0-2}.$ First, we may select $\mathring{h}_0[0]$ to have an amplitude of $1,$ thanks to the equivalence relationships that have been mentioned previously more than once. Additionally, it can be deduced from (\ref{eq:Galois Field One-Dimensional NSM}) that $\Delta \ddot{b}_0[0]$ must be equal to $1,$ because $\sum_l \Delta \ddot{b}_0[l] \ddot{h}_0[k-l]$ is also equal to $1$ for $k=0.$ As a result, we can assume that $\Delta \mathring{b}_0[0]$ has an amplitude of $1,$ without losing generality. 

The aforementioned choices, together with $\Delta \mathring{b}_1[0]$ taken null, guarantee that  $\Delta \bar{s}[0]$ has an amplitude of $1.$ Now, for $0\le l \le k,$ $k < K_{\text{min}}-L_0-1,$ we assume that $\Delta \bar{s}[l]$ has been completely specified, so that its values are in $\{ 0, \pm 1 \},$ by a suitable choice of $\Delta \mathring{b}_0[l]$ and $\Delta \mathring{b}_1[l],$ with the constraint that $\Delta \ddot{b}_0[l]$ is one of those eligible sequences that ensure (\ref{eq:Galois Field One-Dimensional NSM}). At this point, we assume that $\Delta \mathring{b}_0[l]$ and $\Delta \mathring{b}_1[l],$ $k+1 \le l \le K_{\text{min}}-L_0-1,$ have not yet been determined and are set to zero beforehand. As a result, the temporary value of $\Delta \bar{s}[k+1]$ (which hasn't incorporated $\Delta \mathring{b}_0[k+1]$ and $\Delta \mathring{b}_1[k+1]$ yet) is always in the range $[-3,3],$ because $\mathring{h}_0[k+1-l],$ has only three non-null values equal to $1,$ when $l < k+1$ ($\mathring{h}_0[0],$ the multiplicative factor of $\Delta \mathring{b}_0[k+1]$ is not accounted for yet).

The following step is then to specify $\Delta \mathring{b}_0[k+1]$ and $\Delta \mathring{b}_1[k+1],$ so that $\Delta \mathring{b}_0[k+1]$ is in agreement with its binary counterpart, $\Delta \ddot{b}_0[k+1],$ which assures (\ref{eq:Galois Field One-Dimensional NSM}). If $\Delta \ddot{b}_0[k+1]$ is null, $\Delta \mathring{b}_0[k+1]$ is similarly assumed to be null. If, on the one hand, the temporary value of $\Delta \bar{s}[k+1]$ is equal to $2$ or $3$ (respectively, $-2$ or $-3$), then $\Delta \mathring{b}_1[k+1]$ is set to $-1$ (respectively, $+1$), ensuring that the final value of $\Delta \bar{s}[k+1]$ is either $0$ or $1$ (respectively, $0$ or $-1$). If, on the other hand, the temporary value of $\Delta \bar{s}[k+1]$ is in $\{ 0, \pm 1 \}$ then $\Delta \mathring{b}_1[k+1]$ can be set to $0$ to ensure that its final value remains in $\{ 0, \pm 1 \}.$

To sum up, if $\Delta \ddot{b}_0[k+1]$ is not null and hence equal to $1,$ then $\Delta \mathring{b}_0[k+1]$ is set to $-1$ (respectively, $+1$) if the temporary value of $\Delta \bar{s}[k+1]$ is positive (respectively, negative). Otherwise, the value of $\Delta \mathring{b}_0[k+1]$ can be arbitrarily set to $-1$ or $+1.$ Furthermore, if the temporary value of $\Delta \bar{s}[k+1]$ is $+3$ (respectively, $-3$), then $\Delta \mathring{b}_1[k+1]$ is taken to be $-1$ (respectively, $+1$). Otherwise, $\Delta \mathring{b}_1[k+1]$ is assumed to be null. Again, these considerations and choices ensure that the final number $\Delta \bar{s}[k+1]$ is exactly in $\{ 0, \pm 1 \}.$

When we arrive at $k = K_{\text{min}}-L_0-1,$ we supply the last non-zero value of $\Delta \mathring{b}_0[k+1],$ in the same way we did earlier. Subsequent values of $\Delta \mathring{b}_0[k+1],$ with $k \ge K_{\text{min}}-L_0,$ are taken null. Indeed, from the fact that $\mathring{h}_0[k] = \delta[k] \pm \delta[k-k_0] \pm \delta[k-k_1] \pm \delta[k-L_0+1],$ we have that $\ddot{h}_0[k] = \delta[k] + \delta[k-k_0] + \delta[k-k_1] + \delta[k-L_0+1],$ and therefore that $\ddot{h}_0(x) = 1 + x^{k_0} + x^{k_1} + x^{L_0-1}$. Then, using (\ref{eq:Galois Field One-Dimensional NSM Polynomial Form}), we deduce that $\Delta \ddot{b}_0(x)$ is of degree $(K_{\text{min}}-1)-(L_0-1) = K_{\text{min}}-L_0,$ meaning that the sequence $\ddot{b}_0[l]$ is of length $K_{\text{min}}-L_0+1$ and therefore that only values of $\ddot{b}_0[l], 0 \leq l \leq K_{\text{min}}-L_0,$ could take non-null values. The values of $\Delta \mathring{b}_1[k+1],$ $K_{\text{min}}-L_0-1 \le k \le K_{\text{min}}-3,$ are chosen similarly, while $\Delta \mathring{b}_1[k+1]$ is set to zero for $k = K_{\text{min}}-2.$ As a result, the last non-null value of $\Delta \bar{s}[k+1],$ at $k = K_{\text{min}}-2,$ is equal to $+1$ (respectively, $-1$) if the last non-null values of $\Delta \mathring{b}_0[l]$ and $\mathring{h}_0[l],$ are of the same sign (respective, opposite signs).

At this point, we have built input differences sequences $\Delta \mathring{b}_0[k]$ and $\Delta \mathring{b}_1[k]$ that ensure that the modulated difference sequence $\Delta \bar{s}[k]$ in (\ref{eq:Scaled Modulated Sequence Difference One-Dimensional NSM}) has all of its values in $\{ 0, \pm 1 \},$ while further guaranteeing that the binary counterpart, $\Delta \ddot{b}_0[k],$ of $\Delta \mathring{b}_0[k]$ is a solution to (\ref{eq:Galois Field One-Dimensional NSM}). Now, since $\Delta \bar{s}[k]$ is in $\{ 0, \pm 1 \},$ for any value of $k,$ and since, from (\ref{eq:Galois Field One-Dimensional NSM}), we know that the values modulo $2$ of $\Delta \bar{s}[k]$ are exactly equal to $0,$ for $0 < k < K_\text{min}-1,$ we conclude that $\Delta \bar{s}[k]$ is strictly null for this range of values of $k.$ On the other hand, we recall that we established that $\Delta \bar{s}[0]$ is equal to $1,$ whereas $\Delta \bar{s}[K_{\text{min}}-1]$ is either equal to $-1$ or to $+1.$ With the exception of Table~\ref{table:Simple Filters Coefficients L_0=14 L_1=1}, this completes the proof of the presence of at least one error event of length $K_{\text{min}} = 2^{L_0-2},$ for the situations in Tables~\ref{table:Simple Filters Coefficients L_0=9 L_1=1}--\ref{table:Simple Filters Coefficients L_0=15 L_1=1}.

The remaining situation in Table~\ref{table:Simple Filters Coefficients L_0=14 L_1=1}, corresponding to $L_0 = 14,$ is handled in a manner similar to what was described earlier. In this case, and in accordance with the aforementioned table, $\ddot{h}_0(x) = q^n(x)$ is always of the form $(x+1)^2 p^n(x),$ where $p^n(x)$ is a primitive polynomial in $\text{GF}(2)[x]$ of degree $L_0-3.$ As a result, if $K$ is the length of a potential shortest error event of minimal \gls{sed}, $2,$ then $(x+1)^2$ and $p^n(x)$ must each divide $x^{K-1} + 1.$ On the one hand, given that $p^n(x)$ is a primitive polynomial of degree $l \triangleq L_0-3,$ we can show, using elementary algebraic theory arguments, that the only polynomials of the form $x^m + 1$ that are divisible by $p^n(x)$ are those in which $m$ is an integer multiple of $2^l-1.$ On the other hand, in order for $(x+1)^2$ to divide $x^m + 1,$ $1$ must be a double root of $x^m + 1,$ which necessitates an even value for the exponent $m$ (remember that $(x+1)$ must divide the derivative, $m x^{m-1},$ of $x^m + 1,$ meaning that $m x^{m-1},$ must be null and therefore that $m$ must also be null modulo $2$).

In summary, $m$ must be equal to an even integer multiple of $2^l-1$ (notice here that $2^l-1$ is always odd) for any polynomial $x^m + 1$ to be divisible by $q^n(x)=(x+1)^2 p^n(x).$ Consequently, the shortest error event's length, $K,$ is such that $K-1 \ge 2(2^l-1) = 2(2^{L_0-3}-1),$ for any hypothetical \gls{sed} of $2.$ In complete agreement with the experimental value provided in Table~\ref{table:Simple Filters Coefficients L_0=14 L_1=1}, a lower bound on the length of any probable error event with a \gls{msed} of $2$ is thus $K_{\text{min}} = 2^{L_0-2}-1.$ Because $\mathring{h}_0[k]$ has exactly $4$ non-null taps with values in $\{ \pm 1 \},$ the construction of an error event with a minimum Euclidean distance of $2$ and a length equal to the lower bound $K_{\text{min}} = 2^{L_0-2}-1,$ is identical to that given previously for the other treated cases where $L_0 \ne 14.$ This brings the proof to a close.

Now that we've properly characterized the shortest length of error event with \gls{msed}, it's time to characterize all authorized lengths of other error events with the same \gls{msed}. Let then $\Delta \mathring{b}_0[k]$ and $\Delta \mathring{b}_1[k]$ be input sequence differences that result in the modulated sequence difference $\Delta \bar{s}[k] = \pm \delta[k] \pm \delta[k-(K-1)],$ of an error event, with \gls{msed}, of length $K > K_{\text{min}}.$ We recall that the expressions of $\Delta \bar{s}[k],$ its binary counterpart, $\Delta \ddot{s},$ and its polynomial form, $\Delta \ddot{s}(x),$ in (\ref{eq:Scaled Modulated Sequence Difference One-Dimensional NSM}), (\ref{eq:Galois Field One-Dimensional NSM}) and (\ref{eq:Galois Field One-Dimensional NSM Polynomial Form}), respectively, still hold. Let then $\kappa \geq 1$ and $\rho < K_{\text{min}}-1$ respectively denote the quotient and the rest of the Euclidean division of $K-1$ by $K_{\text{min}}-1.$ Then, we can rewrite $x^{K-1}+1$ as $x^{(K_{\text{min}}-1) \kappa + \rho}+1 = (x^{(K_{\text{min}}-1)})^\kappa x^\rho+1 = ((x^{(K_{\text{min}}-1)})^\kappa + 1)x^\rho + x^\rho + 1.$ Now, since $(x^{(K_{\text{min}}-1)})^\kappa + 1$ is divisible by $x^{(K_{\text{min}}-1)} + 1,$ and since $\rho < K_{\text{min}}-1,$ we can affirm that $x^\rho+1$ is the rest of the polynomial division of $x^{K-1}+1$ by $x^{(K_{\text{min}}-1)} + 1.$ Now remember, from (\ref{eq:Galois Field One-Dimensional NSM Polynomial Form}), that both $x^{K-1}+1$ and $x^{(K_{\text{min}}-1)} + 1$ are multiples of $\ddot{h}_0(x) = q^n(x).$ Hence, $x^\rho + 1$ is also a multiple of $\ddot{h}_0(x) = q^n(x).$ But, also remember from the above that $K_{\text{min}}$ is the minimum value of $K$ such that $x^{(K_{\text{min}}-1)} + 1$ is divisible by $q^n(x).$ Since $\rho < K_{\text{min}}-1,$ we conclude that $\rho$ must be null. As a consequence, we can confidently say that $K-1$ must be equal to $\kappa (K_{\text{min}}-1),$ for some $\kappa \geq 1.$ This means that all allowed error events with \gls{msed} must have lengths of the form $K = \kappa (K_{\text{min}}-1) + 1,$ for some $\kappa \geq 1.$ This result is precisely used in Appendix~\ref{app:Characterization Second Minimum Euclidean Distance Rate-2 NSMs Simple Coefficients}.

\begin{figure}[!htbp]
    \centering
    \includegraphics[width=1.0\textwidth]{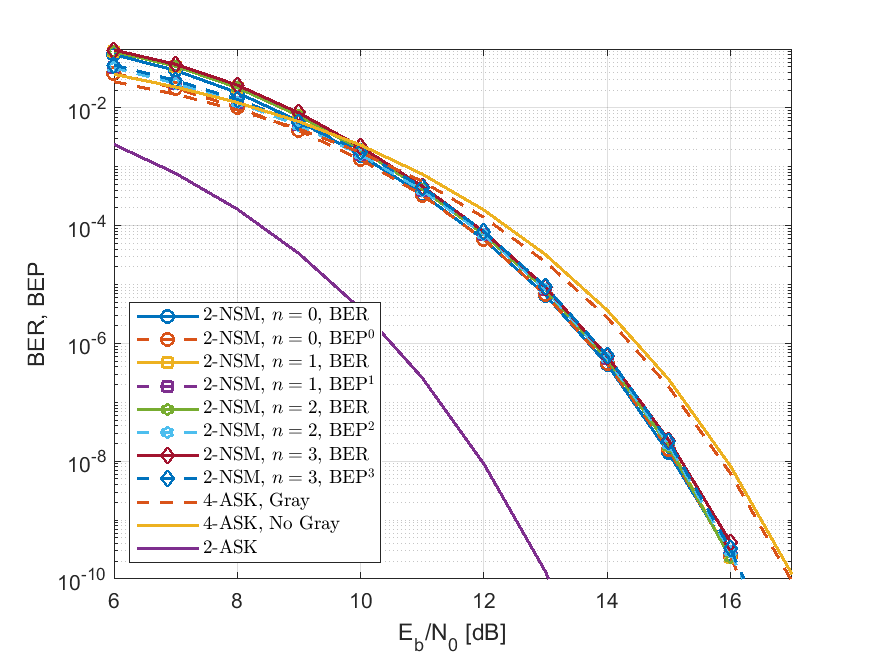}
    \caption{BER of the NSM of rate $2,$ with simplest filter coefficients, given in Table~\ref{table:Simple Filters Coefficients L_0=5 L_1=1}, with $L_0 = 5$ and $L_1 = 1$ ($2\bar{\bm{h}}^0_0=(1,1,-1,0,-1)$, $2\bar{\bm{h}}^1_0=(1,1,-1,0,1)$, $2\bar{\bm{h}}^2_0=(1,1,1,0,-1)$ and $2\bar{\bm{h}}^3_0=(1,1,1,0,1)$). For reference, the BEPs of $2$-ASK and Gray and non-Gray precoded $4$-ASK conventional modulations are shown.}
    \label{fig:BER-BEP-NSM-2-FilterSimpleCoefficients-FilterLength_5}
\end{figure}

\begin{figure}[!htbp]
    \centering
    \includegraphics[width=1.0\textwidth]{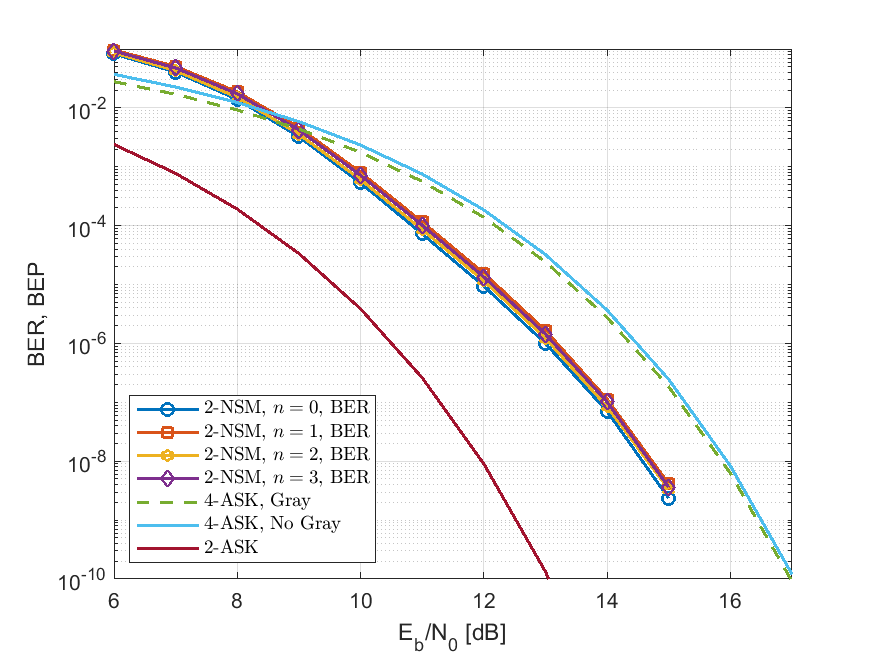}
    \caption{BER of the NSM of rate $2,$ with simplest filter coefficients, given in Table~\ref{table:Simple Filters Coefficients L_0=6 L_1=1}, with $L_0 = 6$ and $L_1 = 1$ ($2\bar{\bm{h}}^0_0=(1,1,0,-1,0,-1)$, $2\bar{\bm{h}}^1_0=(1,1,0,-1,0,1)$, $2\bar{\bm{h}}^2_0=(1,1,0,1,0,-1)$ and $2\bar{\bm{h}}^3_0=(1,1,0,1,0,1)$). For reference, the BEPs of $2$-ASK and Gray and non-Gray precoded $4$-ASK conventional modulations are shown.}
    \label{fig:BER-BEP-NSM-2-FilterSimpleCoefficients-FilterLength_6}
\end{figure}

\begin{figure}[!htbp]
    \centering
    \includegraphics[width=1.0\textwidth]{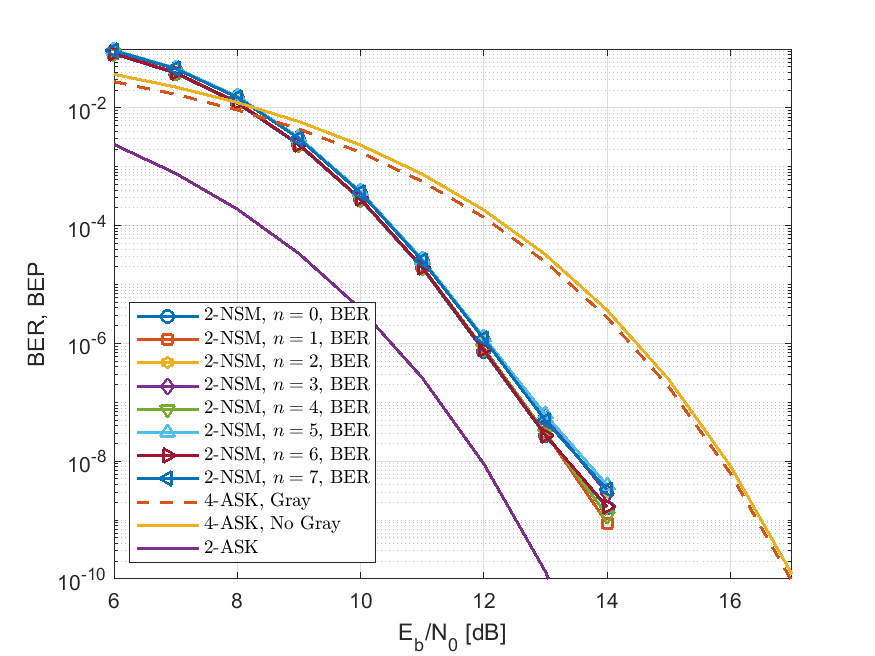}
    \caption{BER of the NSM of rate $2,$ with simplest filter coefficients, given in Table~\ref{table:Simple Filters Coefficients L_0=7 L_1=1}, with $L_0 = 7$ and $L_1 = 1$ ($2\bar{\bm{h}}^0_0=(1,1,-1,0,0,0,-1)$, $2\bar{\bm{h}}^1_0=(1,1,-1,0,0,0,1)$, $2\bar{\bm{h}}^2_0=(1,1,1,0,0,0,-1)$, $2\bar{\bm{h}}^3_0=(1,1,1,0,0,0,1)$, $2\bar{\bm{h}}^4_0=(1,0,-1,1,0,0,-1)$, $2\bar{\bm{h}}^5_0=(1,0,-1,1,0,0,1)$, $2\bar{\bm{h}}^6_0=(1,0,1,1,0,0,-1)$ and $2\bar{\bm{h}}^7_0=(1,0,1,1,0,0,1)$). For reference, the BEPs of $2$-ASK and Gray and non-Gray precoded $4$-ASK conventional modulations are shown.}
    \label{fig:BER-BEP-NSM-2-FilterSimpleCoefficients-FilterLength_7}
\end{figure}

\begin{figure}[!htbp]
    \centering
    \includegraphics[width=1.0\textwidth]{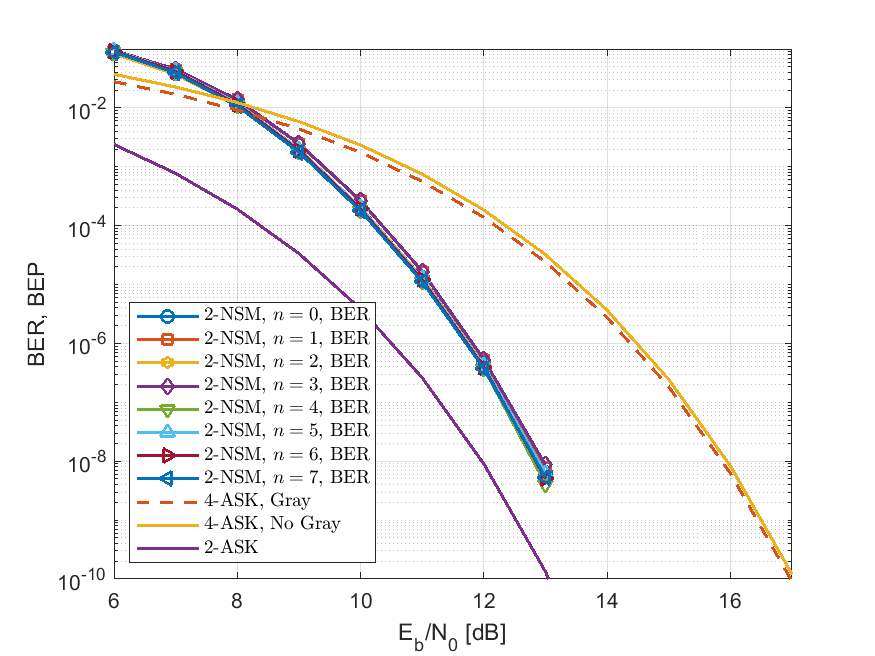}
    \caption{BER of the NSM of rate $2,$ with simplest filter coefficients, given in Table~\ref{table:Simple Filters Coefficients L_0=8 L_1=1}, with $L_0 = 8$ and $L_1 = 1$ ($2\bar{\bm{h}}^0_0=(1,0,-1,0,-1,0,0,1)$, $2\bar{\bm{h}}^1_0=(1,0,-1,0,1,0,0,1)$, $2\bar{\bm{h}}^2_0=(1,0,1,0,-1,0,0,1)$, $2\bar{\bm{h}}^3_0=(1,0,1,0,1,0,0,1)$, $2\bar{\bm{h}}^4_0=(1,1,0,0,0,-1,0,-1)$, $2\bar{\bm{h}}^5_0=(1,1,0,0,0,-1,0,1)$, $2\bar{\bm{h}}^6_0=(1,1,0,0,0,1,0,-1)$ and $2\bar{\bm{h}}^7_0=(1,1,0,0,0,1,0,1)$). For reference, the BEPs of $2$-ASK and Gray and non-Gray precoded $4$-ASK conventional modulations are shown.}
    \label{fig:BER-BEP-NSM-2-FilterSimpleCoefficients-FilterLength_8}
\end{figure}

\begin{figure}[!htbp]
    \centering
    \includegraphics[width=1.0\textwidth]{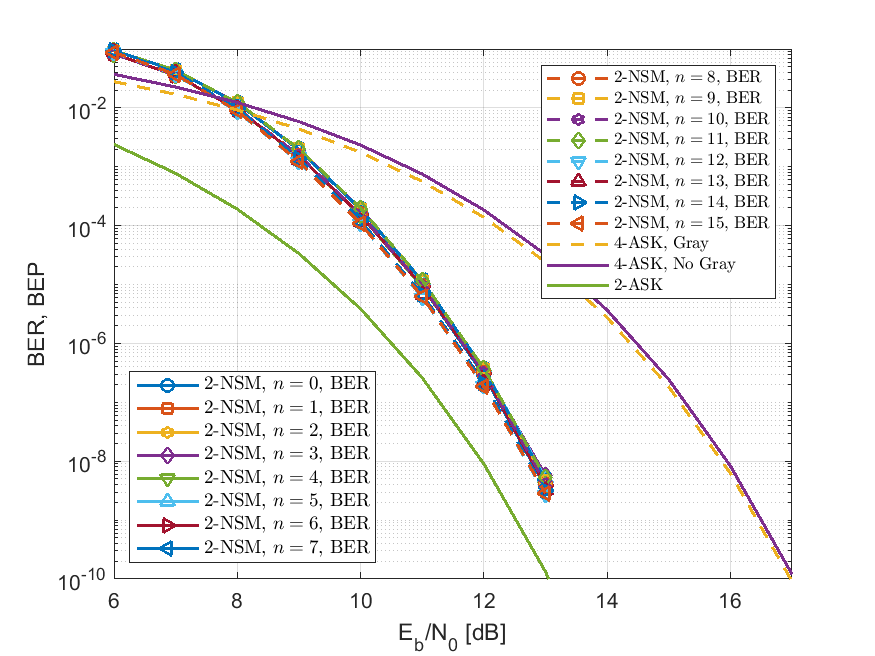}
    \caption{BER of the NSM of rate $2,$ with simplest filter coefficients, given in Table~\ref{table:Simple Filters Coefficients L_0=9 L_1=1}, with $L_0 = 9$ and $L_1 = 1$  ($2\bar{\bm{h}}^0_0=(1,1,-1,0,0,0,0,0,-1)$, $2\bar{\bm{h}}^1_0=(1,1,-1,0,0,0,0,0,1)$,
    $2\bar{\bm{h}}^2_0=(1,1,1,0,0,0,0,0,-1)$,
    $2\bar{\bm{h}}^3_0=(1,1,1,0,0,0,0,0,1)$,
    $2\bar{\bm{h}}^4_0=(1,1,0,0,-1,0,0,0,-1)$, $2\bar{\bm{h}}^5_0=(1,1,0,0,-1,0,0,0,1)$,
    $2\bar{\bm{h}}^6_0=(1,1,0,0,1,0,0,0,-1)$,
    $2\bar{\bm{h}}^7_0=(1,1,0,0,1,0,0,0,1)$, 
    $2\bar{\bm{h}}^8_0=(1,0,0,1,-1,0,0,0,-1)$, $2\bar{\bm{h}}^9_0=(1,0,0,1,-1,0,0,0,1)$, $2\bar{\bm{h}}^{10}_0=(1,0,0,1,1,0,0,0,-1)$, $2\bar{\bm{h}}^{11}_0=(1,0,0,1,1,0,0,0,1)$, $2\bar{\bm{h}}^{12}_0=(1,1,0,0,0,0,-1,0,-1)$, $2\bar{\bm{h}}^{13}_0=(1,1,0,0,0,0,-1,0,1)$, $2\bar{\bm{h}}^{14}_0=(1,1,0,0,0,0,1,0,-1)$ and $2\bar{\bm{h}}^{15}_0=(1,1,0,0,0,0,1,0,1)$). For reference, the BEPs of $2$-ASK and Gray and non-Gray precoded $4$-ASK conventional modulations are shown.}
    \label{fig:BER-BEP-NSM-2-FilterSimpleCoefficients-FilterLength_9}
\end{figure}

\begin{figure}[!htbp]
    \centering
    \includegraphics[width=1.0\textwidth]{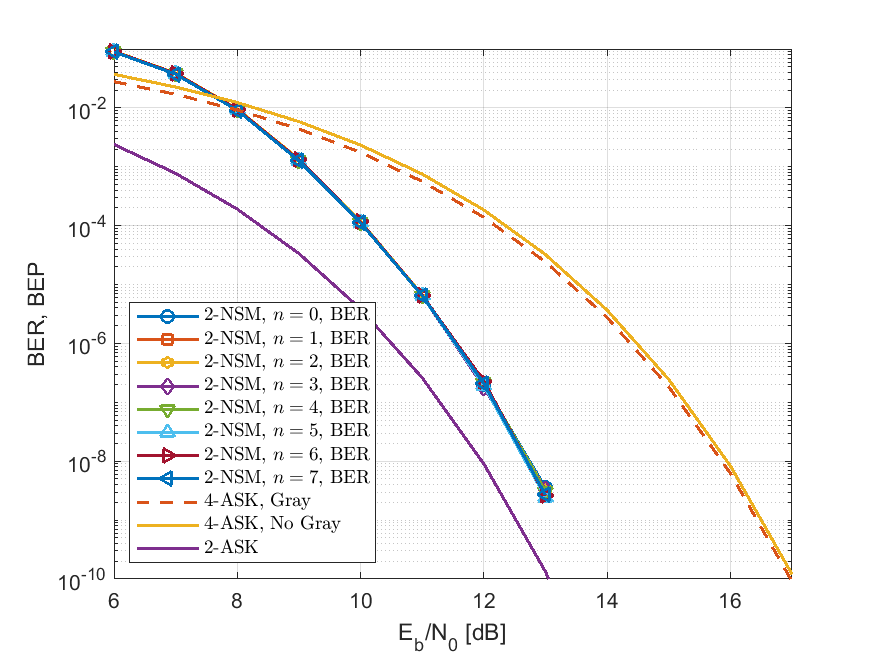}
    \caption{BER of the NSM of rate $2,$ with simplest filter coefficients, given in Table~\ref{table:Simple Filters Coefficients L_0=10 L_1=1}, with $L_0 = 10$ and $L_1 = 1$ ($2\bar{\bm{h}}^0_0=(1,0,0,1,0,-1,0,0,0,-1)$, $2\bar{\bm{h}}^1_0=(1,0,0,1,0,-1,0,0,0,1)$, $2\bar{\bm{h}}^2_0=(1,0,0,1,0,1,0,0,0,-1)$, $2\bar{\bm{h}}^3_0=(1,0,0,1,0,1,0,0,0,1)$, $2\bar{\bm{h}}^4_0=(1,0,-1,0,0,0,-1,0,0,1)$, $2\bar{\bm{h}}^5_0=(1,0,-1,0,0,0,1,0,0,1)$, $2\bar{\bm{h}}^6_0=(1,0,1,0,0,0,-1,0,0,1)$ and $2\bar{\bm{h}}^7_0=(1,0,1,0,0,0,1,0,0,1)$). For reference, the BEPs of $2$-ASK and Gray and non-Gray precoded $4$-ASK conventional modulations are shown.}
    \label{fig:BER-BEP-NSM-2-FilterSimpleCoefficients-FilterLength_10}
\end{figure}

\begin{figure}[!htbp]
    \centering
    \includegraphics[width=1.0\textwidth]{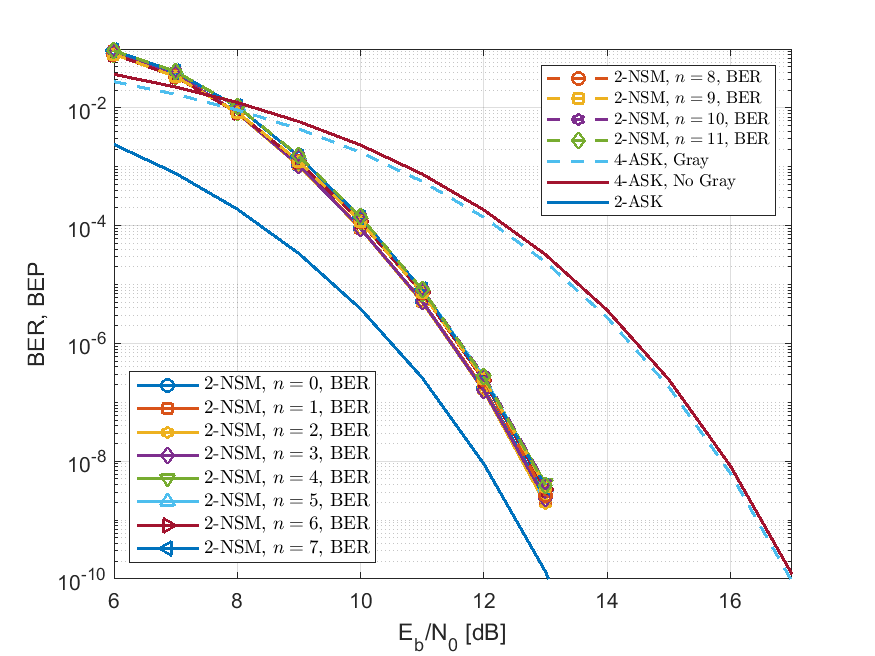}
    \caption{BER of the NSM of rate $2,$ with simplest filter coefficients, given in Table~\ref{table:Simple Filters Coefficients L_0=11 L_1=1}, with $L_0 = 11$ and $L_1 = 1$  ($2\bar{\bm{h}}^0_0=(1,0,-1,1,0,0,0,0,0,0,-1)$, $2\bar{\bm{h}}^1_0=(1,0,-1,1,0,0,0,0,0,0,1)$, $2\bar{\bm{h}}^2_0=(1,0,1,1,0,0,0,0,0,0,-1)$, $2\bar{\bm{h}}^3_0=(1,0,1,1,0,0,0,0,0,0,1)$, 
    $2\bar{\bm{h}}^4_0=(1,0,-1,0,0,1,0,0,0,0,-1)$, $2\bar{\bm{h}}^5_0=(1,0,-1,0,0,1,0,0,0,0,1)$, $2\bar{\bm{h}}^6_0=(1,0,1,0,0,1,0,0,0,0,-1)$, $2\bar{\bm{h}}^7_0=(1,0,1,0,0,1,0,0,0,0,1)$, 
    $2\bar{\bm{h}}^8_0=(1,0,0,1,0,0,-1,0,0,0,-1)$, $2\bar{\bm{h}}^9_0=(1,0,0,1,0,0,-1,0,0,0,1)$, $2\bar{\bm{h}}^{10}_0=(1,0,0,1,0,0,1,0,0,0,-1)$ and $2\bar{\bm{h}}^{11}_0=(1,0,0,1,0,0,1,0,0,0,1)$). For reference, the BEPs of $2$-ASK and Gray and non-Gray precoded $4$-ASK conventional modulations are shown.}
    \label{fig:BER-BEP-NSM-2-FilterSimpleCoefficients-FilterLength_11}
\end{figure}

\begin{figure}[!htbp]
    \centering
    \includegraphics[width=1.0\textwidth]{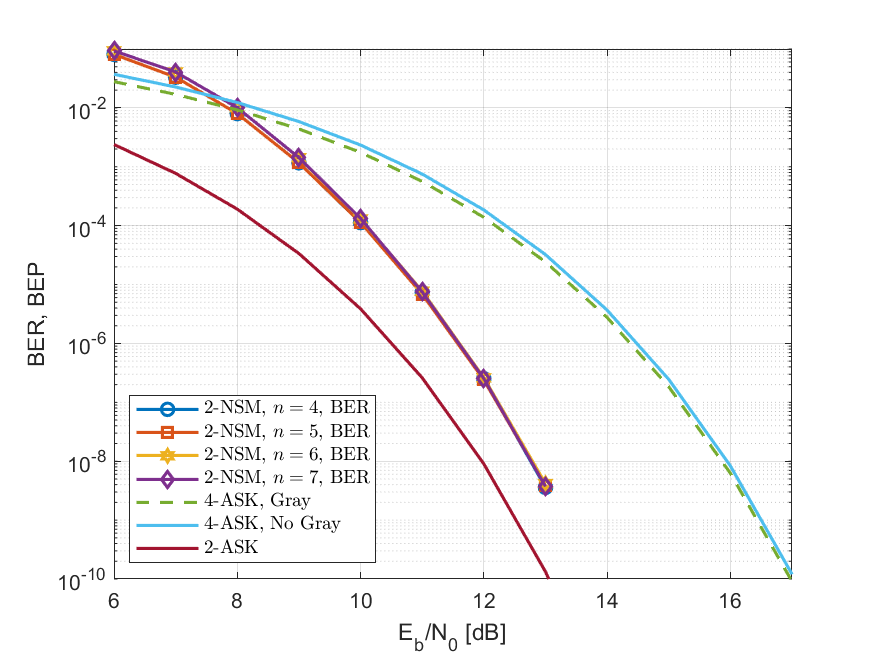}
    \caption{BER of the NSM of rate $2,$ with simplest filter coefficients, given in Table~\ref{table:Simple Filters Coefficients L_0=12 L_1=1}, with $L_0 = 12$ and $L_1 = 1$ ($2\bar{\bm{h}}^4_0=(1,0,-1,0,-1,0,0,0,0,0,0,1)$, $2\bar{\bm{h}}^5_0=(1,0,-1,0,1,0,0,0,0,0,0,1)$, $2\bar{\bm{h}}^6_0=(1,0,1,0,-1,0,0,0,0,0,0,1)$ and $2\bar{\bm{h}}^7_0=(1,0,1,0,1,0,0,0,0,0,0,1)$). For reference, the BEPs of $2$-ASK and Gray and non-Gray precoded $4$-ASK conventional modulations are shown.}
    \label{fig:BER-BEP-NSM-2-FilterSimpleCoefficients-FilterLength_12}
\end{figure}

\begin{figure}[!htbp]
    \centering
    \includegraphics[width=1.0\textwidth]{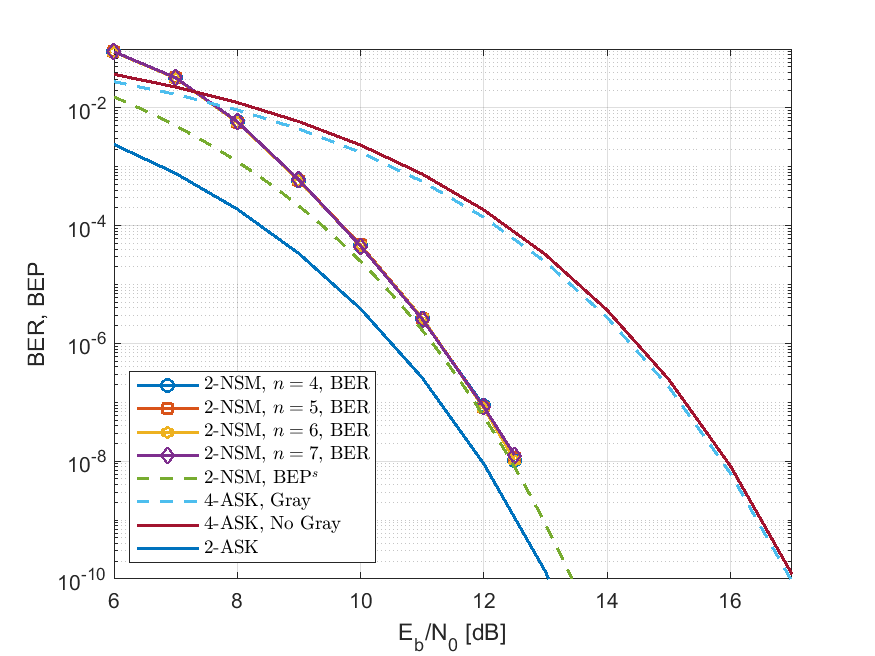}
    \caption{BER of the NSM of rate $2,$ with simplest filter coefficients, given in Table~\ref{table:Simple Filters Coefficients L_0=14 L_1=1}, with $L_0 = 14$ and $L_1 = 1$ ($2\bar{\bm{h}}^4_0=(1,0,0,1,-1,0,0,0,0,0,0,0,0,-1)$, $2\bar{\bm{h}}^5_0=(1,0,0,1,-1,0,0,0,0,0,0,0,0,1)$, $2\bar{\bm{h}}^6_0=(1,0,0,1,1,0,0,0,0,0,0,0,0,-1)$ and $2\bar{\bm{h}}^7_0=(1,0,0,1,1,0,0,0,0,0,0,0,0,1)$). For reference, the approximate BEP contribution, BEP$^s$, of the second category of error events with second minimum square Euclidean distance, $2d_{\text{min}}^2=10.$ Also for reference, the BEPs of $2$-ASK and Gray and non-Gray precoded $4$-ASK conventional modulations are shown.}
    \label{fig:BER-BEP-NSM-2-FilterSimpleCoefficients-FilterLength_14}
\end{figure}

\begin{figure}[!htbp]
    \centering
    \includegraphics[width=1.0\textwidth]{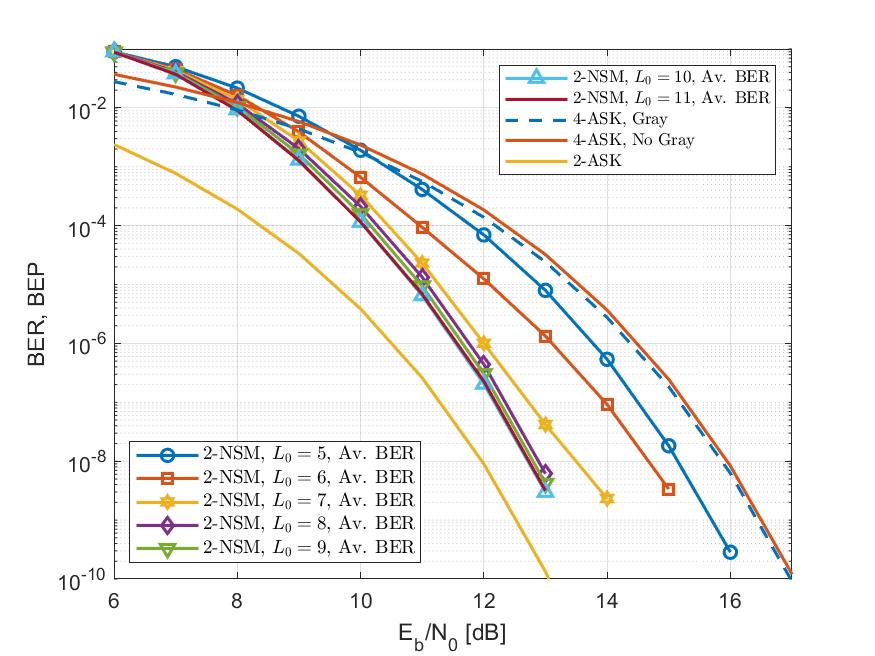}
    \caption{Average BER of each of the NSMs families of rate $2,$ with simplest filter coefficients, given in Tables~\ref{table:Simple Filters Coefficients L_0=5 L_1=1}--\ref{table:Simple Filters Coefficients L_0=11 L_1=1}, corresponding to $L_0 = 5$ to $11$ and $L_1 = 1$. For reference, the BEPs of $2$-ASK and Gray and non-Gray precoded $4$-ASK conventional modulations are shown.}
    \label{fig:BER-BEP-NSM-2-FilterSC-FL_5_6_7_8_9_10_11}
\end{figure}

To push the characterization further, it will be highly interesting to assess the ramifications of raising the first filter's length, $L_0$, from $5$ to $15.$ For this purpose, we have given in Figures~\ref{fig:BER-BEP-NSM-2-FilterSimpleCoefficients-FilterLength_5}--\ref{fig:BER-BEP-NSM-2-FilterSimpleCoefficients-FilterLength_11} the \gls{ber} curves for all the optimal filters stated in Tables~\ref{table:Simple Filters Coefficients L_0=5 L_1=1}--\ref{table:Simple Filters Coefficients L_0=11 L_1=1}, for $L_0$ values ranging from $5$ to $11$. We have also included in Figures~\ref{fig:BER-BEP-NSM-2-FilterSimpleCoefficients-FilterLength_12} and~\ref{fig:BER-BEP-NSM-2-FilterSimpleCoefficients-FilterLength_14} the \gls{ber} curves for only four of the best filters ($\bar{\bm{h}}^n_0,$ $4 \le n < 8$), which are shown in Tables~\ref{table:Simple Filters Coefficients L_0=12 L_1=1} and \ref{table:Simple Filters Coefficients L_0=14 L_1=1}, for $L_0$ values of $12$ and $14$ (only a subset of the optimum filters has been simulated here because simulations are too computationally intensive in this case), respectively. Furthermore, in Figure~\ref{fig:BER-BEP-NSM-2-FilterSC-FL_5_6_7_8_9_10_11}, we have grouped the average \gls{ber} curves, which correspond to the average performance of all best filters, for $L_0$ ranging from $5$ to $11$, to maintain clarity and conciseness and to allow quick and easy inferences about how the \glspl{nsm} behave as $L_0$ increases. In all simulations, we opted for modulated sequences of length $10^3$. We chose not to simulate scenarios $L_0 = 13$ and $15$, because they require a large amount of simulation time, while giving no extra insight. However, we partially simulated a subset of the best filters for the case where $L_0=14,$ since the error event length in this case is fairly specific, being of the form $2^{L_0-2}-1$ rather than $2^{L_0-2}$, as in the other analyzed filter lengths.

First, it is important to notice that the shortest error events with minimum Euclidean distance, for $L_0=12$ and $14,$ have the advantage of having lengths equal to $2^{12-2}=1024$ and $2^{14-2}-1=4095,$ respectively, which are clearly longer than the simulated sequence length of $10^3.$ As a result, $2$-ASK performance is asymptotically achieved in both cases, with an effective \gls{msed} of $10$ (error events with $5$ as \gls{msed} cannot occur here because their lengths exceed $10^3$). Second, the best filters, with $L_0$ lengths ranging from $5$ to $11,$ admit minimum Euclidean distance error event lengths between $2^{5-2}=8$ and $2^{11-2}=512$. This yields an effective minimum Euclidean distance of $5$, as these lengths are shorter than the simulated sequence length of $10^3.$

As expected, Figures~\ref{fig:BER-BEP-NSM-2-FilterSimpleCoefficients-FilterLength_12} and~\ref{fig:BER-BEP-NSM-2-FilterSimpleCoefficients-FilterLength_14} show that the \gls{ber} curves, for $L_0 = 12$ or $14,$ have a tendency to converge towards the $2$-ASK curve, as the \gls{snr} increases. This supports the previous claim that the \gls{msed}, which establishes the \gls{ber} performance at high \gls{snr}, is the same for both values of $L_0$ as it is for $2$-ASK. Furthermore, Figures~\ref{fig:BER-BEP-NSM-2-FilterSimpleCoefficients-FilterLength_12} and~\ref{fig:BER-BEP-NSM-2-FilterSimpleCoefficients-FilterLength_14} demonstrate a slight \gls{ber} benefit for the scenarios where $L_0 = 14$ with respect to those with $L_0=12.$ This outcome makes sense, because the multiplicities of error events with lowest \glspl{sed}, which matter at moderate to high \glspl{snr}, should decrease with a bigger filter memory, $L_0-1.$ Last but not least, these figures reveal a slight differences in performance, in terms of \gls{ber}, at low \gls{snr}, for the group of filters under a common filter length, $L_0,$ which is particularly apparent for $L_0=12.$ This behavior might be explained by the fact that, while all of the resulting \glspl{nsm} have the same minimum Euclidean distance, they differ in the multiplicities of their lowest \glspl{sed}, which matter at low \glspl{snr}.

To quantify the practical gains achieved by these sets of filters, we notice, from Figures~\ref{fig:BER-BEP-NSM-2-FilterSimpleCoefficients-FilterLength_12} and~\ref{fig:BER-BEP-NSM-2-FilterSimpleCoefficients-FilterLength_14}, that the \gls{ber} curves exhibit approximate losses of $0.8$ and $0.5$ dB, with respect to $2$-ASK, at a \gls{ber} target of $10^{-8},$  for $L_0$ equal to $12$ and $14,$ respectively. With regard to conventional $4$-ASK, these losses translate into approximate gains of $3.18$ and $3.48$ dB, respectively.

As also expected, the average \gls{ber} performance for $L_0 = 5,$ in Figure~\ref{fig:BER-BEP-NSM-2-FilterSC-FL_5_6_7_8_9_10_11}, supports the asymptotic gain of $10\log_{10}(5/4) \approx 0.9691$ dB, which corresponds to its effective \gls{msed}, $d_{\text{min}}^2 = 5.$ To acquire a better understanding of this scenario, we considered it useful to accurately characterize it in Appendix~\ref{app:Multiplicities Minimum Euclidean Distance Rate-2 NSMs Simple Coefficients}. Indeed, given that the shortest length of error events with \gls{msed}, $2^{L_0-2}=8,$ is the smallest of all the scenarios considered, this scenario may be fully characterized.

Through Table~\ref{table:Characteristics First Minimum Euclidean Distance Simple Filters Coefficients L_0=5 L_1=1}, Appendix~\ref{app:Multiplicities Minimum Euclidean Distance Rate-2 NSMs Simple Coefficients} reveals that all eligible minimum squared distance error event sequences, for which  $\Delta \ddot{b}^n_0(x) = (x^7 + 1)/q^n(x) = x^3 + x + 1,$ are actually all admissible as error events, for $L_0=5.$ This indicates that, when they are not null, non-boundary components of $\Delta \bar{\bm{s}}_0$ are always in the set $\{0,\pm 2\}$ and can therefore be immediately eliminated by a sequence $\Delta \mathring{\bm{b}}_1$ with non-boundary entries in the set $\{\pm 1\}$ (note that components of $\Delta \bar{\bm{s}}_0$ may occur within the range $[-4,4]$ and cannot be adjusted for if they fall outside the set $\{0,\pm 2\}$). In full agreement with (\ref{eq:Galois Field One-Dimensional NSM}), this yields an error sequence $\Delta \bar{\bm{s}}$ with null non-boundary elements.

Moreover, Appendix~\ref{app:Multiplicities Minimum Euclidean Distance Rate-2 NSMs Simple Coefficients} shows that the best filters, under filter length $L_0=5,$ are expected to exhibit a little difference in performance, in terms of \gls{ber}, at high \gls{snr}, which is in perfect agreement with Figure~\ref{fig:BER-BEP-NSM-2-FilterSimpleCoefficients-FilterLength_5}. This result comes from the fact that, while all of the resulting \glspl{nsm} have the same \gls{msed}, they differ in their associated multiplicities, which matters at high \glspl{snr}. For the same reason, this observation comes to light at high \gls{snr}, when $L_0 = 6$ or $7.$

One of the most important insights to take away from Figures~\ref{fig:BER-BEP-NSM-2-FilterSimpleCoefficients-FilterLength_5}--\ref{fig:BER-BEP-NSM-2-FilterSimpleCoefficients-FilterLength_14} \textemdash and especially from the summary, in Figure~\ref{fig:BER-BEP-NSM-2-FilterSC-FL_5_6_7_8_9_10_11}, of Figures~\ref{fig:BER-BEP-NSM-2-FilterSimpleCoefficients-FilterLength_5}--\ref{fig:BER-BEP-NSM-2-FilterSimpleCoefficients-FilterLength_11} \textemdash is that, for moderate \glspl{snr}, the \gls{ber} curves of the \glspl{nsm} under consideration tend to shift toward the \gls{ber} curve of $2$-ASK, as $L_0$ rises. On the one hand, part of this trend can be explained by the fact that the probability of occurrence of error events with \gls{msed}, $d_{\text{min}}^2=5,$ declines as $L_0$ increases. Indeed, as will be demonstrated next, these error events have an exponentially decreasing compound multiplicity, in the \gls{bep} upperbound term, $0.5 \operatorname{erfc} (\sqrt{0.5 E_b/N_0})$, and can thus only show up at high \glspl{snr}. Therefore, the higher $L_0$, the higher the \glspl{snr} at which their effect is observable and perceptible. On the other hand, part of this behavior is due to the fact that, when $L_0$ increases, the likelihood of occurrence of error events with the next \gls{msed}, $10$ (corresponding to the \gls{msed} of $2$-ASK), remains unchanged and does not vanish. Indeed, as demonstrated in Appendix~\ref{app:Characterization Second Minimum Euclidean Distance Rate-2 NSMs Simple Coefficients} and showcased in (\ref{eq:Contribution Second Minimum Squared Distance BEP}), these error events maintain a constant multiplicity in their \gls{bep} upperbound term, $0.5 \operatorname{erfc} (\sqrt{E_b/N_0}).$ To ensure the validity of this fixed multiplicity, we presented the curve of the associated contribution to the \gls{bep}, in Figure~\ref{fig:BER-BEP-NSM-2-FilterSimpleCoefficients-FilterLength_14} and validated its tightness, as lowerbound, to the simulated \gls{ber} for $L_0=14.$

To this point, it is significant to note that, despite their lack of error-correcting capabilities, \glspl{nsm} perform similarly to turbo-codes based on \glspl{pccc} \cite{Berrou93, Berrou96, Divsalar96, Giulietti04, Yeh03, Perez96, Vucetic00}. First, \gls{pccc} turbo-codes do not prohibit error events with very low Hamming weights from occurring, but rather make their occurrence probabilities decrease dramatically as the coded sequence size grows. Likely, the \gls{nsm} discussed here do not prohibit error events with a \gls{msed}, immediately below that of $2$-ASK (half that of $2$-ASK), from occurring, but rather make their multiplicity drop exponentially, as the filter length, $L_0$, rises. This fact will be explicitly proven next by upper bounding the multiplicity of the \gls{msed} and demonstrating its extraordinarily fast exponential decay to zero, as $L_0$ increases. Second, \gls{pccc} turbo-codes exhibit relatively low multiplicities for error events with low Hamming distances, making their impact on \gls{ber} curves only apparent at high \glspl{snr}, resulting in curve flattening (also referred to as error floor, error flare or plateau effect). Likely, albeit not as noticeable as that of \gls{pccc} turbo-codes, a plateau effect is to be expected for \glspl{nsm} with simple rational filter coefficients. In accordance with Figure~\ref{fig:BER-BEP-NSM-2-FilterSC-FL_5_6_7_8_9_10_11}, for $L_0 = 6$ and $7,$ this error floor begins to show while going from low to high \glspl{snr}. The linked \gls{ber} curves exhibit two consecutive inflections, either entirely or partially, which cause them to go from concave to convex and back to concave. This plateau effect does not show up for $L_0=5$, since, as shown in Appendix~\ref{app:Multiplicities Minimum Euclidean Distance Rate-2 NSMs Simple Coefficients}, the multiplicity of error events with \gls{msed}, are not too small. Additionally, they are absent for $L_0 \geq 8,$ since the multiplicity of the error events with minimal \gls{sed} is below the range of simulated \glspl{ber}, for these filter lengths.

This is being said, the most noteworthy finding is that, if \gls{nsm} filters are constructed properly, the extremely low multiplicities of the related \gls{msed} prevent it from affecting the \gls{ber} at practical \gls{snr} ranges. Therefore, within desired working ranges of the \gls{snr}, this \gls{msed} fails to prevent these \glspl{nsm} from achieving near 2-ASK performance at practical \glspl{ber}.

The cumulative error probability multiplicity of all error events with \gls{msed} cannot be computed precisely. To prove its exponential fading towards zero, as $L_0$ increases, we propose to upper bound it. Adopting a similar approach to that of Appendix~\ref{app:Multiplicities Minimum Euclidean Distance Rate-2 NSMs Simple Coefficients}, our starting point is the assessment of the partial \gls{tf}, $T_K(N),$ encompassing the contributions of all error events with \gls{msed} and common length, $K.$ Let then $\Delta \mathring{b}_0[k]$ and $\Delta \mathring{b}_1[k]$ be input sequence differences and $\Delta \bar{s}[k]$ be the modulated sequence difference associated to one of these error events. Let also $w_0$ and $w_1$ be the number of non-null components of $\Delta \mathring{b}_0[k]$ and $\Delta \mathring{b}_1[k],$ respectively. Before proceeding, notice that $w_0$ is also the Hamming weight of $\Delta \ddot{b}_0[k],$ the binary counterpart of $\Delta \mathring{b}_0[k],$ which remains unchanged for each error event length $K,$ as shown below. The contribution of such an error event to the partial \gls{tf} $T_K(N)$ is therefore given by $N^{w_0+w_1}.$ Consequently, following the considerations of Appendix~\ref{app:Tight Estimate BEP Rate 2} and more particularly the expression of the multiplicity as a function of the \gls{tf} in (\ref{eq:Multiplicity and Distance Rate-2 Modulation}), we deduce that the contribution to the error probability compound multiplicity of all error events with \gls{msed} is equal to $(w_0+w_1)(\tfrac{1}{2})^{w_0+w_1}.$ Now, let us write $w_1$ as $w_1 = w_{10} + w_{11},$ where $w_{10}$ is the total number of non-null components of $\Delta \mathring{b}_1[k],$ in front of non-null components of $\Delta \mathring{b}_0[k].$ Then, $w_{11}$ is lower bounded by $0$ and upper bounded by $K-w_0.$ Hence, the previous contribution to the error probability compound multiplicity can be upper bounded by $(w_0+w_{10}+(K-w_0))(\tfrac{1}{2})^{w_0+w_{10}} = (w_{10}+K)(\tfrac{1}{2})^{w_0+w_{10}}.$ This upper bound results from an evaluation of $N^{1-K+w_0} \partial N^{w_{10}+K} / \partial N$ on $N=1/2.$ Now consider an arbitrary position $k,$ where $\Delta \mathring{b}_0[k]$ is not null. Then we have the following cases. On the one hand, if previous contributions of the input sequences differences $\Delta \mathring{b}_0[l]$ and $\Delta \mathring{b}_1[l],$ $0 \leq l < k,$ to the modulated difference sequence sample $\Delta \bar{s}[k]$ is equal to $\pm 3,$ then $(\Delta \mathring{b}_0[k], \Delta \mathring{b}_1[k])$ must be equal to $\mp (1, 1),$ in order that the final value of $\Delta \bar{s}[k]$ be null, when $0<k<K-1.$ On the other hand, if previous contributions of the previous input sequences differences to the modulated difference sequence sample $\Delta \bar{s}[k]$ is equal to $\pm 1,$ then $(\Delta \mathring{b}_0[k], \Delta \mathring{b}_1[k])$ must be equal to $\mp (1, 0)$ or $\mp (-1, 1).$ The contribution of the former case to the term $N^{w_0+w_{10}}$ above is equal to $N^2,$ while the contribution of the latter case to this term is either equal to $N$ or to $N^2.$ Hence, the sum of all terms $N^{w_0+w_{10}}$ over all eligible error events of common length $K,$ and therefore common weight $w_0,$ can be replaced by the product $(N+N^2)^{w_0},$ in order to preserve the upper bounding procedure. Hence, the upper bound of the cumulative error probability multiplicity contribution, of error events with common length $K,$ is upper bounded by the evaluation of $N^{1-K+w_0} \partial (2(1+N)^{w_0}N^K) / \partial N$ on $N=1/2.$ The $2$ factor in the previous expression comes from the fact that all four values of $(\Delta \mathring{b}_0[0], \Delta \mathring{b}_1[0])$ in the set $\{\pm (1, 0), \pm (-1, 1)\}$ are allowed at the start of the error event. After some calculus, we can state that the contribution of eligible error events, of common length $K$ and weight $w_0,$ to the error probability multiplicity of error events with \gls{msed}, is upper bounded by $(\tfrac{3}{4})^{w_0-1}(\tfrac{1}{2}w_0+\tfrac{3}{2}K) = (\tfrac{3}{4})^{w_0}(\tfrac{2}{3}w_0+2K).$

At this stage, it is time to assess the weight, $w_0,$ for each of the permissible lengths, $K = \kappa (K_{\text{min}}-1) + 1,$ $\kappa > 0,$ of error events with \gls{msed} $d_{\text{min}}^2 = 5.$ For this purpose, we start by fully characterizing the case of error events with minimum length, $K = K_{\text{min}},$ corresponding to $\kappa = 1,$ then deducing the characteristics of other error events length, when $\kappa > 1.$ We first characterize the case where  $\ddot{h}_0(x) = q^n(x) = (x+1)p^n(x),$ with $p^n(x)$ being a primitive polynomial of degree $L_0-2,$ then move to the case where $\ddot{h}_0(x) = q^n(x) = (x+1)^2 p^n(x),$ with $p^n(x)$ being a primitive polynomial of degree $L_0-3.$ Our objective in both cases is to determine the Hamming weight of $\Delta \ddot{b}_0(x)$ in (\ref{eq:Galois Field One-Dimensional NSM Polynomial Form}), when $K = K_{\text{min}}.$ For the two forms of $\ddot{h}_0(x),$ we can write $\Delta \ddot{b}_0(x) = (x^{K_{\text{min}}-1} + 1)/\ddot{h}_0(x) = (x^{K_{\text{min}}-1} + 1)/q^n(x),$ which encourages us to start by studying and characterizing $1/q^n(x).$

For the first case, where $q^n(x) = (x+1)p^n(x),$ we can use the decomposition into partial fractions of $1/q^n(x)$ and write
\begin{equation} \label{eq: Partial Fractions Decomposition Case 1/(x+1)p^n(x)}
\frac{1}{q^n(x)} = \frac{g(x)}{p^n(x)} + \frac{1}{x+1},
\end{equation}
where $g(x)$ is a polynomial of degree $L_0-3$ at most (the degree of polynomial $g(x)$ must be strictly lower than $L_0-2,$ degree of polynomial $p^n(x),$ since $q^n(x)$ is of degree $L_0-1$). Notice that the numerator in the partial fraction involving $(x+1)$ must be equal to $1,$ unique non-null value in the binary Galois field $\text{GF}(2)=\{0, 1\}.$ We know from \cite{Viterbi95} that
\begin{equation} \label{eq: Partial Fractions Case 1/(x+1)p^n(x)}
\frac{g(x)}{p^n(x)} = m^n(x)(1 + x^{K_{\text{min}}-1} + x^{2(K_{\text{min}}-1)} + x^{3(K_{\text{min}}-1)} + \cdots) = m^n(x) \sum_{k \geq 0} x^{k (K_{\text{min}}-1)},
\end{equation}
where $m^n(x)$ is the polynomial corresponding to the maximum-length sequence (m-sequence) associated to primitive polynomial $p^n(x).$ From (\ref{eq: Partial Fractions Case 1/(x+1)p^n(x)}), one can see that partial fraction $g(x)/p^n(x)$ corresponds to a periodization of the m-sequence of length $K_{\text{min}}-1$ associated to polynomial $m^n(x).$ Hence, from the fact that $1/(x+1) = 1+x+x^2+x^3+\cdots = \sum_{k \geq 0} x^k,$ we conclude that $1/q^n(x)$ corresponds to the repetition of a sequence, with polynomial $\ddot{m}^n(x),$ whose elements are the one-complements of the elements of the m-sequence. Hence, we can rewrite (\ref{eq: Partial Fractions Decomposition Case 1/(x+1)p^n(x)}) as
\begin{equation} \label{eq: Developed Partial Fractions Decomposition Case 1/(x+1)p^n(x)}
\frac{1}{q^n(x)} =  \ddot{m}^n(x)(1 + x^{K_{\text{min}}-1} + x^{2(K_{\text{min}}-1)} + x^{3(K_{\text{min}}-1)} + \cdots) = \ddot{m}^n(x) \sum_{k \geq 0} x^{k (K_{\text{min}}-1)},
\end{equation}
which tells us that $1/q^n(x)$ is just a periodization of the polynomial $\ddot{m}^n(x)$ corresponding to the one complements of the m-sequence associated to primitive polynomial $p^n(x).$ Using (\ref{eq: Developed Partial Fractions Decomposition Case 1/(x+1)p^n(x)}), we obtain
\begin{equation} \label{eq: First Input Differences Sequence Case 1/(x+1)p^n(x)}
\Delta \ddot{b}_0(x) = \frac{x^{K_{\text{min}}-1} + 1}{q^n(x)} = \ddot{m}^n(x) \sum_{k \geq 1} x^{k (K_{\text{min}}-1)} + \ddot{m}^n(x) \sum_{k \geq 0} x^{k (K_{\text{min}}-1)} = \ddot{m}^n(x),
\end{equation}
which tells us that the sought first input differences sequence, $\Delta \mathring{b}_0[k],$ admits $\Delta \ddot{b}_0[k],$ with polynomial $\Delta \ddot{b}_0(x) = \ddot{m}^n(x),$ for its binary counterpart. 

Up to this point, given that $\Delta \ddot{b}_0(x)$ obeys to (\ref{eq:Galois Field One-Dimensional NSM Polynomial Form}), we need to ensure that $\ddot{m}^n(x)$ is indeed a polynomial of degree $(K_{\text{min}}-1) - (L_0-1) = K_{\text{min}} - L_0,$ meaning that $\Delta \ddot{b}_0[k],$ of length $K_{\text{min}}-1,$ must end with $(K_{\text{min}}-2)-(K_{\text{min}} - L_0) = L_0-2$ zeros. This is conceivable, since, from \cite{Viterbi95}, the m-sequence corresponding to $m^n(x)$ has a maximum run of $L_0-2$ ones, which, when complemented to one, leads to a maximum run of $L_0-2$ zeros in sequence $\Delta \ddot{b}_0[k].$ Notice that the specific position, at the end of the sequence, of this run of ones in the m-sequence, is determined by polynomial $g(x),$ in (\ref{eq: Partial Fractions Case 1/(x+1)p^n(x)}), which specifies the initial conditions of the m-sequence generating linear feedback shift register. 

With that in mind, we can now determine the Hamming weights, $w_0,$ of $\Delta \ddot{b}_0[k],$ corresponding to error events with \gls{msed} and lengths $K = \kappa (K_{\text{min}}-1) + 1,$ $\kappa > 0.$ First, notice, from \cite{Viterbi95} that the m-sequence, of length $K_{\text{min}}-1,$ corresponding to polynomial $m^n(x)$ has exactly $((K_{\text{min}}-1)-1)/2 = K_{\text{min}}/2-1$ null components and $((K_{\text{min}}-1)+1)/2 = K_{\text{min}}/2$ components equal to $1.$ As a consequence, the binary version, $\Delta \ddot{b}_0[k],$ of the first input differences sequence, corresponding to polynomial $\Delta \ddot{b}_0(x) = \ddot{m}^n(x)$ has exactly $K_{\text{min}}/2$ null components and $K_{\text{min}}/2-1$ components equal to $1.$ From this, we conclude that the Hamming weights, $w_0,$ corresponding to the least error event length, $K = K_{\text{min}},$ is equal to $K_{\text{min}}/2-1.$

For error events with length $K =  \kappa (K_{\text{min}}-1) + 1,$ $\kappa > 1,$ notice that $x^{K-1} + 1 = (x^{K_{\text{min}}-1} + 1)\sum_{k=0}^{\kappa-1} x^{k (K_{\text{min}}-1)}.$ Hence, from (\ref{eq:Galois Field One-Dimensional NSM Polynomial Form}), we can state that $\Delta \ddot{b}_0(x) = \ddot{m}^n(x) \sum_{k=0}^{\kappa-1} x^{k (K_{\text{min}}-1)}.$ Now, the fact that polynomial $\ddot{m}^n(x)$ is of degree at most $K_{\text{min}}-2,$ we can affirm that the different shifts of $\ddot{m}^n(x) x^{k (K_{\text{min}}-1)}$ by $k (K_{\text{min}}-1)$ do not overlap. Hence, the Hamming weight of $\Delta \ddot{b}_0[k]$ for an error event with length $K =  \kappa (K_{\text{min}}-1) + 1$ is equal to $w_0 = \kappa (K_{\text{min}}/2-1),$ $K_{\text{min}}/2-1$ being the Hamming weight of the shortest length Error event with \gls{msed} derived above.

Based on the above results regarding the lengths and Hamming weights of the binary counterparts of the input differences sequence corresponding to error events with \gls{msed}, we can upper bound their error probability multiplicity by 
\begin{equation} \label{Upper Bound Error Probability Multiplicity - First Case}
\mu < \sum_{\kappa > 0}(\tfrac{3}{4})^{\kappa (K_{\text{min}}/2-1)}
(\tfrac{2}{3}\kappa (\tfrac{1}{2}K_{\text{min}}-1) + 2(\kappa (K_{\text{min}}-1) + 1)) <
\frac{7 K_{\text{min}}}{3} \frac{(\tfrac{3}{4})^{K_{\text{min}}/2-1}}{(1-(\tfrac{3}{4})^{K_{\text{min}}/2-1})^2}.
\end{equation}
Given that for the first case at hand, we have $K_{\text{min}} = 2^{L_0-2},$ we find that the multiplicity of error events with \gls{msed} is upper bounded by around $\num{4e-6}$ for $L_0 = 9,$ $\num{8.1e-14}$ for $L_0 = 10$ and $\num{1.6e-29}$ for $L_0 = 11.$ These results tell us that the impact of all error events with \gls{msed} must not show up at practical ranges of the \gls{ber}, above $\num{e-10},$ whenever $L_0 \geq 10.$

For the second case, where $q^n(x)=(x+1)^2 p^n(x)$, we can relay on the partial fraction decomposition of $1/q^n(x)$ and write
\begin{equation} \label{eq: Partial Fractions Decomposition Case 1/(x+1)^2p^n(x)}
\frac{1}{q^n(x)} = \frac{g_1(x)}{p^n(x)} + \frac{g_2(x)}{(x+1)^2},
\end{equation}
where $g_1(x)$ is a polynomial of degree $L_0-4$ at most (the degree of polynomial $g_1(x)$ must be strictly lower than $L_0-3,$ degree of polynomial $p^n(x)$), while $g_2(x)$ is a polynomial of degree $1$ at most. Notice that the numerator, $g_2(x),$ in the partial fraction involving $(x+1)^2,$ must be either equal to $x$ or to $1.$ Since $(x+1)^2$ must figure in the polynomial decomposition of $q^n(x),$ polynomial $g_2(x)$ is forbidden to take $0$ and $x+1$ as possible values (the last value is not allowed since it simplifies with the denominator leading to the appearance of partial fraction $1/(x+1)$ instead of partial fraction $1/(x+1)^2$). Since $p^n(x)$ is a primitive polynomial of degree $L_0-3,$ it generates an m-sequence, $m^n(x),$ of length $L = 2^{L_0-3}-1.$ Given that the length of the error event with \gls{msed} is given by $K_{\text{min}} = 2^{L_0-2}-1,$ we deduce that $K_{\text{min}} = 2L+1.$

On the one hand, we know from \cite{Viterbi95} that
\begin{equation} \label{eq: Partial Fractions Case 1/(x+1)^2p^n(x)}
\frac{g(x)}{p^n(x)} = m^n(x)(1 + x^L + x^{2L} + x^{3L} + \cdots) = m^n(x) \sum_{k \geq 0} x^{kL}.
\end{equation}
From (\ref{eq: Partial Fractions Case 1/(x+1)^2p^n(x)}), one can see that partial fraction $g_1(x)/p^n(x)$ corresponds to a periodization of the m-sequence, with polynomial $m^n(x),$ with odd period $L = 2^{L_0-3}-1.$ On the other hand, we recall that $1/(x+1)^2 = 1/(x^2+1),$ since we are working in $\text{GF}(2).$ Hence $1/(x+1)^2 = 1+x^2+x^4+\cdots = \sum_{k \geq 0} x^{2k},$ and therefore $x/(x+1)^2 = x+x^3+x^5+\cdots = \sum_{k \geq 0} x^{2k+1},$ correspond to periodic sequence of even period, $2,$ resulting from a periodization of binary words $10$ and $01,$ respectively. Hence, whatever the permissible values of polynomials $g_1(x)$ and $g_2(x),$ we can conclude that $1/q^n(x)$ corresponds to a periodic sequence with the least common multiple of $L= 2^{L_0-3}-1$ and $2$ as period. Let $\ddot{m}_0^n(x)$ (respectively, $\ddot{m}_1^n(x)$) denote the polynomial corresponding to the element-wise addition of the m-sequence of polynomial $m^n(x)$ and the alternating binary sequence $101010 \cdots 01$ (respectively, $010101 \cdots 10$) of length $L.$ Then, $1/q^n(x)$ can be seen as the repetition, with period $2L = 2(2^{L_0-3}-1) = 2^{L_0-2}-2,$ of polynomial $\ddot{m}_0^n(x) + x^L \ddot{m}_1^n(x),$ when $g_2(x) = 1,$ and polynomial $\ddot{m}_1^n(x) + x^L \ddot{m}_0^n(x),$ when $g_2(x) = x.$ Hence, noticing that $2L = K_{\text{min}}-1,$ we can rewrite (\ref{eq: Partial Fractions Decomposition Case 1/(x+1)^2p^n(x)}) as
\begin{align} \label{eq: Developed Partial Fractions Decomposition Case 1/(x+1)^2p^n(x) 1}
\frac{1}{q^n(x)} & = (\ddot{m}_i^n(x)+x^L\ddot{m}_{1-i}^n(x))(1 + x^{K_{\text{min}}-1} + x^{2(K_{\text{min}}-1)} + x^{3(K_{\text{min}}-1)} + \cdots) \\
 \label{eq: Developed Partial Fractions Decomposition Case 1/(x+1)^2p^n(x) 2} & = (\ddot{m}_i^n(x)+x^L\ddot{m}_{1-i}^n(x)) \sum_{k \geq 0} x^{k (K_{\text{min}}-1)} ,
\end{align}
where $i=0$ if $g_2(x)=1$ and $i=1$ if $g_2(x)=x.$ Using (\ref{eq: Developed Partial Fractions Decomposition Case 1/(x+1)^2p^n(x) 2}), we obtain
\begin{align} \label{eq: First Input Differences Sequence Case 1/(x+1)^2p^n(x) 1}
\Delta \ddot{b}_0(x) & = \frac{x^{K_{\text{min}}-1} + 1}{q^n(x)} \\
& = (\ddot{m}_i^n(x)+x^L\ddot{m}_{1-i}^n(x)) \sum_{k \geq 1} x^{k (K_{\text{min}}-1)} + (\ddot{m}_i^n(x)+x^L\ddot{m}_{1-i}^n(x)) \sum_{k \geq 0} x^{k (K_{\text{min}}-1)} \\
\label{eq: First Input Differences Sequence Case 1/(x+1)^2p^n(x) 2} & = \ddot{m}_i^n(x)+x^L\ddot{m}_{1-i}^n(x),
\end{align}
which tells us that the sought first input differences sequence, $\Delta \mathring{b}_0[k],$ admits $\Delta \ddot{b}_0[k],$ with polynomial $\Delta \ddot{b}_0(x) = \ddot{m}_i^n(x)+x^L\ddot{m}_{1-i}^n(x),$ for its binary counterpart, $i$ being equal to $0$ or $1,$ depending on whether $g_2(x) = 1$ or $x.$

Again, as above, based on $\Delta \ddot{b}_0(x)$ obeys to (\ref{eq:Galois Field One-Dimensional NSM Polynomial Form}), we need to make certain that $\ddot{m}^n(x)$ is a polynomial of degree $(K_{\text{min}}-1) - (L_0-1) = K_{\text{min}} - L_0,$ meaning that $\Delta \ddot{b}_0[k],$ of length $K_{\text{min}}-1,$ must end with $(K_{\text{min}}-2)-(K_{\text{min}} - L_0) = L_0-2$ zeros. To see this, remember that the m-sequence with polynomial $m^n(x)$ and length $L$ is generated by a linear feedback shift register of length $L_0-3.$ Hence, this m-sequence contains necessary words $1010 \cdots 01$ and $0101 \cdots 10$ of common length $L_0-3.$ Hence, either the merged word $1010 \cdots 010$ or $01010 \cdots 01,$ of length $L_0-2,$ is an integral part of the m-sequence. Moreover, to prevent the sequence of alternating zeros and ones from being reproduced indefinitely, if eligible, the first (second, respectively) merged word should appear in extend form as $11010 \cdots 0100$ (respectively, $001010 \cdots 011$). Hence, on the one hand, if the contribution of the alternating sequence of zeros and ones, coming from fraction $g_2(x)/(x+1)^2,$ in front of one of these extended sequences, is $01010 \cdots 0101,$ then the final extended word that appears in $\ddot{b}_0[k]$ is $10000 \cdots 0001$ (respectively, $01111 \cdots 1110$) for the first (respectively, second) extended word, $11010 \cdots 0100$ (respectively, $001010 \cdots 011$), and hence presents exactly a run of $L_0-2$ zeros (respectively, ones). On the other hand, if the contribution of the alternating sequence of zeros and ones, coming from fraction $g_2(x)/(x+1)^2,$ in front of one of these extended sequences, is $10101 \cdots 1010,$ then the final extended word that appears in $\ddot{b}_0[k]$ is $01111 \cdots 1110$ (respectively, $10000 \cdots 0001$) for the first (respectively, second) extended word, $11010 \cdots 0100$ (respectively, $001010 \cdots 011$), and hence presents exactly a run of $L_0-2$ ones (respectively, zeros). In either case, we are sure that, with appropriate choices of polynomials $g_1(x)$ and $g_2(x),$ a run of $L_0-2$ zeros can be guaranteed at the end of sequence $\Delta \ddot{b}_0[k],$ of length $K_{\text{min}}-1.$

As in the first case, We can now determine the Hamming weights, $w_0,$ of $\Delta \ddot{b}_0[k],$ corresponding to error events with \gls{msed} and lengths $K = \kappa (K_{\text{min}}-1) + 1,$ $\kappa > 0.$ First, recall, from (\ref{eq: First Input Differences Sequence Case 1/(x+1)^2p^n(x) 2}), that the binary version, $\Delta \ddot{b}_0[k],$ of the input differences sequences $\Delta \mathring{b}_0[k],$ corresponding to the error events with \gls{msed} and shortest length, $K_{\text{min}},$ is such that $\Delta \ddot{b}_0(x) = \ddot{m}_i^n(x)+x^L\ddot{m}_{1-i}^n(x),$ where $i = 0$ or $1.$ Hence, this input sequence is composed of two consecutive m-sequences, of common length $L$ and polynomial $m^n(x),$ with one of them complemented to one in its even indices elements and the other complemented to one in its odd indices elements. The resulting Hamming weight of $\Delta \mathring{b}_0[k],$ is therefore equal to that of a sequence composed of two consecutive m-sequences with only one of them fully complemented to one. Hence, the weight, $w_0,$ of $\Delta \mathring{b}_0[k],$ for the error event with least length, $K = K_{\text{min}},$ is exactly equal to $L = 2^{L_0-3}-1 = (K_{\text{min}}-1)/2.$

Following similar steps as in the first case, for error events of the second case with length $K =  \kappa (K_{\text{min}}-1) + 1,$ $\kappa > 1,$ we can state that the Hamming weight of $\Delta \ddot{b}_0[k],$ for an error event with length $K =  \kappa (K_{\text{min}}-1) + 1,$ is equal to $w_0 = \kappa (K_{\text{min}}-1)/2,$ $(K_{\text{min}}-1)/2$ being the Hamming weight of the shortest length Error event with \gls{msed} derived above.

Based on the above results for error events, of the second case, with \gls{msed}, we can upper bound the error probability multiplicity by 
\begin{equation} \label{Upper Bound Error Probability Multiplicity - Second Case}
\mu < \sum_{\kappa > 0}(\tfrac{3}{4})^{\kappa (K_{\text{min}}-1)/2}
(\tfrac{2}{3}\kappa \tfrac{1}{2} (K_{\text{min}}-1) + 2(\kappa (K_{\text{min}}-1) + 1)) <
\frac{7 K_{\text{min}}}{3} \frac{(\tfrac{3}{4})^{(K_{\text{min}}-1)/2}}{(1-(\tfrac{3}{4})^{(K_{\text{min}}-1)/2})^2}.
\end{equation}
Given that for the second case, we have $K_{\text{min}} = 2^{L_0-2}-1,$ we find that the multiplicity of \gls{msed} is upper bounded by around $\num{1.7e-252}$ for $L_0 = 14.$ These results tell us that the impact of all error events with \gls{msed} must not show up at practical ranges of the \gls{ber}, above $\num{e-10},$ for the second case too.

\begin{table}[!htbp]
\caption{Maximum length minimum Euclidean distance guaranteeing NSMs of rate $2$ with extremely simple filters' coefficients, for $L_0=5$ and $L_1=1$ ($\bar{\bm{h}}_1 = (1)$ and $\eta = \eta_0 = \eta_1 = 5/2$).}
\label{table:Simple Filters Coefficients L_0=5 L_1=1}
\centering
\begin{tabular}{|c|c|} 
\hline
\multicolumn{2}{|l|}{\# of extremely simple $h_0[k]$ filters $ = 48$} \\ \hline
\multicolumn{2}{|l|}{\# of non equivalent $h_0[k]$ filters $ = 8$} \\ \hline
\multicolumn{2}{|l|}{\# of best non equivalent $h_0[k]$ filters $ = 4$} \\ \hline
\multicolumn{2}{|l|}{\gls{msed}, $d_{\text{min}}^2 = 5$} \\ \hline
\multicolumn{2}{|l|}{Asymptotic gain with respect to $4$-ASK $ = 10 \log_{10}(5/4) \approx 0.9691$ dB} \\ \hline
\multicolumn{2}{|l|}{Length of shortest error event with minimum Euclidean distance $= 2^{L_0-2} = 8$} \\ [0.5ex] 
\hline\hline
$n$ & $0 - 3$ \\ \hline
$2\bar{\bm{h}}^n_0 \in \mathbb{Z}^{L_0}$ & $(1,1,\pm 1,0,\pm 1)$ \\ \hline
$2|\bar{\bm{h}}^n_0| \in \text{GF}(2)^{L_0}$ & $(1,1,1,0,1)$ \\ \hline
$q^n(x) \in \text{GF}(2)[x]$ & $x^4 + x^2 + x + 1 = (x + 1)(x^3 + x^2 + 1)$  \\ [1ex] 
 \hline
\end{tabular}
\end{table} 

\begin{table}[!htbp]
\caption{Maximum length minimum Euclidean distance guaranteeing NSMs of rate $2$ with extremely simple filters' coefficients, for $L_0=6$ and $L_1=1$ ($\bar{\bm{h}}_1 = (1)$ and $\eta = \eta_0 = \eta_1 = 5/2$).}
\label{table:Simple Filters Coefficients L_0=6 L_1=1}
\centering
\begin{tabular}{|c|c|} 
\hline
\multicolumn{2}{|l|}{\# of extremely simple $h_0[k]$ filters $ = 96$} \\ \hline
\multicolumn{2}{|l|}{\# of non equivalent $h_0[k]$ filters $ = 14$} \\ \hline
\multicolumn{2}{|l|}{\# of best non equivalent $h_0[k]$ filters $ = 4$} \\ \hline
\multicolumn{2}{|l|}{\gls{msed}, $d_{\text{min}}^2 = 5$} \\ \hline
\multicolumn{2}{|l|}{Asymptotic gain with respect to $4$-ASK $ = 10 \log_{10}(5/4) \approx 0.9691$ dB} \\ \hline
\multicolumn{2}{|l|}{Length of shortest error event with minimum Euclidean distance $= 2^{L_0-2} = 16$} \\ [0.5ex] 
\hline\hline
$n$ & $0 - 3$ \\ \hline
$2\bar{\bm{h}}^n_0 \in \mathbb{Z}^{L_0}$ & $(1,1,0,\pm 1,0,\pm 1)$ \\ \hline
$2|\bar{\bm{h}}^n_0| \in \text{GF}(2)^{L_0}$ & $(1,1,0,1,0,1)$ \\ \hline
$q^n(x) \in \text{GF}(2)[x]$ & $x^5 + x^3 + x + 1 = (x + 1)(x^4 + x^3 + 1)$  \\ [1ex] 
 \hline
\end{tabular}
\end{table}

\begin{table}[!htbp]
\caption{Maximum length minimum Euclidean distance guaranteeing NSMs of rate $2$ with extremely simple filters' coefficients, for $L_0=7$ and $L_1=1$ ($\bar{\bm{h}}_1 = (1)$ and $\eta = \eta_0 = \eta_1 = 5/2$).}
\label{table:Simple Filters Coefficients L_0=7 L_1=1}
\centering
\begin{tabular}{|c|c|} 
\hline
\multicolumn{2}{|l|}{\# of extremely simple $h_0[k]$ filters $ = 160$} \\ \hline
\multicolumn{2}{|l|}{\# of non equivalent $h_0[k]$ filters $ = 26$} \\ \hline
\multicolumn{2}{|l|}{\# of best non equivalent $h_0[k]$ filters $ = 8$} \\ \hline
\multicolumn{2}{|l|}{\gls{msed}, $d_{\text{min}}^2 = 5$} \\ \hline
\multicolumn{2}{|l|}{Asymptotic gain with respect to $4$-ASK $ = 10 \log_{10}(5/4) \approx 0.9691$ dB} \\ \hline
\multicolumn{2}{|l|}{Length of shortest error event with minimum Euclidean distance $= 2^{L_0-2} = 32$} \\ [0.5ex] 
\hline\hline
$n$ & $0 - 3$ \\ \hline
$2\bar{\bm{h}}^n_0 \in \mathbb{Z}^{L_0}$ & $(1,1,\pm 1,0,0,0,\pm 1)$ \\ \hline
$2|\bar{\bm{h}}^n_0| \in \text{GF}(2)^{L_0}$ & $(1,1,1,0,0,0,1)$ \\ \hline
$q^n(x) \in \text{GF}(2)[x]$ & $x^6 + x^2 + x + 1 = (x + 1)(x^5 + x^4 + x^3 + x^2 + 1)$  \\ [0.5ex] 
\hline\hline
$n$ & $4 - 7$ \\ \hline
$2\bar{\bm{h}}^n_0 \in \mathbb{Z}^{L_0}$ & $(1,0,\pm 1,1,0,0,\pm 1)$ \\ \hline
$2|\bar{\bm{h}}^n_0| \in \text{GF}(2)^{L_0}$ & $(1,0,1,1,0,0,1)$ \\ \hline
$q^n(x) \in \text{GF}(2)[x]$ & $x^6 + x^3 + x^2 + 1 = (x + 1)(x^5 + x^4 + x^3 + x + 1)$ \\  [1ex] 
 \hline
\end{tabular}
\end{table}

\begin{table}[!htbp]
\caption{Maximum length minimum Euclidean distance guaranteeing NSMs of rate $2$ with extremely simple filters' coefficients, for $L_0=8$ and $L_1=1$ ($\bar{\bm{h}}_1 = (1)$ and $\eta = \eta_0 = \eta_1 = 5/2$).}
\label{table:Simple Filters Coefficients L_0=8 L_1=1}
\centering
\begin{tabular}{|c|c|} 
\hline
\multicolumn{2}{|l|}{\# of extremely simple $h_0[k]$ filters $ = 240$} \\ \hline
\multicolumn{2}{|l|}{\# of non equivalent $h_0[k]$ filters $ = 33$} \\ \hline
\multicolumn{2}{|l|}{\# of best non equivalent $h_0[k]$ filters $ = 8$} \\ \hline
\multicolumn{2}{|l|}{\gls{msed}, $d_{\text{min}}^2 = 5$} \\ \hline
\multicolumn{2}{|l|}{Asymptotic gain with respect to $4$-ASK $ = 10 \log_{10}(5/4) \approx 0.9691$ dB} \\ \hline
\multicolumn{2}{|l|}{Length of shortest error event with minimum Euclidean distance $= 2^{L_0-2} = 64$} \\ [0.5ex] 
\hline\hline
$n$ & $0 - 3$ \\ \hline
$2\bar{\bm{h}}^n_0 \in \mathbb{Z}^{L_0}$ & $(1,0,\pm 1,0,\pm 1,0,0,1)$ \\ \hline
$2|\bar{\bm{h}}^n_0| \in \text{GF}(2)^{L_0}$ & $(1,0,1,0,1,0,0, 1)$ \\ \hline
$q^n(x) \in \text{GF}(2)[x]$ & $x^7 + x^4 + x^2 + 1 = (x + 1)(x^6 + x^5 + x^4 + x + 1)$  \\ [0.5ex] 
\hline\hline
$n$ & $4 - 7$ \\ \hline
$2\bar{\bm{h}}^n_0 \in \mathbb{Z}^{L_0}$ & $(1,1,0,0,0,\pm 1,0,\pm 1)$ \\ \hline
$2|\bar{\bm{h}}^n_0| \in \text{GF}(2)^{L_0}$ & $(1,1,0,0,0,1,0,1)$ \\ \hline
$q^n(x) \in \text{GF}(2)[x]$ & $x^7 + x^5 + x + 1 = (x + 1)(x^6 + x^5 + 1)$ \\  [1ex] 
 \hline
\end{tabular}
\end{table}

\begin{table}[!htbp]
\caption{Maximum length minimum Euclidean distance guaranteeing NSMs of rate $2$ with extremely simple filters' coefficients, for $L_0=9$ and $L_1=1$ ($\bar{\bm{h}}_1 = (1)$ and $\eta = \eta_0 = \eta_1 = 5/2$).}
\label{table:Simple Filters Coefficients L_0=9 L_1=1}
\centering
\begin{tabular}{|c|c|} 
\hline
\multicolumn{2}{|l|}{\# of extremely simple $h_0[k]$ filters $ = 336$} \\ \hline
\multicolumn{2}{|l|}{\# of non equivalent $h_0[k]$ filters $ = 54$} \\ \hline
\multicolumn{2}{|l|}{\# of best non equivalent $h_0[k]$ filters $ = 16$} \\ \hline
\multicolumn{2}{|l|}{\gls{msed}, $d_{\text{min}}^2 = 5$} \\ \hline
\multicolumn{2}{|l|}{Asymptotic gain with respect to $4$-ASK $ = 10 \log_{10}(5/4) \approx 0.9691$ dB} \\ \hline
\multicolumn{2}{|l|}{Length of shortest error event with minimum Euclidean distance $= 2^{L_0-2} = 128$} \\ [0.5ex] 
\hline\hline
$n$ & $0 - 3$ \\ \hline
$2\bar{\bm{h}}^n_0 \in \mathbb{Z}^{L_0}$ & $(1,1,\pm 1,0,0,0,0,0,\pm 1)$ \\ \hline
$2|\bar{\bm{h}}^n_0| \in \text{GF}(2)^{L_0}$ & $(1,1,1,0,0,0,0,0, 1)$ \\ \hline
$q^n(x) \in \text{GF}(2)[x]$ & $x^8 + x^2 + x + 1 = (x + 1)(x^7 + x^6 + x^5 + x^4 + x^3 + x^2 + 1)$  \\ [0.5ex] 
\hline\hline
$n$ & $4 - 7$ \\ \hline
$2\bar{\bm{h}}^n_0 \in \mathbb{Z}^{L_0}$ & $(1,1,0,0,\pm 1,0,0,0,\pm 1)$ \\ \hline
$2|\bar{\bm{h}}^n_0| \in \text{GF}(2)^{L_0}$ & $(1,1,0,0,1,0,0,0, 1)$ \\ \hline
$q^n(x) \in \text{GF}(2)[x]$ & $x^8 + x^4 + x + 1 = (x + 1)(x^7 + x^6 + x^5 + x^4 + 1)$  \\ [0.5ex] 
\hline\hline
$n$ & $8 - 11$ \\ \hline
$2\bar{\bm{h}}^n_0 \in \mathbb{Z}^{L_0}$ & $(1,0,0,1,\pm 1,0,0,0,\pm 1)$ \\ \hline
$2|\bar{\bm{h}}^n_0| \in \text{GF}(2)^{L_0}$ & $(1,0,0,1,1,0,0,0, 1)$ \\ \hline
$q^n(x) \in \text{GF}(2)[x]$ & $x^8 + x^4 + x^3 + 1 = (x + 1)(x^7 + x^6 + x^5 + x^4 + x^2 + x + 1)$  \\ [0.5ex] 
\hline\hline
$n$ & $12 - 15$ \\ \hline
$2\bar{\bm{h}}^n_0 \in \mathbb{Z}^{L_0}$ & $(1,1,0,0,0,0,\pm 1,0,\pm 1)$ \\ \hline
$2|\bar{\bm{h}}^n_0| \in \text{GF}(2)^{L_0}$ & $(1,1,0,0,0,0,1,0, 1)$ \\ \hline
$q^n(x) \in \text{GF}(2)[x]$ & $x^8 + x^6 + x + 1 = (x + 1)(x^7 + x^6 + 1)$  \\  [1ex] 
 \hline
\end{tabular}
\end{table}

\begin{table}[!htbp]
\caption{Maximum length minimum Euclidean distance guaranteeing NSMs of rate $2$ with extremely simple filters' coefficients, for $L_0=10$ and $L_1=1$ ($\bar{\bm{h}}_1 = (1)$ and $\eta = \eta_0 = \eta_1 = 5/2$).}
\label{table:Simple Filters Coefficients L_0=10 L_1=1}
\centering
\begin{tabular}{|c|c|} 
\hline
\multicolumn{2}{|l|}{\# of extremely simple $h_0[k]$ filters $ = 448$} \\ \hline
\multicolumn{2}{|l|}{\# of non equivalent $h_0[k]$ filters $ = 60$} \\ \hline
\multicolumn{2}{|l|}{\# of best non equivalent $h_0[k]$ filters $ = 8$} \\ \hline
\multicolumn{2}{|l|}{\gls{msed}, $d_{\text{min}}^2 = 5$} \\ \hline
\multicolumn{2}{|l|}{Asymptotic gain with respect to $4$-ASK $ = 10 \log_{10}(5/4) \approx 0.9691$ dB} \\ \hline
\multicolumn{2}{|l|}{Length of shortest error event with minimum Euclidean distance $= 2^{L_0-2} = 256$} \\ [0.5ex] 
\hline\hline
$n$ & $0 - 3$ \\ \hline
$2\bar{\bm{h}}^n_0 \in \mathbb{Z}^{L_0}$ & $(1,0,0, 1,0,\pm 1,0,0,0,\pm 1)$ \\ \hline
$2|\bar{\bm{h}}^n_0| \in \text{GF}(2)^{L_0}$ & $(1,0,0,1,0,1,0,0,0, 1)$ \\ \hline
$q^n(x) \in \text{GF}(2)[x]$ & $x^9 + x^5 + x^3 + 1 = (x + 1)(x^8 + x^7 + x^6 + x^5 + x^2 + x + 1)$  \\ [0.5ex] 
\hline\hline
$n$ & $4 - 7$ \\ \hline
$2\bar{\bm{h}}^n_0 \in \mathbb{Z}^{L_0}$ & $(1,0,\pm 1,0,0,0,\pm 1,0,0,1)$ \\ \hline
$2|\bar{\bm{h}}^n_0| \in \text{GF}(2)^{L_0}$ & $(1,0,1,0,0,0,1,0,0, 1)$ \\ \hline
$q^n(x) \in \text{GF}(2)[x]$ & $x^9 + x^6 + x^2 + 1 = (x + 1)(x^8 + x^7 + x^6 + x + 1)$ \\  [1ex] 
 \hline
\end{tabular}
\end{table}

\begin{table}[!htbp]
\caption{Maximum length minimum Euclidean distance guaranteeing NSMs of rate $2$ with extremely simple filters' coefficients, for $L_0=11$ and $L_1=1$ ($\bar{\bm{h}}_1 = (1)$ and $\eta = \eta_0 = \eta_1 = 5/2$).}
\label{table:Simple Filters Coefficients L_0=11 L_1=1}
\centering
\begin{tabular}{|c|c|} 
\hline
\multicolumn{2}{|l|}{\# of extremely simple $h_0[k]$ filters $ = 576$} \\ \hline
\multicolumn{2}{|l|}{\# of non equivalent $h_0[k]$ filters $ = 92$} \\ \hline
\multicolumn{2}{|l|}{\# of best non equivalent $h_0[k]$ filters $ = 12$} \\ \hline
\multicolumn{2}{|l|}{\gls{msed}, $d_{\text{min}}^2 = 5$} \\ \hline
\multicolumn{2}{|l|}{Asymptotic gain with respect to $4$-ASK $ = 10 \log_{10}(5/4) \approx 0.9691$ dB} \\ \hline
\multicolumn{2}{|l|}{Length of shortest error event with minimum Euclidean distance $= 2^{L_0-2} = 512$} \\ [0.5ex] 
\hline\hline
$n$ & $0 - 3$ \\ \hline
$2\bar{\bm{h}}^n_0 \in \mathbb{Z}^{L_0}$ & $(1,0,\pm 1,1,0,0,0,0,0,0,\pm 1)$ \\ \hline
$2|\bar{\bm{h}}^n_0| \in \text{GF}(2)^{L_0}$ & $(1,0,1,1,0,0,0,0,0,1)$ \\ \hline
$q^n(x) \in \text{GF}(2)[x]$ & $x^{10} + x^3 + x^2 + 1 = (x + 1)(x^9 + x^8 + x^7 + x^6 + x^5 + x^4 + x^3 + x + 1)$  \\ [0.5ex] 
\hline\hline
$n$ & $4 - 7$ \\ \hline
$2\bar{\bm{h}}^n_0 \in \mathbb{Z}^{L_0}$ & $(1,0,\pm 1,0,0,1,0,0,0,0,\pm 1)$ \\ \hline
$2|\bar{\bm{h}}^n_0| \in \text{GF}(2)^{L_0}$ & $(1,0,1,0,0,1,0,0,0,0, 1)$ \\ \hline
$q^n(x) \in \text{GF}(2)[x]$ & $x^{10} + x^5 + x^2 + 1 = (x + 1)(x^9 + x^8 + x^7 + x^6 + x^5 + x + 1)$  \\ [0.5ex] 
\hline\hline
$n$ & $8 - 11$ \\ \hline
$2\bar{\bm{h}}^n_0 \in \mathbb{Z}^{L_0}$ & $(1,0,0,1,0,0,\pm 1,0,0,0,\pm 1)$ \\ \hline
$2|\bar{\bm{h}}^n_0| \in \text{GF}(2)^{L_0}$ & $(1,0,0,1,0,0,1,0,0,0, 1)$ \\ \hline
$q^n(x) \in \text{GF}(2)[x]$ & $x^{10} + x^6 + x^3 + 1 = (x + 1)(x^9 + x^8 + x^7 + x^6 + x^2 + x + 1)$ \\  [1ex] 
 \hline
\end{tabular}
\end{table}

\begin{table}[!htbp]
\caption{Maximum length minimum Euclidean distance guaranteeing NSMs of rate $2$ with extremely simple filters' coefficients, for $L_0=12$ and $L_1=1$ ($\bar{\bm{h}}_1 = (1)$ and $\eta = \eta_0 = \eta_1 = 5/2$).}
\label{table:Simple Filters Coefficients L_0=12 L_1=1}
\centering
\begin{tabular}{|c|c|} 
\hline
\multicolumn{2}{|l|}{\# of extremely simple $h_0[k]$ filters $ = 720$} \\ \hline
\multicolumn{2}{|l|}{\# of non equivalent $h_0[k]$ filters $ = 95$} \\ \hline
\multicolumn{2}{|l|}{\# of best non equivalent $h_0[k]$ filters $ = 8$} \\ \hline
\multicolumn{2}{|l|}{\gls{msed}, $d_{\text{min}}^2 = 5$} \\ \hline
\multicolumn{2}{|l|}{Asymptotic gain with respect to $4$-ASK $ = 10 \log_{10}(5/4) \approx 0.9691$ dB} \\ \hline
\multicolumn{2}{|l|}{Length of shortest error event with minimum Euclidean distance $= 2^{L_0-2} = 1024$} \\ [0.5ex] 
\hline\hline
$n$ & $0 - 3$ \\ \hline
$2\bar{\bm{h}}^n_0 \in \mathbb{Z}^{L_0}$ & $(1,1,0,\pm 1, 0,0,0,0,0,0,0,\pm 1)$ \\ \hline
$2|\bar{\bm{h}}^n_0| \in \text{GF}(2)^{L_0}$ & $(1,1,0,1,0,0,0,0,0,0,0,1)$ \\ \hline
$q^n(x) \in \text{GF}(2)[x]$ & $x^{11} + x^3 + x + 1 = (x + 1)(x^{10} + x^9 + x^8 + x^7 + x^6 + x^5 + x^4 + x^3 + 1)$  \\ [0.5ex] 
\hline\hline
$n$ & $4 - 7$ \\ \hline
$2\bar{\bm{h}}^n_0 \in \mathbb{Z}^{L_0}$ & $(1,0,\pm 1,0,\pm 1, 0,0,0,0,0,0,1)$ \\ \hline
$2|\bar{\bm{h}}^n_0| \in \text{GF}(2)^{L_0}$ & $(1,0,1,0,1,0,0,0,0,0,0,1)$ \\ \hline
$q^n(x) \in \text{GF}(2)[x]$ & $x^{11} + x^4 + x^2 + 1 = (x + 1)(x^{10} + x^9 + x^8 + x^7 + x^6 + x^5 + x^4 + x + 1)$ \\  [1ex] 
 \hline
\end{tabular}
\end{table}

\begin{table}[!htbp]
\caption{Maximum length minimum Euclidean distance guaranteeing NSMs of rate $2$ with extremely simple filters' coefficients, for $L_0=13$ and $L_1=1$ ($\bar{\bm{h}}_1 = (1)$ and $\eta = \eta_0 = \eta_1 = 5/2$).}
\label{table:Simple Filters Coefficients L_0=13 L_1=1}
\centering
\begin{tabular}{|c|c|} 
\hline
\multicolumn{2}{|l|}{\# of extremely simple $h_0[k]$ filters $ = 880$} \\ \hline
\multicolumn{2}{|l|}{\# of non equivalent $h_0[k]$ filters $ = 140$} \\ \hline
\multicolumn{2}{|l|}{\# of best non equivalent $h_0[k]$ filters $ = 4$} \\ \hline
\multicolumn{2}{|l|}{\gls{msed}, $d_{\text{min}}^2 = 5$} \\ \hline
\multicolumn{2}{|l|}{Asymptotic gain with respect to $4$-ASK $ = 10 \log_{10}(5/4) \approx 0.9691$ dB} \\ \hline
\multicolumn{2}{|l|}{Length of shortest error event with minimum Euclidean distance $= 2^{L_0-2} = 2048$} \\ [0.5ex] 
\hline\hline
$n$ & $0 - 3$ \\ \hline
$2\bar{\bm{h}}^n_0 \in \mathbb{Z}^{L_0}$ & $(1,0,\pm 1,0,0,0,0,1, 0,0,0,0,\pm 1)$ \\ \hline
$2|\bar{\bm{h}}^n_0| \in \text{GF}(2)^{L_0}$ & $(1,0,1,0,0,0,0,1,0,0,0,0,1)$ \\ \hline
$q^n(x) \in \text{GF}(2)[x]$ & $x^{12} + x^7 + x^2 + 1 = (x + 1)(x^{11} + x^{10} + x^9 + x^8 + x^7 + x + 1)$ \\  [1ex] 
 \hline
\end{tabular}
\end{table}

\begin{table}[!htbp]
\caption{Maximum length minimum Euclidean distance guaranteeing NSMs of rate $2$ with extremely simple filters' coefficients, for $L_0=14$ and $L_1=1$ ($\bar{\bm{h}}_1 = (1)$ and $\eta = \eta_0 = \eta_1 = 5/2$).}
\label{table:Simple Filters Coefficients L_0=14 L_1=1}
\centering
\begin{tabular}{|c|c|} 
\hline
\multicolumn{2}{|l|}{\# of extremely simple $h_0[k]$ filters $ = 1056$} \\ \hline
\multicolumn{2}{|l|}{\# of non equivalent $h_0[k]$ filters $ = 138$} \\ \hline
\multicolumn{2}{|l|}{\# of best non equivalent $h_0[k]$ filters $ = 16$} \\ \hline
\multicolumn{2}{|l|}{\gls{msed}, $d_{\text{min}}^2 = 5$} \\ \hline
\multicolumn{2}{|l|}{Asymptotic gain with respect to $4$-ASK $ = 10 \log_{10}(5/4) \approx 0.9691$ dB} \\ \hline
\multicolumn{2}{|l|}{Length of shortest error event with minimum Euclidean distance $= 2^{L_0-2}-1 = 4095$} \\ [0.5ex] 
\hline\hline
$n$ & $0 - 3$ \\ \hline
$2\bar{\bm{h}}^n_0 \in \mathbb{Z}^{L_0}$ & $(1,1,\pm 1,0,0,0,0,0,0,0,0,0,0,\pm 1)$ \\ \hline
$2|\bar{\bm{h}}^n_0| \in \text{GF}(2)^{L_0}$ & $(1,1,1,0,0,0,0,0,0,0,0,0,0, 1)$ \\ \hline
$q^n(x) \in \text{GF}(2)[x]$ & $x^{13} + x^2 + x + 1 = (x + 1)^2 (x^{11} + x^9 + x^7 + x^5 + x^3 + x + 1)$ \\ [0.5ex] 
\hline\hline
$n$ & $4 - 7$ \\ \hline
$2\bar{\bm{h}}^n_0 \in \mathbb{Z}^{L_0}$ & $(1,0,0,0,0,1, \pm 1,0,0,0,0,0,0,\pm 1)$ \\ \hline
$2|\bar{\bm{h}}^n_0| \in \text{GF}(2)^{L_0}$ & $(1,0,0,0,0,1,1,0,0,0,0,0,0,1)$ \\ \hline
$q^n(x) \in \text{GF}(2)[x]$ & $x^{13} + x^6 + x^5 + 1 = (x + 1)^2 (x^{11} + x^9 + x^7 + x^5 + x^4 + x^2 + 1)$ \\ [0.5ex] 
\hline\hline
$n$ & $8 - 11$ \\ \hline
$2\bar{\bm{h}}^n_0 \in \mathbb{Z}^{L_0}$ & $(1,0,0,0,\pm 1,0,0,1,0,0,0,0,0,\pm 1)$ \\ \hline
$2|\bar{\bm{h}}^n_0| \in \text{GF}(2)^{L_0}$ & $(1,0,0,0,1,0,0,1,0,0,0,0,0,1)$ \\ \hline
$q^n(x) \in \text{GF}(2)[x]$ & $x^{13} + x^7 + x^4 + 1 = (x + 1)^2 (x^{11} + x^9 + x^7 + x^2 + 1)$ \\ [0.5ex] 
\hline\hline
$n$ & $12 - 15$ \\ \hline
$2\bar{\bm{h}}^n_0 \in \mathbb{Z}^{L_0}$ & $(1,0,\pm 1,0,0,0,0,0,0,1,0,0,0,\pm 1)$ \\ \hline
$2|\bar{\bm{h}}^n_0| \in \text{GF}(2)^{L_0}$ & $(1,0,1,0,0,0,0,0,0,1,0,0,0,1)$ \\ \hline
$q^n(x) \in \text{GF}(2)[x]$ & $x^{13} + x^9 + x^2 + 1 = (x + 1)^2 (x^{11} + x^9 + 1)$ \\  [1ex] 
 \hline
\end{tabular}
\end{table}

\begin{table}[!htbp]
\caption{Maximum length minimum Euclidean distance guaranteeing NSMs of rate $2$ with extremely simple filters' coefficients, for $L_0=15$ and $L_1=1$ ($\bar{\bm{h}}_1 = (1)$ and $\eta = \eta_0 = \eta_1 = 5/2$).}
\label{table:Simple Filters Coefficients L_0=15 L_1=1}
\centering
\begin{tabular}{|c|c|} 
\hline
\multicolumn{2}{|l|}{\# of extremely simple $h_0[k]$ filters $ = 1248$} \\ \hline
\multicolumn{2}{|l|}{\# of non equivalent $h_0[k]$ filters $ = 198$} \\ \hline
\multicolumn{2}{|l|}{\# of best non equivalent $h_0[k]$ filters $ = 28$} \\ \hline
\multicolumn{2}{|l|}{\gls{msed}, $d_{\text{min}}^2 = 5$} \\ \hline
\multicolumn{2}{|l|}{Asymptotic gain with respect to $4$-ASK $ = 10 \log_{10}(5/4) \approx 0.9691$ dB} \\ \hline
\multicolumn{2}{|l|}{Length of shortest error event with minimum Euclidean distance $= 2^{L_0-2} = 8192$} \\ [0.5ex] 
\hline\hline
$n$ & $0 - 3$ \\ \hline
$2\bar{\bm{h}}^n_0 \in \mathbb{Z}^{L_0}$ & $(1,1,\pm 1,0,0,0,0,0,0,0,0,0,0,0,\pm 1)$ \\ \hline
$2|\bar{\bm{h}}^n_0| \in \text{GF}(2)^{L_0}$ & $(1,1,1,0,0,0,0,0,0,0,0,0,0,0,1)$ \\ \hline
\multirow{2}{*}{$q^n(x) \in \text{GF}(2)[x]$} & $x^{14} + x^2 + x + 1 = (x + 1)(x^{13} + x^{12} + x^{11} + x^{10} + x^9 + x^8 + x^7 + x^6 + x^5$ \\
& $+ x^4 + x^3 + x^2 + 1)$ \\ [0.5ex] 
\hline\hline
$n$ & $4 - 7$ \\ \hline
$2\bar{\bm{h}}^n_0 \in \mathbb{Z}^{L_0}$ & $(1,0,0,1,\pm 1,0,0,0,0,0,0,0,0,0,\pm 1)$ \\ \hline
$2|\bar{\bm{h}}^n_0| \in \text{GF}(2)^{L_0}$ & $(1,0,0,1,1,0,0,0,0,0,0,0,0,0,1)$ \\ \hline
\multirow{2}{*}{$q^n(x) \in \text{GF}(2)[x]$} & $x^{14} + x^4 + x^3 + 1 = (x + 1)(x^{13} + x^{12} + x^{11} + x^{10} + x^9 + x^8 + x^7 + x^6 + x^5$ \\
& $+ x^4 + x^2 + x + 1)$ \\ [0.5ex] 
\hline\hline
$n$ & $8 - 11$ \\ \hline
$2\bar{\bm{h}}^n_0 \in \mathbb{Z}^{L_0}$ & $(1,0,\pm 1,0,0,1,0,0,0,0,0,0,0,0,\pm 1)$ \\ \hline
$2|\bar{\bm{h}}^n_0| \in \text{GF}(2)^{L_0}$ & $(1,0,1,0,0,1,0,0,0,0,0,0,0,0,1)$ \\ \hline
\multirow{2}{*}{$q^n(x) \in \text{GF}(2)[x]$} & $x^{14} + x^5 + x^2 + 1 = (x + 1)(x^{13} + x^{12} + x^{11} + x^{10} + x^9 + x^8 + x^7 + x^6 + x^5 $ \\
& $+ x + 1)$ \\ [0.5ex]
\hline\hline
$n$ & $12 - 15$ \\ \hline
$2\bar{\bm{h}}^n_0 \in \mathbb{Z}^{L_0}$ & $(1,0,\pm 1,0,0,0,0,1,0,0,0,0,0,0,\pm 1)$ \\ \hline
$2|\bar{\bm{h}}^n_0| \in \text{GF}(2)^{L_0}$ & $(1,0,1,0,0,0,0,1,0,0,0,0,0,0,1)$ \\ \hline
$q^n(x) \in \text{GF}(2)[x]$ & $x^{14} + x^7 + x^2 + 1 = (x + 1)(x^{13} + x^{12} + x^{11} + x^{10} + x^9 + x^8 + x^7 + x + 1)$ \\ [0.5ex]
\hline\hline
$n$ & $16 - 19$ \\ \hline
$2\bar{\bm{h}}^n_0 \in \mathbb{Z}^{L_0}$ & $(1,0,0,0,0,1,0,0,\pm 1,0,0,0,0,0,\pm 1)$ \\ \hline
$2|\bar{\bm{h}}^n_0| \in \text{GF}(2)^{L_0}$ & $(1,0,0,0,0,1,0,0,1,0,0,0,0,0,1)$ \\ \hline
$q^n(x) \in \text{GF}(2)[x]$ & $x^{14} + x^8 + x^5 + 1 = (x + 1)(x^{13} + x^{12} + x^{11} + x^{10} + x^9 + x^8 + x^4 + x^3 + x^2$ \\
& $+ x + 1)$ \\ [0.5ex] 
\hline\hline
$n$ & $20 - 23$ \\ \hline
$2\bar{\bm{h}}^n_0 \in \mathbb{Z}^{L_0}$ & $(1,0,0,0,\pm 1,0,0,0,0,1,,0,0,0,0,\pm 1)$ \\ \hline
$2|\bar{\bm{h}}^n_0| \in \text{GF}(2)^{L_0}$ & $(1,0,0,0,1,0,0,0,0,1,0,0,0,0,1)$ \\ \hline
$q^n(x) \in \text{GF}(2)[x]$ & $x^{14} + x^9 + x^4 + 1 = (x + 1)(x^{13} + x^{12} + x^{11} + x^{10} + x^9 + x^3 + x^2 + x + 1)$ \\ [0.5ex] 
\hline\hline
$n$ & $24 - 27$ \\ \hline
$2\bar{\bm{h}}^n_0 \in \mathbb{Z}^{L_0}$ & $(1,0,\pm 1,0,0,0,0,0,0,0,0,1,0,0,\pm 1)$ \\ \hline
$2|\bar{\bm{h}}^n_0| \in \text{GF}(2)^{L_0}$ & $(1,0,1,0,0,0,0,0,0,0,0,1,0,0,1)$ \\ \hline
$q^n(x) \in \text{GF}(2)[x]$ & $x^{14} + x^{11} + x^2 + 1 = (x + 1)(x^{13} + x^{12} + x^{11} + x + 1)$ \\ [1ex] 
 \hline
\end{tabular}
\end{table}

\begin{table}[!htbp]
\caption{Characteristics of eligible error events input sequences differences, with MSED, $d_{\text{min}}^2=5,$ and associated NSMs of rate $2$ with extremely simple filters' coefficients, for $L_0=5$ and $L_1=1$ ($\bar{\bm{h}}_1 = (1)$, $\eta = \eta_0 = \eta_1 = 5/2,$ $p^n(x) = x^3 + x^2 + 1$, $q^n(x) = x^4 + x^2 + x + 1 = (x + 1)(x^3 + x^2 + 1),$  $\Delta \ddot{b}^n_0(x) = (x^7 + 1)/q^n(x) = x^3 + x + 1,$ and $\Delta \bar{\bm{s}}_0 \triangleq \Delta \mathring{\bm{b}}_0 \circledast \mathring{\bm{h}}_0$).}
\label{table:Characteristics First Minimum Euclidean Distance Simple Filters Coefficients L_0=5 L_1=1}
\centering
\begin{tabular}{|l|l|} 
\hline
\multicolumn{2}{|c|}{$\mathring{\bm{h}}_0 = \mathring{\bm{h}}^0_0 = (1,1,-1,0,-1)$} \\  [0.5ex] 
\hline
$\Delta \mathring{\bm{b}}_0 = \pm (1,-1,0,-1)$ & $\Delta \mathring{\bm{b}}_0 = \pm (1,-1,0,1)$ \\
$\Delta \bar{\bm{s}}_0 = \pm (1,0,-2,0,-2,2,0,1)$ & $\Delta \bar{\bm{s}}_0 = \pm (1,0,-2,2,0,0,0,-1)$ \\
$\Delta \mathring{\bm{b}}_1 = \pm (0\lor-1,0,1,0,1,-1,0,0\lor-1)$ & $\Delta \mathring{\bm{b}}_1 = \pm (0\lor-1,0,1,-1,0,0,0,0\lor1)$ \\
$\Delta \bar{\bm{s}} = \pm (\pm 1,0,0,0,0,0,0,\pm 1)$ & $\Delta \bar{\bm{s}} = \pm (\pm 1,0,0,0,0,0,0,\mp 1)$ \\ \hline
 $\Delta \mathring{\bm{b}}_0 = \pm (1,1,0,-1)$ & 
$\Delta \mathring{\bm{b}}_0 = \pm (1,1,0,1)$ \\
$\Delta \bar{\bm{s}}_0 = \pm (1,2,0,-2,-2,0,0,1)$ & $\Delta \bar{\bm{s}}_0 = \pm (1,2,0,0,0,-2,0,-1)$ \\
$\Delta \mathring{\bm{b}}_1 = \pm (0\lor-1,-1,0,1,1,0,0,0\lor-1)$ & $\Delta \mathring{\bm{b}}_1 = \pm (0\lor-1,-1,0,0,0,1,0,0\lor1)$ \\
$\Delta \bar{\bm{s}} = \pm (\pm 1,0,0,0,0,0,0,\pm 1)$ & $\Delta \bar{\bm{s}} = \pm (\pm 1,0,0,0,0,0,0,\mp 1)$ \\ [0.5ex]
\hline\hline
\multicolumn{2}{|c|}{$\mathring{\bm{h}}_0 = \mathring{\bm{h}}^1_0 = (1,1,-1,0,1)$} \\ [0.5ex] 
\hline
$\Delta \mathring{\bm{b}}_0 = \pm (1,-1,0,-1)$ & $\Delta \mathring{\bm{b}}_0 = \pm (1,-1,0,1)$ \\
$\Delta \bar{\bm{s}}_0 = \pm (1,0,-2,0,0,0,0,-1)$ & $\Delta \bar{\bm{s}}_0 = \pm (1,0,-2,2,2,-2,0,1)$ \\
$\Delta \mathring{\bm{b}}_1 = \pm (0\lor-1,0,1,0,0,0,0,0\lor1)$ & $\Delta \mathring{\bm{b}}_1 = \pm (0\lor-1,0,1,-1,-1,1,0,0\lor-1)$ \\
$\Delta \bar{\bm{s}} = \pm (\pm 1,0,0,0,0,0,0,\mp 1)$ & $\Delta \bar{\bm{s}} = \pm (\pm 1,0,0,0,0,0,0,\pm 1)$ \\ \hline
$\Delta \mathring{\bm{b}}_0 = \pm (1,1,0,-1)$ & $\Delta \mathring{\bm{b}}_0 = \pm (1,1,0,1)$ \\
$\Delta \bar{\bm{s}}_0 = \pm (1,2,0,-2,0,2,0,-1)$  & $\Delta \bar{\bm{s}}_0 = \pm (1,2,0,0,2,0,0,1)$ \\
$\Delta \mathring{\bm{b}}_1 = \pm (0\lor-1,-1,0,1,0,-1,0,0\lor1)$ & $\Delta \mathring{\bm{b}}_1 = \pm (0\lor-1,-1,0,0,-1,0,0,0\lor-1)$ \\
$\Delta \bar{\bm{s}} = \pm (\pm 1,0,0,0,0,0,0,\mp 1)$ & $\Delta \bar{\bm{s}} = \pm (\pm 1,0,0,0,0,0,0,\pm 1)$ \\ [0.5ex]
\hline\hline
\multicolumn{2}{|c|}{$\mathring{\bm{h}}_0 = \mathring{\bm{h}}^2_0 = (1,1,1,0,-1)$} \\ [0.5ex] 
\hline
$\Delta \mathring{\bm{b}}_0 = \pm (1,-1,0,-1)$ & $\Delta \mathring{\bm{b}}_0 = \pm (1,-1,0,1)$ \\
$\Delta \bar{\bm{s}}_0 = \pm (1,0,0,-2,-2,0,0,1)$ & $\Delta \bar{\bm{s}}_0 = \pm (1,0,0,0,0,2,0,-1)$ \\
$\Delta \mathring{\bm{b}}_1 = \pm (0\lor-1,0,0,1,1,0,0,0\lor-1)$ & $\Delta \mathring{\bm{b}}_1 = \pm (0\lor-1,0,0,0,0,-1,0,0\lor1)$ \\
$\Delta \bar{\bm{s}} = \pm (\pm 1,0,0,0,0,0,0, \pm1)$ & $\Delta \bar{\bm{s}} = \pm (\pm 1,0,0,0,0,0,0,\mp 1)$ \\ \hline
$\Delta \mathring{\bm{b}}_0 = \pm (1,1,0,-1)$ & $\Delta \mathring{\bm{b}}_0 = \pm (1,1,0,1)$ \\
$\Delta \bar{\bm{s}}_0 = \pm (1,2,2,0,-2,-2,0,1)$ & $\Delta \bar{\bm{s}}_0 = \pm (1,2,2,2,0,0,0,-1)$ \\
$\Delta \mathring{\bm{b}}_1 = \pm (0\lor-1,-1,-1,0,1,1,0,0\lor-1)$ & $\Delta \mathring{\bm{b}}_1 = \pm (0\lor-1,-1,-1,-1,0,0,0,0\lor1)$ \\
$\Delta \bar{\bm{s}} = \pm (\pm 1,0,0,0,0,0,0,\pm 1)$ & $\Delta \bar{\bm{s}} = \pm (\pm 1,0,0,0,0,0,0,\mp 1)$ \\ [0.5ex]
\hline\hline
\multicolumn{2}{|c|}{$\mathring{\bm{h}}_0 = \mathring{\bm{h}}^3_0 = (1,1,1,0,1)$} \\  [0.5ex] 
\hline
$\Delta \mathring{\bm{b}}_0 = \pm (1,-1,0,-1)$ & $\Delta \mathring{\bm{b}}_0 = \pm (1,-1,0,1)$ \\
$\Delta \bar{\bm{s}}_0 = \pm (1,0,0,-2,0,-2,0,-1)$ & $\Delta \bar{\bm{s}}_0 = \pm (1,0,0,0,2,0,0,1)$ \\
$\Delta \mathring{\bm{b}}_1 = \pm (0\lor-1,0,0,1,0,1,0,0\lor1)$ & $\Delta \mathring{\bm{b}}_1 = \pm (0\lor-1,0,0,0,-1,0,0,0\lor-1)$ \\
$\Delta \bar{\bm{s}} = \pm (\pm 1,0,0,0,0,0,0,\mp 1)$ & $\Delta \bar{\bm{s}} = \pm (\pm 1,0,0,0,0,0,0,\pm 1)$ \\ \hline
$\Delta \mathring{\bm{b}}_0 = \pm (1,1,0,-1)$ & $\Delta \mathring{\bm{b}}_0 = \pm (1,1,0,1)$ \\
$\Delta \bar{\bm{s}}_0 = \pm (1,2,2,0,0,0,0,-1)$ & $\Delta \bar{\bm{s}}_0 = \pm (1,2,2,2,2,2,0,1)$ \\
$\Delta \mathring{\bm{b}}_1 = \pm (0\lor-1,-1,-1,0,0,0,0,0\lor1)$ & $\Delta \mathring{\bm{b}}_1 = \pm (0\lor-1,-1,-1,-1,-1,-1,0,0\lor-1)$ \\
$\Delta \bar{\bm{s}} = \pm (\pm 1,0,0,0,0,0,0,\mp 1)$ & $\Delta \bar{\bm{s}} = \pm (\pm 1,0,0,0,0,0,0,\pm 1)$ \\ [1ex] 
\hline
\end{tabular}
\end{table} 

\begin{table}[!htbp]
\caption{Building blocks of error events, with MSED, $d_{\text{min}}^2=5,$ associated to the rate-$2$ NSMs with $L_0=5, L_1=1$ and $\mathring{\bm{h}}_0 = \mathring{\bm{h}}^0_0 = (1,1,-1,0,-1),$ deduced from Table~\ref{table:Characteristics First Minimum Euclidean Distance Simple Filters Coefficients L_0=5 L_1=1}.}
\label{table:Building Bricks Simple Filters Coefficients L_0=5 L_1=1 h_0=h^0_0}
\small
\centering
\begin{tabular}{|l|l|l|l|l|} 
\hline
Type & $\Delta \bar{\bm{s}}$ & $\Delta \mathring{\bm{b}}_0$ & $\Delta \mathring{\bm{b}}_1$ & Contribution \\ [0.5ex] \hline\hline
\multirow{16}{*}{Primary}
   & \multirow{4}{*}{$(\bm{1},0,0,0,0,0,0,1)$} & $(1,-1,0,-1)$ & $(0,0,1,0,1,-1,0,\bm{0})$ & \multirow{8}{*}{\shortstack{$2(N^5+2N^6+N^7)=$ \\ $2N^5(1+N)^2$}} \\
   &     & $-(1,-1,0,1)$ & $(1,0,-1,1,0,0,0,\bm{0})$ &    \\
   &     & $(1,1,0,-1)$ & $(0,-1,0,1,1,0,0,\bm{0})$ &    \\ 
   &     & $-(1,1,0,1)$ & $(1,1,0,0,0,-1,0,\bm{0})$ &    \\ \cline{2-4}
   & \multirow{4}{*}{$(\bm{1},0,0,0,0,0,0,-1)$} & $-(1,-1,0,-1)$ & $(1,0,-1,0,-1,1,0,\bm{0})$ &   \\ 
   &     & $(1,-1,0,1)$ & $(0,0,1,-1,0,0,0,\bm{0})$ &    \\
   &     & $-(1,1,0,-1)$ & $(1,1,0,-1,-1,0,0,\bm{0})$ &    \\
   &     & $(1,1,0,1)$ & $(0,-1,0,0,0,1,0,\bm{0})$ &    \\ \cline{2-5}
   & \multirow{4}{*}{$(\bm{-1},0,0,0,0,0,0,1)$} & $(1,-1,0,-1)$ & $(-1,0,1,0,1,-1,0,\bm{0})$ &  \multirow{8}{*}{\shortstack{$2(N^5+2N^6+N^7)=$ \\ $2N^5(1+N)^2$}} \\
   &    & $-(1,-1,0,1)$ & $(0,0,-1,1,0,0,0,\bm{0})$ &    \\ 
   &    & $(1,1,0,-1)$ & $(-1,-1,0,1,1,0,0,\bm{0})$ &    \\ 
   &    & $-(1,1,0,1)$ & $(0,1,0,0,0,-1,0,\bm{0})$ &    \\ \cline{2-4}   
   & \multirow{4}{*}{$(\bm{-1},0,0,0,0,0,0,-1)$} & $-(1,-1,0,-1)$ & $(0,0,-1,0,-1,1,0,\bm{0})$ &   \\
   &     & $(1,-1,0,1)$ & $(-1,0,1,-1,0,0,0,\bm{0})$ &    \\
   &     & $-(1,1,0,-1)$ & $(0,1,0,-1,-1,0,0,\bm{0})$ &    \\ 
   &     & $(1,1,0,1)$ & $(-1,-1,0,0,0,1,0,\bm{0})$ &    \\ [1ex] \hline
\multirow{16}{*}{Secondary}
   & \multirow{4}{*}{$(\bm{1},0,0,0,0,0,0,1)$} & $-(1,-1,0,-1)$ & $(1,0,-1,0,-1,1,0,\bm{1})$ & \multirow{8}{*}{\shortstack{$2(N^6+2N^7+N^8)=$ \\ $2N^6(1+N)^2$}} \\ 
   &     & $(1,-1,0,1)$ & $(0,0,1,-1,0,0,0,\bm{1})$ &    \\
   &     & $-(1,1,0,-1)$ & $(1,1,0,-1,-1,0,0,\bm{1})$ &    \\
   &     & $(1,1,0,1)$ & $(0,-1,0,0,0,1,0,\bm{1})$ &    \\ \cline{2-4}
   & \multirow{4}{*}{$(\bm{1},0,0,0,0,0,0,-1)$} & $(1,-1,0,-1)$ & $(0,0,1,0,1,-1,0,\bm{-1})$ &  \\
   &     & $-(1,-1,0,1)$ & $(1,0,-1,1,0,0,0,\bm{-1})$ &    \\
   &     & $(1,1,0,-1)$ & $(0,-1,0,1,1,0,0,\bm{-1})$ &    \\ 
   &     & $-(1,1,0,1)$ & $(1,1,0,0,0,-1,0,\bm{-1})$ &    \\ \cline{2-5}
   & \multirow{4}{*}{$(\bm{-1},0,0,0,0,0,0,1)$} & $-(1,-1,0,-1)$ & $(0,0,-1,0,-1,1,0,\bm{1})$ & \multirow{8}{*}{\shortstack{$2(N^6+2N^7+N^8)=$ \\ $2N^6(1+N)^2$}} \\
   &     & $(1,-1,0,1)$ & $(-1,0,1,-1,0,0,0,\bm{1})$ &    \\
   &     & $-(1,1,0,-1)$ & $(0,1,0,-1,-1,0,0,\bm{1})$ &    \\ 
   &     & $(1,1,0,1)$ & $(-1,-1,0,0,0,1,0,\bm{1})$ &    \\  \cline{2-4}
   & \multirow{4}{*}{$(\bm{-1},0,0,0,0,0,0,-1)$} & $(1,-1,0,-1)$ & $(-1,0,1,0,1,-1,0,\bm{-1})$ &   \\
   &    & $-(1,-1,0,1)$ & $(0,0,-1,1,0,0,0,\bm{-1})$ &    \\ 
   &    & $(1,1,0,-1)$ & $(-1,-1,0,1,1,0,0,\bm{-1})$ &    \\ 
   &    & $-(1,1,0,1)$ & $(0,1,0,0,0,-1,0,\bm{-1})$ &    \\  [1ex] 
\hline
\end{tabular}
\end{table} 

\begin{table}[!htbp]
\caption{Building blocks of error events, with MSED, $d_{\text{min}}^2=5,$ associated to the rate-$2$ NSMs with $L_0=5, L_1=1$ and $\mathring{\bm{h}}_0 = \mathring{\bm{h}}^1_0 = (1,1,-1,0,1),$ deduced from Table~\ref{table:Characteristics First Minimum Euclidean Distance Simple Filters Coefficients L_0=5 L_1=1}.}
\label{table:Building Bricks Simple Filters Coefficients L_0=5 L_1=1 h_0=h^1_0}
\centering
\small
\begin{tabular}{|l|l|l|l|l|} 
\hline
Type & $\Delta \bar{\bm{s}}$ & $\Delta \mathring{\bm{b}}_0$ & $\Delta \mathring{\bm{b}}_1$ & Contribution \\ [0.5ex] \hline\hline
\multirow{16}{*}{Primary}
   & \multirow{4}{*}{$(\bm{1},0,0,0,0,0,0,1)$} & $-(1,-1,0,-1)$ & $(1,0,-1,0,0,0,0,\bm{0})$ & \multirow{8}{*}{\shortstack{$N^4+2N^5+2N^6$ \\ $+2N^7+N^8=$ \\ $N^4(1+N)^2(1+N^2)$}} \\
   &     & $(1,-1,0,1)$ & $(0,0,1,-1,-1,1,0,\bm{0})$ &    \\
   &     & $-(1,1,0,-1)$ & $(1,1,0,-1,0,1,0,\bm{0})$ &    \\ 
   &     & $(1,1,0,1)$ & $(0,-1,0,0,-1,0,0,\bm{0})$ &    \\ \cline{2-4}
   & \multirow{4}{*}{$(\bm{1},0,0,0,0,0,0,-1)$} & $(1,-1,0,-1)$ & $(0,0,1,0,0,0,0,\bm{0})$ &   \\ 
   &     & $-(1,-1,0,1)$ & $(1,0,-1,1,1,-1,0,\bm{0})$ &    \\
   &     & $(1,1,0,-1)$ & $(0,-1,0,1,0,-1,0,\bm{0})$ &    \\
   &     & $-(1,1,0,1)$ & $(1,1,0,0,1,0,0,\bm{0})$ &    \\ \cline{2-5}
   & \multirow{4}{*}{$(\bm{-1},0,0,0,0,0,0,1)$} & $-(1,-1,0,-1)$ & $(0,0,-1,0,0,0,0,\bm{0})$ &  \multirow{8}{*}{\shortstack{$N^4+2N^5+2N^6$ \\ $+2N^7+N^8=$ \\ $N^4(1+N)^2(1+N^2)$}} \\
   &    & $(1,-1,0,1)$ & $(-1,0,1,-1,-1,1,0,\bm{0})$ &    \\ 
   &    & $-(1,1,0,-1)$ & $(0,1,0,-1,0,1,0,\bm{0})$ &    \\ 
   &    & $(1,1,0,1)$ & $(-1,-1,0,0,-1,0,0,\bm{0})$ &    \\ \cline{2-4}   
   & \multirow{4}{*}{$(\bm{-1},0,0,0,0,0,0,-1)$} & $(1,-1,0,-1)$ & $(-1,0,1,0,0,0,0,\bm{0})$ &   \\
   &     & $-(1,-1,0,1)$ & $(0,0,-1,1,1,-1,0,\bm{0})$ &    \\
   &     & $(1,1,0,-1)$ & $(-1,-1,0,1,0,-1,0,\bm{0})$ &    \\ 
   &     & $-(1,1,0,1)$ & $(0,1,0,0,1,0,0,\bm{0})$ &    \\ [1ex] \hline
\multirow{16}{*}{Secondary}   
   & \multirow{4}{*}{$(\bm{1},0,0,0,0,0,0,1)$} & $(1,-1,0,-1)$ & $(0,0,1,0,0,0,0,\bm{1})$ & \multirow{8}{*}{\shortstack{$N^5+2N^6+2N^7$ \\ $+2N^8+N^9=$ \\ $N^5(1+N)^2(1+N^2)$}}  \\ 
   &     & $-(1,-1,0,1)$ & $(1,0,-1,1,1,-1,0,\bm{1})$ &    \\
   &     & $(1,1,0,-1)$ & $(0,-1,0,1,0,-1,0,\bm{1})$ &    \\
   &     & $-(1,1,0,1)$ & $(1,1,0,0,1,0,0,\bm{1})$ &    \\ \cline{2-4}
   & \multirow{4}{*}{$(\bm{1},0,0,0,0,0,0,-1)$} & $-(1,-1,0,-1)$ & $(1,0,-1,0,0,0,0,\bm{-1})$ &  \\
   &     & $(1,-1,0,1)$ & $(0,0,1,-1,-1,1,0,\bm{-1})$ &    \\
   &     & $-(1,1,0,-1)$ & $(1,1,0,-1,0,1,0,\bm{-1})$ &    \\ 
   &     & $(1,1,0,1)$ & $(0,-1,0,0,-1,0,0,\bm{-1})$ &    \\ \cline{2-5}
   & \multirow{4}{*}{$(\bm{-1},0,0,0,0,0,0,1)$} & $(1,-1,0,-1)$ & $(-1,0,1,0,0,0,0,\bm{1})$ &  \multirow{8}{*}{\shortstack{$N^5+2N^6+2N^7$ \\ $+2N^8+N^9=$ \\ $N^5(1+N)^2(1+N^2)$}}  \\
   &     & $-(1,-1,0,1)$ & $(0,0,-1,1,1,-1,0,\bm{1})$ &    \\
   &     & $(1,1,0,-1)$ & $(-1,-1,0,1,0,-1,0,\bm{1})$ &    \\ 
   &     & $-(1,1,0,1)$ & $(0,1,0,0,1,0,0,\bm{1})$ &    \\ \cline{2-4}
   & \multirow{4}{*}{$(\bm{-1},0,0,0,0,0,0,-1)$} & $-(1,-1,0,-1)$ & $(0,0,-1,0,0,0,0,\bm{-1})$ &  \\
   &    & $(1,-1,0,1)$ & $(-1,0,1,-1,-1,1,0,\bm{-1})$ &    \\ 
   &    & $-(1,1,0,-1)$ & $(0,1,0,-1,0,1,0,\bm{-1})$ &    \\ 
   &    & $(1,1,0,1)$ & $(-1,-1,0,0,-1,0,0,\bm{-1})$ &    \\ [1ex] \hline
\end{tabular}
\end{table} 

\begin{table}[!htbp]
\caption{Building blocks of error events, with MSED, $d_{\text{min}}^2=5,$ associated to the rate-$2$ NSMs with $L_0=5, L_1=1$ and $\mathring{\bm{h}}_0 = \mathring{\bm{h}}^2_0 = (1,1,1,0,-1),$ deduced from Table~\ref{table:Characteristics First Minimum Euclidean Distance Simple Filters Coefficients L_0=5 L_1=1}.}
\label{table:Building Bricks Simple Filters Coefficients L_0=5 L_1=1 h_0=h^2_0}
\small
\centering
\begin{tabular}{|l|l|l|l|l|} 
\hline
Type & $\Delta \bar{\bm{s}}$ & $\Delta \mathring{\bm{b}}_0$ & $\Delta \mathring{\bm{b}}_1$ & Contribution \\ [0.5ex] \hline\hline
\multirow{16}{*}{Primary}
   & \multirow{4}{*}{$(\bm{1},0,0,0,0,0,0,1)$} & $(1,-1,0,-1)$ & $(0,0,0,1,1,0,0,\bm{0})$ & \multirow{8}{*}{\shortstack{$N^4+2N^5+2N^6$ \\ $+2N^7+N^8=$ \\ $N^4(1+N)^2(1+N^2)$}} \\
   &     & $-(1,-1,0,1)$ & $(1,0,0,0,0,1,0,\bm{0})$ &    \\
   &     & $(1,1,0,-1)$ & $(0,-1,-1,0,1,1,0,\bm{0})$ &    \\ 
   &     & $-(1,1,0,1)$ & $(1,1,1,1,0,0,0,\bm{0})$ &    \\ \cline{2-4}
   & \multirow{4}{*}{$(\bm{1},0,0,0,0,0,0,-1)$} & $-(1,-1,0,-1)$ & $(1,0,0,-1,-1,0,0,\bm{0})$ &   \\ 
   &     & $(1,-1,0,1)$ & $(0,0,0,0,0,-1,0,\bm{0})$ &    \\
   &     & $-(1,1,0,-1)$ & $(1,1,1,0,-1,-1,0,\bm{0})$ &    \\
   &     & $(1,1,0,1)$ & $(0,-1,-1,-1,0,0,0,\bm{0})$ &    \\ \cline{2-5}
   & \multirow{4}{*}{$(\bm{-1},0,0,0,0,0,0,1)$} & $(1,-1,0,-1)$ & $(-1,0,0,1,1,0,0,\bm{0})$ &  \multirow{8}{*}{\shortstack{$N^4+2N^5+2N^6$ \\ $+2N^7+N^8=$ \\ $N^4(1+N)^2(1+N^2)$}} \\
   &    & $-(1,-1,0,1)$ & $(0,0,0,0,0,1,0,\bm{0})$ &    \\ 
   &    & $(1,1,0,-1)$ & $(-1,-1,-1,0,1,1,0,\bm{0})$ &    \\ 
   &    & $-(1,1,0,1)$ & $(0,1,1,1,0,0,0,\bm{0})$ &    \\ \cline{2-4}   
   & \multirow{4}{*}{$(\bm{-1},0,0,0,0,0,0,-1)$} & $-(1,-1,0,-1)$ & $(0,0,0,-1,-1,0,0,\bm{0})$ &   \\
   &     & $(1,-1,0,1)$ & $(-1,0,0,0,0,-1,0,\bm{0})$ &    \\
   &     & $-(1,1,0,-1)$ & $(0,1,1,0,-1,-1,0,\bm{0})$ &    \\ 
   &     & $(1,1,0,1)$ & $(-1,-1,-1,-1,0,0,0,\bm{0})$ &    \\ [1ex] \hline
\multirow{16}{*}{Secondary}   
   & \multirow{4}{*}{$(\bm{1},0,0,0,0,0,0,1)$} & $-(1,-1,0,-1)$ & $(1,0,0,-1,-1,0,0,\bm{1})$ & \multirow{8}{*}{\shortstack{$N^5+2N^6+2N^7$ \\ $+2N^8+N^9=$ \\ $N^5(1+N)^2(1+N^2)$}} \\ 
   &     & $(1,-1,0,1)$ & $(0,0,0,0,0,-1,0,\bm{1})$ &    \\
   &     & $-(1,1,0,-1)$ & $(1,1,1,0,-1,-1,0,\bm{1})$ &    \\
   &     & $(1,1,0,1)$ & $(0,-1,-1,-1,0,0,0,\bm{1})$ &    \\ \cline{2-4}
   & \multirow{4}{*}{$(\bm{1},0,0,0,0,0,0,-1)$} & $(1,-1,0,-1)$ & $(0,0,0,1,1,0,0,\bm{-1})$ &  \\
   &     & $-(1,-1,0,1)$ & $(1,0,0,0,0,1,0,\bm{-1})$ &    \\
   &     & $(1,1,0,-1)$ & $(0,-1,-1,0,1,1,0,\bm{-1})$ &    \\ 
   &     & $-(1,1,0,1)$ & $(1,1,1,1,0,0,0,\bm{-1})$ &    \\ \cline{2-5}
   & \multirow{4}{*}{$(\bm{-1},0,0,0,0,0,0,1)$} & $-(1,-1,0,-1)$ & $(0,0,0,-1,-1,0,0,\bm{1})$ & \multirow{8}{*}{\shortstack{$N^5+2N^6+2N^7$ \\ $+2N^8+N^9=$ \\ $N^5(1+N)^2(1+N^2)$}} \\
   &     & $(1,-1,0,1)$ & $(-1,0,0,0,0,-1,0,\bm{1})$ &    \\
   &     & $-(1,1,0,-1)$ & $(0,1,1,0,-1,-1,0,\bm{1})$ &    \\ 
   &     & $(1,1,0,1)$ & $(-1,-1,-1,-1,0,0,0,\bm{1})$ &    \\ \cline{2-4}
   & \multirow{4}{*}{$(\bm{-1},0,0,0,0,0,0,-1)$} & $(1,-1,0,-1)$ & $(-1,0,0,1,1,0,0,\bm{-1})$ &   \\
   &    & $-(1,-1,0,1)$ & $(0,0,0,0,0,1,0,\bm{-1})$ &    \\ 
   &    & $(1,1,0,-1)$ & $(-1,-1,-1,0,1,1,0,\bm{-1})$ &    \\ 
   &    & $-(1,1,0,1)$ & $(0,1,1,1,0,0,0,\bm{-1})$ &    \\ [1ex] \hline
   \end{tabular}
\end{table} 

\begin{table}[!htbp]
\caption{Building blocks of error events, with MSED, $d_{\text{min}}^2=5,$ associated to the rate-$2$ NSMs with $L_0=5, L_1=1$ and $\mathring{\bm{h}}_0 = \mathring{\bm{h}}^3_0 = (1,1,1,0,1),$ deduced from Table~\ref{table:Characteristics First Minimum Euclidean Distance Simple Filters Coefficients L_0=5 L_1=1}.}
\label{table:Building Bricks Simple Filters Coefficients L_0=5 L_1=1 h_0=h^3_0}
\small
\centering
\begin{tabular}{|l|l|l|l|l|} 
\hline
Type & $\Delta \bar{\bm{s}}$ & $\Delta \mathring{\bm{b}}_0$ & $\Delta \mathring{\bm{b}}_1$ & Contribution \\ [0.5ex] \hline\hline
\multirow{16}{*}{Primary}
   & \multirow{4}{*}{$(\bm{1},0,0,0,0,0,0,1)$} & $-(1,-1,0,-1)$ & $(1,0,0,-1,0,-1,0,\bm{0})$ & \multirow{8}{*}{\shortstack{$N^4+3N^5+2N^6$ \\ $+N^8+N^9=$ \\ $N^4(1+2N+N^4)(1+N)$}} \\
   &     & $(1,-1,0,1)$ & $(0,0,0,0,-1,0,0,\bm{0})$ &    \\
   &     & $-(1,1,0,-1)$ & $(1,1,1,0,0,0,0,\bm{0})$ &    \\ 
   &     & $(1,1,0,1)$ & $-(0,1,1,1,1,1,0,\bm{0})$ &    \\ \cline{2-4}
   & \multirow{4}{*}{$(\bm{1},0,0,0,0,0,0,-1)$} & $(1,-1,0,-1)$ & $(0,0,0,1,0,1,0,\bm{0})$ &   \\ 
   &     & $-(1,-1,0,1)$ & $(1,0,0,0,1,0,0,\bm{0})$ &    \\
   &     & $(1,1,0,-1)$ & $(0,-1,-1,0,0,0,0,\bm{0})$ &    \\
   &     & $-(1,1,0,1)$ & $(1,1,1,1,1,1,0,\bm{0})$ &    \\ \cline{2-5}
   & \multirow{4}{*}{$(\bm{-1},0,0,0,0,0,0,1)$} & $-(1,-1,0,-1)$ & $(0,0,0,-1,0,-1,0,\bm{0})$ &  \multirow{8}{*}{\shortstack{$N^4+3N^5+2N^6$ \\ $+N^8+N^9=$ \\ $N^4(1+2N+N^4)(1+N)$}} \\
   &    & $(1,-1,0,1)$ & $(-1,0,0,0,-1,0,0,\bm{0})$ &    \\ 
   &    & $-(1,1,0,-1)$ & $(0,1,1,0,0,0,0,\bm{0})$ &    \\ 
   &    & $(1,1,0,1)$ & $-(1,1,1,1,1,1,0,\bm{0})$ &    \\ \cline{2-4}   
   & \multirow{4}{*}{$(\bm{-1},0,0,0,0,0,0,-1)$} & $(1,-1,0,-1)$ & $(-1,0,0,1,0,1,0,\bm{0})$ &   \\
   &     & $-(1,-1,0,1)$ & $(0,0,0,0,1,0,0,\bm{0})$ &    \\
   &     & $(1,1,0,-1)$ & $(-1,-1,-1,0,0,0,0,\bm{0})$ &    \\ 
   &     & $-(1,1,0,1)$ & $(0,1,1,1,1,1,0,\bm{0})$ &    \\ [1ex] \hline
\multirow{16}{*}{Secondary}   
   & \multirow{4}{*}{$(\bm{1},0,0,0,0,0,0,1)$} & $(1,-1,0,-1)$ & $(0,0,0,1,0,1,0,\bm{1})$ & \multirow{8}{*}{\shortstack{$N^5+3N^6+2N^7$ \\ $+N^9+N^{10}=$ \\ $N^5(1+2N+N^4)(1+N)$}} \\ 
   &     & $-(1,-1,0,1)$ & $(1,0,0,0,1,0,0,\bm{1})$ &    \\
   &     & $(1,1,0,-1)$ & $(0,-1,-1,0,0,0,0,\bm{1})$ &    \\
   &     & $-(1,1,0,1)$ & $(1,1,1,1,1,1,0,\bm{1})$ &    \\ \cline{2-4}
   
   & \multirow{4}{*}{$(\bm{1},0,0,0,0,0,0,-1)$} & $-(1,-1,0,-1)$ & $(1,0,0,-1,0,-1,0,\bm{-1})$ &  \\
   &     & $(1,-1,0,1)$ & $(0,0,0,0,-1,0,0,\bm{-1})$ &    \\
   &     & $-(1,1,0,-1)$ & $(1,1,1,0,0,0,0,\bm{-1})$ &    \\ 
   &     & $(1,1,0,1)$ & $-(0,1,1,1,1,1,0,\bm{1})$ &    \\ \cline{2-5}
   & \multirow{4}{*}{$(\bm{-1},0,0,0,0,0,0,1)$} & $(1,-1,0,-1)$ & $(-1,0,0,1,0,1,0,\bm{1})$ & \multirow{8}{*}{\shortstack{$N^5+3N^6+2N^7$ \\ $+N^9+N^{10}=$ \\ $N^5(1+2N+N^4)(1+N)$}} \\
   &     & $-(1,-1,0,1)$ & $(0,0,0,0,1,0,0,\bm{1})$ &    \\
   &     & $(1,1,0,-1)$ & $(-1,-1,-1,0,0,0,0,\bm{1})$ &    \\ 
   &     & $-(1,1,0,1)$ & $(0,1,1,1,1,1,0,\bm{1})$ &    \\ \cline{2-4}
& \multirow{4}{*}{$(\bm{-1},0,0,0,0,0,0,-1)$} & $-(1,-1,0,-1)$ & $(0,0,0,-1,0,-1,0,\bm{-1})$ &  \\
   &    & $(1,-1,0,1)$ & $(-1,0,0,0,-1,0,0,\bm{-1})$ &    \\ 
   &    & $-(1,1,0,-1)$ & $(0,1,1,0,0,0,0,\bm{-1})$ &    \\ 
   &    & $(1,1,0,1)$ & $-(1,1,1,1,1,1,0,\bm{1})$ &    \\ [1ex] \hline
   
\end{tabular}
\end{table} 

\begin{table}[!htbp]
\caption{Characteristics of a subset of the eligible error events input sequences differences, with second MSED, $2d_{\text{min}}^2=10,$ inherited from error events with MSED, for NSMs of rate $2$ with extremely simple filters' coefficients, with $L_0=5$ and $L_1=1$ ($\bar{\bm{h}}_1 = (1)$, $\eta = \eta_0 = \eta_1 = 5/2,$ $p^n(x) = x^3 + x^2 + 1$, $q^n(x) = x^4 + x^2 + x + 1 = (x + 1)(x^3 + x^2 + 1),$  $\Delta \ddot{b}^n_0(x) = (x^3 + x + 1)(x + 1) = x^4 + x^3 + x^2 + 1,$ and $\Delta \ddot{\bm{s}}_0 = (x^7 + 1)(x + 1) = x^8 + x^7 + x + 1$) (Due to space constraints, only a fraction of the available values for vectors $\Delta \mathring{\bm{b}}_1,$ and $\Delta \bar{\bm{s}}$ are given. For the same reason, only the results for filters $\mathring{\bm{h}}_0 = \mathring{\bm{h}}^0_0$ and $\mathring{\bm{h}}^1_0$ are displayed. Similar results are obtained for filters $\mathring{\bm{h}}_0 = \mathring{\bm{h}}^2_0$ and $\mathring{\bm{h}}^3_0$.).}
\label{table:Characteristics Second Minimum Euclidean Distance Simple Filters Coefficients L_0=5 L_1=1 (x + 1)}
\centering
\begin{tabular}{|l|l|} 
\hline
\multicolumn{2}{|c|}{$\mathring{\bm{h}}_0 = \mathring{\bm{h}}^0_0 = (1,1,-1,0,-1)$} \\  [0.5ex] 
\hline
$\Delta \mathring{\bm{b}}_0 = \pm (1,0,-1,-1,-1)$ & $\Delta \mathring{\bm{b}}_0 = \pm (1,0,1,-1,-1)$ \\
$\Delta \bar{\bm{s}}_0 = \pm (1,1,-2,-2,-2,0,2,1,1)$ & $\Delta \bar{\bm{s}}_0 = \pm (1,1,0,0,\bm{-4},0,0,1,1)$ \\
$\Delta \mathring{\bm{b}}_1 = \pm (0,0,1,1,1,0,-1,0,0)$ & $\Delta \mathring{\bm{b}}_1$ not eligible \\
$\Delta \bar{\bm{s}} = \pm (1,1,0,0,0,0,0,1,1)$ & $\Delta \bar{\bm{s}}$ not allowed \\ \hline

$\Delta \mathring{\bm{b}}_0 = \pm (1,0,-1,1,-1)$ & 
$\Delta \mathring{\bm{b}}_0 = \pm (1,0,1,1,-1)$ \\
$\Delta \bar{\bm{s}}_0 = \pm (1,1,-2,0,0,-2,2,-1,1)$ & $\Delta \bar{\bm{s}}_0 = \pm (1,1,0,2,-2,-2,0,-1,1)$ \\
$\Delta \mathring{\bm{b}}_1 = \pm (0,0,1,0,0,1,-1,0,0)$ & $\Delta \mathring{\bm{b}}_1 = \pm (0,0,0,-1,1,1,0,0,0)$ \\
$\Delta \bar{\bm{s}} = \pm (1,1,0,0,0,0,0,-1,1)$ & $\Delta \bar{\bm{s}} = \pm ( 1,1,0,0,0,0,0,-1,1)$ \\ \hline 

$\Delta \mathring{\bm{b}}_0 = \pm (1,0,-1,-1,1)$ & $\Delta \mathring{\bm{b}}_0 = \pm (1,0,1,-1,1)$ \\
$\Delta \bar{\bm{s}}_0 = \pm (1,1,-2,-2,0,2,0,1,-1)$ & $\Delta \bar{\bm{s}}_0 = \pm (1,1,0,0,-2,2,-2,1,-1)$ \\
$\Delta \mathring{\bm{b}}_1 = \pm (0,0,1,1,0,-1,0,0,0)$ & $\Delta \mathring{\bm{b}}_1 = \pm (0,0,0,0,1,-1,1,0,0)$ \\
$\Delta \bar{\bm{s}} = \pm (1,1,0,0,0,0,0,1,-1)$ & $\Delta \bar{\bm{s}} = \pm (1,1,0,0,0,0,0,1,-1)$ \\ \hline

$\Delta \mathring{\bm{b}}_0 = \pm (1,0,-1,1,1)$ & $\Delta \mathring{\bm{b}}_0 = \pm (1,0,1,1,1)$ \\
$\Delta \bar{\bm{s}}_0 = \pm (1,1,-2,0,2,0,0,-1,-1)$ & $\Delta \bar{\bm{s}}_0 = \pm (1,1,0,2,0,0,-2,-1,-1)$ \\
$\Delta \mathring{\bm{b}}_1 = \pm (0,0,1,0,-1,0,0,0,0)$ & $\Delta \mathring{\bm{b}}_1 = \pm (0,0,0,-1,0,0,1,0,0)$ \\
$\Delta \bar{\bm{s}} = \pm (1,1,0,0,0,0,0,-1,-1)$ & $\Delta \bar{\bm{s}} = \pm (1,1,0,0,0,0,0,-1,-1)$ \\ [0.5ex]
\hline\hline

\multicolumn{2}{|c|}{$\mathring{\bm{h}}_0 = \mathring{\bm{h}}^1_0 = (1,1,-1,0,1)$} \\  [0.5ex] 
\hline
$\Delta \mathring{\bm{b}}_0 = \pm (1,0,-1,-1,-1)$ & $\Delta \mathring{\bm{b}}_0 = \pm (1,0,1,-1,-1)$ \\
$\Delta \bar{\bm{s}}_0 = \pm (1,1,-2,-2,0,0,0,-1,-1)$ & $\Delta \bar{\bm{s}}_0 = \pm (1,1,0,0,-2,0,2,-1,-1)$ \\
$\Delta \mathring{\bm{b}}_1 = \pm (0,0,1,1,0,0,0,0,0)$ & $\Delta \mathring{\bm{b}}_1 = \pm (0,0,0,0,1,0,-1,0,0)$ \\
$\Delta \bar{\bm{s}} = \pm (1,1,0,0,0,0,0,-1,-1)$ & $\Delta \bar{\bm{s}} = \pm (1,1,0,0,0,0,0,-1,-1)$ \\ \hline

$\Delta \mathring{\bm{b}}_0 = \pm (1,0,-1,1,-1)$ & 
$\Delta \mathring{\bm{b}}_0 = \pm (1,0,1,1,-1)$ \\
$\Delta \bar{\bm{s}}_0 = \pm (1,1,-2,0,2,-2,0,1,-1)$ & $\Delta \bar{\bm{s}}_0 = \pm (1,1,0,2,0,-2,2,1,-1)$ \\
$\Delta \mathring{\bm{b}}_1 = \pm (0,0,1,0,-1,1,0,0,0)$ & $\Delta \mathring{\bm{b}}_1 = \pm (0,0,0,-1,0,1,-1,0,0)$ \\
$\Delta \bar{\bm{s}} = \pm (1,1,0,0,0,0,0,1,-1)$ & $\Delta \bar{\bm{s}} = \pm ( 1,1,0,0,0,0,0,1,-1)$ \\ \hline 
  
$\Delta \mathring{\bm{b}}_0 = \pm (1,0,-1,-1,1)$ & $\Delta \mathring{\bm{b}}_0 = \pm (1,0,1,-1,1)$ \\
$\Delta \bar{\bm{s}}_0 = \pm (1,1,-2,-2,2,2,-2,-1,1)$ & $\Delta \bar{\bm{s}}_0 = \pm (1,1,0,0,0,2,0,-1,1)$ \\
$\Delta \mathring{\bm{b}}_1 = \pm (0,0,1,1,-1,-1,1,0,0)$ & $\Delta \mathring{\bm{b}}_1 = \pm (0,0,0,0,0,-1,0,0,0)$ \\
$\Delta \bar{\bm{s}} = \pm (1,1,0,0,0,0,0,-1,1)$ & $\Delta \bar{\bm{s}} = \pm (1,1,0,0,0,0,0,-1,1)$ \\ \hline
  
$\Delta \mathring{\bm{b}}_0 = \pm (1,0,-1,1,1)$ & $\Delta \mathring{\bm{b}}_0 = \pm (1,0,1,1,1)$ \\
$\Delta \bar{\bm{s}}_0 = \pm (1,1,-2,0,\bm{4},0,-2,1,1)$ & $\Delta \bar{\bm{s}}_0 = \pm (1,1,0,2,2,0,0,1,1)$ \\
 $\Delta \mathring{\bm{b}}_1$ not eligible & $\Delta \mathring{\bm{b}}_1 = \pm (0,0,0,-1,-1,0,0,0,0)$ \\
$\Delta \bar{\bm{s}}$ not allowed & $\Delta \bar{\bm{s}} = \pm (1,1,0,0,0,0,0,1,1)$     \\ [1ex] \hline
\end{tabular}
\end{table} 

\begin{table}[!htbp]
\caption{Characteristics of a subset of the eligible error events input sequences differences, with second MSED, $2d_{\text{min}}^2=10,$ inherited from error events with MSED, for NSMs of rate $2$ with extremely simple filters' coefficients, with $L_0=5$ and $L_1=1$ ($\bar{\bm{h}}_1 = (1)$, $\eta = \eta_0 = \eta_1 = 5/2,$ $p^n(x) = x^3 + x^2 + 1$, $q^n(x) = x^4 + x^2 + x + 1 = (x + 1)(x^3 + x^2 + 1),$  $\Delta \ddot{b}^n_0(x) = (x^3 + x + 1)(x^2 + 1) = x^5 + x^2 + x + 1,$ and $\Delta \ddot{\bm{s}}_0 = (x^7 + 1)(x^2 + 1) = x^9 + x^7 + x^2 + 1$) (Due to space constraints, only a fraction of the available values for vectors $\Delta \mathring{\bm{b}}_1,$ and $\Delta \bar{\bm{s}}$ are given. For the same reason, only the results for filters $\mathring{\bm{h}}_0 = \mathring{\bm{h}}^0_0$ and $\mathring{\bm{h}}^1_0$ are displayed.).}
\label{table:Characteristics Second Minimum Euclidean Distance Simple Filters Coefficients L_0=5 L_1=1 (x^2 + 1)}
\centering
\begin{tabular}{|l|l|} 
\hline
\multicolumn{2}{|c|}{$\mathring{\bm{h}}_0 = \mathring{\bm{h}}^0_0 = (1,1,-1,0,-1)$} \\  [0.5ex] 
\hline
$\Delta \mathring{\bm{b}}_0 = \pm (1,-1,-1,0,0,-1)$ & $\Delta \mathring{\bm{b}}_0 = \pm (1,1,-1,0,0,-1)$ \\
$\Delta \bar{\bm{s}}_0 = \pm (1,0,\bm{-3},0,0,0,0,1,0,1)$ & $\Delta \bar{\bm{s}}_0 = \pm (1,2,-1,-2,0,-2,0,1,0,1)$ \\
$\Delta \mathring{\bm{b}}_1$ not eligible & $\Delta \mathring{\bm{b}}_1 = \pm (0,-1,0,1,0,1,0,0,0,0)$  \\
$\Delta \bar{\bm{s}}$ not allowed & $\Delta \bar{\bm{s}} = \pm (1,0,-1,0,0,0,0,1,0,1)$  \\ \hline

$\Delta \mathring{\bm{b}}_0 = \pm (1,-1,1,0,0,-1)$ & 
$\Delta \mathring{\bm{b}}_0 = \pm (1,1,1,0,0,-1)$ \\
$\Delta \bar{\bm{s}}_0 = \pm (1,0,-1,2,-2,0,-2,1,0,1)$ & $\Delta \bar{\bm{s}}_0 = \pm (1,2,1,0,-2,-2,-2,1,0,1)$ \\
$\Delta \mathring{\bm{b}}_1 = \pm (0,0,0,-1,1,0,1,0,0,0)$ & $\Delta \mathring{\bm{b}}_1 = \pm (0,-1,0,0,1,1,1,0,0,0)$ \\
$\Delta \bar{\bm{s}} = \pm (1,0,-1,0,0,0,0,1,0,1)$ & $\Delta \bar{\bm{s}} = \pm ( 1,0,1,0,0,0,0,1,0,1)$ \\ \hline 

$\Delta \mathring{\bm{b}}_0 = \pm (1,-1,-1,0,0,1)$ & $\Delta \mathring{\bm{b}}_0 = \pm (1,1,-1,0,0,1)$ \\
$\Delta \bar{\bm{s}}_0 = \pm (1,0,\bm{-3},0,0,2,2,-1,0,-1)$ & $\Delta \bar{\bm{s}}_0 = \pm (1,2,-1,-2,0,0,2,-1,0,-1)$ \\
$\Delta \mathring{\bm{b}}_1$ not eligible & $\Delta \mathring{\bm{b}}_1 = \pm (0,-1,0,1,0,0,-1,0,0,0)$ \\
$\Delta \bar{\bm{s}}$ not allowed & $\Delta \bar{\bm{s}} = \pm (1,0,-1,0,0,0,0,-1,0,-1)$ \\ \hline
    
$\Delta \mathring{\bm{b}}_0 = \pm (1,-1,1,0,0,1)$ & $\Delta \mathring{\bm{b}}_0 = \pm (1,1,1,0,0,1)$ \\
$\Delta \bar{\bm{s}}_0 = \pm (1,0,-1,2,-2,-2,0,-1,0,-1)$ & $\Delta \bar{\bm{s}}_0 = \pm (1,2,1,0,-2,0,0,-1,0,-1)$ \\
$\Delta \mathring{\bm{b}}_1 = \pm (0,0,0,-1,1,-1,0,0,0,0)$ & $\Delta \mathring{\bm{b}}_1 = \pm (0,-1,0,0,1,0,0,0,0,0)$ \\
$\Delta \bar{\bm{s}} = \pm (1,0,-1,0,0,0,0,-1,0,-1)$ & $\Delta \bar{\bm{s}} = \pm (1,0,1,0,0,0,0,-1,0,-1)$ \\ [0.5ex]
\hline\hline
    
\multicolumn{2}{|c|}{$\mathring{\bm{h}}_0 = \mathring{\bm{h}}^1_0 = (1,1,-1,0,1)$} \\  [0.5ex] 
\hline
$\Delta \mathring{\bm{b}}_0 = \pm (1,-1,-1,0,0,-1)$ & $\Delta \mathring{\bm{b}}_0 = \pm (1,1,-1,0,0,-1)$ \\
$\Delta \bar{\bm{s}}_0 = \pm (1,0,\bm{-3},0,2,-2,-2,1,0,-1)$ & $\Delta \bar{\bm{s}}_0 = \pm (1,2,-1,-2,2,0,-2,1,0,-1)$ \\
$\Delta \mathring{\bm{b}}_1$ not eligible & $\Delta \mathring{\bm{b}}_1 = \pm (0,-1,0,1,-1,0,1,0,0,0)$ \\
$\Delta \bar{\bm{s}}$ not allowed & $\Delta \bar{\bm{s}} = \pm (1,0,-1,0,0,0,0,1,0,-1)$ \\ \hline
    
$\Delta \mathring{\bm{b}}_0 = \pm (1,-1,1,0,0,-1)$ & 
$\Delta \mathring{\bm{b}}_0 = \pm (1,1,1,0,0,-1)$ \\
$\Delta \bar{\bm{s}}_0 = \pm (1,0,-1,2,0,-2,0,1,0,-1)$ & $\Delta \bar{\bm{s}}_0 = \pm (1,2,1,0,0,0,0,1,0,-1)$ \\
$\Delta \mathring{\bm{b}}_1 = \pm (0,0,0,-1,0,1,0,0,0,0)$ & $\Delta \mathring{\bm{b}}_1 = \pm (0,-1,0,0,0,0,0,0,0,0)$ \\
$\Delta \bar{\bm{s}} = \pm (1,0,-1,0,0,0,0,1,0,-1)$ & $\Delta \bar{\bm{s}} = \pm ( 1,0,1,0,0,0,0,1,0,-1)$ \\ \hline 

$\Delta \mathring{\bm{b}}_0 = \pm (1,-1,-1,0,0,1)$ & $\Delta \mathring{\bm{b}}_0 = \pm (1,1,-1,0,0,1)$ \\
$\Delta \bar{\bm{s}}_0 = \pm (1,0,\bm{-3},0,2,0,0,-1,0,1)$ & $\Delta \bar{\bm{s}}_0 = \pm (1,2,-1,-2,2,2,0,-1,0,1)$ \\
$\Delta \mathring{\bm{b}}_1$ not eligible & $\Delta \mathring{\bm{b}}_1 = \pm (0,-1,0,1,-1,-1,0,0,0,0)$ \\
$\Delta \bar{\bm{s}}$ not allowed & $\Delta \bar{\bm{s}} = \pm (1,0,-1,0,0,0,0,-1,0,1)$ \\ \hline

$\Delta \mathring{\bm{b}}_0 = \pm (1,-1,1,0,0,1)$ & $\Delta \mathring{\bm{b}}_0 = \pm (1,1,1,0,0,1)$ \\
$\Delta \bar{\bm{s}}_0 = \pm (1,0,-1,2,0,0,2,-1,0,1)$ & $\Delta \bar{\bm{s}}_0 = \pm (1,2,1,0,0,2,2,-1,0,1)$ \\
$\Delta \mathring{\bm{b}}_1 = \pm (0,0,0,-1,0,0,-1,0,0,0)$ & $\Delta \mathring{\bm{b}}_1 = \pm (0,-1,0,0,0,-1,-1,0,0,0)$ \\
$\Delta \bar{\bm{s}} = \pm (1,0,-1,0,0,0,0,-1,0,1)$ & $\Delta \bar{\bm{s}} = \pm (1,0,1,0,0,0,0,-1,0,1)$   \\ [1ex] \hline
\end{tabular}
\end{table} 

\begin{table}[!htbp]
\caption{Characteristics of a subset of the eligible error events input sequences differences, with second MSED, $2d_{\text{min}}^2=10,$ inherited from error events with MSED, for NSMs of rate $2$ with extremely simple filters' coefficients, with $L_0=5$ and $L_1=1$ ($\bar{\bm{h}}_1 = (1)$, $\eta = \eta_0 = \eta_1 = 5/2,$ $p^n(x) = x^3 + x^2 + 1$, $q^n(x) = x^4 + x^2 + x + 1 = (x + 1)(x^3 + x^2 + 1),$  $\Delta \ddot{b}^n_0(x) = (x^3 + x + 1)(x^2 + 1) = x^5 + x^2 + x + 1,$ and $\Delta \ddot{\bm{s}}_0 = (x^7 + 1)(x^3 + 1) = x^{10} + x^7 + x^3 + 1$) (Due to space constraints, only a fraction of the available values for vectors $\Delta \mathring{\bm{b}}_1,$ and $\Delta \bar{\bm{s}}$ are given. For the same reason, only the results for filters $\mathring{\bm{h}}_0 = \mathring{\bm{h}}^0_0$ and $\mathring{\bm{h}}^1_0$ are displayed.).}
\label{table:Characteristics Second Minimum Euclidean Distance Simple Filters Coefficients L_0=5 L_1=1 (x^3 + 1)}
\centering
\begin{tabular}{|l|l|} 
\hline
\multicolumn{2}{|c|}{$\mathring{\bm{h}}_0 = \mathring{\bm{h}}^0_0 = (1,1,-1,0,-1)$} \\  [0.5ex] 
\hline
$\Delta \mathring{\bm{b}}_0 = \pm (1,-1,0,0,-1,0,-1)$ & $\Delta \mathring{\bm{b}}_0 = \pm (1,1,0,0,-1,0,-1)$ \\
$\Delta \bar{\bm{s}}_0 = \pm (1,0,-2,1,-2,0,0,-1,2,0,1)$ & $\Delta \bar{\bm{s}}_0 = \pm (1,2,0,-1,-2,-2,0,-1,2,0,1)$ \\
$\Delta \mathring{\bm{b}}_1 = \pm (0,0,1,0,1,0,0,0,-1,0,0)$ & $\Delta \mathring{\bm{b}}_1 = \pm (0,-1,0,0,1,1,0,0,-1,0,0)$  \\
$\Delta \bar{\bm{s}} = \pm (1,0,0,1,0,0,0,-1,0,0,1)$ & $\Delta \bar{\bm{s}} = \pm (1,0,0,-1,0,0,0,-1,0,0,1)$  \\ \hline

$\Delta \mathring{\bm{b}}_0 = \pm (1,-1,0,0,1,0,-1)$ & 
$\Delta \mathring{\bm{b}}_0 = \pm (1,1,0,0,1,0,-1)$ \\
$\Delta \bar{\bm{s}}_0 = \pm (1,0,-2,1,0,2,-2,-1,0,0,1)$ & $\Delta \bar{\bm{s}}_0 = \pm (1,2,0,-1,0,0,-2,-1,0,0,1)$ \\
$\Delta \mathring{\bm{b}}_1 = \pm (0,0,1,0,0,-1,1,0,0,0,0)$ & $\Delta \mathring{\bm{b}}_1 = \pm (0,-1,0,0,0,0,1,0,0,0,0)$ \\
$\Delta \bar{\bm{s}} = \pm (1,0,0,1,0,0,0,-1,0,0,1)$ & $\Delta \bar{\bm{s}} = \pm (1,0,0,-1,0,0,0,-1,0,0,1)$ \\ \hline

$\Delta \mathring{\bm{b}}_0 = \pm (1,-1,0,0,-1,0,1)$ & $\Delta \mathring{\bm{b}}_0 = \pm (1,1,0,0,-1,0,1)$ \\
$\Delta \bar{\bm{s}}_0 = \pm (1,0,-2,1,-2,0,2,1,0,0,-1)$ & $\Delta \bar{\bm{s}}_0 = \pm (1,2,0,-1,-2,-2,2,1,0,0,-1)$ \\
$\Delta \mathring{\bm{b}}_1 = \pm (0,0,1,0,1,0,-1,0,0,0,0)$ & $\Delta \mathring{\bm{b}}_1 = \pm (0,-1,0,0,1,1,-1,0,0,0,0)$ \\
$\Delta \bar{\bm{s}} = \pm (1,0,0,1,0,0,0,1,0,0,-1)$ & $\Delta \bar{\bm{s}} = \pm (1,0,0,-1,0,0,0,1,0,0,-1)$ \\ \hline
    
$\Delta \mathring{\bm{b}}_0 = \pm (1,-1,0,0,1,0,1)$ & $\Delta \mathring{\bm{b}}_0 = \pm (1,1,0,0,1,0,1)$ \\
$\Delta \bar{\bm{s}}_0 = \pm (1,0,-2,1,0,2,0,1,-2,0,-1)$ & $\Delta \bar{\bm{s}}_0 = \pm (1,2,0,-1,0,0,0,1,-2,0,-1)$ \\
$\Delta \mathring{\bm{b}}_1 = \pm (0,0,1,0,0,0,-1,0,0,1,0,0)$ & $\Delta \mathring{\bm{b}}_1 = \pm (0,-1,0,0,0,0,0,0,1,0,0)$ \\
$\Delta \bar{\bm{s}} = \pm (1,0,0,1,0,0,0,1,0,0,-1)$ & $\Delta \bar{\bm{s}} = \pm (1,0,0,-1,0,0,0,1,0,0,-1)$ \\ [0.5ex]
\hline\hline
    
\multicolumn{2}{|c|}{$\mathring{\bm{h}}_0 = \mathring{\bm{h}}^1_0 = (1,1,-1,0,1)$} \\  [0.5ex] 
\hline
$\Delta \mathring{\bm{b}}_0 = \pm (1,-1,0,0,-1,0,-1)$ & $\Delta \mathring{\bm{b}}_0 = \pm (1,1,0,0,-1,0,-1)$ \\
$\Delta \bar{\bm{s}}_0 = \pm (1,0,-2,1,0,-2,0,-1,0,0,-1)$ & $\Delta \bar{\bm{s}}_0 = \pm (1,2,0,-1,0,0,0,-1,0,0,-1)$ \\
$\Delta \mathring{\bm{b}}_1 = \pm (0,0,1,0,0,1,0,0,0,0,0)$ & $\Delta \mathring{\bm{b}}_1 = \pm (0,-1,0,0,0,0,0,0,0,0,0)$ \\
$\Delta \bar{\bm{s}} = \pm (1,0,0,1,0,0,0,-1,0,0,-1)$ & $\Delta \bar{\bm{s}} = \pm (1,0,0,-1,0,0,0,-1,0,0,-1)$ \\ \hline
    
$\Delta \mathring{\bm{b}}_0 = \pm (1,-1,0,0,1,0,-1)$ & 
$\Delta \mathring{\bm{b}}_0 = \pm (1,1,0,0,1,0,-1)$ \\
$\Delta \bar{\bm{s}}_0 = \pm (1,0,-2,1,2,0,-2,-1,2,0,-1)$ & $\Delta \bar{\bm{s}}_0 = \pm (1,2,0,-1,2,2,-2,-1,2,0,-1)$ \\
$\Delta \mathring{\bm{b}}_1 = \pm (0,0,1,0,-1,0,1,0,-1,0,0)$ & $\Delta \mathring{\bm{b}}_1 = \pm (0,-1,0,0,-1,-1,1,0,-1,0,0)$ \\
$\Delta \bar{\bm{s}} = \pm (1,0,0,1,0,0,0,-1,0,0,-1)$ & $\Delta \bar{\bm{s}} = \pm ( 1,0,0,-1,0,0,0,-1,0,0,-1)$ \\ \hline 
    
$\Delta \mathring{\bm{b}}_0 = \pm (1,-1,0,0,-1,0,1)$ & $\Delta \mathring{\bm{b}}_0 = \pm (1,1,0,0,-1,0,1)$ \\
$\Delta \bar{\bm{s}}_0 = \pm (1,0,-2,1,0,-2,2,1,-2,0,1)$ & $\Delta \bar{\bm{s}}_0 = \pm (1,2,0,-1,0,0,2,1,-2,0,1)$ \\
$\Delta \mathring{\bm{b}}_1 = \pm (0,0,1,0,0,1,-1,0,1,0,0)$ & $\Delta \mathring{\bm{b}}_1 = \pm (0,-1,0,0,0,0,-1,0,1,0,0)$ \\
$\Delta \bar{\bm{s}} = \pm (1,0,0,1,0,0,0,1,0,0,1)$ & $\Delta \bar{\bm{s}} = \pm (1,0,0,-1,0,0,0,1,0,0,1)$ \\ \hline

$\Delta \mathring{\bm{b}}_0 = \pm (1,-1,0,0,1,0,1)$ & $\Delta \mathring{\bm{b}}_0 = \pm (1,1,0,0,1,0,1)$ \\
$\Delta \bar{\bm{s}}_0 = \pm (1,0,-2,1,2,0,0,1,0,0,1)$ & $\Delta \bar{\bm{s}}_0 = \pm (1,2,0,-1,2,2,0,1,0,0,1)$ \\
$\Delta \mathring{\bm{b}}_1 = \pm (0,0,1,0,-1,0,0,0,0,0,0)$ & $\Delta \mathring{\bm{b}}_1 = \pm (0,-1,0,0,-1,-1,0,0,0,0,0)$ \\
$\Delta \bar{\bm{s}} = \pm (1,0,0,1,0,0,0,1,0,0,1)$ & $\Delta \bar{\bm{s}} = \pm (1,0,0,-1,0,0,0,1,0,0,1)$   \\ [1ex] \hline

\end{tabular}
\end{table} 

\paragraph*{\textbf{Extended and generalized NSM constructions}}

In Section~\ref{Rate-2 guaranteeing, minimum Euclidean distance approaching NSMs with real filters' coefficients}, we provided a detailed explanation of the procedure used to identify the optimal \glspl{nsm} with real filter taps. This process has proven to be particularly challenging and of limited efficiency when selecting the optimal filters, $\mathring{h}_m[k]$, for $m = 0, 1$, especially in cases where $L_m \geq 4,$ $m = 0, 1$, or when $L_1 = 1$ and $L_0 \geq 9$. Under these conditions, the search space becomes so vast that the optimization algorithm consistently converges to local maxima of the \gls{msed}. To make the optimization process more manageable, it is essential to reduce the size of the search space—particularly as the values of $L_m$, $m = 0, 1$, increase. Recall that significant performance improvements are only achievable when $L_m$ is large. In line with this observation, we began this section by constraining the search space to scaled filters, $\mathring{h}_0[k],$ with exactly four non-zero components, each belonging to the simple bipolar set $\{\pm 1\}.$ The corresponding scaled filter, $\mathring{h}_1[k],$ with length $L_1 = 1$, is uniquely defined as $\mathring{h}_1[k] = 2 \, \delta[k]$, so that its norm matches that of $\mathring{h}_0[k].$

By narrowing the search space, the optimization procedure became significantly more tractable and highly effective, enabling the discovery of high-quality \glspl{nsm} with large filter lengths, reaching up to $L_0 = 15$. The best \glspl{nsm} obtained under the imposed constraints on $\mathring{h}_m[k]$, $m = 0, 1,$ were thoroughly analyzed both theoretically and experimentally at the beginning of this section. A striking finding was that none of these optimized \glspl{nsm} achieve the \gls{msed} of $2$-ASK. Instead, they exhibit a performance degradation of approximately $3$ dB with respect to its bit error probability. Nevertheless, they still yield a clear asymptotic coding gain of $10 \log{10}(5/4) \approx 0.9691$ dB over $4$-ASK. Fortunately, the input sequence differences associated with the error events corresponding to this reduced \gls{msed} tend to have significantly greater lengths and weights as the filter length, $L_0,$ increases. As a result, their multiplicity decreases exponentially toward zero. Consequently, the second \gls{msed}—which coincides with that of $2$-ASK—becomes the dominant factor in determining the bit error probability performance at moderate \glspl{snr}, and progressively governs performance at high \glspl{snr} as well, as $L_0$ increases. Therefore, although the optimized \glspl{nsm} exhibited only half the \gls{msed} of $2$-ASK, this does not prevent them from achieving comparable performance within practical and relevant \gls{snr} ranges.

The family of \glspl{nsm} discussed at the beginning of this section performs well at moderate to high \gls{snr} levels, when $L_0$ is sufficiently large, specifically greater than $8.$ These \glspl{nsm} are derived from a compact filter pattern $\bm{\pi}_0=(1,1,1,1),$ which defines the structure of the search space. Candidate filters, $\mathring{h}_0[k],$ are formed by appending $L_0-4$ zeros to $\bm{\pi}_0,$ followed by random permutations and sign changes applied to the non-zero elements. Additionally, it is required that both $\mathring{h}_0[0]$ and $\mathring{h}_0[L_0-1]$ be non-zero, ensuring the filter spans its full length, $L_0.$ The use of $\bm{\pi}_0=(1,1,1,1)$ ensures that the non-zero components of $\mathring{h}_0[k]$ take values from the bipolar set $\{\pm 1\}.$ When combined with the specific expression, $\mathring{h}_1[k] = 2,\ \delta[k],$ of second filter, this leads to tightness constraints that significantly increase the multiplicity of error events at the second \gls{msed}. This increase in multiplicity explains why these optimal \glspl{nsm} underperform compared to $2$-ASK at moderate \glspl{snr}. Moreover, the use of $\bm{\pi}_0=(1,1,1,1)$ invariably leads to error events with \gls{msed} equal to half that of $2$-ASK, regardless of the value of $L_0.$ However, when $L_0 \le 7,$ these error events are short in length, and avoiding them entirely requires using data packets with correspondingly small sizes.

To address the limitations associated with the compact filter pattern $\bm{\pi}_0=(1,1,1,1),$ alternative compact patterns with positive integer coefficients can be explored. Specifically, for the case $L_1=1,$ we consider three representative candidate patterns: $\bm{\pi}_0 = (1,1,1,1,1,1,1,1,1),$ $(2,2,2,2,2,2,1)$ and $(3,3,3,2,2,1).$ These patterns have integer squared Euclidean norms $\| \bm{\pi}_0 \| = 3,$ $5$ and $6,$ respectively, leading to fully specified filters $\mathring{h}_1[k] = 3 \, \delta[k],$ $5 \, \delta[k]$ and $6 \, \delta[k].$  As a preliminary observation, only the last pattern, $\bm{\pi}_0 = (3,3,3,2,2,1),$ satisfies the tightness property, characterized by $\| \bm{\pi}_0 \|_\infty = 3 = \| \bm{\pi}_0 \|/2,$ where $\| \cdot \|_\infty$ denotes the infinity norm. Consequently, the multiplicity of error events, with the \gls{msed} of $2$-ASK, will be higher for this pattern compared to the other two. However, it is worth noting that this tightness is slightly less pronounced than that of the original pattern $\bm{\pi}_0 = (1,1,1,1),$ which involves four filter taps, whereas the pattern $\bm{\pi}_0 = (3,3,3,2,2,1),$ involves only three.

\begin{figure}[!htbp]
    \centering
    \includegraphics[width=1.0\textwidth]{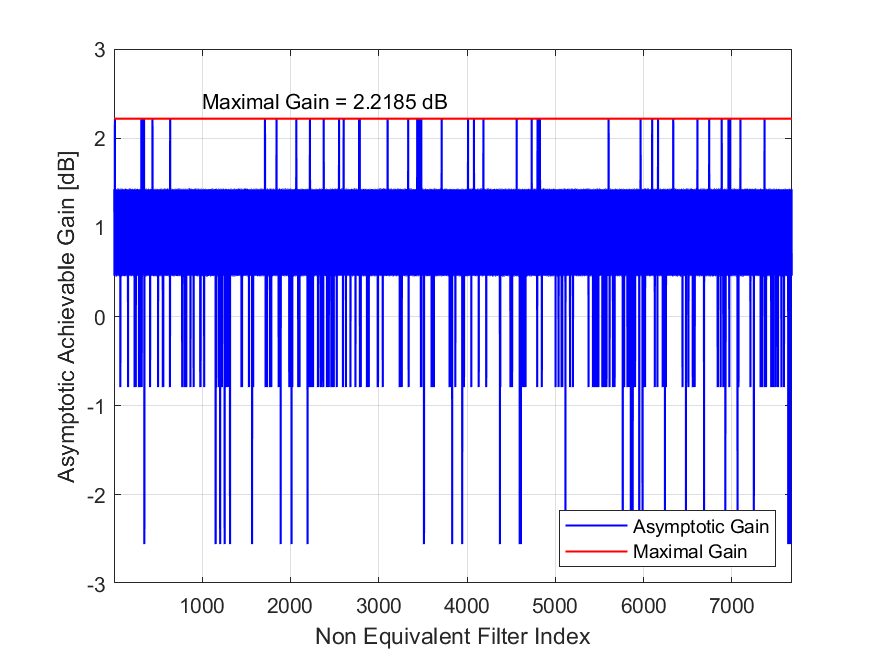}
    \caption{Asymptotic achievable gains of rate-$2$ NSMs, with non-equivalent rational filter coefficients, with compacted filter pattern $\bm{\pi}_0 = (1, 1, 1, 1, 1, 1, 1, 1, 1),$ for $\mathring{\bm{h}}_0,$ and $L_0 = 12.$ The total number of filter $\mathring{\bm{h}}_0$ candidates is $61440.$ The total number of non-equivalent filter candidates is $7680.$ The indices of these filter candidates are shown in the abscissa. The total number of best non-equivalent filter candidates is $46.$ The asymptotic gain in dB is determined as $10\, \log_{10}((d_{\text{min}}^2/(2\cdot9))/(4/5)) = 10\, \log_{10}(5 \, d_{\text{min}}^2/72)).$ The maximum explored trellis depth is $K=1200.$ Achievable MSEDs, $d_{\text{min}}^2,$ are $4 \cdot (6, 5, 4, 3, 2),$ corresponding to gains $(2.2185, 1.4267, 0.4576, -0.7918, -2.5527)$ dB, approximately.} 
    \label{fig:GaindB1DRate2SimpleFilterCoeff_P_111111111_L0_12}
\end{figure}

\begin{figure}[!htbp]
    \centering
    \includegraphics[width=1.0\textwidth]{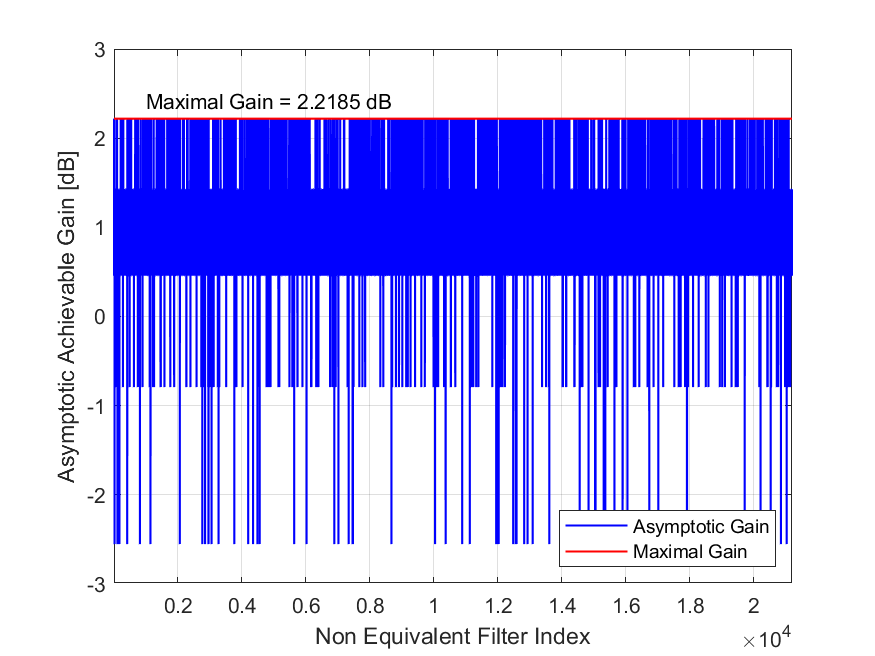}
    \caption{Asymptotic achievable gains of rate-$2$ NSMs, with non-equivalent rational filter coefficients, with compacted filter pattern $\bm{\pi}_0 = (1, 1, 1, 1, 1, 1, 1, 1, 1),$ for $\mathring{\bm{h}}_0,$ and $L_0 = 13.$ The total number of filter $\mathring{\bm{h}}_0$ candidates is $168960.$ The total number of non-equivalent filter candidates is $21200.$ The indices of these filter candidates are shown in the abscissa. The total number of best non-equivalent filter candidates is $872.$ The asymptotic gain in dB is determined as $10\, \log_{10}(5 \, d_{\text{min}}^2/72)).$ The maximum explored trellis depth is $K=2400.$ Achievable MSEDs, $d_{\text{min}}^2,$ are $4 \cdot (6, 5, 4, 3, 2),$ corresponding to gains $(2.2185, 1.4267, 0.4576, -0.7918, -2.5527)$ dB, approximately.} 
    \label{fig:GaindB1DRate2SimpleFilterCoeff_P_111111111_L0_13}
\end{figure}

Considering the candidate compact pattern $\bm{\pi}_0 = (1,1,1,1,1,1,1,1,1),$ Figures~\ref{fig:GaindB1DRate2SimpleFilterCoeff_P_111111111_L0_12} and~\ref{fig:GaindB1DRate2SimpleFilterCoeff_P_111111111_L0_13} present the asymptotically achievable gains in \gls{bep} performance—relative to $4$-ASK—for all non-equivalent scaled filters $\mathring{h}_0[k],$ corresponding to $L_0=12$ and $13,$ respectively. First, observe that the total average symbol energy is given by $\| \mathring{h}_0[k] \|^2 + \| \mathring{h}_1[k] \|^2 = 2 \, \| \bm{\pi}_0 \|^2 = 2 \cdot 9 = 18.$ In contrast, $4$-ASK, which uses a quaternary symbol alphabet $\{ \pm 1, \pm 3 \},$ has an average symbol energy of $5$ and a \gls{msed} of $4.$ Therefore, the asymptotic coding gain achieved with the proposed construction can be expressed as $10\, \log_{10}((d_{\text{min}}^2/18)/(4/5)) = 10\, \log_{10}(5 \, d_{\text{min}}^2/72)).$ Secondly, note that the gains shown in Figures~\ref{fig:GaindB1DRate2SimpleFilterCoeff_P_111111111_L0_12} and~\ref{fig:GaindB1DRate2SimpleFilterCoeff_P_111111111_L0_13}, expressed in dB, correspond to normalized \glspl{msed}, $d_{\text{min}}^2/4,$ taking values $6,5,4,3$ and $2.$ Thirdly, observe that the total number of candidate filters, $\mathring{h}_0[k],$ whether equivalent or not, increases when transitioning from $L_0 = 12$ to $L_0 = 13.$ A similar trend holds for the number of best-performing filters—those yielding the highest gains—although the maximum asymptotic gain achieved remains unchanged between the two cases. The lowest observed normalized \gls{msed}, $\| \Delta \bar{s}[k] \|^2/4=2,$ arises from a modulated sequence differences $\Delta \bar{s}[k],$ containing only two nonzero symbols from the set $\{ \pm 2 \}.$ This behavior mirrors that of the best \glspl{nsm} previously analyzed for the compact filter pattern $\bm{\pi}_0 = (1,1,1,1).$ Fourthly, it is important to mention that the \glspl{msed} are computed based on error events whose lengths do not exceed the maximum trellis depth explored—$K = 1200$ for $L_0 = 12$ and $K = 2400$ for $L_0=13.$ While longer error events could, in principle, yield even smaller \glspl{msed}, this limitation does not affect the overall conclusions. Indeed, the maximum gain achieved by the best codes in both cases, $10\, \log_{10}(5/3) \approx 2.2185,$ remains significantly below the maximum gain of $10\, \log_{10}(5/2) \approx 3.9794,$ attained by $2$-ASK relative to $4$-ASK. Consequently, we conclude that the compact filter pattern $\bm{\pi}_0 = (1,1,1,1,1,1,1,1,1)$ cannot deliver performance comparable to that of $2$-ASK, when the packet size is sufficiently large.

\begin{figure}[!htbp]
    \centering
    \includegraphics[width=1.0\textwidth]{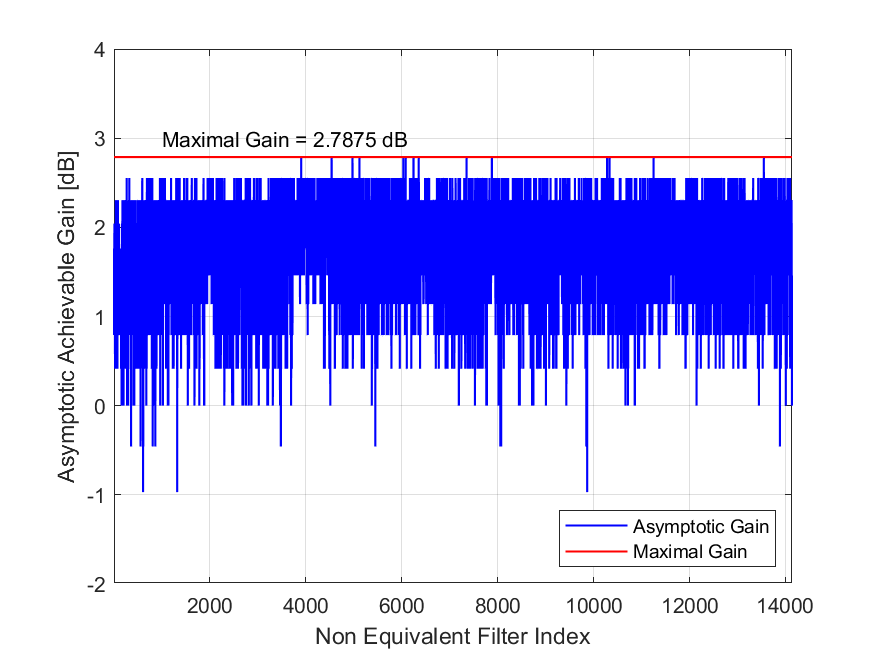}
    \caption{Asymptotic achievable gains of rate-$2$ NSMs, with non-equivalent rational filter coefficients, with compacted filter pattern $\bm{\pi}_0 = (2, 2, 2, 2, 2, 2, 1),$ for $\mathring{\bm{h}}_0,$ and $L_0 = 11.$ The total number of filter $\mathring{\bm{h}}_0$ candidates is $112896.$ The total number of non-equivalent filter candidates is $14136.$ The indices of these filter candidates are shown in the abscissa. The total number of best non-equivalent filter candidates is $15.$ The asymptotic gain in dB is determined as $10\, \log_{10}((d_{\text{min}}^2/(2 \cdot 25))/(4/5)) = 10\, \log_{10}(d_{\text{min}}^2/40)).$ The maximum explored trellis depth is $K=100.$} 
    \label{fig:GaindB1DRate2SimpleFilterCoeff_P_2222221_L0_11}
\end{figure}

\begin{figure}[!htbp]
    \centering
    \includegraphics[width=1.0\textwidth]{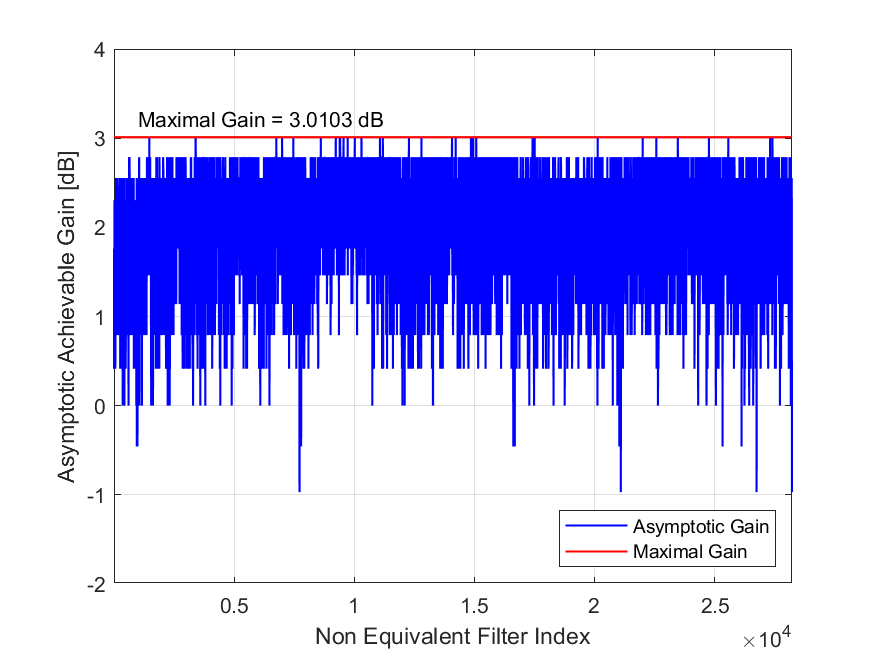}
    \caption{Asymptotic achievable gains of rate-$2$ NSMs, with non-equivalent rational filter coefficients, with compacted filter pattern $\bm{\pi}_0 = (2, 2, 2, 2, 2, 2, 1),$ for $\mathring{\bm{h}}_0,$ and $L_0 = 12.$ The total number of filter $\mathring{\bm{h}}_0$ candidates is $225792.$ The total number of non-equivalent filter candidates is $28224.$ The indices of these filter candidates are shown in the abscissa. The total number of best non-equivalent filter candidates is $33.$ The asymptotic gain in dB is determined as $10\, \log_{10}(d_{\text{min}}^2/40)).$ The maximum explored trellis depth is $K=100.$} 
    \label{fig:GaindB1DRate2SimpleFilterCoeff_P_2222221_L0_12}
\end{figure}

\begin{figure}[!htbp]
    \centering
    \includegraphics[width=1.0\textwidth]{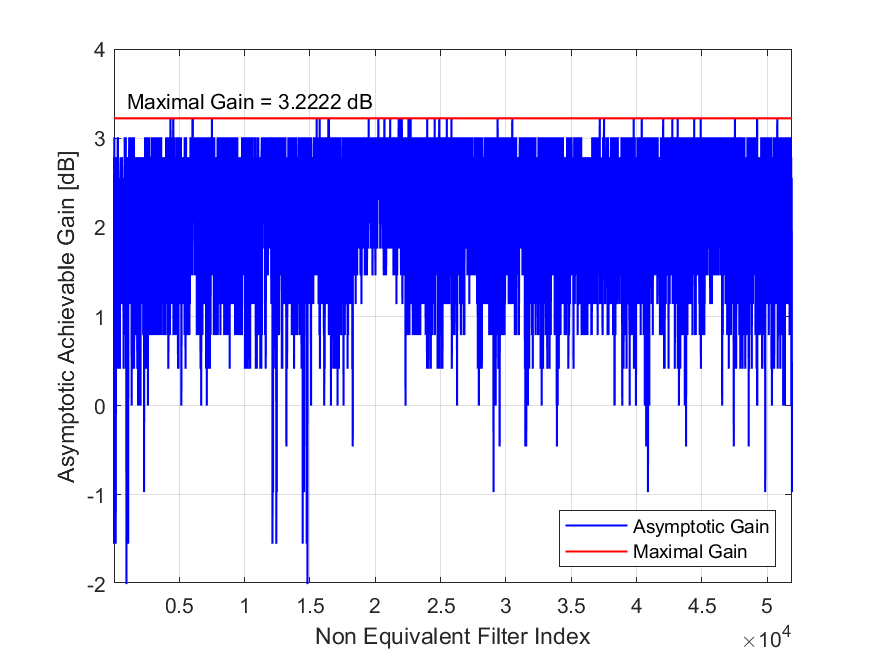}
    \caption{Asymptotic achievable gains of rate-$2$ NSMs, with non-equivalent rational filter coefficients, with compacted filter pattern $\bm{\pi}_0 = (2, 2, 2, 2, 2, 2, 1),$ for $\mathring{\bm{h}}_0,$ and $L_0 = 13.$ The total number of filter $\mathring{\bm{h}}_0$ candidates is $413952.$ The total number of non-equivalent filter candidates is $51896.$ The indices of these filter candidates are shown in the abscissa. The total number of best non-equivalent filter candidates is $39.$ The asymptotic gain in dB is determined as $10\, \log_{10}(d_{\text{min}}^2/40)).$ The maximum explored trellis depth is $K=100.$} 
    \label{fig:GaindB1DRate2SimpleFilterCoeff_P_2222221_L0_13}
\end{figure}

Figures~\ref{fig:GaindB1DRate2SimpleFilterCoeff_P_2222221_L0_11}, \ref{fig:GaindB1DRate2SimpleFilterCoeff_P_2222221_L0_12} and~\ref{fig:GaindB1DRate2SimpleFilterCoeff_P_2222221_L0_13} illustrate the asymptotic gains relative to $4$-ASK, for the compact filter pattern $\bm{\pi}_0 = (2,2,2,2,2,2,1),$ corresponding to $L_0=11,$ $12$ and $13,$ respectively. As before, the number of filter candidates—whether equivalent or not—increases rapidly with $L_0.$ For this pattern, the average symbol energy is given by $\| \mathring{h}_0[k] \|^2 + \| \mathring{h}_1[k] \|^2 = 2 \, \| \bm{\pi}_0 \|^2 = 2 \cdot 25 = 50.$ Consequently, the asymptotic gain in dB is $10\, \log_{10}((d_{\text{min}}^2/50)/(4/5)) = 10\, \log_{10}(d_{\text{min}}^2/40)).$ Unlike the case with the pattern $\bm{\pi}_0 = (1,1,1,1,1,1,1,1,1),$ here the best achieved gains increase with $L_0.$ These gains are derived from \glspl{msed} corresponding to error events of length at most $K=100.$ As with previous cases, longer error events beyond this trellis depth may lead to lower \glspl{msed}, and thus lower gains. However, even accounting for this potential degradation, the observed asymptotic gains—approximately $2.7875,$ $3.0103$ and $3.2222$ dB for for $L_0=11,$ $12$ and $13,$ respectively—remain below the required gain of about $10\, \log_{10}(5/2) \approx 3.9794,$ dB needed to match $2$-ASK performance. Therefore, this pattern is ultimately not of practical interest, as it already exhibits \glspl{msed}—within a limited trellis depth—that are inferior to those achieved by $2$-ASK.

\begin{figure}[!htbp]
    \centering
    \includegraphics[width=1.0\textwidth]{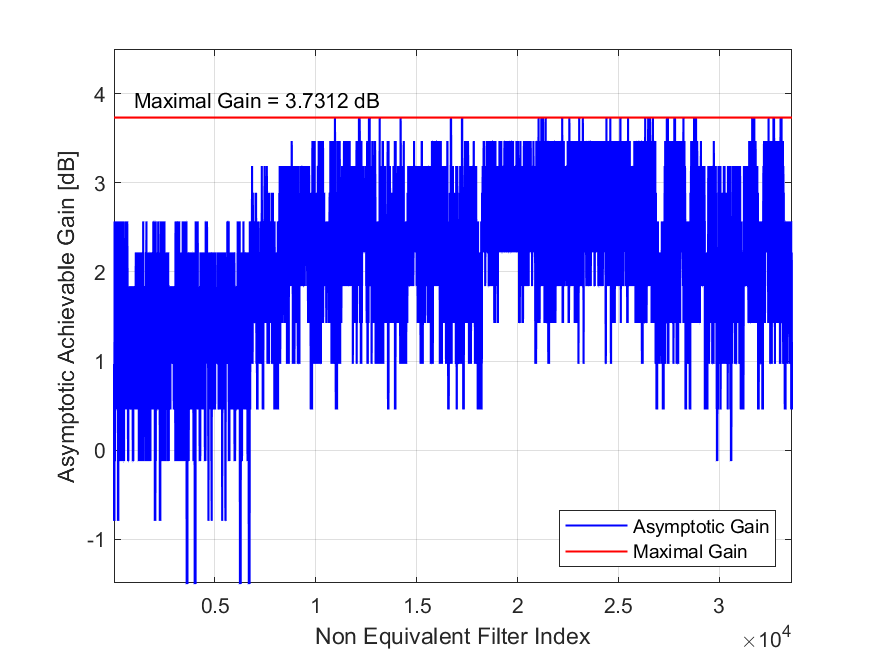}
    \caption{Asymptotic achievable gains of rate-$2$ NSMs, with non-equivalent rational filter coefficients, with compacted filter pattern $\bm{\pi}_0 = (3, 3, 3, 2, 2, 1),$ for $\mathring{\bm{h}}_0,$ and $L_0 = 10.$ The total number of filter $\mathring{\bm{h}}_0$ candidates is $268800.$ The total number of non-equivalent filter candidates is $33600.$ The indices of these filter candidates are shown in the abscissa. The total number of best non-equivalent filter candidates is $100.$ The asymptotic gain in dB is determined as $10\, \log_{10}((d_{\text{min}}^2/(2 \cdot 36))/(4/5)) = 10\, \log_{10}(5 \, d_{\text{min}}^2/288)).$ The maximum explored trellis depth is $K=300.$} 
    \label{fig:GaindB1DRate2SimpleFilterCoeff_P_333221_L0_10}
\end{figure}

\begin{figure}[!htbp]
    \centering
    \includegraphics[width=1.0\textwidth]{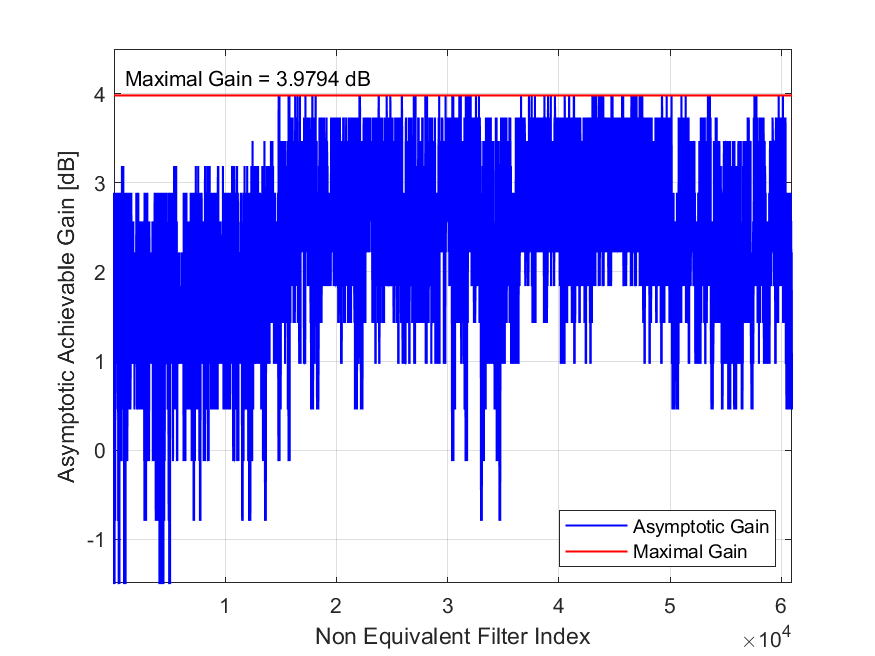}
    \caption{Asymptotic achievable gains of rate-$2$ NSMs, with non-equivalent rational filter coefficients, with compacted filter pattern $\bm{\pi}_0 = (3, 3, 3, 2, 2, 1),$ for $\mathring{\bm{h}}_0,$ and $L_0 = 11.$ The total number of filter $\mathring{\bm{h}}_0$ candidates is $483840.$ The total number of non-equivalent filter candidates is $60960.$ The indices of these filter candidates are shown in the abscissa. The total number of best filter candidates is $1648.$ The total number of best non-equivalent filter candidates is $206.$ The asymptotic gain in dB is determined as $10\, \log_{10}(5 \, d_{\text{min}}^2/288)).$ The maximum explored trellis depth is $K=600.$} 
    \label{fig:GaindB1DRate2SimpleFilterCoeff_P_333221_L0_11}
\end{figure}

\begin{figure}[!htbp]
    \centering
    \includegraphics[width=1.0\textwidth]{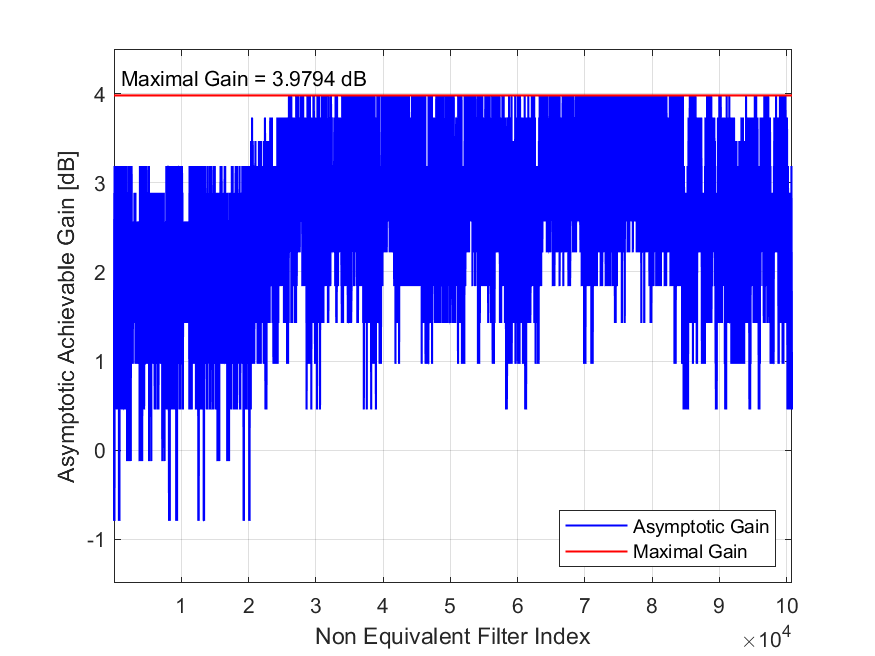}
    \caption{Asymptotic achievable gains of rate-$2$ NSMs, with non-equivalent rational filter coefficients, with compacted filter pattern $\bm{\pi}_0 = (3, 3, 3, 2, 2, 1),$ for $\mathring{\bm{h}}_0,$ and $L_0 = 12.$ The total number of filter $\mathring{\bm{h}}_0$ candidates is $806400.$ The total number of non-equivalent filter candidates is $100800.$ The indices of these filter candidates are shown in the abscissa. The total number of best filter candidates is $19744.$ The total number of best non-equivalent filter candidates is $2468.$ The asymptotic gain in dB is determined as $10\, \log_{10}(5 \, d_{\text{min}}^2/288)).$ The maximum explored trellis depth is $K=1200.$} 
    \label{fig:GaindB1DRate2SimpleFilterCoeff_P_333221_L0_12}
\end{figure}

For the compact filter pattern $\bm{\pi}_0 = (3,3,3,2,2,1),$ Figures~\ref{fig:GaindB1DRate2SimpleFilterCoeff_P_333221_L0_10}, \ref{fig:GaindB1DRate2SimpleFilterCoeff_P_333221_L0_11} and~\ref{fig:GaindB1DRate2SimpleFilterCoeff_P_333221_L0_12} illustrate the asymptotically achieved gains as a function of the indices of candidate non-equivalent filters $\mathring{h}_0[k],$ for $L_0 = 10,$ $11,$ and $12.$ For this pattern, the average symbol energy is given by $\| \mathring{h}_0[k] \|^2 + \| \mathring{h}_1[k] \|^2 = 2 \, \| \bm{\pi}_0 \|^2 = 2 \cdot 36 = 72.$ Consequently, the asymptotic gain in dB is $10\, \log_{10}((d_{\text{min}}^2/72)/(4/5)) = 10\, \log_{10}(5 \, d_{\text{min}}^2/288).$ We observe that the best achievable gain increases from approximately $3.7312$ dB, at $L_0 = 10,$ to the theoretical maximum of $10 \log{10}(5/2) \approx 3.9794$ dB, at $L_0 = 11,$ indicating that the performance of $2$-ASK can be asymptotically approached from this point onward.

Furthermore, while this maximum gain is achieved by $206$ non-equivalent filters for $L_0 = 11,$ the number of such optimal filters increases significantly to $2468$ for $L_0 = 12.$ This growth is not only numerical but also structural, in the sense that the percentage of good filters—those achieving the \gls{msed} of $2$-ASK—increases markedly with $L_0$. Specifically, for $L_0 = 11,$ there are a total of $483840$ candidate filters $\mathring{h}_0[k],$ of which $1{,}648$ achieve the $2$-ASK minimum distance. These correspond to $206$ good filters among the $60960$ non-equivalent candidates, yielding a percentage of approximately $0.34\%$. In contrast, for $L_0 = 12,$ we have $806400$ candidate filters with $19744$ achieving the target distance, including $2468$ good filters among the $100800$ non-equivalent ones—raising the percentage to about $2.45\%$. This represents a more than $7$-fold increase in the proportion of optimal filters when transitioning from $L_0 = 11$ to $L_0 = 12$.

This trend suggests a broader phenomenon: we conjecture that the percentage of good filters continues to increase with $L_0$ and that this percentage rapidly approaches $100\%$ as soon as $L_0$ exceeds moderate values, likely around $14$ or $15$. In other words, almost all candidate filters $\mathring{h}_0[k]$ are expected to become good, achieving the $2$-ASK \gls{msed}, except for a small pathological subset whose proportion rapidly tends to zero. This is a crucial insight, as it implies that once the filter length $L_0$ exceeds a moderate threshold, near-optimal performance in terms of \gls{bep} is attained regardless of the specific filter choice. Although this observation has been detailed for the particular pattern $\bm{\pi}_0 = (3,3,3,2,2,1),$ we conjecture that the same behavior will hold for an increasingly wide range of filter patterns as $L_0$ becomes sufficiently large. In other words, the asymptotic performance of $2$-ASK is expected to become universally achievable for a broad class of patterns, as long as $L_0$ is large enough to mitigate the influence of pattern-specific anomalies.

While the above conjectures are grounded in empirical observations based on limited trellis depths, it is important to validate their reliability through deeper evaluations. To this end, we revisit the numerical results obtained for $L_0 = 11$ and $L_0 = 12$ and strengthen their validity by increasing the trellis exploration depths. These results were initially obtained using limited trellis exploration depths of $K = 600,$ for $L_0 = 11,$ and $K = 1200,$ for $L_0 = 12.$ To confirm the robustness of these findings, we extended the trellis depth to $K = 10^5$ for the $206$ best candidates at $L_0 = 11,$ and to $K = 10^4$ for the $2468$ candidates at $L_0 = 12.$ In both cases, the same maximum gain—consistent with $2$-ASK performance—was retained across all optimal filters identified with the smaller trellis depths. This confirms that, for practical data packet sizes around $10^4,$ \glspl{nsm} derived from the best filters, with $L_0 \ge 11,$ attain the \gls{msed} of $2$-ASK. However, due to current computational limitations, it remains an open question whether the $2$-ASK's asymptotic performance can still be guaranteed for arbitrarily large packet sizes under the same filter pattern $\bm{\pi}_0 = (3,3,3,2,2,1)$ and $L_0 \ge 11.$

\begin{figure}[!htbp]
    \centering
    \includegraphics[width=1.0\textwidth]{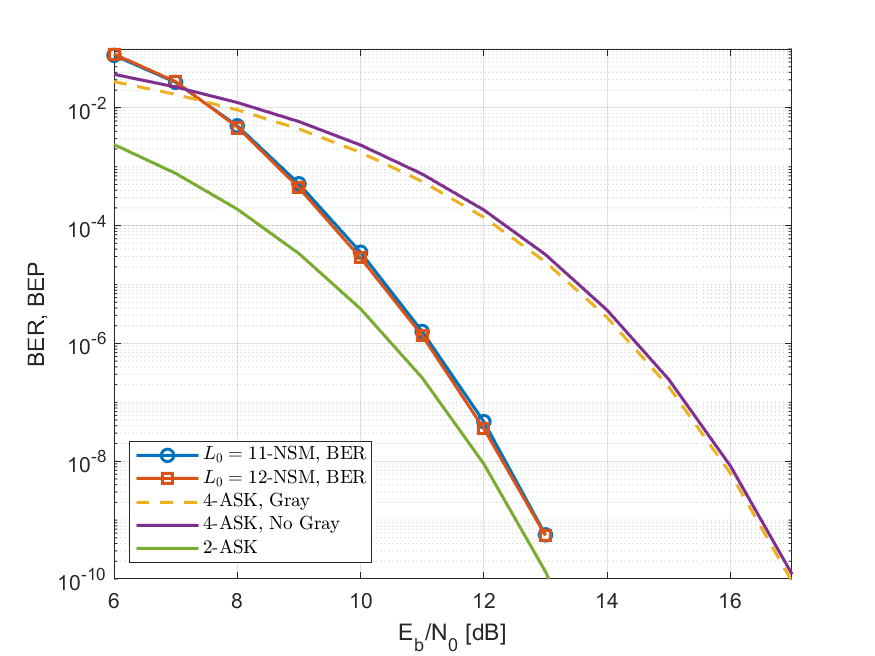}
    \caption{BER of NSMs of rate $2,$ with rational filter coefficients, with compacted filter pattern $\bm{\pi}_0 = (3, 3, 3, 2, 1, 1),$ with $L_0 = 11$ and $12,$ and $L_1 = 1$ ($\mathring{\bm{h}}_0 = 6 \, \bar{\bm{h}}_0 = (3, 2, 0, 0, 3, -2, 0, 0, 0, -1, -3),$ for $L_0=11,$ and $\mathring{\bm{h}}_0 = 6 \, \bar{\bm{h}}_0 = (3, 1, 0, 0, -2, 3, 0, 0, 2, 0, 0, -3),$ for $L_0=12.$). For reference, the BEPs of $2$-ASK and Gray and non-Gray precoded $4$-ASK conventional modulations are shown.}
    \label{fig:BER-BEP-NSM-2-Pattern_333221}
\end{figure}

Given that the pattern $\bm{\pi}_0 = (3,3,3,2,2,1)$ produces \glspl{nsm} that asymptotically achieve $2$-ASK performance for $L_0 \ge 11$, we sought to efficiently characterize the projected distance spectrum of each of the top-performing non-equivalent codes—$206$ for $L_0 = 11$ and $2468$ for $L_0 = 12$. The procedure for determining the projected distance spectrum follows the same principles as that established earlier in Subsection~\ref{Rate-2 guaranteeing, minimum Euclidean distance approaching NSMs with real filters' coefficients} for rate-$2$ \glspl{nsm} with real-tapped filters. The main difference here is that scaled versions of the rational-tapped filters are used, resulting in equivalent integer-tapped representations. The projected distance spectrum is computed as follows: for each finite-length first input sequence difference $\Delta \bar{b}_0[k]$, we determine the corresponding partial modulated sequence $\Delta \bar{s}_0[k] = \Delta \bar{b}_0[k] \circledast \mathring{h}_0[k]$. The \gls{sed} is then obtained by minimizing the \gls{sen} of $\Delta \bar{s}[k] = \Delta \bar{s}_0[k] + \Delta \bar{s}_1[k]$ over all possible second input sequence differences $\Delta \bar{b}_1[k]$, where $\Delta \bar{s}_1[k] = \Delta \bar{b}_1[k] \circledast \mathring{h}_1[k] = \mathring{h}_1[0] \Delta \bar{b}_1[k]$. The projected \gls{sed} for each $\Delta \bar{b}_0[k]$ is thus given by $\sum_k \min_{\Delta \bar{b}_1[k] \in \{0, \pm 2\}}(\Delta \bar{s}_0[k]+\mathring{h}_1[0] \Delta \bar{b}_1[k])^2.$ Based on the projected distance spectrum analysis, we selected two representative filter candidates for further evaluation: $\mathring{\bm{h}}_0 = (3, 2, 0, 0, 3, -2, 0, 0, 0, -1, -3),$ for $L_0 = 11,$ and $\mathring{\bm{h}}_0 = (3, 1, 0, 0, -2, 3, 0, 0, 2, 0, 0, -3),$ for $L_0 = 12.$ These filters were then assessed through simulations to evaluate their \gls{ber} performance. As shown in Figure~\ref{fig:BER-BEP-NSM-2-Pattern_333221}, both \glspl{nsm} exhibit similar performance and closely approach that of $2$-ASK, despite a relatively high multiplicity of error events at the \gls{msed}—an effect associated with the tightness property.

Compared to the best-performing \glspl{nsm} under the compact filter pattern $\bm{\pi}_0 = (1,1,1,1)$ (shown in Figures~\ref{fig:BER-BEP-NSM-2-FilterSimpleCoefficients-FilterLength_11} and~\ref{fig:BER-BEP-NSM-2-FilterSimpleCoefficients-FilterLength_12}, for $L_0 = 11$ and $L_0 = 12,$ respectively), the $(3,3,3,2,2,1)$ pattern offers a clear performance advantage. This improvement cannot be attributed solely to a modest reduction in the multiplicity of minimum-distance error events; rather, it likely stems from a broader enhancement of the entire distance spectrum, particularly among low-weight error events. This comparison highlights the value of exploring alternative filter patterns that not only maintain the $2$-ASK’s \gls{msed}, for practical trellis depths and packet lengths, but also alleviate the limitations imposed by tightness, thereby improving overall \gls{nsm} performance.

\subsubsection{Two-Dimensional NSMs}
\label{ssec:Two-Dimensional Rate-2 NSMs}

\paragraph*{\textbf{Foundational NSMs inspired by rate-5/4 designs}}

\begin{figure}[!htbp]
    \centering
    \includegraphics[width=0.8\textwidth]{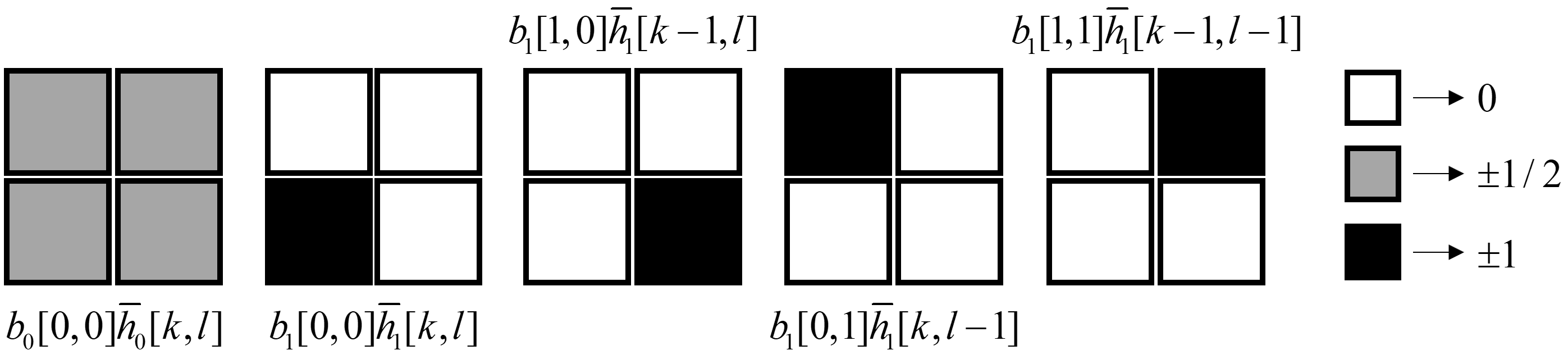}
    \caption{Two-dimensional interpretation of the rate-$5/4$ block NSM proposed in Subsection~\ref{ssec:Minimum Euclidean Distance Guaranteeing 5/4-NSM}.}
    \label{fig:Two-Dimensional-NSM-Illustration-Rate-5-4}
\end{figure}

This family of \glspl{nsm}, like the \glspl{nsm} proposed in Subsection~\ref{sssec:One-dimensional designed NSMs}, employs normalized filters with simple non-null fractional coefficients in the set $\{\pm 1/2, \pm 1 \}.$ They originate from a \gls{2d} interpretation of the \gls{1d} illustrative block \gls{nsm} of rate $5/4,$ which has been thoroughly specified and characterized in Subsection~\ref{ssec:Minimum Euclidean Distance Guaranteeing 5/4-NSM}. This \gls{2d} description, as seen in Figure~\ref{fig:Two-Dimensional-NSM-Illustration-Rate-5-4}, entails replacing the \gls{1d} filters, $\bar{h}_m[k] = \delta[k-m],$ $m=0,1,2$ and $3,$ and $\bar{h}_4[k] = \tfrac{1}{2}(\delta[k]+\delta[k-1]+\delta[k-2]+\delta[k-3]),$ presented in Subsection~\ref{ssec:Minimum Euclidean Distance Guaranteeing 5/4-NSM}, with their respective \gls{2d} counterparts, $\bar{h}_1[k,l] \triangleq \delta[k] \delta[l], \bar{h}_1[k-1,l], \bar{h}_1[k,l-1],\bar{h}_1[k-1,l-1]$ and $\bar{h}_0[k,l] \triangleq \tfrac{1}{2} (\delta[k]+\delta[k-1])(\delta[l]+\delta[l-1]).$ As a consequence, the $5$ \gls{1d} bipolar symbols, $b_m[l],$ $0 \le m \le 4,$ of the $l$-th input block are replaced by the $5$ \gls{2d} input bipolar symbols $b_1[0,0], b_1[1,0], b_1[0,1], b_1[1,1]$ and $b_0[0,0],$ while the $4$ \gls{1d} symbols, $s[4l+m],$ $0 \le m \le 3,$ of the $l$-th modulated block are replaced by the $4$ \gls{2d} modulated symbols $s[0,0], s[1,0], s[0,1]$ and $s[1,1].$ As an outcome, the relationship between the \gls{2d} modulated symbols and the input bipolar symbols may be written as $s[k,l] = b_0[0,0] \bar{h}_0[k,l] + b_1[0,0] \bar{h}_1[k,l] + b_1[1,0] \bar{h}_1[k-1,l] + b_1[0,1] \bar{h}_1[k,l-1] + b_1[1,1] \bar{h}_1[k-1,l-1],$ with the individual contributions of all five input bipolar symbols explicitly displayed in Figure~\ref{fig:Two-Dimensional-NSM-Illustration-Rate-5-4}.

In matrix form, the \gls{2d} modulated symbols, associated to the rate-$5/4$ \gls{nsm} proposed in Subsection~\ref{ssec:Minimum Euclidean Distance Guaranteeing 5/4-NSM}, can be expressed as
\begin{multline}
\begin{pmatrix}
s[0,0] & s[0,1] \\
s[1,0] & s[1,1]
\end{pmatrix} =
b_0[0,0]
\begin{pmatrix}
\tfrac{1}{2} & \tfrac{1}{2} \\
\tfrac{1}{2} & \tfrac{1}{2}
\end{pmatrix} +
b_1[0,0]
\begin{pmatrix}
1 & 0 \\
0 & 0
\end{pmatrix} \\ +
b_1[1,0]
\begin{pmatrix}
0 & 0 \\
1 & 0
\end{pmatrix} +
b_1[0,1]
\begin{pmatrix}
0 & 1 \\
0 & 0
\end{pmatrix} +
b_1[1,1]
\begin{pmatrix}
0 & 0 \\
0 & 1
\end{pmatrix}.  
\end{multline}

Armed with this \gls{2d} understanding of the rate-$5/4$ \gls{nsm}, we begin by defining and characterizing, in terms of minimal Euclidean distance, an infinite \gls{2d} extent \gls{nsm}, of exact rate $2$. As an outcome, this \gls{nsm} assists in the formulation of \glspl{nsm} with finite increasing extent, which have the advantage of achieving rates approaching gradually $2$, while permitting practical implementation.

In mathematical form, the modulated symbols of the infinite extent \gls{nsm} of rate $2$ can be stated as
\begin{equation} \label{eq:Mod Seq Two-Dimensional NSM}
    s[k,l] = \sum_{i,j} b_0[i,j] \bar{h}_0[k-i,l-j] + \sum_{i,j} b_1[i,j] \bar{h}_1[k-i,l-j],
\end{equation}
which amounts to summing the contributions of two \gls{2d} convolutions, involving input data bipolar symbols, $b_m[k,l]$, and \gls{2d} filters, $\bar{h}_m[k,l]$, $m=0,1$.

With the simple formulation of the rate-$2$ \gls{nsm} in (\ref{eq:Mod Seq Two-Dimensional NSM}), the first and most important goal next is to demonstrate that the resulting minimal Euclidean distance is exactly that of $2$-ASK. In order to accomplish this, we introduce, as in Subsection~\ref{sssec:One-dimensional designed NSMs}, two \gls{2d} input sequences, $(b_0^n[k,l], b_1^n[k,l]),$ $n=0,1,$ and their associated modulated sequences $s^n[k,l],$ $n=0,1.$ We then consider the associated \gls{2d} input sequences differences $\Delta \bar{b}_0[k,l] \triangleq \bar{b}_0^1[k,l]-\bar{b}_0^0[k,l]$ and $\Delta \bar{b}_1[k,l] \triangleq \bar{b}_1^1[k,l]-\bar{b}_1^0[k,l],$ which take their values in the ternary alphabet $\{0, \pm 2\},$ as well as the modulated sequence difference $\Delta s[k,l] \triangleq s^1[k,l]-s^0[k,l].$

Given (\ref{eq:Mod Seq Two-Dimensional NSM}), we may write
\begin{equation}
    \Delta s[k,l] = \sum_{i,j} \Delta \bar{b}_0[i,j] h_0[k-i,l-j] + \sum_{i,j} \Delta \bar{b}_1[i,j] h_1[k-i,l-j].
\end{equation}
In order to simplify the determination of the minimum Euclidean distance of the proposed \gls{nsm}, we consider the non-scaled version $\Delta \bar{s}[k,l] \triangleq \Delta s[k,l]/\sqrt{\eta},$ where we recall that $\eta=5/2$ is a generic parameter that allows the proposed \gls{nsm} to be compared to conventional $4$-ASK, on a common basis of average symbol energy. As a results, we can write $\Delta \bar{s}[k,l] = \Delta \bar{s}_0[k,l] + \Delta \bar{s}_1[k,l],$ where
\begin{equation}
    \Delta \bar{s}_m[k,l] \triangleq \sum_{i,j} \Delta \bar{b}_m[i,j] \bar{h}_m[k-i,l-j], m = 0,1.
\end{equation}

\begin{figure}[!htbp]
    \centering
    \includegraphics[width=0.9\textwidth]{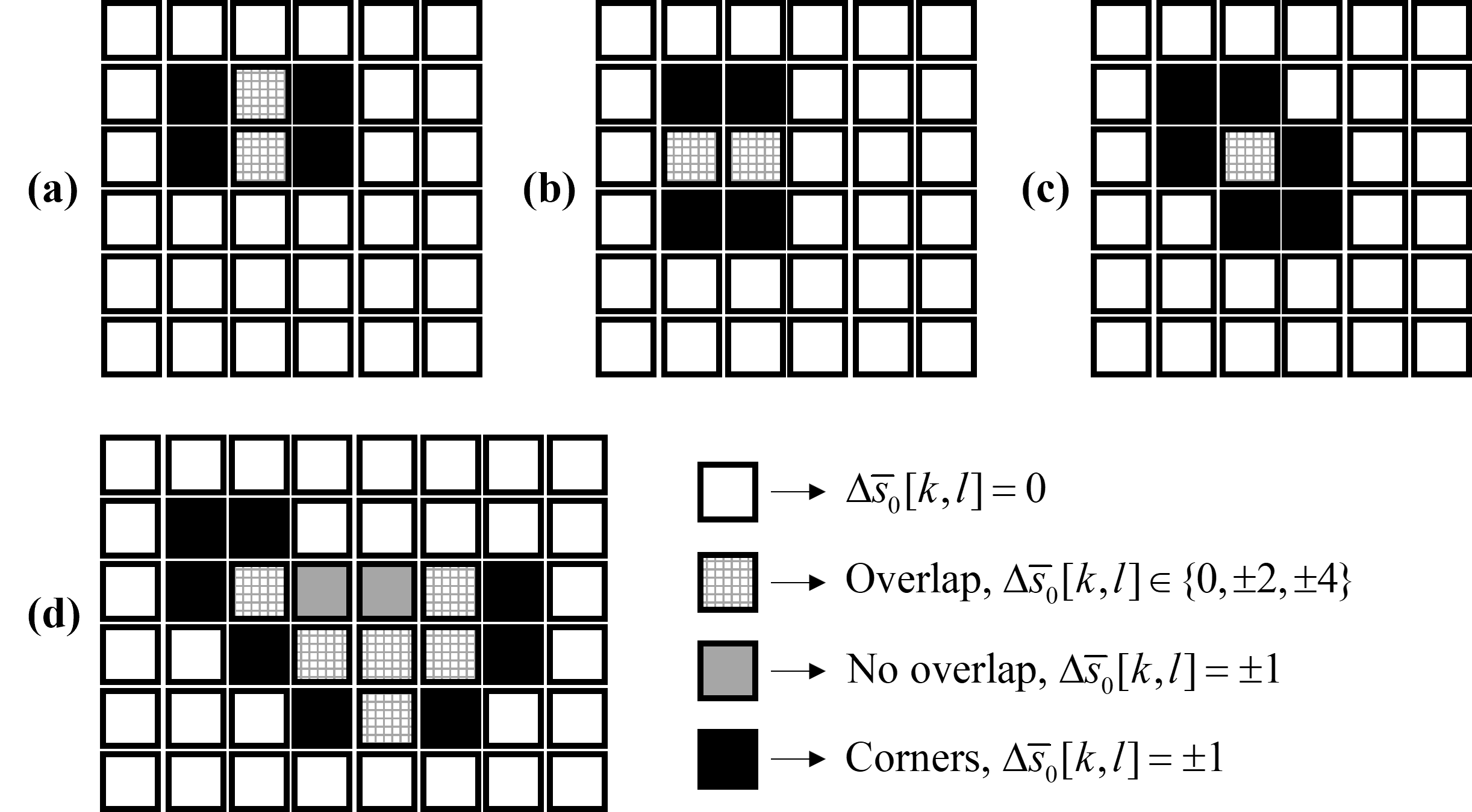}
    \caption{Two-dimensional footprints of $\Delta \bar{s}_0[k,l],$ illustrating the notion of corners: (a)
    $\Delta \bar{b}_0[k,l] = \pm \delta[k]\delta[l] \pm \delta[k-1]\delta[l],$
(b) $\Delta \bar{b}_0[k,l] = \pm \delta[k]\delta[l] \pm \delta[k]\delta[l-1],$
(c) $\Delta \bar{b}_0[k,l] = \pm \delta[k]\delta[l] \pm \delta[k-1]\delta[l-1],$
(d) $\Delta \bar{b}_0[k,l] = \pm \delta[k]\delta[l] \pm \delta[k-1]\delta[l-1]\pm \delta[k-3]\delta[l-1] \pm \delta[k-4]\delta[l-1] \pm \delta[k-2]\delta[l-2] \pm \delta[k-2]\delta[l-3].$} \label{fig:Two-Dimensional-NSM-Minimum-Euclidean-Distance}
\end{figure}

We assume that both, $\Delta \bar{b}_0[k,l]$ and $\Delta \bar{b}_1[k,l],$ are not simultaneously null and contain a finite number of non-null coefficients, all of which must belong to the set $\{ \pm 2 \}.$ On the one hand, if $\Delta \bar{s}_0[k,l]$ is a null \gls{2d} sequence, then the \gls{2d} sequence $\Delta \bar{s}_1[k,l],$ and hence the \gls{2d} sequence $\Delta \bar{s}[k,l],$ has at least one non-zero coefficient in the set $\{ \pm 2 \}.$ As a result, the \gls{sen} of \gls{2d} sequence $\Delta \bar{s}[k,l]$ must be at least equal to $4.$ On the other hand, if the \gls{2d} sequence $\Delta \bar{s}_0[k,l]$ is non-null, it must contain a finite number of non-null coefficients. Figure~\ref{fig:Two-Dimensional-NSM-Minimum-Euclidean-Distance} depicts multiple illustrative footprints of \gls{2d} $\Delta \bar{s}_0[k,l],$ that identify positions in the \gls{2d} grid, $\mathbb{Z}^2,$ where it is potentially non-null. As is readily apparent, each potential footprint of the \gls{2d} sequence $\Delta \bar{s}_0[k,l]$ provides at least four corners, where the footprints of the various underlying \gls{2d} shifts of filter $\bar{h}_0[k,l]$ do not overlap. Because of this, we can be certain that $\Delta \bar{s}_0[k,l],$ takes values from the set $\{ \pm 1 \}$ in at least four distinct locations on the \gls{2d} grid. Given that $\Delta \bar{s}_1[k,l]$ gets values from the set $\{0, \pm 2 \},$ it follows that $\Delta \bar{s}[k,l]$  will undoubtedly take values in the set $\{ \pm 1, \pm 3 \}$ for these four locations. As a result, the \gls{sen} of $\Delta \bar{s}[k,l]$ must at least be equal to $4.$ To demonstrate that this lower bound is met and that the proposed \gls{nsm}'s \gls{msed} is exactly equal to $4,$ consider input sequence differences of the form $(\Delta \bar{b}_0[k,l] = \pm 2 \delta[k-i]\delta[l-j],\Delta \bar{b}_1[k,l] = 0)$ or $(\Delta \bar{b}_0[k,l] = 0,\Delta \bar{b}_1[k,l]=\pm 2 \delta[k-i]\delta[l-j]),$ for an arbitrary position $(i,j)$ in the \gls{2d} grid. Indeed, in such a circumstance, the \gls{sen} of $\Delta \bar{s}[k,l]$ is exactly equal to $4,$ because it is either equal to $\pm 2 \bar{h}_0[k-i,l-j]$ or $\pm 2 \bar{h}_1[k-i,l-j].$

At this point, we believe it is wise and enlightening to highlight some additional aspects of the infinite \gls{2d} extent \glspl{nsm} of rate $2,$ before proposing and evaluating a number of realistic finite \gls{2d} extent \glspl{nsm}. First, the \gls{2d} \gls{nsm} at hand is degenerate, much like the \gls{1d} \gls{nsm} of rate $2$ presented for illustration in Subsection~\ref{ssec:Modulation of Rate 2}. To demonstrate this, all that needs to be done is to note how error events with quincunx alternating-sign input data differences sequences $\Delta \bar{b}_0[k,l] = \pm 2 \sum_{i=0}^{I-1} \sum_{j=0}^{J-1} (-1)^{i+j} \delta[k-i]\delta[l-j]$ and $\Delta \bar{b}_1[k,l] = 0,$ with increasing rectangular extent, as $I$ and $J$ increase to infinity, result in output modulated sequences differences, $\Delta \bar{s}[k,l] = \pm (\delta[k] + (-1)^{I-1}\delta[k-I])(\delta[l] + (-1)^{J-1}\delta[l-J]),$ of \gls{msed} of $4.$ Fortunately, identical argumentation as in Subsection~\ref{ssec:Modulation of Rate 2} and Appendices~\ref{app:Tight Estimate BEP Rate 5/4} and~\ref{app:Tight Estimate BEP Rate 2} show that such error events occur with probability $(\tfrac{1}{2})^{I+J},$ which vanishes exponentially as the \gls{2d} extent extends uniformly towards infinity.

Second, with reference to the normalized filter, $\bar{h}_0[k,l],$ notice that the associated footprint, which in the present instance is the most compact, can be changed to any larger irregular, and therefore non-rectilinear, footprint in order to reduce the multiplicity of error events of low Euclidean distance. However, keep in mind that practical considerations drive the use of the most compact filter presented thus far, as it enables the highest rates when the \gls{nsm} \gls{2d} extent is constrained to be finite.

Still referring to the normalized filters, $\bar{h}_m[k,l],$ $m=0,1,$ observe that there are a variety of transformations that might result in equivalent filters and, as a result, make it possible to find the optimal filter more quickly, subject to certain extent restrictions. These transformations are, in part, identical or generalizations of the equivalence transformations introduced in Subsection~\ref{ssec:Modulation of Rate 2}, for the \gls{1d} case. To begin with, using $-\bar{h}_m[k,l]$ instead of $\bar{h}_m[k,l]$ for $m = 0$ or $m = 1,$ has no effect on the resulting \gls{nsm}'s properties. Additionally, following a similar argumentation as \ref{ssec:Modulation of Rate 2}, we can show that the three alternate sign transformations, whereby $\bar{h}_0[k,l]$ is replaced by $\tilde{h}_0[k,l] = (-1)^k \bar{h}_0[k,l],$ $\tilde{h}_0[k,l] = (-1)^l \bar{h}_0[k,l]$ or $\tilde{h}_0[k,l] = (-1)^{k+l} \bar{h}_0[k,l],$ preserve the characteristics of the resulting \gls{nsm}. Moreover, to any arbitrary shift in time for the \gls{1d} scenario, which preserves the \gls{nsm} characteristics, corresponds a translation in the grid for the \gls{2d} scenario at hand. Similarly, to the time reversal operation in one dimension, which maintains \gls{1d} \gls{nsm} characteristics, corresponds symmetries operations in two dimensions, with regard to the vertical, horizontal and two diagonal axes, whereby $\bar{h}_0[k,l]$ is replaced by $\bar{h}_0[-k,l],$ $\bar{h}_0[k,-l],$ $\bar{h}_0[l,k]$ and $\bar{h}_0[-l,-k],$ respectively. To these geometrical transformations, we can add rotations in the grid with phases in the set $\{\pm \pi/2, \pi \}.$

Using the prior set of characteristics-preserving transformations and the $16$ choices, $\tfrac{1}{2}(\pm \delta[k]\delta[l] \pm \delta[k-1]\delta[l] \pm \delta[k]\delta[l-1] \pm \delta[k-1]\delta[l-1]),$ for filter $\bar{h}_0[k,l],$ presented at the beginning of Section~\ref{Rate 2 Approaching NSM Simple Rational Coefficients}, we end up with just two equivalence class representatives, namely the previously considered filter, $\bar{h}_0[k,l] = \tfrac{1}{2}(\delta[k] + \delta[k-1])(\delta[l] + \delta[l-1]),$ and the new filter, $\bar{h}_0[k,l] = \tfrac{1}{2}(\delta[k]\delta[l] + \delta[k-1]\delta[l] + \delta[k]\delta[l-1] - \delta[k-1]\delta[l-1]).$ Experimental results, in terms of \gls{ber}, as well as theoretical results, in terms of \gls{msed} error events multiplicities, for the rate-$3/2$ \gls{1d} \glspl{nsm} inherited from these two \gls{2d} \gls{nsm} equivalence classes, show a slight superiority for the second equivalence class. This second equivalence class, with $\bar{h}_0[k,l] = \tfrac{1}{2}(\delta[k]\delta[l] + \delta[k-1]\delta[l] + \delta[k]\delta[l-1] - \delta[k-1]\delta[l-1]),$ offers the same \gls{msed} as the first equivalence class, with $\bar{h}_0[k,l] = \tfrac{1}{2}(\delta[k] + \delta[k-1])(\delta[l] + \delta[l-1]).$ However, the former equivalence class presents a reduced multiplicity, for the error events with \gls{msed}, with respect to the latter equivalence class.

To gain a basic understanding of this multiplicity concern, we consider \gls{msed} error events with first input sequences difference, $\Delta \bar{b}_0[k,l],$ such that only $\Delta \bar{b}_0[0,0]$ and $\Delta \bar{b}_0[0,1]$ are non-null, with values in the set $\{ \pm 2 \}.$ For the first equivalence class, if we take $\Delta \bar{b}_0[0,0] = \Delta \bar{b}_0[0,1] = \pm 2,$ then we need to chose the second input sequences difference, $\Delta \bar{b}_1[k,l],$ such that only $\Delta \bar{b}_1[0,1]$ and $\Delta \bar{b}_1[1,1]$ are non-null, and are given by $\Delta \bar{b}_1[0,1] = \Delta \bar{b}_1[1,1] = \mp 2.$ The occurrence probability of these elementary error events is equal to $2(\tfrac{1}{2})^4 = (\tfrac{1}{2})^3.$ Alternatively, if we take $\Delta \bar{b}_0[0,0] = - \Delta \bar{b}_0[0,1] = \pm 2,$ then we need to chose the second input sequences difference to be null. The occurrence of these elementary error events is $2(\tfrac{1}{2})^2 = \tfrac{1}{2}.$ Hence, the cumulative multiplicity for the first equivalence class is $(\tfrac{1}{2})^3 + \tfrac{1}{2} = \tfrac{5}{8}.$ Now, for the second equivalence class, if we take $\Delta \bar{b}_0[0,0] = \Delta \bar{b}_0[0,1] = \pm 2,$ then we need to chose the second input sequences difference, $\Delta \bar{b}_1[k,l],$ such that only $\Delta \bar{b}_1[0,1]$ is non-null, and is given by $\Delta \bar{b}_1[0,1] = \mp 2.$ The occurrence probability of these elementary error events is equal to $2(\tfrac{1}{2})^3 = (\tfrac{1}{2})^2.$ Alternatively, if we take $\Delta \bar{b}_0[0,0] = - \Delta \bar{b}_0[0,1] = \pm 2,$ then we need to chose the second input sequences difference, $\Delta \bar{b}_1[k,l],$ such that only $\Delta \bar{b}_1[1,1]$ is non-null, and is given by $\Delta \bar{b}_1[1,1] = \pm 2.$ The occurrence of these elementary error events is again $2(\tfrac{1}{2})^3 = (\tfrac{1}{2})^2.$ Hence, the cumulative multiplicity for the second equivalence class is $(\tfrac{1}{2})^2 + (\tfrac{1}{2})^2 = \tfrac{1}{2}.$ Comparing the cumulative multiplicities of $\tfrac{5}{8},$ for the first equivalence class, and $\tfrac{1}{2},$ for the second equivalence class, goes in the direction of the slight superiority of the latter class, as first pointed out above.  This is further supported  by the superior performance of the \gls{1d} filter $\mathring{h}_0[k] = 2 \bar{h}_0[k] = \delta[k] + \delta[k-1] + \delta[k-2] - \delta[k-3],$ studied in Section~\ref{Rate-3/2 Approaching NSMs} for rate-$3/2$ \glspl{nsm}, compared to filter $\mathring{h}_0[k] = 2 \bar{h}_0[k] = \delta[k] + \delta[k-1] + \delta[k-2] + \delta[k-3],$ as shown in Table~\ref{table:NSMs Rate-3/2 Filter Pattern (1, 1, 1, 1)}. Indeed filters $\bar{h}_0[k] = \tfrac{1}{2}(\delta[k] + \delta[k-1] + \delta[k-2] - \delta[k-3])$ and $\bar{h}_0[k] = \tfrac{1}{2}(\delta[k] + \delta[k-1] + \delta[k-2] + \delta[k-3])$ are the \gls{1d} counterparts of \gls{2d} filters $\bar{h}_0[k,l] = \tfrac{1}{2}(\delta[k]\delta[l] + \delta[k-1]\delta[l] + \delta[k]\delta[l-1] - \delta[k-1]\delta[l-1])$ and $\bar{h}_0[k,l] = \tfrac{1}{2}(\delta[k] + \delta[k-1])(\delta[l] + \delta[l-1]),$ respectively. 

For the sake of illustration, while keeping the presentation as straightforward as possible, we proceed with the first equivalence class of filters, even though it is marginally less efficient than the second equivalence class of filters. For realistic implementations, we must consider \glspl{nsm} with limited extents. When we consider rectangular extents of width $I \ge 2$ and height $J \ge 2,$ only input sequences $\bar{b}_0[k,l],$ $0 \le k < I-1$ and $0 \le l < J-1,$ and $\bar{b}_1[k,l],$ $0 \le k < I$ and $0 \le l < J,$ are allowed to take non-null values in the set $\{ \pm 1 \}.$ In sum, $IJ +(I-1)(J-1)$ input data bipolar symbols can be modulated into a \gls{2d} normalized sequence, $\bar{s}[k,l],$ $0 \le k < I$ and $0 \le l < J,$ comprising exactly $IJ$ symbols. As expected, the offered rate, given by $\rho = (IJ +(I-1)(J-1))/(IJ) = 1 + (1-1/I)(1-1/J),$ approaches $2$ as the extent rises in size towards infinity, along both grid axes.

Next, for illustration purposes, we look at extents that are square-shaped, with $I = J \ge 2.$ In this particular situation, the offered rate is given by $\rho = 1+(1-1/I)^2.$ The scenario where $I=2$ provides a rate of $5/4$ and corresponds to the illustrative \gls{nsm} that was introduced and thoroughly discussed in \ref{ssec:Minimum Euclidean Distance Guaranteeing 5/4-NSM}. As a consequence, we will now concentrate on the scenarios where $I \ge 3.$ We first investigate the scenario where $I=3,$ to show how demodulation can be accomplished with reduced complexity, by generalizing the demodulation technique described at the end of Subsection~\ref{ssec:Minimum Euclidean Distance Guaranteeing 5/4-NSM} for the rate $5/4$ \gls{nsm}. We then wrap up this subsection by showcasing some simulation results that demonstrate how well \glspl{nsm} perform in terms of \gls{ber}, for $I \in \{ 3, 4, 5 \},$ and their corresponding increasing rates, $\rho \in \{13/9, 25/16, 41/25\}.$

\begin{figure}[!htbp]
    \centering
    \includegraphics[width=0.8\textwidth]{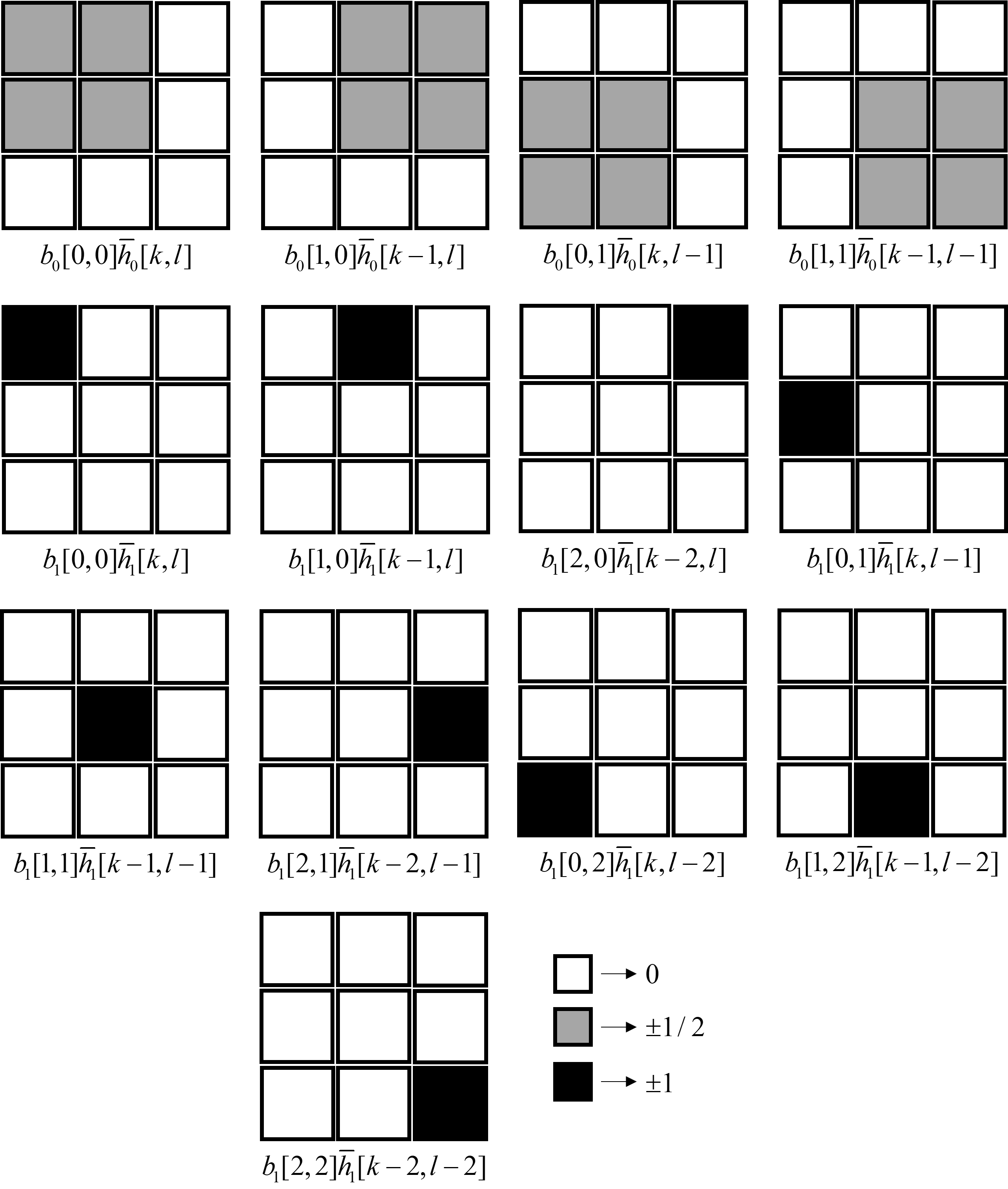}
    \caption{Interpretation in two dimensions of the 2D rate-$13/9$ block NSM associated with $I=3.$}
    \label{fig:Two-Dimensional-NSM-Illustration-Rate-13-9}
\end{figure}

When $I=3,$ the individual contributions of all non-null input bipolar symbols are plainly displayed in Figure~\ref{fig:Two-Dimensional-NSM-Illustration-Rate-13-9}. We have $9$ input bipolar data symbols from $\bar{b}_1[k,l],$ corresponding to coordinates $(k,l) \in \{ 0, 1, 2 \}^2,$ in the \gls{2d} grid $\mathbb{Z}^2.$ For flexible notations, we introduce the $9 \times 1$ row vector, $\bar{\bm{b}_1},$ which arises from reading these symbols row-wise. We also have $4$ input bipolar data symbols from $\bar{b}_0[k,l],$ with coordinates $(k,l) \in \{ 0, 1 \}^2,$ in the \gls{2d} grid $\mathbb{Z}^2.$ Always with the goal of keeping notations simple, we introduce the $4 \times 1$ row vector, $\bar{\bm{b}_0},$ which arises from reading these symbols row-wise. For more concise notations, these distinct input bipolar data vectors can then be aggregated into a unique $13 \times 1$ row vector, $\bar{\bm{b}} \triangleq [\bar{\bm{b}_1} \bar{\bm{b}_0}].$ This vector of input symbols yields $9$ modulated symbols, $\bar{s}[k,l],$ with coordinates in the set $\{ 0, 1, 2 \}^2.$ Also to keep notation simple, we denote the $9 \times 1$ row vector aggregating these modulated symbols, when read row-wise, by $\bar{\bm{s}}.$ 
This analysis of the involved vectors allows us to confirm that the proposed \gls{nsm} offered rate for $I=3$ is indeed $\rho = 1+(1-1/I)^2 = 13/9$.

Similar to the considerations at the beginning of Subsection~\ref{ssec:Minimum Euclidean Distance Guaranteeing 5/4-NSM}, we can relate $\bar{\bm{s}}$ to $\bar{\bm{b}}$ by $\bar{\bm{s}} = \bar{\bm{b}} \bm{G},$ where $\bm{G} \triangleq [\bm{I} \bm{G}_0^T]^T$ is now a $13 \times 9$ \gls{nsm} generating analog matrix, $\bm{I}$ is the $9 \times 9$ identity matrix and
\begin{equation} \label{eq: Generating Matrix Rate 13/9 2D NSM}
    \bm{G}_0 =
    \begin{pmatrix}
    \tfrac{1}{2} & \tfrac{1}{2} & 0 & \tfrac{1}{2} & \tfrac{1}{2} & 0 & 0 & 0 & 0 \\
    0 & \tfrac{1}{2} & \tfrac{1}{2} & 0 & \tfrac{1}{2} & \tfrac{1}{2} & 0 & 0 & 0 \\
    0 & 0 & 0 & \tfrac{1}{2} & \tfrac{1}{2} & 0 & \tfrac{1}{2} & \tfrac{1}{2} & 0 \\
    0 & 0 & 0 & 0 & \tfrac{1}{2} & \tfrac{1}{2} & 0 & \tfrac{1}{2} & \tfrac{1}{2}
    \end{pmatrix}
\end{equation}
is the $4 \times 9$ sub-matrix associated to non-null input data symbols $\bar{b}_0[k,l],$ and $(\cdot)^T$ is the transpose operator.

Because the current generating matrix, $\bm{G},$ is structurally identical to that in (\ref{eq:Generating Matrix NSM Rate 5/4}), we may simplify the \gls{ml} detection method by proceeding in two phases, as in the case of the rate-$5/4$ \gls{nsm} of Section~\ref{ssec:Minimum Euclidean Distance Guaranteeing 5/4-NSM}. With this goal in mind, we begin by rewriting the modulated vector as $\bar{\bm{s}} = \bar{\bm{b}}_1 + \bar{\bm{s}}_0,$ where $\bar{\bm{s}}_0 \triangleq \bar{\bm{b}}_0 \bm{G}_0$ is the contribution to the modulated vector from input bipolar symbols $\bar{b}_0[k,l].$ As in (\ref{eq:ML Detection Rate-5/4 NSM}) for the rate-$5/4$ \gls{nsm}, the most complex one-shot \gls{ml} detection algorithm, can be stated as
\begin{equation} \label{eq:ML Detection Rate-13/9 NSM}
    \hat{\bar{\bm{b}}} = \arg \min_{\bar{\bm{b}}} \| \bar{\bm{r}} - \bar{\bm{s}} \|^2,
\end{equation}
where $\bar{\bm{r}}$ stands for the vector received on a Gaussian channel, as a result of the transmission of modulated vector $\bar{\bm{s}}.$ This algorithm can be replaced with the following simpler two-step algorithm.

Following (\ref{eq:First Step Detection Algorithm Rate-5/4 NSM}), we begin by deciding on $\bar{\bm{b}}_0$ via
\begin{equation}
    \hat{\bar{\bm{b}}}_0 = \arg \min_{\bar{\bm{b}}_0} \left( \min_{\bar{\bm{b}}_1} \| \bar{\bm{r}} - \bar{\bm{s}} \|^2 \right).
\end{equation}
The simplified nature of this approach stems from the fact that the minimization of $\| \bar{\bm{r}} - \bar{\bm{s}} \|^2$ with respect to $\bar{\bm{b}}_1$ is straightforward and leads to a closed form expression as a function of $\bar{\bm{b}}_0,$ as in (\ref{eq:Closed Form Expression Metric Rate-5/4 NSM}). More specifically, because the input data vector $\bar{\bm{b}}_1$ involves components in the bipolar set $\{ \pm 1 \},$ we may write
\begin{equation}
\min_{\bar{\bm{b}}_1} \| \bar{\bm{r}} - \bar{\bm{s}} \|^2 = \min_{\bar{\bm{b}}_1} \| \bar{\bm{r}} - \bar{\bm{s}}_0 - \bar{\bm{b}}_1\|^2 = \sum_{k=0}^8  \left( |\bar{r}[k]-\bar{s}_0[k]|-1 \right)^2, 
\end{equation}
where $\bar{r}[k]$ and $\bar{s}_0[k],$ $0 \le k \le 8,$ are the components of vectors $\bar{\bm{r}}$ and $\bar{\bm{s}}_0,$ respectively. 
Summing up, we can write the decision on $\bar{\bm{b}}_0$ as
\begin{equation} \label{eq:Closed Form Expression Metric Rate-13/9 NSM}
    \hat{\bar{\bm{b}}}_0 = \arg \min_{\bar{\bm{b}}_0} \sum_{k=0}^8  \left( |\bar{r}[k]-\bar{s}_0[k]|-1 \right)^2. 
\end{equation}

As in (\ref{eq:Last Step ML Decision Rate-5/4 NSM}) for the rate-$5/4$ \gls{nsm}, after deciding on $\bar{\bm{b}}_0,$ we proceed to the decision on the $9$ components, $\bar{b}_1[k],$ of $\bar{\bm{b}}_1,$ as
\begin{equation} \label{eq:Last Step ML Decision Rate-13/9 NSM}
    \hat{\bar{b}}_1[k] = \operatorname{sgn} \left( \bar{r}[k]-\hat{\bar{s}}_0[k] \right), \; 0 \le k \le 8,
\end{equation}
where $\hat{\bar{s}}_0[k] \triangleq \hat{\bar{\bm{b}}}_0 \bm{G}_0.$

Before going to the \gls{ber} simulation results for multiple finite squared-extent \glspl{nsm}, it is useful to note that (\ref{eq:Closed Form Expression Metric Rate-13/9 NSM}), which needs the computation of $16$ metrics and the determination of the related minimum, is the most involved step in the proposed two-step \gls{ml} algorithm. Notice that the one-shot \gls{ml} detection approach, on the other hand, necessitates the computation of $2^{13} = 8192$ metrics and the determination of their minimum value. 

For alternative values of the parameter $I,$ which specifies the square extent of the suggested \gls{nsm}'s practical implementations, vector $\bar{\bm{b}}$ has $I^2+(I-1)^2$ dimensions, but vector $\bar{\bm{b}}_0$ has only $(I-1)^2$ dimensions. As a result, the suggested two-step \gls{ml} detection approach requires only $2^{(I-1)^2}$ metric computations and related comparisons, whereas the one-shot \gls{ml} detection algorithm requires $2^{I^2+(I-1)^2}$ metric computations and related comparisons. As a result, the complexity of the one-shot \gls{ml} detection algorithm grows faster than the square of the complexity of the two-step proposed \gls{ml} detection algorithm.

\begin{figure}[!htbp]
    \centering
    \includegraphics[width=1.0\textwidth]{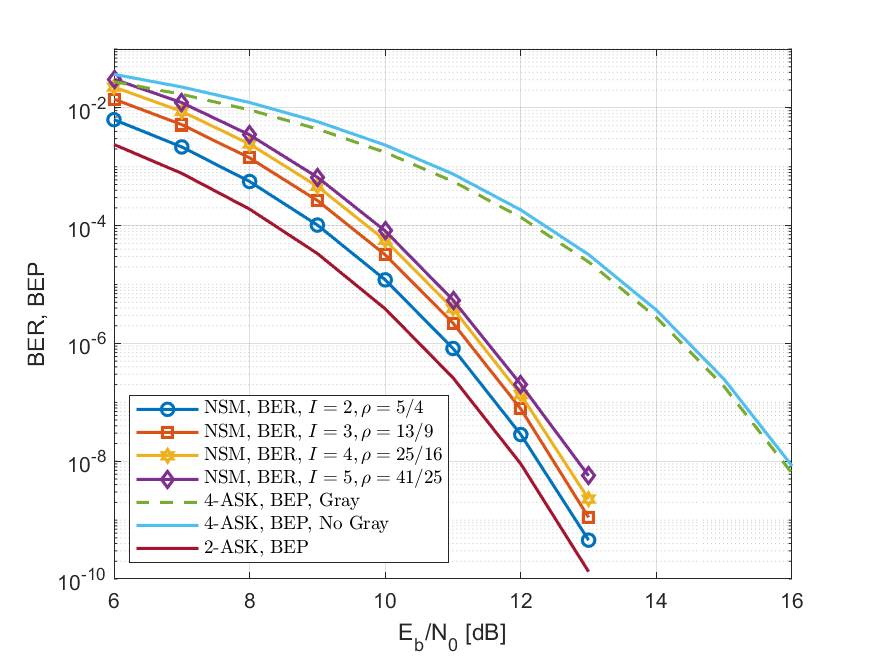}
    \caption{BER of the 2D NSMs of rates $\rho = 5/4,13/9, 25/16$ and $41/25,$ corresponding to $I=2, 3, 4$ and $5,$ respectively. For reference, the BEPs of $2$-ASK and Gray and non-Gray precoded $4$-ASK conventional modulations are presented.}
    \label{fig:BER-BEP-Two-Dimensional-NSM}
\end{figure}

To wrap up this subsection, we display in Figure~\ref{fig:BER-BEP-Two-Dimensional-NSM} the \gls{ber} curves related to the \gls{2d} \glspl{nsm} that were previously reported. For these \glspl{nsm}, the \gls{2d} extent parameter, $I,$ belongs to the set $\{2, 3, 4, 5\}$ and the consequent rate, $\rho,$ to the set $\{ 5/4, 13/9, 25/16, 41/25 \}.$ It should be noted that Subsection~\ref{ssec:Minimum Euclidean Distance Guaranteeing 5/4-NSM} provides a full analysis of the rate-$5/4$ \gls{nsm}, and that Figure~\ref{fig:BER-BEP-NSM-5_4} is the first to show both its \gls{ber} and a \gls{bep} approximation, with the latter being derived in Appendix~\ref{app:Tight Estimate BEP Rate 5/4} and given in (\ref{eq:BEP Estimate Rate 5/4}). As shown in Figure~\ref{fig:BER-BEP-Two-Dimensional-NSM}, all \gls{2d} \gls{nsm} \gls{ber} curves perform admirably with regard to $4$-ASK, even for low values of the \gls{snr}. However, despite being close to the \gls{bep} curve of $2$-ASK, they exhibit a monotonic performance degradation as the extent parameter $I$ increases from $2$ to $5.$ This tendency is expected because, as previously mentioned, degeneracy is exacerbated and amplified as the extent, $I,$ increases. Next, w provide a more extensive set of \gls{2d} \glspl{nsm} that have the advantage of avoiding this degeneracy. 

\paragraph*{\textbf{Extended and generalized NSM constructions}}

In the previous subsection, we conducted a detailed analysis of rate-$2$ \gls{2d} \glspl{nsm} characterized by a compact footprint for the filter $h_0[k,l].$ We found that this family of \glspl{nsm} achieves the \gls{msed} of $2$-ASK. Moreover, filter $h_0[k,l]$ was shown to fall into two distinct equivalence classes, each associated with different multiplicities for the \gls{msed}. Importantly, all \glspl{nsm} within a given equivalence class exhibit identical performance in terms of binary error probability—implying that, overall, there are only two distinct performance profiles for this family of \glspl{nsm}. Additionally, due to the tightness property discussed earlier, the \gls{msed} error events exhibit certain degeneracies, which result in a notable increase in their multiplicities in the binary error probability bounds. This increase affects both equivalence classes, though the degeneracy—and hence the resulting multiplicity—is more pronounced in one class than in the other.

To reduce multiplicity, two complementary approaches can be considered, either independently or in combination. The first approach maintains the same filter, $h_0[k,l],$ footprint size—as in the previous cases where a $2\times2$ footprint, corresponding to a size of $4,$ is used—but replaces the compact footprint with an irregular, non-compact one. This irregularity makes it more difficult for many of the unnecessary or superfluous minimum Euclidean distance error events to occur. While this method asymptotically achieves a rate of $2,$ in practice it results in a significant reduction in the effective rate, when applied to a finite subset of the \gls{2d} grid. As a result, this approach should be avoided in practical scenarios with finite extent, particularly when the extent is kept small to limit detection complexity.

The second approach avoids the main drawback of the first by preserving the compact nature of the filter $h_0[k,l],$ footprint. Instead of introducing footprint irregularity, this method reduces the multiplicity of minimum Euclidean distance error events by increasing the size of the compact footprint. While enlarging the footprint leads to a decrease in \gls{nsm} rate for a given fixed \gls{2d} grid extent, this reduction remains moderate and acceptable, due to the retained compactness. More importantly, this approach offers significant potential for mitigating—and in some cases, entirely eliminating—the degeneracies introduced by the tightness phenomenon.

In what follows, we specifically adopt the second approach to reduce—or ideally eliminate—the degeneracy responsible for the increased multiplicity of minimum Euclidean distance error events. Our ultimate objective is to ensure that the dominant term in the binary error probability bound aligns with that of $2$-ASK, thereby achieving equivalent performance at moderate to high \glspl{snr}. To this end, we concentrate on $3 \times 3$ square \gls{2d} patterns, which strike a favorable balance between maintaining a reasonable \gls{nsm} rate—when applied to finite \gls{2d} grid extents—and effectively lowering the multiplicity of the dominant error term. As will be demonstrated through numerical and simulation results, there is generally no need to consider larger pattern sizes, since the $3 \times 3$ configuration already enables us, in most cases, to match the leading error probability term of $2$-ASK.

We define our $3 \times 3$ square-shaped footprint filters, $h_0[k,l],$ as being zero outside the \gls{2d} region, $0 \le k,l \le 2,$ and specified within this region by a $1 \times 9$ filter pattern vector $\bm{\pi}_0.$ Starting from a suitably chosen admissible pattern, $\bm{\pi}_0,$ we generate multiple candidate scaled filters, $\mathring{h}_0[k,l],$ by mapping the nine components of $\bm{\pi}_0$ to the nine corresponding positions of the \gls{2d} footprint. This is done through random permutations of the components of $\bm{\pi}_0,$ along with random sign changes, producing a diverse set of candidate scaled filters, $\mathring{h}_0[k,l].$

For a candidate $1 \times 9$ filter pattern $\bm{\pi}_0$ to be admissible, several conditions must be satisfied. First, its components—assumed, without loss of generality, to be non-negative—must be integers, since the taps of any scaled candidate filter, $\mathring{h}_0[k,l],$ are required to be integer-valued as well.

Second, to ensure the possibility of achieving $2$-ASK performance asymptotically, the Euclidean norms of filters $\mathring{h}_m[k,l],$ $m=0,1,$ must be equal. Given that the Euclidean norm of $\mathring{h}_0[k,l],$ is identical to that of $\bm{\pi}_0,$ and assuming that filter $\mathring{h}_0[k,l],$ is given by $\mathring{h}_m[k,l] = \| \bm{\pi}_0 \| \delta[k] \delta[l],$ it follows that the norm of, $\| \bm{\pi}_0 \|$ of candidate filter pattern vector $\bm{\pi}_0$ must also be integer.

Furthermore, in order for the \gls{nsm} associated with candidate filter $\mathring{h}_0[k,l]$ to preserve the \gls{msed} of $2$-ASK, it is necessary that the maximum component of $\bm{\pi}_0$— which corresponds to its infinity norm, $\|\bm{\pi}_0 \|_\infty$— of $\bm{\pi}_0,$ be less than or equal to half the Euclidean norm, $\|\bm{\pi}_0 \|.$

Lastly, to prevent the degeneracy effects associated with the tightness property—which were shown to be responsible for the large multiplicity of minimum Euclidean distance error events in previously studied \gls{2d} \glspl{nsm}—it is strongly advisable to enforce the strict inequality $\|\bm{\pi}_0 \|_\infty < \|\bm{\pi}_0 \|/2.$ Enforcing this strict inequality is crucial to prevent the degeneracy that leads to a dramatic increase in the multiplicity of minimum-distance error events.

It is worth highlighting that, in light of the pattern-based framework introduced above, the basic rate-$2$ \glspl{nsm} studied previously can now be recognized as being associated with the specific pattern $\bm{\pi}_0=(1,1,1,1).$ For this pattern, the strict inequality, $\|\bm{\pi}_0 \|_\infty < \|\bm{\pi}_0 \|/2,$ does not hold; instead, equality is met. This tightness condition is precisely what accounts for the high degeneracy and the resulting large multiplicity of minimum-distance error events observed in those basic \glspl{nsm}.

For illustrative purposes, we consider candidate pattern vectors, $\bm{\pi}_0,$ with non-negative integer components, ranging from $0$ to $4.$ This restricted range is chosen solely for simplicity and does not represent a fundamental constraint. By allowing a broader range of integer values, we can generate a significantly larger set of candidate patterns, each capable of producing multiple \gls{2d} candidate scaled filters $\mathring{h}_0[k,l].$ Increasing the pool of candidate filters enhances our chances of identifying better ones and, consequently, constructing more powerful \glspl{nsm}. As will be shown through numerical and simulation results, even within this limited range, there exist promising patterns that can lead to highly effective \glspl{nsm} through optimal selection of their associated filters $\mathring{h}_0[k,l].$

When restricting ourselves to the range $[0,4]$ and ensuring that all admissibility constraints are satisfied, we identify three candidate pattern vectors, $\bm{\pi}_0,$ namely $(1,1,1,1,1,1,1,1,1),$ $(3,3,3,3,3,1,1,1,1)$ and $(4,4,4,4,3,3,3,3,0).$

For the first candidate pattern, $\bm{\pi}_0=(1,1,1,1,1,1,1,1,1),$ we have $\| \bm{\pi}_0 \| = 3,$ and $\| \bm{\pi}_0 \|_\infty = 1,$ so the strict non-tightness inequality, $\|\bm{\pi}_0 \|_\infty < \|\bm{\pi}_0 \|/2,$ is met. In this case, the scaled filter $\mathring{h}_1[k,l]$ is fully determined as $\mathring{h}_1[k,l]=3 \, \delta[k]\delta[l],$ while the scaled filter $\mathring{h}_0[k,l]$ has nine non-zero taps taking values in the bipolar set $\{\pm 1 \}.$ For the second candidate pattern $\bm{\pi}_0=(3,3,3,3,3,1,1,1,1),$ we find $\| \bm{\pi}_0 \| = 7,$ while $\| \bm{\pi}_0 \|_\infty = 3,$ again satisfying the strict non-tightness inequality. Here, $\mathring{h}_1[k,l]=7 \,\delta[k]\delta[l]$ and $\mathring{h}_0[k,l]$ consists of five non-zero taps in the set $\{\pm 3\},$ along with four non-zero taps in the set $\{\pm 1\}.$ Finally, for $\bm{\pi}_0=(4,4,4,4,3,3,3,3,0),$ we obtain $\| \bm{\pi}_0 \| = 10$ and $\| \bm{\pi}_0 \|_\infty = 4,$ so the non-tightness condition is still satisfied. The corresponding scaled filter, $\mathring{h}_1[k,l],$ is given by $\mathring{h}_1[k,l]=10 \, \delta[k]\delta[l],$ while the other scaled filter, $\mathring{h}_0[k,l]$ includes four non-zero values in the set $\{\pm 4\}$ and four additional non-zero values in the set $\{\pm 3\}.$

Tables~\ref{table:2D NSMs Filter Pattern (1, 1, 1, 1, 1, 1, 1, 1, 1)}, \ref{table:2D NSMs Filter Pattern (3, 3, 3, 3, 3, 1, 1, 1, 1)} and~\ref{table:2D NSMs Filter Pattern (4, 4, 4, 4, 3, 3, 3, 3, 0)} provide a full characterization of the patterns $\bm{\pi}_0=(1,1,1,1,1,1,1,1,1),$ $(3,3,3,3,3,1,1,1,1)$ and $(4,4,4,4,3,3,3,3,0),$ respectively. As indicated at the beginning of each table, the corresponding numbers of candidate filters, $\mathring{h}_0[k,l]$ are $512,$ for Table~\ref{table:2D NSMs Filter Pattern (1, 1, 1, 1, 1, 1, 1, 1, 1)}, $64512,$ for Table~\ref{table:2D NSMs Filter Pattern (3, 3, 3, 3, 3, 1, 1, 1, 1)}, and $161280,$ for Table~\ref{table:2D NSMs Filter Pattern (4, 4, 4, 4, 3, 3, 3, 3, 0)}. Typically, optimizing \glspl{nsm} performance requires evaluating each candidate filter, $\mathring{h}_0[k,l],$ individually—assessing its distance spectrum, \gls{tf}, or an upper bound on its error probability. However, given the large number of candidates—particularly in the latter two cases—this exhaustive search becomes highly time- and computation-intensive. To mitigate this complexity, we introduce equivalence relations between candidate filters that yield identical performance in terms of the relevant performance metrics. By grouping filters into such equivalence classes, we can focus the analysis on a significantly smaller set of class representatives. This approach not only preserves the accuracy of the optimization process but also dramatically reduces the computational burden, as the number of equivalence classes is typically smaller by at least an order of magnitude.

\begin{table}[H]

\caption{Selected 2D NSMs, with simple filters' coefficients, with common filter $\bm{h}_0$ pattern $\bm{\pi}_0 = (1, 1, 1, 1, 1, 1, 1, 1, 1),$ arranged in a decreasing performance order, starting from the best one. Relative degradation in performance, due to distance reduction or/and multiplicity increase, when moving from one filter to the next, is emphasized in bold. 
} \label{table:2D NSMs Filter Pattern (1, 1, 1, 1, 1, 1, 1, 1, 1)}

\centering
\begin{tabular}{|c|c|} 
\hline

\multicolumn{2}{|l|}{\# of simple $h_0[k,l]$ filters $ = 512$, \# of non equivalent $h_0[k,l]$ filters $ = 19$} \\ \hline
\multicolumn{2}{|l|}{\# of non equivalent $h_0[k,l]$ filters with ($\dot{T}_H(D)=\dot{T}_V(D)$) $ = 10$} \\ \hline
\multicolumn{2}{|l|}{\# of non equivalent $h_0[k,l]$ filters with ($\dot{T}_H(D)=\dot{T}_V(D)$) \& ($\dot{T}_{MD}(D)=\dot{T}_{SD}(D)$) $ = 4$} \\ \hline
\multicolumn{2}{|l|}{$\mathring{h}_1[k,l] = 3 \, \delta[k]\delta[l]$} \\ \hline
\multicolumn{2}{|l|}{\gls{msed}, $d_\text{min}^2 = 36$} \\ [0.5ex]
\hline\hline

\multicolumn{2}{|c|}{Filter~\#$1$} \\ \hline
\multirow{2}{*}{$\mathring{h}_0[k,l]$} & $\delta[k]\delta[l]+\delta[k-1]\delta[l]-\delta[k-2]\delta[l]+\delta[k]\delta[l-1] -\delta[k-1]\delta[l-1]$  \\
& $- \delta[k-2]\delta[l-1] - \delta[k]\delta[l-2] - \delta[k-1]\delta[l-2] + \delta[k-2]\delta[l-2]$ \\
\hline

\end{tabular}
\end{table}

\begin{table}[H]

\captionsetup{list=no} 
\caption*{Table~\ref{table:2D NSMs Filter Pattern (1, 1, 1, 1, 1, 1, 1, 1, 1)} (Continued).}
\centering
\begin{tabular}{|c|c|} 
\hline

\multicolumn{2}{|c|}{Filter~\#$4$} \\ \hline
$\mathring{h}_0[k,l]$ & $(\delta[k]+\delta[k-1]+\delta[k-2])(\delta[l]+\delta[l-1] +\delta[l-2])$  \\ \hline
$\mathring{\bm{h}}_0$ & $\begin{pmatrix}
1 & 1 & 1 \\
1 & 1 & 1 \\
1 & 1 & 1
\end{pmatrix}$ \\ [0.5ex] \hline
\multirow{2}{*}{$\dot{T}_{H,V}(D)$} & 
$\frac{\bm{16}}{\bm{9}}\,D^{24}+\frac{599}{36}\,D^{36}+\frac{28087}{432}\,D^{48}+\frac{5701409}{20736}\,D^{60}+\frac{9873844279}{9331200}\,D^{72}+\frac{377124913061}{99532800}\,D^{84}$ \\
& $+\frac{723691319633083}{53747712000}\,D^{96}+\frac{40447523750736953}{859963392000}\,D^{108}+\frac{520011991761469387}{3224862720000}\,D^{120} + \cdots$ \\ [0.5ex] 
\hline
\multirow{2}{*}{$\dot{T}_{MD,SD}(D)$} & 
$6\,D^{36}+D^{40}+9\,D^{48}+\frac{33}{4}\,D^{52}+\frac{3}{16}\,D^{56}+27\,D^{60}+\frac{247}{8}\,D^{64}+\frac{147}{32}\,D^{68}+\frac{19381}{256}\,D^{72}$ \\
& $+\frac{1591}{16}\,D^{76}+\frac{5165}{128}\,D^{80}+\frac{36487}{256}\,D^{84}+\frac{1336999}{4096}\,D^{88}+\frac{12579}{64}\,D^{92}+\frac{797359}{2048}\,D^{96} + \cdots$ \\ [1ex] 
 \hline

\end{tabular}
\end{table}

\begin{table}[H]
\caption{Selected 2D NSMs, with simple filters' coefficients, with common filter $\bm{h}_0$ pattern $\bm{\pi}_0 = (3, 3, 3, 3, 3, 1, 1, 1, 1),$ arranged in a decreasing performance order, starting from the best one. Relative degradation in performance, due to distance reduction or/and multiplicity increase, when moving from one filter to the next, is emphasized in bold.}
\label{table:2D NSMs Filter Pattern (3, 3, 3, 3, 3, 1, 1, 1, 1)}
\centering
\begin{tabular}{|c|c|} 
\hline

\multicolumn{2}{|l|}{\# of simple $h_0[k,l]$ filters $ = 64512$, \# of non equivalent $h_0[k,l]$ filters $ = 1118$} \\ \hline
\multicolumn{2}{|l|}{\# of non equivalent $h_0[k,l]$ filters with ($\dot{T}_H(D)=\dot{T}_V(D)$) $ = 100$} \\ \hline
\multicolumn{2}{|l|}{\# of non equivalent $h_0[k,l]$ filters with ($\dot{T}_H(D)=\dot{T}_V(D)$) \& ($\dot{T}_{MD}(D)=\dot{T}_{SD}(D)$) $ = 8$} \\ \hline
\multicolumn{2}{|l|}{$\mathring{h}_1[k,l] = 7\, \delta[k]\delta[l]$} \\ \hline
\multicolumn{2}{|l|}{\gls{msed}, $d_\text{min}^2 = 196$} \\ [0.5ex]
\hline\hline

\multicolumn{2}{|c|}{Filter~\#$1$} \\ \hline
\multirow{2}{*}{$\mathring{h}_0[k,l]$} & $3\,\delta[k]\delta[l]+\delta[k-1]\delta[l]-3\,\delta[k-2]\delta[l]+\delta[k]\delta[l-1] -3\,\delta[k-1]\delta[l-1]$  \\
& $- \delta[k-2]\delta[l-1] - 3\,\delta[k]\delta[l-2] - \delta[k-1]\delta[l-2] + 3\,\delta[k-2]\delta[l-2]$ \\
\hline

\end{tabular}
\end{table}

\begin{table}[H]
\captionsetup{list=no} 
\caption*{Table~\ref{table:2D NSMs Filter Pattern (3, 3, 3, 3, 3, 1, 1, 1, 1)} (Continued).}
\centering
\begin{tabular}{|c|c|} 
\hline

\multicolumn{2}{|c|}{Filter~\#$4$} \\ \hline
\multirow{2}{*}{$\mathring{h}_0[k,l]$} & $3\,\delta[k]\delta[l]+\delta[k-1]\delta[l]+3\,\delta[k-2]\delta[l]+\delta[k]\delta[l-1] +3\,\delta[k-1]\delta[l-1]$  \\
& $+\delta[k-2]\delta[l-1] + 3\,\delta[k]\delta[l-2] + \delta[k-1]\delta[l-2] + 3\,\delta[k-2]\delta[l-2]$ \\ \hline
$\mathring{\bm{h}}_0$ & $\begin{pmatrix}
3 & 1 & 3 \\
1 & 3 & 1 \\
3 & 1 & 3
\end{pmatrix}$ \\ [0.5ex] \hline
\multirow{2}{*}{$\dot{T}_{H,V}(D)$} & 
$4\,D^{196}+5\,D^{224}+D^{240}+D^{\bm{248}}+\frac{15}{2}\,D^{252}+\frac{1}{2}\,D^{264}+\frac{9}{2}\,D^{268}+3\,D^{276}+5\,D^{280}$ \\
& $+\frac{17}{16}\,D^{284}+\frac{27}{8}\,D^{292}+\frac{15{}{2}}\,D^{296}+3\,D^{304}+\frac{35}{16}\,D^{308}+\frac{17}{4}\,D^{312}+\frac{5}{8}\,D^{316}+\frac{63}{8}\,D^{320} + \cdots$ \\ [0.5ex] 
\hline
\multirow{2}{*}{$\dot{T}_{MD,SD}(D)$} & 
$6\,D^{196}+5\,D^{224}+D^{232}+\frac{15}{2}\,D^{252}+\frac{9}{2}\,D^{260}+\frac{1}{2}D^{272}+5\,D^{280}+\frac{15}{2}\,D^{288}+\frac{15}{8}\,D^{300}$ \\
& $+\frac{25}{16}\,D^{308}+\frac{25}{4}\,D^{316}+D^{320}+\frac{3}{4}\,D^{324}+\frac{45}{16}\,D^{328}+\frac{67}{16}\,D^{336}+\frac{3}{4}\,D^{340}+\frac{45}{16}\,D^{344} + \cdots$ \\
\hline

 \end{tabular}
\end{table}

\begin{table}[H]

\caption{Selected 2D NSMs, with simple filters' coefficients, with common filter $\bm{h}_0$ pattern $\bm{\pi}_0 = (4, 4, 4, 4, 3, 3, 3, 3, 0),$ arranged in a decreasing performance order, starting from the best one. Relative degradation in performance, due to distance reduction or/and multiplicity increase, when moving from one filter to the next, is emphasized in bold.}
\label{table:2D NSMs Filter Pattern (4, 4, 4, 4, 3, 3, 3, 3, 0)}
\centering
\begin{tabular}{|c|c|} 
\hline

\multicolumn{2}{|l|}{\# of simple $h_0[k,l]$ filters $ = 161280$, \# of non equivalent $h_0[k,l]$ filters $ = 2618$} \\ \hline
\multicolumn{2}{|l|}{\# of non equivalent $h_0[k,l]$ filters with ($\dot{T}_H(D)=\dot{T}_V(D)$) $ = 80$} \\ \hline
\multicolumn{2}{|l|}{\# of non equivalent $h_0[k,l]$ filters with ($\dot{T}_H(D)=\dot{T}_V(D)$) \& ($\dot{T}_{MD}(D)=\dot{T}_{SD}(D)$) $ = 10$} \\ \hline
\multicolumn{2}{|l|}{$\mathring{h}_1[k,l] = 10\, \delta[k]\delta[l]$} \\ \hline
\multicolumn{2}{|l|}{\gls{msed}, $d_\text{min}^2 = 400$} \\ [0.5ex]
\hline\hline

\multicolumn{2}{|c|}{Filter~\#$1$} \\ \hline
\multirow{2}{*}{$\mathring{h}_0[k,l]$} & $4\,\delta[k]\delta[l]+3\,\delta[k-1]\delta[l]+4\,\delta[k-2]\delta[l]+3\,\delta[k]\delta[l-1] - 3\,\delta[k-2]\delta[l-1]$  \\
& $ - 4\,\delta[k]\delta[l-2] + 3\,\delta[k-1]\delta[l-2] - 4\,\delta[k-2]\delta[l-2]$ \\
\hline

\end{tabular}
\end{table}

\begin{table}[H]

\captionsetup{list=no} 
\caption*{Table~\ref{table:2D NSMs Filter Pattern (4, 4, 4, 4, 3, 3, 3, 3, 0)} (Continued).}
\centering
\begin{tabular}{|c|c|} 
\hline

\multicolumn{2}{|c|}{Filter~\#$4$} \\ \hline
\multirow{2}{*}{$\mathring{h}_0[k,l]$} & $3\,\delta[k]\delta[l]+4\,\delta[k-1]\delta[l]+3\,\delta[k-2]\delta[l]+4\,\delta[k]\delta[l-1] - 4\,\delta[k-2]\delta[l-1]$  \\
& $ - 3\,\delta[k]\delta[l-2] + 4\,\delta[k-1]\delta[l-2] - 3\,\delta[k-2]\delta[l-2]$ \\ \hline
$\mathring{\bm{h}}_0$ & $\begin{pmatrix}
3 & 4 & -3 \\
4 & 0 & 4 \\
3 & -4 & -3
\end{pmatrix}$ \\ [0.5ex] \hline
\multirow{2}{*}{$\dot{T}_{H,V}(D)$} & 
$4\,D^{400}+5\,D^{480}+\frac{3}{8}\,D^{\bm{496}}+\frac{1}{8}\,D^{512}+\frac{5}{128}\,D^{528}+\frac{195}{256}\,D^{544}+\frac{22535}{2048}\,D^{560}+\frac{897}{1024}\,D^{576}$ \\
& $+\frac{9225}{32768}\,D^{592}+\frac{5637}{65536}\,D^{608}+\frac{2176011}{524288}\,D^{624}+\frac{11669251\,D^{640}}{524288}+\frac{29967373\,D^{656}}{8388608} + \cdots$ \\ [0.5ex] 
\hline
\multirow{2}{*}{$\dot{T}_{MD,SD}(D)$} & 
$6\,D^{400}+4\,D^{480}+10\,D^{560}+18\,D^{640}+\frac{5}{8}\,D^{648}+\frac{3}{8}\,D^{712}+\frac{457}{16}\,D^{720}+\frac{5}{2}\,D^{728}+\frac{7}{32}\,D^{736}$ \\
& $+\frac{9}{4}\,D^{792}+\frac{299}{4}\,D^{800}+\frac{177}{16}\,D^{808}+\frac{1}{2}\,D^{816}+\frac{9}{128}\,D^{824}+\frac{9}{128}\,D^{864}+\frac{151}{16}\,D^{872} + \cdots$ \\
\hline

 \end{tabular}
\end{table}

To define equivalence classes of normalized $\bar{h}_0[k,l]$ filters that yield identical \gls{nsm} performance, we revisit the \gls{2d} transformations introduced earlier in this subsection for illustrative rate-$2$ \glspl{nsm}. These transformations, which preserve binary error probability, form the basis of the equivalence relation. First, sign inversion—replacing $\bar{h}_0[k,l]$ with $-\bar{h}_0[k,l]$—has no impact on performance. Additionally, the three alternate sign transformations, $\tilde{h}_0[k,l] = (-1)^k \bar{h}_0[k,l],$ $\tilde{h}_0[k,l] = (-1)^l \bar{h}_0[k,l]$ and $\tilde{h}_0[k,l] = (-1)^{k+l} \bar{h}_0[k,l],$ produce equivalent \glspl{nsm}. Translations, $\bar{h}_0[k-\Delta k,l-\Delta l],$ of $\bar{h}_0[k,l]$ by $(\Delta k, \Delta l),$ across the \gls{2d} grid also preserve performance. Furthermore, symmetry operations with respect to vertical, horizontal, and diagonal axes—resulting in transformed filters $\bar{h}_0[-k,l],$ $\bar{h}_0[k,-l],$ $\bar{h}_0[l,k]$ and $\bar{h}_0[-l,-k],$—do not alter \gls{nsm} characteristics. Finally, rotations by $\pm \pi/2$ and $\pi$ in the grid maintain equivalence. These transformations collectively define the equivalence classes, enabling performance assessment to be limited to a smaller set of representative filters.
   
These equivalence-preserving transformations significantly reduce the number of candidate filters that must be considered during \glspl{nsm} optimization. As shown at the beginning of Tables~\ref{table:2D NSMs Filter Pattern (1, 1, 1, 1, 1, 1, 1, 1, 1)}, \ref{table:2D NSMs Filter Pattern (3, 3, 3, 3, 3, 1, 1, 1, 1)} and~\ref{table:2D NSMs Filter Pattern (4, 4, 4, 4, 3, 3, 3, 3, 0)}, applying the equivalence relation reduces the number of distinct filter candidates to be examined and characterized from $512$ to just $19,$ for pattern $\bm{\pi}_0=(1,1,1,1,1,1,1,1,1)$ (Table~\ref{table:2D NSMs Filter Pattern (1, 1, 1, 1, 1, 1, 1, 1, 1)}), from $64512$ to $1118,$ for pattern $\bm{\pi}_0=(3,3,3,3,3,1,1,1,1)$ (Table~\ref{table:2D NSMs Filter Pattern (3, 3, 3, 3, 3, 1, 1, 1, 1)}), and from $161280$ to $2618$ for pattern $\bm{\pi}_0=(4,4,4,4,3,3,3,3,0)$ (Table~\ref{table:2D NSMs Filter Pattern (4, 4, 4, 4, 3, 3, 3, 3, 0)}). This reduction is substantial—often more than an order of magnitude—and plays a crucial role in making the filter optimization process computationally feasible.

\begin{figure}[!htbp]
    \centering
    \includegraphics[width=1.0\textwidth]{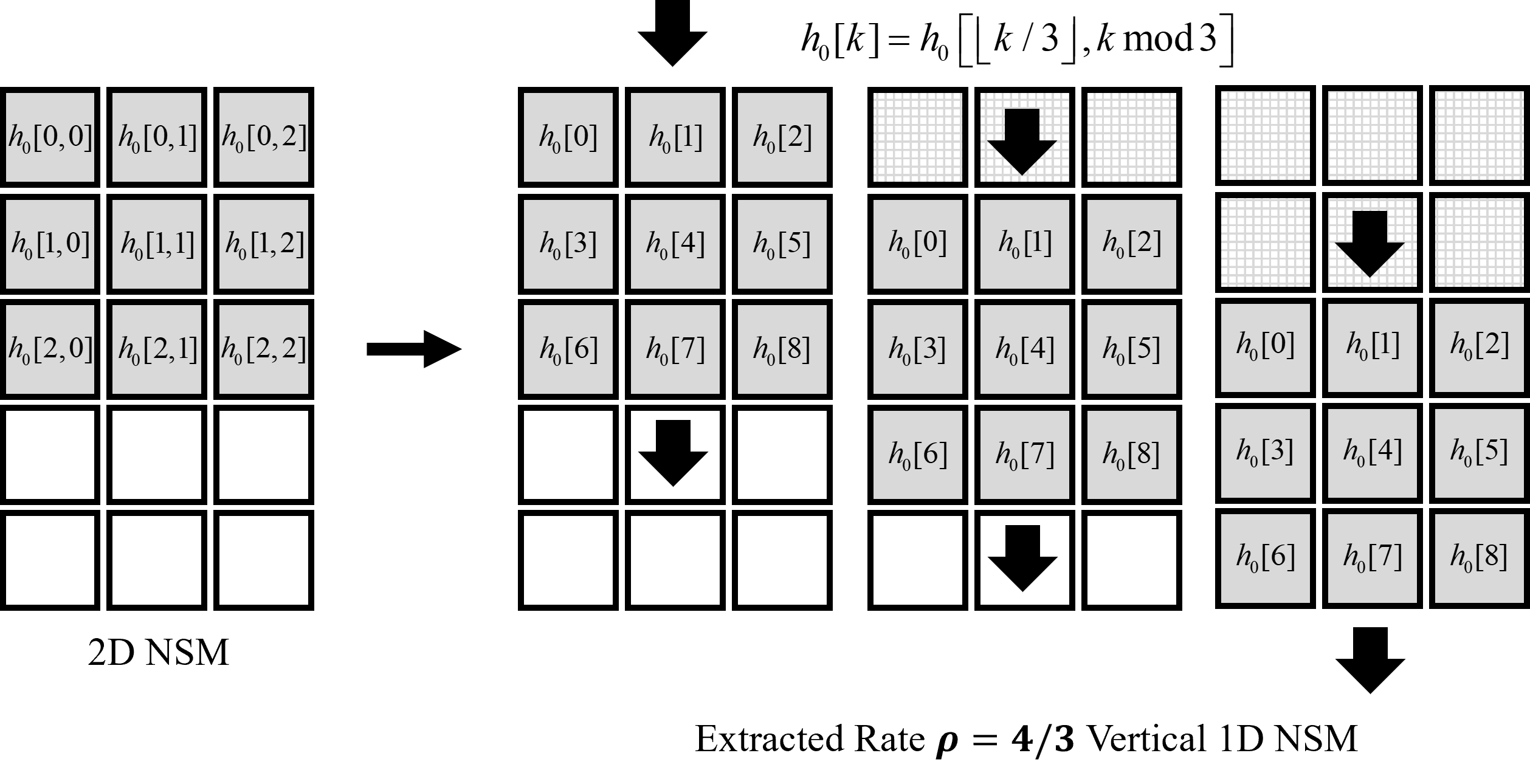}
    \caption{Successive footprints of rate $\rho = 4/3$ vertical (V) 1D NSM, extracted from rate $\rho = 2$ 2D NSM. One-dimensional filter, $h_0[k],$ expressed as $h_0[k]=h_0[\lfloor k/3 \rfloor, k \bmod{3}],$ as a function of 2D filter $h_0[k,l].$ The RTF resulting from the extracted vertical 1D NSM is denoted by $\dot{T}_V(D).$}
    \label{fig:Extracted Vertical 1D NSM}
\end{figure}

\begin{figure}[!htbp]
    \centering
    \includegraphics[width=1.0\textwidth]{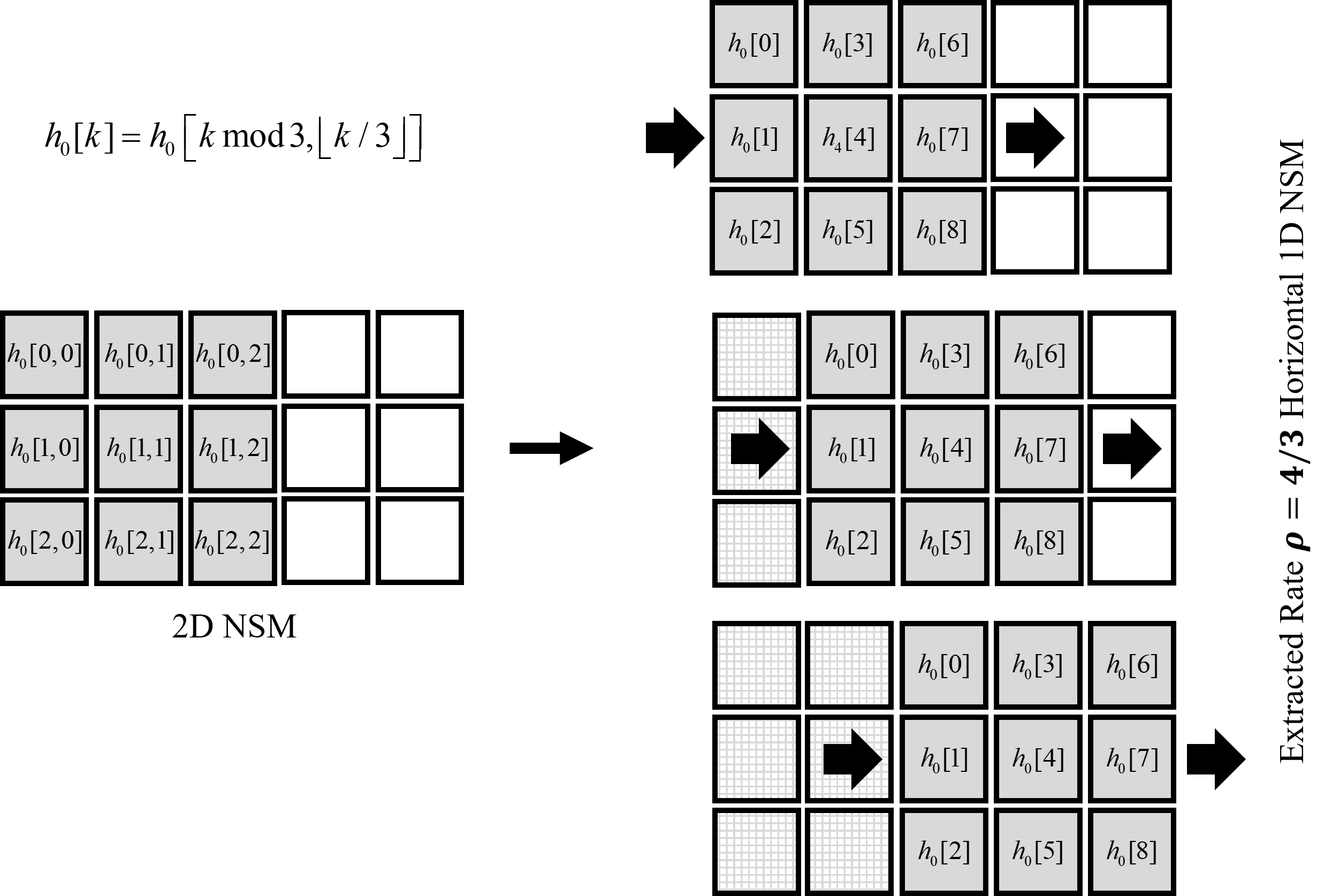}
    \caption{Successive footprints of rate $\rho = 4/3$ horizontal (H) 1D NSM, extracted from rate $\rho = 2$ 2D NSM. One-dimensional filter, $h_0[k],$ expressed as $h_0[k]=h_0[k \bmod{3}, \lfloor k/3 \rfloor],$ as a function of 2D filter $h_0[k,l].$ The RTF resulting from the extracted vertical 1D NSM is denoted by $\dot{T}_H(D).$}
    \label{fig:Extracted Horizontal 1D NSM}
\end{figure}

\begin{figure}[!htbp]
    \centering
    \includegraphics[width=1.0\textwidth]{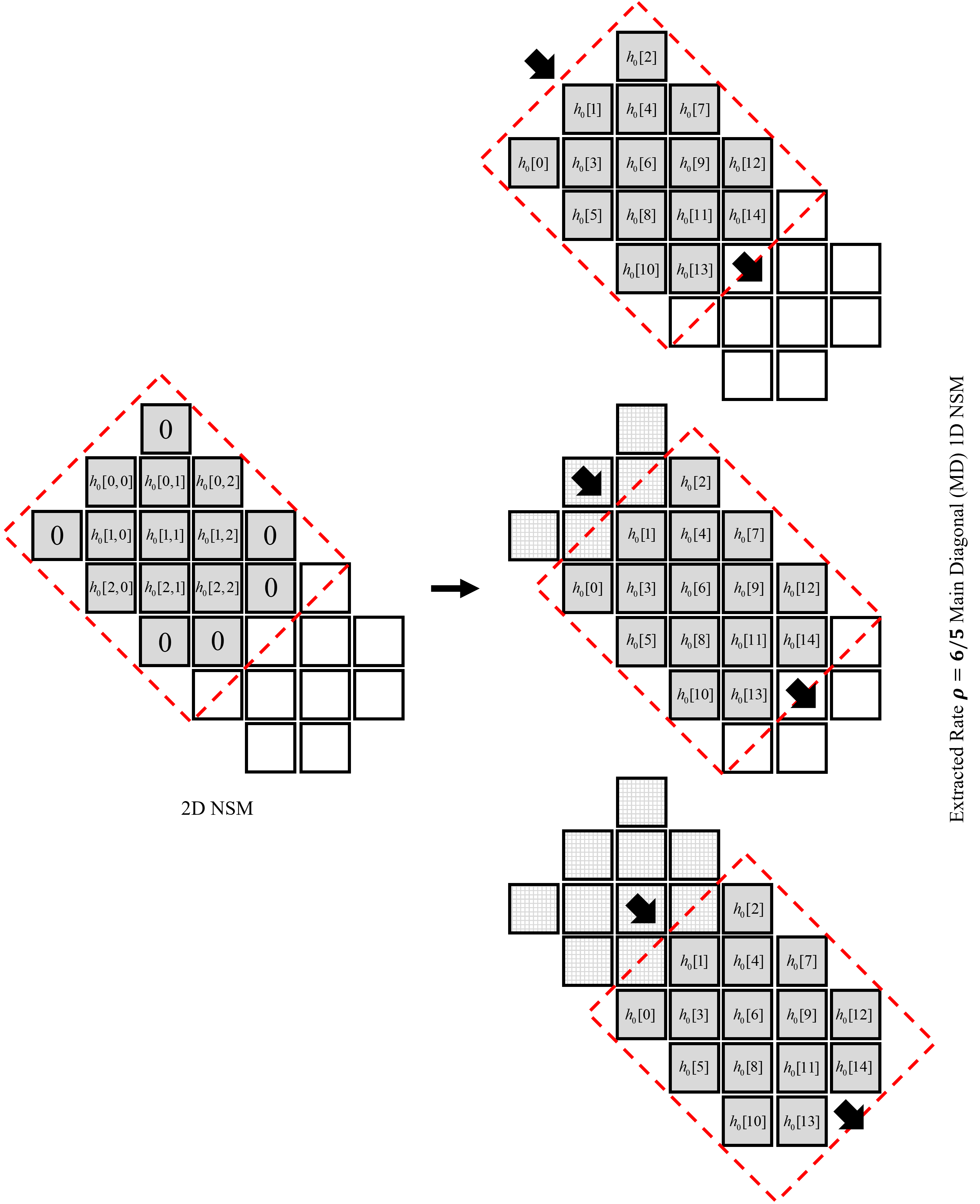}
    \caption{Successive footprints of rate $\rho = 6/5$ MD 1D NSM, extracted from rate $\rho = 2$ 2D NSM. Expressed as a function of 2D filter $h_0[k,l],$ 1D filter, $h_0[k],$ is such that $h_0[2k+3l+1]=h_0[k,l],$ for $0 \le k, l \le 2,$ and $h_0[k]=0,$ otherwise. The RTF resulting from the extracted MD 1D NSM is denoted by $\dot{T}_{MD}(D).$}
    \label{fig:Extracted Main Diagonal 1D NSM}
\end{figure}

\begin{figure}[!htbp]
    \centering
    \includegraphics[width=1.0\textwidth]{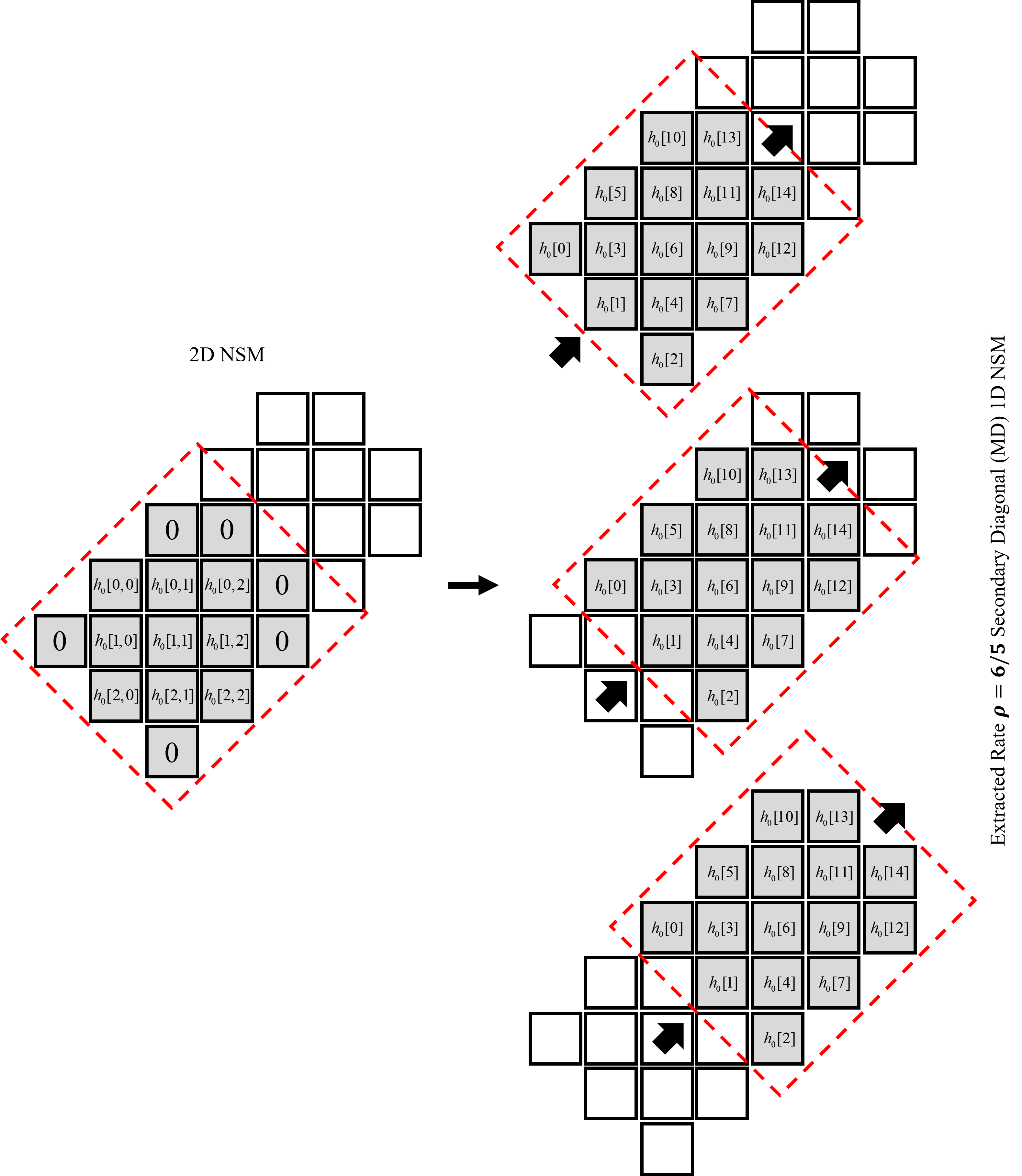}
    \caption{Successive footprints of rate $\rho = 6/5$ SD 1D NSM, extracted from rate $\rho = 2$ 2D NSM. Expressed as a function of 2D filter $h_0[k,l],$ 1D filter, $h_0[k],$ is such that $h_0[-2k+3l+5]=h_0[k,l],$ for $0 \le k, l \le 2,$ and $h_0[k]=0,$ otherwise. The RTF resulting from the extracted MD 1D NSM is denoted by $\dot{T}_{SD}(D).$}
    \label{fig:Extracted Secondary Diagonal 1D NSM}
\end{figure}

The \gls{2d} nature of the considered \glspl{nsm} prevents the use of trellis representations, rendering traditional analytical tools—such as \gls{tf}-based distance spectrum computation—inapplicable. To overcome this limitation, we systematically derive four structured \gls{1d} \glspl{nsm} from each candidate parent \gls{2d} \gls{nsm}, with each child \gls{1d} \gls{nsm} defined by a specific constraint on input sequences symbols, $\bar{b}_m[k,l],$ $m=0,1,$ positions, $k$ and $l.$ As illustrated in Figure~\ref{fig:Extracted Vertical 1D NSM}, the \emph{Vertical} (V) \gls{1d} \gls{nsm} restricts \gls{2d} non-zero input symbols to $l=0,$ yielding \gls{1d} scaled filter $\mathring{h}_0[n]=\mathring{h}_0[\lfloor n/3 \rfloor, n \bmod{3}].$ In Figure~\ref{fig:Extracted Horizontal 1D NSM}, the \emph{Horizontal} (H) \gls{1d} \gls{nsm} isolates \gls{2d} input symbols at $k=0,$ resulting in \gls{1d} scaled filters $\mathring{h}_0[n]=\mathring{h}_0[n \bmod{3}, \lfloor n/3 \rfloor].$ Figures~\ref{fig:Extracted Main Diagonal 1D NSM} and~\ref{fig:Extracted Secondary Diagonal 1D NSM} illustrate the \gls{md} and \gls{sd} \gls{1d} \glspl{nsm}, which retain inputs along $k=l$ and $k=-l,$ respectively. For the \gls{md} \gls{nsm}, the \gls{1d} filter $h_0[n]$ is defined by $h_0[n]=h_0[k,l],$ with $n=2k+3l+1,$ for $0 \le k, l \le 2,$ and $h_0[n]=0,$ otherwise. Similarly, for the \gls{sd} \gls{nsm}, $h_0[n]=h_0[k,l],$ with $n=-2k+3l+5,$ under the same index constraints, and zero elsewhere. The \gls{1d} filter $\mathring{h}_1[n] = \| \bm{\pi}_0 \| \delta[n]$ is shared across all \gls{1d} child \glspl{nsm}. However, the Vertical (V) and Horizontal (H) \gls{1d} child \glspl{nsm} operate at rate $\rho=4/3,$ while the \gls{md} and \gls{sd} \gls{1d} child \glspl{nsm} operate at a slightly reduced rate of $\rho=6/5.$ Crucially, each of these four \gls{1d} \glspl{nsm} admits a trellis representation, enabling the use of \gls{tf} techniques to partially but rigorously assess the performance of the original \gls{2d} parent \gls{nsm}.

The optimization process—aimed at identifying the best equivalence classes through their representative scaled filters $\mathring{h}_0[k,l]$—proceeds through the following steps. For each candidate \gls{2d} \gls{nsm} with filter $\mathring{h}_0[k,l],$ we derive four associated \gls{1d} child \glspl{nsm}, each with its own scaled filter $\mathring{h}_0[k].$ Since the other filter, $\mathring{h}_1[k] = \| \bm{\pi}_0 \| \delta[k]$ is the same across all four child \glspl{nsm}, we have complete information on the filters for the Vertical (V), Horizontal (H), \gls{md}, and \gls{sd} \glspl{nsm}.

To analyze these \gls{1d} child \glspl{nsm}, we apply Algorithm~\ref{alg:Simplified One-Shot Two-Step Reduced Transfer Function T(D) Rate-(Q+1)/Q NSM}—a simplified and faster version of Algorithm~\ref{alg:One-Shot Two-Step Reduced Transfer Function T(D) Rate-(Q+1)/Q NSM}—to compute truncated versions of their \glspl{rtf}. Although Algorithm~\ref{alg:One-Shot Two-Step Reduced Transfer Function T(D) Rate-(Q+1)/Q NSM} could also be used, it is more computationally intensive. Both algorithms are considered two-step processes, as they first compute the full \gls{tf} $T(N,D),$ then derive the \gls{rtf} $\dot{T}(D),$ followed by its truncated version $\dot{T}(D;P).$ We denote the \glspl{rtf} of the V, H, \gls{md}, and \gls{sd} child \glspl{nsm} as $\dot{T}_V(D),$ $\dot{T}_H(D),$ $\dot{T}_{MD}(D)$ and $\dot{T}_{SD}(D),$ respectively.

\begin{algorithm}
\caption{One-shot, \emph{two-step}, symbolic determination of \gls{rtf}, $\dot{T}(D),$ of a rate-$(Q+1)/Q$ \gls{nsm} and its truncated version $\dot{T}(D;P),$ \emph{with} prior computation of TF $T(N,D)$}\label{alg:One-Shot Two-Step Reduced Transfer Function T(D) Rate-(Q+1)/Q NSM}
\begin{algorithmic}[1]
\Require $\mathring{\bm{h}}_0 = (\mathring{h}_0[0], \mathring{h}_0[1], \ldots, \mathring{h}_0[L_0-1]),$ $\mathring{\bm{h}}_1 = (\mathring{h}_1[0]),$ $Q \ge 1$ and $Q \mid L_0$, $P$ \Comment{$L_0$ must be a multiple of $Q$, $P$ is the Taylor series truncation order}
\Ensure $\dot{T}(D),$ $\dot{T}(D;P)$

\State $k \gets 0$
\State $M \gets L_0/Q$ 
\State $\Sigma \gets \{ \bm{\sigma} = (\Delta \bar{b}_0[k-(M-2)], \ldots, \Delta \bar{b}_0[k-1], \Delta \bar{b}_0[k]), \Delta \bar{b}_0[k-m] \in \{0, \pm 2\}, 0 \le m < M-1 \}$ \Comment{$\Sigma$ is the set of states of the NSM state diagram}

\For{$\Delta \bar{b}_0[k-m] \in \{0, \pm 2\}, m=0,1,\ldots,M-1,$}

\State $\bm{\sigma}[k-1] \gets (\Delta \bar{b}_0[k-(M-1)], \ldots, \Delta \bar{b}_0[k-2], \Delta \bar{b}_0[k-1])$
\State $\bm{\sigma}[k] \gets (\Delta \bar{b}_0[k-(M-2)], \ldots, \Delta \bar{b}_0[k-1], \Delta \bar{b}_0[k])$
\State $i_0 \gets \tfrac{1}{2} |\Delta \bar{b}_0[k]|$ \Comment{Weight of the current input difference corresponding to filter $h_0[k]$}
\State $\Delta \mathring{\bm{s}}_0 = (\Delta \mathring{s}_0[kQ], \Delta \mathring{s}_0[kQ+1], \ldots, \Delta \mathring{s}_0[kQ+Q-1]) \gets \sum_{m=0}^{M-1} \Delta \bar{b}_0[k-m] (\mathring{h}_0[mQ], \mathring{h}_0[mQ+1], \ldots, \mathring{h}_0[mQ+Q-1])$
\State Let $ N, D $ be two symbolic variables
\State $L_{\bm{\sigma}[k-1]\bm{\sigma}[k]}(N,D) \gets 0$

\For{$\Delta \bar{b}_1[kQ+l] \in \{0, \pm 2\}, l=0,1,\ldots,Q-1,$}
 \State $\Delta \mathring{\bm{s}} \gets \Delta \mathring{\bm{s}}_0 + \mathring{h}_1[0](\Delta \bar{b}_1[kQ], \Delta \bar{b}_1[kQ+1], \ldots, \Delta \bar{b}_1[kQ+Q-1])$
 \State $i_1 \gets \tfrac{1}{2} \sum_{l=0}^{Q-1} |\Delta \bar{b}_1[kQ+l]|$ \Comment{Weight of the $Q$ current inputs differences corresponding to filter $h_1[k]$}
 \State $j \gets \norm{\Delta \mathring{\bm{s}}}^2$ \Comment{Squared Euclidean norm of the $Q$ current outputs differences}
 \State $L_{\bm{\sigma}[k-1]\bm{\sigma}[k]}(N,D) \gets L_{\bm{\sigma}[k-1]\bm{\sigma}[k]}(N,D) + N^{i_1} D^j$
\EndFor
\State $L_{\bm{\sigma}[k-1]\bm{\sigma}[k]}(N,D) \gets N^{i_0} L_{\bm{\sigma}[k-1]\bm{\sigma}[k]}(N,D)$
\EndFor

\State $L_{\bm{0}\bm{0}}(N,D) \gets L_{\bm{0}\bm{0}}(N,D)-1$ \Comment{Branch with null input and output differences and label $N^0 D^0 = 1$ should be discarded to keep true error events only} \label{step:last two-step algorithm common step}

\State $\bm{L}(N,D) \gets (L_{\bm{\sigma}^\prime \bm{\sigma}}(N,D))_{(\bm{\sigma}^\prime, \bm{\sigma}) \in \Sigma \times \Sigma}$ \Comment{$\bm{L}(N,D)$ is a $|\Sigma| \times |\Sigma|$ square matrix, with $|\Sigma| = 3^{M-1}$}

\State Let $\bm{T}^e(N,D) = (T_{\bm{\sigma}}^e(N,D))_{\bm{\sigma} \in \Sigma}$ be a symbolic $1 \times|\Sigma|$ row vector \Comment{$\bm{T}^e(N,D)$ is a version of $\bm{T}(N,D),$ which focuses on the ending state $\bm{\sigma}=\bm{0}_e$}

\State $\bm{T}^s(N,D) \gets \bm{T}^e(N,D)$ \Comment{$\bm{T}^s(N,D)$ is a version of $\bm{T}(N,D),$ which focuses on the starting state $\bm{\sigma}=\bm{0}_s$}

\State $T^s_{\bm{0}}(N,D) \gets 1$ \Comment{For the assessment of all error events, we set $T_{\bm{0}}(N,D) = 1$ when the zero state is $\bm{\sigma}=\bm{0}_s$}

\State $\bm{T}^e(N,D) \gets  \text{Solve} \left( \bm{T}^e(N,D)  = \bm{T}^s(N,D)  \bm{L}(N,D)  \right)$

\State $T(N,D) \gets T^e_{\bm{0}}(N,D)$ \Comment{The symbolic expression of the transfer function, $T(N,D),$ when the zero state is seen as the to ending state $\bm{\sigma} = \bm{0}_e$}

\State $\dot{T}(D) \gets \left. N \tfrac{\partial }{\partial N} T(N,D)\right|_{N=1/2}$

\State $\dot{T}(D;P) \gets \text{Taylor} \left(\dot{T}(D), P\right)$ \Comment{Truncated Taylor series of the reduced transfer function $\dot{T}(D)$}

\end{algorithmic}
\end{algorithm}

To narrow down the number of candidate equivalence classes, we impose an isotropy condition on the \gls{2d} grid. Specifically, we retain only those \gls{2d} \gls{nsm} candidates whose Horizontal and Vertical child \glspl{nsm} share the same \gls{rtf}—i.e., $\dot{T}_V(D)=\dot{T}_H(D).$ This constraint introduces a form of \emph{partial cardinal isotropy}. The number of candidate \glspl{nsm} satisfying this condition is summarized at the beginning of Tables~\ref{table:2D NSMs Filter Pattern (1, 1, 1, 1, 1, 1, 1, 1, 1)}, \ref{table:2D NSMs Filter Pattern (3, 3, 3, 3, 3, 1, 1, 1, 1)} and \ref{table:2D NSMs Filter Pattern (4, 4, 4, 4, 3, 3, 3, 3, 0)} for the respective patterns $(1,1,1,1,1,1,1,1,1),$ $(3,3,3,3,3,1,1,1,1)$ and $(4,4,4,4,3,3,3,3,0).$ Thanks to this isotropy requirement, the search space is drastically reduced—from $19$ to $10$ candidates for pattern $(1,1,1,1,1,1,1,1,1),$ from $1118$ to $100$ for $(3,3,3,3,3,1,1,1,1),$ and from $2618$ to just $80$ for $(4,4,4,4,3,3,3,3,0).$ To further refine the set of candidates, we introduce an additional constraint: the \glspl{rtf} of the Main and Secondary Diagonal \glspl{nsm} must also be equal—i.e., $\dot{T}_{MD}(D)=\dot{T}_{SD}(D).$ This condition introduces a form of \emph{partial diagonal isotropy}. Applying both isotropy conditions ultimately yields only $4,$ $8,$ and $10$ optimal \gls{2d} equivalence class candidates for patterns $(1,1,1,1,1,1,1,1,1),$ $(3,3,3,3,3,1,1,1,1)$ and $(4,4,4,4,3,3,3,3,0),$ respectively.

Tables~\ref{table:2D NSMs Filter Pattern (1, 1, 1, 1, 1, 1, 1, 1, 1)}, \ref{table:2D NSMs Filter Pattern (3, 3, 3, 3, 3, 1, 1, 1, 1)} and~\ref{table:2D NSMs Filter Pattern (4, 4, 4, 4, 3, 3, 3, 3, 0)} present particularly interesting subsets of equivalence classes that satisfy both cardinal and diagonal isotropy constraints. Each class is represented by its \gls{2d} scaled filter $\mathring{h}_0[k,l],$ its matrix form $\mathring{\bm{h}}_0 \triangleq (\mathring{h}_0[k,l])_{0\le k, l \le 2},$ and two associated \glspl{rtf}: the common cardinal \gls{tf} $\dot{T}_{H,V}(D),$ shared by the Horizontal and Vertical directions, and the common diagonal \gls{tf} $\dot{T}_{MD,SD}(D),$ shared by the Main and Secondary Diagonals. Whenever possible, these selected equivalence classes are ranked from best to worst based on their \glspl{rtf}. The ranking prioritizes $\dot{T}_{H,V}(D)$ over $\dot{T}_{MD,SD}(D).$ A candidate is said to “outperforms” another if it has better cardinal and diagonal \glspl{rtf}, or if one of the two is strictly better while the other is equal. However, in cases where one class has a better cardinal function but a worse diagonal function (or vice versa), no clear ranking can be established. Such a tie in performance metrics makes the comparison locally undecidable. An example of this ambiguity is found in Table~\ref{table:2D NSMs Filter Pattern (3, 3, 3, 3, 3, 1, 1, 1, 1)}, for pattern $(3,3,3,3,3,1,1,1,1),$ where Filter~\#$1$ and Filter~\#$2$ cannot be definitively ranked against each other.

Up to this point, we have used the notion of \gls{rtf} ranking, without explicitly defining it. We now clarify this concept. Given two \glspl{rtf} expressed symbolically in the symbolic parameter $D,$ we compare them term by term. A \gls{rtf} is considered better if it has, at the first point of difference, either a smaller exponent in $D,$ or the same exponent but with a lower multiplicity. The earliest differing terms in the series dominate the comparison, as they play the most significant role in determining performance.

An examination of Tables~\ref{table:2D NSMs Filter Pattern (1, 1, 1, 1, 1, 1, 1, 1, 1)}, \ref{table:2D NSMs Filter Pattern (3, 3, 3, 3, 3, 1, 1, 1, 1)} and \ref{table:2D NSMs Filter Pattern (4, 4, 4, 4, 3, 3, 3, 3, 0)} reveals that the cardinal and diagonal child \gls{1d} \glspl{nsm} derived from the best \gls{2d} filters, $\mathring{h}_0[k,l]$—particularly Filter~\#$1$—attain the \gls{msed} of $2$-ASK. Specifically, this minimum distance, represented by $4 \, \| \bm{\pi}_0\|^2,$ takes the values $4 \cdot 3^2 = 36,$ $4 \cdot 7^2 = 196$ and $4 \cdot 10^2 = 400,$ for the patterns $(1,1,1,1,1,1,1,1,1),$ $(3,3,3,3,3,1,1,1,1)$ and $(4,4,4,4,3,3,3,3,0),$ respectively, corresponding to patterns Euclidean norms of $3,$ $7,$ and $10.$ These variations in $2$-ASK’s \gls{msed} are directly linked to the norm of the scaled filter $\mathring{h}_0[k,l],$ and are precisely the values achieved by the cardinal and diagonal child \gls{1d} \glspl{nsm} associated with the most effective \gls{2d} filters.

Given that these basic \gls{1d} \glspl{nsm} exhibit the optimal \gls{msed} properties of $2$-ASK, it is reasonable to hope that similar properties might extend to the corresponding \gls{2d} \glspl{nsm}. While simulation results—specifically, \gls{ber} performance at medium to high \glspl{snr}—cannot offer formal proof, they can provide valuable insight. The upcoming results will help assess whether this expectation is likely justified, though a degree of uncertainty inevitably remains.

At this stage, it is useful to highlight an important observation. For all considered filter patterns $\bm{\pi}_0,$ the best \gls{2d} \glspl{nsm} achieve the same \gls{msed} as $2$-ASK. However, achieving comparable \gls{bep} performance also depends on the multiplicities of dominant error events in the \glspl{rtf} of the associated cardinal and diagonal child \gls{1d} \glspl{nsm}. For patterns $\bm{\pi}_0 = (3,3,3,3,3,1,1,1,1)$ and $\bm{\pi}_0 = (4,4,4,4,3,3,3,3,0),$ the dominant term in the \gls{rtf}—the one with the smallest exponent in the symbolic parameter $D$—has multiplicities of $4$ and $6$ for the cardinal and diagonal \glspl{nsm}, respectively. When normalized by the numerators, $4$ and $6,$ of their respective rates ($\rho = 4/3$ and $\rho = 6/5$), these yield an effective multiplicity of $1,$ which exactly matches the leading term in the \gls{bep} of $2$-ASK. 

In the specific case of the pattern $\bm{\pi}_0=(1,1,1,1,1,1,1,1,1),$ the dominant term in the \gls{rtf} appears with multiplicities of $303/64$ and $6$ for the cardinal and diagonal \gls{1d} \glspl{nsm}, respectively. After normalization by their corresponding rates, the effective multiplicities become $1$ for the cardinal \glspl{nsm} and $303/256 \approx 1.18$ for the diagonal \glspl{nsm}. This leads to an average effective multiplicity strictly between $1$ and $303/256$ for the underlying \gls{2d} \gls{nsm}, resulting in a \gls{bep} upper bound that is asymptotically higher than that of $2$-ASK. Therefore, although the diagonal \gls{1d} \glspl{nsm} still achieve an effective multiplicity of $1,$ matching $2$-ASK \gls{bep} asymptotically, it is the increased effective multiplicity of the cardinal \glspl{nsm}—equal to $303/256$—that causes the overall performance degradation for pattern $\bm{\pi}_0 = (1,1,1,1,1,1,1,1,1).$ In contrast, the patterns $\bm{\pi}_0 = (3,3,3,3,3,1,1,1,1)$ and $\bm{\pi}_0 = (4,4,4,4,3,3,3,3,0)$ lead to best \gls{2d} \glspl{nsm} whose average effective multiplicities in the \gls{bep} upper bounds match that of $2$-ASK, thereby achieving its asymptotic performance.

At this point in the discussion, it is important to understand why the leading term in the \gls{rtf} of the cardinal \gls{1d} \glspl{nsm}—derived from the best \gls{2d} \gls{nsm} (Filter~\#$1$) for the pattern $\bm{\pi}_0=(1,1,1,1,1,1,1,1,1),$—yields an effective multiplicity of $303/256$ in the corresponding \gls{bep} upper bound, which is strictly greater than $1,$ with respect to the \gls{bep} of $2$-ASK. First, note that this pattern—like the others considered—was specifically chosen to avoid the tightness property, ensuring that the increased multiplicity cannot be attributed to any tightness-related artifact. The origin of this higher multiplicity lies in the presence of non-standard error events with \gls{msed}. Unlike standard events, which correspond to first input sequence differences with only a single non-zero component, these non-standard events involve first input differences with more than one non-zero component. 

In the context of the pattern $\bm{\pi}_0=(1,1,1,1,1,1,1,1,1),$ we are able to construct configurations where the first input sequence difference contains two consecutive non-zero components in either the Horizontal or Vertical direction, and the second input sequence difference has three non-zero components—while still resulting in error events of \gls{msed}. The matrix representation $\mathring{\bm{h}}_0,$ of \gls{2d} filter $\mathring{h}_0[k,l],$ (associated with Filter~\#$1$, in Table~\ref{table:2D NSMs Filter Pattern (1, 1, 1, 1, 1, 1, 1, 1, 1)}) is symmetric. This symmetry (which follows directly from our cardinal isotropic requirement) ensures that the \glspl{rtf} in the Horizontal and Vertical directions are identical, allowing us to focus on the Horizontal case without loss of generality. Specifically, we take the first input sequence difference as $\Delta \bar{b}_0[k,l]= \pm 2\, \delta[k](\delta[k]+\delta[k-1])$ or $\Delta \bar{b}_0[k,l]= \pm 2\, \delta[k](\delta[k]-\delta[k-1]),$ and the second input difference as $\Delta \bar{b}_1[k,l] = \mp (\delta[k]\delta[l-1]-\delta[k-1]\delta[l-2]-\delta[k-2]\delta[l-1]),$ or $\Delta \bar{b}_1[k,l] = \mp (-\delta[k]\delta[l-2]-\delta[k-1]\delta[l-1]+\delta[k-2]\delta[l-2]),$ respectively.

One can verify that, when the rescaled version 
\begin{equation} \label{eq:Scaled Mod Seq Two-Dimensional NSM-1}
    \mathring{s}[k,l] = \sum_{i,j} \bar{b}_0[i,j] \mathring{h}_0[k-i,l-j] + \sum_{i,j} \bar{b}_1[i,j] \mathring{h}_1[k-i,l-j],
\end{equation}
of the original modulated sequence, as defined in (\ref{eq:Mod Seq Two-Dimensional NSM}), is applied, these configurations result in a scaled modulated sequence difference with a \gls{msed} of $36.$

While the optimized \gls{nsm} filters were designed for infinite \gls{2d} grid extents, practical implementations require us to work with finite grids. To this end, we focus on \gls{2d} rectangular grids with dimensions $I \ge 3$ and $J \ge 3.$ Within this practical framework, only the input sequences $\bar{b}_0[k,l],$ defined for $0 \le k < I-2,$ $0 \le l < J-2,$ and $\bar{b}_1[k,l],$ defined for $0 \le k < I,$ $0 \le l < J,$ are permitted to take values in the bipolar set $\{ \pm 1 \}.$ These sequences carry a total of $(I-2)(J-2)+IJ$ input data symbols, which are mapped onto a \gls{2d} normalized output sequence $\bar{s}[k,l],$ defined for $0 \le k < I,$ $0 \le l < J,$ consisting of exactly $IJ$ symbols. The resulting \gls{nsm} rate, $\rho = ((I-2)(J-2)+IJ)/(IJ) = 1 + (1-2/I)(1-2/J),$ approaches $2$ asymptotically as both $I$ and $J$ increase, reflecting the behavior of increasingly large \gls{2d} grid extents.

To illustrate the performance of the best obtained \gls{2d} \glspl{nsm}, we consider square-shaped extents with $I = J \in \{ 4, 5, 6 \},$ corresponding to increased rates of $\rho = 20/16, 34/25$ and $52/36,$, respectively, since $\rho = 1+(1-2/I)^2.$ For each extent, we simulate the \gls{ber} performance of the best \glspl{nsm} associated with the patterns $\bm{\pi}_0 = (1,1,1,1,1,1,1,1,1),$ $\bm{\pi}_0 = (3,3,3,3,3,1,1,1,1)$ and $\bm{\pi}_0 = (4,4,4,4,3,3,3,3,0),$ using Filter~\#$1$ in all cases. Additionally, for pattern $\bm{\pi}_0 = (3,3,3,3,3,1,1,1,1),$ we also include results for Filter~\#$2$, due to a subtle tradeoff: its diagonal \gls{1d} \glspl{nsm} perform slightly better than those of Filter~\#$1$, although the opposite is true for the cardinal direction.

\begin{figure}[!htbp]
    \centering
    \includegraphics[width=1.0\textwidth]{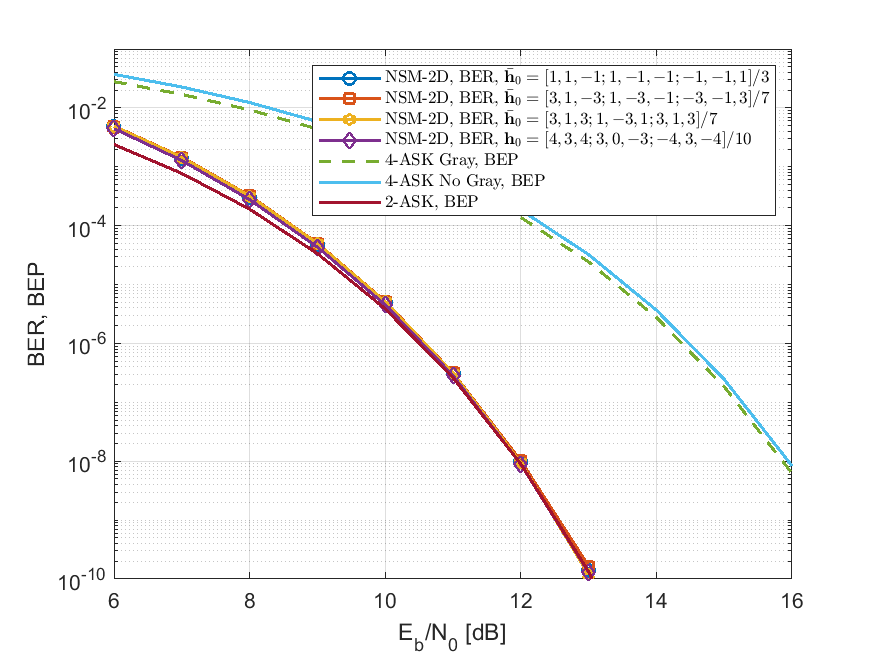}
    \caption{BER of the 2D rate $\rho = 20/16=5/4$ NSMs, corresponding to $I=J=4,$ using best 2D filters $\mathring{\bm{h}}_0 = (1, 1, -1; 1, -1, -1; -1, -1, 1)$ (pattern $\bm{\pi}_0 = (1, 1, 1, 1, 1, 1, 1, 1, 1)$), $\mathring{\bm{h}}_0 = (3, 1, -3; 1, -3, -1; -3, -1, 3)$ and $(3, 1, 3; 1, -3, 1; 3, 1, 3)$ (pattern $\bm{\pi}_0 = (3, 3, 3, 3, 3, 1, 1, 1, 1)$), and $\mathring{\bm{h}}_0 = (4, 3, 4; 3, 0, -3; -4, 3, -4)$ (pattern $\bm{\pi}_0 = (4, 4, 4, 4, 3, 3, 3, 3, 0)$). For reference, the BEPs of $2$-ASK and Gray and non-Gray precoded $4$-ASK conventional modulations are presented.}
    \label{fig:BER_2D_NSM_Rate20_16}
\end{figure}

\begin{figure}[!htbp]
    \centering
    \includegraphics[width=1.0\textwidth]{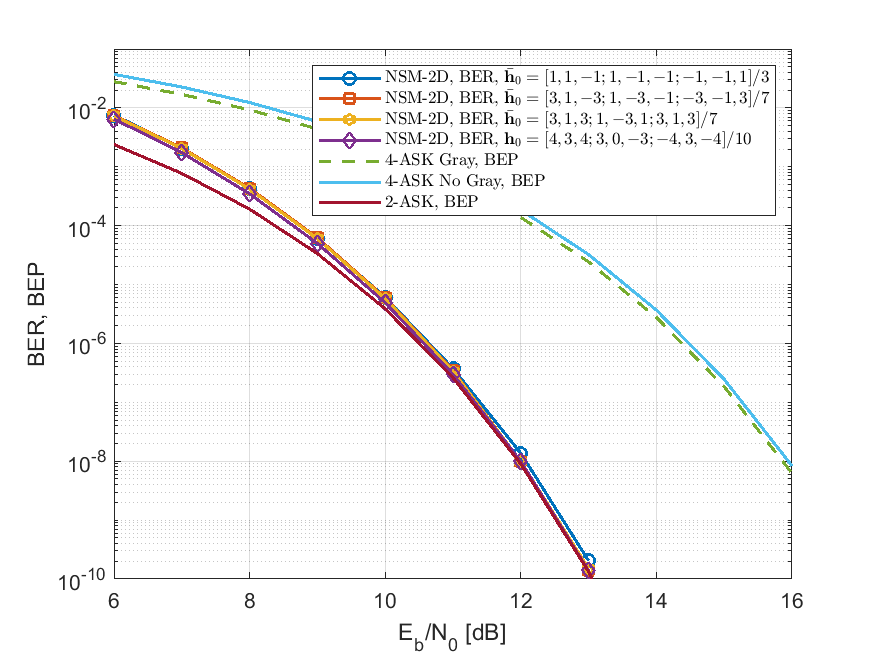}
    \caption{BER of the 2D rate $\rho = 34/25$ NSMs, corresponding to $I=J=5,$ using best 2D filters $\mathring{\bm{h}}_0 = (1, 1, -1; 1, -1, -1; -1, -1, 1)$ (pattern $\bm{\pi}_0 = (1, 1, 1, 1, 1, 1, 1, 1, 1)$), $\mathring{\bm{h}}_0 = (3, 1, -3; 1, -3, -1; -3, -1, 3)$ and $(3, 1, 3; 1, -3, 1; 3, 1, 3)$ (pattern $\bm{\pi}_0 = (3, 3, 3, 3, 3, 1, 1, 1, 1)$), and $\mathring{\bm{h}}_0 = (4, 3, 4; 3, 0, -3; -4, 3, -4)$ (pattern $\bm{\pi}_0 = (4, 4, 4, 4, 3, 3, 3, 3, 0)$). For reference, the BEPs of $2$-ASK and Gray and non-Gray precoded $4$-ASK conventional modulations are presented.}
    \label{fig:BER_2D_NSM_Rate34_25}
\end{figure}

\begin{figure}[!htbp]
    \centering
    \includegraphics[width=1.0\textwidth]{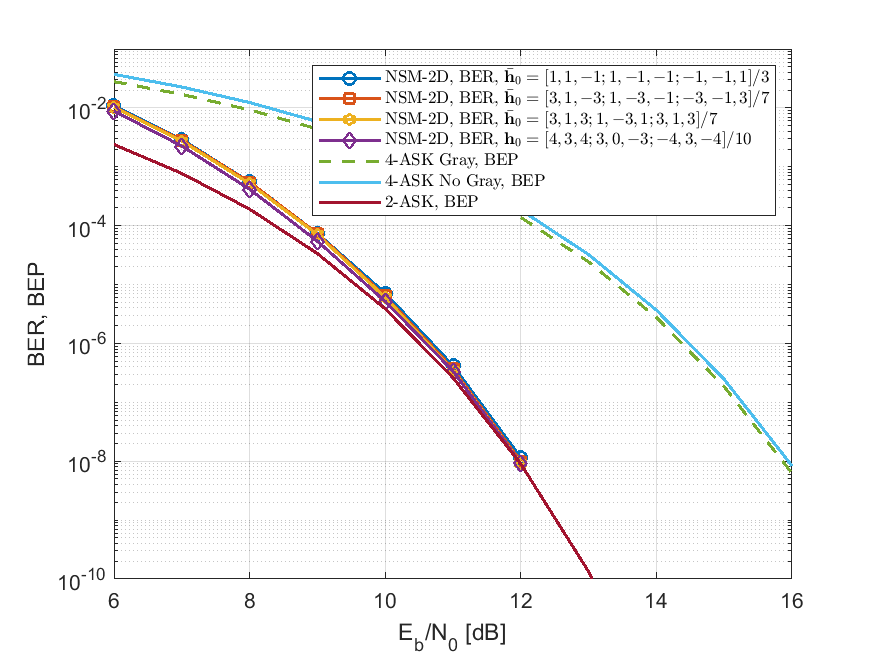}
    \caption{BER of the 2D rate $\rho = 52/36=13/9$ NSMs, corresponding to $I=J=6,$ using best 2D filters $\mathring{\bm{h}}_0 = (1, 1, -1; 1, -1, -1; -1, -1, 1)$ (pattern $\bm{\pi}_0 = (1, 1, 1, 1, 1, 1, 1, 1, 1)$), $\mathring{\bm{h}}_0 = (3, 1, -3; 1, -3, -1; -3, -1, 3)$ and $(3, 1, 3; 1, -3, 1; 3, 1, 3)$ (pattern $\bm{\pi}_0 = (3, 3, 3, 3, 3, 1, 1, 1, 1)$), and $\mathring{\bm{h}}_0 = (4, 3, 4; 3, 0, -3; -4, 3, -4)$ (pattern $\bm{\pi}_0 = (4, 4, 4, 4, 3, 3, 3, 3, 0)$). For reference, the BEPs of $2$-ASK and Gray and non-Gray precoded $4$-ASK conventional modulations are presented.}
    \label{fig:BER_2D_NSM_Rate52_36}
\end{figure}

The simulation results, presented in Figures~\ref{fig:BER_2D_NSM_Rate20_16}, \ref{fig:BER_2D_NSM_Rate34_25} and~\ref{fig:BER_2D_NSM_Rate52_36}, show a slight degradation in \gls{ber} performance as the extent increases from $4 \times 4$ to $6 \times 6.$ This behavior is expected, as larger extents allow a greater number of low-distance error events to occur, which slightly worsens performance at low to moderate \glspl{snr}.

Among the three patterns, $\bm{\pi}_0 = (4,4,4,4,3,3,3,3,0),$ shows a marginal but consistent advantage over the others—in line with theoretical expectations. Although $\bm{\pi}_0 = (1,1,1,1,1,1,1,1,1)$ has the best normalized second \gls{msed} (equal to $48/36=4/3$), it suffers from a higher effective multiplicity for the error events of \gls{msed}, particularly due to its cardinal \gls{1d} \glspl{nsm}. As a result, it performs slightly worse than the other two patterns, especially at high \glspl{snr}.

For the intermediate pattern $\bm{\pi}_0 = (3,3,3,3,3,1,1,1,1),$ the normalized second \gls{msed} is $224/196=8/7,$ whereas for the pattern $\bm{\pi}_0 = (4,4,4,4,3,3,3,3,0),$ it increases to $480/400=6/5.$ This higher distance already offers a performance benefit. In addition, although both patterns have the same original multiplicity of $5$ for this second minimum distance in the cardinal \gls{1d} \glspl{nsm}, the diagonal multiplicity is smaller for the pattern $\bm{\pi}_0 = (4,4,4,4,3,3,3,3,0),$—only $4$ compared to $5$ for $\bm{\pi}_0 = (3,3,3,3,3,1,1,1,1).$ This results in a cumulative effect: a combination of increased normalized distance and reduced multiplicity in the diagonal \glspl{nsm}, which helps explain the improved performance observed for the pattern $\bm{\pi}_0 = (4,4,4,4,3,3,3,3,0).$

Most importantly, across all extents and patterns, especially for the latter two, the \gls{ber} curves align very closely with that of $2$-ASK in the moderate to high \gls{snr} range. This tight performance suggests that exploring more complex pattern vectors, different pattern shapes, or larger pattern sizes may not yield significant gains. Instead, maintaining simple structures like those explored here allows us to keep the offered rates high while minimizing detection complexity at the receiver.

\subsubsection{Three-Dimensional NSMs}
\label{ssec:Three-Dimensional Rate-2 NSMs}

Up to this point, we have focused on a family of \gls{1d} \glspl{nsm} characterized by simple filter coefficients. In Section~\ref{ssec:Two-Dimensional Rate-2 NSMs}, we extend this framework to \gls{2d} \glspl{nsm}, still maintaining simple filter coefficients. This transition from the \gls{1d} to the \gls{2d} setting can be viewed as analogous to the shift from time-domain or frequency-domain \gls{ftn} \cite{Anderson13} to \gls{mftn} \cite{Rusek05, Anderson13}.

We now extend our framework from \gls{1d} and \gls{2d} \glspl{nsm} to \gls{3d} \glspl{nsm}. Unlike the transition from one dimension to two dimensions, this move to three dimensions has no direct analogue in the \gls{ftn} framework. \gls{ftn} begins with an analog modulated signal and relies on carefully chosen transmit and receive waveforms. The key idea is to compress these waveforms—whether in time (time-domain \gls{ftn}), in frequency (frequency-domain \gls{ftn}), or jointly in time and frequency (\gls{mftn})—thereby introducing structured \gls{isi}. This results in \gls{ftn} signals being interpreted as a form of \gls{prs}, where the \gls{isi} arises from the specific waveform pair used. However, this indirect method of introducing \gls{isi} to improve spectral efficiency is fundamentally tied to the structure of the modulated signal itself. As such, it imposes significant limitations and does not permit a natural extension to \gls{3d} or higher-dimensional \glspl{nsm}. At best, it corresponds to the \gls{1d} and \gls{2d} \glspl{nsm} we have already defined. In contrast, our approach enables the construction of \gls{3d} and higher-dimensional \glspl{nsm}, without these inherent constraints.

Extending the construction of \gls{2d} \glspl{nsm} to the \gls{3d} case, while maintaining a modulation rate of $\rho = 2,$ the rescaled \gls{3d} modulated sequence can be expressed as 
\begin{equation} \label{eq:Scaled Mod Seq Two-Dimensional NSM-2}
    \mathring{s}[l,m,n] = \sum_{i,j,k} \bar{b}_0[i,j,k] \mathring{h}_0[l-i,m-j,n-k] + \sum_{i,j,k} \bar{b}_1[i,j,k] \mathring{h}_1[l-i,m-j,n-k],
\end{equation}
In this formulation, both the rescaled filters $\mathring{h}_0[l,m,n]$ and $\mathring{h}_1[l,m,n]$, which have integer taps, and the input sequences, $\bar{b}_0[l,m,n]$ and $\bar{b}_1[l,m,n]$, which are bipolar-valued, are now fully \gls{3d}.

For illustrative purposes, we consider finite \gls{3d} grids of size $I \times J \times K,$ along with compact $2 \times 2 \times 2$ \gls{3d} filters, $\mathring{h}_0[l,m,n],$ defined such that $\mathring{h}_0[l,m,n] = 0,$ except when $0 \le l,m,n \le 1.$ This compact structure brings both advantages and limitations. Compact filters offer a key advantage: they support higher \gls{nsm} rates—specifically, $\rho = (IJK + (I-1)(J-1)(K-1))/IJK = 1 +(1-1/I)(1-1/J)(1-1/K)$—compared to non-compact alternatives on the same \gls{3d} grid. This advantage is particularly noticeable when the \gls{3d} grid extent dimensions $I,J$ and $K$ are small. Although the proposed \gls{3d} \glspl{nsm} can asymptotically achieve the optimal rate of $2$ as these dimensions grow, practical implementations are constrained by the need to keep complexity as low as possible. Therefore, it is highly desirable to work with small values of $I,J$ and $K,$ and in this regime, compact generator footprints are especially beneficial—they allow high rates to be achieved without exceeding complexity limits.

Furthermore, if the \gls{msed} of $2$-ASK is achieved by the most compact error events—namely, those with a cubic footprint of size $2 \times 2 \times 2$—then this minimum distance is guaranteed for all other error events as well, even those with larger or non-compact footprints. This implication relies on a structural property of the error patterns. To explain this in more detail, the non-zero values of modulated sequences differences corresponding to error events with the minimum cubic footprint of size $2 \times 2 \times 2$ are exclusively located within the $8$ vertices of the $2 \times 2 \times 2$ cube. The \gls{sed} of such events is determined by the sum of their $8$ non-zero components, at these $8$ vertices, weighted by 4—reflecting the squared magnitude of non-null components of input sequences differences taking their non-null values in the alphabet $\{\pm2\}$). Any arbitrary error event, regardless of its extent or compactness, necessarily includes at least one occurrence at each of these $8$ vertex positions within its footprint. As a result, its \gls{sed} is lower-bounded by the \gls{msed} of the minimal $2 \times 2 \times 2$ footprint events, since it replicates all the key positions that contribute to the minimum distance. Consequently, if the minimal footprint error events—those confined to the $2 \times 2 \times 2$ region—achieve the \gls{msed} of $2$-ASK, then all other error events inherit this property. That is, the \gls{msed} of the overall \gls{nsm} is guaranteed to be at least that of $2$-ASK, as long as the most compact error patterns meet this \gls{msed}. This vertex-based property generalizes an analogous one seen in \gls{2d} settings, discussed at the beginning of Section~\ref{ssec:Two-Dimensional Rate-2 NSMs}. There, in the context of basic rate-$2$ \gls{2d} \glspl{nsm}, it was shown—and illustrated in Figure~\ref{fig:Two-Dimensional-NSM-Minimum-Euclidean-Distance}—that any valid error footprint reproduce at least one time each of the four corners of a compact $2 \times 2$ square, thus inheriting the \gls{msed}, provided the most compact footprint meets it.

On the downside, compact filters can increase the multiplicities associated with the upper bounds of the binary error probability, particularly for error events corresponding to the second \gls{msed} and higher. 

As in the case of the \gls{2d} \glspl{nsm} discussed in Subsection~\ref{ssec:Two-Dimensional Rate-2 NSMs}, each filter $\mathring{h}_0[l,m,n]$ must must conform to a $1 \times 8$ filter pattern, $\bm{\pi}_0,$ composed of non-negative integers and characterized by an integer-valued norm. The second filter, $\mathring{h}_1[l,m,n],$ is defined simply as $\mathring{h}_1[l,m,n]=\| \bm{\pi}_0 \|\,\delta[l]\delta[m]\delta[n].$ To ensure the desired \gls{msed}—matching that of $2$-ASK—we impose a key constraint on $\bm{\pi}_0$: no component may exceed half of its norm. This can be expressed as $\| \bm{\pi}_0 \|_\infty \le \| \bm{\pi}_0 \|/2.$ This inequality serves as both a necessary and sufficient condition for ensuring that the \gls{msed} associated with the minimal \gls{3d} footprint—specifically, the $2 \times 2 \times 2$ cube shared by  $\mathring{h}_0[l,m,n]$—matches that of $2$-ASK. Once established for these compact error events, the same \gls{msed} extends to all other error patterns, regardless of their shape or extent. Notably, when the inequality becomes an equality, the the filters $\mathring{h}_p[l,m,n],$ $p=0,1,$ are said to exhibit tightness—a condition best avoided, as it tends to increase the multiplicity of minimum-distance error events.

For \gls{nsm} optimization, our objective is to find the most effective filter, $\mathring{h}_0[l,m,n],$ within a given eligible pattern $\bm{\pi}_0,$ one that minimizes the upper bound on the binary error probability. The pool of candidate filters includes all $\mathring{h}_0[l,m,n],$ filters whose non-zero components are generated through permutations and arbitrary sign changes of the components of $\bm{\pi}_0.$ However, directly computing the full \gls{tf}, $T(N,D),$ or even its reduced version $\dot{T}(D),$ for an infinite-grid-extent \gls{3d} \gls{nsm} built from a given filter is computationally intractable. To overcome this, we generalize the \gls{1d} \gls{nsm} extraction method previously applied to the \gls{2d} \glspl{nsm} in Section~\ref{ssec:Two-Dimensional Rate-2 NSMs}. In the \gls{3d} context, \gls{1d} \glspl{nsm} can be extracted along three types of paths: \gls{f2f}, \gls{e2e}, and \gls{v2v}. These extracted \gls{1d} \glspl{nsm} allow for tractable computation of their (reduced) transfer functions. For practical and computational efficiency, we restrict our analysis to \gls{f2f} and \gls{e2e} extractions, as \gls{v2v} paths will ultimately prove unnecessary for our purposes.

\begin{figure}[!htbp]
    \centering
    \includegraphics[width=1.0\textwidth]{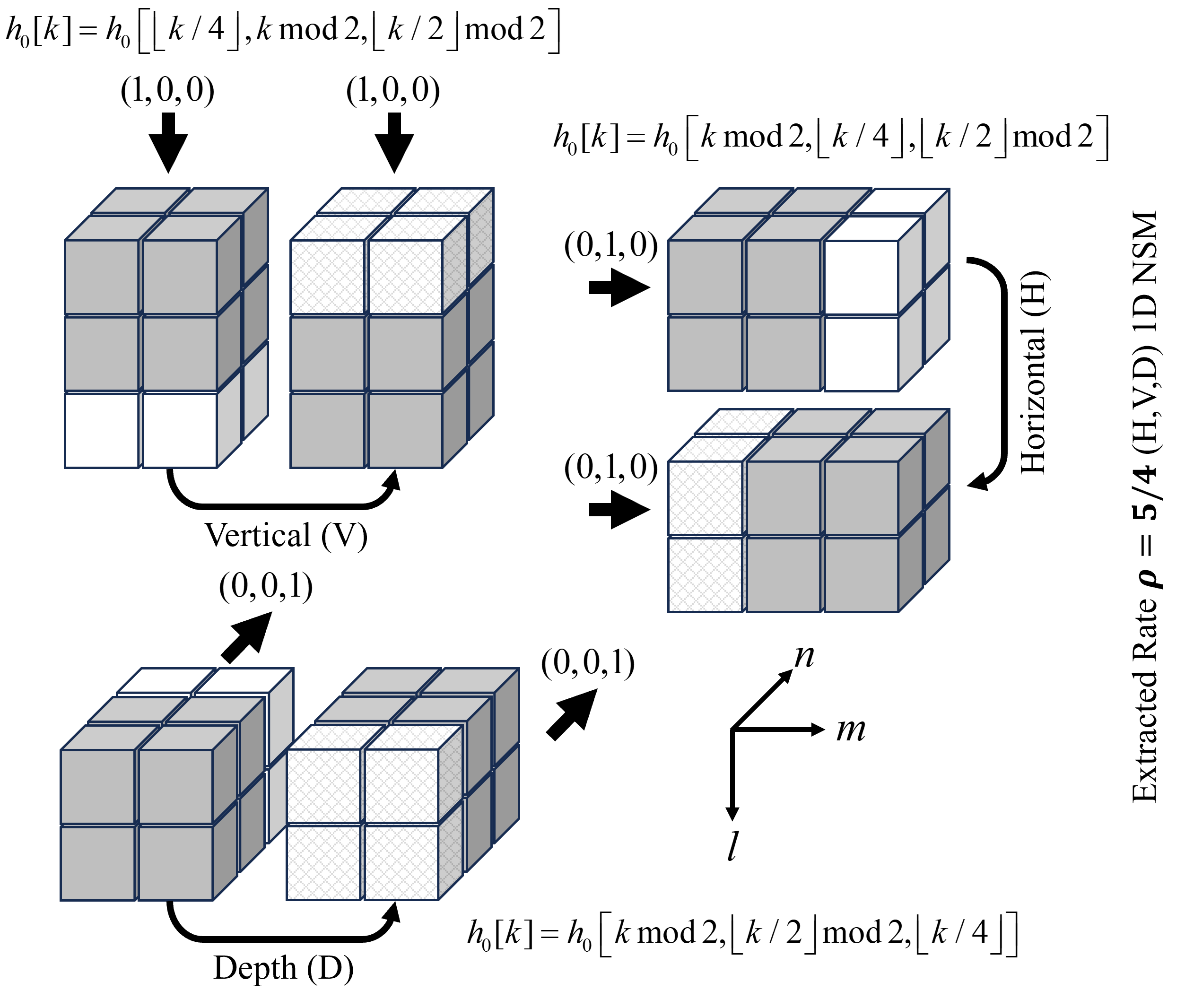}
    \caption{Footprints of rate $\rho = 5/4$ Vertical (V), Horizontal (H) and Depth (D) 1D NSM, extracted from rate $\rho = 2$ 3D NSM. Vertical 1D filter, $h_0[k],$ expressed as $h_0[k]=h_0[\lfloor k/4 \rfloor, k \bmod{2}, \lfloor k/2 \rfloor \bmod{2}],$ as a function of 3D filter $h_0[k,l,m].$ Horizontal 1D filter, $h_0[k],$ expressed as $h_0[k]=h_0[k \bmod{2}, \lfloor k/4 \rfloor, \lfloor k/2 \rfloor \bmod{2}],$ as a function of 3D filter $h_0[k,l,m].$ Depth 1D filter, $h_0[k],$ expressed as $h_0[k]=h_0[k \bmod{2}, \lfloor k/2 \rfloor \bmod{2}, \lfloor k/4 \rfloor],$ as a function of 3D filter $h_0[k,l,m].$ The \glspl{rtf} resulting from the extracted V, H and D 1D NSM are denoted by $\dot{T}_V(D),$ $\dot{T}_H(D)$ and $\dot{T}_D(D),$ respectively}
    \label{fig:3D Extracted VHD 1D NSM}
\end{figure}

\gls{f2f} directions in the \gls{3d} grid are defined by the direction vectors, $(1,0,0),$ $(0,1,0)$ and $(0,0,1),$ corresponding to the Vertical (V), Horizontal (H), and Depth (D) axes, respectively. The V and H directions are already present in the \gls{2d} framework, while the D direction is introduced in the \gls{3d} context. As illustrated in Figure~\ref{fig:3D Extracted VHD 1D NSM}, extracting \gls{1d} filters along these directions yields three \glspl{nsm} of rate $\rho=5/4,$ with \gls{1d} filters given by $h_0[k]=h_0[\lfloor k/4 \rfloor, k \bmod{2}, \lfloor k/2 \rfloor \bmod{2}],$ $h_0[k]=h_0[k \bmod{2}, \lfloor k/4 \rfloor, \lfloor k/2 \rfloor \bmod{2}]$ and $h_0[k]=h_0[k \bmod{2}, \lfloor k/2 \rfloor \bmod{2}, \lfloor k/4 \rfloor],$ for the V, H, and D directions, respectively. The corresponding \glspl{rtf} are denoted $\dot{T}_V(D),$ $\dot{T}_H(D)$ and $\dot{T}_D(D).$

\begin{figure}[!htbp]
    \centering
    \includegraphics[width=1.0\textwidth]{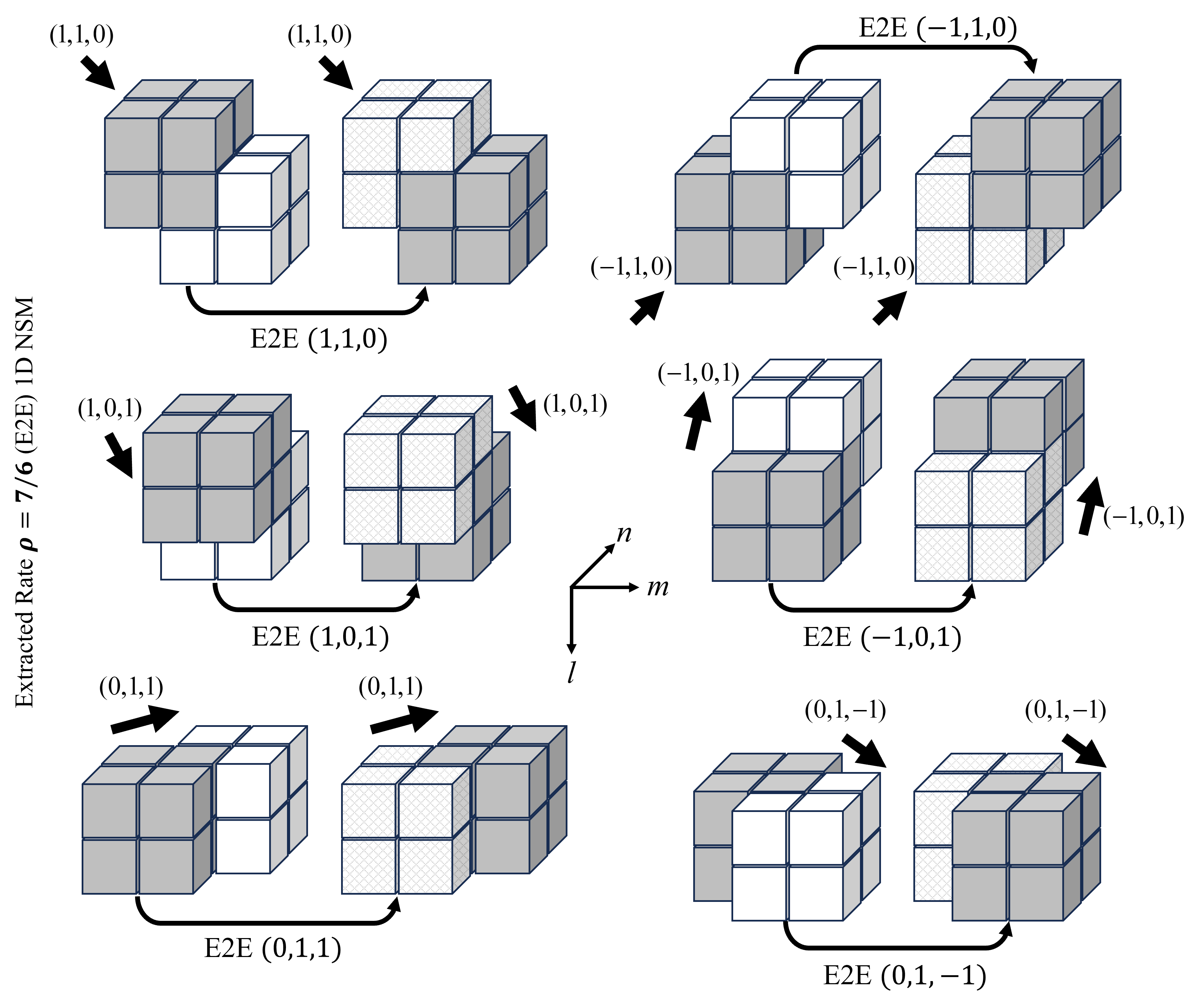}
    \caption{Footprints of rate $\rho = 7/6$ E2E 1D NSM, extracted from rate $\rho = 2$ 3D NSM. The RTFs resulting from the extracted E2E 1D NSMs are denoted by $\dot{T}_{(k,l,m)}(D),$ for E2E directions $(k,l,m)$ in $\{(\pm 1, 1,0), (\pm 1,0,1), (0,1,\pm 1) \}.$}
    \label{fig:3D Extracted E2E 1D NSM}
\end{figure}

The \gls{e2e} extraction involves six distinct directions, defined by the vectors $(\pm1,1,0),$ $(\pm1,0,1)$ and $(0,1,\pm1).$ As shown in Figure~\ref{fig:3D Extracted E2E 1D NSM}, the \gls{1d} \glspl{nsm} extracted along these directions all share a common rate of $\rho=7/6.$ Although giving the explicit forms of the \gls{1d} filters $h_0[p],$ as a function of the original \gls{3d} filters $h_0[l,m,n],$ for all six directions is quite involved, it is still instructive to provide expressions for two representative cases, namely the directions $(\pm1,1,0),$ to help grasp how such extractions are defined. For direction $(1,1,0),$ the filter is given by $h_0[p] = h_0[l,m,n],$ with $p=2(k+2l)+m=4l+2k+m,$ for $0 \le l,m,n \le 1,$ and $h_0[p]=0,$ otherwise. For direction $(-1,1,0),$ it is defined as $h_0[p] = h_0[l,m,n],$ with $p=2(-k+2l+1)+m=4l-2k+m+2,$ for $0 \le l,m,n \le 1,$ and $h_0[p]=0,$ elsewhere. For each of the $6$ directions $(i,j,k) \in \{(\pm1,1,0), (\pm1,0,1), (0,1,\pm1) \},$ the corresponding extracted \gls{1d} \gls{nsm} is associated with a \gls{rtf}, denoted by $\dot{T}_{(i,j,k)}(D).$

For a given eligible pattern $\bm{\pi}_0,$ the number of associated \gls{3d} filter candidates, $h_0[l,m,n],$ can be prohibitively large. This leads to significant computational complexity in identifying the optimal filters for constructing the best \gls{3d} \glspl{nsm}, as evaluating each candidate requires computing the corresponding \glspl{rtf}: $\dot{T}_V(D),$ $\dot{T}_H(D),$ $\dot{T}_D(D)$ and $\dot{T}_{(i,j,k)}(D),$ $(i,j,k) \in \{(\pm1,1,0), (\pm1,0,1), (0,1,\pm1) \}.$ This evaluation process is both time-consuming and computationally expensive.

To simplify the optimization, as in the \gls{2d} case, we introduce an equivalence relation among candidate \gls{3d} \glspl{nsm}. This relation groups together \gls{3d} filter $h_0[l,m,n]$ candidates that yield the same underlying \gls{nsm} characteristics and, consequently, identical performance in terms of binary error probability. It relies on transformations of the \gls{3d} filters $h_0[l,m,n]$ that preserve the \gls{nsm} structure. Among these transformations is the octahedral group—the group of symmetries of the cube—which preserves the cubic $2 \times 2 \times 2$ footprint of $h_0[l,m,n]$ and hence the properties of the generated \gls{nsm}. This group includes $48$ orthogonal transformations: $24$ proper rotations (with determinant $+1$) and $24$ improper symmetries, comprising reflections and rotoreflections (with determinant $-1$). Additional admissible transformations include sign inversion, where the filter $h_0[l,m,n]$ is replaced by its negative $-h_0[l,m,n],$ and translations on the \gls{3d} grid, where the filter is shifted by a translation vector $(\Delta l, \Delta m, \Delta n),$ resulting in $h_0[l-\Delta l,m-\Delta m,n-\Delta n].$ Another class of transformations involves sign-alternating modifications, where the filter is transformed into one of seven equivalent forms defined by $\tilde{h}_0[l,m,n]=(-1)^{\lambda l + \mu m + \nu n} h_0[l,m,n],$ $(\lambda, \mu, \nu) \in \{0,1\}^3,$ $(\lambda, \mu, \nu) \ne (0,0,0).$

To streamline the optimization process, we introduce the concept of isotropy and apply two fundamental isotropy conditions: \gls{f2f} \emph{isotropy} and \gls{e2e} \emph{isotropy}. In \gls{f2f} isotropy, the \glspl{rtf} $\dot{T}_V(D),$ $\dot{T}_H(D)$ and $\dot{T}_D(D)$ are constrained to be identical, collectively represented by a single common function $\dot{T}_{F2F}(D).$ In \gls{e2e} isotropy, the \glspl{rtf} $\dot{T}_{(i,j,k)}(D)$—where  $(i,j,k) \in \{(\pm1,1,0), (\pm1,0,1), (0,1,\pm1) \}$—are also required to be identical and equal to a shared function $\dot{T}_{E2E}(D).$ We begin by enforcing \gls{f2f} isotropy across all representatives $h_0[l,m,n],$ within each equivalence class, which significantly reduces the pool of eligible classes for optimal filter selection. Applying the second constraint of \gls{f2f} isotropy further refines the set by eliminating additional candidates. Among the remaining equivalence classes that satisfy both \gls{f2f} and \gls{f2f} isotropy, a final ranking is performed based on their distance spectra, evaluated using the corresponding \glspl{rtf} $\dot{T}_{F2F}(D)$ and $\dot{T}_{E2E}(D).$

To illustrate the construction, we focus on a class of \gls{3d} \glspl{nsm} with $2 \times 2 \times 2$ filter footprints, examining in detail those associated with the filter patterns $\bm{\pi}_0 = (1,1,1,1,0,0,0,0),$ $(4,4,4,4,3,3,3,3),$ $(12,12,12,12,5,5,5,5)$ and $(15,15,15,15,8,8,8,8).$ The corresponding second \gls{3d} rescaled filters are defined as $\mathring{h}_1[l,m,n]=\|\bm{\pi}_0\| \, \delta[l]\delta[m]\delta[n],$ with $\bm{\pi}_0$ having respective norms $\|\bm{\pi}_0\| = 2, 10, 26$ and $34.$ Each of these patterns satisfies the necessary and sufficient condition $\| \bm{\pi}_0 \|_\infty \le \| \bm{\pi}_0 \|/2,$ which ensures that the resulting \glspl{nsm} can achieve the \gls{msed} of $2$-ASK. However, this inequality is strictly satisfied by all the listed patterns except $\bm{\pi}_0=(1,1,1,1,0,0,0,0),$ which attains equality. This distinction has a direct impact on the multiplicity of \gls{msed} error events, as confirmed by the data in Tables~\ref{table:3D NSMs Filter Pattern (1, 1, 1, 1, 0, 0, 0, 0)}--\ref{table:3D NSMs Filter Pattern (15, 15, 15, 15, 8, 8, 8, 8)}. These tables detail the characteristics of the corresponding rescaled filters $\mathring{h}_0[l,m,n],$ in terms of their \glspl{rtf} $\dot{T}_{F2F}(D)$ and $\dot{T}_{E2E}(D).$ For the patterns that strictly satisfy the inequality, the multiplicities of \gls{msed} events are $5$ and $7,$ respectively, for the \gls{f2f} and \gls{e2e} \glspl{rtf}. When normalized by the number of bipolar input sequences—determined by the associated \gls{1d} extracted \gls{f2f} and \gls{e2e} \gls{nsm} rates $\rho=5/4$ and $\rho=7/6$—these multiplicities yield a normalized value of $1$ for the leading term in the binary error probability. This leading term corresponds precisely to the (binary) error probability of $2$-ASK. Crucially, this implies that the performance of the proposed \glspl{nsm}—except for those based on the pattern $\bm{\pi}_0=(1,1,1,1,0,0,0,0)$—can asymptotically match that of $2-ASK,$ and do so rapidly, even in the moderate \gls{snr} range.

\begin{table}[H]

\caption{“Isotropic” 3D NSMs, with simple filters' coefficients, with common filter $\bm{h}_0$ pattern $\bm{\pi}_0 = (1, 1, 1, 1, 0, 0, 0, 0),$ arranged in a decreasing performance order, starting from the best one. Relative degradation in performance, due to distance reduction or/and multiplicity increase, when moving from one filter to the next, is emphasized in bold. 
} \label{table:3D NSMs Filter Pattern (1, 1, 1, 1, 0, 0, 0, 0)}

\centering
\begin{tabular}{|c|c|} 
\hline

\multicolumn{2}{|l|}{\# of simple $h_0[k,l,m]$ filters $ = 1120$} \\ \hline
\multicolumn{2}{|l|}{\# of non equivalent $h_0[k,l,m]$ filters $ = 9$} \\ \hline
\multicolumn{2}{|l|}{\# of non equivalent $h_0[k,l,m]$ filters with ($\dot{T}_{(1,0,0)}(D)=\dot{T}_{(0,1,0)}(D)=\dot{T}_{(0,0,1)}(D)$) $ = 4$} \\ \hline
\multicolumn{2}{|l|}{\# of non equivalent $h_0[k,l,m]$ filters with ($\dot{T}_{(1,0,0)}(D)=\dot{T}_{(0,1,0)}(D)=\dot{T}_{(0,0,1)}(D)$)} \\
\multicolumn{2}{|l|}{\& ($\dot{T}_{(1,1,0)}(D)=\dot{T}_{(1,-1,0)}(D)=\dot{T}_{(1,0,1)}(D)=\dot{T}_{(1,0,-1)}(D) =\dot{T}_{(0,1,1)}(D)=\dot{T}_{(0,1,-1)}(D)$) $ = 2$} \\ [0.5ex] \hline

\multicolumn{2}{|l|}{$\mathring{h}_1[k,l,m] = 2 \, \delta[k]\delta[l]\delta[l]$} \\ \hline
\multicolumn{2}{|l|}{\gls{msed}, $d_\text{min}^2 = 4\cdot2^2 = 16$} \\ [0.5ex]
\hline\hline

\multicolumn{2}{|c|}{Filter~\#$1$} \\ \hline
$\mathring{h}_0[k,l,m]$ & $\delta[k]\delta[l]\delta[m]+\delta[k-1]\delta[l-1]\delta[m]+\delta[k]\delta[l-1]\delta[m-1]+\delta[k-1]\delta[l]\delta[m-1]$  \\ \hline
$\mathring{\bm{h}}_0$ & Front slice: $\mathring{\bm{h}}_0[:,:,0]=\begin{pmatrix}
1 & 0 \\
0 & 1
\end{pmatrix}$, Back slice: $\mathring{\bm{h}}_0[:,:,1]=\begin{pmatrix}
0 & 1 \\
1 & 0
\end{pmatrix}$ \\ [0.5ex] \hline
\multirow{2}{*}{$\dot{T}_{F2F}(D)$} & 
$\frac{253}{16}\,D^{16}+\frac{25485}{128}\,D^{32}+\frac{6770775}{4096}\,D^{48}+\frac{194839405}{16384}\,D^{64}+\frac{83564731953}{1048576}\,D^{80}+\frac{4286309642343}{8388608}\,D^{96}$ \\
& $+\frac{853295774758027}{268435456}\,D^{112}+\frac{10387369294825229\,D^{128}}{536870912}+\frac{7959726382139001701\,D^{144}}{68719476736} + \cdots$ \\ [0.5ex] 
\hline
\multirow{2}{*}{$\dot{T}_{E2E}(D)$} & 
$\frac{285}{16}\,D^{16}+\frac{9477}{256}\,D^{24}+\frac{1049817}{4096}\,D^{32}+\frac{49491081}{65536}\,D^{40}+\frac{2987037429}{1048576}\,D^{48}+\frac{155891474829}{16777216}\,D^{56}$ \\
& $+\frac{8181013894161}{268435456}\,D^{64}+\frac{419413681702161}{4294967296}\,D^{72}+\frac{21237071167490349}{68719476736}\,D^{80} + \cdots$ \\ [0.5ex] 
 \hline\hline

\multicolumn{2}{|c|}{Filter~\#$2$} \\ \hline
$\mathring{h}_0[k,l,m]$ & $\delta[k]\delta[l]\delta[m]+\delta[k-1]\delta[l-1]\delta[m]+\delta[k]\delta[l-1]\delta[m-1]-\delta[k-1]\delta[l]\delta[m-1]$  \\ \hline
$\mathring{\bm{h}}_0$ & Front slice: $\mathring{\bm{h}}_0[:,:,0]=\begin{pmatrix}
1 & 0 \\
0 & 1
\end{pmatrix}$, Back slice: $\mathring{\bm{h}}_0[:,:,1]=\begin{pmatrix}
0 & 1 \\
-1 & 0
\end{pmatrix}$ \\ [0.5ex] \hline
\multirow{2}{*}{$\dot{T}_{F2F}(D)$} & 
$\frac{253}{16}\,D^{16}+\frac{25485}{128}\,D^{32}+\frac{6770775}{4096}\,D^{48}+\frac{194839405}{16384}\,D^{64}+\frac{83564731953}{1048576}\,D^{80}+\frac{4286309642343}{8388608}\,D^{96}$ \\
& $+\frac{853295774758027}{268435456}\,D^{112}+\frac{10387369294825229}{536870912}\,D^{128}+\frac{7959726382139001701}{68719476736}\,D^{144} + \cdots$ \\ [0.5ex] 
\hline
\multirow{2}{*}{$\dot{T}_{E2E}(D)$} & 
$\frac{285}{16}\,D^{16}+\frac{9477}{256}\,D^{24}+\frac{1049817}{4096}\,D^{32}+\frac{49491081}{65536}\,D^{40}+\frac{2987037429}{1048576}\,D^{48}+\frac{155891474829}{16777216}\,D^{56}$ \\
& $+\frac{8181013894161}{268435456}\,D^{64}+\frac{419413681702161}{4294967296}\,D^{72}+\frac{21237071167490349}{68719476736}\,D^{80} + \cdots$ \\ [1ex] 
 \hline

\end{tabular}
\end{table}

\begin{table}[H]
\caption{“Isotropic” 3D NSMs, with simple filters' coefficients, with common filter $\bm{h}_0$ pattern $\bm{\pi}_0 = (4, 4, 4, 4, 3, 3, 3, 3),$ arranged in a decreasing performance order, starting from the best one. Relative degradation in performance, due to distance reduction or/and multiplicity increase, when moving from one filter to the next, is emphasized in bold. 
} \label{table:3D NSMs Filter Pattern (4, 4, 4, 4, 3, 3, 3, 3)}

\centering
\begin{tabular}{|c|c|} 
\hline

\multicolumn{2}{|l|}{\# of simple $h_0[k,l,m]$ filters $ = 17920$} \\ \hline
\multicolumn{2}{|l|}{\# of non equivalent $h_0[k,l,m]$ filters $ = 52$} \\ \hline
\multicolumn{2}{|l|}{\# of non equivalent $h_0[k,l,m]$ filters with ($\dot{T}_{(1,0,0)}(D)=\dot{T}_{(0,1,0)}(D)=\dot{T}_{(0,0,1)}(D)$) $ = 12$} \\ \hline
\multicolumn{2}{|l|}{\# of non equivalent $h_0[k,l,m]$ filters with ($\dot{T}_{(1,0,0)}(D)=\dot{T}_{(0,1,0)}(D)=\dot{T}_{(0,0,1)}(D)$)} \\
\multicolumn{2}{|l|}{\& ($\dot{T}_{(1,1,0)}(D)=\dot{T}_{(1,-1,0)}(D)=\dot{T}_{(1,0,1)}(D)=\dot{T}_{(1,0,-1)}(D) =\dot{T}_{(0,1,1)}(D)=\dot{T}_{(0,1,-1)}(D)$) $ = 2$} \\ [0.5ex] \hline

\multicolumn{2}{|l|}{$\mathring{h}_1[k,l,m] = 10 \, \delta[k]\delta[l]\delta[l]$} \\ \hline
\multicolumn{2}{|l|}{\gls{msed}, $d_\text{min}^2 = 4\cdot10^2 = 400$} \\ [0.5ex]
\hline\hline

\multicolumn{2}{|c|}{Filter~\#$1$} \\ \hline

\multirow{3}{*}{$\mathring{h}_0[k,l,m]$} & $4 \,\delta[k]\delta[l]\delta[m]+3\,\delta[k]\delta[l-1]\delta[m]+3\,\delta[k-1]\delta[l]\delta[m]-4\,\delta[k-1]\delta[l-1]\delta[m]$ \\
& $3 \,\delta[k]\delta[l]\delta[m-1]-4\,\delta[k]\delta[l-1]\delta[m-1]-4\,\delta[k-1]\delta[l]\delta[m-1]$ \\
& $-3\delta[k-1]\delta[l-1]\delta[m-1]$ \\\hline

$\mathring{\bm{h}}_0$ & Front slice: $\mathring{\bm{h}}_0[:,:,0]=\begin{pmatrix}
4 & 3 \\
3 & -4
\end{pmatrix}$, Back slice: $\mathring{\bm{h}}_0[:,:,1]=\begin{pmatrix}
3 & -4 \\
-4 & -3
\end{pmatrix}$ \\ [0.5ex] \hline
\multirow{2}{*}{$\dot{T}_{F2F}(D)$} & 
$5\,D^{400}+5\,D^{480}+\frac{183}{16}\,D^{560}+\frac{733}{32}\,D^{640}+\frac{9453}{256}\,D^{720}+\frac{539}{8}\,D^{800}+\frac{340371}{4096}\,D^{880}$ \\
& $+\frac{1008019}{8192}\,D^{960}+\frac{12489785}{65536}\,D^{1040}+\frac{19863347}{65536}\,D^{1120}+\frac{509983967}{1048576}\,D^{1200} + \cdots$ \\ [0.5ex] 
\hline
\multirow{2}{*}{$\dot{T}_{E2E}(D)$} & 
$7\,D^{400}+4\,D^{480}+\frac{17}{2}\,D^{560}+D^{600}+14\,D^{640}+5\,D^{680}+\frac{269}{16}\,D^{720}+\frac{117}{8}\,D^{760}+\frac{223}{4}\,D^{800}$ \\
& $+\frac{511}{16}\,D^{840}+\frac{341}{8}\,D^{880}+\frac{893}{16}\,D^{920}+\frac{4759}{64}\,D^{960}+\frac{747}{8}\,D^{1000}+\frac{4127}{32}\,D^{1040} + \cdots$ \\ [0.5ex] 
 \hline\hline

\multicolumn{2}{|c|}{Filter~\#$2$} \\ \hline
\multirow{3}{*}{$\mathring{h}_0[k,l,m]$} & $4 \,\delta[k]\delta[l]\delta[m]+3\,\delta[k]\delta[l-1]\delta[m]+3\,\delta[k-1]\delta[l]\delta[m]+4\,\delta[k-1]\delta[l-1]\delta[m]$ \\
& $3 \,\delta[k]\delta[l]\delta[m-1]+4\,\delta[k]\delta[l-1]\delta[m-1]+4\,\delta[k-1]\delta[l]\delta[m-1]$ \\
& $+3\delta[k-1]\delta[l-1]\delta[m-1]$ \\\hline

$\mathring{\bm{h}}_0$ & Front slice: $\mathring{\bm{h}}_0[:,:,0]=\begin{pmatrix}
4 & 3 \\
3 & 4
\end{pmatrix}$, Back slice: $\mathring{\bm{h}}_0[:,:,1]=\begin{pmatrix}
3 & 4 \\
4 & 3
\end{pmatrix}$ \\ [0.5ex] \hline
\multirow{2}{*}{$\dot{T}_{F2F}(D)$} & 
$5\,D^{400}+D^{\bm{416}}+\frac{3}{4}\,D^{432}+\frac{1}{2}\,D^{448}+\frac{5}{16}\,D^{464}+\frac{67}{16}\,D^{480}+\frac{199}{64}\,D^{496}+\frac{33}{16}\,D^{512}+\frac{329}{256}\,D^{528}$ \\
& $+\frac{245}{256}\,D^{544}+\frac{9387}{1024}\,D^{560}+\frac{3299}{512}\,D^{576}+\frac{17037}{4096}\,D^{592}+\frac{10455}{4096}\,D^{608}+\frac{31983}{16384}\,D^{624} + \cdots$ \\ [0.5ex] 
\hline
\multirow{2}{*}{$\dot{T}_{E2E}(D)$} & 
$7\,D^{400}+4\,D^{480}+\frac{17}{2}\,D^{560}+D^{600}+14\,D^{640}+5\,D^{680}+\frac{269}{16}\,D^{720}+\frac{117}{8}\,D^{760}+\frac{223}{4}\,D^{800}$ \\
& $+\frac{511}{16}\,D^{840}+\frac{341}{8}\,D^{880}+\frac{893}{16}\,D^{920}+\frac{4759}{64}\,D^{960}+\frac{747}{8}\,D^{1000}+\frac{4127}{32}\,D^{1040} + \cdots$ \\ [1ex] 
 \hline

\end{tabular}
\end{table}

\begin{table}[H]
\caption{“Isotropic” 3D NSMs, with simple filters' coefficients, with common filter $\bm{h}_0$ pattern $\bm{\pi}_0 = (12, 12, 12, 12, 5, 5, 5, 5),$ arranged in a decreasing performance order, starting from the best one. Relative degradation in performance, due to distance reduction or/and multiplicity increase, when moving from one filter to the next, is emphasized in bold. 
} \label{table:3D NSMs Filter Pattern (12, 12, 12, 12, 5, 5, 5, 5)}

\centering
\begin{tabular}{|c|c|} 
\hline

\multicolumn{2}{|l|}{\# of simple $h_0[k,l,m]$ filters $ = 17920$} \\ \hline
\multicolumn{2}{|l|}{\# of non equivalent $h_0[k,l,m]$ filters $ = 52$} \\ \hline
\multicolumn{2}{|l|}{\# of non equivalent $h_0[k,l,m]$ filters with ($\dot{T}_{(1,0,0)}(D)=\dot{T}_{(0,1,0)}(D)=\dot{T}_{(0,0,1)}(D)$) $ = 12$} \\ \hline
\multicolumn{2}{|l|}{\# of non equivalent $h_0[k,l,m]$ filters with ($\dot{T}_{(1,0,0)}(D)=\dot{T}_{(0,1,0)}(D)=\dot{T}_{(0,0,1)}(D)$)} \\
\multicolumn{2}{|l|}{\& ($\dot{T}_{(1,1,0)}(D)=\dot{T}_{(1,-1,0)}(D)=\dot{T}_{(1,0,1)}(D)=\dot{T}_{(1,0,-1)}(D) =\dot{T}_{(0,1,1)}(D)=\dot{T}_{(0,1,-1)}(D)$) $ = 2$} \\ [0.5ex] \hline

\multicolumn{2}{|l|}{$\mathring{h}_1[k,l,m] = 26 \, \delta[k]\delta[l]\delta[l]$} \\ \hline
\multicolumn{2}{|l|}{\gls{msed}, $d_\text{min}^2 = 4\cdot26^2 = 2704$} \\ [0.5ex]
\hline\hline

\multicolumn{2}{|c|}{Filter~\#$1$} \\ \hline

\multirow{3}{*}{$\mathring{h}_0[k,l,m]$} & $12 \,\delta[k]\delta[l]\delta[m]+5\,\delta[k]\delta[l-1]\delta[m]+5\,\delta[k-1]\delta[l]\delta[m]-12\,\delta[k-1]\delta[l-1]\delta[m]$ \\
& $5 \,\delta[k]\delta[l]\delta[m-1]-12\,\delta[k]\delta[l-1]\delta[m-1]-12\,\delta[k-1]\delta[l]\delta[m-1]$ \\
& $-5\delta[k-1]\delta[l-1]\delta[m-1]$ \\\hline

$\mathring{\bm{h}}_0$ & Front slice: $\mathring{\bm{h}}_0[:,:,0]=\begin{pmatrix}
12 & 5 \\
5 & -12
\end{pmatrix}$, Back slice: $\mathring{\bm{h}}_0[:,:,1]=\begin{pmatrix}
5 & -12 \\
-12 & -5
\end{pmatrix}$ \\ [0.5ex] \hline
\multirow{2}{*}{$\dot{T}_{F2F}(D)$} & 
$5\,D^{2704}+4\,D^{2912}+\frac{9}{2}\,D^{3120}+2\,D^{3328}+\frac{5}{16}\,D^{3536}+D^{3744}+\frac{5}{2}\,D^{3952}+\frac{9}{4}\,D^{4160}$ \\
& $+\frac{39}{8}\,D^{4368}+\frac{121}{8}\,D^{4576}+\frac{327}{16}\,D^{4784}+\frac{59}{4}\,D^{4992}+\frac{243}{32}\,D^{5200}+\frac{159}{8}\,D^{5408} + \cdots$ \\ [0.5ex] 
\hline
\multirow{2}{*}{$\dot{T}_{E2E}(D)$} & 
$7\,D^{2704}+4\,D^{2912}+\frac{9}{2}\,D^{3120}+2\,D^{3328}+\frac{5}{16}\,D^{3536}+D^{4056}+\frac{9}{2}\,D^{4264}+4\,D^{4368}$ \\
& $+\frac{33}{4}\,D^{4472}+12\,D^{4576}+\frac{37}{4}\,D^{4680}+12\,D^{4784}+\frac{15}{2}\,D^{4888}+5\,D^{4992}+\frac{157}{32}\,D^{5096} + \cdots$ \\ [0.5ex] 
 \hline\hline

\multicolumn{2}{|c|}{Filter~\#$2$} \\ \hline
\multirow{3}{*}{$\mathring{h}_0[k,l,m]$} & $12 \,\delta[k]\delta[l]\delta[m]+5\,\delta[k]\delta[l-1]\delta[m]+5\,\delta[k-1]\delta[l]\delta[m]+12\,\delta[k-1]\delta[l-1]\delta[m]$ \\
& $5 \,\delta[k]\delta[l]\delta[m-1]+12\,\delta[k]\delta[l-1]\delta[m-1]+12\,\delta[k-1]\delta[l]\delta[m-1]$ \\
& $+5\delta[k-1]\delta[l-1]\delta[m-1]$ \\\hline

$\mathring{\bm{h}}_0$ & Front slice: $\mathring{\bm{h}}_0[:,:,0]=\begin{pmatrix}
12 & 5 \\
5 & 12
\end{pmatrix}$, Back slice: $\mathring{\bm{h}}_0[:,:,1]=\begin{pmatrix}
5 & 12 \\
12 & 5
\end{pmatrix}$ \\ [0.5ex] \hline
\multirow{2}{*}{$\dot{T}_{F2F}(D)$} & 
$5\,D^{2704}+4\,D^{2912}+\frac{9}{2}\,D^{3120}+2\,D^{3328}+D^{\bm{3488}}+\frac{5}{16}\,D^{3536}+3\,D^{3696}+3\,D^{3904}$ \\
& $+\frac{3}{16}\,D^{4000}+\frac{5}{4}\,D^{4112}+\frac{7}{16}\,D^{4208}+\frac{3}{4}\,D^{4272}+\frac{3}{16}\,D^{4320}+4\,D^{4368}+\frac{3}{8}\,D^{4416} + \cdots$ \\ [0.5ex] 
\hline
\multirow{2}{*}{$\dot{T}_{E2E}(D)$} & 
$7\,D^{2704}+4\,D^{2912}+\frac{9}{2}\,D^{3120}+2\,D^{3328}+\frac{5}{16}\,D^{3536}+D^{4056}+\frac{9}{2}\,D^{4264}+4\,D^{4368}$ \\
& $+\frac{33}{4}\,D^{4472}+12\,D^{4576}+\frac{37}{4}\,D^{4680}+12\,D^{4784}+\frac{15}{2}\,D^{4888}+5\,D^{4992}+\frac{157}{32}\,D^{5096} + \cdots$ \\ [1ex] 
 \hline

\end{tabular}
\end{table}

\begin{table}[H]
\caption{“Isotropic” 3D NSMs, with simple filters' coefficients, with common filter $\bm{h}_0$ pattern $\bm{\pi}_0 = (15, 15, 15, 15, 8, 8, 8, 8),$ arranged in a decreasing performance order, starting from the best one. Relative degradation in performance, due to distance reduction or/and multiplicity increase, when moving from one filter to the next, is emphasized in bold. 
} \label{table:3D NSMs Filter Pattern (15, 15, 15, 15, 8, 8, 8, 8)}

\centering
\begin{tabular}{|c|c|} 
\hline

\multicolumn{2}{|l|}{\# of simple $h_0[k,l,m]$ filters $ = 17920$} \\ \hline
\multicolumn{2}{|l|}{\# of non equivalent $h_0[k,l,m]$ filters $ = 52$} \\ \hline
\multicolumn{2}{|l|}{\# of non equivalent $h_0[k,l,m]$ filters with ($\dot{T}_{(1,0,0)}(D)=\dot{T}_{(0,1,0)}(D)=\dot{T}_{(0,0,1)}(D)$) $ = 12$} \\ \hline
\multicolumn{2}{|l|}{\# of non equivalent $h_0[k,l,m]$ filters with ($\dot{T}_{(1,0,0)}(D)=\dot{T}_{(0,1,0)}(D)=\dot{T}_{(0,0,1)}(D)$)} \\
\multicolumn{2}{|l|}{\& ($\dot{T}_{(1,1,0)}(D)=\dot{T}_{(1,-1,0)}(D)=\dot{T}_{(1,0,1)}(D)=\dot{T}_{(1,0,-1)}(D) =\dot{T}_{(0,1,1)}(D)=\dot{T}_{(0,1,-1)}(D)$) $ = 2$} \\ [0.5ex] \hline

\multicolumn{2}{|l|}{$\mathring{h}_1[k,l,m] = 34 \, \delta[k]\delta[l]\delta[l]$} \\ \hline
\multicolumn{2}{|l|}{\gls{msed}, $d_\text{min}^2 = 4\cdot34^2=4624$} \\ [0.5ex]
\hline\hline

\multicolumn{2}{|c|}{Filter~\#$1$} \\ \hline

\multirow{3}{*}{$\mathring{h}_0[k,l,m]$} & $15 \,\delta[k]\delta[l]\delta[m]+8\,\delta[k]\delta[l-1]\delta[m]+8\,\delta[k-1]\delta[l]\delta[m]-15\,\delta[k-1]\delta[l-1]\delta[m]$ \\
& $8 \,\delta[k]\delta[l]\delta[m-1]-15\,\delta[k]\delta[l-1]\delta[m-1]-15\,\delta[k-1]\delta[l]\delta[m-1]$ \\
& $-8\delta[k-1]\delta[l-1]\delta[m-1]$ \\\hline

$\mathring{\bm{h}}_0$ & Front slice: $\mathring{\bm{h}}_0[:,:,0]=\begin{pmatrix}
15 & 8 \\
8 & -15
\end{pmatrix}$, Back slice: $\mathring{\bm{h}}_0[:,:,1]=\begin{pmatrix}
8 & -15 \\
-15 & -8
\end{pmatrix}$ \\ [0.5ex] \hline
\multirow{2}{*}{$\dot{T}_{F2F}(D)$} & 
$5\,D^{4624}+4\,D^{5168}+\frac{9}{2}\,D^{5712}+D^{5984}+2\,D^{6256}+\frac{5}{2}\,D^{6528}+\frac{5}{16}\,D^{6800}+\frac{25}{4}\,D^{7072}$ \\
& $+\frac{7}{16}\,D^{7344}+\frac{127}{8}\,D^{7616}+D^{7888}+\frac{161}{8}\,D^{8160} + \cdots$ \\ [0.5ex] 
\hline
\multirow{2}{*}{$\dot{T}_{E2E}(D)$} & 
$7\,D^{4624}+4\,D^{5168}+\frac{9}{2}\,D^{5712}+2\,D^{6256}+\frac{5}{16}\,D^{6800}+D^{6936}+4\,D^{7072}+\frac{9}{2}\,D^{7480}$ \\
& $+12\,D^{7616}+\frac{33}{4}\,D^{8024}+12\,D^{8160} + \cdots$ \\ [0.5ex] 
 \hline\hline

\multicolumn{2}{|c|}{Filter~\#$2$} \\ \hline

\multirow{3}{*}{$\mathring{h}_0[k,l,m]$} & $15 \,\delta[k]\delta[l]\delta[m]+8\,\delta[k]\delta[l-1]\delta[m]+8\,\delta[k-1]\delta[l]\delta[m]+15\,\delta[k-1]\delta[l-1]\delta[m]$ \\
& $8 \,\delta[k]\delta[l]\delta[m-1]+15\,\delta[k]\delta[l-1]\delta[m-1]+15\,\delta[k-1]\delta[l]\delta[m-1]$ \\
& $+8\delta[k-1]\delta[l-1]\delta[m-1]$ \\\hline

$\mathring{\bm{h}}_0$ & Front slice: $\mathring{\bm{h}}_0[:,:,0]=\begin{pmatrix}
15 & 8 \\
8 & 15
\end{pmatrix}$, Back slice: $\mathring{\bm{h}}_0[:,:,1]=\begin{pmatrix}
8 & 15 \\
15 & 8
\end{pmatrix}$ \\ [0.5ex] \hline
\multirow{2}{*}{$\dot{T}_{F2F}(D)$} & 
$5\,D^{4624}+4\,D^{5168}+D^{\bm{5408}}+\frac{9}{2}\,D^{5712}+3\,D^{5952}+\frac{3}{4}\,D^{6192}+2\,D^{6256}+3\,D^{6496}$ \\
& $+\frac{3}{16}\,D^{6560}+2\,D^{6736}+\frac{5}{16}\,D^{6800}+\frac{1}{2}\,D^{6976}+\frac{5}{4}\,D^{7040}+4\,D^{7072}+\frac{7}{16}\,D^{7104} + \cdots$ \\ [0.5ex] 
\hline
\multirow{2}{*}{$\dot{T}_{E2E}(D)$} & 
$7\,D^{4624}+4\,D^{5168}+\frac{9}{2}\,D^{5712}+2\,D^{6256}+\frac{5}{16}\,D^{6800}+D^{6936}+4\,D^{7072}+\frac{9}{2}\,D^{7480}$ \\
& $+12\,D^{7616}+\frac{33}{4}\,D^{8024}+12\,D^{8160} + \cdots$ \\ [1ex] 
 \hline

\end{tabular}
\end{table}

\begin{figure}[!htbp]
    \centering
    \includegraphics[width=0.8\textwidth]{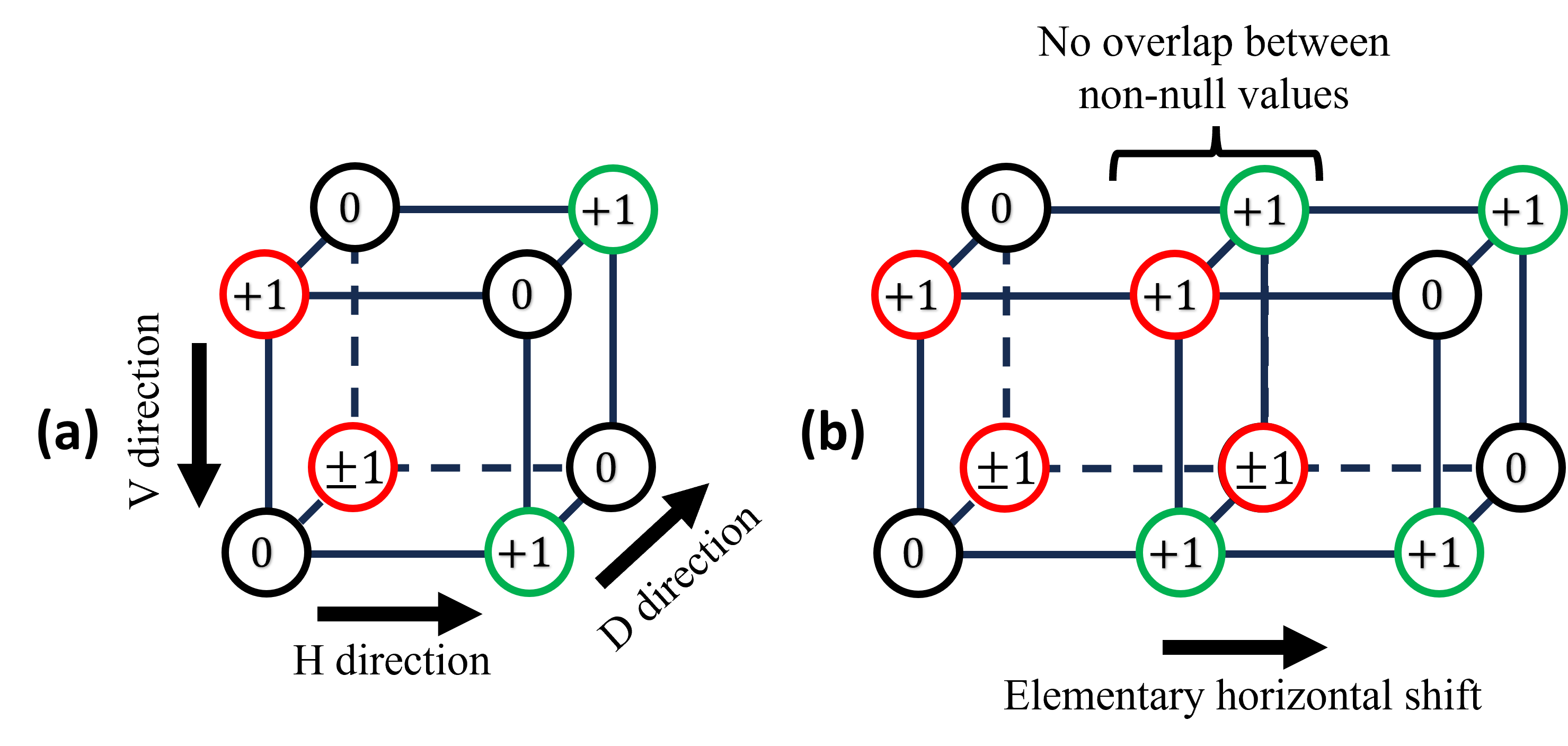}
    \caption{Illustration of the absence of overlap between non-zero taps in the vertically (V), horizontally (H), and depth (D) shifted versions of the footprints of isotropic filters associated with the filter pattern $\bm{\pi}_0 = (1,1,1,1,0,0,0,0).$ (a) 3D footprints of filters $\mathring{\bm{h}}_0 = ((1, 0; 0, 1) \mathbin{::} (0, 1; \pm 1, 0))$ (Filters \# $1$ and \# $2$ in Table~\ref{table:3D NSMs Filter Pattern (1, 1, 1, 1, 0, 0, 0, 0)}). (b) Elementary horizontal shift of the filters' footprint showing no overlap between non-zero taps in the shifted footprints.}
    \label{fig:No-Overlap 3D Pattern (11110000)}
\end{figure}

Unlike the other patterns considered, the filter pattern $\bm{\pi}_0 = (1,1,1,1,0,0,0,0)$ exhibits several distinctive and noteworthy features that merit further examination. First, the multiplicities of the dominant terms in the \glspl{rtf}, $\dot{T}_{F2F}(D)$ and $\dot{T}_{E2E}(D),$ are $\tfrac{253}{16}$ and $\tfrac{285}{16},$ respectively. When normalized by the number of input sequences for the corresponding extracted \gls{f2f} and \gls{e2e} \gls{1d} \glspl{nsm}—$5$ and $7,$ respectively—these values yield effective multiplicities of approximately $\tfrac{253}{80} \approx 3.16$ and $\tfrac{285}{112} \approx 2.54.$ Both exceed $1,$ indicating that at high \gls{snr}, the binary error probability of the \glspl{nsm} associated with this pattern (as detailed in Table~\ref{table:3D NSMs Filter Pattern (1, 1, 1, 1, 0, 0, 0, 0)}) is more than $2.5$ times greater than that of $2$-ASK. Notably, the multiplicity of the \gls{msed} error events for the \gls{f2f}-extracted \gls{1d} \gls{nsm} results in a leading term in its upper-bound probability given by $\tfrac{253}{80} \tfrac{1}{2} \operatorname{erfc} (\sqrt{ E_b/N_0}) = \tfrac{253}{160} \operatorname{erfc} (\sqrt{ E_b/N_0}),$ which matches the approximate bit error probability given in Equation (\ref{eq:BEP Estimate Rate 5/4}) for the illustrative rate-$5/4$ \gls{nsm} studied in Subsection~\ref{ssec:Minimum Euclidean Distance Guaranteeing 5/4-NSM}. This connection can be intuitively explained as follows: as shown in Figure~\ref{fig:No-Overlap 3D Pattern (11110000)}, two consecutive axial shifts of the footprint of $\mathring{h}_0[l,m,n],$ in the V, H or D direction, do not overlap in their non-zero components when considering Filters \# 1 and \# 2 from Table~\ref{table:3D NSMs Filter Pattern (1, 1, 1, 1, 0, 0, 0, 0)}. As a result, in the \gls{f2f}-extracted \gls{1d} \gls{nsm}, each shifted footprint behaves as if it operates independently, unaffected by neighboring shifts. For each such footprint, the corresponding output depends on a single input from the first input sequence $\bar{b}_0[i,j,k]$ along with four inputs from the second input sequence $\bar{b}_1[i,j,k],$ precisely those that overlap with the non-zero taps of the shifted version of $\mathring{h}_0[l,m,n],$ linked to that primary input. In this local configuration, the underlying structure effectively mimics the rate-$5/4$ block \gls{nsm}, as described in Subsection~\ref{ssec:Minimum Euclidean Distance Guaranteeing 5/4-NSM}.

The absence of overlap between \gls{f2f} shifts of the isotropic filters $\mathring{h}_0[l,m,n],$ $\bm{\pi}_0 = (1,1,1,1,0,0,0,0),$ as shown in Table~\ref{table:3D NSMs Filter Pattern (1, 1, 1, 1, 0, 0, 0, 0)}, leads to a significantly large second \gls{sed}. Specifically, the first and second \glspl{sed} are $16$ and $32,$ yielding a normalized second distance of $2$—indicating a substantial improvement. In contrast, for the other patterns, $\bm{\pi}_0 = (4,4,4,4,3,3,3,3),$ $(12,12,12,12,5,5,5,5)$ and $(15,15,15,15,8,8,8,8),$ the normalized increases are marginal: $480/400=6/5,$ $2912/2704=14/13$ and $5168/4624=19/17,$ respectively. This is due to the considerable overlap between consecutive \gls{f2f} shifts in any spatial direction (Vertical, Horizontal, or Depth), which reduces the relative gain in \gls{sed}. Moreover, this partial overlap introduces memory into the extracted \gls{1d} \gls{f2f} \glspl{nsm}, resulting in a two-state detection trellis. While the \gls{rtf} $\dot{T}_{F2F}(D)$ reflects this trellis structure, it is unnecessary for computing the distance spectrum in the case of pattern $\bm{\pi}_0 = (1,1,1,1,0,0,0,0).$ Due to the complete lack of overlap, in this case, there is no memory with respect to the first input $\bar{b}_0[l,m,n],$ with $(m,n)=(0,0)$ for the V direction, $(l,n)=(0,0)$ for the H direction and $(l,m)=(0,0)$ for the D direction. As a result, the additional terms in the \gls{rtf}—beyond the main term corresponding to the rate-$4/3$ block \gls{nsm} described in Subsection~\ref{ssec:Minimum Euclidean Distance Guaranteeing 5/4-NSM}—do not contribute to the distance spectrum and can be considered superfluous.

To reduce the overlap between consecutive shifts of the generator footprints for the patterns $\bm{\pi}_0 = (4,4,4,4,3,3,3,3),$ $(12,12,12,12,5,5,5,5)$ and $(15,15,15,15,8,8,8,8),$ one possible approach is to move to four or higher dimensions. For illustration purposes, we focus here on the \gls{4d} case, considering a $2 \times 2 \times 2 \times 2$ hypercube of size $16.$ In this \gls{4d} setting, the first filter, $\mathring{h}_0[m,n,p,q],$ occupies a hypercubic footprint with only $8$ non-zero taps. These taps are derived by applying permutations and sign changes to the components of the original $1 \times 8$ patterns $\bm{\pi}_0.$ To prevent overlap during \gls{f2f} shifts in four dimensions—i.e., shifts that involve changing only one of the indices $m$, $n$, $p$, or $q$—the $8$ non-zero taps of $\mathring{h}_0[m,n,p,q]$ can be placed at positions in a subset $P_4 \subset \{0,1\}^4$ that avoids such conflicts. One possible choice for $P_4$ is to build it hierarchically from lower-dimensional patterns: $P_2 = \{(0,0), (1,1)\},$ $P_3 = \{(\bm{r},0), (\bm{r},0)+(0,1,1) \bmod 2, \bm{r} \in P_2 \} = \{(0,0,0), (1,1,0), (0,1,1), (1,0,1)\}$ and $P_4 = \{(\bm{r},0), (\bm{r},0)+(0,0,1,1) \bmod 2, \bm{r} \in P_3 \} = \{(0,0,0,0), (1,1,0,0), (0,1,1,0), (1,0,1,0), (0,0,1,1), (1,1,1,1), (0,1,0,1), (1,0,0,1)\}.$ For reference, the \gls{3d} positions $P_3$ match the non-zero tap locations of the filter $\mathring{h}_0[l,m,n]$ shown in Figure~\ref{fig:No-Overlap 3D Pattern (11110000)}. Alternatively, the non-zero taps of $\mathring{h}_0[m,n,p,q]$ can occupy the complementary positions of this specific $P_4$ within the full set $\{0,1\}^4,$ i.e., $\{0,1\}^4 \setminus P_4.$ This alternative configuration also ensures that \gls{f2f} shifts do not result in overlapping taps.

It is clear that moving from three dimensions to four dimensions helps improve the distance spectrum properties of the resulting \gls{nsm}. Starting from a $1 \times 8$ pattern in three dimensions, this transition allows us to extend it to a $1 \times 16$ pattern in four dimensions by appending four zero components. In the same spirit, a \gls{3d} pattern such as $\bm{\pi}_0 = (1,1,1,1,0,0,0,0),$ which suffers from tightness issues in the \gls{3d} setting, can be replaced by the fully populated \gls{4d} pattern $\bm{\pi}_0 = (1,1,1,1,1,1,1,1,1,1,1,1,1,1,1,1),$ with all $16$ components equal to $1,$ leading to a better-performing configuration. Based on this design choice, we define the \gls{4d} filter $\mathring{h}_0[m,n,p,q]$ to occupy a $2 \times 2 \times 2 \times 2$ hyper-cubic footprint, with all values taken from the set $\{ \pm 1 \}$—a construction equivalent to applying permutations and arbitrary sign changes to the components of the \gls{4d} pattern $\bm{\pi}_0$. By setting the second filter as $\mathring{h}_1[m,n,p,q] = \| \bm{\pi}_0 \| \, \delta[m] \delta[n] \delta[p] \delta[q] = 4 \, \delta[m] \delta[n] \delta[p] \delta[q],$ we obtain a \gls{4d} \gls{nsm} with rational taps that preserves the \gls{sed} of $2$-ASK. This performance is ensured by the fact that, in any \gls{4d} error event, all $16$ vertices of the $2 \times 2 \times 2 \times 2$ hypercube appear at least once. Moreover, the construction asymptotically achieves rate $2$ and avoids the tightness problem, since in this case $\| \bm{\pi}_0 \|_\infty=1<\| \bm{\pi}_0 \|/2=2.$ The only drawback is that, for a finite \gls{4d} extent, the \gls{nsm} rate, $\rho,$ approaches $2$ more slowly than it does in the \gls{3d} case, as the extent increases. As a result, moving from \gls{3d} to \gls{4d} \glspl{nsm}—while maintaining the same offered rate—comes with an expected increase in complexity.

\begin{figure}[!htbp]
    \centering
    \includegraphics[width=1.0\textwidth]{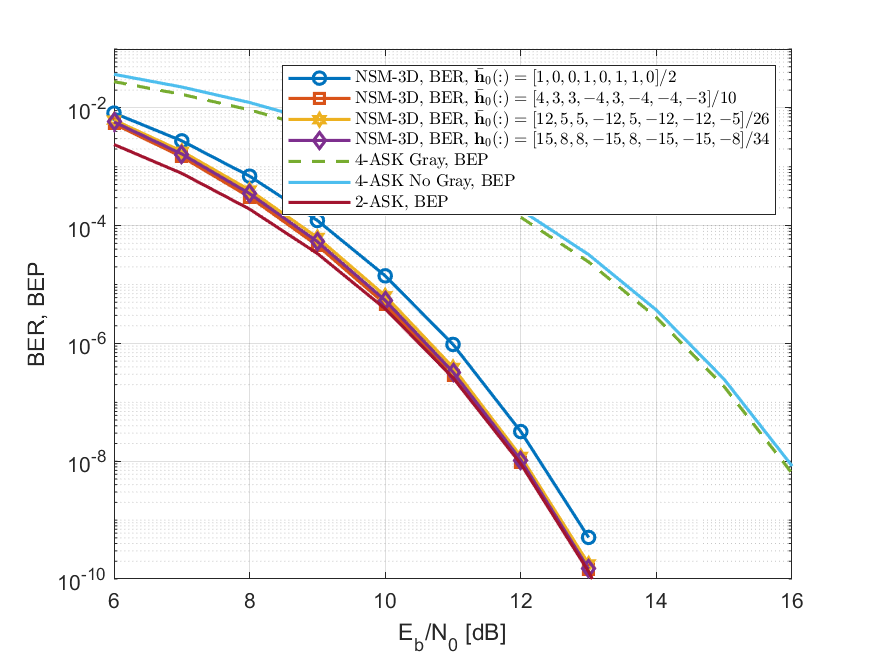}
    \caption{BER of the 3D rate $\rho = 35/27$ NSMs, corresponding to $I=J=K=3,$ using best isotropic 3D filters $\mathring{\bm{h}}_0 = ((1, 0; 0, 1) \mathbin{::} (0, 1; 1, 0))$ (pattern $\bm{\pi}_0 = (1, 1, 1, 1, 0, 0, 0, 0)$), $\mathring{\bm{h}}_0 = ((4, 3; 3, -4) \mathbin{::} (3, -4; -4, -3))$ (pattern $\bm{\pi}_0 = (4, 4, 4, 4, 3, 3, 3, 3)$), $\mathring{\bm{h}}_0 = ((12, 5; 5, -12) \mathbin{::} (5, -12; -12, -5))$ (pattern $\bm{\pi}_0 = (12, 12, 12, 12, 5, 5, 5, 5)$), and $\mathring{\bm{h}}_0 = ((15, 8; 8, -15) \mathbin{::} (8, -15; -15, -8))$ (pattern $\bm{\pi}_0 = (15, 15, 15, 15, 8, 8, 8, 8)$). For reference, the BEPs of $2$-ASK and Gray and non-Gray precoded $4$-ASK conventional modulations are presented.}
    \label{fig:BER_3D_NSM_Rate35_27}
\end{figure}

\begin{figure}[!htbp]
    \centering
    \includegraphics[width=1.0\textwidth]{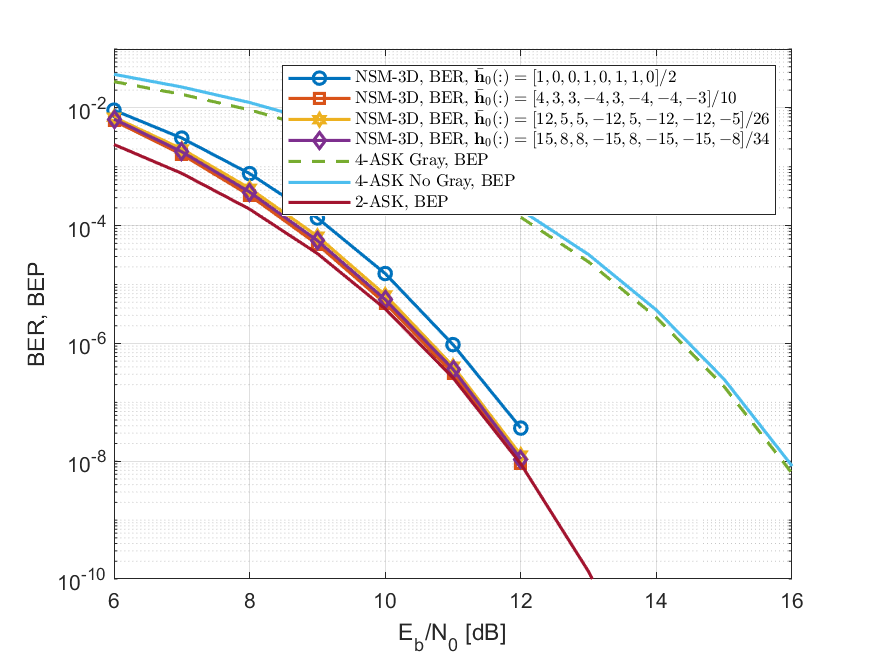}
    \caption{BER of the 3D rate $\rho = 48/36$ NSMs, corresponding to $I=4, J=K=3,$ using best isotropic 3D filters $\mathring{\bm{h}}_0 = ((1, 0; 0, 1) \mathbin{::} (0, 1; 1, 0))$ (pattern $\bm{\pi}_0 = (1, 1, 1, 1, 0, 0, 0, 0)$), $\mathring{\bm{h}}_0 = ((4, 3; 3, -4) \mathbin{::} (3, -4; -4, -3))$ (pattern $\bm{\pi}_0 = (4, 4, 4, 4, 3, 3, 3, 3)$), $\mathring{\bm{h}}_0 = ((12, 5; 5, -12) \mathbin{::} (5, -12; -12, -5))$ (pattern $\bm{\pi}_0 = (12, 12, 12, 12, 5, 5, 5, 5)$), and $\mathring{\bm{h}}_0 = ((15, 8; 8, -15) \mathbin{::} (8, -15; -15, -8))$ (pattern $\bm{\pi}_0 = (15, 15, 15, 15, 8, 8, 8, 8)$). For reference, the BEPs of $2$-ASK and Gray and non-Gray precoded $4$-ASK conventional modulations are presented.}
    \label{fig:BER_3D_NSM_Rate48_36}
\end{figure}

\begin{figure}[!htbp]
    \centering
    \includegraphics[width=1.0\textwidth]{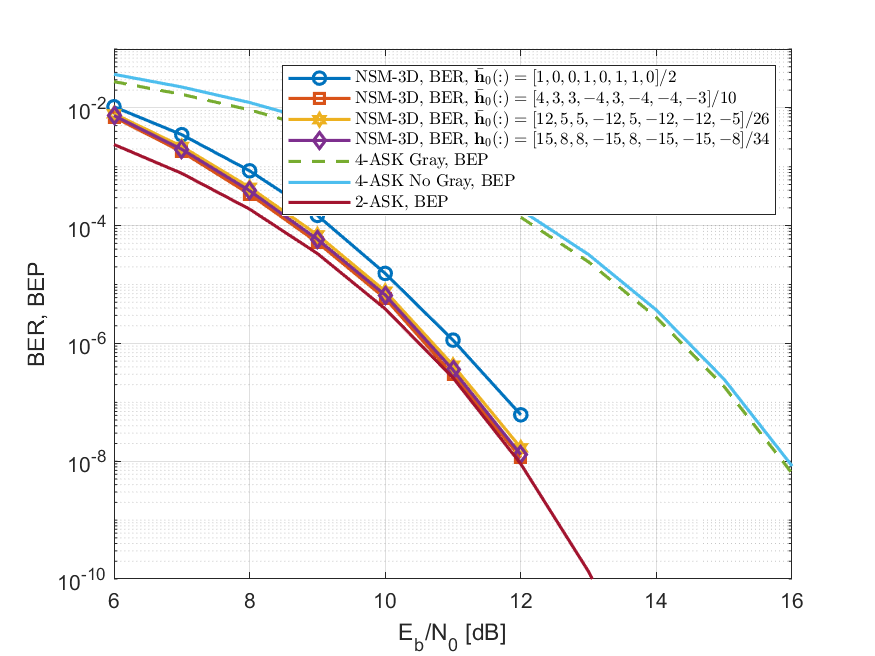}
    \caption{BER of the 3D rate $\rho = 66/48$ NSMs, corresponding to $I=J=4, K=3,$ using best isotropic 3D filters $\mathring{\bm{h}}_0 = ((1, 0; 0, 1) \mathbin{::} (0, 1; 1, 0))$ (pattern $\bm{\pi}_0 = (1, 1, 1, 1, 0, 0, 0, 0)$), $\mathring{\bm{h}}_0 = ((4, 3; 3, -4) \mathbin{::} (3, -4; -4, -3))$ (pattern $\bm{\pi}_0 = (4, 4, 4, 4, 3, 3, 3, 3)$), $\mathring{\bm{h}}_0 = ((12, 5; 5, -12) \mathbin{::} (5, -12; -12, -5))$ (pattern $\bm{\pi}_0 = (12, 12, 12, 12, 5, 5, 5, 5)$), and $\mathring{\bm{h}}_0 = ((15, 8; 8, -15) \mathbin{::} (8, -15; -15, -8))$ (pattern $\bm{\pi}_0 = (15, 15, 15, 15, 8, 8, 8, 8)$). For reference, the BEPs of $2$-ASK and Gray and non-Gray precoded $4$-ASK conventional modulations are presented.}
    \label{fig:BER_3D_NSM_Rate66_48}
\end{figure}

For the analysis of the most effective \gls{3d} \glspl{nsm} associated with each considered filter pattern—namely $\bm{\pi}_0 = (1,1,1,1,0,0,0,0),$ $(4,4,4,4,3,3,3,3),$ $(12,12,12,12,5,5,5,5)$ and $(15,15,15,15,8,8,8,8)$—we present, in Figures~\ref{fig:BER_3D_NSM_Rate35_27}--\ref{fig:BER_3D_NSM_Rate66_48}, the \gls{ber} curves of the \glspl{nsm} built from the best-performing filters, identified as Filter~\#$1$ in Tables~\ref{table:3D NSMs Filter Pattern (1, 1, 1, 1, 0, 0, 0, 0)}-- \ref{table:3D NSMs Filter Pattern (15, 15, 15, 15, 8, 8, 8, 8)}. These results correspond to finite \gls{3d} grid sizes of $(I,J,K)=(3,3,3),$ $(4,3,3)$ and $(4,4,3),$ offering respective \gls{nsm} rates of $\rho=\tfrac{35}{27},$ $\tfrac{48}{36}$ and $\tfrac{66}{48}.$ For comparison, the \gls{ber} performances of $2$-ASL $4$-ASK, with and without Gray precoding, are also included as benchmarks. As expected, all \gls{3d} \glspl{nsm}—except those associated with the pattern $\bm{\pi}_0 = (1,1,1,1,0,0,0,0)$—closely track the performance of $2$-ASK, even at moderate \gls{snr} levels. In contrast, the \gls{ber} curves corresponding to the $\bm{\pi}_0 = (1,1,1,1,0,0,0,0)$ pattern remain consistently parallel to the $2$-ASK curve, but at a noticeable distance, indicating a persistent performance gap. These results are in line with the anticipated limitations stemming from the tightness phenomenon previously discussed.

To conclude this section, it's worth emphasizing the crucial role played by the equivalence relation in significantly narrowing down the pool of candidate $\mathring{h}_0[l,m,n]$ filter representatives. For example, in the case of the pattern $\bm{\pi}_0 = (1,1,1,1,0,0,0,0),$ the total number of filter candidates was reduced from $1120$ to just $9$ non-equivalent ones. For the remaining patterns, the reduction was even more dramatic—from $17920$ initial candidates down to only $52$ distinct equivalence classes. This corresponds to a reduction factor of approximately $124$ for the $\bm{\pi}_0 = (1,1,1,1,0,0,0,0),$ pattern, and about $344$ for the others. By further enforcing both \gls{f2f} and \gls{e2e} isotropy constraints, the set of candidates was limited to only two per pattern, making the selection of the best filter straightforward. Of course, alternative selection strategies could have been considered without relying on isotropy conditions—for instance, by employing a single weighted criterion based on the \glspl{rtf} $\dot{T}_{F2F}(D)$ and $\dot{T}_{E2E}(D),$ allowing for a clear and consistent ranking of the non-equivalent candidates from best to worst.




\section{Rate-3 NSMs} \label{sec:Rate-3 NSMs}

We begin this section with a brief experimental analysis of high-performing rate-$3$ \glspl{nsm} that use use real filter taps, focusing on their asymptotic gain relative to $64$-QAM. These rate-$3$ \glspl{nsm} serve as alternatives to the $64$-QAM, much like rate-$2$ \glspl{nsm} are used as alternatives to $16$-QAM. We then consider rate-$3$ \glspl{nsm} with rational filter taps, introducing a specific \gls{nsm} that attains the \gls{msed} of $2$-ASK and thus asymptotically matches its performance as the \gls{snr} increases. The discussion in this case centers on this particular \gls{nsm}, from which we draw several interesting insights and conclusions.

\subsection{Good Rate-3 NSMs with Real Filter Coefficients} \label{Sseq:Good rate-3 NSMs with real filters' coefficients}

\begin{figure}[!htbp]
    \centering
    \includegraphics[width=0.75\textwidth]{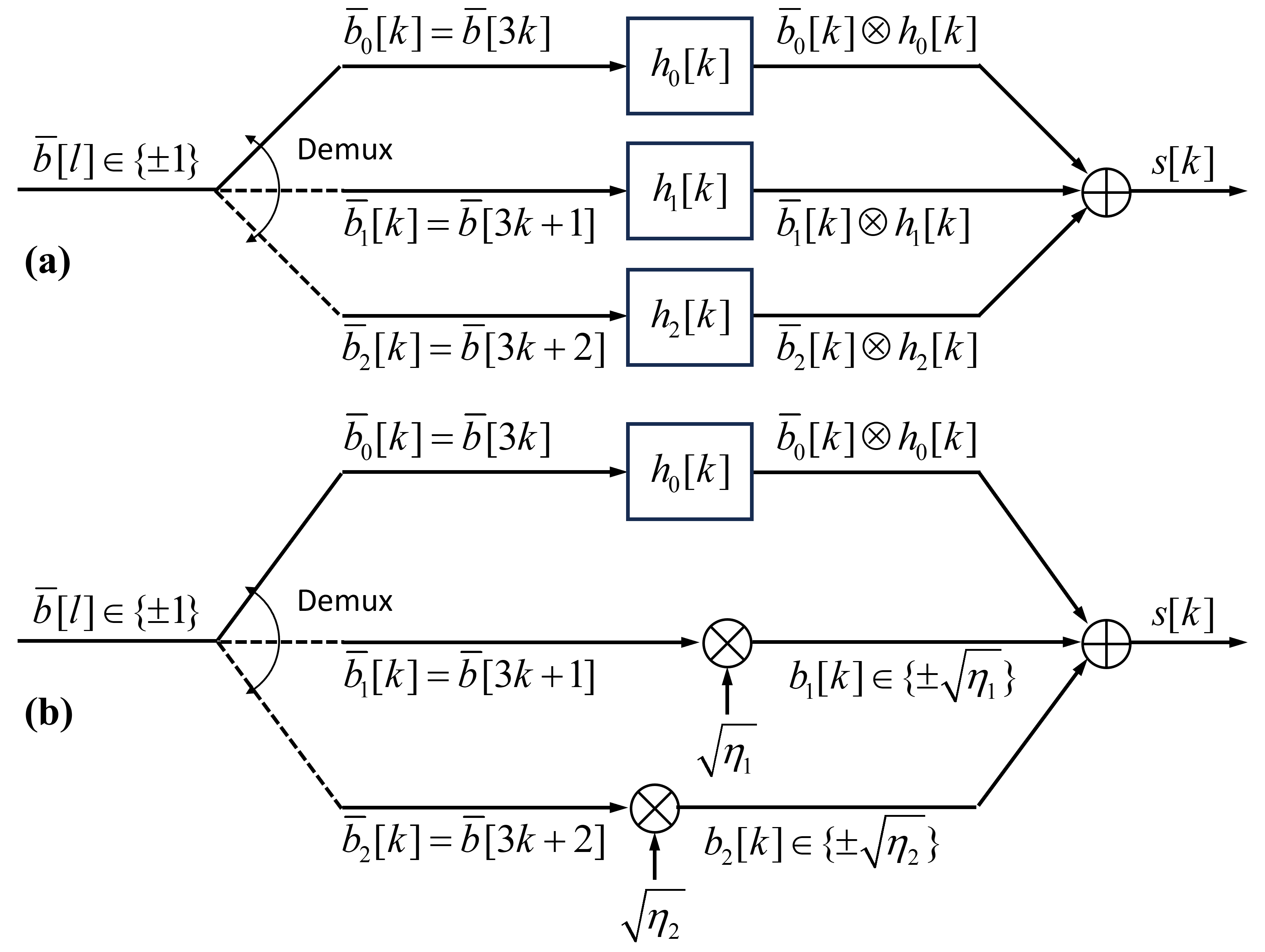}
    \caption{ Block diagram of the rate-$3$ Nyquist signaling modulator: (a) General form, (b) Specific form.}
    \label{fig:BlockDiagramRate3NSMs}
\end{figure}

The set of rate-$3$ \glspl{nsm} is constructed using three bipolar input sequences, $\bar{b}_m[k],$ $m=0,1,2,$ and three finite-length filters, $h_m[k]$, each of length $L_m$ and with \gls{sen} $\eta_m = \|h_m[k]\|^2$. As shown in Figure~\ref{fig:BlockDiagramRate3NSMs}(a), the output modulated sequence is given by the sum 
\begin{equation} \label{eq:Modulated Signal Rate-3 NSM}
s[k] = \sum_{m=0}^2 \sum_l \bar{b}_m[l] h_m[k-l] = \sum_{m=0}^2 \bar{b}_m[k] \circledast h_m[k].
\end{equation}
The average symbol energy of the resulting \gls{nsm} is thus $\eta_0 + \eta_1 + \eta_2.$

To enable a simple and direct comparison with $8$-ASK (or equivalently, $64$-QAM), we normalize this average symbol energy to $21,$ which is the average symbol energy of $8$-ASK. It's worth noting that $8$-ASK can itself be viewed as a rate-$3$ \gls{nsm} with filters given by $h_m[k] = 2^m \, \delta[k]$, for $m = 0, 1, 2$. In this case, the squared norms are $\eta_m = 2^{2m} = 4^m$, resulting in a total average symbol energy of $\eta_0+\eta_1+\eta_2=1+4+16=21.$

To determine the \gls{msed} of these rate-$3$ \glspl{nsm}, we analyze error events based on the differences between input sequences. Specifically, for $m = 0, 1, 2,$ we define $\Delta \bar{b}_m[k] \triangleq \bar{b}_m^1[k]-\bar{b}_m^0[k],$ where $\bar{b}_m^0[k]$ and $\bar{b}_m^1[k]$ are two bipolar input sequences. These difference sequences, $\Delta \bar{b}_m[k],$ take values from the ternary alphabet $\{0, \pm 2 \}.$ Let $s^0[k]$ and $s^1[k]$ be the modulated sequences corresponding to the input sequences triplets $b_m^0[k],$ $m = 0, 1, 2,$ and $b_m^1[k]$, $m = 0, 1, 2,$ respectively. The difference between the modulated sequences, defined as $\Delta s[k] \triangleq s^1[k] - s^0[k],$ is given by
\begin{equation} \label{eq:Modulated Signal Difference Rate-3 NSM}
\Delta s[k] = \sum_{m=0}^2 \sum_l \Delta \bar{b}_m[l] h_m[k-l].
\end{equation}
The \gls{msed}, $d_{\text{min}}^2,$ is obtained by minimizing the \gls{sen}, $\|\Delta s[k] \|^2,$ over all error events for which not all input differences $\Delta \bar{b}_m[k],$ $m = 0, 1, 2,$ are zero. For fixed filter lengths, $L_m,$ $m = 0, 1, 2,$ and real-valued filters $h_m[k],$ $m = 0, 1, 2,$ the goal is to design the filters to maximize $d_{\text{min}}^2$. As with the rate-$2$ \glspl{nsm} discussed in Section~\ref{Background}, it is insightful to derive an upper bound on the \gls{sed} for these rate-$3$ \glspl{nsm}. This upper bound helps us understand how to distribute the average symbol energy, specified by the filters contributions $\eta_m$, across the three input sequences $\bar{b}_m[k],$ $m = 0, 1, 2,$ to maximize $d_{\text{min}}^2.$

To extend the upper bound on $d_{\text{min}}^2$ derived in (\ref{Upper-Bound Minimum Squared Euclidean Distance}) for rate-$2$ \glspl{nsm} to the rate-$3$ case, we examine specific error events in which only one of the three input sequence differences is nonzero—say, $\bar{b}_n[k] = \pm 2 \, \delta[k],$ for some $n \in \{0, 1, 2\}$. In such a scenario, the resulting modulated sequence difference is simply $\Delta s[k] = \pm 2 \, h_n[k],$ and its squared norm is $\|\Delta s[k] \|^2 = 4 \, \|h[k] \|^2 = 4 \, \eta_n.$ This implies that the \gls{msed}, $d_{\text{min}}^2,$ is upper-bounded by the smallest of these three values. Therefore, we conclude that
\begin{equation} \label{eq:Upper Bound Minimum Distance Rate-3 NSM}
d_{\text{min}}^2 \le 4 \, \min (\eta_0, \eta_1, \eta_2).
\end{equation}

Since the filters energy contributions $\eta_0, \eta_1,$ and $\eta_2$ are constrained to sum to $21$—aligned with the average symbol energy of $8$-ASK—their minimum value, $\min (\eta_0, \eta_1, \eta_2),$ is upper bounded by $7.$ This upper bound is achieved only when $\eta_0 = \eta_1 = \eta_2 = 7$, meaning that to maximize the \gls{msed}, $d_{\text{min}}^2$, the average symbol energy must be evenly distributed among the three input sequences $\bar{b}_m[k].$ This perfectly balanced configuration, or any close configuration, where the $\eta_m$ values are all near $7,$ is highly promising for maximizing $d_{\text{min}}^2.$ However, it also carries a major drawback: it increases the risk of destructive interference between the contributions of the individual input sequences $\bar{b}_m[k],$ to the modulated sequence, $s[k],$ potentially reducing the actual \gls{msed} to well below the theoretical maximum. To effectively explore this critical—but promising—region and still achieve a high \gls{msed}, $d_{\text{min}}^2,$ the only viable strategy is to significantly increase the lengths $L_m$ of filters $h_m[k]$, $m=0,1,2,$ within the limits of practical complexity. As observed in the analysis of rate-$2$ \glspl{nsm} in Subsection~\ref{Rate-2 guaranteeing, minimum Euclidean distance approaching NSMs with real filters' coefficients}, the upper bound on $d_{\text{min}}^2$—equal to $10$ and achievable when $\eta_0 = \eta_1 = 5/2$—was attained only when $L_0 \geq 10$, with $L_1$ fixed at $1.$ In all cases where $L_0 < 10$ (still with $L_1 = 1$), this maximum value was never reached, despite extensive and intensive optimization efforts.

A desirable and ultimate rate-$3$ \gls{nsm}, aiming for a performance comparable to that of $2$-ASK (which operates at rate $1$), should behave as if it consists of three independent and non-interfering multiplexed $2$-ASK streams. If the energy allocated to a $2$-ASK within each stream is $\eta$, then a rate-$3$ \gls{nsm} should have a cumulative average symbol energy of $3,\eta$. Similarly to rate-$2$ \glspl{nsm}, achieving $2$-ASK performance corresponds to the configuration $\eta_0 = \eta_1 = \eta_2 = \eta$, which maximizes the upper bound on the \gls{msed}, $d_{\text{min}}^2$, as given in (\ref{eq:Upper Bound Minimum Distance Rate-3 NSM}). Therefore, to attain performance comparable to $2$-ASK, we must maintain comparable values for $\eta_0$, $\eta_1$, and $\eta_2$, a situation where destructive interference among the bipolar input sequences is maximized. To effectively combat this interference, it becomes essential to increase the filters lengths $L_0$, $L_1$, and $L_2$.

To achieve the \gls{msed} of $2$-ASK by a rate-$3$ \gls{nsm}, it is necessary to ensure that $\eta_0 = \eta_1 = \eta_2 = 7$, assuming alignment with the average symbol energy of $8$-ASK, for ease of comparison. Since the input sequences are bipolar, taking values from the alphabet ${\pm1}$, the ideal \gls{msed} in this case is $d_{\text{min}} = 4 \times 7 = 28$. Consequently, the asymptotic \gls{snr} gap between any arbitrary rate-$3$ \gls{nsm} with \gls{msed} $d_{\text{min}}^2$ and $2$-ASK can be calculated as $-10 \, \log_{10}(d_{\text{min}}^2/28)$, expressed in decibels (dB).

It is important to note that for $8$-ASK, we have $\eta_0 = 1$, $\eta_1 = 4$, and $\eta_2 = 16$, with the \gls{msed} given by $d_{\text{min}}^2 = 4 \, \min(\eta_0, \eta_1, \eta_2) = 4$. As a result, the asymptotic \gls{snr} gain achieved by any rate-$3$ \gls{nsm}, relative to $8$-ASK (which is equivalent to $64$-QAM), at identical spectral efficiency, is calculated as $10 \, \log_{10}(d_{\text{min}}^2/4)$, expressed in dB.

To optimize rate-$3$ \glspl{nsm}, we sought the best filters $h_m[k],$ $m=0,1,2,$ under the conditions $L_2=1,$ $1 \le L_1 \le 5,$ and $\max(L_1,2) \le L_0 \le 11-L_1.$ For each configuration, a simulated annealing-like algorithm \cite{Aarts88} was employed to maximize the \gls{msed}, $d_{\text{min}}^2.$ Additionally, Algorithm~\ref{alg:d_min^2 Rate-3 NSM L_0>1, L_1>1, L_2>1} was repeatedly applied to compute $d_{\text{min}}^2$ for the admissible candidate filters of each configuration.

\begin{algorithm}[h]
\caption{Determination of the MSED, $d_{\text{min}}^2,$ of a rate-$3$ NSM with $L_0>1,$ $L_1>1$ and $L_2>1$} \label{alg:d_min^2 Rate-3 NSM L_0>1, L_1>1, L_2>1}
\begin{algorithmic}
\Require $h_0[k],$ $h_1[k],$ $h_2[k],$ $L_0>1,$ $L_1>1,$ $L_2>1$
\Ensure $d_{\text{min}}^2,$ $\text{Degeneracy}$
\State $\infty \gets 10^{6}$ \Comment{Plays the role of infinity}
\State $k \gets 0$
\For {$\Delta \bar{b}_m[k-l] \in \{0,\pm 2\}, l=0,1,\ldots,L_m-1, m=0,1,2$}
\State $\bm{\sigma}[k-1] \gets \Delta \bar{b}_0[k-(L_0-1)] \ldots \Delta \bar{b}_0[k-1], \Delta \bar{b}_1[k-(L_1-1)] \ldots \Delta \bar{b}_1[k-1], \Delta \bar{b}_2[k-(L_2-1)] \ldots \Delta \bar{b}_2[k-1]$
\State $\bm{\sigma}[k] \gets \Delta \bar{b}_0[k-(L_0-2)] \ldots \Delta \bar{b}_0[k], \Delta \bar{b}_1[k-(L_1-2)] \ldots \Delta \bar{b}_1[k], \Delta \bar{b}_2[k-(L_2-2)] \ldots \Delta \bar{b}_2[k]$
\State $\Delta s(\bm{\sigma}[k-1], \bm{\sigma}[k])  \gets \sum_{l=0}^{L_0-1} h_0[l] \Delta \bar{b}_0[k-l] + \sum_{l=0}^{L_1-1} h_1[l] \Delta \bar{b}_1[k-l] + \sum_{l=0}^{L_2-1} h_2[l] \Delta \bar{b}_2[k-l]$
\State $m(\bm{\sigma}[k-1], \bm{\sigma}[k])  \gets (\Delta s(\bm{\sigma}[k-1], \bm{\sigma}[k]))^2$
\EndFor

\State Follows Algorithm~\ref{alg:d_min^2 Rate-2 NSM L_0>1, L_1>1} from Step \ref{step:start determination minimum squared Euclidean distance algorithm common step 2}

\end{algorithmic}
\end{algorithm}

For the same value of $L_0,$ the optimized \glspl{nsm} with $L_1=1$ consistently achieved better \glspl{msed} than those with $L_1>1.$ Ideally, higher minimum distances should have be obtained, as $L_1$ increases for a fixed $L_0,$ since larger filters offer greater design flexibility. However, in practice, the optimization algorithm often became trapped in local maxima of the achievable minimum distance, when $L_1>1.$ This difficulty mainly stems from the increased size of the search space as the filter sizes grow, particularly when it comes to filter $h_1[k].$ Consequently, we chose to exclude all scenarios with $L_1>1$ and focused instead on the more tractable cases where $L_1=L_2=1$ and $L_0>1.$ Figure~\ref{fig:BlockDiagramRate3NSMs}(b) illustrates the structure of the modulator, as determined under these simplifying and restrictive constraints. For simplicity, we also used Algorithm~\ref{alg:d_min^2 Rate-3 NSM L_0>1, L_1=L_2=1}—specifically tailored for these settings—to compute the \gls{msed}, $d_{\text{min}}^2,$ for each triplet of candidate generators $h_m[k],$ $m=0,1,2.$

\begin{algorithm}[h]
\caption{Determination of the MSED, $d_{\text{min}}^2,$ of a rate-$3$ NSM with $L_0>1$ and $L_1=L_2=1$} \label{alg:d_min^2 Rate-3 NSM L_0>1, L_1=L_2=1}
\begin{algorithmic}
\Require $h_0[k],$ $h_1[0],$ $h_2[0],$ $L_0>1$
\Ensure $d_{\text{min}}^2,$ $\text{Degeneracy}$
\State $\infty \gets 10^{6}$
\State $k \gets 0$
\For {$\Delta \bar{b}_0[k-l] \in \{0,\pm 2\}, l=0,1,\ldots,L_0-1$}
\State $\bm{\sigma}[k-1] \gets \Delta \bar{b}_0[k-(L_0-1)] \ldots \Delta \bar{b}_0[k-1]$
\State $\bm{\sigma}[k] \gets \Delta \bar{b}_0[k-(L_0-2)] \ldots \Delta \bar{b}_0[k]$

\For {$\Delta \bar{b}_1[k] \in \{0,\pm 2\}$}
\For {$\Delta \bar{b}_2[k] \in \{0,\pm 2\}$}
\State $\Delta s(\bm{\sigma}[k-1], \bm{\sigma}[k], \Delta \bar{b}_1[k], \Delta \bar{b}_2[k])  \gets \sum_{l=0}^{L_0-1} h_0[l] \Delta \bar{b}_0[k-l] + h_1[0] \Delta \bar{b}_1[k] + h_2[0] \Delta \bar{b}_2[k]$
\EndFor
\EndFor

\State $m(\bm{\sigma}[k-1], \bm{\sigma}[k])  \gets 
\min_{(\Delta \bar{b}_1[k], \Delta \bar{b}_2[k]) \in \{ 0, \pm 2 \}^2}
(\Delta s(\bm{\sigma}[k-1], \bm{\sigma}[k], \Delta \bar{b}_1[k], \Delta \bar{b}_2[k]))^2$
\EndFor
\State $m(\bm{0}, \bm{0})  \gets 
\min_{(\Delta \bar{b}_1[k], \Delta \bar{b}_2[k]) \in \{0, \pm 2 \}^2 \setminus \{(0,0) \} }
(\Delta s(\bm{0}, \bm{0}, \Delta \bar{b}_1[k]), \Delta \bar{b}_2[k]))^2$

\State Follows Algorithm~\ref{alg:d_min^2 Rate-2 NSM L_0>1, L_1>1} from Step \ref{step:start determination minimum squared Euclidean distance algorithm common step}

\end{algorithmic}
\end{algorithm}

Table~\ref{table:Best Filters Numerical Form Rate-3 NSM} presents the characteristics of several notable rate-$3$ \glspl{nsm} for $2 \leq L_0 \leq 8$, with $L_1 = L_2 = 1$. These characteristics include the underlying normalized filters, $\hat{\bar{h}}_m[k]$ for $m = 0,1,2$, the corresponding average symbol energy contributions $\hat{\eta}_m[k]$, the achieved \gls{msed} $d_{\text{min}}^2$, the theoretical upper bound (in dB) on the achievable gain, given by $10 \, \log_{10}(\min(\hat{\eta}_0, \hat{\eta}_1, \hat{\eta}_2)),$ and the actual gain in dB achieved relative to $8$-ASK. From these characteristics, the non-normalized filters $\hat{h}_m[k],$ $m=1,2,3,$ can be fully determined using the relation $\hat{h}_m[k] = \sqrt{\hat{\eta}_m} \, \hat{\bar{h}}_m[k]$, where $\hat{\bar{h}}_m[k] = \delta[k]$ for $m=1,2$. We restricted our analysis to $L_0 \leq 8$, although promising results were also obtained for $L_0 = 9$ and $L_0 = 10$, which show additional asymptotic \gls{snr} gains compared to $8$-ASK. However, we chose not to include these cases, as the consistency of the filters for these higher values of $L_0$ remains uncertain.

\begin{table}[H]
\caption{Characteristics of good rate-$3$ NSMs, in numerical form, with $L_0$ ranging from $2$ to $8$ ($L_1=L_2=1$ and $\bar{\hat{h}}_1[k]=\bar{\hat{h}}_2[k]=\delta[k]$). $\star$: Filter $h_0[k]$ resulting from rate-$2$ NSMs optimization. $\dagger$: Filter $h_0[k]$ resulting from rate-$3$ NSMs optimization.}
\label{table:Best Filters Numerical Form Rate-3 NSM}
\centering
\begin{tabular}{|c||c|c|} 
\hline

$L_0$ & $2^{\star\dagger}$ & $3^{\star\dagger}$ \\ \hline
$\bar{\hat{h}}_0[0]$ & $0.707106781186547$ & $0.653281482438188$ \\
$\bar{\hat{h}}_0[1]$ & $0.707106781186547$ & $0.382683432365090$ \\
$\bar{\hat{h}}_0[2]$ & - & $0.653281482438188$ \\ \hline
$\hat{\eta}_0$ & $1.909090909090910$ & $2.202288850017553$ \\ \hline
$\hat{\eta}_1$ & $3.818181818181818$ & $3.759542229996490$ \\ \hline
$\hat{\eta}_2$ & $15.272727272727273$ & $15.038168919985958$ \\ \hline
$\hat{d}_{\text{min}}^2$ & $7.636363636363629$ & $8.809155400070200$ \\ \hline
$10 \, \log_{10}(\min(\hat{\eta}_0,\hat{\eta}_1,\hat{\eta}_2))$ & $2.808266095756943$ & $3.428742800043420$ \\ \hline
Gain [dB] & $2.808266095756938$ & $3.428742800043413$ \\ [0.5ex] 
 \hline\hline

$L_0$ & $4^{\star}$ & $\bm{4^\dagger}$ \\ \hline
$\bar{\hat{h}}_0[0]$ & $0.627963030199554$ & $0.561516668266344$ \\
$\bar{\hat{h}}_0[1]$ & $0.398112608509063$ & $0.561516668266344$ \\
$\bar{\hat{h}}_0[2]$ & $0.229850421690492$ & $\pm 0.232587819494474$ \\
$\bar{\hat{h}}_0[3]$ & $0.627963030199554$ & $\pm 0.561516668266344$  \\ \hline
$\hat{\eta}_0$ & $2.363068255485155$ & $2.874342226955987$ \\ \hline
$\hat{\eta}_1$ & $3.727386348902969$ & $3.625131554608803$  \\ \hline
$\hat{\eta}_2$ & $14.909545395611875$ & $14.500526218435210$ \\ \hline
$\hat{d}_{\text{min}}^2$ & $9.452273021940615$ & $9.738160085207779$ \\ \hline
$10 \, \log_{10}(\min(\hat{\eta}_0,\hat{\eta}_1,\hat{\eta}_2))$ & $3.734762660896059$ & $\bm{4.585384751517942}$ \\ \hline
Gain [dB] & $3.734762660896056$ & $\bm{3.864169182894832}$ \\ [0.5ex] 
 \hline\hline

$L_0$ & $5^\star$ & $\bm{5^\dagger}$ \\ \hline
$\bar{\hat{h}}_0[0]$ & $0.589216898252358$ & $0.561516668266344$ \\
$\bar{\hat{h}}_0[1]$ & $0.498472935623522$ & $0.561516668266344$ \\
$\bar{\hat{h}}_0[2]$ & $0$ & $0$ \\
$\bar{\hat{h}}_0[3]$ & $-0.239105888841342$ & $\pm 0.232587819494474$ \\
$\bar{\hat{h}}_0[4]$ & $0.589216898252358$ & $\pm 0.561516668266344$ \\\hline
$\hat{\eta}_0$ & $2.643660587657938$ & $2.874342226955987$ \\ \hline
$\hat{\eta}_1$ & $3.671267882468412$ & $3.625131554608803$ \\ \hline
$\hat{\eta}_2$ & $14.685071529873650$ & $14.500526218435210$ \\ \hline
$\hat{d}_{\text{min}}^2$ & $10.574642350631722$ & $10.982108573202868$ \\ \hline
$10 \, \log_{10}(\min(\hat{\eta}_0,\hat{\eta}_1,\hat{\eta}_2))$ & $4.222056965135653$ & $\bm{4.585384751517942}$ \\ \hline
Gain [dB] & $4.222056965135641$ & $\bm{4.386257416631158}$ \\ [1ex] 
\hline
\end{tabular}
\end{table}

\begin{table}[H]
\captionsetup{list=no} 
\caption*{Table~\ref{table:Best Filters Numerical Form Rate-3 NSM}
 (Continued).}
\centering
\begin{tabular}{|c|c|c|} 
\hline

$L_0$ & $6^{\star\dagger}$ & $7^\star$ \\ \hline
$\bar{\hat{h}}_0[0]$ & $0.554700196225229$ & $0.542805974620653$ \\
$\bar{\hat{h}}_0[1]$ & $0$ & $0.542805974620653$ \\
$\bar{\hat{h}}_0[2]$ & $0.277350098112615$ & $0$ \\
$\bar{\hat{h}}_0[3]$ & $0.554700196225229$ & $0.284912667360959$ \\
$\bar{\hat{h}}_0[4]$ & $0$ & $0$ \\
$\bar{\hat{h}}_0[5]$ & $0.554700196225229$ & $0.186841627389698$ \\
$\bar{\hat{h}}_0[6]$ & - & $0.542805974620653$ \\ \hline
$\hat{\eta}_0$ & $2.935483847206641$ & $3.046672179299205$ \\ \hline
$\hat{\eta}_1$ & $3.612903230558672$ & $3.590665564140159$ \\ \hline
$\hat{\eta}_2$ & $14.451612922234688$ & $14.362662256560636$ \\ \hline
$\hat{d}_{\text{min}}^2$ & $11.741935307223184$ & $12.186688717196777$ \\ \hline
$10 \, \log_{10}(\min(\hat{\eta}_0,\hat{\eta}_1,\hat{\eta}_2))$ & $4.676796919532168$ & $4.838257268250200$ \\ \hline
Gain [dB] & $4.676796919532168$ & $4.838257268250185$ \\ [0.5ex] 
 \hline\hline

$L_0$ & $7^\dagger$ & $8^{\star\dagger}$ \\ \hline
$\bar{\hat{h}}_0[0]$ & $0.490290337845460$ & $0.516339056518458$  \\
$\bar{\hat{h}}_0[1]$ & $0.490290337845460$  & $0.269675949396294$ \\
$\bar{\hat{h}}_0[2]$ & $\pm 0.490290337845460$ & $0.516339056518458$ \\
$\bar{\hat{h}}_0[3]$ & $\pm 0.196116135138184$ & $0$ \\
$\bar{\hat{h}}_0[4]$ & $0$ & $0$ \\
$\bar{\hat{h}}_0[5]$ & $0$ & $0$ \\
$\bar{\hat{h}}_0[6]$ & $\pm 0.490290337845460$ & $0.357010950053487$ \\
$\bar{\hat{h}}_0[7]$ & - & $0.516339056518458$  \\ \hline
$\hat{\eta}_0$ & $3.615894039735100$ & $3.316424607889151$ \\ \hline
$\hat{\eta}_1$ & $3.476821192052980$ & $3.536715078422170$ \\ \hline
$\hat{\eta}_2$ & $13.907284768211920$ & $14.146860313688679$ \\ \hline
$\hat{d}_{\text{min}}^2$ & $12.238410596026483$ & $13.265698431556606$ \\ \hline
$10 \, \log_{10}(\min(\hat{\eta}_0,\hat{\eta}_1,\hat{\eta}_2))$ & $\bm{5.411823561127874}$ & $5.206701289523861$ \\ \hline
Gain [dB] & $\bm{4.856650282629558}$ & $5.206701289523861$ \\ [1ex] 
\hline

\end{tabular}
\end{table}

In Table~\ref{table:Best Filters Numerical Form Rate-3 NSM}, each column is labeled with a dagger ($\dagger$), an asterisk ($\star$), or both ($\dagger\star$). Columns marked with a dagger ($\dagger$) correspond to fully optimized rate-$3$ \glspl{nsm}. Those marked with both a dagger and an asterisk ($\dagger\star$) represent a subset of these \glspl{nsm} that, as a result of full rate-$3$ \gls{nsm} optimization, happen to share the same normalized filters, $\bar{h}_0[k],$ as fully optimized rate-$2$ \glspl{nsm}, with identical filter lengths, $L_0.$ In contrast, columns marked only with a dagger ($\dagger$) correspond to fully optimized rate-$3$ \glspl{nsm} whose optimal normalized filters, $\bar{h}_0[k],$ differ from those of rate-$2$ \glspl{nsm}, despite having the same filter lengths, $L_0.$

The numerical results in Table~\ref{table:Best Filters Numerical Form Rate-3 NSM}, highlight a key observation: for all fully optimized rate-$3$ \glspl{nsm}, with $2 \le L_0 \le 8,$ corresponding to columns marked with either a dagger ($\dagger$) or both a dagger and an asterisk ($\dagger\star$), the three filters $h_m[k],$ $m=0,1,2,$ consistently satisfy an extensive set of tightness conditions. This behavior closely mirrors what was previously observed for the two filters $h_m[k],$ $m=0,1,$, in fully optimized rate-$2$ \glspl{nsm}. Specifically, in the rate-$2$ case, the tightness conditions required that $\eta_0 + \eta_1 = 5,$ along with structural constraints such as $h_1[0] = 2 \, h_0[0] = 2 \, h_1[L_0-1],$ In comparison, the tightness conditions for rate-$3$ \glspl{nsm} are both more involved and more comprehensive: they require $\eta_0 + \eta_1 + \eta_2 = 21,$ and include at least the relations $h_2[0] = 2 \, h_1[0] = 4 \, h_0[0] = 4 \, h_0[L_0-1],$ thereby imposing coordinated structure across all three filters.

The observation that fully optimized rate-$3$ \glspl{nsm}—represented by columns marked exclusively with a dagger ($\dagger$)—produce normalized filters, $\bar{h}_0[k],$ different from those of the fully optimized rate-$2$ \glspl{nsm} (despite having the same normalized filter lengths, $L_0$), while still satisfying the full set of tightness conditions, motivated us to explore an alternative design. Specifically, we evaluated a new class of rate-$3$ \glspl{nsm} that retain the same filter lengths, $L_0,$ but adopt the normalized filters, $\bar{h}_0[k],$ from the corresponding rate-$2$ \glspl{nsm}, instead of using the one obtained through full rate-$3$ optimization. Importantly, these \glspl{nsm} still enforce the complete set of tightness constraints, namely $h_2[0] = 2 \, h_1[0] = 4 \, h_0[0]= 4 \, h_0[L_0-1],$ across all three filters. These correspond precisely to the remaining columns in Table~\ref{table:Best Filters Numerical Form Rate-3 NSM}, identified by the asterisk symbol ($\star$).

Assuming filter lengths, $L_1=L_2=1,$ and the partial tightness constraint, $h_2[0] = 2 \, h_1[0],$ which is satisfied by all considered rate-$3$ \gls{nsm} filters, the contribution of the bipolar input sequences $\bar{b}_m[k],$ for $m=1,2,$ to the modulated signal in (\ref{eq:Modulated Signal Rate-3 NSM}), namely
\begin{equation}
s_{12}[k]= h_1[0]\bar{b}_1[k]+h_2[0]\bar{b}_2[k] = h_1[0](\bar{b}_1[k]+2\bar{b}_2[k]),
\end{equation}
is equivalent to a non-Gray-coded $4$-ASK modulation. Therefore, the overall modulated signal $s[k],$ in (\ref{eq:Modulated Signal Rate-3 NSM}), can be interpreted as the superposition of the contribution from the first input sequence, $\bar{b}_0[k],$ represented by $s_0[k] = \bar{b}_0[k] \circledast h_0[k],$ and a $4$-ASK modulation.

The restriction of the filter lengths to $L_1=L_2=1$ is precisely what gives rise to the partial tightness constraint $h_2[0] = 2 \, h_1[0].$ This constraint emerges as the only way to prevent the contributions $s_1[k] = \bar{b}_1[k] \circledast h_1[k]= h_1[0] \bar{b}_1[k]$ and $s_2[k] = \bar{b}_2[k] \circledast h_2[k] = h_2[0] \bar{b}_2[k],$ to the modulated signal, $s[k]$, in (\ref{eq:Modulated Signal Rate-3 NSM}), from destructively interfering, as in conventional $4$-ASK. This constraint prevents the resulting rate-$3$ \gls{nsm} from asymptotically matching the performance of $2$-ASK, since the corresponding \gls{msed}, $d_{\text{min}}^2,$ will always remain strictly less than $28$—the threshold required to attain such asymptotic performance. Specifically, given that $h_2[0] = 2 \, h_1[0],$ it follows that $\eta_2 = 4 \,\eta_1.$ With the constraint $\eta_0 + \eta_1 + \eta_2 = 21,$ we can deduce that $\eta_0 = 21 - 5 \,\eta_1,$ and since $\eta_1 < \eta_2,$ we find that the \gls{msed} is bounded above by
\begin{equation}
d_{\text{min}}^2 \le 4 \, \min (\eta_0 = 21-5 \, \eta_1, \eta_1).    
\end{equation}
The upper bound in this equation is maximized when $\eta_0 = 21-5 \, \eta_1 = \eta_1,$ implying that the optimal value occurs at $\eta_0 = \eta_1 = 7/2.$ This yields a maximum upper bound on the \gls{msed} of $4 \cdot 7/2 = 14,$ corresponding to a maximum asymptotic \gls{snr} gain of $10 \log_{10}(7/2) \approx 5.441$ dB compared to $8$-ASK, and a minimum \gls{snr} gap of approximately $10 \, \log_{10}(7/(7/2)) \approx 3.01,$ dB relative to $2$-ASK. The only way to reduce this \gls{snr} gap is to allow at least the filter $h_1[k]$ to have an unconstrained length $L_1 > 1.$

Table~\ref{table:Best Filters Numerical Form Rate-3 NSM} complements the preceding discussion by presenting the best achievable gains for \glspl{nsm} with $L_1=L_2=1$ and $2 \le L_0 \le 8,$ all satisfying the partial tightness condition. The data supports earlier observations, showing a gain relative to $8$-ASK that rises from approximately $2.808$ dB, at $L_0=2,$ to about $5.207$ dB, at $L_0=8.$ Remarkably, at $L_0=8,$ the gain comes within $0.234$ dB of the theoretical maximum for this class of configurations.

The upper bound on the \gls{msed}, given in (\ref{eq:Upper Bound Minimum Distance Rate-3 NSM}), is exactly achieved only when the normalized filters, $\bar{h}_0[k],$ from optimized rate-$2$ \glspl{nsm} are used. The improvement brought by the fully optimized normalized filters for rate-$3$ \glspl{nsm} builds directly on this fact. Specifically, for filter lengths $L_0 = 4, 5,$ and $7$, the first components $\bar{h}_0[0]$ in the rate-$3$ optimized filters, $\bar{h}_0[k],$ are smaller than those in their rate-$2$ counterparts. This reduction leads to lower values of $\eta_1$ and $\eta_2$—the energy contributions from the second and third input sequences $\bar{b}_1[k]$ and $\bar{b}_2[k]$—in the total average symbol energy budget $\eta_0 + \eta_1 + \eta_2 = 21,$ with a consistently more pronounced drop in $\eta_2$, reflecting the tightness condition $\eta_2 = 4\eta_1$. This decrease creates room to significantly increase $\eta_0$, the portion of the average symbol energy associated with the first input sequence $\bar{b}_0[k]$. Since the upper bound in (\ref{eq:Upper Bound Minimum Distance Rate-3 NSM}) is determined by the smallest of the $\eta_m$ values, $m=0,1,2,$ and $\eta_0$ always turns out to be the minimum, this raises the upper bound accordingly. However, the reduction in $\bar{h}_0[0]$ also intensifies self-interference in the signal component $s_0[k] = \bar{b}_0[k] \circledast h_0[k]$, which represents the contribution from $\bar{b}_0[k]$ to the overall modulated signal $s[k],$ as described in (\ref{eq:Modulated Signal Rate-3 NSM}). As a result, when the contribution $s_0[k]$ is examined in isolation—under the assumption that the second and third input sequences, $\bar{b}_1[k]$ and $\bar{b}_2[k]$, remain unchanged—it exhibits an \gls{msed} strictly below $4\eta_0.$ This explains why the upper bound in (\ref{eq:Upper Bound Minimum Distance Rate-3 NSM}) is never reached for these configurations. Crucially, this result shows that the decrease in \gls{sed} for $s_0[k]$ is more than compensated by the substantial increase in its energy contribution $\eta_0$ to the total symbol energy—an essential finding that underpins the overall performance gain achieved through the full optimization of the rate-$3$ filters, $h_0[k]$.

Before concluding this section, it is worth discussing the fully optimized rate-$3$ \glspl{nsm}, for $L_0 = 4$ and $5,$ as they support the preceding analysis. As shown in Table~\ref{table:Best Filters Numerical Form Rate-3 NSM}, the contributions, $\eta_m,$ $m=0,1,2,$ from the three input sequences to the average symbol energy are identical in both cases. Additionally, for the filter $h_0[k],$ the first and last coefficients, $h_0[0]$ and $h_0[L_0-1],$ are equal up to a sign, and all non-zero coefficients have the same magnitude, differing only by sign. Although the theoretical upper bound on the asymptotic gain is the same in both configurations, the case with $L_0 = 5$ comes closer to achieving it. This improvement results from reduced self-interference in the signal component $s_0[k],$ enabled by increasing the length of $h_0[k]$ by one. This extension is achieved by inserting a zero after the first two non-zero coefficients in the $L_0 = 4$ configuration.

\subsection{Exploring Rate-3 NSMs with Rational Filter Coefficients}

As previously noted, optimizing rate-$3$ \glspl{nsm} with real-valued filter taps becomes particularly difficult when $L_1 > 1,$ even when we take $L_2=1.$ We recall that all solutions found using optimization methods resembling simulated annealing consistently get trapped in local maxima, which, paradoxically, led to better results in the more constrained case of $L_1 = 1$. To overcome this issue efficiently and without significant performance loss, we can restrict the filter taps, $h_m[k]$, for $m = 0, 1, 2$, to rational values. This approach substantially speeds up the search for good \glspl{nsm} even when $L_1$ and $L_2$ are greater than one, and also simplifies the implementation. Rational-valued filters, when appropriately scaled to integers, facilitate more precise and efficient modulation and demodulation operations.

To clearly understand and internalize this concept, we now construct, step-by-step, a layered rate-$3$ \gls{nsm}. The modulated sequence $s[k]$, of assumed finite length $Q$, is formed as the sum of three partial modulated sequences $s_m[k]$, $m = 0,1,2$, each generated from a corresponding input sequence $\bar{b}_m[k]$ of finite length. In particular, the third partial modulated sequence, $s_2[k],$ is produced using a simple \gls{1d} convolution with filter $h_2[k] = 4 \, \delta[k]$, resulting in $s_2[k] = \bar{b}_2[k] \circledast h_2[k]=4 \, \bar{b}_m[k],$ which shows that $\bar{b}_2[k]$ is simply scaled and retains length $Q$.

We construct the partial modulated signal $s_1[k],$ using a \gls{2d} structural framework similar to the one introduced in Section~\ref{ssec:Two-Dimensional Rate-2 NSMs} for rate-$2$ \glspl{nsm}. We assume that the sequence length, $Q,$ can be factored as $Q=IJ,$ where $I, J \ge 2,$ and are chosen to be as close as possible to each other in value, so as to maximize spectral efficiency. Starting with the \gls{1d} input sequence $\bar{b}_1[k],$ we reshape it into an equivalent \gls{2d} sequence, $\bar{b}_1[m,n],$ defined over an $(I-1) \times (J-1)$ rectangular grid. This implies that $\bar{b}_1[k]$ has length $(I-1)(J-1).$ We then filter $\bar{b}_1[m,n]$ with a \gls{2d} filter selected from the class of non-equivalent $2 \times 2$ footprint filters $h_1[i,j]=2 (\delta[i]\delta[j] + \delta[i-1]\delta[j] + \delta[i]\delta[j-1] \pm \delta[i-1]\delta[j-1]).$ These filters are scaled versions (by a factor of $2$) of the \gls{2d} filters $h_0[k,l]= \delta[k]\delta[l] + \delta[k-1]\delta[l] + \delta[k]\delta[l-1] \pm \delta[k-1]\delta[l-1],$ discussed in Section~\ref{ssec:Two-Dimensional Rate-2 NSMs} in the context of \gls{2d} rate-$2$ \glspl{nsm}. To help reduce the multiplicity of \gls{msed} error events in the rate-$3$ \gls{nsm} being constructed, we specifically choose the filter with a negative sign on the cross term, namely $h_1[i,j]=2 (\delta[i]\delta[j] + \delta[i-1]\delta[j] + \delta[i]\delta[j-1] - \delta[i-1]\delta[j-1]).$ The resulting convolved sequence $s_1[m,n]=\bar{b}_1[m,n] \circledast  h_1[m,n],$ spanning an $I \times J$ rectangular region, is finally reshaped into a \gls{1d} sequence $s_1[k],$ of length $IJ=Q.$

The initial \gls{1d} input sequence, $\bar{b}_0[k],$ is modulated within a \gls{4d} framework, as introduced at the end of Section~\ref{ssec:Three-Dimensional Rate-2 NSMs} in the context of \gls{3d} rate-$2$ \glspl{nsm}. We assume that the length of the resulting modulated sequence, $s[k],$ denoted by $Q$, can be factored as $Q = KLMN$, with integers $K, L, M, N \ge 2$. To maximize spectral efficiency, these factors are chosen to be as close in value as possible. The sequence $\bar{b}_0[k]$ is then mapped onto an equivalent \gls{4d} array, $\bar{b}_0[m,n,p,q],$ defined over a $(K-1) \times (L-1) \times (M-1) \times (N-1)$ hyper-rectangular grid. Consequently, the original sequence is assumed to contain exactly $(K-1)(L-1)(M-1)(N-1)$ elements. The transformed sequence $\bar{b}_0[m,n,p,q]$ is then convolved with a \gls{4d}, $2 \times 2 \times 2 \times 2$ hyper-cubic filter $h_0[i,j,k,l]$, whose non-zero entries belong to the bipolar set ${\pm 1}$ for indices $0 \le i,j,k,l \le 1$. This convolution yields a \gls{4d} output sequence $s_0[m,n,p,q] = \bar{b}_0[m,n,p,q] \circledast h_0[i,j,k,l]$, which spans a $K \times L \times M \times N$ hyper-rectangular grid. The resulting 4D sequence is then converted into a \gls{1d} sequence $s_0[k]$ of length $Q = KLMN$. To date, no systematic study has been conducted on the equivalence classes of \gls{4d} filters $h_0[i,j,k,l]$—those arising through transformations that preserve modulation performance properties—and there is no known optimal class in terms of minimizing the \gls{sed} multiplicity for the target rate-$3$ \gls{nsm}. However, any of the candidate \gls{4d} filters considered yields the same \gls{msed} for the resulting modulation.

The modulated sequence $s[k]$, of total length $Q = IJ = KLMN$, is constructed as the sum of three component sequences: $s[k] = s_0[k] + s_1[k] + s_2[k].$ The total number of bipolar input symbols is then $Q+(I-1)(J-1)+(K-1)(L-1)(M-1)(N-1),$ accounting respectively for the contributions from $s_2[k]$, $s_1[k]$, and $s_0[k]$. This leads to an overall offered \gls{nsm} rate $\rho = (Q + (I-1)(J-1)+(K-1)(L-1)(M-1)(N-1))/Q = 1 + (1-1/I)(1-1/J) + (1-1/K)(1-1/L)(1-1/M)(1-1/N).$ Assuming a square footprint in two dimensions for $s_1[m,n]$ and a hyper-cubic footprint in four dimensions for $s_0[m,n,p,q]$, we set $I = J$ and $K = L = M = N$. This implies $I = K^2$ and thus $Q = I^2 = K^4$. Substituting into the rate expression yields: $\rho=1+(1-1/K^2)^2+(1-1/K)^4.$ As expected, this rate approaches $3$ in the limit as $K \to \infty$—that is, as the modulated sequence size $Q$ tends to infinity. Unfortunately, the asymptotic rate is reached only very slowly. For instance, with a modulated sequence, $s[k],$ of length $Q = K^4 = 10^4,$ the achieved rate is approximately $\rho = 2.636,$ which remains well below the target value of $3,$ even though the modulated sequence size is already quite large.

To demonstrate that the resulting asymptotic rate-$3$ \gls{nsm} achieves the \gls{msed} of $2$-ASK, we begin by observing that $\eta_0 = \eta_1 = \eta_2 = 16$, where these values represent the squared norms of the respective filters: $\eta_0 = \|h_0[i,j,k,l]\|^2$, $\eta_1 = \|h_1[i,j]\|^2$, and $\eta_2 = \|h_2[k]\|^2$. According to the discussion immediately following equation (\ref{eq:Upper Bound Minimum Distance Rate-3 NSM}), this implies that the \gls{msed} of the \gls{nsm} must satisfy $d_{\text{min}}^2 = 4 \cdot 16 = 64,$ which matches the \gls{msed} of $2$-ASK. This condition is indeed satisfied. To verify this, observe that the partial sum $s_{12}[k] = s_1[k] + s_2[k]$, when scaled by a factor of $1/2$, is exactly equivalent to the modulated output of one of the rate-$2$ \glspl{nsm} discussed in Section~\ref{ssec:Two-Dimensional Rate-2 NSMs}, specifically those generated by filter pattern $\bm{\pi}_0 = (1, 1, 1, 1)$. It has already been established that such rate-$2$ \glspl{nsm} attain an \gls{msed} of $16.$ Now, turning back to the constructed rate-$3$ \gls{nsm}, consider error events characterized by input sequence differences $\Delta \bar{b}_m[k]$, $m = 0, 1, 2$, under the assumption that $\Delta \bar{b}_0[k]$ is identically zero. In this case, the resulting modulated sequence differences, $\Delta s[k],$ is exactly equal to the partial sequence differences, $\Delta s_{12}[k]$. From the previously established connection with the rate-$2$ \glspl{nsm}, discussed in Section~\ref{ssec:Two-Dimensional Rate-2 NSMs}, we know that these error events satisfy $\| \Delta s[k] \|^2 = \| \Delta s_{12}[k] \|^2 \ge 4 \cdot 16 = 64.$ Therefore, this class of error events always results in \glspl{sed} that are bounded below by the \gls{msed} of $2$-ASK.

To complete the analysis, we now consider error events in which the first input sequence difference, $\Delta \bar{b}_0[k]$, is nonzero. In this case, we examine the modulo-$2$ version of the scaled modulated sequence differences, $\Delta s[k]/2,$ defined as $\Delta \mathring{s}[k] \triangleq (\Delta s[k]/2) \bmod 2.$ Since each $\Delta \bar{b}_m[k]$, $m = 0, 1, 2$, takes values in the ternary set $\{0, \pm 2\}$, the scaled sequence $\Delta s[k]/2$ is guaranteed to consist of integer components. Therefore, the \gls{sen} of $\Delta s[k]/2$ is always greater than or equal to that of $\Delta \mathring{s}[k],$ with the modulo-$2$ values being interpreted as integers in the set $\{0, 1\}.$ This implies that the \gls{sed}, $\|\Delta s[k]\|^2,$ of the error event, is lower bounded by four times the \gls{sen} of $\Delta \mathring{s}[k].$ As discussed at the end of Section~\ref{ssec:Three-Dimensional Rate-2 NSMs} in the context of 4D rate-$2$ \glspl{nsm} with filter pattern $\bm{\pi}_0 = (1,1,1,1,1,1,1,1,1,1,1,1,1,1,1,1),$ the sequence $\Delta \mathring{s}[k]$, when reinterpreted in four dimensions, reproduces, modulo $2,$ all $16$ vertices of the \gls{4d} filter $h_0[m,n,p,q]$ at least once. As a result, we can assert that $\| \Delta \mathring{s}[k] \|^2 \ge \| h_0[m,n,p,q] \|^2 = \eta_0 = 16.$ Consequently, the \gls{sed} of this class of error events is also lower bounded by $4 \cdot 16 = 64$.

In summary, regardless of the error event, the corresponding \gls{sed} is always bounded below by $64,$ which also serves as the upper bound for the \gls{msed}, $d_{\text{min}}^2$, of the constructed asymptotic rate-$3$ \gls{nsm}. This confirms that the constructed \gls{nsm}, while asymptotically achieving a rate of $3,$ consistently guarantees the \gls{msed} of $2$-ASK. The result provides a straightforward example demonstrating that it is indeed possible to design asymptotically rate-$3$ achieving \glspl{nsm} using simple rational-valued filter coefficients. Moreover, this construction can naturally be extended to higher-rate \glspl{nsm}, by exploring various higher-dimensional embeddings, beyond the two- and four-dimensional embeddings employed here.

Before concluding this section, it is both insightful and informative to observe the following. First, when the constructed rate-$3$ \gls{nsm} is reduced to a rate-$2$ \gls{nsm} by eliminating the first input sequence, $\bar{b}_0[k]$ (i.e., setting $\bar{b}_0[k]$ to a null sequence), the result is equivalent to one of the rate-$2$ \glspl{nsm} described in detail in Section~\ref{ssec:Two-Dimensional Rate-2 NSMs}. On the other hand, if we transform the constructed rate-$3$ \gls{nsm} into a rate-$2$ \gls{nsm} by eliminating the second input sequence, $\bar{b}_1[k]$ (i.e., setting $\bar{b}_1[k]$ to a null sequence), we obtain one of the \gls{4d} rate-$2$ \glspl{nsm} with the filter pattern $\bm{\pi}_0 = (1,1,1,1,1,1,1,1,1,1,1,1,1,1,1,1),$ which were thoroughly characterized in Section~\ref{ssec:Three-Dimensional Rate-2 NSMs}. Clearly, both transformations of the rate-$3$ \gls{nsm} achieve the \gls{msed} of $2$-ASK. However, the second transformation results in a significantly lower multiplicity for the error events corresponding to the \gls{msed}.



\section{Rate-3/2 NSMs with Rational Filter Coefficients} \label{Rate-3/2 Approaching NSMs}

\subsection{A Basic Rate-3/2 NSM Inspired by a Previous 2D Rate-2 NSM} \label{ssec:Basic Rate-3/2 NSM}

The new family of \glspl{nsm} presented in this subsection is directly derived from the \gls{2d} \glspl{nsm} proposed in Subsection~\ref{ssec:Two-Dimensional Rate-2 NSMs}. It corresponds to the specific case in which one of the \gls{2d} grid's rectangular extents, $I$ and $J,$ is equal to its smallest value, $2.$ As a result, it can be classified as \gls{1d} and, as we will see, allows for trellis representations, which greatly simplify detection at the receiver. Other than that, it ensures a minimum squared Euclidean distance of $4,$ as in $2$-ASK.

Since the two extents, $I$ and $J,$ play symmetrical roles, we can take $J=2.$ As a result, the offered \glspl{nsm} rates, which are expressed in Subsection~\ref{ssec:Two-Dimensional Rate-2 NSMs} as $\rho = 1 + (1-1/I)(1-1/J),$ for arbitrary values of $I$ and $J,$ greater than or equal to $2,$ are now given by $\rho = 1 + (1-1/I)/2 = 3/2 -1/2I.$ These rates go to $3/2$ as the one dimensional extent $I$ goes to infinity. As a result, when $I$ goes to infinity, these \glspl{nsm} approach conventional $8$-PSK and $8$-QAM in terms of spectrum efficiency.

\begin{figure}[!htbp]
    \centering
    \includegraphics[width=0.8\textwidth]{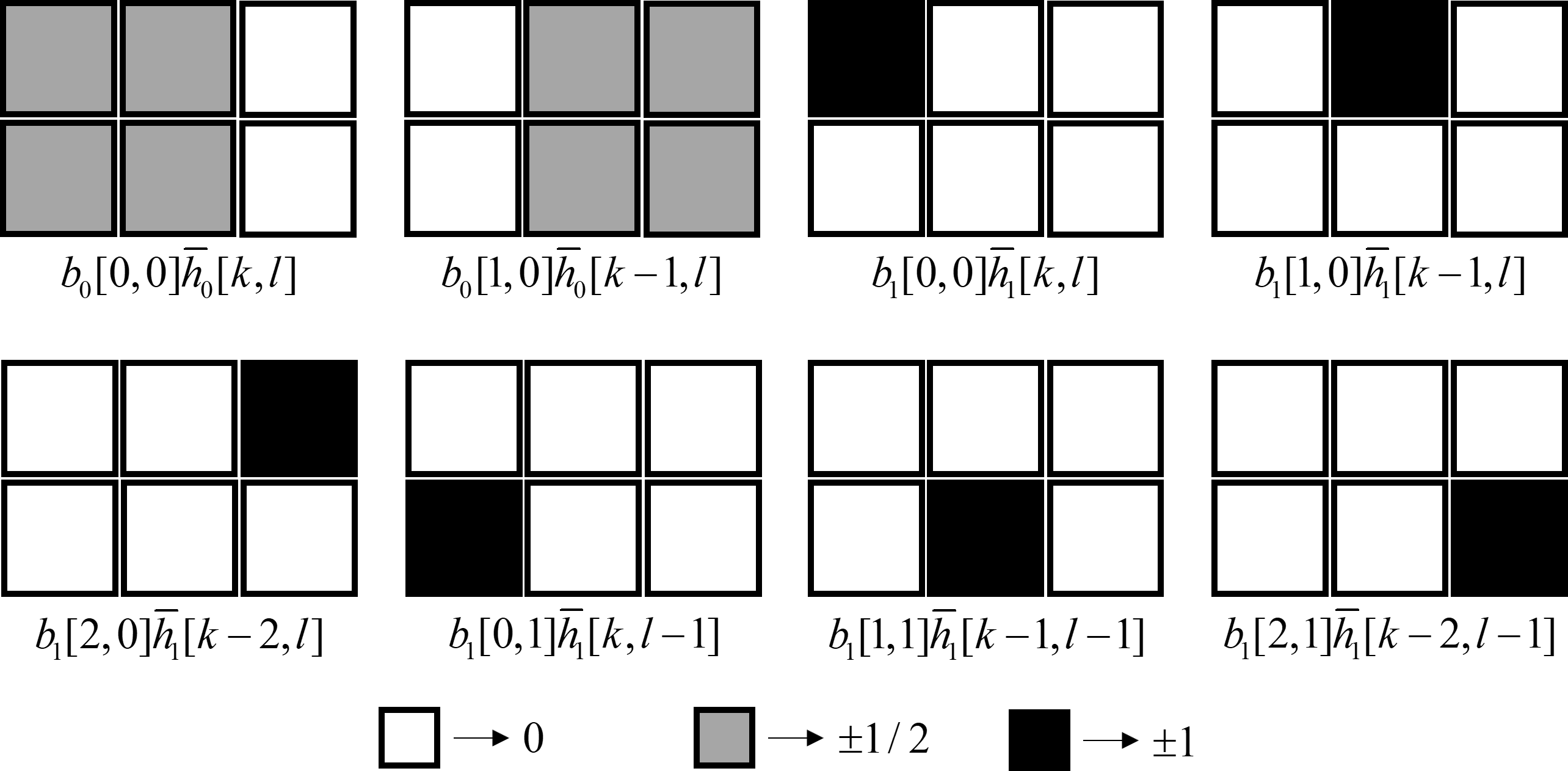}
    \caption{Interpretation in two dimensions of the 1D rate-$4/3$ block NSM associated with $I=3.$}
    \label{fig:One-Dimensional-NSM-Illustration-Rate-4-3}
\end{figure}

\begin{figure}[!htbp]
    \centering
    \includegraphics[width=0.8\textwidth]{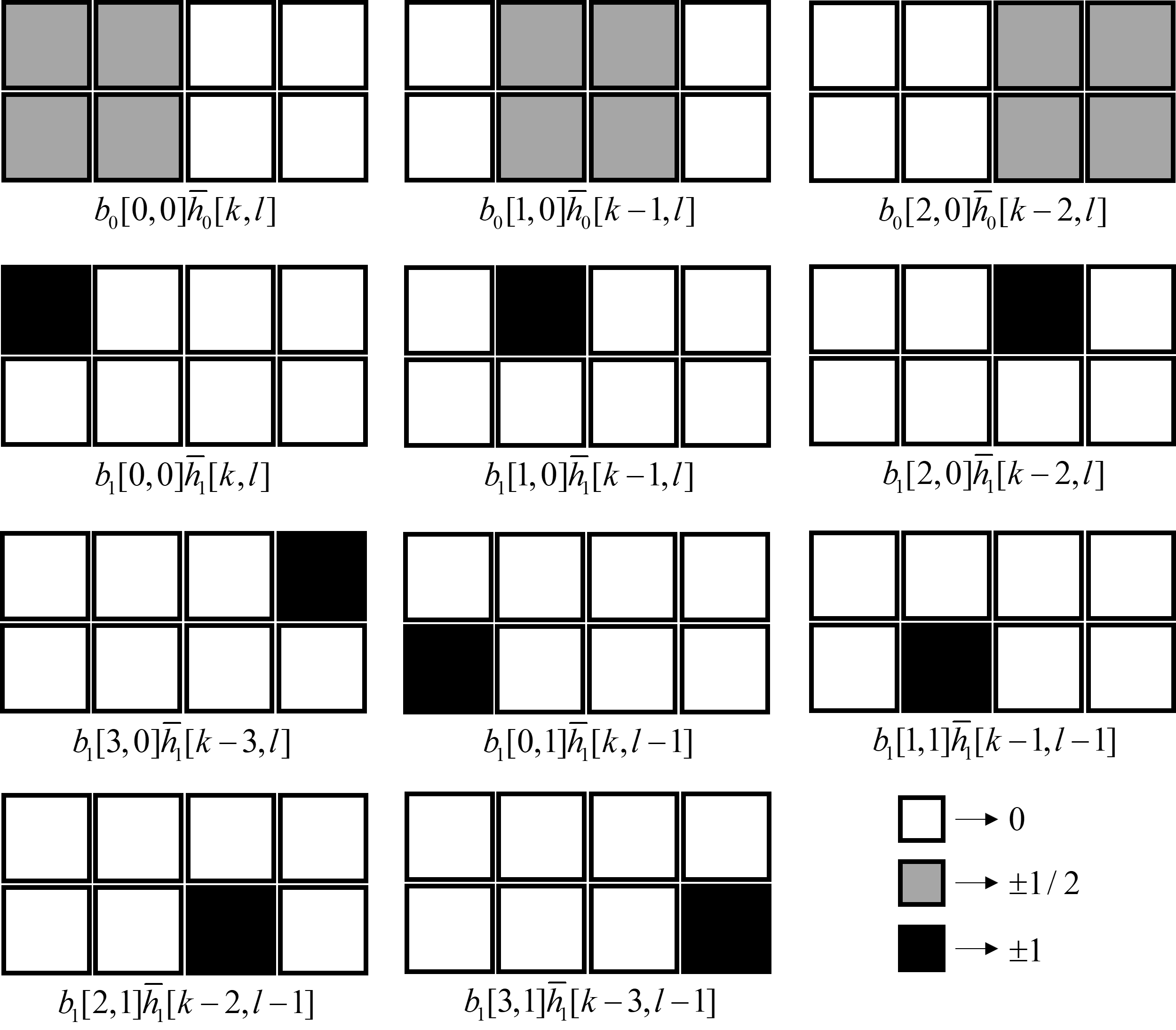}
    \caption{Interpretation in two dimensions of the 1D rate-$11/8$ block NSM associated with $I=4.$}
    \label{fig:One-Dimensional-NSM-Illustration-Rate-11-8}
\end{figure}

The smallest rate offered by this family of \glspl{nsm}, which is achieved for $I=2,$ is $\rho = 5/4.$ It corresponds to the rate-$5/4$ \gls{nsm} that has first been proposed in Subsection~\ref{ssec:Minimum Euclidean Distance Guaranteeing 5/4-NSM}, for illustration purpose, and that has also been at the origin of the new family of \gls{2d} \glspl{nsm} proposed in Subsection~\ref{ssec:Two-Dimensional Rate-2 NSMs}. The \gls{2d} layout of this \gls{nsm} has already been depicted in Figure~\ref{fig:Two-Dimensional-NSM-Illustration-Rate-5-4}, as part of the discussions about \gls{2d} \glspl{nsm}, in Subsection~\ref{ssec:Two-Dimensional Rate-2 NSMs}. For the sake of illustration and completeness, Figures~\ref{fig:One-Dimensional-NSM-Illustration-Rate-4-3} and~\ref{fig:One-Dimensional-NSM-Illustration-Rate-11-8} show the \gls{2d} layouts of the \glspl{nsm} with rates $\rho=4/3$ and $11/8,$ corresponding to $I=3$ and $4,$ respectively.

In what follows, we characterize the resulting \glspl{nsm} as much as we can, just as we did for the rate-$2$ \gls{nsm} presented for illustration in Subsection~\ref{ssec:Modulation of Rate 2}. Reading the \gls{2d} input data symbols $\bar{b}_0[k,l]$ and $\bar{b}_1[k,l],$ which take their values in the bipolar set $\{ \pm 1 \},$ column-wise and considering the infinite extent \gls{nsm} of rate $3/2,$ corresponding to $I = \infty$ and $J=2,$ we can write the normalized modulated symbols, as in (\ref{eq:Mod Seq FTN}), as
\begin{equation}  \label{eq:Normalized Modulated Symbols Rate 3/2 MSN}
    \bar{s}[k] = \sum_{m=0}^2 \sum_l \bar{b}_m[l] \bar{h}_m[k-2l],
\end{equation}
where $\bar{b}_0[k] \triangleq \bar{b}_0[0,k],$ $\bar{b}_m[k] \triangleq \bar{b}_1[m-1,k],$ $m=1,2,$ $\bar{h}_0[k] \triangleq \tfrac{1}{2}(\delta[k]+\delta[k-1]+\delta[k-2]+\delta[k-3])$ and $\bar{h}_m[k] \triangleq \delta[k-(m-1)],$ $m=1,2.$ 

To keep the expressions for the labels in the trellis and state diagrams, as well as the \gls{tf}, associated with the proposed rate $3/2$ \gls{nsm}, as simple as possible, we use, as in Subsection~\ref{ssec:Modulation of Rate 2}, the rescaled versions, $\mathring{h}_m[k] \triangleq 2 \bar{h}_m[k],$ of the filters $\bar{h}_m,$ $0 \le m \le 2,$ as well as the rescaled version, $\mathring{s}[k] \triangleq 2 \bar{s}[k],$ of the normalized modulated symbols, as expressed in (\ref{eq:Normalized Modulated Symbols Rate 3/2 MSN}). The rescaled modulated symbols, $\mathring{s}[k],$ are now expressed as
\begin{equation}
    \mathring{s}[k] = \sum_{m=0}^2 \sum_l \bar{b}_m[l] \mathring{h}_m[k-2l],
\end{equation}
using the modified filters $\mathring{h}_0[k] = \delta[k]+\delta[k-1]+\delta[k-2]+\delta[k-3],$ $\mathring{h}_m[k] = 2 \delta[k-(m-1)],$ $m=1,2,$ which have the merit of having simple integer taps.

\begin{figure}[!htbp]
    \centering
    \includegraphics[width=0.9\textwidth]{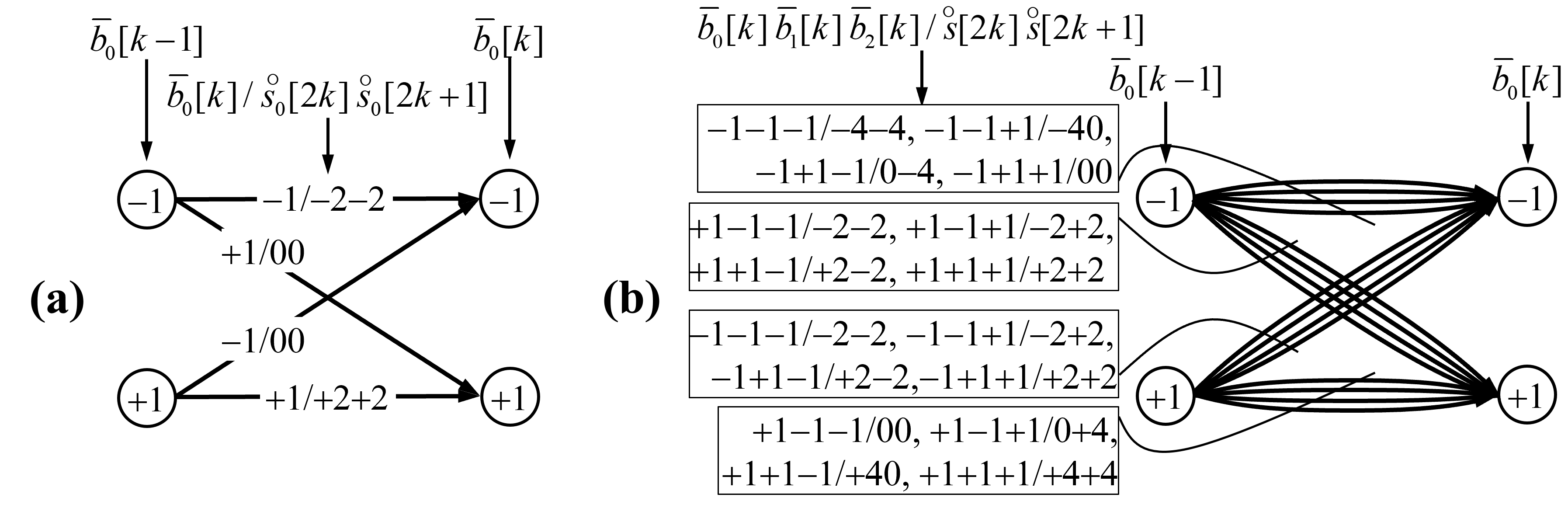}
    \caption{Trellises of filter $\mathring{h}_0[k]$ and the rate-$3/2$ NSM: (a) Filter $\mathring{h}_0[k]$, (b) Rate-$3/2$ modulation.}
    \label{fig:Trellis Filter h_0 Rate-3/2 Modulation}
\end{figure}

To begin characterizing the rate $3/2$ \gls{nsm}, we build its trellis in two phases. To begin, in Figure~\ref{fig:Trellis Filter h_0 Rate-3/2 Modulation}(a), we build up the trellis corresponding to the individual contribution, $\mathring{s}_0[k] \triangleq \sum_l \bar{b}_0[l] \mathring{h}_0[k-2l],$ of input data sequence $\bar{b}_0[k],$ to the modulated sequence $\mathring{s}[k].$ Then, in Figure~\ref{fig:Trellis Filter h_0 Rate-3/2 Modulation}(b), we construct the final and full \gls{nsm} modulation trellis, by incorporating the individual contributions of input data sequences $\bar{b}_m[k],$ $m=1,2.$ For both trellises, the states delimiting the $k$-th section are the admissible values of $\bar{b}_0[k-1],$ to the left and $\bar{b}_0[k],$ to the right.

As depicted in Figure~\ref{fig:Trellis Filter h_0 Rate-3/2 Modulation}(a), each branch in the $k$-th section of the first trellis, connecting state $\bar{b}_0[k-1]$ to state $\bar{b}_0[k],$ has $\bar{b}_0[k]$ as input label and $\mathring{s}_0[2k] \, \mathring{s}_0[2k+1]$ as output label, where
\begin{equation} \label{eq:MathringS_0[k] Rate 3/2}
  \mathring{s}_0[2k] = \mathring{s}_0[2k+1] = \sum_l \bar{b}_0[l] \mathring{h}_0[2k-2l] = \sum_l \bar{b}_0[l] \mathring{h}_0[(2k+1)-2l] = \bar{b}_0[k] + \bar{b}_0[k-1].  
\end{equation}

Each branch in the second trellis, depicted in Figure~\ref{fig:Trellis Filter h_0 Rate-3/2 Modulation}(b), has an extended input label of $\bar{b}_0[k] \, \bar{b}_1[k] \, \bar{b}_2[k]$ and an updated output label of $\mathring{s}[2k] \, \mathring{s}[2k+1],$ where $\mathring{s}[2k] = \mathring{s}_0[2k] + 2 \bar{b}_1[k] = \bar{b}_0[k] + \bar{b}_0[k-1] + 2 \bar{b}_1[k]$ and $\mathring{s}[2k+1] = \mathring{s}_0[2k+1] + 2 \bar{b}_2[k] = \bar{b}_0[k] + \bar{b}_0[k-1] + 2 \bar{b}_2[k].$ As a result, the second lattice is formed by quadrupling each branch of the first trellis, leading to four “parallel” branches, corresponding to the four possible values of the pair $(\bar{b}_1[k], \bar{b}_2[k])$ in the set $\{ (\pm 1, \pm 1) \}.$

\begin{figure}[!htbp]
    \centering
    \includegraphics[width=1\textwidth]{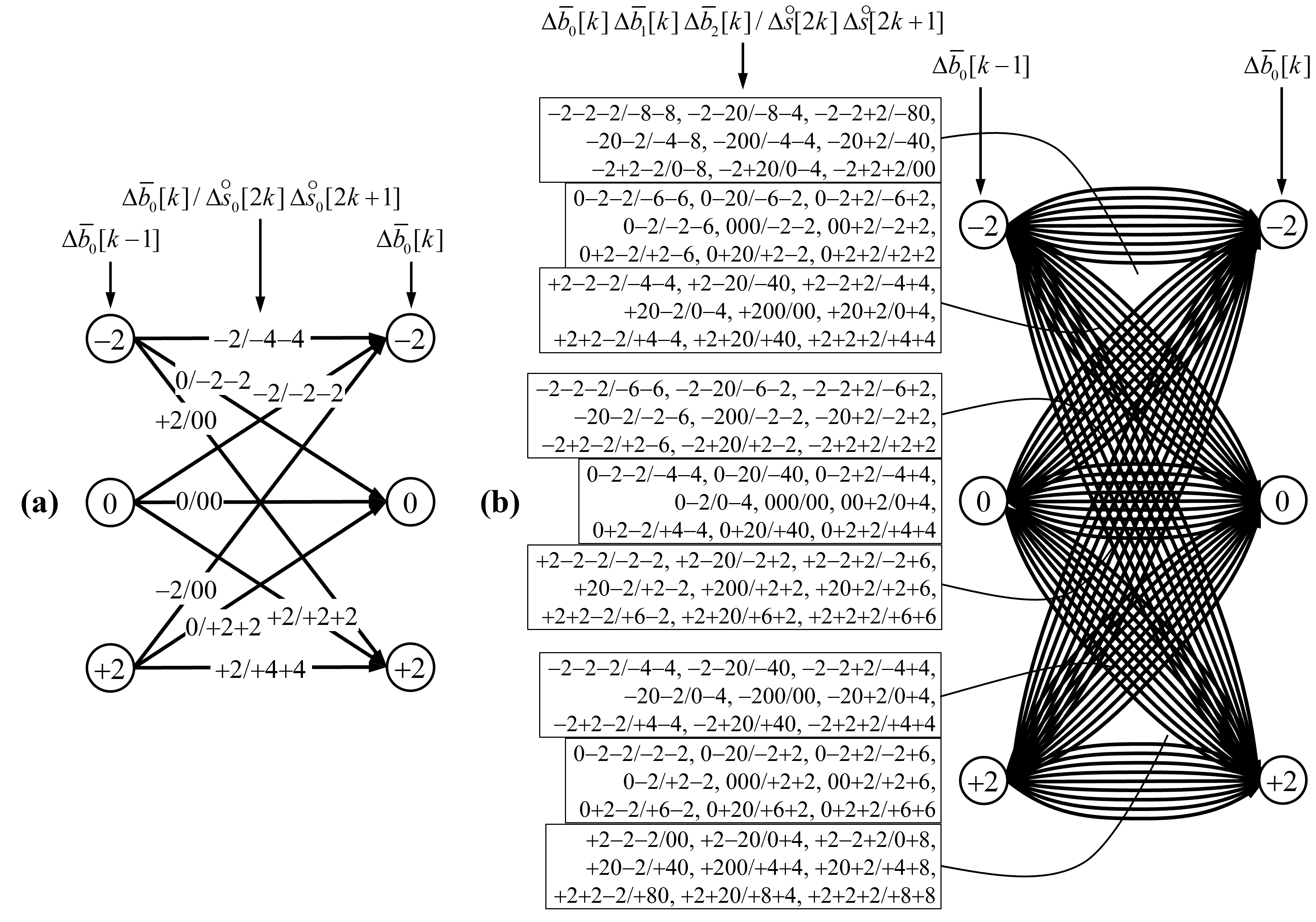}
    \caption{Trellises of the input/output sequences differences of $\mathring{h}_0[k]$ and the rate-$3/2$ NSM: (a) Filter $\mathring{h}_0[k]$, (b) Rate-$3/2$ modulation.}
    \label{fig:Trellis Input Difference Filter h_0 Rate-3/2 Modulation}
\end{figure}

As for the rate-$2$ \gls{nsm} studied in Subsection~\ref{ssec:Modulation of Rate 2}, it is instructive to characterize, in a finer way, the degenerate nature of the rate $3/2$ at hand, which is inherited from the \gls{2d} \glspl{nsm} proposed in Subsection~\ref{ssec:Two-Dimensional Rate-2 NSMs}. For this purpose, we show in Figures~\ref{fig:Trellis Input Difference Filter h_0 Rate-3/2 Modulation}(a) and~\ref{fig:Trellis Input Difference Filter h_0 Rate-3/2 Modulation}(b) the trellises of the differences in input and output sequences associated to $\mathring{s}_0[k]$ and $\mathring{s}[k],$ respectively.

The first trellis, depicted in Figure~\ref{fig:Trellis Input Difference Filter h_0 Rate-3/2 Modulation}(a), which corresponds to $\mathring{s}_0[k],$ provides a \emph{de facto} description of filter $\mathring{h}_0[k] = \delta[k] + \delta[k-1] + \delta[k-2] + \delta[k-3].$ Its $k$-th section is delimited by three states to the left and three other states to the right, corresponding respectively to the values that $\Delta \bar{b}_0[k-1]$ and $\Delta \bar{b}_0[k]$ take in the ternary set $\{ 0, \pm 2 \}.$ The branch in the $k$-th section of this trellis, which connects state $\Delta \bar{b}_0[k-1] \in \{ 0, \pm 2 \}$ to state $\Delta \bar{b}_0[k] \in \{ 0, \pm 2 \},$ has $\Delta \bar{b}_0[k]$ as input difference label and $\Delta \mathring{s}_0[2k] \, \Delta \mathring{s}_0[2k+1]$ as output difference label, where, from (\ref{eq:MathringS_0[k] Rate 3/2}), $\Delta \mathring{s}_0[2k] = \Delta \mathring{s}_0[2k+1] = \sum_l \Delta \bar{b}_0[l] \mathring{h}_0[2k-2l] = \sum_l \Delta \bar{b}_0[l] \mathring{h}_0[(2k+1)-2l] = \Delta \bar{b}_0[k] + \Delta \bar{b}_0[k-1].$

According to (\ref{eq:MathringS_0[k] Rate 3/2}), filter $\mathring{h}_0[k]$ acts as two parallel “duobinary” channels, with respect to the partial outputs $\mathring{s}_0[2k]$ and $\mathring{s}_0[2k+1],$ with even and odd indices, respectively, taken separately. As a result, borrowing from Subsection~\ref{ssec:Modulation of Rate 2}, we can claim that this filter is inherently degenerate and that error events of minimum Euclidean distance, with length $K \ge 1,$ have $\Delta \bar{b}_0[k] = \sum_{l=0}^{K-1} (-1)^l 2 \delta[k-l]$ or $\Delta \bar{b}_0[k] = -\sum_{l=0}^{K-1} (-1)^l 2 \delta[k-l],$ as input sequences differences and  $\Delta \mathring{s}_0[2k] = \Delta \mathring{s}_0[2k+1] = 2 (\delta[k] + (-1)^{K-1}\delta[k-K])$ or $\Delta \mathring{s}_0[2k] = \Delta \mathring{s}_0[2k+1] = -2 (\delta[k] + (-1)^{K-1}\delta[k-K]),$ respectively, as output sequence differences. Similarly to the “duobinary” channel, which was discussed in detail in Subsection~\ref{ssec:Modulation of Rate 2}, these error events are infinite in number, resulting in a minimum squared Euclidean distance of $16$ (twice that of the “duobinary” channel). Fortunately, the occurrence probability of any such error event, of length $K,$ which is equal to $(\tfrac{1}{2})^K,$ decreases exponentially as $K$ increases.

To construct a section of the trellis of the rate-$3/2$ \gls{nsm}, shown in Figure~\ref{fig:Trellis Input Difference Filter h_0 Rate-3/2 Modulation}(b), we replicate each branch in the trellis of $\Delta\mathring{s}_0[2k] \, \Delta\mathring{s}_0[2k+1],$ shown in Figure~\ref{fig:Trellis Input Difference Filter h_0 Rate-3/2 Modulation}(a), nine times, with one replica for each of the nine values of the pair $(\Delta \bar{b}_1[k], \Delta \bar{b}_2[k])$ in the set $\{0, \pm 2 \}^2.$ More precisely, each branch in the trellis of $\Delta\mathring{s}_0[2k] \, \Delta\mathring{s}_0[2k+1],$ with label $\Delta \bar{b}_0[k]/\Delta\mathring{s}_0[2k] \, \Delta\mathring{s}_0[2k+1],$ leads to $9$ branches with labels $\Delta \bar{b}_0[k] \, \Delta \bar{b}_1[k] \, \Delta \bar{b}_2[k] /\Delta \mathring{s}[2k] \, \Delta \mathring{s}[2k+1],$ corresponding to the $9$ values of $(\Delta \bar{b}_1[k], \Delta \bar{b}_2[k]).$ The output label, $\Delta \mathring{s}[2k] \, \Delta \mathring{s}[2k+1],$ is such that $\Delta \mathring{s}[2k] = \Delta \bar{b}_0[k-1] + \Delta \bar{b}_0[k] + 2 \Delta \bar{b}_1[k]$ and $\Delta \mathring{s}[2k+1] = \Delta \bar{b}_0[k-1] + \Delta \bar{b}_0[k] + 2 \Delta \bar{b}_2[k].$

In addition to the first category of degeneracy acquired from $\mathring{s}_0[2k],$ and its underlying filter $\mathring{h}_0[k],$  the terms $2 \Delta \bar{b}_1[k]$ and $2 \Delta \bar{b}_2[k]$, which appear in $\Delta \mathring{s}[2k]$ and $\Delta \mathring{s}[2k+1],$ respectively, introduce a second category of degeneracy. Indeed, when we transition from state $\Delta \bar{b}_0[k-1] = -2,$ (respectively, $+2$) to state $\Delta \bar{b}_0[k] = -2,$ (respectively, $+2$), with $\Delta \bar{b}_1[k] = \Delta \bar{b}_2[k] = +2,$ (respectively, $-2$), the output sequence differences, $\Delta \mathring{s}[2k]$ and $\Delta \mathring{s}[2k+1],$ are both null. Before continuing, it is important to point out that both forms of degeneracy have been identified in Appendix~\ref{app:Tight Estimate BEP Rate 3/2}, in a different way, with the second category being categorized as less consequential in terms of likelihood of occurrence. Indeed, whereas the likelihood of crossing a branch of the first category is equal to $\tfrac{1}{2}$ and is at the root of the previously indicated exponential decrease, $(\tfrac{1}{2})^K,$ the probability of crossing a branch of the second category is only $(\tfrac{1}{2})^3=\tfrac{1}{8}.$

\begin{figure}[!htbp]
    \centering
    \includegraphics[width=1.0\textwidth]{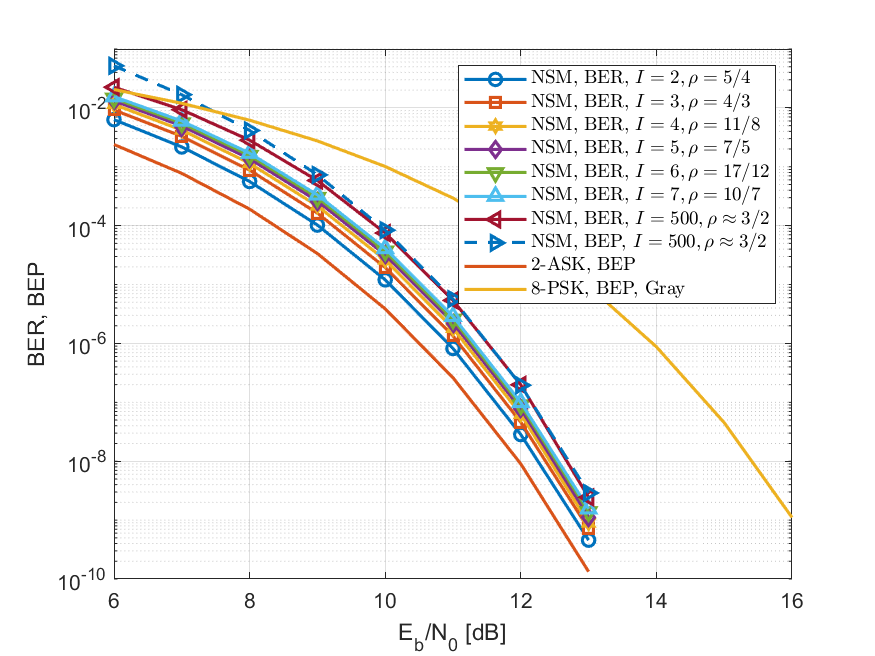}
    \caption{BER of the 1D NSMs of rates $\rho=5/4, 4/3, 11/8, 7/5, 17/12, 10/7$ and $3/2-1/1000,$ corresponding to $I=2, 3, 4, 5, 6, 7$ and $500,$ respectively. For validation the BEP approximation of the NSM of rate $\rho = 1499/1000,$ given in (\ref{eq:Tight Estimate BEP Rate 3-2}), is shown. Also for reference and comparison, the BEPs of $2$-ASK and Gray-precoded $8$-PSK conventional modulations are presented.}
    \label{fig:Figure One Dimensional NSM-BER}
\end{figure}

To round up the characterization of the rate-$3/2$ \gls{nsm}, we show, in Figure~\ref{fig:Figure One Dimensional NSM-BER}, many simulation results in terms of \gls{ber}, for situations where $I = 2, 3, 4, 5, 6, 7,$ and $500.$ The latter value of $I$ results in $1000$ modulated symbols and an effective rate of $\rho = 1499/1000$, which is quite close to the asymptotic rate of $3/2.$ The Viterbi algorithm \cite{Forney73} is used in these simulations to achieve maximum likelihood detection, based on the trellis structure, a section of which is depicted in Figure~\ref{fig:Trellis Filter h_0 Rate-3/2 Modulation}(b).

For purposes of verification and better comprehension, Figure~\ref{fig:Figure One Dimensional NSM-BER} also depicts an approximate \gls{bep} for the asymptotic case where $I = \infty$. The utilized expressed for the approximation \gls{bep}, 
\begin{equation} \label{eq:Tight Estimate BEP Rate 3-2}
        \text{BEP} \approx \frac{65}{3} \frac{1}{2} \operatorname{erfc} \left( \sqrt{ \frac{E_b}{N_0}} \right) = \frac{65}{6} \operatorname{erfc} \left( \sqrt{ \frac{E_b}{N_0}} \right),
\end{equation}
is the result of the full derivation of the \gls{tf} of the rate-$3/2$ \gls{nsm}, addressed in Appendix~\ref{app:Tight Estimate BEP Rate 3/2}.

For comparison, the \gls{bep} curves for $2$-ASK and $8$-PSK are provided, in Figure~\ref{fig:Figure One Dimensional NSM-BER}, as benchmarks. On the one hand, because the rate-$3/2$ \gls{nsm} at hand has the same minimum Euclidean distance as $2$-ASK, the \gls{bep} of $2$-ASK can be used to measure visually the severity of degeneracy. On the other hand, because $8$-PSK has the same spectrum efficiency as the suggested rate-$3/2$ \gls{nsm}, it is employed to assess the performance improvement brought at various practical \gls{ber} target values. The \gls{bep} provided for $8$-ASK results from the union bound approximation $\text{BEP} \approx \tfrac{1}{3} \operatorname{erfc} \left( \sin{(\pi/8)} \sqrt{3 E_b/N_0} \right).$

A first observation, based on the \gls{nsm}'s \gls{ber} curves, shows a constant decline in performance as the \gls{nsm} extent, $I,$ increases from $2$ to $500$. This is to be expected, because decreasing the \gls{nsm} extent by choosing small values of $I$ lowers the detrimental effect of degeneracy. A second observation reveals a nearly perfect match between the simulated \gls{ber}, at an extent of $I=500$ and a rate $\rho=1499/1000,$ and the approximation \gls{bep}, provided in (\ref{eq:Tight Estimate BEP Rate 3-2}), at an infinite extent and an asymptotic rate $\rho=3/2.$

In terms of the performance enhancement offered by the proposed \gls{nsm}, when $I$ is large ($I \ge 500$), we may report gains in $E_b/N_0$ of roughly $2.4$ and $2.7$ dB, at target \glspl{ber}/\glspl{bep} of $10^{-6}$ and $10^{-8},$ respectively. It is crucial to note that the asymptotic gain that the rate-$3/2$ \gls{nsm} may reach at high \gls{snr}, which is equal to $-10 \log_{10}(3\sin^2(\pi/8)),$ or around $3.57$ dB, is far from being realized at moderate \glspl{snr}. This is clearly the outcome of the substantial degeneracy, caused by the prohibitive multiplicity factor of $65/3,$ of the \gls{bep} approximation for the rate-$3/2$ \gls{nsm}, in comparison to that of $2$-ASK.

Finally, for completeness,  we need to underline the fact that, apart from $\mathring{h}_0[k] = \delta[k] + \delta[k-1] + \delta[k-2] + \delta[k-3],$ all other filters $\mathring{h}_0[k],$ of length $L_0=4,$ with coefficients in the bipolar set $\{ \pm 1 \},$ achieve the minimum squared Euclidean distance of $2$-ASK. However, when considering the squared Euclidean distance spectrum, and consequently the multiplicity of the minimum squared Euclidean distance, they are not identical. As in Section~\ref{ssec:Two-Dimensional Rate-2 NSMs}, we discuss here the equivalence transformations that can be operated on filter  $\mathring{h}_0[k],$ while preserving its squared Euclidean distance spectrum.

First off, as has been argued numerous times before, the \gls{nsm} properties in terms of the distance spectrum are maintained when $\mathring{h}_0[k]$ is substituted with its opposite, $-\mathring{h}_0[k],$ or its time reverse $\mathring{h}_0[L_0-1-k],$ $L_0=4$ being the filter length. Now, note that, by replacing the input bipolar sequence $\bar{b}_0[k]$ with the alternate bipolar sequence $\tilde{b}_0[k] \triangleq (-1)^k \bar{b}_0[k]$ and the partial output sequence $\mathring{s}_0[k]$ with the new output alternate sequence  $\tilde{\mathring{s}}_0[k] \triangleq (-1)^{\lfloor (k-l)/2 \rfloor} \mathring{s}_0[k],$ for an arbitrary index shift, $l,$ where $\lfloor \cdot \rfloor$ is the floor function, we get a new \gls{nsm} system with filter $\tilde{\mathring{h}}_0[k] \triangleq (-1)^{\lfloor (k-l)/2 \rfloor} \mathring{h}_0[k],$ while preserving the distance spectrum (notice here that there are only $4$ such “polyphase” alternating transformations, with half of them obtained by a sign change of the other half).  Furthermore, by replacing the partial output sequence $\mathring{s}_0[k]$ with the new scrambled output sequence  $\tilde{\mathring{s}}_0[k] \triangleq r[k] \mathring{s}_0[k],$ where $r[k] = r[k \bmod 2],$ is a periodic scrambling bipolar sequence with period $2,$ that repeats indefinitely the length-$2$ pattern $r[0]r[1],$ we gain a new \gls{nsm} system with scrambled filter $\tilde{\mathring{h}}_0[k] \triangleq r[k] \mathring{h}_0[k],$ while keeping the distance spectrum. In addition to these transformations, shifts in time can change the detection trellis structure (number of states and branch labels), while preserving distance spectrum. Finally permutations of consecutive even and odd indexed samples, leading to the new output sequence $\tilde{\mathring{s}}_0[k] \triangleq \mathring{s}_0[k+1 - 2(k \bmod 2)],$ preserve spectrum distance, while leading to the equivalent \gls{nsm} with filter $\tilde{\mathring{h}}_0[k] \triangleq \mathring{h}_0[k+1 - 2(k \bmod 2)].$

The previous transformations, applied to filter $\mathring{h}_0[k],$ define an equivalence relation among filters, resulting in only two equivalence classes: $\mathring{h}_0[k] = \delta[k] + \delta[k-1] + \delta[k-2] + \delta[k-3],$ which has already been studied in detail, and $\mathring{h}_0[k] = \delta[k] + \delta[k-1] + \delta[k-2] - \delta[k-3],$ which is new. Although not covered in detail in this section, filter $\mathring{h}_0[k] = \delta[k] + \delta[k-1] + \delta[k-2] - \delta[k-3]$ performs better than the thoroughly studied filter, $\mathring{h}_0[k] = \delta[k] + \delta[k-1] + \delta[k-2] + \delta[k-3],$ in terms of error events minimum Squared distance multiplicity. This superiority of the former filter will be evident from the results in the next Section, particularly those in Table~\ref{table:NSMs Rate-3/2 Filter Pattern (1, 1, 1, 1)}, on \gls{rtf} $\dot{T}(D).$ That said, this superiority was expected, given that the \gls{2d} filter $\bar{h}_0[k,l] = \tfrac{1}{2} \mathring{h}_0[k,l] =  \tfrac{1}{2}(\delta[k]\delta[l] + \delta[k-1]\delta[l] + \delta[k]\delta[l-1] - \delta[k-1]\delta[l-1]),$ is superior to the \gls{2d} filter $\bar{h}_0[k,l] = \tfrac{1}{2} \mathring{h}_0[k,l] = \tfrac{1}{2}(\delta[k] + \delta[k-1])(\delta[l] + \delta[l-1]),$ as stated in Section~\ref{ssec:Two-Dimensional Rate-2 NSMs}. Indeed, notice that the present filters, $\mathring{h}_0[k] = \delta[k] + \delta[k-1] + \delta[k-2] - \delta[k-3]$ and $\mathring{h}_0[k] = \delta[k] + \delta[k-1] + \delta[k-2] + \delta[k-3],$ are the \gls{1d} counterparts of \gls{2d} filters $\mathring{h}_0[k,l] = \delta[k]\delta[l] + \delta[k-1]\delta[l] + \delta[k]\delta[l-1] - \delta[k-1]\delta[l-1],$ is superior to the \gls{2d} filter $\mathring{h}_0[k,l] =  (\delta[k] + \delta[k-1])(\delta[l] + \delta[l-1]),$ respectively.


\begin{table}[H]
\caption{Selected rate-$3/2$ NSMs, with simple filters' coefficients, with common filter $\bm{h}_0$ pattern $\bm{\pi}_0 = (1, 1, 1, 1),$ arranged in a decreasing performance order, starting from the best one. Relative degradation in performance due to distance reduction or/and multiplicity increase when moving from one filter to the next is emphasized in bold.}
\label{table:NSMs Rate-3/2 Filter Pattern (1, 1, 1, 1)}
\centering
\begin{tabular}{|c|c|} 
\hline
\multicolumn{2}{|l|}{\# of simple $h_0[k]$ filters $ = 16$} \\ \hline
\multicolumn{2}{|l|}{\# of non equivalent $h_0[k]$ filters $ = 2$} \\ \hline
\multicolumn{2}{|l|}{$\mathring{h}_1[k] = 2 \delta[k]$ and $\mathring{h}_2[k] = 2 \delta[k-1]$} \\ \hline
\multicolumn{2}{|l|}{Minimum squared Euclidean distance, $d_\text{min}^2 = 16$} \\ \hline
\multicolumn{2}{|l|}{Number of trellis states $ = 2$} \\  [0.5ex]
\hline\hline
\multicolumn{2}{|c|}{Filter \# $1$} \\ \hline
$\mathring{h}_0[k]$ & $\delta[k]+\delta[k-1]+\delta[k-2]-\delta[k-3]$ \\ \hline
$\mathring{\bm{h}}_0$ & $(1, 1, 1, -1)$ \\ \hline
$\dot{T}(D)$ & $\tfrac{367}{8} D^{16}+245 D^{32} + \tfrac{10395}{8} D^{40} + \tfrac{24003}{4} D^{64} + \tfrac{106047}{4} D^{80} + 112779 D^{96} + \cdots$ \\ \hline
\multirow{2}{*}{\gls{bep} approximation} & $\frac{367}{48} \operatorname{erfc}\left( \sqrt{\frac{E_b}{N_0}} \right) + \frac{245}{6} \operatorname{erfc}\left( \sqrt{2 \frac{E_b}{N_0}} \right) + \frac{3465}{16} \operatorname{erfc}\left( \sqrt{\frac{5}{2} \frac{E_b}{N_0}} \right) $ \\
& $+ \frac{8001}{2} \operatorname{erfc}\left( \sqrt{4 \frac{E_b}{N_0}} \right) + \frac{35349}{8} \operatorname{erfc}\left( \sqrt{5 \frac{E_b}{N_0}} \right)$ \\ [0.5ex] 
\hline\hline
\multicolumn{2}{|c|}{Filter \# $2$} \\ \hline
$\mathring{h}_0[k]$ & $\delta[k]+\delta[k-1]+\delta[k-2]+\delta[k-3]$ \\ \hline
$\mathring{\bm{h}}_0$ & $(1, 1, 1, 1)$ \\ \hline
$\dot{T}(D)$ & $\bm{65} D^{16} + 488 D^{32} + 3300 D^{48} + 19800 D^{64} + 112480 D^{80} + 615198 D^{96} + \cdots$ \\ \hline
\multirow{2}{*}{BEP approximation} & $\frac{65}{6} \operatorname{erfc}\left( \sqrt{\frac{E_b}{N_0}} \right) + \frac{244}{3} \operatorname{erfc}\left( \sqrt{2 \frac{E_b}{N_0}} \right) + 550 \operatorname{erfc}\left( \sqrt{3 \frac{E_b}{N_0}} \right) $ \\
& $+ 3300 \operatorname{erfc}\left( \sqrt{4 \frac{E_b}{N_0}} \right) + \frac{56240}{3} \operatorname{erfc}\left( \sqrt{5 \frac{E_b}{N_0}} \right)$ \\  [1ex] 
 \hline
\end{tabular}
\end{table}

\subsection{A Broader Collection of Rate-3/2 NSMs}
\label{A broader collection of rate-3/2 NSMs}

Up to this point, we have thoroughly analyzed the rate-$3/2$ \gls{nsm} with bipolar input sequences $\bar{b}_m,$ $m=0,1,2,$ and rescaled filters $\mathring{h}_0[k]=\delta[k] + \delta[k-1] + \delta[k-2] \pm \delta[k-3]$ and $\mathring{h}_m[k]=2 \delta[k-(m-1)],$ $m=1,2.$ We have introduced a variety of equivalence transformations---including global sign inversion, time reversal, alternating sign changes, scrambling with a periodic bipolar sequence of period $2,$ and permutations of consecutive even and odd indexed taps of filter $\bar{h}_0[k]$ or its rescaled form, $\mathring{h}_0[k],$ that preserve the \gls{nsm}'s \gls{tf}, $T(N,D),$ or its reduced form, $\dot{T}(D).$ We have demonstrated that all length-$4$ filters, $\mathring{h}_0[k],$ with bipolar taps in $\{ \pm 1\},$ are equivalent to one of the two canonical filters $\mathring{h}_0[k]=\delta[k] + \delta[k-1] + \delta[k-2] \pm \delta[k-3].$ Furthermore, we showed that, although both filters yield the same minimum squared Euclidean distance, filter $\mathring{h}_0[k]=\delta[k] + \delta[k-1] + \delta[k-2] - \delta[k-3]$ outperforms filter $\mathring{h}_0[k]=\delta[k] + \delta[k-1] + \delta[k-2] + \delta[k-3],$ due to the lower multiplicity of this distance.

In vector form, all candidate filters within the previous \gls{nsm} family are represented by filter candidates, $\mathring{\bm{h}}_0 = (\pm 1, \pm 1, \pm 1, \pm 1),$ corresponding to base pattern $\bm{\pi}_0 = (1,1,1,1).$ Building on this structure of the rate-$3/2$ \glspl{nsm}, we now extend our analysis by introducing new filter $\mathring{h}_0[k]$ candidates, associated with filter patterns $(1,1,1,1,0,0),$ $(3,3,3,3,3,2)$ and $(4,4,4,4,4,1),$ all defined for filter length $L_0=6$. In particular, the pattern $(1,1,1,1,0,0),$ also accommodates the special case, $L_0=4,$ when the two zero components are positioned at either end of the filter. Additionally, we consider the extended patterns $(2,2,2,1,1,1,1,0),$ $(2,2,2,2,2,2,1,0),$ $(3,3,3,3,2,2,2,1),$ $(3,3,3,3,3,3,3,1)$ and $(4,4,4,4,3,3,3,3),$ corresponding to filter length $L_0=8.$

For each previously mentioned filter pattern, $\bm{\pi}_0,$ we generate all candidate filters, $\mathring{\bm{h}}_0,$ by applying arbitrary sign change operations and permutations to the components of the pattern. The number of potential candidate filters can be extremely large, making the search for the optimal ones quite challenging. To mitigate this complexity, we can effectively leverage the equivalence relation introduced in Subsection~\ref{ssec:Basic Rate-3/2 NSM} and summarized earlier. This relation, based on transformations that preserve the \gls{nsm}’s \gls{tf}---and thus its distance spectrum---allows us to limit our search to representative filters from each equivalence class, without sacrificing optimality.

The pattern vectors, $\bm{\pi}_0,$ introduced earlier are not chosen arbitrarily, but are subject to several constrains, that ensure the generation of good \glspl{nsm}. First, their components must be non-negative integers. Second, their Euclidean norms, $\| \bm{\pi}_0 \|,$ must also be integers. This requirement arises from the fact that filters $\mathring{\bm{h}}_m,$ $m=1,2,$ which are designed to have the same norm (and therefore the same energy contribution to the modulated signal) as $\mathring{\bm{h}}_0,$ are now defined as $\mathring{h}_m[k] = \| \mathring{\bm{h}}_0 \| \delta[k-(m-1)] = \| \bm{\pi}_0 \| \delta[k-(m-1)],$ $m=1,2.$ Third, to preserve the minimum squared Euclidean distance guaranteed by $2$-ASK, the maximum component of $\bm{\pi}_0,$ which corresponds to the infinity norm, $\| \bm{\pi}_0 \|_\infty,$ of $\bm{\pi}_0,$ must not be greater than half $\| \bm{\pi}_0 \|.$ This condition is already satisfied by the exploratory \glspl{nsm}, discussed in Subsection~\ref{ssec:Basic Rate-3/2 NSM}. In fact, it holds with equality, as $\| \bm{\pi}_0 \|_\infty = 1 = \| \bm{\pi}_0 \|/2.$ Based on our experience with rate-$2$ \glspl{nsm}, it is recommended that this last constraint be made as strict as possible. Doing so helps avoid the tightness effect, which is known to significantly increase the multiplicity of the minimum Euclidean distance in the \gls{nsm}’s distance spectrum.

Tables~\ref{table:NSMs Rate-3/2 Filter Pattern (1, 1, 1, 1)}--\ref{table:NSMs Rate-3/2 Filter Pattern (4, 4, 4, 4, 3, 3, 3, 3)} present, for each pattern, the total number of filter $\mathring{h}_0[k]$ candidates as well as the number of non-equivalent candidates. The latter corresponds to the number of equivalent classes defined by the equivalence relation based on transfer-function-preserving transformations. We also explicitly provide the expressions for the other scaled filters $\mathring{h}_m[k],$ $m=0,1.$ Furthermore, we report the largest minimum squared Euclidean distance, achieved by filter candidates in the best-performing equivalent class. In addition, the number of states in the demodulator detection trellis is specified. Since the input sequences are of bipolar type, this number of states is given by $2^{L_0/2-1},$ where the exponent, $L_0/2-1,$ represents the memory depth associated with the first input sequence, $\bar{b}_0[k].$


\begin{table}[H]
\caption{Selected rate-$3/2$ NSMs, with simple filters' coefficients, with common filter $\bm{h}_0$ pattern $\bm{\pi}_0 = (1, 1, 1, 1, 0, 0),$ arranged in a decreasing performance order, starting from the best one. Relative degradation in performance due to distance reduction or/and multiplicity increase when moving from one filter to the next is emphasized in bold. Notice that filters \# $7,$ $8$ and $9$ are equivalent to filter \# $2$ in Table~\ref{table:NSMs Rate-3/2 Filter Pattern (1, 1, 1, 1)}.}
\label{table:NSMs Rate-3/2 Filter Pattern (1, 1, 1, 1, 0, 0)}
\centering
\begin{tabular}{|c|c|} 
\hline
\multicolumn{2}{|l|}{\# of simple $h_0[k]$ filters $ = 240$} \\ \hline
\multicolumn{2}{|l|}{\# of non equivalent $h_0[k]$ filters $ = 9$} \\ \hline
\multicolumn{2}{|l|}{$\mathring{h}_1[k] = 2 \delta[k]$ and $\mathring{h}_2[k] = 2 \delta[k-1]$} \\ \hline
\multicolumn{2}{|l|}{Minimum squared Euclidean distance, $d_\text{min}^2 = 16$} \\ \hline
\multicolumn{2}{|l|}{Number of trellis states $ = 2$ or $4$} \\  [0.5ex]
\hline\hline
\multicolumn{2}{|c|}{Filter \# $1$} \\ \hline
$\mathring{h}_0[k]$ & $\delta[k]+\delta[k-1]+\delta[k-2]+\delta[k-5]$ \\ \hline
$\mathring{\bm{h}}_0$ & $(1, 1, 1, 0, 0, 1)$ \\ \hline
$\dot{T}(D)$ & $\tfrac{221}{16} D^{16}+\tfrac{11743461}{73984} D^{24} + \tfrac{786552761}{1183744} D^{32} + \cdots$ \\ \hline
BEP approximation & $\frac{221}{96} \operatorname{erfc}\left( \sqrt{\frac{E_b}{N_0}} \right) + \frac{3914487}{147968} \operatorname{erfc}\left( \sqrt{\frac{3}{2} \frac{E_b}{N_0}} \right) + \frac{786552761}{7102464} \operatorname{erfc}\left( \sqrt{2 \frac{E_b}{N_0}} \right)$ \\ [0.5ex] 
\hline\hline
\multicolumn{2}{|c|}{Filter \# $2$} \\ \hline
$\mathring{h}_0[k]$ & $\delta[k]+\delta[k-1]+\delta[k-2]-\delta[k-5]$ \\ \hline
$\mathring{\bm{h}}_0$ & $(1, 1, 1, 0, 0, -1)$ \\ \hline
$\dot{T}(D)$ & $\tfrac{221}{16} D^{16}+\tfrac{\bm{5299101}}{\bm{30976}} D^{24} + \tfrac{356482385}{495616} D^{32} + \cdots$ \\ [0.5ex] 
\hline\hline
\multicolumn{2}{|c|}{Filter \# $3$} \\ \hline
$\mathring{h}_0[k]$ & $\delta[k]+\delta[k-1]+\delta[k-2]-\delta[k-4]$ \\ \hline
$\mathring{\bm{h}}_0$ & $(1, 1, 1, 0, -1, 0)$ \\ \hline
$\dot{T}(D)$ & $\tfrac{\bm{793}}{\bm{32}} D^{16}+\tfrac{34911}{256} D^{24} + \tfrac{24949951}{32768} D^{32} + \cdots$ \\ [0.5ex] 
\hline\hline
\multicolumn{2}{|c|}{$\vdots$} \\ [0.5ex] 
\hline\hline
\multicolumn{2}{|c|}{Filter \# $7$} \\ \hline
$\mathring{h}_0[k]$ & $\delta[k]+\delta[k-1]+\delta[k-2]+\delta[k-3]$ \\ \hline
$\mathring{\bm{h}}_0$ & $(1, 1, 1, 1, 0, 0)$ \\ \hline
$\dot{T}(D)$ & $\bm{65} D^{16}+ 2342 D^{32} + \tfrac{250431}{4} D^{48} + \cdots$  \\ [0.5ex] 
\hline\hline
\multicolumn{2}{|c|}{Filters \# $8$ \& $9$} \\ \hline
$\mathring{h}_0[k]$ & $\delta[k]+\delta[k-1]+\delta[k-4]+\delta[k-5]$ \& $\delta[k]+\delta[k-1]-\delta[k-4]-\delta[k-5]$ \\ \hline
$\mathring{\bm{h}}_0$ & $(1, 1, 0, 0, 1, 1)$ \& $(1, 1, 0, 0, -1, -1)$ \\ \hline
$\dot{T}(D)$ & $65 D^{16}+ \bm{6734} D^{32} + 379661 D^{48} + \cdots$ \\ [1ex] 
 \hline
\end{tabular}
\end{table}


\begin{table}[H]
\caption{Selected rate-$3/2$ NSMs, with simple filters' coefficients, with common filter $\bm{h}_0$ pattern $\bm{\pi}_0 = (3, 3, 3, 3, 3, 2),$ arranged in a decreasing performance order, starting from the best one. Relative degradation in performance due to distance reduction or/and multiplicity increase when moving from one filter to the next is emphasized in bold. Notice that all filters do not achieve the minimum Euclidean distance of $2$-ASK. to be equivalent to $2$-ASK, in terms of minimum Euclidean distance, $d_\text{min}^2,$ must be equal to $196.$}
\label{table:NSMs Rate-3/2 Filter Pattern (3, 3, 3, 3, 3, 2)}
\centering
\begin{tabular}{|c|c|} 
\hline
\multicolumn{2}{|l|}{\# of simple $h_0[k]$ filters $ = 384$} \\ \hline
\multicolumn{2}{|l|}{\# of non equivalent $h_0[k]$ filters $ = 14$} \\ \hline
\multicolumn{2}{|l|}{$\mathring{h}_1[k] = 7 \delta[k]$ and $\mathring{h}_2[k] = 7 \delta[k-1]$} \\ \hline
\multicolumn{2}{|l|}{Best minimum squared Euclidean distance, $d_\text{min}^2 = 168$} \\
\multicolumn{2}{|l|}{(far below $196,$ the equivalent minimum squared Euclidean distance of $2$-ASK)} \\ \hline
\multicolumn{2}{|l|}{Number of trellis states $ = 4$} \\  [0.5ex]
\hline\hline
\multicolumn{2}{|c|}{Filter \# $1$} \\ \hline
$\mathring{h}_0[k]$ & $3 \delta[k] + 3 \delta[k-1] + 3 \delta[k-2] -2 \delta[k-3] -3 \delta[k-4] -3 \delta[k-5]$ \\ \hline
$\mathring{\bm{h}}_0$ & $(3, 3, 3, -2, -3, -3)$ \\ \hline
$\dot{T}(D)$ & $D^{168}+\tfrac{1}{4} D^{192} + \tfrac{11}{2} D^{196} + \cdots$ \\ [0.5ex] 
\hline\hline
\multicolumn{2}{|c|}{Filter \# $2$} \\ \hline
$\mathring{h}_0[k]$ & $3 \delta[k] + 3 \delta[k-1] + 3 \delta[k-2] + 2 \delta[k-3] -3 \delta[k-4] -3 \delta[k-5]$ \\ \hline
$\mathring{\bm{h}}_0$ & $(3, 3, 3, 2, -3, -3)$ \\ \hline
$\dot{T}(D)$ & $D^{168}+\tfrac{\bm{5}}{\bm{16}} D^{172} + \cdots$ \\ [0.5ex] 
\hline\hline
\multicolumn{2}{|c|}{$\vdots$} \\ [0.5ex] 
\hline\hline
\multicolumn{2}{|c|}{Filter \# $13$} \\ \hline
$\mathring{h}_0[k]$ & $3 \delta[k] + 3 \delta[k-1] + 3 \delta[k-2] -3 \delta[k-3] + 2 \delta[k-4] -3 \delta[k-5]$ \\ \hline
$\mathring{\bm{h}}_0$ & $(3, 3, 3, -3, 2, -3)$ \\ \hline
$\dot{T}(D)$ & $\tfrac{3}{4} D^{\bm{132}}+ \tfrac{27}{64} D^{148} + \cdots$  \\ [0.5ex] 
\hline\hline
\multicolumn{2}{|c|}{Filter \# $14$} \\ \hline
$\mathring{h}_0[k]$ & $3 \delta[k] + 3 \delta[k-1] + 3 \delta[k-2] + 3 \delta[k-3] + 3 \delta[k-4] + 2 \delta[k-5]$ \\ \hline
$\mathring{\bm{h}}_0$ & $(3, 3, 3, 3, 3, 2)$ \\ \hline
$\dot{T}(D)$ & $D^{\bm{128}}+ \tfrac{1}{2} D^{136} + \cdots$ \\ [1ex] 
 \hline
\end{tabular}
\end{table}


\begin{table}[H]
\caption{Selected rate-$3/2$ NSMs, with simple filters' coefficients, with common filter $\bm{h}_0$ pattern $\bm{\pi}_0 = (4, 4, 4, 4, 4, 1),$ arranged in a decreasing performance order, starting from the best one. Relative degradation in performance due to distance reduction or/and multiplicity increase when moving from one filter to the next is emphasized in bold.}
\label{table:NSMs Rate-3/2 Filter Pattern (4, 4, 4, 4, 4, 1)}
\centering
\begin{tabular}{|c|c|} 
\hline
\multicolumn{2}{|l|}{\# of simple $h_0[k]$ filters $ = 384$} \\ \hline
\multicolumn{2}{|l|}{\# of non equivalent $h_0[k]$ filters $ = 14$} \\ \hline
\multicolumn{2}{|l|}{$\mathring{h}_1[k] = 9 \delta[k]$ and $\mathring{h}_2[k] = 9 \delta[k-1]$} \\ \hline
\multicolumn{2}{|l|}{Best minimum squared Euclidean distance, $d_\text{min}^2 = 384$} \\ \hline
\multicolumn{2}{|l|}{Number of trellis states $ = 4$} \\  [0.5ex]
\hline\hline
\multicolumn{2}{|c|}{Filter \# $1$} \\ \hline
$\mathring{h}_0[k]$ & $4 \delta[k] + 4 \delta[k-1] + 4 \delta[k-2] - \delta[k-3] -4 \delta[k-4] -4 \delta[k-5]$ \\ \hline
$\mathring{\bm{h}}_0$ & $(4, 4, 4, -1, -4, -4)$ \\
\hline
\end{tabular}
\end{table}


\begin{table}[H]
\caption{Selected rate-$3/2$ NSMs, with simple filters' coefficients, with common filter $\bm{h}_0$ pattern $\bm{\pi}_0 = (2, 2, 2, 1, 1, 1, 1, 0),$ arranged in a decreasing performance order, starting from the best one. Relative degradation in performance due to distance reduction or/and multiplicity increase when moving from one filter to the next is emphasized in bold.}
\label{table:NSMs Rate-3/2 Filter Pattern (2, 2, 2, 1, 1, 1, 1, 0)}
\centering
\begin{tabular}{|c|c|} 
\hline
\multicolumn{2}{|l|}{\# of simple $h_0[k]$ filters $ = 35840$} \\ \hline
\multicolumn{2}{|l|}{\# of non equivalent $h_0[k]$ filters $ = 864$} \\ \hline
\multicolumn{2}{|l|}{$\mathring{h}_1[k] = 4 \delta[k]$ and $\mathring{h}_2[k] = 4 \delta[k-1]$} \\ \hline
\multicolumn{2}{|l|}{Best minimum squared Euclidean distance, $d_\text{min}^2 = 64$} \\ \hline
\multicolumn{2}{|l|}{Number of trellis states $ = 8$} \\  [0.5ex]
\hline\hline
\multicolumn{2}{|c|}{Filter \# $1$} \\ \hline
$\mathring{h}_0[k]$ & $2 \delta[k] + 2 \delta[k-1] + \delta[k-2] + \delta[k-4] + \delta[k-5] - \delta[k-6] + 2 \delta[k-7]$ \\ \hline
$\mathring{\bm{h}}_0$ & $(2, 2, 1, 0, 1, 1, -1, 2)$ \\
\hline
\end{tabular}
\end{table}


\begin{table}[H]
\caption{Selected rate-$3/2$ NSMs, with simple filters' coefficients, with common filter $\bm{h}_0$ pattern $\bm{\pi}_0 = (2, 2, 2, 2, 2, 2, 1, 0),$ arranged in a decreasing performance order, starting from the best one. Relative degradation in performance due to distance reduction or/and multiplicity increase when moving from one filter to the next is emphasized in bold.}
\label{table:NSMs Rate-3/2 Filter Pattern (2, 2, 2, 2, 2, 2, 1, 0)}
\centering
\begin{tabular}{|c|c|} 
\hline
\multicolumn{2}{|l|}{\# of simple $h_0[k]$ filters $ = 7168$} \\ \hline
\multicolumn{2}{|l|}{\# of non equivalent $h_0[k]$ filters $ = 176$} \\ \hline
\multicolumn{2}{|l|}{$\mathring{h}_1[k] = 5 \delta[k]$ and $\mathring{h}_2[k] = 5 \delta[k-1]$} \\ \hline
\multicolumn{2}{|l|}{Best minimum squared Euclidean distance, $d_\text{min}^2 = 100$} \\ \hline
\multicolumn{2}{|l|}{Number of trellis states $ = 8$} \\  [0.5ex]
\hline\hline
\multicolumn{2}{|c|}{Filter \# $1$} \\ \hline
$\mathring{h}_0[k]$ & $2 \delta[k] + 2 \delta[k-1] + \delta[k-2] + \delta[k-4] + 2 \delta[k-5] - 2 \delta[k-6] + 2 \delta[k-7]$ \\ \hline
$\mathring{\bm{h}}_0$ & $(2, 2, 2, 0, 1, 2, -2, 2)$ \\
\hline
\end{tabular}
\end{table}


\begin{table}[H]
\caption{Selected rate-$3/2$ NSMs, with simple filters' coefficients, with common filter $\bm{h}_0$ pattern $\bm{\pi}_0 = (3, 3, 3, 3, 2, 2, 2, 1),$ arranged in a decreasing performance order, starting from the best one. Relative degradation in performance due to distance reduction or/and multiplicity increase when moving from one filter to the next is emphasized in bold.}
\label{table:NSMs Rate-3/2 Filter Pattern (3, 3, 3, 3, 2, 2, 2, 1)}
\centering
\begin{tabular}{|c|c|} 
\hline
\multicolumn{2}{|l|}{\# of simple $h_0[k]$ filters $ = 71680$} \\ \hline
\multicolumn{2}{|l|}{\# of non equivalent $h_0[k]$ filters $ = 2240$} \\ \hline
\multicolumn{2}{|l|}{$\mathring{h}_1[k] = 7 \delta[k]$ and $\mathring{h}_2[k] = 7 \delta[k-1]$} \\ \hline
\multicolumn{2}{|l|}{Best minimum squared Euclidean distance, $d_\text{min}^2 = 196$} \\ \hline
\multicolumn{2}{|l|}{Number of trellis states $ = 8$} \\  [0.5ex]
\hline\hline
\multicolumn{2}{|c|}{Filter \# $1$} \\ \hline
\multirow{2}{*}{$\mathring{h}_0[k]$} & $3 \delta[k] + 3 \delta[k-1] + 2 \delta[k-2] - \delta[k-3] + 2 \delta[k-4] + 2 \delta[k-5]$ \\
& $- 3 \delta[k-6] + 3 \delta[k-7]$ \\
\hline
\end{tabular}
\end{table}


\begin{table}[H]
\caption{Selected rate-$3/2$ NSMs, with simple filters' coefficients, with common filter $\bm{h}_0$ pattern $\bm{\pi}_0 = (3, 3, 3, 3, 3, 3, 3, 1),$ arranged in a decreasing performance order, starting from the best one. Relative degradation in performance due to distance reduction or/and multiplicity increase when moving from one filter to the next is emphasized in bold.}
\label{table:NSMs Rate-3/2 Filter Pattern (3, 3, 3, 3, 3, 3, 3, 1)}
\centering
\begin{tabular}{|c|c|} 
\hline
\multicolumn{2}{|l|}{\# of simple $h_0[k]$ filters $ = 2048$} \\ \hline
\multicolumn{2}{|l|}{\# of non equivalent $h_0[k]$ filters $ = 64$} \\ \hline
\multicolumn{2}{|l|}{$\mathring{h}_1[k] = 8 \delta[k]$ and $\mathring{h}_2[k] = 8 \delta[k-1]$} \\ \hline
\multicolumn{2}{|l|}{Best minimum squared Euclidean distance, $d_\text{min}^2 = 256$} \\ \hline
\multicolumn{2}{|l|}{Number of trellis states $ = 8$} \\  [0.5ex]
\hline\hline
\multicolumn{2}{|c|}{Filter \# $1$} \\ \hline
\multirow{2}{*}{$\mathring{h}_0[k]$} & $3 \delta[k] + 3 \delta[k-1] + 3 \delta[k-2] - \delta[k-3] - 3 \delta[k-4] - 3 \delta[k-5]$ \\
& $+ 3 \delta[k-6] - 3 \delta[k-7]$ \\
\hline
\end{tabular}
\end{table}


\begin{table}[H]
\caption{Selected rate-$3/2$ NSMs, with simple filters' coefficients, with common filter $\bm{h}_0$ pattern $\bm{\pi}_0 = (4, 4, 4, 4, 3, 3, 3, 3),$ arranged in a decreasing performance order, starting from the best one. Relative degradation in performance due to distance reduction or/and multiplicity increase when moving from one filter to the next is emphasized in bold.}
\label{table:NSMs Rate-3/2 Filter Pattern (4, 4, 4, 4, 3, 3, 3, 3)}
\centering
\begin{tabular}{|c|c|} 
\hline
\multicolumn{2}{|l|}{\# of simple $h_0[k]$ filters $ = 17920$} \\ \hline
\multicolumn{2}{|l|}{\# of non equivalent $h_0[k]$ filters $ = 596$} \\ \hline
\multicolumn{2}{|l|}{$\mathring{h}_1[k] = 10 \delta[k]$ and $\mathring{h}_2[k] = 10 \delta[k-1]$} \\ \hline
\multicolumn{2}{|l|}{Best minimum squared Euclidean distance, $d_\text{min}^2 = 400$} \\ \hline
\multicolumn{2}{|l|}{Number of trellis states $ = 8$} \\  [0.5ex]
\hline\hline
\multicolumn{2}{|c|}{Filter \# $1$} \\ \hline
\multirow{2}{*}{$\mathring{h}_0[k]$} & $4 \delta[k] + 4 \delta[k-1] + 3 \delta[k-2] - 3 \delta[k-3] + 3 \delta[k-4] + 3 \delta[k-5]$ \\
& $- 4 \delta[k-6] + 4 \delta[k-7]$ \\
\hline
\end{tabular}
\end{table}

Alongside the summary information provided at the beginning of each Table, we also present the leading terms of the \gls{rtf}, $\dot{T}(D),$ for selected filter representatives from various equivalence classes, beginning with the best-performing equivalence class and progressing toward the least effective equivalence class. The \glspl{rtf} and the leading terms of their Taylor series expansions for all filters representatives were computed symbolically using Algorithm~\ref{alg:One-Shot Reduced Transfer Function T(D) Rate-(Q+1)/Q NSM}, as described in Appendix~\ref{app:Symbolic Determination Reduced Transfer Function Rate-(Q+1)/Q NSM}, with parameter $Q=2$. Alternatively, Algorithms~\ref{alg:Simplified One-Shot Reduced Transfer Function T(D) Rate-(Q+1)/Q NSM} and~\ref{alg:Simplified Iterative Reduced Transfer Function T(D) Rate-(Q+1)/Q NSM}, also described in Appendix~\ref{app:Symbolic Determination Reduced Transfer Function Rate-(Q+1)/Q NSM}, could have been used. In fact, these algorithms will be applied to the rate-$4/3$ and rate-$5/4$ MSNs, corresponding to $Q=3$ and $Q=4,$ respectively, as will be discussed in Sections~\ref{Rate-4/3 Approaching NSMs} and~\ref{Rate-5/4 Approaching NSMs}.

Equivalence classes are ranked based on the dominant terms in their \glspl{rtf}. To illustrate performance degradation between successive non-equivalent filter candidates, we highlight in bold the specific cause—either a drop in squared Euclidean distance or an increase in its multiplicity. Importantly, this degradation always affects a later term in the \gls{tf}, while earlier terms remain unchanged, marking it as the most significant decline in performance.

Finally, for each pattern, we provide an approximate upper bound on the \gls{bep} for the best non-equivalent filter candidates (referred to as Filter~\#1 in the tables). Filter pattern $\bm{\pi}_0= (3,3,3,3,3,2),$ is excluded from this, since it never achieves the performance of $2$-ASK. This \gls{bep} bound is known to be tight at moderate to high \glspl{snr}, as will soon be confirmed through \gls{ber} simulations.

Table~\ref{table:NSMs Rate-3/2 Filter Pattern (1, 1, 1, 1)} presents the characteristics of pattern $\bm{\pi}_0= (1,1,1,1),$ which was thoroughly analyzed in Subsection~\ref{ssec:Basic Rate-3/2 NSM}. Both Filter \# 1 and Filter \# 2 were studied in detail, with Filter \# 2 receiving the most in-depth and comprehensive characterization. The first few terms of the \gls{rtf} $\dot{T}(D)$ for Filter \# 2 align perfectly with the terms obtained from its Taylor series expansion, in (\ref{eq:Multiplicity and Distance Rate-3/2 Modulation}). The derivation of the exact expression for $\dot{T}(D)$ and its Taylor series expansion can be found in Appendix~\ref{app:Tight Estimate BEP Rate 3/2}.

Table~\ref{table:NSMs Rate-3/2 Filter Pattern (1, 1, 1, 1, 0, 0)} presents the characteristics of pattern $\bm{\pi}_0= (1,1,1,1,0,0),$ which is simply an extension of pattern, $\bm{\pi}_0= (1,1,1,1),$ in Table~\ref{table:NSMs Rate-3/2 Filter Pattern (1, 1, 1, 1)}, by appending two zeros. This extended pattern gives rise to $9$ equivalence classes, including the two original equivalence classes of pattern $\bm{\pi}_0= (1,1,1,1),$ as special cases. For example Filter \# $1$ in Table~\ref{table:NSMs Rate-3/2 Filter Pattern (1, 1, 1, 1)}, $\mathring{\bm{h}}_0=(1,1,1,1),$ derived from pattern $\bm{\pi}_0= (1,1,1,1),$ reappears in Table~\ref{table:NSMs Rate-3/2 Filter Pattern (1, 1, 1, 1, 0, 0)}, as Filter \# $7,$ in the form $\mathring{\bm{h}}_0=(1,1,1,1,0,0),$ now associated with pattern $\bm{\pi}_0= (1,1,1,1,0,0).$

As previously noted, with the exception of pattern $\bm{\pi}_0= (3,3,3,3,3,2),$ all considered patterns achieve the minimum squared Euclidean distance of $2$-ASK, in at least one equivalence class, typically represented by Filter \# $1$ as the best candidate. However, only three patterns---$(4,4,4,4,4,1),$ with $L_0=6,$ and $(2,2,2,2,2,2,1,0)$ and $(3,3,3,3,2,2,2,1),$ with $L_0=8$---feature the \gls{bep} of $2$-ASK as leading term in their \gls{bep} upper bound expressions. Notably, the best filter associated with the last of these patterns achieves the highest second minimum squared Euclidean distance, normalized to $8/7,$ with respect to $d_{\text{min}}^2.$ In contrast, except for patterns $(1,1,1,1),$ with $L_0=4,$ and $(1,1,1,1,0,0),$ with $L_0 \in \{4,6\},$ where all candidate filters attain the $2$-ASK minimum squared Euclidean distance, the remaining patterns include at least one filter class whose performance falls strictly below that benchmark.

\begin{figure}[!htbp]
    \centering
    \includegraphics[width=1.0\textwidth]{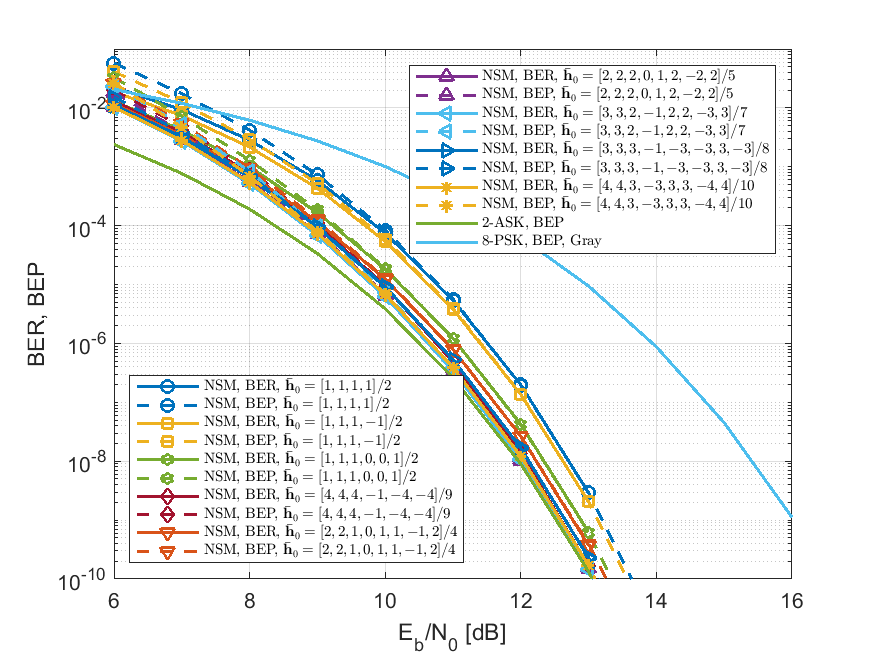}
    \caption{BER of the best rate $3/2$ NSMs, with rational filter coefficients, as specified in Tables~\ref{table:NSMs Rate-3/2 Filter Pattern (1, 1, 1, 1)}--\ref{table:NSMs Rate-3/2 Filter Pattern (4, 4, 4, 4, 3, 3, 3, 3)}, for filter patterns $(1,1,1,1),$ $(1,1,1,1,0,0),$ $(4,4,4,4,4,1),$ $(2,2,2,1,1,1,1,0),$ $(2,2,2,2,2,2,1,0),$ $(3,3,3,3,2,2,2,1),$ $(3,3,3,3,3,3,3,1,),$ and $(4,4,4,4,3,3,3,3).$ For reference the BEP upper bounds from Tables~\ref{table:NSMs Rate-3/2 Filter Pattern (1, 1, 1, 1)}--\ref{table:NSMs Rate-3/2 Filter Pattern (4, 4, 4, 4, 3, 3, 3, 3)}, along with the BEP of $2$-ASK, are shown.}
    \label{fig:BER-BEP-Non Trivial NSM-3_2}
\end{figure}

Figure~\ref{fig:BER-BEP-Non Trivial NSM-3_2} presents the \gls{ber} simulation results for the best $\mathring{h}_0[k]$ filters, across all considered filter patterns $\bm{\pi}_0,$ excluding pattern $\bm{\pi}_0= (3,3,3,3,3,2).$ For reference, the figure also includes the \gls{bep} upper bounds from Tables~\ref{table:NSMs Rate-3/2 Filter Pattern (1, 1, 1, 1)}--\ref{table:NSMs Rate-3/2 Filter Pattern (4, 4, 4, 4, 3, 3, 3, 3)} (excluding Table~\ref{table:NSMs Rate-3/2 Filter Pattern (3, 3, 3, 3, 3, 2)}), as well as the \gls{bep} of $2$-ASK, used as performance benchmark. The results show excellent alignment between simulated \gls{ber} curves and the corresponding \gls{bep} upper bounds at moderate to high \glspl{snr}. This confirms that the best $\mathring{h}_0[k]$ filters perform very closely to $2$-ASK, in accordance with predictions based on the \gls{bep} bounds. Notably, filter, $\mathring{\bm{h}}_0 = (3,3,2,-1,2,2,-3,3),$ derived from pattern $\bm{\pi}_0=(3,3,3,3,2,2,2,1),$ outperforms all others, which is consistent with its superior second normalized minimum squared Euclidean distance, $8/7$ (relative to $d_{\text{min}}^2$). This strong performance aligns well with theoretical expectations.

\section{Rate-4/3 NSMs with Rational Filter Coefficients} \label{Rate-4/3 Approaching NSMs}

\subsection{A Basic Rate-4/3 NSM Inspired by a Previous Rate-3/2 NSM} \label{A basic rate-4/3 NSM inspired by a previous basic rate-3/2 NSM}

Following the presentation and discussion in Section~\ref{ssec:Basic Rate-3/2 NSM}, we can write the normalized modulated symbols, as in (\ref{eq:Normalized Modulated Symbols Rate 3/2 MSN}), as
\begin{equation}  \label{eq:Normalized Modulated Symbols Rate 4/3 MSN}
    \bar{s}[k] = \sum_{m=0}^3 \sum_l \bar{b}_m[l] \bar{h}_m[k-3l],
\end{equation}
where $\bar{b}_0[k] \triangleq \bar{b}_0[0,k],$ $\bar{b}_m[k] \triangleq \bar{b}_1[m-1,k],$ $m=1,2,3,$ $\bar{h}_0[k] \triangleq \tfrac{1}{2}(\delta[k]+\delta[k-1]+\delta[k-2]+\delta[k-3])$ and $\bar{h}_m[k] \triangleq \delta[k-(m-1)],$ $m=1,2,3.$ 

To keep the expressions for the labels in the trellis and state diagrams, as well as the \gls{tf}, associated with the proposed rate $4/3$ \gls{nsm}, as simple as possible, we use, as in Section~\ref{ssec:Basic Rate-3/2 NSM}, the rescaled versions, $\mathring{h}_m[k] \triangleq 2 \bar{h}_m[k],$ of the filters $\bar{h}_m,$ $0 \le m \le 2,$ as well as the rescaled version, $\mathring{s}[k] \triangleq 2 \bar{s}[k],$ of the normalized modulated symbols, as expressed in (\ref{eq:Normalized Modulated Symbols Rate 4/3 MSN}). The rescaled modulated symbols, $\mathring{s}[k],$ are now expressed as
\begin{equation} \label{eq:Rescaled Modulated Symbols Rate 4/3 MSN}
    \mathring{s}[k] = \sum_{m=0}^3 \sum_l \bar{b}_m[l] \mathring{h}_m[k-3l],
\end{equation}
using the modified filters $\mathring{h}_0[k] = \delta[k]+\delta[k-1]+\delta[k-2]+\delta[k-3],$ $\mathring{h}_m[k] = 2 \delta[k-(m-1)],$ $m=1,2,3$ which have the merit of having simple integer taps.

Unlike the basic rate-$3/2$ \gls{nsm} studied in Section~\ref{ssec:Basic Rate-3/2 NSM}, we derive the \gls{tf} for the basic rate-$4/3$ \gls{nsm} at hand using a modified trellis structure. Compared to the trellis in Figure~\ref{fig:Trellis Input Difference Filter h_0 Rate-3/2 Modulation}(b), this new structure, with its subsections and intermediate substates, simplifies the derivation process. This simplification arises from the memoryless contributions of the inputs $\bar{b}_m[l], m=1,2,3,$ to the modulated sequence $\mathring{s}[k],$ which result in trellis branches formed as Cartesian products of sub-branches within the subsections---a property linked to the memoryless nature of filters $\mathring{h}_m[k] = 2 \delta[k-(m-1)],$ $m=1,2,3.$

\begin{figure}[!htbp]
    \centering
    \includegraphics[width=1\textwidth]{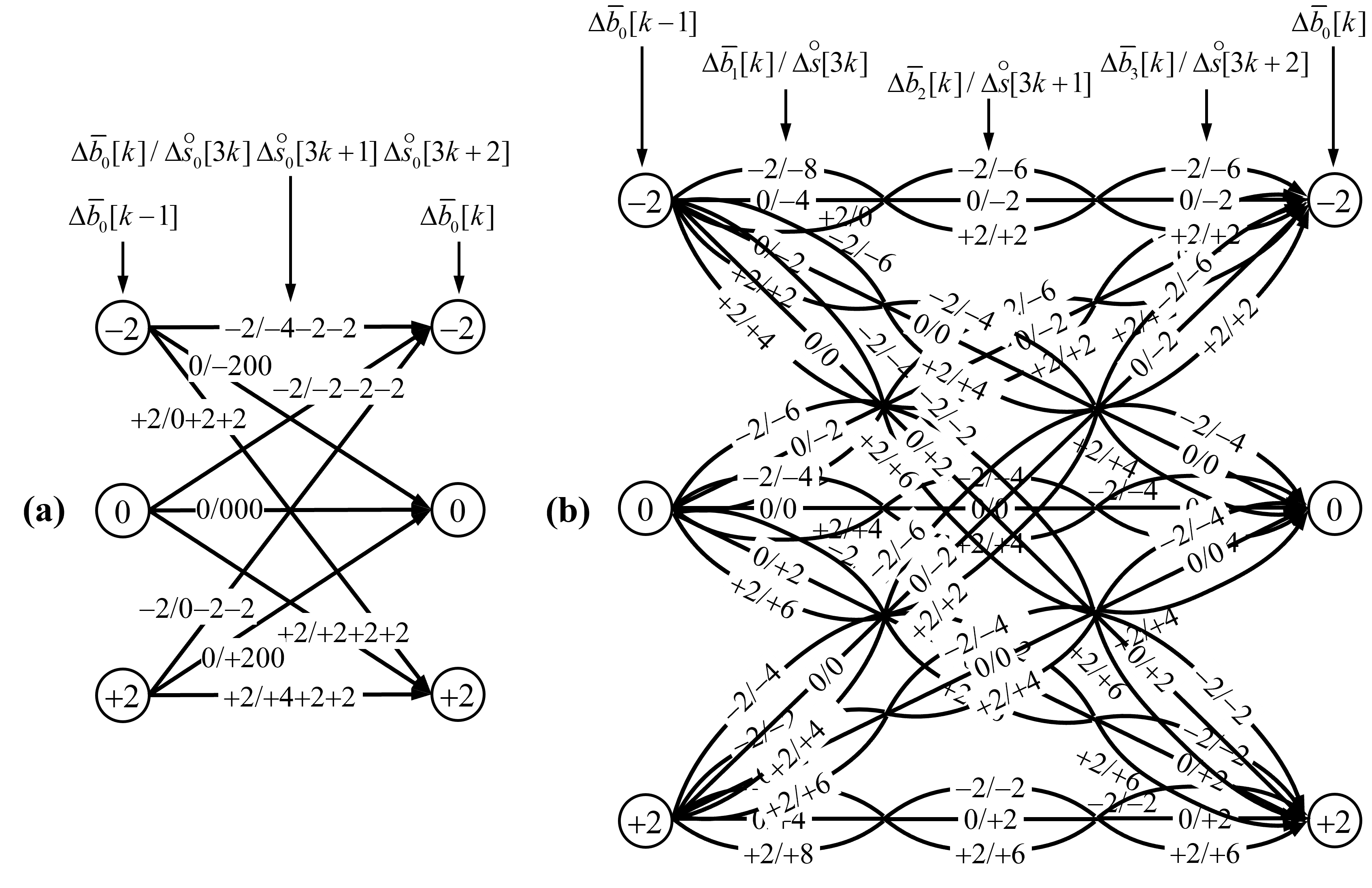}
    \caption{Trellises of the input/output sequences differences of $\mathring{h}_0[k]$ and the rate-$4/3$ NSM: (a) Filter $\mathring{h}_0[k]$, (b) Rate-$4/3$ modulation. Introduction of intermediate states and subsections in the trellis in (b) to simplify TF derivation.}
    \label{fig:Trellis Input Difference Filter h_0 Rate-4/3 Modulation}
\end{figure}

We start by showing in Figures~\ref{fig:Trellis Input Difference Filter h_0 Rate-4/3 Modulation}(a) and~\ref{fig:Trellis Input Difference Filter h_0 Rate-4/3 Modulation}(b) the trellises of the differences in input and output sequences associated to $\mathring{s}_0[k]$ and $\mathring{s}[k],$ respectively. The first trellis, depicted in Figure~\ref{fig:Trellis Input Difference Filter h_0 Rate-4/3 Modulation}(a), which corresponds to $\mathring{s}_0[k],$ provides a description of filter $\mathring{h}_0[k] = \delta[k] + \delta[k-1] + \delta[k-2] + \delta[k-3].$ Its $k$-th section is delimited by three states to the left and three other states to the right, corresponding respectively to the values that $\Delta \bar{b}_0[k-1]$ and $\Delta \bar{b}_0[k]$ take in the ternary set $\{ 0, \pm 2 \}.$ The branch in the $k$-th section of this trellis, which connects state $\Delta \bar{b}_0[k-1] \in \{ 0, \pm 2 \}$ to state $\Delta \bar{b}_0[k] \in \{ 0, \pm 2 \},$ has $\Delta \bar{b}_0[k]$ as input difference label and $\Delta \mathring{s}_0[3k] \, \Delta \mathring{s}_0[3k+1]  \, \Delta \mathring{s}_0[3k+2]$ as output difference label, where, $\Delta \mathring{s}_0[3k] = \sum_l \Delta \bar{b}_0[l] \mathring{h}_0[3k-3l] = \Delta \bar{b}_0[k] + \Delta \bar{b}_0[k-1]$
and $\Delta \mathring{s}_0[3k+1] = \Delta \mathring{s}_0[3k+2] = \sum_l \Delta \bar{b}_0[l] \mathring{h}_0[(3k+1)-3l] = \sum_l \Delta \bar{b}_0[l] \mathring{h}_0[(3k+2)-3l] = \Delta \bar{b}_0[k].$

To construct a section of the trellis for the rate-$4/3$ \gls{nsm}, as shown in Figure~\ref{fig:Trellis Input Difference Filter h_0 Rate-4/3 Modulation}(b), the three consecutive output sequence differences, $\Delta\mathring{s}_0[3k],$ $\Delta\mathring{s}_0[3k+1]$ and  $\Delta\mathring{s}_0[3k+2],$ from a branch in the trellis of Figure~\ref{fig:Trellis Input Difference Filter h_0 Rate-4/3 Modulation}(a), are each replicated three times. These replicas correspond, respectively, to the three values $\Delta \bar{b}_1[k],$ $\Delta \bar{b}_2[k])$ and $\Delta \bar{b}_3[k])$ in the ternary set $\{0, \pm 2 \}.$ More specifically, each section of the trellis in Figure~\ref{fig:Trellis Input Difference Filter h_0 Rate-4/3 Modulation}(a), with branches labeled as $\Delta \bar{b}_0[k]/\mathring{s}_0[3k] \, \mathring{s}_0[3k+1] \, \mathring{s}_0[3k+2],$ expands into three consecutive subsections. Each subsection contains three parallel sub-branches, carrying partial labels $\Delta \bar{b}_1[k]/\mathring{s}[3k],$ $\Delta \bar{b}_2[k]/\mathring{s}[3k+1]$ and $\Delta \bar{b}_3[k]/\mathring{s}[3k+2],$ corresponding to the three values of $\Delta \bar{b}_1[k], \Delta \bar{b}_2[k]$ and $\bar{b}_3[k],$ respectively. The output labels, $\Delta \mathring{s}[3k],$ $\Delta \mathring{s}[3k+1]$ and $\Delta \mathring{s}[3k+2]$ are such that $\Delta \mathring{s}[3k] = \Delta \bar{b}_0[k-1] + \Delta \bar{b}_0[k] + 2 \Delta \bar{b}_1[k],$ $\Delta \mathring{s}[3k+1] = \Delta \bar{b}_0[k] + 2 \Delta \bar{b}_2[k]$ and $\Delta \mathring{s}[3k+2] = \Delta \bar{b}_0[k] + 2 \Delta \bar{b}_3[k].$

\begin{figure}[!htbp]
    \centering
    \includegraphics[width=0.6\textwidth]{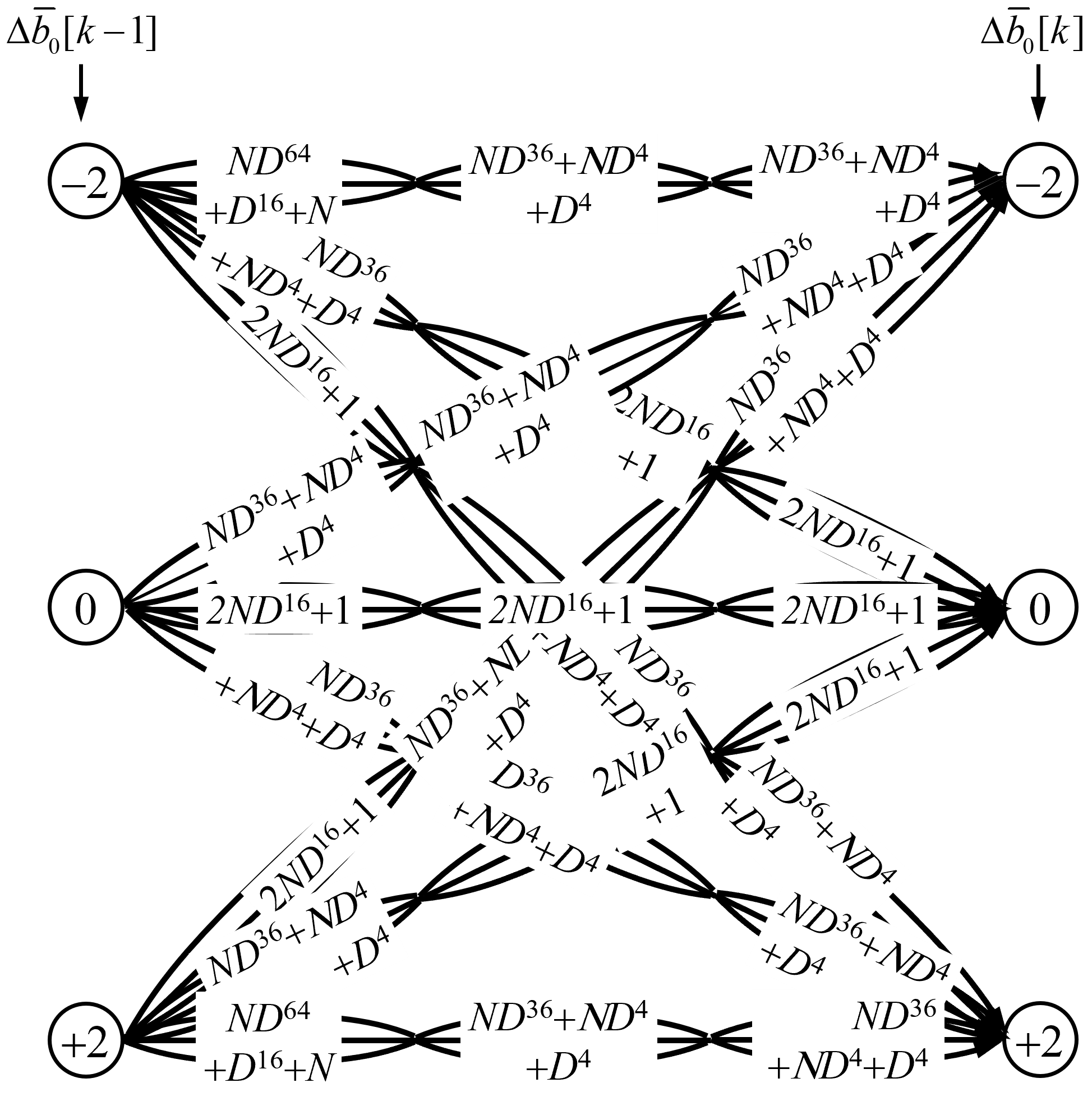}
    \caption{Trellis, with $N^iD^j$ labels, of the input/output sequences differences of the rate-$4/3$ NSM. Introduction of intermediate states and subsections in the trellis to simplify TF derivation.}
    \label{fig:Trellis Labels Filter h_0 Rate-4/3 Modulation}
\end{figure}

To derive the basic rate-$4/3$ \gls{nsm} \gls{tf}, we rename the branches in the trellis diagram shown in Figure~\ref{fig:Trellis Labels Filter h_0 Rate-4/3 Modulation} by assigning labels in the form $N^i D^j$ to the sub-branches. This renaming is based on the trellis from Figure~\ref{fig:Trellis Input Difference Filter h_0 Rate-4/3 Modulation}(b). We chose to work with the trellis diagram instead of the state diagram to maintain the clear structure of subsections, sub-branches, and intermediate states. In this process, the original sub-branch labels $\Delta \bar{b}_m[k]/\mathring{s}_0[3k+m-1],$ $m=1,2,3$ from Figure~\ref{fig:Trellis Input Difference Filter h_0 Rate-4/3 Modulation}(b) are replaced in Figure~\ref{fig:Trellis Labels Filter h_0 Rate-4/3 Modulation} with labels of the form $N^i D^j,$ where $i = |\Delta \bar{b}_m[k]|/2$ and $j = (\mathring{s}[3k+m-1])^2.$

\begin{figure}[!htbp]
    \centering
    \includegraphics[width=0.9\textwidth]{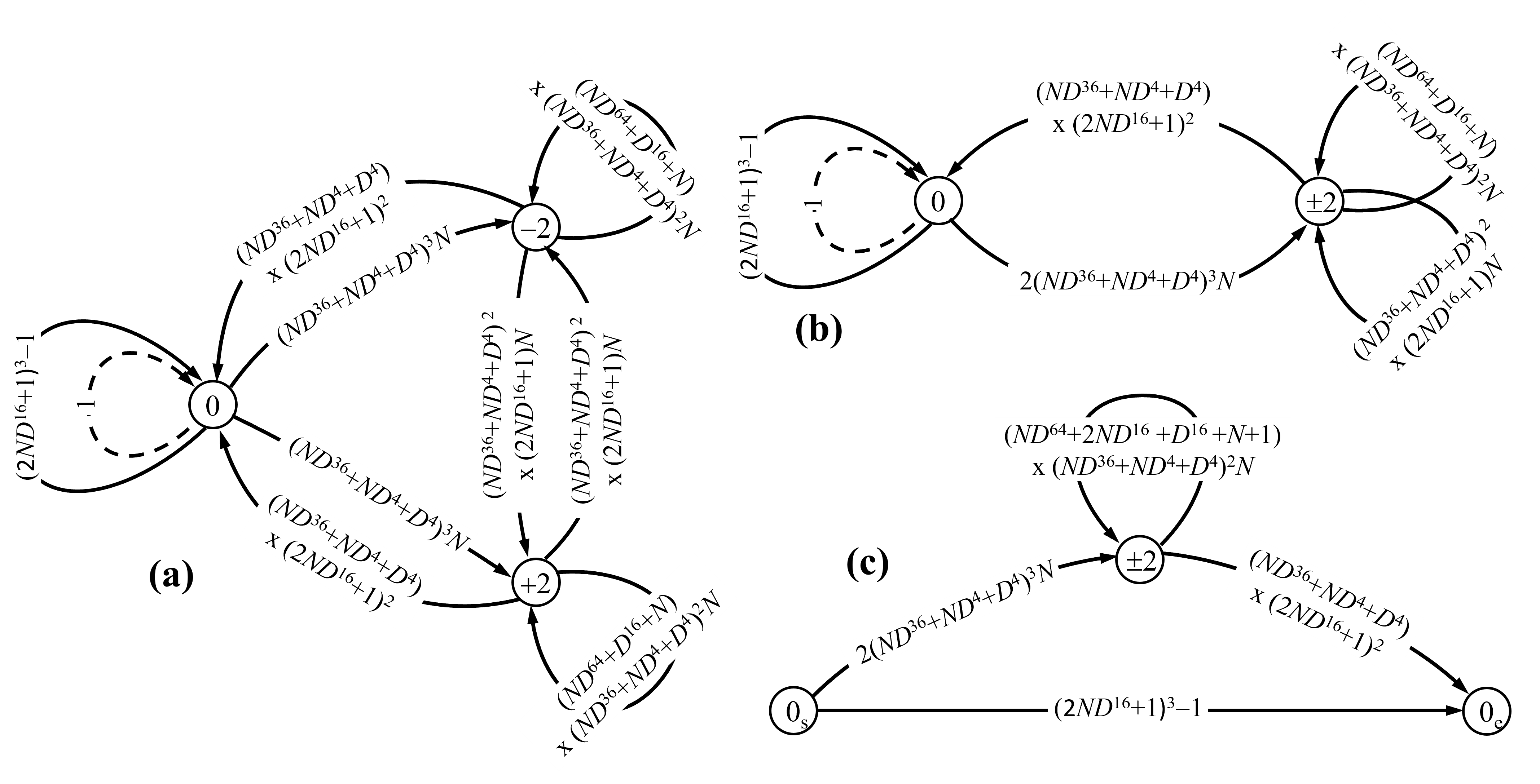}
    \caption{State diagrams, with $N^iD^j$ labels, of the input/output sequences differences of the rate-$4/3$ NSM: (a) Simplified state diagram, aggregating $N^i D^i$ labels with identical starting and ending states, (b) Simplified state diagram with state merging, (c) Modified state diagram for TF computation. Branch with zero Input/Output difference shown with dashed line.}
    \label{fig:State Diagram Input/output Difference Rate-4/3 Modulation}
\end{figure}

To compute the overall label for each transition from state $\Delta \bar{b}_0[k-1]$ to state $\Delta \bar{b}_0[k],$ in the state diagram of Figure~\ref{fig:State Diagram Input/output Difference Rate-4/3 Modulation}(a), we first add the labels $N^i D^j$ of the parallel sub-branches within each of the three consecutive subsections. Then, we multiply the contributions from these three subsections and multiply the final result by $N^i,$ where $i=|\Delta \bar{b}_0[k]|/2.$

This simplified approach, used in Algorithms~\ref{alg:Simplified One-Shot Reduced Transfer Function T(D) Rate-(Q+1)/Q NSM} and~\ref{alg:Simplified Iterative Reduced Transfer Function T(D) Rate-(Q+1)/Q NSM} (described in Appendix~\ref{app:Symbolic Determination Reduced Transfer Function Rate-(Q+1)/Q NSM}), leverages the division of each trellis section into three subsections, with elementary distinct output sequence differences, and computes the aggregate label for each state transition by multiplying the contributions from the subsections. Instead of handling $9$ parallel branches between consecutive states $\Delta \bar{b}_0[k-1]$ and $\Delta \bar{b}_0[k],$ as in a conventional trellis, we work with three subsections, each containing three parallel sub-branches. This effectively breaks the global problem into simpler partial problems, as the $9$ parallel branches in a standard trellis represent the Cartesian product of three parallel sub-branches from three consecutive subsections. As seen in Algorithms~\ref{alg:Simplified One-Shot Reduced Transfer Function T(D) Rate-(Q+1)/Q NSM} and~\ref{alg:Simplified Iterative Reduced Transfer Function T(D) Rate-(Q+1)/Q NSM}, this simplification becomes more evident when dealing with more general \glspl{nsm} with rates $\rho=(Q+1)/Q,$ particularly when $Q$ exceeds $3,$ which is the case for the current \gls{nsm} of rate $\rho=4/3.$ 

The aggregation of subsections contributions in the trellis of Figure~\ref{fig:Trellis Labels Filter h_0 Rate-4/3 Modulation} results in the state diagram shown in Figure~\ref{fig:State Diagram Input/output Difference Rate-4/3 Modulation}(a). Similar to Figure~\ref{fig:State Diagram Input/output Difference Rate-3/2 Modulation}(c) for the rate-$3/2$ \gls{nsm} discussed in detail in Appendix~\ref{app:Tight Estimate BEP Rate 3/2}, the "null" branch, labeled $N^0D^0=1,$ is excluded from the \gls{tf} expression and is kept separate from the other branches that connect to and from the same $0$ state.

To further simplify the \gls{tf} derivation, we merge states $-2$ and $+2$ from Figure~\ref{fig:State Diagram Input/output Difference Rate-4/3 Modulation}(a) into a single state, labeled $\pm 2.$ as shown in Figure~\ref{fig:State Diagram Input/output Difference Rate-4/3 Modulation}(b). This merging, similar to the approach in Figure~\ref{fig:State Diagram Input/output Difference Rate-3/2 Modulation}(d) for the rate-$3/2$ \gls{nsm} detailed in Appendix~\ref{app:Tight Estimate BEP Rate 3/2}, reduces the complexity of the state diagram and streamlines the \gls{tf} computation. 

This technique has been generalized in Appendix~\ref{app:Symbolic Determination Reduced Transfer Function Rate-(Q+1)/Q NSM} for any rate-$(Q+1)/Q$ \gls{nsm} and applied in Algorithms~\ref{alg:Simplified One-Shot Reduced Transfer Function T(D) Rate-(Q+1)/Q NSM} and~\ref{alg:Simplified Iterative Reduced Transfer Function T(D) Rate-(Q+1)/Q NSM} to simplify the symbolic determination of the \gls{nsm} \gls{tf}.

The final step before deriving the \gls{tf}, as shown in Figure~\ref{fig:State Diagram Input/output Difference Rate-4/3 Modulation}(c), involves removing the "null" branch labeled $1$ and splitting the $0$ state into two separate states: a starting state labeled $0_\text{s}$ and an ending state labeled $0_\text{e}.$

Using the state diagram in Figure~\ref{fig:State Diagram Input/output Difference Rate-4/3 Modulation}(c) and a method similar to the one applied in Appendix~\ref{app:Tight Estimate BEP Rate 3/2} to derive the \gls{tf} of the rate-$3/2$ \gls{nsm}, we obtain the desired \gls{tf},
\begin{equation}
T(N,D) = (1+2ND^{16})^3-1 + \frac{2N (1+2ND^{16})^2 (D^4+ND^4+ND^{36})^4}{1 - N (1+N+D^{16}+2ND^{16}+ND^{64}) (D^4+ND^4+ND^{36})^2}.
\end{equation}
Expanding this \gls{tf}, while focusing on the main terms with the lowest powers of $D$ corresponding to the \gls{nsm}'s \gls{msed}, yields
\begin{equation} \label{eq:Expansion Transfer Function Rate-4/3 Modulation}
T(N,D) = \left(6\,N+2\,N\,{\left(1+N\right)}^4\right)\,D^{16}+2\,N^2\,{\left(1+N\right)}^7\,D^{24} + \cdots.
\end{equation}
As with the rate-$3/2$ \gls{nsm} thoroughly analyzed in Appendix~\ref{app:Tight Estimate BEP Rate 3/2}, this expansion reinforces the hypothesis that the rate-$4/3$ \gls{nsm}'s \gls{msed}, when using normalized $\bar{h}_m[k],$ $m \in \{0,1,2,3\},$ equals 4. This conclusion is based on the fact that the smallest \gls{sed} with scaled filters $\mathring{h}_m[k],$ $m \in \{0,1,2,3\},$ (scaled by a factor of 2) is 16. Consequently, the proposed rate-$4/3$ \gls{nsm} possesses the highly desirable property of perfectly achieving the \gls{msed} of $2$-ASK.

In order to derive an expansion of the \gls{rtf}, limited to its major terms, for the suggested rate-$4/3$ \gls{nsm}, we apply the same rules as in Appendices~\ref{app:Tight Estimate BEP Rate 2} and~\ref{app:Tight Estimate BEP Rate 3/2} and obtain,
\begin{equation} \label{eq:Multiplicity and Distance Rate-4/3 Modulation}
\dot{T}(D) = \left. N \cfrac{\partial T(N,D)}{\partial N} \right|_{N=1/2} =
\frac{237}{16}\,D^{16}+\frac{9477}{256}\,D^{24}+\frac{536793}{4096}\,D^{32}+\frac{22617225}{65536}\,D^{40}+ \cdots.
\end{equation}
This expansion perfectly aligns with the result obtained directly in Table~\ref{table:NSMs Rate-4/3 Filter Pattern (1, 1, 1, 1, 0, 0)} for the same \gls{nsm}, using scaled filter \# $1,$ $\mathring{h}_0[k] = \delta[k] + \delta[k-1] + \delta[k-2] + \delta[k-3],$ and symbolic calculus through Algorithms~\ref{alg:Simplified One-Shot Reduced Transfer Function T(D) Rate-(Q+1)/Q NSM} and~\ref{alg:Simplified Iterative Reduced Transfer Function T(D) Rate-(Q+1)/Q NSM}. Notably, the latter approach, based on symbolic calculus and advocated in Appendix~\ref{app:Symbolic Determination Reduced Transfer Function Rate-(Q+1)/Q NSM}, eliminates the need for a more intricate two-step procedure. In contrast, in the alternative solution implemented here, the more advanced two-argument \gls{tf} $T(N,D)$ is initially computed as an auxiliary function.


\begin{table}[H]
\caption{Selected rate-$4/3$ NSMs, with simple filters' coefficients, with common filter $\bm{h}_0$ pattern $\bm{\pi}_0 = (1, 1, 1, 1, 0, 0),$ arranged in a decreasing performance order, starting from the best one. Relative degradation in performance due to distance reduction or/and multiplicity increase when moving from one filter to the next is emphasized in bold.}
\label{table:NSMs Rate-4/3 Filter Pattern (1, 1, 1, 1, 0, 0)}
\centering
\begin{tabular}{|c|c|} 
\hline
\multicolumn{2}{|l|}{\# of simple $h_0[k]$ filters $ = 240$} \\ \hline
\multicolumn{2}{|l|}{\# of non equivalent $h_0[k]$ filters $ = 3$} \\ \hline
\multicolumn{2}{|l|}{$\mathring{h}_1[k] = 2 \delta[k],$ $\mathring{h}_2[k] = 2 \delta[k-1]$ and $\mathring{h}_3[k] = 2 \delta[k-2]$} \\ \hline
\multicolumn{2}{|l|}{Minimum squared Euclidean distance, $d_\text{min}^2 = 16$} \\ \hline
\multicolumn{2}{|l|}{Number of trellis states $ = 2$} \\  [0.5ex]
\hline\hline
\multicolumn{2}{|c|}{Filter \# $1$} \\ \hline
$\mathring{h}_0[k]$ & $\delta[k] + \delta[k-1] + \delta[k-2] + \delta[k-3]$ \\ \hline
$\mathring{\bm{h}}_0$ & $(1, 1, 1, 1, 0, 0)$ \\ \hline
\multirow{2}{*}{$\dot{T}(D)$} & $\tfrac{237}{16} D^{16} + \tfrac{9477}{256} D^{24} + \tfrac{536793}{4096}D^{32} + \tfrac{22617225}{65536} D^{40} + \tfrac{975928053}{1048576} D^{48}$ \\
& $ + \tfrac{40273427853}{16777216} D^{56} + \cdots$ \\ \hline
\multirow{2}{*}{BEP approximation} & $\frac{237}{128} \operatorname{erfc}\left( \sqrt{\frac{E_b}{N_0}} \right)
+ \tfrac{9477}{2048}\operatorname{erfc}\left( \sqrt{\tfrac{3}{2} \frac{E_b}{N_0}} \right)
+ \tfrac{536793}{32768} \operatorname{erfc}\left( \sqrt{2 \frac{E_b}{N_0}} \right)$ \\
& $+ \tfrac{22617225}{524288} \operatorname{erfc}\left( \sqrt{\tfrac{5}{2} \frac{E_b}{N_0}} \right) + \frac{975928053}{8388608
} \operatorname{erfc}\left( \sqrt{3 \frac{E_b}{N_0}} \right)$  \\ [0.5ex] 
\hline\hline
\multicolumn{2}{|c|}{Filter \# $2$} \\ \hline
$\mathring{h}_0[k]$ & $\delta[k] + \delta[k-1] + \delta[k-3] - \delta[k-4]$ \\ \hline
$\mathring{\bm{h}}_0$ & $(1, 1, 0, 1, -1, 0)$ \\ \hline
$\dot{T}(D)$ & $\tfrac{\bm{375}}{\bm{8}} D^{16} + \tfrac{903}{2} D^{32} + \tfrac{13323}{4} D^{48} + 22302 D^{64} + \cdots$ \\ [0.5ex] 
\hline\hline
\multicolumn{2}{|c|}{Filter \# $3$} \\ \hline
$\mathring{h}_0[k]$ & $\delta[k] + \delta[k-1] + \delta[k-3] + \delta[k-4]$ \\ \hline
$\mathring{\bm{h}}_0$ & $(1, 1, 0, 1, 1, 0)$ \\ \hline
$\dot{T}(D)$ & $\bm{66} D^{16} + \tfrac{1713}{2} D^{32} + 8483 D^{48} + \tfrac{226718}{3} D^{64} + \cdots$ \\ [1ex] 
 \hline
\end{tabular}
\end{table}

\subsection{A Broader Collection of Rate-4/3 NSMs} 
\label{A broader collection of rate-4/3 NSMs}

As in Section~\ref{A broader collection of rate-3/2 NSMs}, we propose and evaluate the performance of several practical rate-$4/3$ \glspl{nsm} with rational filters coefficients. These \glspl{nsm} use four normalized filters, $\bar{h}_m[k],$ $m \in \{0,1,2,3\},$ with rational coefficients, such that their scaled versions, $\mathring{h}_m[k],$ $m \in \{0,1,2,3\},$ share a common norm and possess integer coefficients. Specifically, the normalized filters $\bar{h}_m[k],$ $m \in \{1,2,3\},$ are explicitly defined as $\bar{h}_m[k]=\delta[k-m+1].$ Meanwhile, the scaled version, $\mathring{h}_0[k],$ of $\bar{h}_0[k],$ is expressed in its vectorized form, $\mathring{\bm{h}}_0,$ using a pattern vector, $\bm{\pi}_0,$ of the same dimension, with non-negative integer components.

Based on this framework, candidate scaled filters $\mathring{\bm{h}}_0$ are generated by arbitrarily permuting the components of $\bm{\pi}_0$ and optionally changing their signs. Consequently, we note that $\| \mathring{\bm{h}}_0 \| = \| \bm{\pi}_0 \|$ and $\mathring{h}_m[k]=\| \bm{\pi}_0 \|\delta[k-m+1].$

As a result, the norm $\| \bm{\pi}_0 \|,$ of pattern vector $\bm{\pi}_0$ must be an integer, since the scaled filters, $\mathring{h}_m[k],$ $m \in \{1,2,3\},$ are required to have integer coefficients. Additionally, to guarantee that the resulting \gls{nsm} achieves the minimum Euclidean distance of the $2$-ASK, the maximum component of $\bm{\pi}_0$ --- which is also equal to its infinity norm, $\| \bm{\pi}_0 \|_\infty$ --- must not exceed half of its Euclidean norm, $\| \bm{\pi}_0 \|.$ When $\| \bm{\pi}_0 \|_\infty = \| \bm{\pi}_0 \|/2,$ the tightness property arises, a condition frequently emphasized in this text. However, this property  has the drawback of significantly increasing the multiplicity of error events with the minimum Euclidean distance.

To improve the performance of the resulting \glspl{nsm}, it is generally advisable to select pattern vectors with $\| \bm{\pi}_0 \|_\infty$ values as small as possible compared to $\| \bm{\pi}_0 \|/2.$ Unfortunately, achieving this desirable trait comes at a cost: it increases the length, $L_0,$ of normalized filter $\bar{\bm{h}}_0,$ which, in turn, raises the detection complexity at the receiver. This increased complexity is due to the exponential growth in the number of detection trellis states, driven by the additional memory, relative to input sequence $b_0[k],$ required as a result of the longer filter length, $L_0.$

Most of the time, arbitrary permutations and sign changes, applied to any pattern vector $\bm{\pi}_0$, produce a substantial collection of potential candidate filter vectors $\mathring{\bm{h}}_0.$ Similar to rate-$3/2$ \glspl{nsm}, we introduce specific transformations to vectors $\mathring{\bm{h}}_0,$ that preserve the \gls{tf} $T(N,D),$ and, by extension, the \gls{rtf} $\dot{T}(D).$ These transformations establish an equivalence relation among filters, allowing us to group them into equivalence classes. By focusing on a smaller set of representative filters from these classes, we simplify the analysis in terms of \glspl{rtf} and corresponding bounds on error probability.

Similar to rate-$3/2$ \glspl{nsm}, we consider several equivalence transformations that preserve the key properties of rate-$4/3$ \glspl{nsm}. These transformations include sign change, time reversal, sign alternation, scrambling, time shifts, and permutations. Sign alternation yields equivalent filters $\tilde{\mathring{h}}_0[k] \triangleq (-1)^{\lfloor (k-l)/3 \rfloor} \mathring{h}_0[k],$ $l$ being an arbitrary integer, with only six possible “polyphase” alternating transformations, half of which can be obtained by applying a sign change change to the other half.

Scrambling produces an equivalent scrambled filter, $\tilde{\mathring{h}}_0[k] \triangleq r[k] \mathring{h}_0[k],$ where $r[k] = r[k \bmod 3],$ is a periodic bipolar scrambling sequence of period $3,$ repeating the length-$3$ pattern $r[0]r[1]r[2].$ Time shifting may alter the structure of the  detection trellis (affecting the number of states and branch labels), but it doesn't impact the \gls{nsm} performance.

Permutations, $\pi(\cdot),$ which operate on the ternary set, $\{0,1,2\},$ transform each consecutive triplet of output samples, $\mathring{s}_0[3k] \mathring{s}_0[3k+1] \mathring{s}_0[3k+2],$ into a new triplet, $\mathring{s}_0[3k+\pi(0)]  \mathring{s}_0[3k+\pi(1)] \mathring{s}_0[3k+\pi(2)].$ They result in an equivalent \gls{nsm} with filter $\tilde{\mathring{h}}_0[k] \triangleq \mathring{h}_0[k - (k \bmod 3)+\pi(k \bmod 3)].$

In Tables~\ref{table:NSMs Rate-4/3 Filter Pattern (1, 1, 1, 1, 0, 0)}--\ref{table:NSMs Rate-4/3 Filter Pattern (3, 3, 3, 3, 3, 3, 3, 1, 0)}, we present the characteristics of rate-$4/3$ \glspl{nsm}, with rational filters coefficients, specified by their patterns $\bm{\pi}_0.$ We consider two sets of patterns. The first set includes patterns $(1,1,1,1,0,0),$ $(3,3,3,3,3,2)$ and $(4,4,4,4,4,1),$ which yield candidate filters, $\mathring{h}_0[k],$ of common length, $L_0=6,$ and detection trellises with $2$ states. The second set comprises patterns $(1,1,1,1,1,1,1,1,1),$ $(3,3,3,3,2,2,2,1,0)$ and $(3,3,3,3,3,3,3,1,0),$ resulting in candidate filters, $\mathring{h}_0[k],$ of length, $L_0=9,$ and detection trellises with $4$ states.

For each proposed filter pattern, we provide the total number of possible candidate filters $\mathring{h}_0[k]$ and the count of non-equivalent filters representatives among them. For completeness, we also specify the additional filters, expressed as $\mathring{h}_m[k] = \| \bm{\pi}_0 \| \delta[k-(m-1)] = 2 \delta[k-(m-1)],$ $m = 1,2,3.$ For several of the non-equivalent filters, we list the first main terms of the \gls{rtf}, $\dot{T}(D),$ corresponding to their underlying \gls{nsm}. These terms are derived using symbolic calculus and Taylor series expansions, through either Algorithm~\ref{alg:Simplified One-Shot Reduced Transfer Function T(D) Rate-(Q+1)/Q NSM} or Algorithm~\ref{alg:Simplified Iterative Reduced Transfer Function T(D) Rate-(Q+1)/Q NSM}.

The selected non-equivalent filters candidates are ranked from the best to the worst, based on their distance spectrum, relying on the most significant terms of their \glspl{rtf}, $\dot{T}(D).$ To highlight performance degradation when transitioning from one of the selected non-equivalent filters candidates to the next, we emphasize in bold the specific cause: either a reduction in the \gls{sed} or an increase in the multiplicity of certain error events. Crucially, this degradation always affects a subsequent term of the \gls{rtf}, while the preceding terms remain unchanged, ensuring that it represents the most impactful decline when moving from one \gls{nsm} to the next.

Finally, for the best non-equivalent filter candidates—referred to as Filter \# $1$ in the tables—we provide an approximate upper bound on the \gls{bep}, for each considered pattern. This bound is known for its tight accuracy at moderate to high \gls{snr} values, which will soon be validated through \gls{ber} simulations.

For all filter patterns analyzed, the leading term in the \gls{bep} upper bound indicates that the best filters asymptotically achieve the performance of $2$-ASK. In most cases, the multiplicity of this term relative to $2$-ASK \gls{bep} is $1,$ suggesting that the performance of these filters closely matches that of $2$-ASK. This will be further confirmed next through \gls{ber} simulation results.

In Table~\ref{table:NSMs Rate-4/3 Filter Pattern (1, 1, 1, 1, 0, 0)}, we examine pattern $\bm{\pi}_0=(1,1,1,1,0,0),$ which generates $240$ candidate $\mathring{h}_0[k]$ filters. These candidates can be grouped into $3$ non-equivalent filters, represented in vector form as $\mathring{\bm{h}}_0 = (1, 1, 1, 1, 0, 0),$ $(1, 1, 0, 1, -1, 0)$ and $(1, 1, 0, 1, 1, 0).$ Among these, the best representative is $\mathring{\bm{h}}_0 = (1, 1, 1, 1, 0, 0),$ corresponding to the \gls{nsm} system thoroughly analyzed in Section~\ref{A basic rate-4/3 NSM inspired by a previous basic rate-3/2 NSM}. The first terms of the corresponding \gls{rtf}, $\dot{T}(D),$ listed in Table~\ref{table:NSMs Rate-4/3 Filter Pattern (1, 1, 1, 1, 0, 0)} and derived using symbolic calculus and Taylor series expansions, perfectly match the results obtained manually through mathematical calculations, in (\ref{eq:Multiplicity and Distance Rate-4/3 Modulation}).

The first term of the associated tight upper bound of the \gls{bep}, expressed as $\tfrac{237}{128} \operatorname{erfc} (\sqrt{E_b/N_0}),$ is derived from the leading term, $\tfrac{237}{16} D^{16},$ of the \gls{rtf}. This derivation follows two key observations. First, the exponent of $D$ corresponds exactly to the \gls{msed}, $d_{\text{min}}^2=4 \|\mathring{\bm{h}}_0 \|^2 = 4 \|\bm{\pi}_0 \|^2=16,$ which matches the \gls{msed} of $2$-ASK. Consequently, the probability of the associated error events equals the \gls{bep} of $2$-ASK, $\tfrac{1}{2}\operatorname{erfc} (\sqrt{E_b/N_0}),$ multiplied by the error events multiplicity, $\tfrac{237}{16}.$ Second, to obtain a bound on the \gls{bep}, the resulting error event probability is divided by $4,$ which accounts for the $4$ simultaneous input sequences, $\bar{b}_m[k],$ $m \in \{0,1,2,3\}.$

As previously mentioned, the best filter,  $\mathring{\bm{h}}_0 = (1, 1, 1, 1, 0, 0),$ for pattern $\bm{\pi}_0=(1,1,1,1,0,0),$ asymptotically matches the performance of $2$-ASK. However, it is important to emphasize that the corresponding leading term in the \gls{bep} is $\tfrac{237}{64}$ times that of $2$-ASK. This increase in multiplicity is anticipated and is partly attributed to the degeneracy of filter $\mathring{h}_0[k]$ relative to filters $\mathring{h}_m[k],$ $m \in \{1,2,3\},$ whose non-zero coefficients are twice the $\mathring{h}_0[k].$

The two remaining non-equivalent filter representatives, $(1, 1, 0, 1, -1, 0)$ and $(1, 1, 0, 1, 1, 0),$ referred to as Filters \# $2$ and \# $3,$ each correspond to two decoupled time-multiplexed streams, associated to \gls{nsm} systems of rates $3/2$ and $1,$ respectively. The rate-$3/2$ \gls{nsm} has $\mathring{b}_0[k] \mathring{b}_1[k] \mathring{b}_2[k]/\mathring{s}_0[3k] \mathring{s}_1[3k+1]$ as its trellis input/output branch label and uses either $(1, 1, 1, -1)$ or $(1, 1, 1, 1)$ as its specific $\mathring{\bm{h}}_0$ filter. The rate-$1$ \gls{nsm}, on the other hand, is trivial, with a memoryless input/output correspondence labeled $\mathring{b}_3[k]/\mathring{s}_0[3k+2]].$ 

As established in Section~\ref{ssec:Basic Rate-3/2 NSM}, the rate-$3/2$ \gls{nsm}, with $\mathring{\bm{h}}_0=(1, 1, 1, -1),$ outperforms the one with $\mathring{\bm{h}}_0=(1, 1, 1, 1).$ This superiority is inherited here by the rate-$4/3$ \gls{nsm} with $\mathring{\bm{h}}_0=(1, 1, 0, 1, -1, 0),$ which is shown to be superior to the rate-$4/3$ \gls{nsm} with $\mathring{\bm{h}}_0=(1, 1, 0, 1, 1, 0),$ as evidenced by their corresponding \glspl{rtf}, listed in Table~\ref{table:NSMs Rate-4/3 Filter Pattern (1, 1, 1, 1, 0, 0)}.

Except for pattern $\bm{\pi}_0=(1,1,1,1,0,0),$ which has been studied in detail, none of other considered patterns exhibit tightness properties. This is a desirable feature, as it helps prevent high multiplicities in the leading term in the \gls{bep} upper bound. As shown in Tables~\ref{table:NSMs Rate-4/3 Filter Pattern (3, 3, 3, 3, 3, 2)}--\ref{table:NSMs Rate-4/3 Filter Pattern (3, 3, 3, 3, 3, 3, 3, 1, 0)}, this property ensures a multiplicity of $1,$ for the leading \gls{bep} upper bound term, in most cases. However it  fails to do so for pattern $\bm{\pi}_0=(1,1,1,1,1,1,1,1,1).$


\begin{table}[H]
\caption{Selected rate-$4/3$ NSMs, with simple filters' coefficients, with common filter $\bm{h}_0$ pattern $\bm{\pi}_0 = (3, 3, 3, 3, 3, 2),$ arranged in a decreasing performance order, starting from the best one. Relative degradation in performance due to distance reduction or/and multiplicity increase when moving from one filter to the next is emphasized in bold.}
\label{table:NSMs Rate-4/3 Filter Pattern (3, 3, 3, 3, 3, 2)}
\centering
\begin{tabular}{|c|c|} 
\hline
\multicolumn{2}{|l|}{\# of simple $h_0[k]$ filters $ = 384$} \\ \hline
\multicolumn{2}{|l|}{\# of non equivalent $h_0[k]$ filters $ = 3$} \\ \hline
\multicolumn{2}{|l|}{$\mathring{h}_1[k] = 7 \delta[k],$ $\mathring{h}_2[k] = 7 \delta[k-1]$ and $\mathring{h}_3[k] = 7 \delta[k-2]$} \\ \hline
\multicolumn{2}{|l|}{Minimum squared Euclidean distance, $d_\text{min}^2 = 196$} \\ \hline
\multicolumn{2}{|l|}{Number of trellis states $ = 2$} \\  [0.5ex]
\hline\hline
\multicolumn{2}{|c|}{Filter \# $1$} \\ \hline
$\mathring{h}_0[k]$ & $3 \delta[k] + 3 \delta[k-1] + 3 \delta[k-2] + 3 \delta[k-3] + 3 \delta[k-4] - 2 \delta[k-5]$ \\ \hline
$\mathring{\bm{h}}_0$ & $(3, 3, 3, 3, 3, -2)$ \\
\hline
\end{tabular}
\end{table}


\begin{table}[H]
\caption{Selected rate-$4/3$ NSMs, with simple filters' coefficients, with common filter $\bm{h}_0$ pattern $\bm{\pi}_0 = (4, 4, 4, 4, 4, 1),$ arranged in a decreasing performance order, starting from the best one. Relative degradation in performance due to distance reduction or/and multiplicity increase when moving from one filter to the next is emphasized in bold.}
\label{table:NSMs Rate-4/3 Filter Pattern (4, 4, 4, 4, 4, 1)}
\centering
\begin{tabular}{|c|c|} 
\hline
\multicolumn{2}{|l|}{\# of simple $h_0[k]$ filters $ = 384$} \\ \hline
\multicolumn{2}{|l|}{\# of non equivalent $h_0[k]$ filters $ = 3$} \\ \hline
\multicolumn{2}{|l|}{$\mathring{h}_1[k] = 9 \delta[k],$ $\mathring{h}_2[k] = 9 \delta[k-1]$ and $\mathring{h}_3[k] = 9 \delta[k-2]$} \\ \hline
\multicolumn{2}{|l|}{Minimum squared Euclidean distance, $d_\text{min}^2 = 324$} \\ \hline
\multicolumn{2}{|l|}{Number of trellis states $ = 2$} \\  [0.5ex]
\hline\hline
\multicolumn{2}{|c|}{Filter \# $1$} \\ \hline
$\mathring{h}_0[k]$ & $4 \delta[k] + 4 \delta[k-1] + 4 \delta[k-2] + 4 \delta[k-3] + 4 \delta[k-4] - \delta[k-5]$ \\ \hline
$\mathring{\bm{h}}_0$ & $(4, 4, 4, 4, 4, -1)$ \\
\hline
\end{tabular}
\end{table}


\begin{table}[H]
\caption{Selected rate-$4/3$ NSMs, with simple filters' coefficients, with common filter $\bm{h}_0$ pattern $\bm{\pi}_0 = (1, 1, 1, 1, 1, 1, 1, 1, 1),$ arranged in a decreasing performance order, starting from the best one. Relative degradation in performance due to distance reduction or/and multiplicity increase when moving from one filter to the next is emphasized in bold.}
\label{table:NSMs Rate-4/3 Filter Pattern (1, 1, 1, 1, 1, 1, 1, 1, 1)}
\centering
\begin{tabular}{|c|c|} 
\hline
\multicolumn{2}{|l|}{\# of simple $h_0[k]$ filters $ = 512$} \\ \hline
\multicolumn{2}{|l|}{\# of non equivalent $h_0[k]$ filters $ = 8$} \\ \hline
\multicolumn{2}{|l|}{$\mathring{h}_1[k] = 3 \delta[k],$ $\mathring{h}_2[k] = 3 \delta[k-1]$ and $\mathring{h}_3[k] = 3 \delta[k-2]$} \\ \hline
\multicolumn{2}{|l|}{Best minimum squared Euclidean distance, $d_\text{min}^2 = 36$} \\ \hline
\multicolumn{2}{|l|}{Number of trellis states $ = 4$} \\  [0.5ex]
\hline\hline
\multicolumn{2}{|c|}{Filter \# $1$} \\ \hline
\multirow{2}{*}{$\mathring{h}_0[k]$} & $\delta[k] + \delta[k-1] + \delta[k-2] + \delta[k-3] + \delta[k-4] - \delta[k-5]$ \\
& $+ \delta[k-6] - \delta[k-7] + \delta[k-8]$ \\
\hline
\end{tabular}
\end{table}


\begin{table}[H]
\caption{Selected rate-$4/3$ NSMs, with simple filters' coefficients, with common filter $\bm{h}_0$ pattern $\bm{\pi}_0 = (3, 3, 3, 3, 2, 2, 2, 1, 0),$ arranged in a decreasing performance order, starting from the best one. Relative degradation in performance due to distance reduction or/and multiplicity increase when moving from one filter to the next is emphasized in bold.}
\label{table:NSMs Rate-4/3 Filter Pattern (3, 3, 3, 3, 2, 2, 2, 1, 0)}
\centering
\begin{tabular}{|c|c|} 
\hline
\multicolumn{2}{|l|}{\# of simple $h_0[k]$ filters $ = 645120$} \\ \hline
\multicolumn{2}{|l|}{\# of non equivalent $h_0[k]$ filters $ = 2294$} \\ \hline
\multicolumn{2}{|l|}{$\mathring{h}_1[k] = 7 \delta[k],$ $\mathring{h}_2[k] = 7 \delta[k-1]$ and $\mathring{h}_3[k] = 7 \delta[k-2]$} \\ \hline
\multicolumn{2}{|l|}{Best minimum squared Euclidean distance, $d_\text{min}^2 = 196$} \\ \hline
\multicolumn{2}{|l|}{Number of trellis states $ = 4$} \\  [0.5ex]
\hline\hline
\multicolumn{2}{|c|}{Filter \# $1$} \\ \hline
\multirow{2}{*}{$\mathring{h}_0[k]$} & $3 \delta[k] + 3 \delta[k-1] + 2 \delta[k-2] + 2 \delta[k-3] + 3 \delta[k-5] - 2 \delta[k-6]$ \\
& $- 3 \delta[k-7] + \delta[k-8]$ \\
\hline
\end{tabular}
\end{table}


\begin{table}[H]
\caption{Selected rate-$4/3$ NSMs, with simple filters' coefficients, with common filter $\bm{h}_0$ pattern $\bm{\pi}_0 = (3, 3, 3, 3, 3, 1, 1, 1, 1),$ arranged in a decreasing performance order, starting from the best one. Relative degradation in performance due to distance reduction or/and multiplicity increase when moving from one filter to the next is emphasized in bold.}
\label{table:NSMs Rate-4/3 Filter Pattern (3, 3, 3, 3, 3, 1, 1, 1, 1)}
\centering
\begin{tabular}{|c|c|} 
\hline
\multicolumn{2}{|l|}{\# of simple $h_0[k]$ filters $ = 64512$} \\ \hline
\multicolumn{2}{|l|}{\# of non equivalent $h_0[k]$ filters $ = 380$} \\ \hline
\multicolumn{2}{|l|}{$\mathring{h}_1[k] = 7 \delta[k],$ $\mathring{h}_2[k] = 7 \delta[k-1]$ and $\mathring{h}_3[k] = 7 \delta[k-2]$} \\ \hline
\multicolumn{2}{|l|}{Best minimum squared Euclidean distance, $d_\text{min}^2 = 196$} \\ \hline
\multicolumn{2}{|l|}{Number of trellis states $ = 4$} \\  [0.5ex]
\hline\hline
\multicolumn{2}{|c|}{Filter \# $1$} \\ \hline
\multirow{2}{*}{$\mathring{h}_0[k]$} & $3 \delta[k] + 3 \delta[k-1] + \delta[k-2] + 3 \delta[k-3] + \delta[k-4] - \delta[k-5]$ \\
& $- \delta[k-6] + 3 \delta[k-7] + 3 \delta[k-8]$ \\
\hline
\end{tabular}
\end{table}


\begin{table}[H]
\caption{Selected rate-$4/3$ NSMs, with simple filters' coefficients, with common filter $\bm{h}_0$ pattern $\bm{\pi}_0 = (3, 3, 3, 3, 3, 3, 3, 1, 0),$ arranged in a decreasing performance order, starting from the best one. Relative degradation in performance due to distance reduction or/and multiplicity increase when moving from one filter to the next is emphasized in bold.}
\label{table:NSMs Rate-4/3 Filter Pattern (3, 3, 3, 3, 3, 3, 3, 1, 0)}
\centering
\begin{tabular}{|c|c|} 
\hline
\multicolumn{2}{|l|}{\# of simple $h_0[k]$ filters $ = 18432$} \\ \hline
\multicolumn{2}{|l|}{\# of non equivalent $h_0[k]$ filters $ = 82$} \\ \hline
\multicolumn{2}{|l|}{$\mathring{h}_1[k] = 8 \delta[k],$ $\mathring{h}_2[k] = 8 \delta[k-1]$ and $\mathring{h}_3[k] = 8 \delta[k-2]$} \\ \hline
\multicolumn{2}{|l|}{Best minimum squared Euclidean distance, $d_\text{min}^2 = 256$} \\ \hline
\multicolumn{2}{|l|}{Number of trellis states $ = 4$} \\  [0.5ex]
\hline\hline
\multicolumn{2}{|c|}{Filter \# $1$} \\ \hline
\multirow{2}{*}{$\mathring{h}_0[k]$} & $3 \delta[k] + 3 \delta[k-1] + 3 \delta[k-2] + 3 \delta[k-3] + 3 \delta[k-4] + 3 \delta[k-6]$ \\
& $- \delta[k-7] - 3 \delta[k-8]$ \\
\hline
\end{tabular}
\end{table}

To understand why the best filter, expressed in vector form as $\mathring{\bm{h}}_0=(1,1,1,1,1,-1,1,-1,1),$ and associated to pattern $\bm{\pi}_0=(1,1,1,1,1,1,1,1,1),$ does not achieve a multiplicity of $1$ for the leading term in the \gls{bep} upper bound, consider the first input sequence differences $\Delta \bar{b}_0[k] = \pm 2 (\delta[k] - \delta[k-1]).$ In vector form, this sequence contributes partially to the output sequence difference, $\Delta \mathring{s}[k],$ through $\Delta \mathring{\bm{s}}_0 = \pm 2 (1,1,1,0,0,-2,0,-2,2,-1,1,-1).$

By selecting the remaining input sequence differences as $\Delta \bar{b}_1[k] = 0,$ $\Delta \bar{b}_2[k] = \pm 2 \delta[k-2]$ and $\Delta \bar{b}_3[k] = \pm 2 (\delta[k-1] - \delta[k-2]),$ the resulting output sequence difference takes the vector form 
$\Delta \mathring{\bm{s}} = \pm 2 (1,1,1,0,0,1,0,1,-1,-1,1,-1).$ In \gls{sen}, this vector leads to the \gls{msed}, $d_{\text{min}}^2 = 36,$ reported in Table~\ref{table:NSMs Rate-4/3 Filter Pattern (1, 1, 1, 1, 1, 1, 1, 1, 1)}.

\begin{figure}[!htbp]
    \centering
    \includegraphics[width=1.0\textwidth]{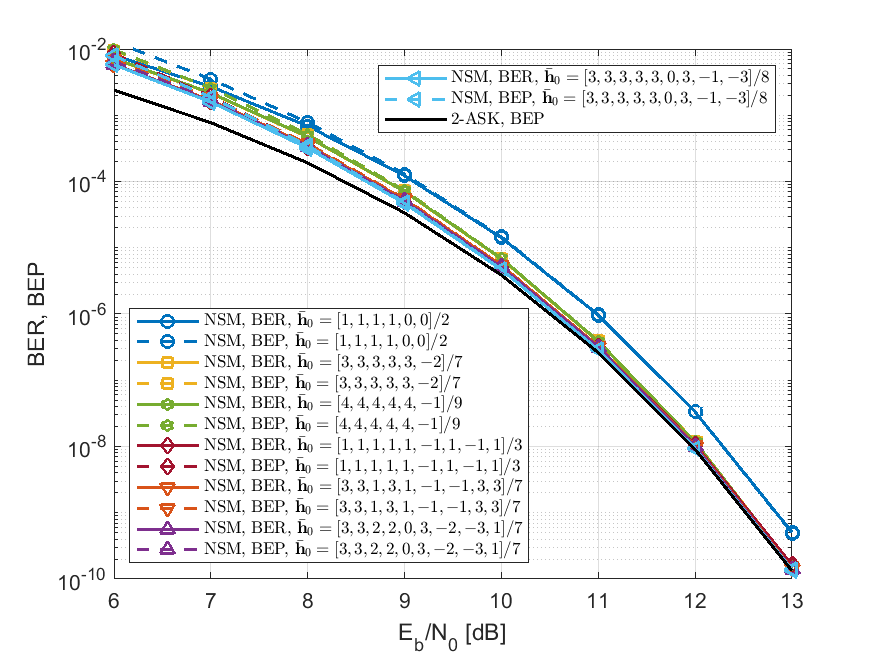}
    \caption{BER of the best rate $4/3$ NSMs, with rational filter coefficients, as specified in Tables~\ref{table:NSMs Rate-4/3 Filter Pattern (1, 1, 1, 1, 0, 0)}--\ref{table:NSMs Rate-4/3 Filter Pattern (3, 3, 3, 3, 3, 3, 3, 1, 0)}, for filter patterns $(1,1,1,1,0,0),$ $(3,3,3,3,3,2),$ $(4,4,4,4,4,1),$ $(1,1,1,1,1,1,1,1,1),$ $(3,3,3,3,2,2,2,1,0)$ and $(3,3,3,3,3,3,3,1,0).$ For reference the BEP upper bounds from Tables~\ref{table:NSMs Rate-4/3 Filter Pattern (1, 1, 1, 1, 0, 0)}--\ref{table:NSMs Rate-4/3 Filter Pattern (3, 3, 3, 3, 3, 3, 3, 1, 0)}, along with the BEP of $2$-ASK, are shown.}
    \label{fig:BER-BEP-Non Trivial NSM-4_3}
\end{figure}

Figure~\ref{fig:BER-BEP-Non Trivial NSM-4_3} shows the \gls{ber} simulation results for the best $\mathring{h}_0[k]$ filters, across all considered filter patterns $\bm{\pi}_0.$ For comparison, the \gls{bep} upper bounds from Tables~\ref{table:NSMs Rate-4/3 Filter Pattern (1, 1, 1, 1, 0, 0)}--\ref{table:NSMs Rate-4/3 Filter Pattern (3, 3, 3, 3, 3, 3, 3, 1, 0)} and the $2$-ASK \gls{bep} (used as the ultimate benchmark) are also included. This figure demonstrates excellent agreement between simulated \glspl{ber} and \gls{bep} upper bounds, for all \glspl{nsm}, using the best $\mathring{h}_0[k]$ filters, at moderate to high \glspl{snr}. It also highlights the performance degradation caused by the degeneracy of the best filter, $\mathring{\bm{h}}_0 = (1, 1, 1, 1, 0, 0),$ corresponding to pattern $\bm{\pi}_0=(1,1,1,1,0,0),$ compared to $2$-ASK and other filters. In contrast, the best filters for the other patterns perform closely to $2$-ASK, as predicted by the \gls{bep} upper bounds in Tables~\ref{table:NSMs Rate-4/3 Filter Pattern (3, 3, 3, 3, 3, 2)}--\ref{table:NSMs Rate-4/3 Filter Pattern (3, 3, 3, 3, 3, 3, 3, 1, 0)}. Notably, the best filter, $\mathring{\bm{h}}_0 = (3, 3, 3, 3, 3, 0, 3, -1, -3),$ corresponding to pattern $\bm{\pi}_0=(3,3,3,3, 3, 3, 3, 1, 0),$ outperforms all others, consistent with its superior second normalized \gls{msed} (normalized by $d_{\text{min}}^2$). With a second normalized \gls{sed} of $5/4$--—higher than that of other \glspl{nsm}—--this result aligns with theoretical expectations.

\section{Rate-5/4 NSMs with Rational Filter Coefficients} \label{Rate-5/4 Approaching NSMs}

Building on Subsection~\ref{A basic rate-4/3 NSM inspired by a previous basic rate-3/2 NSM}, and specifically (\ref{eq:Rescaled Modulated Symbols Rate 4/3 MSN}),
the rescaled modulated symbols can be written as
\begin{equation}  \label{eq:Rescaled Modulated Symbols Rate 5/4 MSN}
 \mathring{s}[k] = \sum_{m=0}^4 \sum_l \bar{b}_m[l] \mathring{h}_m[k-4l],
\end{equation}
where $\bar{b}_m[k],$ $0 \le m \le 4,$ are the $5$ bipolar input sequence to be modulated. Filter $\mathring{h}_0[k]$ has integer taps and length $L_0 > 4.$ Its vector version, $\mathring{\bm{h}}_0,$ is derived from a pattern vector, $\bm{\pi}_0,$ of the same dimension, with non-negative integer components and integer Euclidean norm, using arbitrary components permutations and sign changes. The remaining filters, $\mathring{h}_m[k] \triangleq \| \mathring{\bm{h}}_0 \| \delta[k-(m-1)] = \| \bm{\pi}_0 \| \delta[k-(m-1)],$ $0 < m \le 4,$ are completely specified by the Euclidean norm of pattern vector $\bm{\pi}_0.$ Hence, notice that the choice of $\bar{h}_0[k]$ is the sole degree of freedom in the \gls{nsm} design and should be made carefully, as it determines the performance of the underlying \gls{nsm}.

Apart from the fact that the norm $\| \bm{\pi}_0 \|,$ of pattern vector $\bm{\pi}_0,$ must be an integer, as the scaled filters, $\mathring{h}_m[k],$ $0 < m \le 4,$ are required to have integer coefficients, the maximum component of $\bm{\pi}_0,$ which is equal to its infinity norm, $\| \bm{\pi}_0 \|_\infty,$ must not exceed half of its Euclidean norm, $\| \bm{\pi}_0 \|,$ in order to guarantee that the resulting \gls{nsm} achieves the minimum Euclidean distance of the $2$-ASK modulation. While equality between $\| \bm{\pi}_0 \|/2$ and $\| \bm{\pi}_0 \|_\infty$ doesn't prevent from achieving the minimum Euclidean distance of $2$-ASK, it undesirably increases the multiplicity of error events with the minimum Euclidean distance.

Applying arbitrary permutations and sign changes to the pattern vector $\bm{\pi}_0$, generates a large set of possible filter vector candidates $\mathring{\bm{h}}_0.$ To reduce the complexity of identifying the optimal transformations that yield the best filter $\mathring{h}_0[k],$ we introduce specific transformations---similar to rate-$3/2$ and rate-$4/3$ \glspl{nsm}---that preserve both the \gls{tf} $T(N,D)$ and its reduced form, $\dot{T}(D).$ These transformations define an equivalence relation among filters, grouping them into distinct classes. Instead of evaluating the \gls{rtf} for all candidate filters, we only need to consider a smaller set of filters representatives, one per class, which greatly simplifies the \gls{nsm} optimization process.

As in rate-$3/2$ and rate-$4/3$ \glspl{nsm}, we define a set of equivalence transformations for rate-$5/4$ \glspl{nsm} that preserve their essential properties. These include sign change, time reversal, sign alternation, scrambling, time shifts, and permutations. Sign alternation yields equivalent filters, $\tilde{\mathring{h}}_0[k] \triangleq (-1)^{\lfloor (k-l)/4 \rfloor} \mathring{h}_0[k],$ $l$ being an arbitrary integer, with only eight possible “polyphase” variations, half of which are simply sign-inverted versions of the others.

Permutations, $\pi(\cdot),$ operate on the quaternary set, $\{0,1,2,3\},$ and transform each consecutive quadruplet of output samples, $\mathring{s}_0[4k] \mathring{s}_0[4k+1] \mathring{s}_0[4k+2] \mathring{s}_0[4k+3],$ into a new quadruplet, $\mathring{s}_0[4k+\pi(0)]  \mathring{s}_0[4k+\pi(1)] \mathring{s}_0[4k+\pi(2)] \mathring{s}_0[4k+\pi(3)].$ They result in equivalent \glspl{nsm} with filters $\tilde{\mathring{h}}_0[k] \triangleq \mathring{h}_0[k - (k \bmod 4)+\pi(k \bmod 4)].$

Scrambling generates equivalent scrambled filters, $\tilde{\mathring{h}}_0[k] \triangleq r[k] \mathring{h}_0[k],$ where $r[k] = r[k \bmod 4],$ represents a periodic bipolar scrambling sequences with a period of $4.$ This sequence repeats the length-$4$ pattern $r[0]r[1]r[2]r[3].$ Additionally, while time shifting can modify the structure of the  detection trellis---changing the number of states and branch labels---it doesn't affect the overall \gls{nsm} performance.

Tables~\ref{table:NSMs Rate-5/4 Filter Pattern (2, 2, 2, 1, 1, 1, 1, 0)}--\ref{table:NSMs Rate-5/4 Filter Pattern (2, 1, 1, 1, 1, 1, 1, 1, 1, 1, 1, 1, 1, 0, 0, 0)} summarize the characteristics of several rate-$5/4$ NSMs, with rational filters coefficients, specified by their patterns $\bm{\pi}_0.$ Three sets of patterns are considered. The first set consists of patterns $(2,2,2,1,1,1,1,0),$ $(2,2,2,2,2,2,1,0),$ $(3,3,3,3,2,2,2,1),$ $(3,3,3,3,3,2,0,0),$ $(3,3,3,3,3,3,3,1)$ and $(4,4,4,4,3,3,3,3),$ generating candidate filters, $\mathring{h}_0[k],$ of length, $L_0=8,$ and detection trellises with $2$ states. The second set includes patterns $(1,1,1,1,1,1,1,1,1,0,0,0)$ and $(2,2,1,1,1,1,1,1,1,1,0,0),$ producing candidate filters, $\mathring{h}_0[k],$ of length, $L_0=12,$ and trellises with $4$ states. The third set contains patterns $(1,1,1,1,1,1,1,1,1,1,1,1,1,1,1,1)$ and $(2,1,1,1,1,1,1,1,1,1,1,1,1,0,0,0),$ resulting in candidate filters $\mathring{h}_0[k],$ of length, $L_0=16,$ and trellises with $8$ states.


\begin{table}[H]
\caption{Selected rate-$5/4$ NSMs, with simple filters' coefficients, with common filter $\bm{h}_0$ pattern $\bm{\pi}_0 = (2, 2, 2, 1, 1, 1, 1, 0),$ arranged in a decreasing performance order, starting from the best one. Relative degradation in performance due to distance reduction or/and multiplicity increase when moving from one filter to the next is emphasized in bold.}
\label{table:NSMs Rate-5/4 Filter Pattern (2, 2, 2, 1, 1, 1, 1, 0)}
\centering
\begin{tabular}{|c|c|} 
\hline
\multicolumn{2}{|l|}{\# of simple $h_0[k]$ filters $ = 35840$} \\ \hline
\multicolumn{2}{|l|}{\# of non equivalent $h_0[k]$ filters $ = 19$} \\ \hline
\multicolumn{2}{|l|}{$\mathring{h}_1[k] = 4 \delta[k],$ $\mathring{h}_2[k] = 4 \delta[k-1],$ $\mathring{h}_3[k] = 4 \delta[k-2]$ and $\mathring{h}_4[k] = 4 \delta[k-3]$} \\ \hline
\multicolumn{2}{|l|}{Minimum squared Euclidean distance, $d_\text{min}^2 = 64$} \\ \hline
\multicolumn{2}{|l|}{Number of trellis states $ = 2$} \\  [0.5ex]
\hline\hline
\multicolumn{2}{|c|}{Filter \# $1$} \\ \hline
$\mathring{h}_0[k]$ & $2 \delta[k] + 2 \delta[k-1] + \delta[k-2] + \delta[k-3] + 2 \delta[k-4] + \delta[k-6] - \delta[k-7]$ \\ \hline
$\mathring{\bm{h}}_0$ & $(2, 2, 1, 1, 2, 0, 1, -1)$ \\
\hline
\end{tabular}
\end{table}


\begin{table}[H]
\caption{Selected rate-$5/4$ NSMs, with simple filters' coefficients, with common filter $\bm{h}_0$ pattern $\bm{\pi}_0 = (2, 2, 2, 2, 2, 2, 1, 0),$ arranged in a decreasing performance order, starting from the best one. Relative degradation in performance due to distance reduction or/and multiplicity increase when moving from one filter to the next is emphasized in bold.}
\label{table:NSMs Rate-5/4 Filter Pattern (2, 2, 2, 2, 2, 2, 1, 0)}
\centering
\begin{tabular}{|c|c|} 
\hline
\multicolumn{2}{|l|}{\# of simple $h_0[k]$ filters $ = 7168$} \\ \hline
\multicolumn{2}{|l|}{\# of non equivalent $h_0[k]$ filters $ = 5$} \\ \hline
\multicolumn{2}{|l|}{$\mathring{h}_1[k] = 5 \delta[k],$ $\mathring{h}_2[k] = 5 \delta[k-1],$ $\mathring{h}_3[k] = 5 \delta[k-2]$ and $\mathring{h}_4[k] = 5 \delta[k-3]$} \\ \hline
\multicolumn{2}{|l|}{Minimum squared Euclidean distance, $d_\text{min}^2 = 100$} \\ \hline
\multicolumn{2}{|l|}{Number of trellis states $ = 2$} \\  [0.5ex]
\hline\hline
\multicolumn{2}{|c|}{Filter \# $1$} \\ \hline
$\mathring{h}_0[k]$ & $2 \delta[k] + 2 \delta[k-1] + 2 \delta[k-2] + 2 \delta[k-3] + 2 \delta[k-4] + 2 \delta[k-5] - \delta[k-7]$ \\ \hline
$\mathring{\bm{h}}_0$ & $(2, 2, 2, 2, 2, 2, 0, -1)$ \\
\hline
\end{tabular}
\end{table}


\begin{table}[H]
\caption{Selected rate-$5/4$ NSMs, with simple filters' coefficients, with common filter $\bm{h}_0$ pattern $\bm{\pi}_0 = (3, 3, 3, 3, 2, 2, 2, 1),$ arranged in a decreasing performance order, starting from the best one. Relative degradation in performance due to distance reduction or/and multiplicity increase when moving from one filter to the next is emphasized in bold.}
\label{table:NSMs Rate-5/4 Filter Pattern (3, 3, 3, 3, 2, 2, 2, 1)}
\centering
\begin{tabular}{|c|c|} 
\hline
\multicolumn{2}{|l|}{\# of simple $h_0[k]$ filters $ = 71680$} \\ \hline
\multicolumn{2}{|l|}{\# of non equivalent $h_0[k]$ filters $ = 62$} \\ \hline
\multicolumn{2}{|l|}{$\mathring{h}_1[k] = 7 \delta[k],$ $\mathring{h}_2[k] = 7 \delta[k-1],$ $\mathring{h}_3[k] = 7 \delta[k-2]$ and $\mathring{h}_4[k] = 7 \delta[k-3]$} \\ \hline
\multicolumn{2}{|l|}{Minimum squared Euclidean distance, $d_\text{min}^2 = 196$} \\ \hline
\multicolumn{2}{|l|}{Number of trellis states $ = 2$} \\  [0.5ex]
\hline\hline
\multicolumn{2}{|c|}{Filters \# $1$ \& $2$} \\ \hline
\multirow{3}{*}{$\mathring{h}_0[k]$} & $3 \delta[k] + 3 \delta[k-1] + 3 \delta[k-2] + 2 \delta[k-3] + 3 \delta[k-4] + 2 \delta[k-5]$ \\
 & $+ \delta[k-6] - 2 \delta[k-7]$ \& $3 \delta[k] + 3 \delta[k-1]$ \\
\hline
\end{tabular}
\end{table}


\begin{table}[H]
\caption{Selected rate-$5/4$ NSMs, with simple filters' coefficients, with common filter $\bm{h}_0$ pattern $\bm{\pi}_0 = (3, 3, 3, 3, 3, 2, 0, 0),$ arranged in a decreasing performance order, starting from the best one. Relative degradation in performance due to distance reduction or/and multiplicity increase when moving from one filter to the next is emphasized in bold.}
\label{table:NSMs Rate-5/4 Filter Pattern (3, 3, 3, 3, 3, 2, 0, 0)}
\centering
\begin{tabular}{|c|c|} 
\hline
\multicolumn{2}{|l|}{\# of simple $h_0[k]$ filters $ = 10752$} \\ \hline
\multicolumn{2}{|l|}{\# of non equivalent $h_0[k]$ filters $ = 7$} \\ \hline
\multicolumn{2}{|l|}{$\mathring{h}_1[k] = 7 \delta[k],$ $\mathring{h}_2[k] = 7 \delta[k-1],$ $\mathring{h}_3[k] = 7 \delta[k-2]$ and $\mathring{h}_4[k] = 7 \delta[k-3]$} \\ \hline
\multicolumn{2}{|l|}{Minimum squared Euclidean distance, $d_\text{min}^2 = 196$} \\ \hline
\multicolumn{2}{|l|}{Number of trellis states $ = 2$} \\  [0.5ex]
\hline\hline
\multicolumn{2}{|c|}{Filter \# $1$} \\ \hline
$\mathring{h}_0[k]$ & $3 \delta[k] + 3 \delta[k-1] + 3 \delta[k-2] + 3 \delta[k-3] + 3 \delta[k-4] - 2 \delta[k-7]$ \\ \hline
$\mathring{\bm{h}}_0$ & $(3, 3, 3, 3, 3, 0, 0, -2)$ \\
\hline
\end{tabular}
\end{table}


\begin{table}[H]
\caption{Selected rate-$5/4$ NSMs, with simple filters' coefficients, with common filter $\bm{h}_0$ pattern $\bm{\pi}_0 = (3, 3, 3, 3, 3, 3, 3, 1),$ arranged in a decreasing performance order, starting from the best one. Relative degradation in performance due to distance reduction or/and multiplicity increase when moving from one filter to the next is emphasized in bold.}
\label{table:NSMs Rate-5/4 Filter Pattern (3, 3, 3, 3, 3, 3, 3, 1)}
\centering
\begin{tabular}{|c|c|} 
\hline
\multicolumn{2}{|l|}{\# of simple $h_0[k]$ filters $ = 2048$} \\ \hline
\multicolumn{2}{|l|}{\# of non equivalent $h_0[k]$ filters $ = 4$} \\ \hline
\multicolumn{2}{|l|}{$\mathring{h}_1[k] = 8 \delta[k],$ $\mathring{h}_2[k] = 8 \delta[k-1],$ $\mathring{h}_3[k] = 8 \delta[k-2]$ and $\mathring{h}_4[k] = 8 \delta[k-3]$} \\ \hline
\multicolumn{2}{|l|}{Minimum squared Euclidean distance, $d_\text{min}^2 = 256$} \\ \hline
\multicolumn{2}{|l|}{Number of trellis states $ = 2$} \\  [0.5ex]
\hline\hline
\multicolumn{2}{|c|}{Filter \# $1$} \\ \hline
\multirow{2}{*}{$\mathring{h}_0[k]$} & $3 \delta[k] + 3 \delta[k-1] + 3 \delta[k-2] + 3 \delta[k-3] + 3 \delta[k-4] + 3 \delta[k-5]$ \\
 & $+ 3 \delta[k-6] - \delta[k-7]$ \\
\hline
\end{tabular}
\end{table}


\begin{table}[H]
\caption{Selected rate-$5/4$ NSMs, with simple filters' coefficients, with common filter $\bm{h}_0$ pattern $\bm{\pi}_0 = (4, 4, 4, 4, 3, 3, 3, 3),$ arranged in a decreasing performance order, starting from the best one. Relative degradation in performance due to distance reduction or/and multiplicity increase when moving from one filter to the next is emphasized in bold.}
\label{table:NSMs Rate-5/4 Filter Pattern (4, 4, 4, 4, 3, 3, 3, 3)}
\centering
\begin{tabular}{|c|c|} 
\hline
\multicolumn{2}{|l|}{\# of simple $h_0[k]$ filters $ = 17920$} \\ \hline
\multicolumn{2}{|l|}{\# of non equivalent $h_0[k]$ filters $ = 28$} \\ \hline
\multicolumn{2}{|l|}{$\mathring{h}_1[k] = 10 \delta[k],$ $\mathring{h}_2[k] = 10 \delta[k-1],$ $\mathring{h}_3[k] = 10 \delta[k-2]$ and $\mathring{h}_4[k] = 10 \delta[k-3]$} \\ \hline
\multicolumn{2}{|l|}{Minimum squared Euclidean distance, $d_\text{min}^2 = 400$} \\ \hline
\multicolumn{2}{|l|}{Number of trellis states $ = 2$} \\  [0.5ex]
\hline\hline
\multicolumn{2}{|c|}{Filters \# $1, 2, 3$ \& $4$ } \\ \hline
\multirow{3}{*}{$\mathring{h}_0[k]$} & $4 \delta[k] + 4 \delta[k-1] + 4 \delta[k-2] + 4 \delta[k-3] + 3 \delta[k-4] + 3 \delta[k-5]$ \\
 & $- 3 \delta[k-6] - 3 \delta[k-7]$ \& $\cdots$ \& $4 \delta[k] + 4 \delta[k-1]$ \\
\hline
\end{tabular}
\end{table}


\begin{table}[H]
\caption{Selected rate-$5/4$ NSMs, with simple filters' coefficients, with common filter $\bm{h}_0$ pattern $\bm{\pi}_0 = (1, 1, 1, 1, 1, 1, 1, 1, 1, 0, 0, 0),$ arranged in a decreasing performance order, starting from the best one. Relative degradation in performance due to distance reduction or/and multiplicity increase when moving from one filter to the next is emphasized in bold.}
\label{table:NSMs Rate-5/4 Filter Pattern (1, 1, 1, 1, 1, 1, 1, 1, 1, 0, 0, 0)}
\centering
\begin{tabular}{|c|c|} 
\hline
\multicolumn{2}{|l|}{\# of simple $h_0[k]$ filters $ = 112640$} \\ \hline
\multicolumn{2}{|l|}{\# of non equivalent $h_0[k]$ filters $ = 70$} \\ \hline
\multicolumn{2}{|l|}{$\mathring{h}_1[k] = 3 \delta[k],$ $\mathring{h}_2[k] = 3 \delta[k-1],$ $\mathring{h}_3[k] = 3 \delta[k-2]$ and $\mathring{h}_4[k] = 3 \delta[k-3]$} \\ \hline
\multicolumn{2}{|l|}{Best minimum squared Euclidean distance, $d_\text{min}^2 = 36$} \\ \hline
\multicolumn{2}{|l|}{Number of trellis states $ = 4$} \\  [0.5ex]
\hline\hline
\multicolumn{2}{|c|}{Filter \# $1$ } \\ \hline
\multirow{2}{*}{$\mathring{h}_0[k]$} & $\delta[k] + \delta[k-1] + \delta[k-2] + \delta[k-3] + \delta[k-4] - \delta[k-7]$ \\
 & $ + \delta[k-8] + \delta[k-9] - \delta[k-10] $ \\
\hline
\end{tabular}
\end{table}


\begin{table}[H]
\caption{Selected rate-$5/4$ NSMs, with simple filters' coefficients, with common filter $\bm{h}_0$ pattern $\bm{\pi}_0 = (2, 2, 1, 1, 1, 1, 1, 1, 1, 1, 0, 0),$ arranged in a decreasing performance order, starting from the best one. Relative degradation in performance due to distance reduction or/and multiplicity increase when moving from one filter to the next is emphasized in bold.}
\label{table:NSMs Rate-5/4 Filter Pattern (2, 2, 1, 1, 1, 1, 1, 1, 1, 1, 0, 0)}
\centering
\begin{tabular}{|c|c|} 
\hline
\multicolumn{2}{|l|}{\# of simple $h_0[k]$ filters $ = 3041280$} \\ \hline
\multicolumn{2}{|l|}{\# of non equivalent $h_0[k]$ filters $ = 1171$} \\ \hline
\multicolumn{2}{|l|}{$\mathring{h}_1[k] = 4 \delta[k],$ $\mathring{h}_2[k] = 4 \delta[k-1],$ $\mathring{h}_3[k] = 4 \delta[k-2]$ and $\mathring{h}_4[k] = 4 \delta[k-3]$} \\ \hline
\multicolumn{2}{|l|}{Best minimum squared Euclidean distance, $d_\text{min}^2 = 64$} \\ \hline
\multicolumn{2}{|l|}{Number of trellis states $ = 4$} \\  [0.5ex]
\hline\hline
\multicolumn{2}{|c|}{Filter \# $1$ } \\ \hline
\multirow{2}{*}{$\mathring{h}_0[k]$} & $2 \delta[k] + 2 \delta[k-1] + \delta[k-2] + \delta[k-3] + \delta[k-4] + \delta[k-6]$ \\
 & $ - \delta[k-7] + \delta[k-8] - \delta[k-10] + \delta[k-11]$ \\
\hline
\end{tabular}
\end{table}


\begin{table}[H]
\caption{Selected rate-$5/4$ NSMs, with simple filters' coefficients, with common filter $\bm{h}_0$ pattern $\bm{\pi}_0 = (1, 1, 1, 1, 1, 1, 1, 1, 1, 1, 1, 1, 1, 1, 1, 1),$ arranged in a decreasing performance order, starting from the best one. Relative degradation in performance due to distance reduction or/and multiplicity increase when moving from one filter to the next is emphasized in bold.}
\label{table:NSMs Rate-5/4 Filter Pattern (1, 1, 1, 1, 1, 1, 1, 1, 1, 1, 1, 1, 1, 1, 1, 1)}
\centering
\begin{tabular}{|c|c|} 
\hline
\multicolumn{2}{|l|}{\# of simple $h_0[k]$ filters $ = 65536$} \\ \hline
\multicolumn{2}{|l|}{\# of non equivalent $h_0[k]$ filters $ = 102$} \\ \hline
\multicolumn{2}{|l|}{$\mathring{h}_1[k] = 4 \delta[k],$ $\mathring{h}_2[k] = 4 \delta[k-1],$ $\mathring{h}_3[k] = 4 \delta[k-2]$ and $\mathring{h}_4[k] = 4 \delta[k-3]$} \\ \hline
\multicolumn{2}{|l|}{Best minimum squared Euclidean distance, $d_\text{min}^2 = 64$} \\ \hline
\multicolumn{2}{|l|}{Number of trellis states $ = 8$} \\  [0.5ex]
\hline\hline
\multicolumn{2}{|c|}{Filters \# $1$ \& $2$} \\ \hline
\multirow{5}{*}{$\mathring{h}_0[k]$} & $\delta[k] + \delta[k-1] + \delta[k-2] + \delta[k-3] + \delta[k-4] + \delta[k-5] - \delta[k-6]$ \\
 & $ - \delta[k-7] + \delta[k-8] - \delta[k-9] + \delta[k-10] - \delta[k-11] + \delta[k-12]$ \\
\hline
\end{tabular}
\end{table}


\begin{table}[H]
\caption{Selected rate-$5/4$ NSMs, with simple filters' coefficients, with common filter $\bm{h}_0$ pattern $\bm{\pi}_0 = (2, 1, 1, 1, 1, 1, 1, 1, 1, 1, 1, 1, 1, 0, 0, 0),$ arranged in a decreasing performance order, starting from the best one. Relative degradation in performance due to distance reduction or/and multiplicity increase when moving from one filter to the next is emphasized in bold.}
\label{table:NSMs Rate-5/4 Filter Pattern (2, 1, 1, 1, 1, 1, 1, 1, 1, 1, 1, 1, 1, 0, 0, 0)}
\centering
\begin{tabular}{|c|c|} 
\hline
\multicolumn{2}{|l|}{\# of simple $h_0[k]$ filters $ = 59637760$} \\ \hline
\multicolumn{2}{|l|}{\# of non equivalent $h_0[k]$ filters $ = 17148$} \\ \hline
\multicolumn{2}{|l|}{$\mathring{h}_1[k] = 4 \delta[k],$ $\mathring{h}_2[k] = 4 \delta[k-1],$ $\mathring{h}_3[k] = 4 \delta[k-2]$ and $\mathring{h}_4[k] = 4 \delta[k-3]$} \\ \hline
\multicolumn{2}{|l|}{Best minimum squared Euclidean distance, $d_\text{min}^2 = 64$} \\ \hline
\multicolumn{2}{|l|}{Number of trellis states $ = 8$} \\  [0.5ex]
\hline\hline
\multicolumn{2}{|c|}{Filter \# $1$} \\ \hline
\multirow{2}{*}{$\mathring{h}_0[k]$} & $2 \delta[k] + \delta[k-1] + \delta[k-2] + \delta[k-3] + \delta[k-5] + \delta[k-6] - \delta[k-7]$ \\
 & $ + \delta[k-9] + \delta[k-10] + \delta[k-11] - \delta[k-13] - \delta[k-14] + \delta[k-15]$ \\
\hline
\end{tabular}
\end{table}

For each considered pattern, $\bm{\pi}_0,$ the non-equivalent filters representatives are ranked from the best to the worst, based on their distance spectrum. Moreover, for the best non-equivalent filter representatives---referred to as Filter \# 1, and sometimes as Filters \# 1 \& 2 in the tables---we provide an approximate upper bound on the \gls{bep}.

For most of the filter patterns considered, the leading term in the \gls{bep} upper bound matches the \gls{bep} of $2$-ASK, indicating that the \glspl{nsm} with these best filters asymptotically reach its performance. This holds for patterns $(2,2,2,2,2,2,1,0),$ $(3,3,3,3,2,2,2,1),$ $(3,3,3,3,3,2,0,0),$ $(3,3,3,3,3,3,3,1)$ and $(4,4,4,4,3,3,3,3),$ with $L_0=8,$ $(1,1,1,1,1,1,1,1,1,0,0,0),$ with $L_0=12,$ and $(1,1,1,1,1,1,1,1,1,1,1,$ $1,1,1,1,1),$ with $L_0=16.$ Moreover, even the worst filters representatives maintain the minimum Euclidean distance of $2$-ASK, for all patterns, except those with $L_0=12$ and $L_0=16.$

Among all the patterns considered, pattern $\bm{\pi}_0 = (1,1,1,1,1,1,1,1,1,1,1,1,1,1,1,1)$ yields the best filter representatives for $\mathring{h}_0[k],$ given in vector form as $\mathring{\bm{h}}_0 =
(1,1,1,1,1,1,-1,-1,1,-1,1,-1,1,-1,-1,$ $1)$ and $\mathring{\bm{h}}_0 = (1,1,1,1,1,1,-1,-1,1,-1,1,-1,-1,1,1,-1).$ These two are not equivalent under the previously defined equivalence transformations, but share the same \gls{rtf} $\dot{T}(D).$ In terms of distance spectrum, they offer the lowest multiplicity at the minimum Euclidean distance and the highest second minimum Euclidean distance---whose normalized squared value is $3/2,$ with normalization based on the \gls{msed}, $d_{\text{min}}^2.$

It is noteworthy that in the two best non-equivalent filters representatives, the fourth four-component section shows complimentary pairs, while the first three sections (covering the first eleven components) are identical. Interestingly, stacking the four-component sections of the first representative forms an interleaved version of the $4 \times 4$ Hadamard matrix, with the second and third rows swapped--—an underlying reason for the superior performance of these filters. This structure is reminiscent of the rate-$3/2$ \gls{nsm} in Subsection~\ref{ssec:Basic Rate-3/2 NSM}, where the vector version, $\mathring{\bm{h}}_0 = (1,1,1,-1),$ of the optimal filter, when viewed as two stacked pairs, yields the $2 \times 2$ Hadamard matrix.

\begin{figure}[!htbp]
    \centering
    \includegraphics[width=1.0\textwidth]{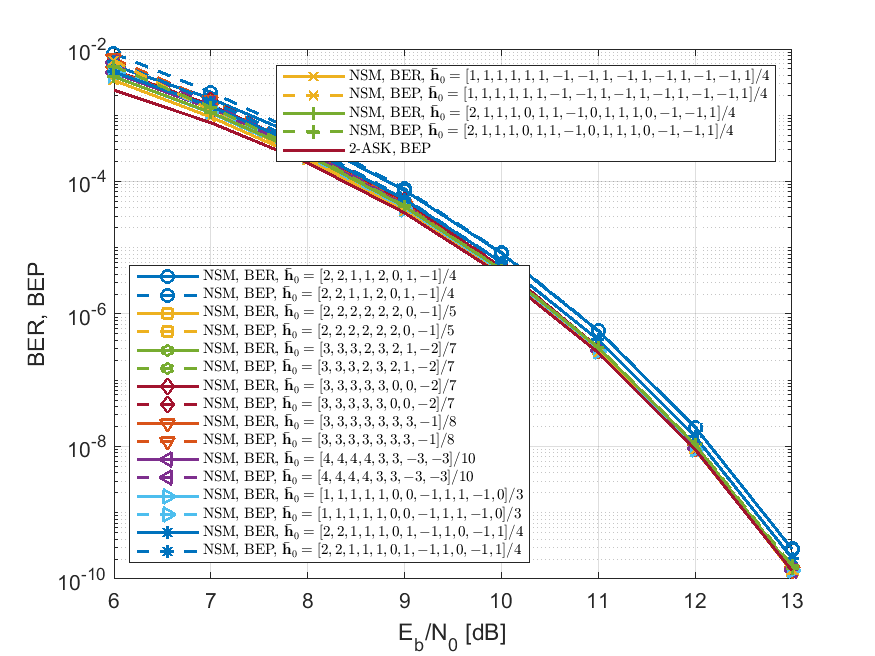}
    \caption{BER of the best rate $5/4$ NSMs, with rational filter coefficients, as specified in Tables~\ref{table:NSMs Rate-5/4 Filter Pattern (2, 2, 2, 1, 1, 1, 1, 0)}--\ref{table:NSMs Rate-5/4 Filter Pattern (2, 1, 1, 1, 1, 1, 1, 1, 1, 1, 1, 1, 1, 0, 0, 0)}, for filter patterns $(2,2,2,1,1,1,1,0),$ $ (2,2,2,2,2,2,1,0),$ $(3,3,3,3,2,2,2,1),$ $(3,3,3,3,3,2,0,0),$ $(3,3,3,3,3,3,3,1),$ $(4,4,4,4,3,3,3,3),$ $(1,1,1,1,1,1,1,1,1,0,0,0),$ $ (2,2,1,1,1,1,1,1,1,1,0,0),$ $ (1,1,1,1,1,1,1,1,1,1,1,1,1,1,1,1)$ and $(2,1,1,1,1,1,1,1,1,1,1,1,1,0,0,0).$ For reference the BEP upper bounds from Tables~\ref{table:NSMs Rate-5/4 Filter Pattern (2, 2, 2, 1, 1, 1, 1, 0)}--\ref{table:NSMs Rate-5/4 Filter Pattern (2, 1, 1, 1, 1, 1, 1, 1, 1, 1, 1, 1, 1, 0, 0, 0)}, along with the BEP of $2$-ASK, are shown.}
    \label{fig:BER-BEP-Non Trivial NSM-5_4}
\end{figure}

Figure~\ref{fig:BER-BEP-Non Trivial NSM-5_4} presents the \gls{ber} simulation results for the best $\mathring{h}_0[k]$ filters, across all considered patterns candidates, $\bm{\pi}_0.$ For reference, it also includes the \gls{bep} upper bounds from Tables~\ref{table:NSMs Rate-5/4 Filter Pattern (2, 2, 2, 1, 1, 1, 1, 0)}--\ref{table:NSMs Rate-5/4 Filter Pattern (2, 1, 1, 1, 1, 1, 1, 1, 1, 1, 1, 1, 1, 0, 0, 0)} and the $2$-ASK \gls{bep}, used as benchmark. This figure shows excellent agreement between simulated \glspl{ber} and the theoretical bounds, at low to high \glspl{snr}, for all \glspl{nsm}, using the best $\mathring{h}_0[k]$ filters. As expected, the first of the two top-performing filters representatives, $\mathring{\bm{h}}_0 = (1,1,1,1,1,1,-1,-1,1,-1,1,-1,1,-1,-1,$ $1),$ corresponding to pattern $\bm{\pi}_0 = (1,1,1,1,1,1,1,1,1,1,1,1,1,1,1,1),$ consistently outperforms the others, reflecting its superior second normalized \gls{msed}.

\section{Unifying Frameworks, Analogies, and Extensions of NSMs} \label{Discussion and conceptual parallels}

\subsection{Gray- and Non-Gray-Coded ASK and QAM as Particular Cases of NSMs} \label{Gray- and non-Gray-coded ASK and QAM as special cases of NSMs}

\begin{figure}[!htbp]
    \centering
    \includegraphics[width=0.85\textwidth]{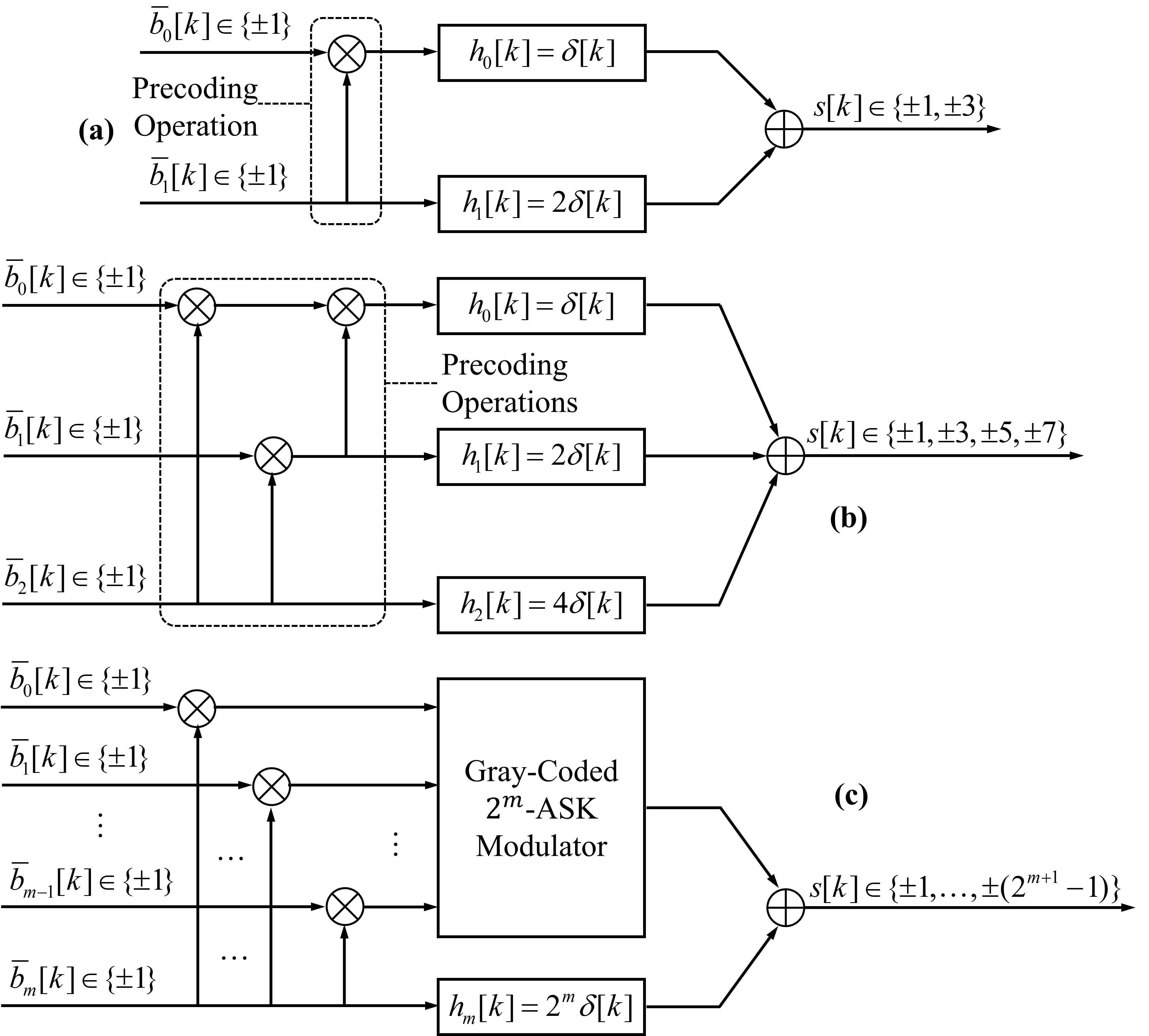}
    \caption{Block diagrams of Gray-coded $2^m$-ASK modulations ($m \ge 1$) as special cases of NSMs with \emph{analog} precoding: (a) Gray-coded $4$-ASK, (b) Gray-coded $8$-ASK, (c) Recursive construction of Gray-coded $2^{m+1}$-ASK from Gray-coded $2^m$-ASK.}
    \label{fig:BlockDiagramGrayCodingASKAnalogDomain}
\end{figure}

As illustrated in Figure~\ref{fig:Block Diagram 4-ASK}, we have demonstrated that $4$-ASK—whether Gray-coded or not—can be regarded as a special case of \glspl{nsm}. Gray coding is implemented through a precoding operation, applied in the \emph{analog} domain, before conventional non-Gray-coded $4$-ASK modulation. An alternative but equivalent representation, highlighting the \gls{nsm}-specific filtering structure for Gray-coded $4$-ASK, is shown in Figure~\ref{fig:BlockDiagramGrayCodingASKAnalogDomain} (a). In this configuration, the Gray-coded output is obtained by \emph{analog-domain} precoding of one of the bipolar input sequences, $\bar{b}_0[k]$ and $\bar{b}_1[k],$ before applying the standard $4$-ASK modulation. For this particular case, the filters associated with the input sequences $\bar{b}_0[k]$ and $\bar{b}_1[k]$ are $h_0[k] = \delta[k]$ and $h_1[k] = 2\,\delta[k]$, respectively.

Building on the detailed analysis of $4$-ASK and its underlying \gls{nsm} structure, Figure~\ref{fig:BlockDiagramGrayCodingASKAnalogDomain} (b) extends this framework to $8$-ASK with Gray coding, illustrating it as a specific instance of \glspl{nsm}. Figure~\ref{fig:BlockDiagramGrayCodingASKAnalogDomain} (c) further generalizes the approach, showing how the block diagram of any $2^{m+1}$-ASK can be recursively derived from that of a $2^m$-ASK. Together, these figures demonstrate that any Gray-coded $2^m$-ASK modulation, $m \ge 1,$ can be interpreted as a non-Gray-coded counterpart—naturally aligned with an \gls{nsm} formulation—preceded by simple precoding operations in the analog domain. In this interpretation, the non-Gray-coded $2^m$-ASK modulator takes $m$ bipolar bipolar input sequences, each passed through a filter $h_l[k] = 2^l \, \delta[k],$ $0 \le l \le m-1,$ and sums their outputs to produce the modulated sequence.

\begin{figure}[!htbp]
    \centering
    \includegraphics[width=0.85\textwidth]{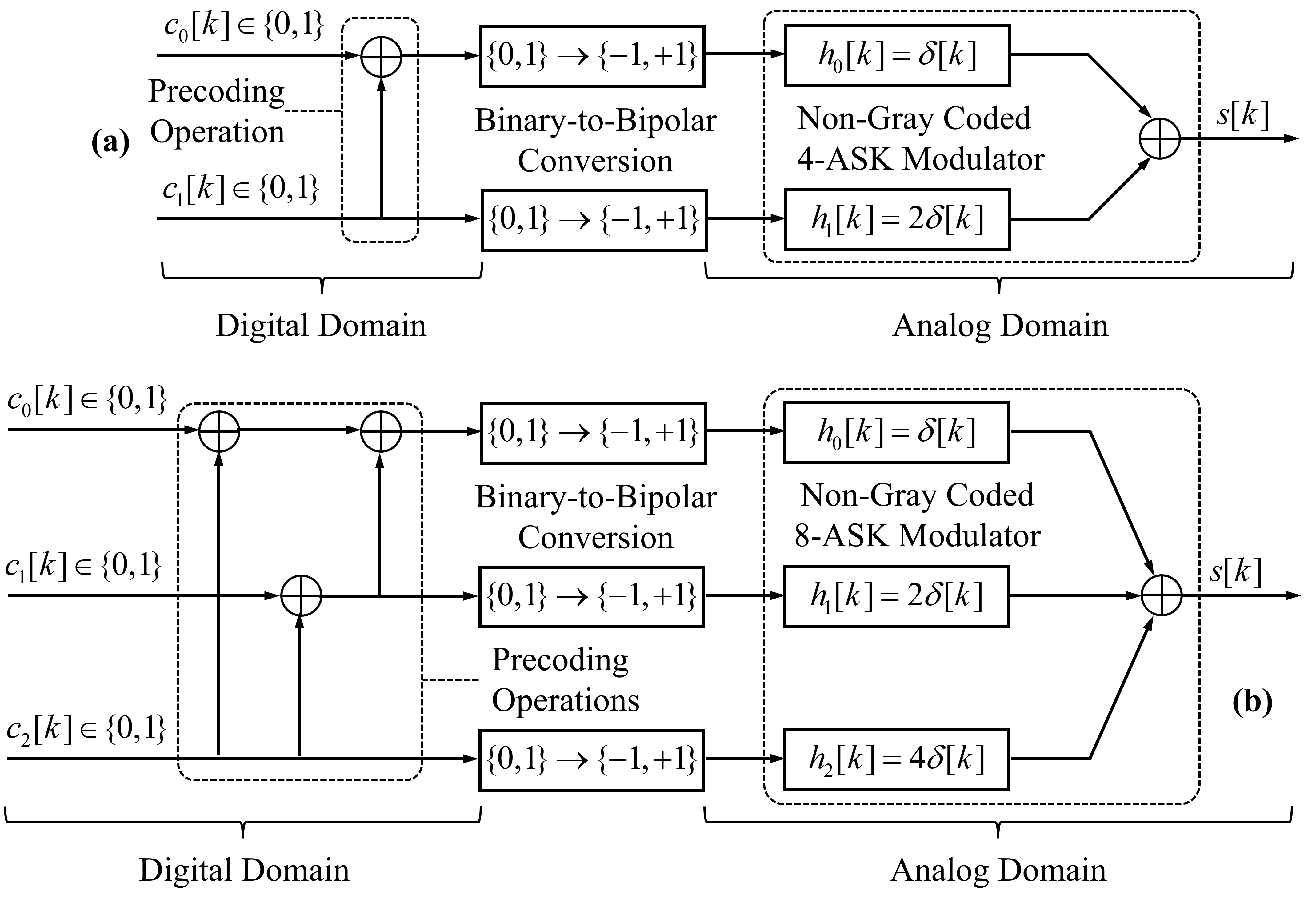}
    \caption{Block diagrams of Gray-coded $2^m$-ASK modulations ($m \ge 1$) as special cases of NSMs with \emph{digital} precoding: (a) Gray-coded $4$-ASK, (b) Gray-coded $8$-ASK.}
    \label{fig:BlockDiagramGrayCodingASKDigitalDomain}
\end{figure}

In Figure~\ref{fig:BlockDiagramGrayCodingASKDigitalDomain}, we revisit the block diagrams of Gray-coded $4$-ASK and $8$-ASK, this time adopting an alternative representation that separates the processing into two distinct stages: \emph{digital} and \emph{analog}. These stages are connected via binary-to-bipolar conversions. In this reformulated view, the analog section contains only the \gls{nsm}-based implementation of the corresponding non-Gray-coded modulation. The analog-domain precoding—previously required to transform a non-Gray-coded modulator into its Gray-coded version—is now shifted entirely to the \emph{digital domain}, where it is realized using simple linear (modulo-$2$) operations. In the earlier representation of Gray-coded $2^m$-ASK (Figure~\ref{fig:BlockDiagramGrayCodingASKAnalogDomain}), a binary-coded data stream from an error correction encoder is first converted to a bipolar sequence and then de-multiplexed into $m$ bipolar input streams $\bar{b}_l[k],$ $0 \le l \le m-1,$ which feed the Gray-coded modulator. By contrast, in the representation shown in Figure~\ref{fig:BlockDiagramGrayCodingASKDigitalDomain}, the encoded binary sequence is first de-multiplexed into $m$ binary streams $c_l[k],$ $0 \le l \le m-1,$ which are then digitally preprocessed using linear operations in the binary domain. These preprocessed sequences are subsequently passed through parallel binary-to-bipolar converters and directly fed into the non-Gray-coded modulator. This shift to digital-domain precoding aligns naturally with most linear error correction encoding schemes, which operate entirely in the binary domain using linear binary operations. As such, the digital Gray-coding stage can be seamlessly integrated into the structure of the error correction encoder itself, resulting in a unified encoding scheme. In this merged structure, the $m$ resulting binary streams are directly converted to bipolar form and used as inputs to the non-Gray-coded modulator.

The purpose of this analysis is to demonstrate that \gls{nsm}-based non-Gray-coded ASK modulators can follow directly after the binary-to-bipolar conversion of encoded data streams, without requiring any analog-domain precoding. This reinterpretation of Gray coding—as a purely digital linear preprocessing step—offers a cleaner, modular design approach and extends naturally to more general \glspl{nsm}, such as those rate-$(Q+1)/Q,$ $1 \le Q \le 4,$ or rate-$3$ \glspl{nsm}, discussed earlier. Consequently, analog-domain precoding can be entirely eliminated from the design of Gray-coded or similarly structured \glspl{nsm}, enabling implementations based solely on linear operations in the binary domain, followed by a pure and standard multi-stream partial response signaling structure, operating linearly in the analog domain.

\subsection{Redesigning NSMs for Selective Fading Channels: Leveraging Diversity}

Most of the \glspl{nsm} that we have searched, optimized, designed and characterized, admitting $m$ bipolar input sequences, bank on $m$ filters, $h_l[k],$ $0 \le l \le m-1,$ with lengths, $L_l,$ such that $L_0 \ge 2,$ while $L_l = 1,$ for $l \ge 1.$ These \glspl{nsm} proved to be good performing and well suited to the Gaussian channel, since they have been designed to maximize the \gls{msed}. Unfortunately, the channel that is encountered in practical systems, such as wireless and radio-mobile systems, is a selective fading channel. This selectivity can be met in time, due to Doppler frequency dispersive channels, in frequency, due to delay time dispersive channels, or both. For these channels, and because of the fact that $L_l = 1,$ for $l \ge 1,$ input streams, carrying bipolar sequences $\bar{b}_l[k],$ $l \ge 1,$ procure no diversity, leading to an overall diversity of order one, at the receiver detector, prior to error correction decoding, even though the input stream carrying bipolar sequences $\bar{b}_0[k],$ procures a diversity order of $2$ or higher, reaching order $L_0,$ when the $L_0$ taps filter $h_0[k]$ are all non-null. In this case, any diversity can only be procured by error-correction-coding with interleaving, prior to \gls{nsm}. Hence, in order to procure additional diversity at the \gls{nsm} level, in addition to the error-correction coding level, we need to increase the lengths of all filters $h_l[k],$ other than filter $h_0[k].$ If $\lambda_l$ denotes the number of non-null taps in filter $h_l[k],$ then, assuming perfect or quasi-perfect interleaving, the diversity order procured at the \gls{nsm} level is equal to $\min_l \, (\lambda_l).$

Most of the \glspl{nsm} we have studied and characterized rely on a bank of filters $h_l[k],$ $0 \le l \le m-1,$ corresponding to the $m$ bipolar input sequences they accept at their input. In these configurations, the first filter, $h_0[k],$ typically has a length $L_0 \ge 2,$ while the remaining filters, $h_l[k],$ $l \ge 1,$ are limited to a single tap, i.e., $L_l = 1.$ These \glspl{nsm} perform well over \gls{awgn} channels, as they are constructed to maximize the \gls{msed} between transmitted sequences. However, in practical communication systems such as wireless and mobile radio, the channel is often subject to selective fading. This selectivity can manifest in time, due to Doppler spread, in frequency, due to multipath delay spread, or simultaneously in both domains. Under such conditions, the unit-length filters for $l \ge 1$ prevent the corresponding bipolar input streams $\bar{b}_l[k]$ from contributing any diversity at the receiver detector level. As a result, the overall diversity order prior to error correction decoding is limited to one, even though the stream $\bar{b}_0[k]$ can provide a diversity order up to $L_0,$ provided that all taps of the filter $h_0[k]$ are simultaneously non-zero.

At this stage, diversity can only be provided through error correction coding, and to make it effective, an interleaver must be inserted between the encoder and the \gls{nsm} modulator. To achieve diversity at the modulation level as well, it is necessary to extend the lengths of the filters, $h_l[k],$ $l \ge 1.$ Letting $\lambda_l$ denote the number of non-zero taps in each filter $h_l[k],$ and assuming ideal or near-ideal interleaving, the diversity order offered by the \gls{nsm} is equal to $\min_l \, (\lambda_l).$ Therefore, increasing the span of all filters—not just $h_0[k]$—is essential to enable diversity in fading channels and improve robustness at the modulation level.

\subsection{Power Spectral Density of NSMs and Whitening via Scrambling}

The $m$ bipolar input sequences, $\bar{b}_l[k],$ $l = 0, 1, \ldots, m-1$, that feed into an \gls{nsm} typically originate from a sequence of processing steps: digital scrambling (elementwise addition of a random binary sequence, using modulo $2$ arithmetic), interleaving, and binary-to-bipolar conversion of the binary-coded output from the error correction encoder. As a result of this processing, the input sequences exhibit white characteristics, meaning their power spectral densities are flat. However, the \gls{nsm} architecture includes a set of $m$ filters, $h_l[k],$, $l = 0, 1, \ldots, m-1,$ with at least one filter having a length $L_l > 1.$ This filtering process introduces spectral coloration, causing the output modulated sequence to have a non-flat, or colored, power spectral density. To mitigate the impact of this spectral coloration on the final transmitted analog signal, analog scrambling is employed prior to pulse shaping. Specifically, the modulated sequence $s[k]$ is multiplied elementwise by a bipolar scrambling sequence $p[k]$ that is spectrally white or nearly white. This yields a scrambled modulated sequence $\tilde{s}[k] \triangleq s[k] p[k]$ that is spectrally white or quasi-white.

In \glspl{nsm} with rates $\rho \geq 2,$ the scrambled modulated sequence is typically both spectrally white and stationary. However, for \glspl{nsm} with rates of the form $(Q+1)/Q,$ where $Q > 1,$ whiteness is still guaranteed after scrambling, but stationarity is not necessarily ensured. This difference originates from the inherent cyclostationary structure of the modulated signal. Specifically, the system effectively upsamples the input sequences by a factor of $Q,$ which increases the sampling rate and introduces a periodic pattern in the statistics of the modulated sequence. The polyphase structure of the modulation, detailed in Section~\ref{sec:Analogy with convolutional error correction coding}, reflects this periodicity and directly follows from the oversampling operation. For example, the rate-$4/3$ \gls{nsm} examined in Section~\ref{sec:Analogy with convolutional error correction coding} clearly reveals this periodic variation in the average energy of the modulated sequence samples, illustrating its cyclostationary nature.

To the above case of cyclostationary modulated signals with periodic variation in average symbol energies, there are exceptions. Some \glspl{nsm} with rates of the form $(Q+1)/Q,$ such as those specified in Tables~\ref{table:NSMs Rate-3/2 Filter Pattern (1, 1, 1, 1)}, \ref{table:NSMs Rate-4/3 Filter Pattern (1, 1, 1, 1, 1, 1, 1, 1, 1)}, and~\ref{table:NSMs Rate-5/4 Filter Pattern (1, 1, 1, 1, 1, 1, 1, 1, 1, 1, 1, 1, 1, 1, 1, 1)}, corresponding to rates $3/2$, $4/3$, and $5/4$ respectively, yield modulated sequences whose components have equal average energies. Despite their cyclostationary nature and polyphase representation, these special cases exhibit stationary behavior after scrambling due to the specific forms of their underlying filters and associated polyphase filters.

\subsection{Transition from Real to Complex Filters for PAPR Reduction in Single-Carrier Systems}

So far, all considered \glspl{nsm} treat and generate the in-phase and quadrature-phase components of the modulated sequence separately. Each of these components is produced by an independent \gls{nsm}, implemented using $m$ real-valued filters, $h_l[k],$ $l=0,1,\ldots,m-1,$ and designed to be spectrally equivalent to a $2^m$-ASK modulation. When these component \glspl{nsm} are combined to form the overall complex modulated sequence, the resulting compound rate is equivalent to that of a $2^{2m}$-QAM.

Proceeding in this way results in a square-shaped footprint (or scatter plot) of the complex modulated sequence in the complex plane. A key drawback of this structure is its tendency to produce a high \gls{papr}, which poses challenges in \gls{sc} systems. This is especially problematic when using nonlinear power amplifiers, which are most power-efficient near saturation—precisely where signal distortion is at its peak. In contrast, \gls{papr} becomes largely irrelevant in systems such as \gls{ofdm}~\cite{Bingham90,Jiang08} or \gls{otfs}~\cite{Hadani17,Raviteja18}, where the transmitted analog signal is the sum of many modulated symbols. By the central limit theorem, the resulting complex envelope approaches a Gaussian distribution, effectively masking any reduction in \gls{papr} achieved at the symbol level.

Focusing on \gls{sc} systems, how can we reduce the \gls{papr} of the modulated sequence, which directly affects the analog signal’s complex envelope? The solution lies in using complex-valued filters $h_l[k],$ $l=0,1,\ldots,m-1,$ instead of real-valued ones. This change provides an additional degree of freedom in filter design, enlarging the optimization space and increasing the chances of producing circular-shaped footprints (or scatter plots) in the complex plane, which can help reduce \gls{papr}.

However, the effectiveness of this approach is limited by two main factors that cause the in-phase and quadrature-phase components to behave in a Gaussian-like manner, regardless of whether real or complex filters are used. When both components exhibit this Gaussian behavior, the resulting complex samples tend to be circularly symmetric Gaussian distributed, producing circular-shaped footprints in the complex plane. The first factor is the growing number of input streams required to achieve high spectral efficiency and data rates. This growth pushes both components closer to Gaussianity. This effect is stronger with complex filters since all input sequences influence both the in-phase and quadrature components. In contrast, with real filters, each component depends on only half of the input streams. The second factor is the increase in filters lengths, which is motivated by the desire to approach the performance of $2$-ASK modulation. Longer filters cause the in-phase and quadrature components to depend on more input bipolar symbols, which accelerates their convergence toward a Gaussian distribution.

As a result, even in \glspl{sc} systems, the benefits of switching from real to complex filters decrease as spectral efficiency rises. Additionally, this change increases receiver complexity by doubling the number of states in the detection trellis. This happens because twice as many input symbols define each component compared to the case of real filters. In summary, using complex filters to reduce \gls{papr} is only beneficial for low-rate modulations with short filters. For high spectral efficiency cases, this approach offers minimal \gls{papr} improvement while significantly increasing detection complexity, making it an unattractive option.

\subsection{Comparison of the 4D Rate-5/4 Block NSM with the Densest 4D Lattice D4} \label{Comparison of rate-5/4 block NSM and the densest four-dimensional lattice D4}

The densest lattice in four dimensions, denoted by $D_4,$ is defined as
\begin{equation}
D_4 = \{\bm{s} = (s[0], s[1], s[2], s[3]) \in \mathbb{Z}^4 \,| \, s[0]+s[1]+s[2]+s[3]=0 \bmod 2  \}.
\end{equation}
As discussed in Section~\ref{ssec:Minimum Euclidean Distance Guaranteeing 5/4-NSM}, the modulated row vectors associated with the rate-$5/4$ block \gls{nsm} are constructed as $\bm{s} = \bm{b}\bm{G},$ where $\bm{b} = (b[0], b[1], b[2], b[3], b[4]) \in \{\pm1\}^5$ and $\bm{G}$ is the generator matrix defined in (\ref{eq:Generating Matrix NSM Rate 5/4}).

To ensure that the modulated vectors $\bm{s}$ align with the $D_4$ lattice, two approaches can be considered. One approach involves scaling the modulated vectors by a factor of two. This results in a set $S \triangleq \{\pm(1\pm2,1\pm2,1\pm2,1\pm2)\},$ which consists of $32$ vectors belonging to $D_4.$ These vectors have a maximum Euclidean norm of $6$ and a maximum infinity norm of $3.$ Now, consider the intersection of $D_4$ with two types of norm balls in $\mathbb{Z}^4.$ First, the Euclidean ball of radius $6$ contains $3337$ lattice points from $D_4.$ Second, the infinity-norm ball (i.e., the \gls{4d} hypercube of radius $3$) contains $1201$ such points. Evidently, the modulated vectors form a very small and sparse subset of these much larger sets. Despite this sparsity, the modulated vectors from the scaled rate-$5/4$ block \gls{nsm} offer a high \gls{msed} of $16.$ This is significantly greater than the \gls{msed} of $2$ in the full $D_4$ lattice, which, for instance, separates the origin, $\bm{0} \triangleq (0,0,0,0),$ from points having exactly two nonzero entries in the bipolar set $\{\pm1\}.$

In a second approach, we express $\bm{b} \in \{\pm1\}^5$ as $\bm{b} = 2\,\bm{b}-\bm{1},$ where $\bm{c} \in \{0,1\}^5$ and $\bm{1} \triangleq (1,1,1,1).$ Substituting this into the modulation formula yields the modulated vectors $\bm{s} = (2\,\bm{c} -\bm{1})\, \bm{G} = \bm{c} \, (2\bm{G}) - \bm{1} \, \bm{G} = \bm{c}\,(2\bm{G}) - \tfrac{3}{2} \, \bm{1},$ where we used the fact that $\bm{1} \, \bm{G} = \tfrac{3}{2} \, \bm{1}.$ This shows that the modulated vectors lie in the coset $D_4-\tfrac{3}{2} \, \bm{1} = D_4+\tfrac{1}{2} \, \bm{1},$ and the rate-$5/4$ block \gls{nsm} can thus be viewed as a subset of this shifted lattice. Let us define the re-centered vectors as $\bm{t} \triangleq \bm{s}+\tfrac{3}{2} \, \bm{1} = \bm{c}\,(2\bm{G}),$ where  $\bm{c} \in \{0,1\}^5.$ These vectors lie in the $D_4$ lattice and represent a centered version of the \gls{nsm} constellation. The components of $\bm{t}$ take values in the set $\{0,1,2,3\},$ so that $\bm{t} \in \{0,1,2,3\}^4 \cap D_4.$ This intersection contains 1$28$ points. However, only $32$ of them are used in the block \gls{nsm}, meaning that $96$ points within this hypercube are not part of the modulation set. Despite this restriction, the modulated vectors $\bm{s},$ or equivalently their centered versions, $\bm{t},$ achieve a \gls{msed} of $4.$ In contrast, the full $D_4$ lattice has a smaller minimum squared distance of $2,$ such as between the origin, $\bm{0},$ and vectors with two nonzero components equal to $1.$

\subsection{Structural Analogies Between NSMs and Convolutional Error-Correcting Codes} \label{sec:Analogy with convolutional error correction coding}

To draw an analogy between convolutional error correction codes and \glspl{nsm}, we take Figure~\ref{fig:mpr_modulations} as a reference for the general structure of any rate-$k/n$ \gls{nsm}. At the input of the \gls{nsm}, we consider a bipolar data stream that is demultiplexed into $k$ parallel bipolar substreams, denoted as $\bar{b}_l[p],$ $l = 0, 1, \ldots, k-1.$ These substreams typically result from error correction coding, followed by interleaving, digital scrambling, and binary-to-bipolar conversion. At the output, the \gls{nsm} generates $n$ parallel modulated substreams, $s_m[p],$ $m = 0, 1, \ldots, n-1,$ which are subsequently multiplexed into a single modulated stream, $s[q].$ Each substream $s_m[p]$ is related to the overall modulated sequence by the expression $s_m[p] = s[np + m].$

Referring again to Figure~\ref{fig:mpr_modulations}, and in line with the expressions provided in (\ref{eq:Normalized Modulated Symbols Rate 3/2 MSN}), (\ref{eq:Normalized Modulated Symbols Rate 4/3 MSN}) and (\ref{eq:Rescaled Modulated Symbols Rate 5/4 MSN}) for rate-$3/2,$ $4/3$ and $5/4$ \glspl{nsm}, respectively, the modulated sequence $s[q]$ can be expressed in the general case as
\begin{equation} \label{eq:Modulated Symbols Rate k/n NSM}
s[q] = \sum_{l=0}^{k-1} \sum_p \bar{b}_l[p] h_l[q-np],
\end{equation}
where $h_l[p],$ $l = 0, 1, \ldots, k - 1,$ are filters of lengths $L_l \ge 1$ that satisfy $h_l[p] = 0$ for $p < 0$ or $p \ge L_l.$

Given the expression $s_m[p] = s[np + m]$ that relates each substream $s_m[p]$ to the modulated sequence $s[q],$ we can use equation (\ref{eq:Modulated Symbols Rate k/n NSM}) to write
\begin{equation}  \label{eq:Modulated Substream Rate k/n NSM}
    s_m[p] = s[np+m] = \sum_{l=0}^{k-1} \sum_i \bar{b}_l[i] h_l[np+m-ni] =  \sum_{l=0}^{k-1} \sum_i \bar{b}_l[i] h_l[n(p-i)+m].
\end{equation}
By defining the polyphase filters as $h_{ml}[p] \triangleq h_l[np + m],$ each modulated substream can be expressed as
\begin{equation}  \label{eq:Polyphase Modulated Substream Rate k/n NSM}
    s_m[p] = \sum_{l=0}^{k-1} \sum_i \bar{b}_l[i] h_{ml}[p-i] = \sum_{l=0}^{k-1} \bar{b}_l[p] \circledast h_{ml}[p].
\end{equation}
It is important to note that the polyphase filters $h_{ml}[p],$ $l = 0, 1, \ldots, k-1$ and $m = 0, 1, \ldots, n-1,$ have lengths at most $\lambda_l \triangleq \lceil L_l / n \rceil,$ $l = 0, 1, \ldots, k-1.$

For illustration, let us examine how the previous expression applies to the best rate-$4/3$ \gls{nsm}, which corresponds to filter pattern $\bm{\pi}_0 = (3, 3, 3, 3, 2, 2, 2, 1, 0)$ and is labeled as Filter \# 1 in Table~\ref{table:NSMs Rate-4/3 Filter Pattern (3, 3, 3, 3, 2, 2, 2, 1, 0)}. Given the rate $\rho = 4/3,$ we identify $k = 4$ and $n = 3,$ corresponding to input bipolar data streams $\bar{b}_l[p],$ $l=0,1,2,3$, and output modulated streams $s_m[p],$ $m=0,1,2.$ In scaled form, the associated filters are $\mathring{h}_0[q] = 3 \delta[q] + 3 \delta[q-1] + 2 \delta[q-2] + 2 \delta[q-3] + 3 \delta[q-5] - 2 \delta[q-6]- 3 \delta[q-7] + \delta[q-8]$ and $\mathring{h}_l[q] = 7\, \delta[q-(l-1)],$ $l=1,2,3.$ As a consequence, the scaled versions of the polyphase filters, $h_{ml}[p],$ $l=0,1,2,3$ and $m=0,1,2,$ that are non-null, are given by $\mathring{h}_{00}[p] = 3\,\delta[p] + 2\,\delta[p-1] - 2\,\delta[p-2],$ $\mathring{h}_{10}[p] = 3\,\delta[p] - 3\,\delta[p-2],$ $\mathring{h}_{20}[p] = 2\,\delta[p] + 3\,\delta[p-1] + \delta[p-2]$ and $\mathring{h}_{01}[p] = \mathring{h}_{12}[p] = \mathring{h}_{23}[p] = 7\,\delta[p].$ 

\begin{figure}[!htbp]
    \centering
    \includegraphics[width=0.75\textwidth]{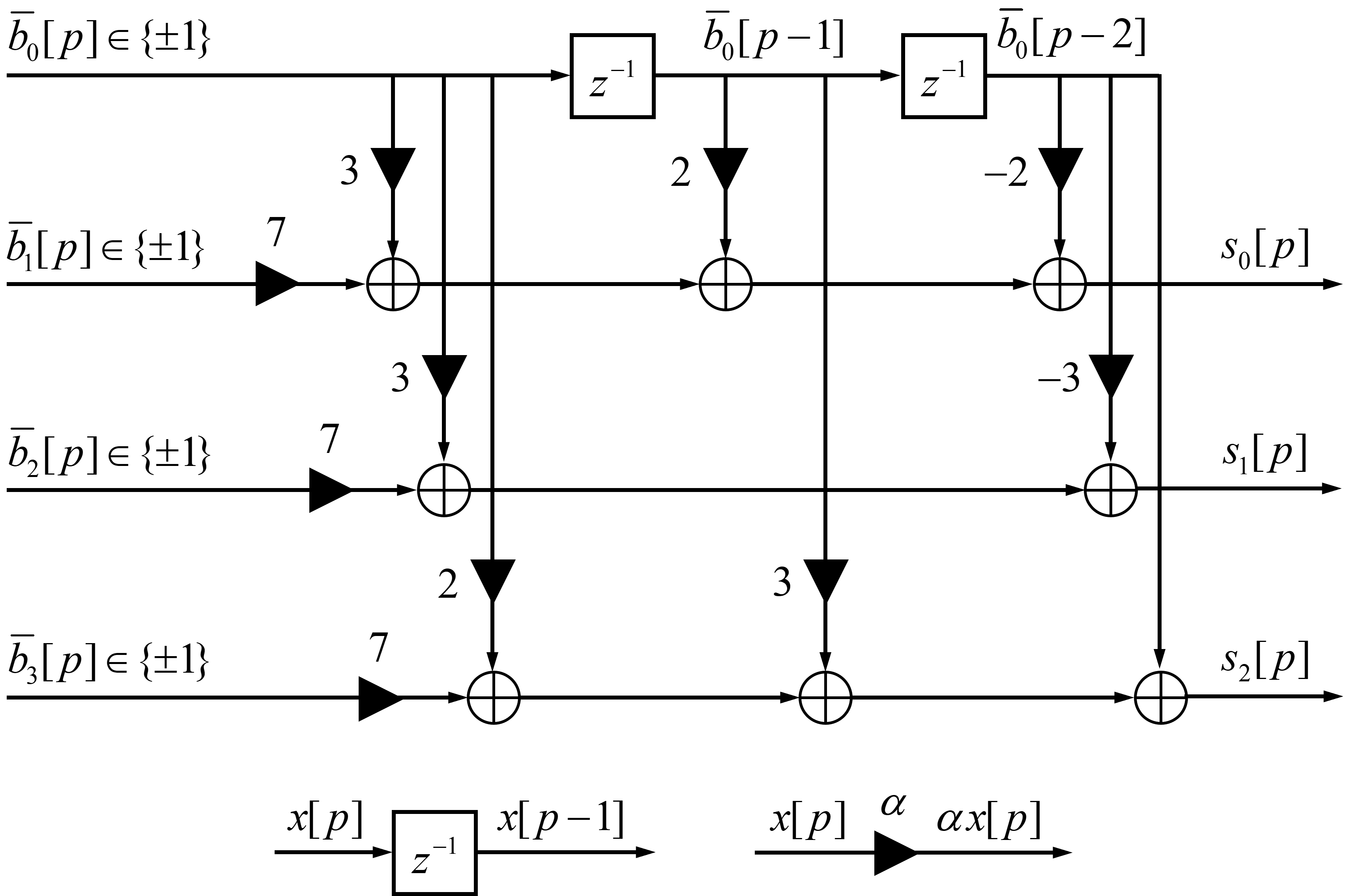}
    \caption{Block diagram showcasing the convolutional structure of the best rate-$4/3$ \gls{nsm}, with first scaled filter, $\mathring{h}_0[q]$, referred to as Filter \# 1, in Table~\ref{table:NSMs Rate-4/3 Filter Pattern (3, 3, 3, 3, 2, 2, 2, 1, 0)}.}
    \label{fig:BlockDiagramConvolutionalStructureNSM}
\end{figure}

The convolutional structure of the considered best rate-$4/3$ \gls{nsm} is illustrated in Figure~\ref{fig:BlockDiagramConvolutionalStructureNSM}, resembling the architecture of a convolutional encoder. This similarity naturally raises the question: what are the key similarities and differences between the \gls{nsm} modulator and the convolutional encoder?

The primary difference lies in their domains of operation. Convolutional coding takes place in the binary domain, using modulo-two arithmetic, which results in binary coded outputs. In contrast, \glspl{nsm} operate using integer arithmetic, producing modulated sequences that typically take more than two distinct values. This distinction also reflects in the achievable rates: convolutional codes, constrained by binary outputs and the injective nature of coding, require rates $\rho = k/n < 1$ (i.e., $k < n$), whereas \glspl{nsm}, with multilevel modulated outputs, can support rates larger than $1$ (i.e., $k > n$).

Another difference concerns the flexibility in code design. Convolutional codes are specified by generator polynomials with binary coefficients, which significantly restricts the set of candidate codes available for optimization. Moreover, some of these polynomial choices may lead to catastrophic codes, whose undesirable properties are well studied in the literature~\cite{Wicker95,Costello04,Johannesson99}. In contrast, \glspl{nsm} are defined by polyphase filters with integer coefficients, offering a much broader design space from which high-performance structures can be selected. As an added benefit, \glspl{nsm} inherently avoid the risk of catastrophic behavior due to their injective convolutional structure.

Furthermore, the average energies of output streams differ between the two. In convolutional codes, binary-to-bipolar conversion yields parallel coded streams with identical average energies. For \glspl{nsm}, however, the average symbol energy generally varies between streams. Specifically, the average energy per modulated stream $s_m[p]$ is given by $E_m \triangleq \sum_{l=0}^{n-1} \| h_{ml}[p] \|^2,$ and for a general set of filters $h_l[q],$ $ l=0,1,\ldots,k-1,$ there is no inherent reason for these energies $E_m$ to be equal across all $m.$ This is confirmed in the rate-$4/3$ \gls{nsm} example discussed earlier, where the average symbol energies of the scaled modulated streams $\mathring{s}_m[p]$, $m = 0, 1, 2$, defined as $\mathring{E}_m \triangleq \sum_{l=0}^{n-1} \| \mathring{h}_{ml}[p] \|^2,$ are found to be unequal:
$\mathring{E}_0 = 3^2 + 2^2 + (-2)^2 + 7^2 = 66$, $\mathring{E}_1 = 3^2 + (-3)^2 + 7^2 = 67$ and $\mathring{E}_2 = 2^2 + 3^2 + 1^2 + 7^2 = 63.$

Despite these differences, the most important similarity between convolutional codes and \glspl{nsm} is their underlying convolutional structure, which relates outputs to inputs through convolution. This structural analogy motivates alternative terms for \glspl{nsm}, such as \emph{convolutional modulations} or \emph{analog codes}, to emphasize their operation in the analog domain. In this light, convolutional codes may be viewed as digital codes, operating within the discrete domain of binary arithmetic.


\subsection{Application of Tail-Biting to One-Dimensional NSMs} \label{sec:One-dimensional NSMs and tail-biting}

Due to the filtered nature of the modulated sequences underlying \glspl{nsm}, whereby, as in (\ref{eq:Modulated Symbols Rate k/n NSM}), the modulated sequence $s[q]$ is obtained as
\begin{equation}
s[q] = \sum_{l=0}^{k-1} \sum_p \bar{b}_l[p] h_l[q-np],
\end{equation}
where $\bar{b}_l[p]$ are bipolar input sequences of finite length $K_l \ge 1,$ and $h_l[p],$ $l = 0, 1, \ldots, k - 1,$ are causal filters of lengths $L_l \ge 1,$ the resulting modulated sequences extend over a total length of $L = \max_l(nK_l + L_l) - 1.$ In this setting, the effective transmission rate is no longer equal to the nominal value $\rho = k/n,$ but instead becomes $\check{\rho} \triangleq (\sum_l K_l)/L.$ This effective rate, $\check{\rho},$ is always strictly less than the theoretical rate, $\rho,$ and only approaches it in the asymptotic regime where the shortest bipolar input sequence length grows without bound, i.e., when $L_\text{min} = \min_l K_l \rightarrow \infty.$ The gap between the theoretical and effective rates can be significant when the input sequences are short and/or when the filters exhibit long impulse responses. Such conditions often arise in practice: for instance, short input sequences are typical when small data packets are transmitted, and longer filters may be preferred to enhance the performance of the \gls{nsm}—approaching that of $2$-ASK—particularly in high-rate scenarios where $\rho$ becomes large.

In order to preserve spectral efficiency and recover the nominal rates of \glspl{nsm}, one can extend the principle of tail-biting—originally developed for convolutional codes~\cite{Ma86}—to the context of \glspl{nsm}. The first step in this approach is to assume that all bipolar input sequences have a common length $K,$ and that the filters involved have a maximum length $L_\text{max} = \max_l L_l,$ which does not exceed $Kn.$ The second step is to apply a cyclic convolution in the multirate setting, where each bipolar input sequence of length $K$ is first upsampled by a factor $n,$ producing a sequence of length $Kn,$ and then cyclically convolved with its corresponding filter. This yields a modulated sequence $s[q]$ of length $Kn,$ given by
\begin{equation}
s[q] = \sum_{l=0}^{k-1} \sum_{p=0}^{K-1} \bar{b}_l[p] h_l[q-np \bmod Kn],
\end{equation}
where $q = 0, 1, \ldots, Kn-1$. The modulo-$Kn$ indexing imposes a cyclic structure on the modulated signal, where the contribution of each input symbol—spread in time through filtering after upsampling—wraps cyclically around the sequence. This construction guarantees that the nominal \gls{nsm} rate is maintained, even in the case of short data packets, as long as the filters do not exceed the length of the modulated sequence.

However, this cyclic multirate structure can also introduce performance vulnerabilities. In particular, it may degrade the \gls{msed} or significantly increase its multiplicity, both of which may negatively impact the system's error performance. To investigate these effects and identify possible countermeasures, we turn our attention to the most fragile \glspl{nsm} studied so far: the rate-$2$ \glspl{nsm} ($k=2$, $n=1$) introduced in Section~\ref{ssec:Modulation of Rate 2}. These \glspl{nsm} share a common second scaled filter $\mathring{h}_1[p] = 2\,\delta[p]$, and differ in their first scaled filter, which is either $\mathring{h}_0[p] = \delta[p] + \delta[p-1]$—representing the “duobinary” channel—or $\mathring{h}_0[p] = \delta[p] - \delta[p-1]$—representing the “dicode” channel. A brief discussion of tail-biting for these \glspl{nsm} was already provided in Section~\ref{ssec:Modulation of Rate 2}. The objective now is to carry out a more detailed analysis of the impact of tail-biting on these two schemes.

Before examining each of the two \glspl{nsm} in detail, we consider a pair of bipolar input sequences $\{\bar{b}_0^m[p], \bar{b}_1^m[p],\, p = 0, 1, \ldots, K-1\},$ $m = 0, 1,$ and their associated scaled modulated sequences $\{\mathring{s}^m[p],\, p = 0, 1, \ldots, K-1\}.$ We define the input sequence differences as $\Delta \bar{b}_0[p] \triangleq \bar{b}_0^1[p] - \bar{b}_0^0[p]$ and $\Delta \bar{b}_1[p] \triangleq \bar{b}_1^1[p] - \bar{b}_1^0[p],$ and the corresponding modulated sequence difference as $\Delta \mathring{s}[p] \triangleq \mathring{s}^1[p] - \mathring{s}^0[p],$ $p = 0, 1, \ldots, K - 1.$ As usual, the input sequence differences $\Delta \bar{b}_0[p]$ and $\Delta \bar{b}_1[p]$ take values in the ternary set $\{0, \pm 2\}.$

First, observe that under tail-biting, some input sequence differences still achieve the same \gls{msed} of $8$ as in the non-tail-bitten \glspl{nsm} discussed in Section~\ref{ssec:Modulation of Rate 2}. For instance, consider $\Delta \bar{b}_0[p] = \pm 2\,\delta[p]$ and $\Delta \bar{b}_1[p] = 0.$ The corresponding scaled modulated sequences difference is then given by $\Delta \mathring{s}[p] = \pm 2\,\mathring{h}_0[p] = \pm 2 (\delta[p] \pm \delta[p-1]),$ which clearly has a \gls{sen} of $8,$ matching the \gls{msed} of the original, non-tail-bitten, \glspl{nsm}. More importantly, the number of such input pairs is reduced in the tail-biting case, since the corresponding error events are now restricted to have durations that do not exceed the finite input sequence length $K.$ As a result, the multiplicity of input differences that achieve the \gls{msed} of the corresponding non-tail-bitten \glspl{nsm} decreases, and falls below the asymptotic value $51/2$ derived in Appendix~\ref{app:Tight Estimate BEP Rate 2}.

Second, in accordance with the branch labels of the \gls{nsm} trellis associated with the “duobinary” filter shown in Figure~\ref{fig:Trellis Input Difference Duobinary Channel Rate-2 Modulation} (b), we note that the scaled modulated sequences difference $\Delta \mathring{s}[p]$ admits values in the set $\{0, \pm 2, \pm 4, \pm 6, \pm 8\}.$ The only cases where the \gls{sen} of $\Delta \mathring{s}[p]$ can fall below $8$ are when this norm is equal to $4$ or $0,$ corresponding respectively to (i) a single nonzero component of $\Delta \mathring{s}[p]$ equal to $\pm 2,$ or (ii) a perfectly zero sequence. The first case cannot occur for any valid input sequences differences. However, the second case, in which $\Delta \mathring{s}[p]$ is identically zero, can arise from nonzero input differences. This more critical situation, in which $\Delta \mathring{s}[p]$ is identically zero, was already briefly noted in Section~\ref{ssec:Modulation of Rate 2}. It corresponds to a complete loss of distinguishability between certain specific pairs of different input sequences that, under tail-biting, produce the same modulated sequence. While the vast majority of input sequence pairs still lead to distinct modulated sequences, these exceptional cases reveal that the mapping from input to modulated sequences is not always injective. As a result, even in the absence of channel noise, such structural ambiguities can cause decoding uncertainty. A detailed investigation of this phenomenon will follow in the next sections, focusing on both \glspl{nsm}.

We begin with the \gls{nsm} associated with the filter $\mathring{h}_0[p] = \delta[p] + \delta[p-1],$ which corresponds to the “duobinary” channel. In this case, the samples of the modulated sequence difference are given explicitly by
\begin{equation}
\Delta \mathring{s}[p] = \Delta \bar{b}_0[p] + \Delta \bar{b}_0[p-1 \bmod K] + 2\,\Delta \bar{b}_1[p],\, p=0,1,\ldots,K-1.
\end{equation}
Imposing that $\Delta \mathring{s}[p]$ be identically zero leads to the constraint $\Delta \bar{b}_0[p] + \Delta \bar{b}_0[p-1 \bmod K] = -2\,\Delta \bar{b}_1[p],\, p=0,1,\ldots,K-1.$ This condition leads to two distinct cases. In the first case, when $\Delta \bar{b}_1[p] = \pm 2,$ the samples $\Delta \bar{b}_0[p]$ and $\Delta \bar{b}_0[p-1 \bmod K],$ of the input sequence differences, must be equal and take the value $\mp 2.$ That is, $\Delta \bar{b}_0[p] = \Delta \bar{b}_0[p-1 \bmod K] = -\Delta \bar{b}_1[p] = \mp 2.$ In the second case, when $\Delta \bar{b}_1[p] = 0,$ the samples $\Delta \bar{b}_0[p]$ and $\Delta \bar{b}_0[p-1 \bmod K],$ of the first input sequence difference, must either both be zero, i.e., $\Delta \bar{b}_0[p] = \Delta \bar{b}_0[p-1 \bmod K] = 0,$ or take opposite signs, i.e., $\Delta \bar{b}_0[p] = -\Delta \bar{b}_0[p-1 \bmod K] = \pm 2.$

In summary, for the first bipolar input sequence difference $\Delta \bar{b}_0[p],$ there are only two admissible relations between any pair of cyclically consecutive samples: (i) $\Delta \bar{b}_0[p] = \Delta \bar{b}_0[p-1 \bmod K] \in \{0, \pm 2\}$ or (ii) $\Delta \bar{b}_0[p] = -\Delta \bar{b}_0[p-1 \bmod K] = \pm 2.$ From these possibilities, we can infer that if $\Delta \bar{b}_0[p] = 0$ at any index $p,$ then the entire sequence must be identically zero. Indeed, only case (iii) allows continuation without violating the cyclic consistency, and applying it recursively around the cycle forces all samples of $\Delta \bar{b}_0[p]$ to be zero. Substituting this into the nullity condition, it follows that all samples of $\Delta \bar{b}_1[p]$ must also be zero. This would imply that both input sequence differences are identically zero, contradicting the assumption that the bipolar input sequences are not simultaneously identical. Therefore, valid sequences $\Delta \bar{b}_0[p \bmod K]$ must be such that consecutive samples (modulo $K$) are either equal or opposite in sign, with all values restricted to $\{\pm 2\}.$

As a result, transitions in polarity between consecutive samples must occur in an even number around the cycle. This necessity arises from the cyclic nature of the indexing, which requires the sequence to return to its starting sign after one full rotation, thereby enforcing that the total count of sign changes be even. If we denote this even number by $2l,$ with $0 \le l \le \lfloor K/2 \rfloor,$ then $\Delta \bar{b}_1[p \bmod K]$ has $2l$ zero-valued samples, while the remaining $K - 2l$ samples take values in $\{\pm 2\}.$ Hence, the total number of nonzero samples in $\Delta \bar{b}_0[p \bmod K]$ and $\Delta \bar{b}_1[p \bmod K]$ combined is $K + (K - 2l) = 2(K - l),$ yielding an occurrence probability of $1/2^{2(K - l)} = 2^{2l}/2^{2K}$.

To count the number of first input sequence differences with exactly $2l$ transitions, we note that such a sequence is completely specified by fixing the value of one sample and selecting $2l$ transition positions among the $K$ indices. This gives $2 \binom{K}{2l}$ valid configurations.

Summing over all possible values of $l,$ the probability that a given pair of bipolar input sequences has at least one different competitor pair yielding the same modulated sequence is:
\begin{equation}
    \pi =  \frac{1}{2^{2K}}\sum_{l=0}^{\lfloor K/2\rfloor} 2 \binom{K}{2l} 2^{2l} = \frac{1}{2^{2K}} \left( (1+2)^K+(1-2)^K \right) = \left(\frac{3}{4}\right)^K + \left(-\frac{1}{4}\right)^K.
\end{equation}

We now turn our attention to the \gls{nsm} defined by the filter $\mathring{h}_0[p] = \delta[p] - \delta[p-1],$ which characterizes the “dicode” channel. In this scenario, the difference between modulated sequences can be explicitly written as
\begin{equation}
\Delta \mathring{s}[p] = \Delta \bar{b}_0[p] - \Delta \bar{b}_0[p-1 \bmod K] + 2\,\Delta \bar{b}_1[p],\, p=0,1,\ldots,K-1.
\end{equation}
Ensuring that $\Delta \mathring{s}[p]$ is identically zero imposes the condition $\Delta \bar{b}_0[p] - \Delta \bar{b}_0[p-1 \bmod K] = -2\,\Delta \bar{b}_1[p],\, p=0,1,\ldots,K-1.$ From this relation, two distinct cases arise. First, when $\Delta \bar{b}_1[p] = \pm 2,$ the consecutive samples $\Delta \bar{b}_0[p]$ and $\Delta \bar{b}_0[p-1 \bmod K]$ must be of opposite signs, specifically, $\Delta \bar{b}_0[p] = -\Delta \bar{b}_0[p-1 \bmod K] = -\Delta \bar{b}_1[p] = \mp 2.$ Second, if $\Delta \bar{b}_1[p] = 0,$ the values of $\Delta \bar{b}_0[p]$ and $\Delta \bar{b}_0[p-1 \bmod K]$ coincide and belong to the set $\{0, \pm 2\},$ i.e., $\Delta \bar{b}_0[p] = \Delta \bar{b}_0[p-1 \bmod K].$

Summarizing these findings, the first input sequence difference $\Delta \bar{b}_0[p]$ can only satisfy one of the following conditions for consecutive samples modulo $K$: either they are equal with values in $\{0,\pm 2\}$ or they are opposite in sign with magnitude $2,$ that is, $\Delta \bar{b}_0[p] = \Delta \bar{b}_0[p-1 \bmod K] \in \{0,\pm 2\}$ or $\Delta \bar{b}_0[p] = -\Delta \bar{b}_0[p-1 \bmod K] = \pm 2.$ Interestingly, these allowed patterns match those found in the “duobinary” case, leading to similar conclusions about the behavior of the first input sequences difference. Notably, if $\Delta \bar{b}_0[p]$ is zero at any sample index, the cyclic constraints and permitted patterns require that $\Delta \bar{b}_0[p]$ be zero for all $p$. Consequently, the relationship between $\Delta \bar{b}_0[p]$ and $\Delta \bar{b}_1[p]$ enforces that $\Delta \bar{b}_1[p]$ is also zero for all $p$. This contradicts the assumption that the pair of bipolar input sequences differ in at least one of their components, either in the first sequence or in the second sequence.

Therefore, as before, consecutive cyclic samples of $\Delta \bar{b}_0[p \bmod K]$ can either share the same sign or have opposite signs within the set $\{\pm 2\}.$ The key distinction for the “dicode” \gls{nsm} lies in the condition that only transitions in polarity of $\Delta \bar{b}_0[p],$ when moving from $p-1 \bmod K$ to $p \bmod K,$ can produce nonzero values in $\Delta \bar{b}_1[p].$ Moreover, the cyclic indexing still mandates an even number of such cyclic polarity transitions within the set $\{0,1,\ldots, K-1\}.$ Denoting this even number as $2l$ with $0 \le l \le \lfloor K/2 \rfloor,$ the count of non-null components in $\Delta \bar{b}_1[p \bmod K]$ is $2l,$ while the number of zero components is $K-2l.$ Consequently, for any sequence $\Delta \bar{b}_0[p \bmod K]$ exhibiting $2l$ transitions, the total number of non-null elements in $\Delta \bar{b}_0[p \bmod K]$ and $\Delta \bar{b}_1[p \bmod K]$ combined equals $K + 2l,$ yielding an occurrence probability of $1/2^{K+2l}.$

Furthermore, the quantity of first input sequence differences with $2l$ transitions remains $2 \binom{K}{2l},$ as previously established for the “duobinary” case. Summarizing, the probability that a given bipolar input sequence set admits at least one distinct competitor producing the same modulated sequence evaluates to  
\begin{equation}
    \pi = \sum_{l=0}^{\lfloor K/2 \rfloor} 2 \binom{K}{2l} \frac{1}{2^{K+2l}} =
    \frac{1}{2^K} \sum_{l=0}^{\lfloor K/2 \rfloor} 2 \binom{K}{2l} \frac{1}{2^{2l}} = \frac{1}{2^{K}} \left( (1+\tfrac{1}{2})^K + (1-\tfrac{1}{2})^K \right) = \left(\frac{3}{4}\right)^K + \left(\frac{1}{4}\right)^K.
\end{equation}

Hence, for both \glspl{nsm} based on the “duobinary” and “dicode” channels, the probability that a given pair of input sequences admits at least one distinct competing pair yielding the same modulated sequence is identical when the sequence length $K$ is even. When $K$ is odd, the “duobinary” case exhibits a slightly lower confusion probability. Most importantly, for large values of $K$, the behavior of both \glspl{nsm} is dominated by the same exponentially decaying term, leading to the tight approximation $\pi \approx (\tfrac{3}{4})^K$ for the confusion probability.

An immediate implication of this approximation is that the probability of confusion—i.e., the inability to uniquely determine the input sequences from the modulated sequence—can be made arbitrarily small by choosing a sufficiently large $K.$ For instance, to ensure a confusion probability $\pi$ below a target threshold $\varepsilon,$ it suffices to select a sequence length approximately given by $K \approx \lceil \ln(\varepsilon)/\ln(3/4) \rceil.$ For a typical target of $\varepsilon = 10^{-10},$ this yields a minimum required sequence length of about $K = 80.$ Fortunately, in most communication scenarios, packet sizes tend to exceed this threshold, which guarantees that the confusion probability remains significantly below $10^{-10}$ in practice.

In contrast, when the sequence length $K$ is smaller than $80,$ the resulting confusion probability $\pi$ may exceed the acceptable level $\varepsilon = 10^{-10},$ potentially leading to repetitive and endless detection errors. To address this situation, a practical countermeasure consists in introducing a scrambling mechanism that is jointly known by both transmitter and receiver, and that applies distinct scrambling patterns across successive (re)transmissions of the same input sequences. Specifically, let $r_0^{(i)}[p]$ and $r_1^{(i)}[p]$ denote pseudo-random bipolar scrambling sequences used during the $i$-th (re)transmission attempt. These sequences multiply the original inputs $\bar{b}_0[p]$ and $\bar{b}_1[p]$ element-wise, producing scrambled versions $\tilde{\bar{b}}_0^{(i)}[p] \triangleq \bar{b}_0[p]\, r_0^{(i)}[p]$ and $\tilde{\bar{b}}_1^{(i)}[p] \triangleq \bar{b}_1[p]\, r_1^{(i)}[p]$ for $p = 0, 1, \ldots, K - 1$. Consequently, any input sequence differences become $\Delta \tilde{\bar{b}}_0^{(i)}[p] \triangleq \Delta \bar{b}_0[p]\, r_0^{(i)}[p]$ and $\Delta \tilde{\bar{b}}_1^{(i)}[p] \triangleq \Delta \bar{b}_1[p]\, r_1^{(i)}[p]$, thereby altering the structure of possible competing sequences at each attempt. This randomized behavior makes confusion events across $I$ (re)transmissions nearly independent, so that the overall probability of failure after $I$ (re)transmissions is approximately $\pi^I.$ Through this scrambling strategy, the confusion probability can be reduced to arbitrarily low values, even when working with short packets.

Before concluding this section, let us recall that our goal has been to evaluate the probability that a non-zero difference between two input sequence pairs results in a zero modulated sequence difference of length $K.$ We have found that this confusion probability is asymptotically characterized by the exponential decay term $(\tfrac{3}{4})^K.$ Interestingly, this same dominant behavior also emerges in a different context: the upper bounds derived in (\ref{Upper Bound Error Probability Multiplicity - First Case}) and (\ref{Upper Bound Error Probability Multiplicity - Second Case}) for the error event multiplicity at \gls{msed} (equal to half the the \gls{sed} of $2$-ASK) in rate-$2$ \glspl{nsm} with rational taps, as discussed in Section~\ref{Rate 2 Approaching NSM Simple Rational Coefficients}. In those cases, the modulated sequence difference $\Delta s[p]$ is nearly zero, except for two non-zero samples (from the set $\{\pm 2\}$) located at positions $p = 0$ and $p = K - 1$, separated by $K - 2$ consecutive zeros. This again highlights that the decay of the occurrence probability of such low-weight modulated differences follows a power law in $(\tfrac{3}{4})^K,$ reinforcing the idea that this behavior is not unique to “duobinary” and “dicode” \gls{nsm} structures, but extends more broadly to other rate-$2$ \glspl{nsm}, particularly those using scaled filters of the form $\mathring{h}_0[p] = \delta[p] \pm \delta[p - p_0] \pm \delta[p - p_1] \pm \delta[p - L_0 + 1],$ where $0 < p_0 < p_1 < L_0 - 1$.

\subsection{Analogies Between NSMs and Overcomplete Frames Using Sigma-Delta Quantization}

Frames generalize orthonormal bases in Hilbert spaces by allowing redundancy while preserving stable signal representations. The foundational theory of frames can be found in~\cite{Christensen03}, with comprehensive applications to signal processing described in~\cite{Kovacevic08} and~\cite{Daubechies03_1}. Frames can be defined in various functional spaces including the finite-dimensional real space $\mathbb{R}^d$, the discrete-time space $\ell^2(\mathbb{Z})$, and the continuous-time space $L^2(\mathbb{R})$. Here, we focus on frames in $\ell^2(\mathbb{Z})$, which is naturally suited for modeling discrete-time signals.

Let $\mathcal{H}$ be a real or complex Hilbert space. A countable collection $\{\bm{h}_l\}_{l \in \Lambda} \subset \mathcal{H}$ is a \emph{frame} for $\mathcal{H}$ if there exist constants $A, B > 0$ such that for all $\bm{x} \in \mathcal{H}$,
\begin{equation}
A \|\bm{x}\|^2 \leq \sum_{l \in \Lambda} |\langle \bm{h}_l, \bm{x} \rangle|^2 \leq B \|\bm{x}\|^2,
\end{equation}
where $\langle \cdot, \cdot \rangle$ denotes the scalar or Hermitian inner product in $\mathcal{H}$, depending on whether the space is real or complex. This ensures that signal energy is stably captured by the frame coefficients, even in redundant settings.

Frames are often categorized by their structural properties. A \emph{tight frame} satisfies $A = B$, and when $A = B = 1$, the frame is called a \emph{Parseval frame}. An \emph{overcomplete frame} contains more elements than a basis, offering non-unique but robust signal representations. It is important to note that overcompleteness is determined by redundancy in the number of frame elements relative to the space dimension, rather than by the frame bounds $A$ and $B$. In finite-dimensional spaces, an overcomplete frame has more elements than the dimension of the space, which implies linear dependence among its elements. In infinite-dimensional spaces, overcompleteness means some elements can be removed without losing the spanning property. In contrast, a system with fewer elements than required to span the space is \emph{undercomplete} and does not satisfy the frame condition.

In many systems modeled by structured expansions, including the \gls{nsm} framework studied here, frame-like elements are built from a finite set of $k$ prototype filters $h_l[p]$, $l = 0, 1, \ldots, k-1$, via regular time shifts:
\begin{equation}
h_{lm}[p] \triangleq h_l[p - mn], \quad l = 0, 1, \ldots, k-1,\; m \in \mathbb{Z},
\end{equation}
where $n$ is a fixed oversampling factor. The family $\{h_{lm}[p]\}_{lm}$ defines a structured and shift-invariant set of functions, naturally implementable in filter banks. These functions resemble structured frames generated by shifts of prototype filters, and \glspl{nsm} modulate bipolar sequences on these functions to generate transmitted signals.

A frame becomes \emph{overcomplete} when its number of generating elements exceeds the minimum required to span the space, either by using additional filters or denser shifts. The oversampling factor $\rho = k/n$ characterizes the degree of redundancy. The condition $\rho > 1$ is necessary for overcompleteness but not sufficient, since other conditions on the set of generators also apply.

Given a frame $\{\bm{h}_l \triangleq (h_l[p])_{p \in \mathbb{Z}}\}_{l \in \Lambda}$ and a signal $\bm{x} \triangleq (x[p])_{p \in \mathbb{Z}}$, one can reconstruct $\bm{x}$ using a dual frame $\{\bm{g}_l \triangleq (g_l[p])_{p \in \mathbb{Z}}\}_{l \in \Lambda}$ through the synthesis formula:
\begin{equation} \label{Synthesis Formula Overcomplete Frames}
x[p] = \sum_{l \in \Lambda} \langle \bm{h}_l, \bm{x} \rangle \, g_l[p].
\end{equation}
Overcompleteness allows for multiple dual frames, offering flexibility in design, especially under quantization or noise.

In coarse quantization settings using a 1-bit alphabet—corresponding to bipolar symbols $\{\pm 1\}$ as employed in both \glspl{nsm} and quantized frame expansions—a sufficiently high redundancy becomes essential. When only coarse quantized coefficients $q_l \in \{\pm 1\}$ are available instead of exact frame coefficients, an approximate reconstruction $\hat{\bm{x}}$ can be obtained:
\begin{equation} \label{Approximate Synthesis Formula}
\hat{x}[p] = \sum_{l \in \Lambda} q_l \, g_l[p].
\end{equation}
This approximation $\hat{\bm{x}}$ serves as a coarse reconstruction of the original signal $\bm{x}$, where the bipolar coefficients $q_l$ replace the inner products $\langle \bm{h}_l, \bm{x} \rangle$. This approximate synthesis formula is structurally analogous to the modulation formula in \glspl{nsm}, where bipolar information sequences modulate the basis functions to generate the transmitted signal.

Sigma-delta quantization is a simple and practical technique that recursively encodes signal projections $\langle \bm{h}_{lm}, \bm{x} \rangle$ into bipolar sequences. It employs a feedback loop to shape the quantization noise and improve the fidelity of $\hat{\bm{x}}$ relative to $\bm{x}$. Various orders of sigma-delta quantization exist. Increasing the order typically reduces quantization error and enhances approximation quality without increasing the alphabet size, although it comes with increased implementation complexity.

In the context of overcomplete frames, the best performance in terms of distortion reduction is achieved by applying maximum likelihood determination of the coarse quantization coefficients of the approximate reconstruction $\hat{\bm{x}}$. Such ML quantization finds the optimal bipolar coefficient vector $\bm{q} = (q_l)_{l \in \Lambda}$ that yields the closest $\hat{\bm{x}}$ to $\bm{x}$ according to some distortion metric, such as \gls{sed}. However, the combinatorial complexity of ML quantization grows exponentially with the number of coefficients, rendering it impractical for many applications. For this reason, sigma-delta quantization with recursive feedback is widely used in practice, trading off optimality for simplicity and stability. Several studies have investigated ML quantization in the context of overcomplete frames~\cite{Boufounos07}, and recent works explore iterative and belief propagation algorithms to approach ML or MAP quantization performance~\cite{Do07,Kaslovsky13}.

Other quantization algorithms for overcomplete frames have been proposed, including those based on iterative coefficient refinement and joint source-channel coding methods. For instance, combinations of iterative quantization and decoding techniques have been studied to improve reconstruction fidelity and robustness by exploiting probabilistic graphical models~\cite{Gupta14}.

Despite the different objectives of the two systems, symbol detection in \glspl{nsm} and signal approximation in sigma-delta quantization, both rely on the ability to operate over coarse alphabets through frame-like representations. The transmission in \glspl{nsm} consists of modulating bipolar information sequences over the set of structured functions $\bm{h}_{lm}$ to generate the transmitted signal. Although in classical frame theory the synthesis functions are typically denoted by $\bm{g}_{lm}$, we use the notation $\bm{h}_{lm}$ throughout this work to denote the synthesis elements used in the \gls{nsm} modulation process. This modulation step is structurally analogous to the approximate synthesis formula (\ref{Approximate Synthesis Formula}) used in sigma-delta quantization, where coarse quantized frame coefficients in $\{\pm 1\}$ are used to reconstruct an approximation $\bm{\hat{x}}$ of the original signal $\bm{x}$. The bipolar sequences in \glspl{nsm} carry encoded information, whereas in sigma-delta systems they represent quantized projection coefficients.

At the receiver side, \glspl{nsm} detect transmitted symbols by comparing the received noisy signal to all possible modulated signals and choosing the most likely bipolar sequence via ML detection. Sigma-delta quantization, instead, uses a simpler recursive decision with feedback to generate bipolar sequences, achieving good practical performance despite not guaranteeing ML optimality.

There is also a useful analogy to draw between \glspl{nsm} and \gls{ldpc} codes. Structured \glspl{nsm}, based on regularly time-shifted prototypes, resemble regular \gls{ldpc} codes, where the parity-check matrix has a uniform structure. Unstructured \glspl{nsm}, on the other hand, do not restrict their modulation functions to regular shifts of a small set of filters. Instead, they exploit increased degrees of freedom by allowing a larger, more flexible set of modulation functions. This design freedom aims to increase the \gls{msed} between signals, reduce the multiplicity of this minimum distance, and potentially lower the binary error probability. Such unstructured \glspl{nsm} are analogous to irregular \gls{ldpc} codes, which break uniformity to improve performance.

Structured \glspl{nsm} typically benefit from simpler encoding and detection algorithms but may suffer from performance limitations due to their regularity, whereas unstructured \glspl{nsm} provide enhanced performance at the cost of higher complexity. This analogy motivates the exploration of unstructured \glspl{nsm}, particularly in scenarios where \gls{ldpc} coding is integrated, as discussed in Section~\ref{A natural and efficient synergy between NSM and LDPC coding}~\cite{Gallager62,Richardson01,Richardson08}.

\subsection{Is Extending NSMs Beyond One Dimension Required?}

In this section, we investigate whether constructing \glspl{nsm} in two or more dimensions offers intrinsic advantages over \gls{1d} approaches. While multidimensional constructions are often preferred for their design clarity and ease of achieving structured properties, we argue that any multidimensional \gls{nsm} can ultimately be expressed as a \gls{1d} block \gls{nsm}. This equivalence is established through the existence of a generating matrix, which acts as a complete modulator description.

Block \gls{nsm} signals operate on a finite set of bipolar data symbols, typically arranged in vectors of fixed length. These are transformed into modulated vectors through a linear mapping defined by a generating matrix, in much the same way that block error correction codes operate. Importantly, this concept applies irrespective of the number of dimensions originally used to conceive the modulation. Whether the modulated symbols are arranged on a \gls{1d} line, a \gls{2d} grid, or higher-dimensional lattices, the modulation process can be fully characterized by an appropriate generator matrix.

A \gls{2d} \gls{nsm}, for instance, may modulate data symbols over a grid-like structure. However, this arrangement can always be \emph{unfolded} (or \emph{unwrapped}) into a \gls{1d} sequence by reading the modulated values row-wise or column-wise, just like scanning the elements of a matrix. This transformation preserves all distance properties relevant to performance, including the \gls{msed} between modulated vectors. Consequently, the unfolded structure admits a generating matrix, making it indistinguishable from any other block \gls{nsm} constructed in one dimension.

To illustrate this principle, we recall the \gls{2d} \gls{nsm} thoroughly studied in Section~\ref{ssec:Two-Dimensional Rate-2 NSMs} and represented in Figure~\ref{fig:Two-Dimensional-NSM-Illustration-Rate-13-9}. This \gls{nsm} modulates $9$ bipolar symbols over a $3 \times 3$ \gls{2d} grid, producing $9$ modulated symbols. As detailed in the referenced section, this scheme offers a modulation rate of $\rho = 13/9$.

After unfolding the \gls{2d} modulated grid column-wise, the modulation process can be fully described using the generating matrix $\bm{G}$ introduced in (\ref{eq: Generating Matrix Rate 13/9 2D NSM}). For consistency and easier visualization in the present discussion, we define a scaled version of this matrix by a factor of two.This scaled matrix, defined as $\mathring{\bm{G}} \triangleq 2 \, \bm{G}$, is given explicitly as
\begin{equation}
\mathring{\bm{G}} = \begin{bmatrix}
    2 & 0 & 0 & 0 & 0 & 0 & 0 & 0 & 0 \\
    0 & 2 & 0 & 0 & 0 & 0 & 0 & 0 & 0 \\
    0 & 0 & 2 & 0 & 0 & 0 & 0 & 0 & 0 \\
    0 & 0 & 0 & 2 & 0 & 0 & 0 & 0 & 0 \\
    0 & 0 & 0 & 0 & 2 & 0 & 0 & 0 & 0 \\
    0 & 0 & 0 & 0 & 0 & 2 & 0 & 0 & 0 \\
    0 & 0 & 0 & 0 & 0 & 0 & 2 & 0 & 0 \\
    0 & 0 & 0 & 0 & 0 & 0 & 0 & 2 & 0 \\
    0 & 0 & 0 & 0 & 0 & 0 & 0 & 0 & 2 \\
    1 & 1 & 0 & 1 & 1 & 0 & 0 & 0 & 0 \\
    0 & 1 & 1 & 0 & 1 & 1 & 0 & 0 & 0 \\
    0 & 0 & 0 & 1 & 1 & 0 & 1 & 1 & 0 \\
    0 & 0 & 0 & 0 & 1 & 1 & 0 & 1 & 1
\end{bmatrix}
\end{equation}
This matrix captures the full \gls{nsm} structure in its unfolded (\gls{1d}) form, demonstrating how the originally \gls{2d} construction admits a conventional \gls{1d} representation via a generating matrix. This reinforces the conclusion that block multidimensional \glspl{nsm} are fundamentally indistinguishable from \gls{1d} \glspl{nsm} when expressed through their generating matrices.

\begin{figure}[!htbp]
    \centering
    \includegraphics[width=0.8\textwidth]{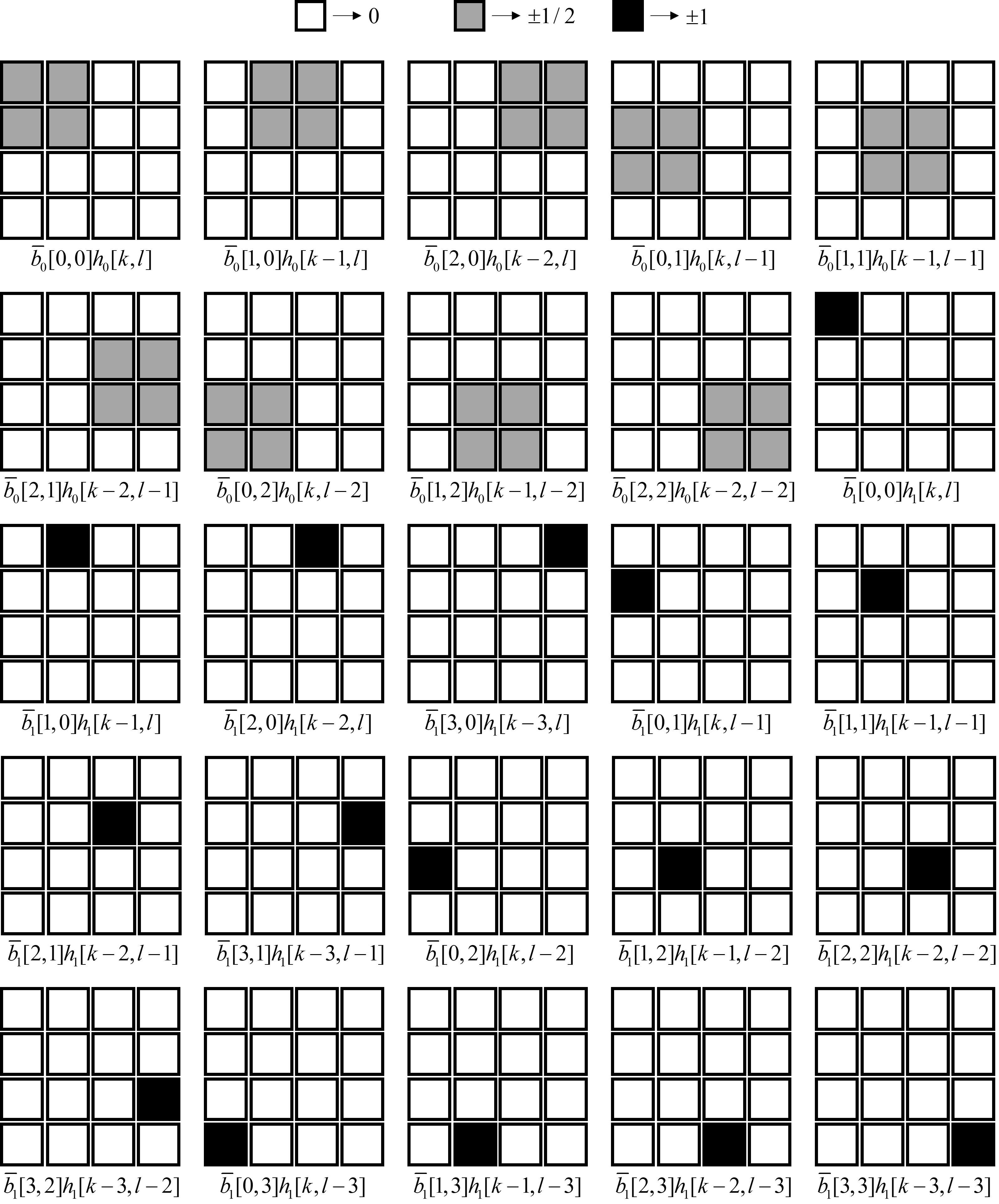}
    \caption{Illustration in two dimensions of the two-dimensional rate-$25/16$ block NSM associated with a $4 \times 4$ grid}
    \label{fig:2D-NSM-Illustration-Rate-25-16}
\end{figure}

To further reinforce the idea that \gls{2d} \glspl{nsm}, once unfolded, yield generating matrices structurally indistinguishable from those of \gls{1d} \glspl{nsm}, we reconsider here an additional \gls{2d} scheme. This second example, previously introduced in Section~\ref{ssec:Two-Dimensional Rate-2 NSMs}, is based on a $4 \times 4$ grid. Although its basic properties were briefly described earlier, neither a visual representation nor its generating matrix were provided. We now address this by illustrating the scheme in Figure~\ref{fig:2D-NSM-Illustration-Rate-25-16}, showing how the $25$ bipolar input symbols are arranged across the $4 \times 4$ \gls{2d} grid.

This \gls{nsm} maps $25$ input symbols into $16$ modulated outputs, resulting in a modulation rate of $\rho = 25/16$. As in the previous example, unfolding the \gls{2d} structure column-wise allows us to describe the modulation process using a generating matrix.As in the rate-$13/9$ case, we consider the scaled version of the generating matrix, whose explicit form is:
\begin{equation}
\mathring{\bm{G}} = \left[ {\scriptstyle \begin{array}{cccccccccccccccc}
    2 & 0 & 0 & 0 & 0 & 0 & 0 & 0 & 0 & 0 & 0 & 0 & 0 & 0 & 0 & 0 \\
    0 & 2 & 0 & 0 & 0 & 0 & 0 & 0 & 0 & 0 & 0 & 0 & 0 & 0 & 0 & 0 \\
    0 & 0 & 2 & 0 & 0 & 0 & 0 & 0 & 0 & 0 & 0 & 0 & 0 & 0 & 0 & 0 \\
    \vdots & \vdots & \vdots & \vdots & \vdots & \vdots & \vdots & \vdots & \vdots & \vdots & \vdots & \vdots & \vdots & \vdots & \vdots & \vdots \\
    0 & 0 & 0 & 0 & 0 & 0 & 0 & 0 & 0 & 0 & 0 & 0 & 0 & 2 & 0 & 0 \\
    0 & 0 & 0 & 0 & 0 & 0 & 0 & 0 & 0 & 0 & 0 & 0 & 0 & 0 & 2 & 0 \\
    0 & 0 & 0 & 0 & 0 & 0 & 0 & 0 & 0 & 0 & 0 & 0 & 0 & 0 & 0 & 2 \\
    1 & 1 & 0 & 0 & 1 & 1 & 0 & 0 & 0 & 0 & 0 & 0 & 0 & 0 & 0 & 0 \\
    0 & 1 & 1 & 0 & 0 & 1 & 1 & 0 & 0 & 0 & 0 & 0 & 0 & 0 & 0 & 0 \\
    0 & 0 & 1 & 1 & 0 & 0 & 1 & 1 & 0 & 0 & 0 & 0 & 0 & 0 & 0 & 0 \\
    0 & 0 & 0 & 0 & 1 & 1 & 0 & 0 & 1 & 1 & 0 & 0 & 0 & 0 & 0 & 0 \\
    0 & 0 & 0 & 0 & 0 & 1 & 1 & 0 & 0 & 1 & 1 & 0 & 0 & 0 & 0 & 0 \\
    0 & 0 & 0 & 0 & 0 & 0 & 0 & 1 & 1 & 0 & 0 & 1 & 1 & 0 & 0 & 0 \\
    0 & 0 & 0 & 0 & 0 & 0 & 0 & 0 & 1 & 1 & 0 & 0 & 1 & 1 & 0 & 0 \\
    0 & 0 & 0 & 0 & 0 & 0 & 0 & 0 & 0 & 1 & 1 & 0 & 0 & 1 & 1 & 0 \\
    0 & 0 & 0 & 0 & 0 & 0 & 0 & 0 & 0 & 0 & 1 & 1 & 0 & 0 & 1 & 1 
\end{array} } \right]
\end{equation}
Once unfolded, this matrix exhibits no visible structure that would distinguish it from a generating matrix associated with a conventional \gls{1d} \gls{nsm}. This second case thus complements the earlier example based on a $3 \times 3$ grid and strengthens the argument that the dimensionality of the modulation design plays no essential role once the underlying block structure is expressed in matrix form.

To complete this comparative illustration, we now revisit a genuinely \gls{1d} \gls{nsm} previously introduced and thoroughly analyzed in Section~\ref{sssec:One-dimensional designed NSMs}. This \gls{nsm} belongs to the class of rate-$2$ schemes constructed using rational-valued filter taps, where the second filter has length $L_1 = 1$ and the first filter length $L_0$ ranges from $5$ to $15.$ Among these, we focus on the configuration with $L_0 = 8$, as it will be used here in a finite-block modulation setting that matches the rate of the previously discussed \gls{2d} \gls{nsm} based on the $4 \times 4$ grid.

Although this \gls{nsm} is originally defined in streaming (infinite-length) form with an asymptotic modulation rate of $\rho = 2,$ we deliberately constrain its operation here to a finite number of input symbols. Specifically, we consider a block of $25$ bipolar input symbols, of which $9$ are filtered through the first filter $h_0[k]$ of length $L_0 = 8,$ and the remaining $16$ through the second filter $h_1[k]$ of length $L_1 = 1.$ This setting yields a total of $16$ modulated symbols, and hence an effective modulation rate of $\rho = 25/16$—identical to that of the \gls{2d} \gls{nsm} previously illustrated in Figure~\ref{fig:2D-NSM-Illustration-Rate-25-16}.

Among the top-performing filters identified for this case, we adopt the one defined by the scaled tap vector $\mathring{\bm{h}}_0 = (1,\ 0,\ 1,\ 0,\ 1,\ 0,\ 0,\ 1)$, listed in Table~\ref{table:Simple Filters Coefficients L_0=8 L_1=1}. The corresponding generating matrix, computed under this finite-block framework and presented below in its scaled form for consistency, shows no distinguishable structural trait when compared to the generating matrices of the earlier \gls{2d} \glspl{nsm}:
\begin{equation}
\mathring{\bm{G}} = \left[ {\small \begin{array}{cccccccccccccccc}
    2 & 0 & 0 & 0 & 0 & 0 & 0 & 0 & 0 & 0 & 0 & 0 & 0 & 0 & 0 & 0 \\
    0 & 2 & 0 & 0 & 0 & 0 & 0 & 0 & 0 & 0 & 0 & 0 & 0 & 0 & 0 & 0 \\
    0 & 0 & 2 & 0 & 0 & 0 & 0 & 0 & 0 & 0 & 0 & 0 & 0 & 0 & 0 & 0 \\
    \vdots & \vdots & \vdots & \vdots & \vdots & \vdots & \vdots & \vdots & \vdots & \vdots & \vdots & \vdots & \vdots & \vdots & \vdots & \vdots \\
    0 & 0 & 0 & 0 & 0 & 0 & 0 & 0 & 0 & 0 & 0 & 0 & 0 & 2 & 0 & 0 \\
    0 & 0 & 0 & 0 & 0 & 0 & 0 & 0 & 0 & 0 & 0 & 0 & 0 & 0 & 2 & 0 \\
    0 & 0 & 0 & 0 & 0 & 0 & 0 & 0 & 0 & 0 & 0 & 0 & 0 & 0 & 0 & 2 \\
    1 & 0 & 1 & 0 & 1 & 0 & 0 & 1 & 0 & 0 & 0 & 0 & 0 & 0 & 0 & 0 \\
    0 & 1 & 0 & 1 & 0 & 1 & 0 & 0 & 1 & 0 & 0 & 0 & 0 & 0 & 0 & 0 \\
    0 & 0 & 1 & 0 & 1 & 0 & 1 & 0 & 0 & 1 & 0 & 0 & 0 & 0 & 0 & 0 \\
    0 & 0 & 0 & 1 & 0 & 1 & 0 & 1 & 0 & 0 & 1 & 0 & 0 & 0 & 0 & 0 \\
    0 & 0 & 0 & 0 & 1 & 0 & 1 & 0 & 1 & 0 & 0 & 1 & 0 & 0 & 0 & 0 \\
    0 & 0 & 0 & 0 & 0 & 1 & 0 & 1 & 0 & 1 & 0 & 0 & 1 & 0 & 0 & 0 \\
    0 & 0 & 0 & 0 & 0 & 0 & 1 & 0 & 1 & 0 & 1 & 0 & 0 & 1 & 0 & 0 \\
    0 & 0 & 0 & 0 & 0 & 0 & 0 & 1 & 0 & 1 & 0 & 1 & 0 & 0 & 1 & 0 \\
    0 & 0 & 0 & 0 & 0 & 0 & 0 & 0 & 1 & 0 & 1 & 0 & 1 & 0 & 0 & 1 
\end{array} } \right]
\end{equation}
This final comparison reinforces the central point of this section: when viewed through the lens of their unfolded generating matrices, \gls{2d} and \gls{1d} \glspl{nsm} exhibit no essential distinction—each can be described as a conventional linear mapping defined by a generating matrix. However, the initial \gls{2d} design retains significant practical value beyond the conception phase. In particular, the \gls{2d} grid reveals the local footprint of each information symbol and its pattern of interference with neighboring symbols. This structural insight becomes especially beneficial in iterative detection processes, where local interactions play a crucial role. It is even more critical when joint detection and decoding are performed, as in turbo-detection or turbo-equalization settings with forward error correction. In such contexts, the \gls{2d} layout facilitates an intuitive and effective exploitation of symbol dependencies, enhancing both detection accuracy and convergence behavior.

To enhance spectral efficiency, we revisit the \gls{2d} \gls{nsm} based on the $3 \times 3$ grid, previously introduced in Section~\ref{ssec:Two-Dimensional Rate-2 NSMs}. In its original configuration, boundary padding prevented the full utilization of the grid. Specifically, only $4$ input symbols were allowed to be modulated via filter $h_0[k,l],$ whose $2 \times 2$ footprint (one unit of memory in both horizontal and vertical directions) could not wrap around the grid borders. An additional $9$ input symbols were modulated via the memoryless filter $h_1[k,l],$ leading to a total of $13$ bipolar input symbols and $9$ output symbols, hence a rate of $\rho = 13/9$.

\begin{figure}[!htbp]
    \centering
    \includegraphics[width=0.8\textwidth]{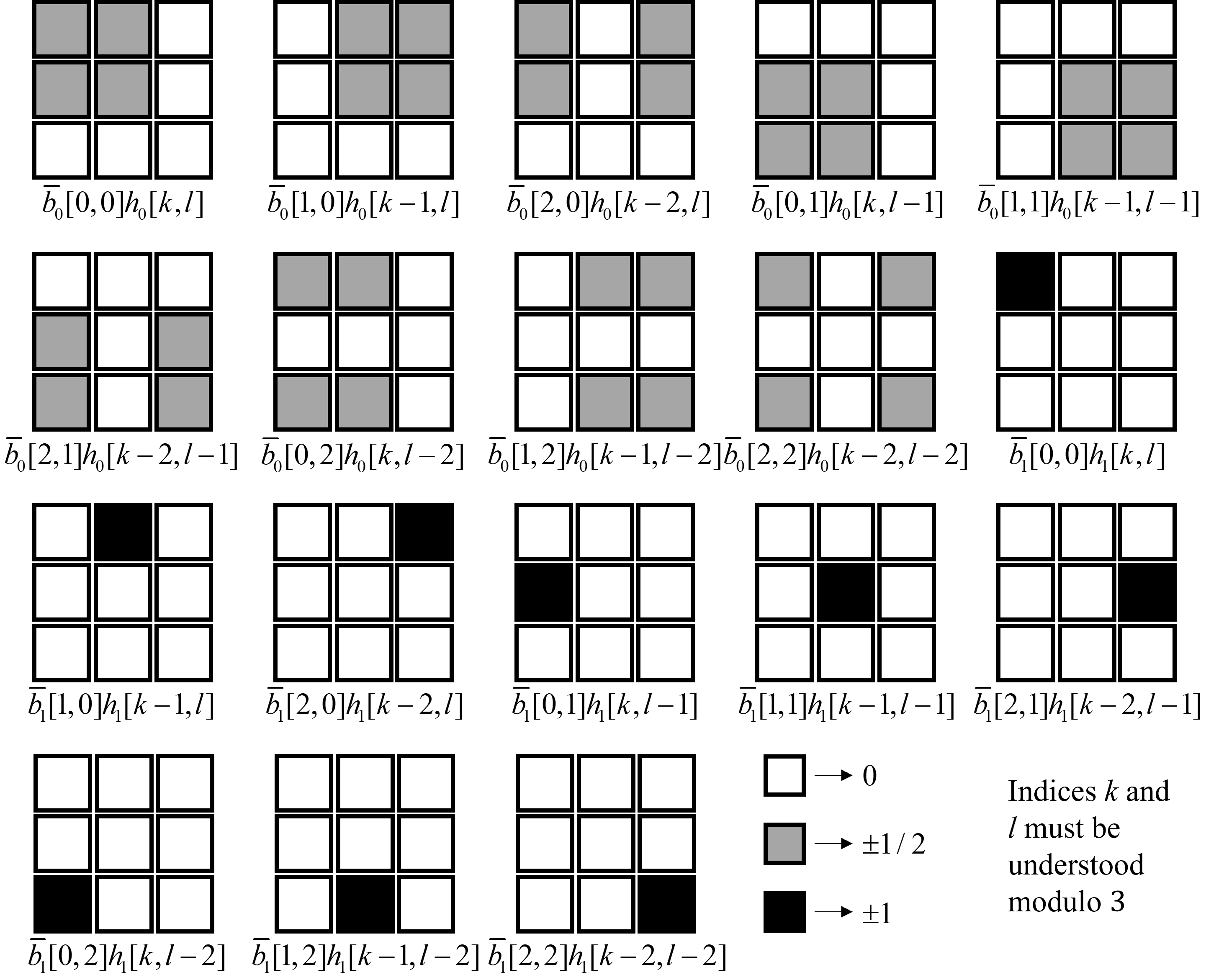}
    \caption{Illustration of the two-dimensional block NSM based on a $3 \times 3$ grid with two-dimensional tail-biting}
    \label{fig:2D-NSM-Illustration-Rate-13-9 Tail-Biting}
\end{figure}

We now extend the concept of tail-biting, previously analyzed in detail for \gls{1d} \glspl{nsm} in Section~\ref{sec:One-dimensional NSMs and tail-biting}, to the \gls{2d} case. This extension, which we refer to as \emph{two-dimensional tail-biting}, applies the principle of \emph{cyclic wrapping} over both spatial dimensions. The $2 \times 2$ footprint of the filter $h_0[k,l]$ is allowed to wrap cyclically across the horizontal and vertical edges of the $3 \times 3$ grid. This eliminates the need for padding and ensures that each filter application maintains full support, regardless of its location. The resulting structure is illustrated in Figure~\ref{fig:2D-NSM-Illustration-Rate-13-9 Tail-Biting}, where the input symbol footprints wrap seamlessly across the grid boundaries in both dimensions.

Thanks to the cyclic wrapping of its $2 \times 2$ footprint over the edges of the $3 \times 3$ grid in both dimensions, each of the $9$ grid positions can now be used to transmit a symbol through the filter $h_0[k,l]$, preserving the full footprint even at the boundaries. Consequently, $9$ bipolar input symbols are modulated via $h_0[k,l]$, each producing a $2 \times 2$ output footprint that wraps cyclically around the $3 \times 3$ grid boundaries, as illustrated in Figure~\ref{fig:2D-NSM-Illustration-Rate-13-9 Tail-Biting}. The number of symbols modulated by the memoryless filter $h_1[k,l]$ remains $9$, as in the case of the non-tail-bitten \gls{2d} rate $13/9$ \gls{nsm} examined above. Thus, the total number of bipolar input symbols increases to $18,$ while the output remains $9,$ yielding a rate of $\rho = \tfrac{18}{9} = 2.$ Notice that, without tail-biting, this rate of $2$ can only be achieved asymptotically by letting the grid size grow unbounded in both dimensions, thereby eliminating boundary overhead.

This demonstrates a key advantage of \gls{2d} tail-biting: it recovers the full spectral efficiency that would otherwise only be achievable asymptotically as the grid size tends to infinity. In the absence of tail-biting, padding losses due to filter memory become increasingly pronounced in higher-dimensional \glspl{nsm}. This is because the proportion of boundary-affected symbols increases with dimensionality due to the decreasing volume-to-surface ratio. Tail-biting completely eliminates this inefficiency by cyclically wrapping the modulation structure, making it especially beneficial in two or more dimensions.

To explicitly represent the \gls{2d} \gls{nsm} using tail-biting over the $3 \times 3$ grid, we now provide the corresponding generator matrix. This matrix reflects the structure described above, where the $9$ input symbols modulated via $h_0[k,l]$ contribute to the $3 \times 3$ output grid through cyclic wrapping of their $2 \times 2$ footprints, while the $9$ symbols modulated via the memoryless filter $h_1[k,l]$ remain unaffected by the wrapping mechanism. For clarity, the generator matrix is scaled and given by:
\begin{equation}
\mathring{\bm{G}} = \begin{bmatrix}
    2 & 0 & 0 & 0 & 0 & 0 & 0 & 0 & 0 \\
    0 & 2 & 0 & 0 & 0 & 0 & 0 & 0 & 0 \\
    0 & 0 & 2 & 0 & 0 & 0 & 0 & 0 & 0 \\
    0 & 0 & 0 & 2 & 0 & 0 & 0 & 0 & 0 \\
    0 & 0 & 0 & 0 & 2 & 0 & 0 & 0 & 0 \\
    0 & 0 & 0 & 0 & 0 & 2 & 0 & 0 & 0 \\
    0 & 0 & 0 & 0 & 0 & 0 & 2 & 0 & 0 \\
    0 & 0 & 0 & 0 & 0 & 0 & 0 & 2 & 0 \\
    0 & 0 & 0 & 0 & 0 & 0 & 0 & 0 & 2 \\
    1 & 1 & 0 & 1 & 1 & 0 & 0 & 0 & 0 \\
    0 & 1 & 1 & 0 & 1 & 1 & 0 & 0 & 0 \\
    1 & 0 & 1 & 1 & 0 & 1 & 0 & 0 & 0 \\
    0 & 0 & 0 & 1 & 1 & 0 & 1 & 1 & 0 \\
    0 & 0 & 0 & 0 & 1 & 1 & 0 & 1 & 1 \\
    0 & 0 & 0 & 1 & 0 & 1 & 1 & 0 & 1 \\
    1 & 1 & 0 & 0 & 0 & 0 & 1 & 1 & 0 \\
    0 & 1 & 1 & 0 & 0 & 0 & 0 & 1 & 1 \\
    1 & 0 & 1 & 0 & 0 & 0 & 1 & 0 & 1
\end{bmatrix}
\end{equation}

The use of \gls{2d} tail-biting comes with a fundamental trade-off: the modulation process is no longer injective. That is, multiple distinct bipolar input vectors may yield the same modulated output, introducing ambiguity at the demodulation stage. This phenomenon is in concordance with the injectivity issue already raised in the \gls{1d} case (see Section~\ref{sec:One-dimensional NSMs and tail-biting}), where it was shown that the probability of such ambiguities decreases exponentially with the modulated block size.

While a rigorous analysis of the injectivity loss in the \gls{2d} case is beyond the scope of this section, the structural parallels with the \gls{1d} setting strongly suggest that similar asymptotic behavior holds. As the grid size increases, the likelihood of distinct bipolar input sequences mapping to the same output becomes negligible. In practice, for moderate-sized blocks, such ambiguities occur very rarely. Therefore, \gls{2d} tail-biting represents a deliberate trade-off: it improves spectral efficiency while introducing only minimal and well-controlled ambiguity.

In conclusion, the exploration of \gls{2d} tail-biting highlights several important insights. First, tail-biting is a powerful method for mitigating inefficiencies introduced by filter memory near the boundaries of finite grids and this benefit becomes increasingly significant in higher dimensions. Second, while \gls{2d} \glspl{nsm} are naturally designed and interpreted in spatial layouts, the underlying modulation process is always expressible through a \gls{1d} generator matrix and thus the dimensionality primarily aids design intuition and optimization. Third, the \gls{2d} structure, particularly when enhanced by cyclic wrapping, provides valuable information about symbol interactions, which can be exploited in iterative detection schemes, especially when combined with error-correction coding in turbo-like architectures. Finally, by enabling full utilization of compact grids, \gls{2d} tail-biting allows practical implementations to achieve the asymptotic modulation rates exactly, making it a valuable tool for the construction of efficient, high-rate \glspl{nsm}.

\subsection{Analog Packing and Digital Packing: Complementarity Between NSMs and Error-Correcting Codes} \label{Analog packing and digital packing}

When considering standard lattice codes, discussions often revolve around sphere packing in the context of error correction and sphere covering in the context of source coding. These two geometric interpretations are dual in nature: enhancing the performance of error-correcting codes corresponds to maximizing the packing radius, while improving source coding corresponds to minimizing the covering radius. Theoretical foundations for optimal packing and covering in high dimensions are comprehensively addressed in the work of Conway and Sloane~\cite{Conway99}.

A typical and powerful approach in constructing lattice codes optimized for the Euclidean distance involves a binary partition chain applied to the lattice $\mathbb{Z}^2$: $\mathbb{Z}^2/\bm{R}\mathbb{Z}^2/\bm{R}^2\mathbb{Z}^2/\bm{R}^3\mathbb{Z}^2/\ldots,$ where $\bm{R}$ is a similitude on the plane that combines a rotation by $\pi/4$ with a scaling by $\sqrt{2}.$ At each level $l,$ the coarser lattice $\bm{R}^{l+1} \mathbb{Z}^2$ is a sublattice of the finer lattice $\bm{R}^l \mathbb{Z}^2,$ partitioning it into two cosets (equivalence classes modulo the sublattice). This binary partition chain is tightly matched to the Euclidean metric: within each coset, the \gls{msed} ($\mathcal{L}_2$ norm) between points doubles from one level to the next. This exponential growth in \gls{sed} is a key geometric feature that supports code constructions aiming to maximize the minimum Euclidean distance under power constraints, which is a fundamental objective in the design of codes for Gaussian noise channels.

This sequence of nested partitions underlies the squaring and cubing constructions introduced by G.D. Forney Jr.~\cite{Forney88-1,Forney88-2}. These constructions combine binary linear codes with the partition chain to form dense and highly structured lattices. For instance, the squaring construction applied to $\mathbb{Z}^2$ yields the \gls{4d} lattice $D_4$ (mentioned in Section~\ref{Comparison of rate-5/4 block NSM and the densest four-dimensional lattice D4}) and can be recursively extended to generate the exceptional eight-dimensional lattice $E_8.$ The Leech lattice $\Lambda_{24},$ celebrated for its remarkable packing and symmetry properties, is obtained using a combination of the squaring and cubing constructions.

For completeness, it is helpful to consider a related hierarchy in the \gls{1d} setting: $\mathbb{Z}/2\mathbb{Z}/2^2\mathbb{Z}/2^3\mathbb{Z}\ldots.$ This binary partition chain exhibits a geometric property analogous to that of the \gls{2d} chain $\mathbb{Z}^2/\bm{R}\mathbb{Z}^2/$ $\bm{R}^2\mathbb{Z}^2/\bm{R}^3\mathbb{Z}^2/\ldots$:the minimum $\mathcal{L}_1$ distance between elements doubles when moving from one level to the next, just as the \gls{sed} doubles at each step in the \gls{2d} hierarchy. In this sense, the \gls{1d} hierarchy is naturally suited to the $\mathcal{L}_1$ norm, just as the \gls{2d} hierarchy is matched to the squared Euclidean ($\mathcal{L}_2$) norm. Siala and Kaleh~\cite{Siala&Kaleh95} have explored this correspondence in their adaptation of the squaring construction to the $\mathcal{L}_1$ setting, applying the same hierarchical coding principles developed for the Euclidean context. When the partition chain is truncated at level $2^m\mathbb{Z},$ the resulting structure corresponds to codes over $\mathbb{Z}_{2^m} \triangleq \mathbb{Z}/2^m\mathbb{Z},$ where the $\mathcal{L}_1$ geometry induces the Lee metric. These lattices therefore provide a natural foundation for designing codes well suited to the Lee metric, especially in digital systems with modular or asymmetric error characteristics.

Beyond their geometric appeal, lattice codes can be understood as a synthesis of digital and analog components, each playing a complementary role. The digital component is embodied in the binary block codes assigned to each level of the hierarchical partitioning, while the analog component is provided by the underlying lattice $\mathbb{Z}^2$ and its nested two-way partitioning. Together, they form a tightly integrated scheme in which digital information is mapped onto analog geometry in a structured and efficient way.

The binary block codes serve to select coset representatives at each level of the partition chain. However, not all levels require the same degree of coding protection. At finer partition levels (i.e., deeper in the hierarchy), where the Euclidean distance between cosets is small, the binary codes must be strong and robust to guard against likely decoding errors. At coarser levels, where cosets are already well separated, weaker codes suffice, and in some cases, no coding is needed at all. As a result, coding strength is distributed unequally across the hierarchy, decreasing as one moves up toward coarser partitions.

This unequal distribution of coding strength aligns naturally with the algebraic structure of the lattice. Since lattices are additive groups closed under integer combinations, the binary block code used at a finer level is a subcode of the one used at the coarser level above. This nesting ensures consistency across levels and reflects the increasing need for error protection as the partitions become finer and the Euclidean distances between cosets decrease. The resulting design is layered: strong binary codes protect densely packed points at finer levels, while weaker codes, or even no coding at all, are sufficient where the lattice geometry already provides large separations.

This layered organization results in an inherently unequal error correction strategy. The binary block codes assigned to finer partition levels are designed to correct small-amplitude errors, since the Euclidean distance between cosets is minimal at those levels. As one moves upward in the hierarchy, the distance between cosets increases, and the need for error correction decreases accordingly. This asymmetry in digital protection is directly matched by an analog asymmetry: finer levels offer closely spaced cosets requiring strong protection, while coarser levels offer well-separated cosets that can be reliably distinguished even without coding. The two asymmetries are thus complementary: where the analog domain is vulnerable, the digital domain compensates; where the geometry itself is robust, the digital structure can be minimal or absent.

Viewed from a broader perspective, the overall lattice code benefits from a \emph{dual form of packing}, combining a digital component, provided by nested binary codes, and an analog component, realized through the geometric structure of the lattice. On the digital side, the binary codes select coset representatives in a way that maximizes separation in \emph{Hamming space}, effectively providing \emph{digital packing}. On the analog side, the hierarchical partitioning of $\mathbb{Z}^2$ introduces \emph{geometric packing} in \emph{Euclidean space}, where lattice points are placed to maximize distance under energy constraints. These two packing mechanisms operate in \emph{tandem}: digital packing combats bit-level noise, while analog packing provides resilience in amplitude and phase. The resulting design is a tight fusion of discrete and continuous techniques, tailored to efficiently manage both bit errors and analog distortions.

An important feature of this hierarchical design is the \emph{unequal energy cost} associated with each binary element, depending on its position in the partition chain. A binary decision made at a coarser level determines large-scale structural shifts in the signal and typically results in a higher variation in amplitude, leading to a greater contribution to the average symbol energy. In contrast, binary elements associated with finer levels induce smaller perturbations in amplitude and therefore have minimal impact on energy. As a result, not all bits contribute equally to the power budget: bits at higher levels carry more energy weight, while those at deeper levels are energetically lightweight.

This phenomenon closely parallels what happens in non-Gray-coded $2^m$-ASK. In such modulations, the $m$ bits specifying a symbol are interpreted in their bipolar form and weighted by increasing powers of two. As a result, coarser bits induce large amplitude swings and dominate the symbol's energy, while finer bits only introduce small fluctuations. The hierarchical bit-to-energy relationship seen in lattice codes mirrors this structure, offering a clear physical intuition for how bit position translates into energy expenditure.

This reasoning, rooted in hierarchical partitioning and unequal error protection, extends naturally beyond block-coded lattice constructions. In particular, Forney and Eyuboglu demonstrated how similar principles can be applied within a trellis-oriented framework using convolutional codes~\cite{Forney91}. Their work on \gls{tcm} exploits multilevel lattice partitions to assign unequal levels of convolutional error protection across the hierarchy, resulting in powerful coding schemes that combine high spectral efficiency with strong error resilience.

While lattice codes offer a compelling combination of analog structure and digital protection, they also present notable practical challenges. One major difficulty lies in labeling: for reliable communication, it is highly desirable that nearby lattice points—those most likely to be confused under noise—be assigned binary labels that differ by as few bits as possible, ideally only one, in the spirit of Gray coding. However, due to the multidimensional and often irregular nature of high-dimensional lattices, constructing such Gray-like mappings is far from straightforward and remains an open design problem in many settings.

A second challenge arises from shaping: to approach the power efficiency limits of the Gaussian channel, lattice codes are often constrained to lie within a hypersphere, a technique known as sphere shaping. This can yield up to 1.53 dB in shaping gain in high dimensions, but enumerating lattice points within a hypersphere, indexing them, and ensuring uniformly distributed mappings for encoding purposes is computationally intensive. Combining this with Gray-like labeling further compounds the complexity, making practical implementations of lattice codes significantly more demanding than their theoretical appeal might suggest.

While lattice codes offer a powerful framework for combining digital and analog communication principles, their practical limitations call for alternative approaches that better align with implementation and performance constraints. An appealing alternative should first and foremost avoid the hierarchical structure of unequal error protection inherent to lattice codes. Instead of relying on multiple layers of binary codes with varying strengths, each associated with a different level of partitioning, the alternative should implement a single, uniformly powerful error-correcting code. This ensures that all coded bits contribute equally to reliability, simplifying the coding architecture and eliminating the asymmetry in protection that characterizes lattice-based schemes.

Closely tied to this uniformity in error correction is the need for all coded bits to have the same impact on the modulated signal. In lattice codes, bits associated with finer partition levels cause only small changes in the resulting waveform, while bits from coarser levels induce larger variations in amplitude. This creates an imbalance in how bits affect the transmitted signal, which complicates both system design and performance analysis. A well-designed alternative should avoid this uneven structure by ensuring that each bit influences the modulated sequence in the same way, with no bit exerting significantly more or less control over the output than others. Furthermore, each bit should carry an equal share of the average symbol energy. The unequal energy allocation found in lattice codes—where some bits dominate the total energy budget while others contribute far less—results in energy inefficiency and places unnecessary constraints on code design. Ensuring that all bits cost the same in energy simplifies the modulation structure and supports a more balanced use of the available power resources.

Another desirable property for a robust alternative is a form of Gray-like protection: the modulated sequences that are closest in Euclidean distance should correspond to binary sequences that differ in only a small number of bits. This ensures that symbol-level errors caused by small noise perturbations are likely to produce only minor bit-level errors, improving the overall reliability of the system. This type of bit-symbol locality is difficult to achieve in high-dimensional lattice codes due to their complex geometric structure, but it remains a valuable goal in any alternative scheme.

Finally, an effective alternative to lattice codes should incorporate some form of shaping to approach power efficiency limits, but without the complexity of explicit sphere shaping. Ideally, the modulated sequence should exhibit an implicit shaping behavior in high-dimensional space—meaning that, over long blocks, it statistically resembles a Gaussian process whose realizations tend to lie within a hypersphere. This type of indirect sphere shaping provides the same energy-concentration benefits as explicit shaping, enabling the signal to approach the Gaussian channel capacity more closely. Crucially, this should be achieved without requiring complex enumeration or indexing of lattice points within a sphere, thereby simplifying implementation while still harnessing the advantages of high-dimensional energy compaction.

Interestingly, nearly all the desirable properties identified—uniform error correction strength, equal bit influence on the modulated signal, balanced energy per bit, and perfect alignment between Hamming and Euclidean distances—are naturally satisfied when using the simple bipolar modulation $2$-ASK as the analog layer of the coded modulation scheme. With its binary-to-bipolar mapping, $2$-ASK ensures that each bit corresponds to a fixed-amplitude symbol with identical energy and influence, making the system symmetric and well-balanced. Furthermore, since each symbol directly encodes a single bit, the nearest Euclidean neighbors differ by exactly one bit, thereby enforcing a trivial but perfect Gray-like mapping. However, all $2$-ASK sequences are composed of symbols drawn uniformly from the set $\{\pm 1\},$ resulting in constant-amplitude signals with no shaping behavior. As such, $2$-ASK—like lattice codes deprived of explicit sphere shaping—fails to achieve any shaping gain, leaving the energy compaction advantages of shaping unexploited.

More importantly, $2$-ASK suffers from a fundamental practical limitation: its spectral efficiency is capped at just one bit per symbol. While it elegantly avoids the structural and energetic imbalances of lattice codes, this low efficiency makes it unsuitable for modern communication systems that demand high throughput. In realistic settings, modulations must support higher spectral efficiencies—well beyond $1$ bit per symbol—while still preserving the desirable structural properties we have outlined: uniform error protection, equal energy distribution per bit, Gray-like behavior, and ideally, high-dimensional shaping gains. This naturally leads to a central question: Is there a way to design a modulation scheme that retains the structural advantages of $2$-ASK, yet offers arbitrary spectral efficiency and introduces a natural, inherent shaping gain—something that lattice codes deprived of sphere shaping do not provide?

Having highlighted the structural limitations of lattice codes and their inability to simultaneously satisfy all desirable features of coded modulation—particularly equal bit influence, energy fairness, Gray-mapping compatibility, and sphere shaping—we now turn our attention to the potential of \glspl{nsm}. We aim to demonstrate, step by step, how \glspl{nsm} can asymptotically fulfill all the requirements previously identified as critical, while also allowing any desirable spectral efficiency. In contrast to $2$-ASK, which naturally resolves many of the limitations of lattice codes but suffers from low spectral efficiency, \glspl{nsm} retain these desirable properties while supporting arbitrarily high spectral efficiencies.

A first essential property of \glspl{nsm}, which directly addresses one of the central limitations of lattice codes, is their ability to naturally induce shaping in high-dimensional spaces. As shown and analyzed in detail in Section~\ref{Link with information theory}, when the filter lengths used in the construction of \glspl{nsm} are sufficiently large, the resulting modulated sequences tend to exhibit a Gaussian-like energy distribution. This behavior is asymptotic and stems from the high-dimensional interaction of independent bipolar input streams through spreading filters. The superposition of many such filtered sequences, each individually normalized, causes the overall signal to become statistically concentrated within a hypersphere—thereby mimicking the sphere shaping effect that must otherwise be explicitly engineered in lattice coding frameworks. This shaping behavior is not only intrinsic to the \gls{nsm} architecture, but also improves energy efficiency and enables the modulated sequence to better approximate capacity-achieving Gaussian input distributions for \gls{awgn} channels. In contrast, pure lattice codes (deprived from explicit sphere shaping layers) fail to provide such a distributional behavior, and the resulting modulated points spread uniformly across an infinite grid, far from the desired sphere-like concentration.

Another key property of \glspl{nsm} is their ability to guarantee equal energy contribution from each coded bit to the overall modulated sequence. This stands in contrast to hierarchical coding schemes such as those based on lattice codes, where bits associated with finer partitions tend to have a much smaller impact on the transmitted signal energy compared to bits at coarser levels. In the \gls{nsm} setting, the structure of the modulated sequence has already been introduced and justified in (\ref{eq:Modulated Symbols Rate k/n NSM}), which we recall here for completeness:
\begin{equation}
s[q] = \sum_{l=0}^{k-1} \sum_p \bar{b}_l[p] h_l[q-np],
\end{equation}
where the $\bar{b}_l[p]$ represent $k$ bipolar streams derived from the coded binary sequence, and $h_l[q]$ are corresponding filters of finite length $L_l.$ Each stream $\bar{b}_l[p]$ is modulated by its associated filter $h_l[q],$ and the contribution of each stream to the average symbol energy is given by the squared norm $\|h_l[q]\|^2.$ Therefore, if all filters are normalized to have equal energy, then each coded bit contributes equally to the average energy of the transmitted modulated sequence. This uniform energy distribution is particularly desirable, as it eliminates the imbalance found in multilevel modulation schemes—such as lattice coding—where some bits may dominate the energy profile while others contribute marginally.

Beyond energy fairness, \glspl{nsm} also exhibit a powerful Gray-like mapping property, which ensures that coded binary sequences differing by only a few bits correspond to modulated sequences that are close in Euclidean distance. To see this, consider two coded binary sequences that differ in a small number of bits. After binary-to-bipolar transformation, these sequences generate two sets of bipolar streams, denoted $\{ \bar{b}_l^m[p] \}$ $m=0,1,$ and $l=0,\ldots,k-1$, where each stream is filtered by the corresponding impulse response $h_l[q].$ As previously recalled, the modulated sequences $s^m[q]$ are then given by:
\begin{equation}
s^m[q] = \sum_{l=0}^{k-1} \sum_p \bar{b}_l^m[p] h_l[q - np].
\end{equation}
The difference between the two modulated sequences, $\Delta s[q] = s^1[q] - s^0[q],$ becomes:
\begin{equation}
\Delta s[q] = \sum_{l=0}^{k-1} \sum_p \Delta \bar{b}_l[p] h_l[q - np],
\end{equation}
where $\Delta \bar{b}_l[p] = \bar{b}_l^1[p] - \bar{b}_l^0[p]$ takes values in $\{0, \pm 2\}.$ Introducing the scaled modulated sequences difference, $\Delta \mathring{s}[q] = \Delta s[q]/2,$ and scaled input sequences differences, $\Delta \mathring{b}_l[p] = \Delta \bar{b}_l[p]/2,$ we obtain:
\begin{equation}
\Delta \mathring{s}[q] = \sum_{l=0}^{k-1} \sum_p \Delta \mathring{b}_l[p] h_l[q - np],
\end{equation}
where the scaled differences $\Delta \mathring{b}_l[p]$ take values in $\{0, \pm 1\}$ and are non-zero only at positions where the original sequences differ. Consequently, $\Delta \mathring{s}[q]$ is a linear combination of a small number of filter responses and their time shifts. If the filter lengths $L_l$ are sufficiently large, these shifted filter responses behave as quasi-orthogonal components.This high-dimensional quasi-orthogonality—well-documented in random matrix theory and asymptotic free probability~\cite{Tao12,Vershynin18,Mingo17}—ensures that the energy of $\Delta \mathring{s}[q]$ grows proportionally with the number of differing bits. In this regime, the \gls{sed} between the modulated sequences is approximately proportional to the Hamming distance between the corresponding binary sequences, establishing a direct and reliable Gray-like relationship.

Moreover, when focusing locally on sequences that differ in only a small number of positions, the corresponding shifted filter responses span a subspace in which the involved components behave as if they were mutually orthogonal. This local structure mimics the behavior of $2$-ASK modulation, where bipolar symbols are projected onto perfectly orthogonal dimensions. In the \gls{nsm} framework, although the full set of shifted filters does not occupy mutually orthogonal subspaces, the limited subset associated with sparse differences exhibits quasi-orthogonality that is sufficient to approximate the benefits of orthogonal signaling locally. This enables effective discrimination between sequences differing by a few bits, similar to what $2$-ASK achieves globally. The global non-orthogonality, on the other hand, reflects the increased spectral efficiency of \gls{nsm} relative to fully orthogonal schemes like $2$-ASK and represents the necessary trade-off to achieve higher rates. Thus, \glspl{nsm} manage to retain desirable local properties of orthogonal signaling, akin to $2$-ASK, while overcoming the spectral efficiency limitations inherent to fully orthogonal schemes.

A remaining concern arises when two coded binary sequences differ in a large number of bits. In this case, the modulated sequences difference $\Delta \mathring{s}[q]$ is a linear combination of many filters $h_l[q]$ and their time shifts. For fixed filter lengths $L_l, l=0,1,\ldots,k-1,$ as the number of combined filters and shifts increases with the Hamming distance $w$ between the coded sequences, the quasi-orthogonality property weakens and eventually is lost. This raises the critical question: could there be situations where two coded sequences differ in many bits, yet their scaled modulated sequence difference $\Delta \mathring{s}[q]$ exhibits a very small \gls{sen}? Such a scenario would represent a vulnerability, implying sensitivity to noise and a significant departure from the desired Gray-like mapping behavior.

To partially address this issue, first consider the \emph{bipolar representation} of the coded sequences. When two binary sequences differ in exactly $w$ bits, their corresponding bipolar sequences differ in exactly $w$ components, each taking values in the bipolar set ${-1, +1}.$ The difference sequence between the bipolar streams therefore contains exactly $w$ non-zero components in the ternary set ${0, \pm 2}.$ Importantly, in accordance with the discussion at the beginning of Appendix~\ref{app:Tight Estimate BEP Rate 5/4}, the probability of occurrence of each of these error events is governed by the number, $w,$ of non-null input sequence differences, and vanishes exponentially as $(\tfrac{1}{2})^w.$ Thus, although large Hamming distance binary pairs exist, the likelihood of encountering their corresponding bipolar difference sequences—which could potentially yield critical low-energy modulated sequence differences—rapidly vanishes as $w$ grows.

Second, even when two sequences differ in a large number $w$ of bits, one might worry that the resulting modulated difference $\Delta \mathring{s}[q]$ could yield a very small \gls{sen} due to destructive interference between the shifted filter responses. In other words, although $\Delta \mathring{s}[q]$ is a linear combination of $w$ time-shifted filters, it could, in principle, have much lower energy than expected, thus weakening the Gray-like property in such high-weight cases.

To address this, we consider a probabilistic model where the filters $h_l[q]$ are treated as random variables, while the bipolar difference pattern is fixed. Assuming all filters have the same energy $\eta = \| h_l[q] \|^2$, the expected \gls{sen} of the modulated difference is $\mathbb{E}[\| \Delta \mathring{s} \|^2] = w \eta.$ Moreover, the variance of this quantity is upper bounded by $(w^2-w) \eta^2 / L_{\min},$ where $L_{\min}$ denotes the minimum filter length. It follows that the standard deviation satisfies $\sigma \leq w \, \eta / \sqrt{L_{\min}},$ and the normalized standard deviation is bounded as $\sigma / \mathbb{E}[\| \Delta \mathring{s} \|^2] \leq 1 / \sqrt{L_{\min}}.$

Although in practice the filters are fixed, when they are sufficiently long and well-spread in high-dimensional space, their deterministic structure mimics the statistical behavior of random filters. This quasi-orthogonality ensures that $\| \Delta \mathring{s} \|^2$ remains highly concentrated around its expected value $w \eta.$ As a result, the probability that the modulated difference exhibits a small \gls{sen} decays rapidly with $L_{\min}.$

In summary, by adequately increasing the filter lengths, the risk of having very small squared Euclidean distances between modulated sequences corresponding to binary sequences differing in many bits can be made arbitrarily small. This preserves the Gray-like robustness of the \gls{nsm} approach, ensuring reliable performance even for large Hamming distance cases.

\subsection{Structural Link Between One-Dimensional NSMs and Ungerboeck’s TCM}

In his seminal 1982 paper, \cite{Ungerboeck82}, Ungerboeck examined the capacity of modulations with progressively larger constellation sizes in both \gls{1d} and \gls{2d} signal spaces. He showed that the most substantial coding gain—relative to uncoded modulation—is obtained by doubling the size of the constellation. Specifically, if the uncoded modulation uses a constellation of size $2^m$ (i.e., carrying $m$ bits per symbol), Ungerboeck proposed expanding the constellation to size $2^{m+1},$ which carries $m+1$ bits per symbol, and encoding the original $m$ information bits using a rate $m/(m+1)$ convolutional code. This efficient strategy was illustrated in Figure~3 of the paper, where capacity curves demonstrate that most of the coding gain can be captured with this simple expansion. A similar conclusion was drawn in the 1998 paper, \cite{Forney98}, by Forney and Ungerboeck, where Figure~2 shows that such constellation expansion—combined with trellis coding—approaches the Shannon capacity with relatively low complexity. Both works emphasize that moving from a $2^m$-ary to a $2^{m+1}$-ary modulation and applying a rate $m/(m+1)$ code captures the majority of potential coding gain.

An essential practical insight into the implementation of this modulation expansion strategy is provided in Figure~11 of Ungerboeck’s 1982 paper, \cite{Ungerboeck82}, where he illustrates equivalent realizations of $8$-PSK and $16$-QASK trellis-coded modulation schemes, using systematic convolutional encoders with feedback. In this architecture, the encoder outputs the original $m$ input bits directly—hence \emph{systematic}—while the additional bit, which enables the expansion from a $2^m$-ary to a $2^{m+1}$-ary constellation, is generated as a recursive function of the $m$ input bits and the encoder’s internal state. This feedback structure defines what are commonly known as \gls{rsc} codes. The recursive nature of these encoders enhances the distance properties of the resulting trellis paths, allowing more efficient use of redundancy for coding gain. Importantly, the systematic nature ensures that the core information bits are preserved and can be accessed directly, which simplifies certain decoder implementations and aligns naturally with multilevel coding frameworks. Ungerboeck's proposed structure thus provides not only theoretical efficiency but also practical realizability in hardware and digital signal processing systems.

This behavior closely mirrors the general structure we have identified in the optimized \glspl{nsm} considered so far. To illustrate this, we begin by examining the cases of rate-$2$ and rate-$3$ \glspl{nsm}. In the best-performing rate-$2$ \glspl{nsm}, the two input bipolar sequences, $\bar{b}_0[k]$ and $\bar{b}_1[k],$ play distinct roles: only $\bar{b}_0[k]$ is processed through a non-trivial filter $h_0[k]$ of length $L_0 > 1,$ while $\bar{b}_1[k]$ is simply scaled by a factor $\sqrt{\eta_1}$ via the trivial filter $h_1[k] = \sqrt{\eta_1} \, \delta[k].$ In effect, $\bar{b}_1[k]$ is incorporated into the modulated sequence without any structural modification, aside from a multiplicative scaling factor. In a similar fashion, the best-performing rate-$3$ \glspl{nsm} involve three input bipolar sequences, $\bar{b}_0[k],$ $\bar{b}_1[k]$ and $\bar{b}_2[k]$, where again only $\bar{b}_0[k]$ is filtered through a non-trivial filter $h_0[k]$ of length $L_0 > 1.$ The remaining sequences, $\bar{b}_1[k]$ and $\bar{b}_2[k],$ are scaled by factors $\sqrt{\eta_1}$ and $\sqrt{\eta_2}$, respectively, via the trivial filters $h_1[k] = \sqrt{\eta_1} \, \delta[k]$ and $h_2[k] = \sqrt{\eta_2} \, \delta[k]$, both of length $1.$ As a result, $\bar{b}_1[k]$ and $\bar{b}_2[k]$ are effectively incorporated into the modulated sequence, preserved in their original form, aside from their respective scalings.

Building on the previous analysis of rate-$2$ and rate-$3$ \glspl{nsm}, which provided insight into structural similarities with Ungerboeck’s \gls{tcm}, we now extend this line of reasoning to the remaining \glspl{nsm}, with rates $\rho = (Q+1)/Q,$ $1 < Q \le 4.$ Adopting the convolutional code-like perspective introduced in Section~\ref{sec:Analogy with convolutional error correction coding}, the global bipolar input sequence—after binary-to-bipolar conversion—is demultiplexed into $(Q+1)$ parallel streams, $\bar{b}_l[k],$ $l=0, 1, \ldots, Q$. The modulated sequence $s[k]$ is formed by multiplexing $Q$ parallel modulated output streams $s_m[k],$ $m=0, 1, \ldots, Q-1.$ Each output stream corresponds to regularly spaced samples of $s[k],$ defined as $s_m[k] \triangleq s[kQ + m].$ Using a general symbolic expression for $s[k]$, inspired by the known forms, in (\ref{eq:Normalized Modulated Symbols Rate 3/2 MSN}), (\ref{eq:Normalized Modulated Symbols Rate 4/3 MSN}) and (\ref{eq:Rescaled Modulated Symbols Rate 5/4 MSN}), for rates $3/2$, $4/3$, and $5/4,$ the normalized modulated sequence can be represented as:
\begin{equation}  \label{eq:Normalized Modulated Symbols Rate (Q+1)/Q MSN}
    \bar{s}[k] = \sum_{q=0}^Q \sum_p \bar{b}_q[p] \bar{h}_q[k-Qp].
\end{equation}
Here, $\bar{h}_0[k]$ is a filter of length $L_0,$ typically a multiple of $Q$, while the other filters take the form $\bar{h}_q[k] = \delta[k - (q-1)].$ From this, the normalized modulated streams $\bar{s}_m[k],$ $m=0,1,\ldots,Q-1,$ can be expressed as:
\begin{equation}  \label{eq:m-th Normalized Stream}
    \bar{s}_m[k] = \sum_p \bar{b}_0[p] \bar{h}_0[m+Q(k-p)] + \sum_{q=1}^Q \sum_p \bar{b}_q[p] \delta[(m-(q-1))+Q(k-p)].
\end{equation}
Noting that the difference $m - (q - 1)$ is strictly between $-Q$ and $Q,$ the term $\delta[(m - (q - 1)) + Q(k - p)]$ vanishes unless both $m = q-1$ and $k = p$ hold simultaneously. This observation allows the expression to simplify as follows:
\begin{equation}  \label{eq:Simplified m-th Normalized Stream}
    \bar{s}_m[k] = \sum_p \bar{b}_0[p] \bar{h}_0[m+Q(k-p)] + \bar{b}_{m+1}[k].
\end{equation}
Introducing the $Q$ polyphase filters, $\bar{h}_0^m[k] \triangleq \bar{h}_0[m+Qk],$ $m=0,1,\ldots,Q-1,$ we rewrite the streams as: 
\begin{equation}  \label{eq:Polyphase Simplified m-th Normalized Stream}
    \bar{s}_m[k] = \sum_p \bar{b}_0[p] \bar{h}_0^m[k-p] + \bar{b}_{m+1}[k] = \bar{b}_0[p] \circledast \bar{h}_0^m[k] + \bar{b}_{m+1}[k].
\end{equation}
This shows that each normalized modulated stream $\bar{s}_m[k]$ includes the input stream $\bar{b}_{m+1}[k]$ directly, without modification, along with a filtered version of the input stream $\bar{b}_0[k],$ via the polyphase filter $\bar{h}_0^m[k].$ The length of each polyphase filter is bounded above by $\lceil L_0 / Q \rceil,$ where $\lceil \cdot \rceil$ denotes the ceiling function.

In summary, for any \gls{nsm} within the family of optimized \glspl{nsm} considered so far, all input streams appear in the modulated sequence $s[k]$ essentially as they are (up to scaling), except for $\bar{b}_0[k],$ which is uniquely filtered before inclusion. This analysis helps to reveal the structural similarity with Ungerboeck’s \gls{tcm}, where the underlying recursive systematic convolutional code outputs all input streams unchanged except for one, which is recursively generated from all inputs.

Continuing along the trend observed for all studied \glspl{nsm} so far, which aligns closely with the structural behavior found in Ungerboeck’s \glspl{tcm}, we now build upon the rate-$2$ and rate-$3$ \glspl{nsm} discussed in Sections~\ref{Rate-2 guaranteeing, minimum Euclidean distance approaching NSMs with real filters' coefficients} and~\ref{Sseq:Good rate-3 NSMs with real filters' coefficients}, and numerically characterized in Tables~\ref{table:Best Filters Numerical Form Rate-2 NSM} and~\ref{table:Best Filters Numerical Form Rate-3 NSM}, respectively. Using the optimal filter $h_0[k]$ of length $L_0 = 8$ presented in both tables, we construct extensions of the corresponding \glspl{nsm} to higher rates—specifically, rate-$4$ and rate-$5$ \glspl{nsm}. The original rate-$2$ and rate-$3$ \glspl{nsm} provided asymptotic gains of approximately $3.837$ dB and $5.206$ dB relative to $4$-ASK and $8$-ASK, respectively. It is important to note that the target extensions, with rates $\rho = 4$ and $\rho = 5,$ correspond, on a spectral efficiency basis, to \gls{1d} modulations $16$-ASK and $32$-ASK, and to \gls{2d} modulations $256$-QAM and $1024$-QAM, respectively. Accordingly, the asymptotic gains achieved by the rate-$4$ and rate-$5$ \glspl{nsm}, which will be specified and characterized next, must be assessed with respect to the corresponding reference modulations, $16$-ASK and $32$-ASK, respectively.

Following the methodology outlined in Section~\ref{Sseq:Good rate-3 NSMs with real filters' coefficients}, we now turn our attention to the specification and analysis of a rate-$4$ \gls{nsm}. In this case, alongside the filter $h_0[k]$ of length $L_0 = 8$—previously presented in Tables~\ref{table:Best Filters Numerical Form Rate-2 NSM} and~\ref{table:Best Filters Numerical Form Rate-3 NSM}—we incorporate filters $h_m[k],$ $m = 1, 2, 3$ of unit length. These filters are related through tightness constraints defined by $h_3[0] = 2\,h_2[0] = 4\,h_1[0] = 8\,h_0[0].$ The normalized form, $\bar{h}_0[k],$ of $h_0[k]$ admits an exact closed-form expression for $\bar{h}_0[0]$, inferred from Section~\ref{Closed-form expressions for L0 = 8 and L1 = 1}, given by $\bar{h}_0[0] = \sqrt{8570} \sqrt{17 \sqrt{29} + 793}/\sqrt{31582880-582760\sqrt{29}},$ which approximates to $\bar{h}_0[0] \approx 0.516339056518458$, as reported in Tables~\ref{table:Best Filters Numerical Form Rate-2 NSM} and~\ref{table:Best Filters Numerical Form Rate-3 NSM}. Let $\eta_m$ for $m = 0,1,2,3$ represent the contributions to the average symbol energy from the bipolar input sequences $\bar{b}_m[k].$ The corresponding filters are then specified as $h_0[k] = \sqrt{\eta_0}\,\bar{h}_0[k]$ and $h_m[k] = \sqrt{\eta_m}\,\delta[k],$ $m=1,2,3.$ The tightness relations among the $\eta_m$ coefficients can be expressed as $\eta_3 = 4\,\eta_2 = 16\,\eta_1 = 64 \, (\bar{h}_0[0])^2 \eta_0.$ At this point, it is important to observe that the total \gls{nsm} energy, $\eta_0 + \eta_1 + \eta_2 + \eta_3,$ corresponds to the average symbol energy, which is matched to that of $16$-ASK and equals $1 + 4 + 16 + 64 = 85.$ Additionally, following reasoning similar to that used for $4$-ASK and $8$-ASK, the \gls{msed}—provisionally assumed as a conjecture—can be expressed as $d_{\text{min}}^2 = 4\,\eta_0$. In conclusion, this leads to a value of $d_{\text{min}}^2 = 765611936652984320/52661142836239673,$ which corresponds to an asymptotic gain, relative to $16$-ASK (or equivalently $256$-QAM), of approximately $10\,\log_{10}(d_{\text{min}}^2/4) \approx 5.604584247771476$ dB.

We now extend the discussion to the rate-$5$ case by incorporating an additional branch corresponding to $m = 4$, which introduces a new filter $h_4[k]$ of unit length. As before, filters $h_m[k]$, $m = 1,2,3,4$, are defined through tightness constraints, now given by $h_4[0] = 2\,h_3[0] = 4\,h_2[0] = 8\,h_1[0] = 16\,h_0[0],$ where $h_0[k]$ is the same length-$8$ filter used in previous constructions. Let $\eta_m$, $m = 0,1,2,3,4$, denote the contributions to the average symbol energy from the respective bipolar input sequences $\bar{b}_m[k].$ These filters are then specified as $h_0[k] = \sqrt{\eta_0}\,\bar{h}_0[k]$ and $h_m[k] = \sqrt{\eta_m}\,\delta[k],$ $m=1,2,3,4.$ In this configuration, the tightness relations among the energy coefficients take the form $\eta_4 = 4\,\eta_3 = 16\,\eta_2 = 64\,\eta_1 = 256 \, (\bar{h}_0[0])^2 \eta_0.$ The total \gls{nsm} energy, given by $\eta_0 + \eta_1 + \eta_2 + \eta_3 + \eta_4$, aligns with the average symbol energy of $32$-ASK and evaluates to $1 + 4 + 16 + 64 + 256 = 341.$ Following the same rationale applied in the previous cases, and under the same conjectured \gls{msed} model, we obtain $d_{\text{min}}^2 = 4\,\eta_0$. Substituting the corresponding value leads to $d_{\text{min}}^2 = 3071454945866678272/206289616809738873,$ which yields an asymptotic gain, relative to $32$-ASK (or equivalently $1024$-QAM), of $10\,\log_{10}(d_{\text{min}}^2/4) \approx 5.708067887079716$ dB.

In summary, the tightness relation $\eta_1 = 4 \, (\bar{h}_0[0])^2 \eta_0$—combined with the specific value $\bar{h}_0[0] \approx 0.516339056518458,$ which is notably close to $1/2$ due to the choice $L_0 = 8$—implies that $\eta_1$ is nearly equal to $\eta_0$. In contrast, standard $2^m$-ASK modulation yields $\eta_1 = 4\,\eta_0$ exactly. This results in the energy coefficients $\eta_l,$ $l = 1, 2, \ldots, m-1,$ in conventional $2^m$-ASK being approximately $1/(\bar{h}_0[0])^2$ times larger than those in the \gls{nsm} setting. Consequently, when $L_0 = 8$ and $m$ increases, the asymptotic gain of the \gls{nsm} over $2^m$-ASK converges to $-10\,\log_{10}((\bar{h}_0[0])^2) \approx 5.742980275450664$ dB.

\subsection{Connection Between NSMs and Information-Theoretic Principles} \label{Link with information theory}

In information theory, the capacity of a memoryless channel is defined as the maximum mutual information between the channel input and output, optimized over all input distributions~\cite{Cover06,MacKay03}. For the \gls{awgn} channel, where the noise is Gaussian and independent of the input, the input distribution that maximizes mutual information is itself Gaussian. When the input $X$ is distributed as $\mathcal{N}(0,P),$ subject to an average power constraint, $E[X^2] \le P,$ this achieves the channel capacity given by $\tfrac{1}{2} \log_2(1+\tfrac{P}{N}),$ bits per channel use, where $N$ is the noise power.

In practical communication systems, the theoretical Gaussian input distribution that achieves capacity in the \gls{awgn} channel is approximated using coded modulation schemes. Lattice codes provide a structured approach to this problem, where shaping techniques are used to emulate the Gaussian input. One such method is sphere shaping, in which codewords are confined to a $d$-dimensional hypersphere. As the dimension $d$ increases, the marginal distribution of the discrete lattice points within the sphere approaches a Gaussian, due to the concentration of measure phenomenon~\cite{Conway99}. However, since practical lattice codes operate over $\mathbb{Z}^d,$ the distribution remains inherently discrete, and a perfect match to the continuous Gaussian distribution is not possible. Despite this, a shaping gain of up to $1.53$ dB, corresponding to the optimal gain of $\pi e/6,$ can be asymptotically approached through such high-dimensional shaping~\cite{Erez04}. To realize this gain in lower dimensions and practical implementations, trellis shaping was proposed, particularly in conjunction with \gls{tcm}, offering a practical method to approximate Gaussian-like behavior using a finite-state machine~\cite{Forney92}. More recent developments, such as \gls{ccdm}, provide further practical tools to approximate Gaussian inputs under power constraints~\cite{Schulte16}.

In our new vision and interpretation of rate $\rho = k/n$ \glspl{nsm}, in Section~\ref{sec:Analogy with convolutional error correction coding}, the $n$ modulated sequence streams, $s_m[p],$ $m = 0, 1, \ldots, n-1,$ the multiplexing of which leads to the modulated sequence $s[q],$ are modeled as the sums of multiple independent component signals, $s_{ml}[p],$ $l = 0, 1, \ldots, k-1,$ each generated by filtering a random bipolar sequence, $\bar{b}_l[p]$—a sequence of independent random variables taking values in ${\pm 1}$ with equal probability. Each bipolar sequence has zero mean and unit variance, and the sequences are assumed independent. Each modulated sequence sample is a weighted sum of these bipolar inputs, where the weights correspond to the magnitudes of the involved filter coefficients, since sign differences are absorbed by the bipolar nature of the inputs. Importantly, in our modulation framework, long filter lengths arise naturally as a consequence of performance optimization: increasing the filter length allows for greater degrees of freedom and improved spectral shaping, which are essential for maximizing efficiency and minimizing interference. As the filter lengths increase, the sum involves a large number of independent, but non-identically distributed, bipolar variables weighted by the filter tap magnitudes. Provided these magnitudes satisfy mild regularity conditions, the generalized \gls{clt} applies, ensuring that each modulated sequence sample, $s[k],$ converges in distribution to a Gaussian random variable. This result explains why such signals exhibit Gaussian-like statistics in the limit of long filters~\cite{{Billingsley95},Feller71,Middleton77}.

While neighboring modulated sequence samples share many common input symbols due to the convolutional structure, the key to asymptotic uncorrelatedness lies in the near-orthogonality of the involved filter impulse responses and their time-shifted versions, as the filter lengths tend to infinity. This property is analogous to the pseudo-random spreading sequences used in spread spectrum and \gls{cdma} systems, where shifted versions have low cross-correlation. Consequently, despite the significant overlap in inputs, the modulated sequence samples, $s[k],$ become asymptotically uncorrelated. In the Gaussian limit, uncorrelatedness implies asymptotic independence of the samples, a fact well-established in the theory of Gaussian processes and linear systems~\cite{Verdu98,Poor94}.

In summary, our proposed \gls{nsm} vision and framework for modulation provides a principled and natural pathway to generate modulated signals with asymptotically Gaussian statistics, aligning seamlessly with the fundamental optimality conditions of information theory. Specifically, since Gaussian inputs maximize the mutual information over \gls{awgn} channels, our approach ensures that the modulated sequence satisfies this requirement without additional shaping complexity. Unlike traditional strategies that rely on sphere-shaped lattice codes or \gls{tcm} with trellis shaping, both of which involve significant design and decoding complexity, our method achieves Gaussianity through the superposition of filtered bipolar sequences with long filter responses. This structure leads to samples that converge in probability density to a true Gaussian distribution, not merely an approximation. In contrast, lattice codes shaped within a high-dimensional sphere remain constrained to integer components in $\mathbb{Z}^d,$ preventing their samples from being perfectly Gaussian even as the dimension grows. Thus, our \gls{nsm} modulation strategy not only simplifies implementation but also provides a theoretically grounded and practically efficient route to capacity-approaching signaling.

\subsection{Structured Synergy Between NSMs and LDPC Coding via Analog LDGM Representation} 
\label{A natural and efficient synergy between NSM and LDPC coding}

The integration of \gls{nsm} with \gls{ldpc} coding yields a structurally coherent and computationally efficient framework for joint modulation and coding. Far from being merely compatible, the two components are mutually reinforcing. The intrinsic structure of \gls{nsm} aligns naturally with the graph-based representation of \gls{ldpc} codes, enabling a unified formulation that supports seamless iterative processing. The iterative detection and decoding process benefits directly from the structured nature of \gls{nsm}, resulting in improved convergence behavior, lower complexity, and a system architecture that is both high-performing and suitable for practical deployment.

\begin{figure}[!htbp]
    \centering
    \includegraphics[width=1.0\linewidth]{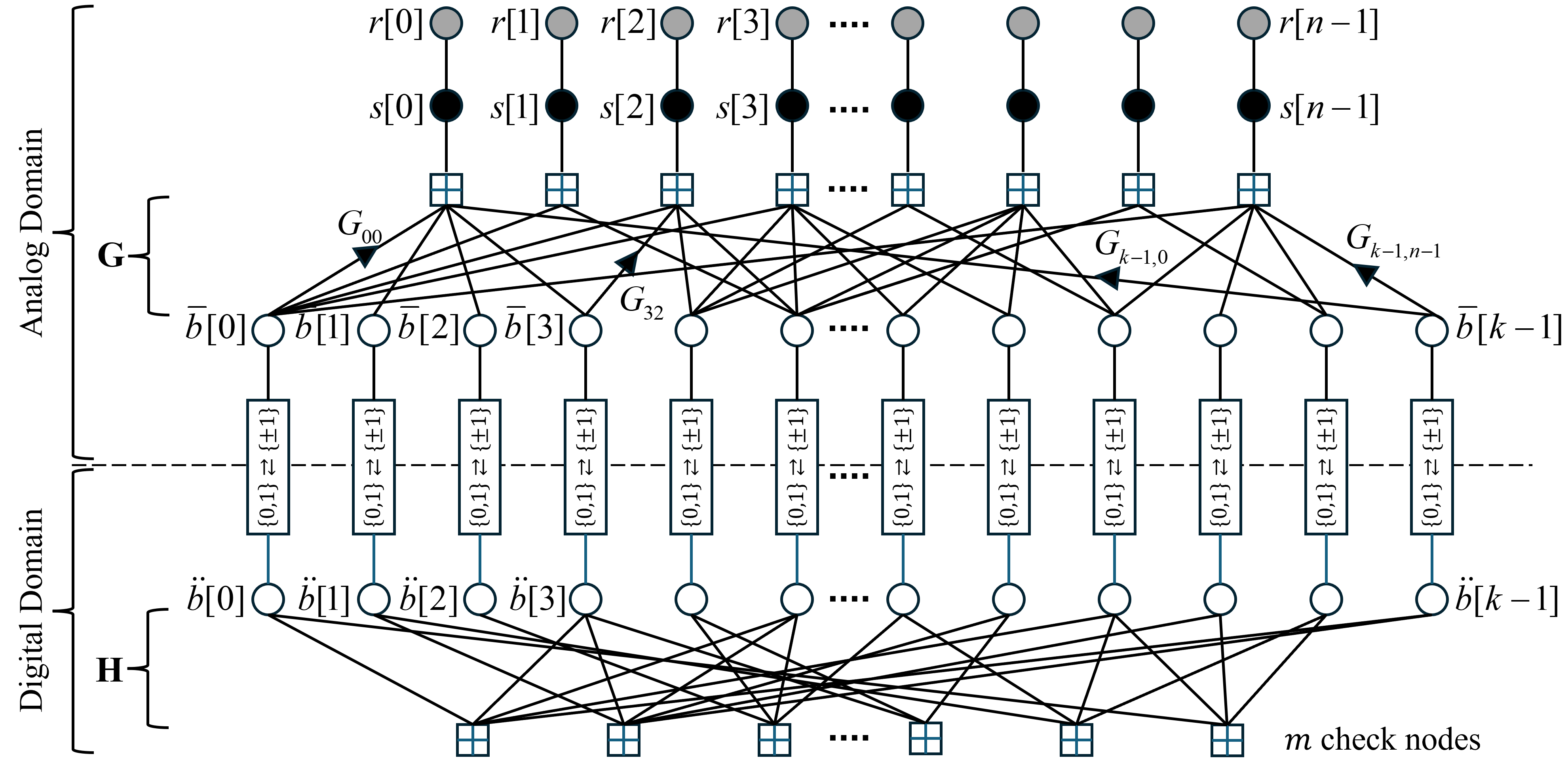}
    \caption{Three-layer factor graph of the proposed analog LDGM–\gls{ldpc} scheme, illustrating the interface between analog-domain modulation and digital-domain \gls{ldpc} decoding. The graph highlights the analog summation nodes, bipolar and binary variable nodes, modulated symbols, and received observations.}
    \label{fig:Factor Graph Analog LDGM NSM-LDPC Code}
\end{figure}

The three-layer factor graph illustrated in Figure~\ref{fig:Factor Graph Analog LDGM NSM-LDPC Code} is inspired by the compound \gls{ldgm}/\gls{ldpc} constructions presented in~\cite{Wainwright07,Wainwright09} (specifically Figure~2). In that construction, an \gls{ldgm} code spans the top and middle layers, while an \gls{ldpc} code connects the middle and bottom layers. We adopt a structurally analogous architecture to model our \gls{nsm}–\gls{ldpc} scheme, extending it to include analog-domain signal processing. The resulting factor graph captures both the real-valued analog detection layer and the binary-valued digital decoding layer, interconnected through a deterministic binary-to-bipolar mapping. This mapping is used in both directions: from digital to analog during modulation, and from analog to digital during demodulation, via its inverse bipolar-to-binary transformation.

In the analog part of the graph, the $n$ modulated components $\bm{s} = (s[0], s[1], \ldots, s[n-1])$ (represented by filled black circles) are linked to $n$ analog summation nodes (represented by squares). These are connected via the generating matrix $\bm{G} \in \mathbb{R}^{k \times n}$ to the bipolar vector $\bar{\bm{b}} = (\bar{b}[0], \bar{b}[1], \ldots, \bar{b}[k-1]) \in \{\pm1\}^k$ (represented by empty circles), which is the bipolar version of the \gls{ldpc}-coded binary vector $\ddot{\bm{b}} \in \text{GF}(2)^k.$ The modulated sequence is defined by the linear relation $\bm{s} = \bar{\bm{b}} \bm{G}.$ The top portion of the graph models the effect of a memoryless noisy channel that maps $\bm{s}$ to the received observation vector $\bm{r} = (r[0], r[1], \ldots, r[n-1])$ (represented by filled gray circles).

Each edge in the analog domain represents a contribution from a coded bit to a transmitted symbol. An edge from $\bar{b}[i]$ to $s[j]$ is labeled with $G_{ij},$ signifying that the term $\bar{b}[i] G_{ij}$ adds to $s[j].$ Likewise, an edge from $s[j]$ to $r[j]$ reflects the effect of the channel. We consider here the case where each received observation $r[j]$ depends solely on a single transmitted symbol $s[j],$ which applies to a wide range of practical scenarios. This includes \gls{awgn} channels and multiplicative fading channels, such as those encountered in cyclic prefix \gls{ofdm} systems. Although the fading coefficients across subcarriers in \gls{ofdm} may be correlated due to limited coherence bandwidth, the system architecture ensures that each $r[j]$ is determined directly by the corresponding $s[j],$ maintaining a symbol-wise dependence.

In more general cases where the received observation depends on multiple transmitted symbols—such as in \gls{sc} systems over multipath fading channels, in doubly dispersive environments modeled by \gls{otfs} modulation, or in \gls{mimo} systems where spatial interference induces symbol coupling—the channel introduces a linear transformation across transmitted symbols. This transformation can be absorbed into the analog generating matrix $\bm{G},$ which then jointly captures both modulation and channel effects. The resulting factor graph structure remains valid and supports iterative detection over such channel-impaired observations without modification to the decoding process.

In the digital part of the graph, the coded binary vector $\ddot{\bm{b}}$ satisfies the \gls{ldpc} parity-check equation $\ddot{\bm{b}} \bm{H}^T = \bm{0},$ where $\bm{H} \in \text{GF}(2)^{m \times k}$ is a sparse binary matrix. A variable node (represented as an empty circle) corresponding to bit $\ddot{b}[j]$ is connected to a check node (represented as a square) corresponding to row $i$ whenever $H_{ij} = 1,$ reflecting the structure of the parity-check constraints.

The interface between the analog and digital domains is established via two inverse mappings. The binary-to-bipolar transformation, given by $\bar{b}[l] = 2\ddot{b}[l] - 1$ for $l = 0, 1, \ldots, k-1$, is used to move from the digital domain (or the digital part of the factor graph) to the analog domain (or the analog part of the factor graph). Conversely, the bipolar-to-binary conversion, defined as $\ddot{b}[l] = (\bar{b}[l] + 1)/2$, is applied to transition from the analog domain back to the digital domain. These mappings are essential for iterative detection and decoding, as they enable alternating updates between analog and digital representations of the codeword.

Notice that the \gls{nsm} has a rate $\rho_G = k/n > 1$, while the \gls{ldpc} code has a rate $\rho_H = 1 - m/k < 1$, assuming that the parity-check matrix $\bm{H}$ has full row rank. Hence, the overall effective rate, obtained by serially concatenating the \gls{ldpc} code with the \gls{nsm}—where the \gls{ldpc} encoder takes $k - m$ information bits in $\text{GF}(2)$ and the \gls{nsm} generates $n$ modulated symbols in $\mathbb{Z}$—is given by $\rho = \rho_G \rho_H = (k - m)/n$.

In the case of a regular \gls{ldpc} code, all columns of the parity-check matrix $\bm{H}$ have a constant Hamming weight $d_v,$ meaning that each variable node is connected to exactly $d_v$ check nodes. Similarly, each row has a constant Hamming weight $d_c,$ so that each check node is connected to exactly $d_c$ variable nodes. The total number of ones in $\bm{H}$ is then equal to $m \, d_c = k \, d_v,$ and the code rate can be expressed alternatively as $\rho_H = 1 - d_v/d_c.$ In both regular and irregular constructions, the sparsity of $\bm{H}$ is a crucial property that enables the use of efficient message-passing algorithms for decoding.

Given an optimal decoder, the best performance would be achieved by codes that most closely resemble random codes—namely, those with the largest $d_v$. However, practical message passing (or belief propagation) algorithms, such as the sum–product decoder, tend to perform poorly on dense graphs, resulting in the best performance being obtained for relatively small values of $d_v.$ This observation is confirmed in~\cite{MacKay03}, where Figure~47.8(b) shows the \gls{bler} as a function of $E_b/N_0$ for $d_v = 3, 4, 5,$ and $6,$ with $k = 816$ and $m = 408,$ corresponding to a code rate $\rho_H = 1/2.$ For \gls{bler} values above $10^{-5}$, the code with $d_v = 3$ outperforms the others with larger $d_v,$ and as $d_v$ increases from $3$ to $6,$ the \gls{bler} degrades within the range of considered $E_b/N_0$ values.

Now, given an optimal \gls{nsm} and in perfect alignment with the discussion at the end of Section~\ref{Analog packing and digital packing}, the best performance would be obtained for random generating matrices, whose rows act as quasi-orthogonal \gls{nsm} filters with lengths that can approach and even reach the modulated sequence length, $n.$ In this setting, any finite number of rows in the generating matrix $\bm{G}$ correspond to quasi-orthogonal vectors in $\mathbb{R}^n,$ offering a $2$-ASK-like local environment for the most detrimental and weakening error events in the \gls{ldpc} code with small Hamming weights. As such, \gls{ldpc} coding can be seen as a structuring mechanism that increases the \gls{msed} of the coded modulation scheme in the same way it does for a pure $2$-ASK modulation.

Practical implementations of the \gls{nsm} demodulator that allow belief propagation from and towards the \gls{ldpc} decoder could employ the \gls{bcjr} algorithm whenever the modulator can be characterized by a trellis with a complexity-acceptable number of states. This occurs, for example, with well-organized \glspl{nsm} having an asymptotic rate $\rho = \kappa/\nu$ (obtained as $n$ goes to infinity), for which the rows of the generating matrix $\bm{G}$ are regular shifts of the corresponding filters $h_i[p],$ $i=0,1,\ldots,\kappa-1,$ provided that the number of such filters is small and their normalized lengths, $L_i/\nu,$ are also small. Additional degrees of freedom are possible by allowing the shifts of each of the $\kappa$ filters $h_i[p]$ to change from one position to another while preserving their lengths, thereby maintaining a trellis structure that remains suitable for trellis-based detection algorithms.

Unfortunately, restricting the filter lengths to small values due to detection complexity constraints limits the performance of the modulation part in the compound \gls{ldpc} coding and \gls{nsm} scheme. Allowing for better-performing \glspl{nsm} requires the normalized filter lengths, $L_i/\nu,$ to be very large. Building on the discussion at the end of Section~\ref{Analog packing and digital packing}, this allows any finite number of rows in the \gls{nsm} generating matrix to be quasi-orthogonal as in the case of perfectly random matrices. When the normalized filter lengths are large, regardless of whether the underlying filter impulse responses are preserved from one shift to the next or not, the \gls{bcjr} algorithm or any variant relying on a modulation trellis structure must be replaced by a simpler message passing algorithm, such as the sum-product algorithm~\cite{MacKay03,Kurkoski02,Richardson08}. Fortunately, employing message passing in the detection part is not only beneficial from a complexity standpoint but also highly desirable since it offers a natural harmonization with \gls{ldpc} decoding, which typically relies on such algorithms.

Proceeding as in \gls{ldpc} codes, and in order to further simplify the implementation of the message passing algorithm at the \gls{nsm} detector level, we require the generating matrix $\bm{G}$ to be sparse, so that the number of edges in the analog part of the factor graph of Figure~\ref{fig:Factor Graph Analog LDGM NSM-LDPC Code} is reduced to its strict minimum. To push the simplification even further, we restrict the non-zero entries of $\bm{G}$ to take values in the rational field $\mathbb{Q}.$ Under this restriction, it is convenient to work with a scaled version $\mathring{\bm{G}}$ of the analog generating matrix $\bm{G},$ where $\mathring{\bm{G}}$ is a $k \times n$ matrix with entries in the integer set $\mathbb{Z}.$ This allows the generation of a scaled modulated vector $\mathring{\bm{s}} = (\mathring{s}[0], \mathring{s}[1], \ldots, \mathring{s}[n-1])$ with integer components, related to the bipolar version of the coded vector $\bar{\bm{b}}$ through the relation $\mathring{\bm{s}} = \bar{\bm{b}} \mathring{\bm{G}}$. The resulting compounded \gls{ldpc} and \gls{nsm} scheme recalls a known class of code constructions combining binary \gls{ldpc} coding and binary \gls{ldgm} coding~\cite{Wainwright07,Wainwright09}, in which the \gls{ldpc} code acts as an outer code and the \gls{ldgm} code as an inner code. It is worth noting, before proceeding, that another class of compound codes known as \gls{mn} codes~\cite{Richardson08} also combines \gls{ldpc} and \gls{ldgm} components, but in a structurally distinct way that bears no direct relation to the compounding architecture described here.

Proceeding as in \gls{ldpc} codes, and in order to further simplify the implementation of the message passing algorithm at the \gls{nsm} detector level, we require the generating matrix $\bm{G}$ to be sparse, so that the number of edges in the analog part of the factor graph of Figure~\ref{fig:Factor Graph Analog LDGM NSM-LDPC Code} is reduced to its strict minimum. To push the simplification even further, we restrict the non-zero entries of $\bm{G}$ to take values in the rational field $\mathbb{Q}.$ Under this restriction, it is convenient to work with a scaled version $\mathring{\bm{G}}$ of the analog generating matrix $\bm{G},$ where $\mathring{\bm{G}}$ is a $k \times n$ matrix with entries in the integer set $\mathbb{Z}.$ This allows the generation of a scaled modulated vector $\mathring{\bm{s}} = (\mathring{s}[0], \mathring{s}[1], \ldots, \mathring{s}[n-1])$ with integer components, related to the bipolar version of the coded vector $\bar{\bm{b}}$ through the relation $\mathring{\bm{s}} = \bar{\bm{b}} \mathring{\bm{G}}$.

The resulting compounded \gls{ldpc} and \gls{nsm} scheme recalls a known class of code constructions combining binary \gls{ldpc} coding and binary \gls{ldgm} coding~\cite{Wainwright07,Wainwright09}, in which the \gls{ldpc} code acts as an outer code and the \gls{ldgm} code as an inner code. It is worth noting, before proceeding, that another class of compound codes known as \gls{mn} codes~\cite{Richardson08} also combines \gls{ldpc} and \gls{ldgm} components, but in a structurally distinct way that bears no direct relation to the compounding architecture described here.

At first glance, requiring $\bm{G}$ to be sparse may appear to contradict the earlier observation, made in Section~\ref{Analog packing and digital packing}, that random generating matrices—whose rows behave as quasi-orthogonal \gls{nsm} filters—are optimal for achieving the asymptotic performance of $2$-ASK. However, there is good reason to believe that this apparent contradiction may be reconciled in light of the findings discussed in Section~\ref{sssec:One-dimensional designed NSMs}. There, we examined \glspl{nsm} associated with the filter pattern $\bm{\pi}_0 = (3,3,3,2,2,1),$ whose filters are inherently sparse due to the small number of non-zero taps. Despite their sparsity, we observed that as the filter length $L_0$ increases, the percentage of filters achieving the asymptotic $2$-ASK performance increases significantly—for example, it was multiplied by more than $7$ when moving from $L_0 = 11$ to $L_0 = 12$, although the absolute percentage at $L_0 = 12$ remains relatively modest at around $2.45\%$. This promising trend motivated the conjecture that the proportion of good filters continues to grow rapidly with $L_0$, and could approach $100\%$ once $L_0$ exceeds moderate values, likely around $14$ or $15$. More importantly, this behavior was conjectured to extend beyond this specific pattern to a broad class of filter patterns as $L_0$ increases, making sparsity progressively less restrictive in terms of performance.

This result is particularly reassuring for the compound \gls{ldgm}–\gls{ldpc} framework considered here. Unlike trellis-based detection algorithms such as the \gls{bcjr} algorithm, whose complexity grows exponentially with $L_0$, message passing algorithms (e.g., the sum-product algorithm) are not sensitive to the absolute filter length but rather to the number of non-zero entries in the filters—i.e., the size of the pattern $\bm{\pi}_0,$ which is typically kept small. This means that large values of $L_0$ can be used to secure performance guarantees without incurring prohibitive complexity. Consequently, the imposition of sparsity on $\bm{G}$ is not detrimental; on the contrary, it enables practical implementation while still allowing the system to benefit from the asymptotic distance properties of well-designed \glspl{nsm}.

With respect to the first category of \gls{ldpc} and \gls{ldgm} compounding, up to three key differences arise when comparing the \gls{nsm}-based construction to conventional binary \gls{ldgm} coding. The first difference is that the binary \gls{ldgm} scheme operates in the digital binary domain $\text{GF}(2)$, while the \gls{nsm} scheme operates in the analog integer domain $\mathbb{Z}$. The second difference concerns the relationship between the input and output lengths of the generating matrix. For \gls{ldgm} codes, we typically have $k < n$ due to the addition of parity bits, whereas the \gls{nsm} scheme is characterized by $k > n$, reflecting the presence of multiple streams in \gls{nsm} and the fact that its rate $\rho_G = k/n$ is strictly greater than $1.$ The third difference concerns the per-component discrimination power. Binary \gls{ldgm} lacks such discrimination at the component level because it operates over $\text{GF}(2),$ where all computations are performed modulo $2,$ restricting each output symbol to only two possible values when considered individually. In contrast, the \gls{nsm} scheme benefits from operating over $\mathbb{Z}$, which allows each modulated symbol $\mathring{s}[j]$, for $j = 0, 1, \ldots, n - 1,$ to take on a much larger number of values. This enhanced per-component discrimination power is a direct consequence of the fact that the analog \gls{ldgm} \gls{nsm} operates at a rate $\rho_G = k/n$ greater than $1,$ whereas binary \gls{ldgm} codes typically operate at rates strictly less than $1.$ In the binary \gls{ldgm} case, meaningful discrimination is achieved only at the level of the full codeword by appending $n - k$ redundancy bits to the $k$ binary inputs of the \gls{ldgm} encoder. In contrast, the analog \gls{ldgm} \gls{nsm} provides strong per-component discrimination, enabling $k$ bipolar input symbols to be compressed into a smaller number $n$ of modulated outputs. In light of these structural and operational differences, and by analogy with the digital \gls{ldgm} structure, it is appropriate to refer to the \gls{nsm} scheme as an analog \gls{ldgm} \gls{nsm} scheme.

As in \gls{ldpc} coding, it is possible to introduce regularization constraints on the \gls{nsm} generating matrix $\mathring{\bm{G}},$ whose entries lie in $\mathbb{Z}.$ First, observe that the average power of the $j$-th scaled modulated symbol $\mathring{s}[j]$ is equal to the \gls{sen} $\|\mathring{\bm{g}}_j\|^2$ of the $j$-th column $\mathring{\bm{g}}_j$ of the scaled generating matrix $\mathring{\bm{G}}.$ Ensuring that the modulated sequence power is evenly distributed among all modulated components $\mathring{s}[j],$ $j = 0, 1, \ldots, n - 1,$ therefore requires that all columns of $\mathring{\bm{G}}$ have the same \gls{sen}. This uniform column-norm constraint, whether applied to the scaled matrix $\mathring{\bm{G}}$ or to its unscaled counterpart $\bm{G},$ is analogous to the constraint in \gls{ldpc} coding that enforces a constant Hamming weight $d_v$ across all columns of the parity-check matrix $\bm{H}.$

Second, observe that the contribution of a coded bit $\ddot{b}[i],$ or its bipolar counterpart $\bar{b}[i],$ $i = 0, 1, \ldots, k - 1,$ to the total energy of the scaled modulated vector $\mathring{\bm{s}}$ is given by the \gls{sen} $\|\mathring{\bm{h}}_i\|^2$ of the $i$-th row $\mathring{\bm{h}}_i$ of matrix $\mathring{\bm{G}}.$ In scenarios where the system employs turbo-equalization, involving iterative message passing between \gls{ldpc} decoding and \gls{nsm} detection, and aims to asymptotically match the performance of an \gls{ldpc}-coded $2$-ASK system at moderate to high \glspl{snr}, it becomes essential for all coded bits $\ddot{b}[i],$ or equivalently their bipolar versions $\bar{b}[i],$ to contribute equally to the overall energy of the modulated sequence $\mathring{\bm{s}},$ or its unscaled version $\bm{s}.$ This design requirement leads to the constraint that all rows of $\mathring{\bm{G}}$ must have identical squared Euclidean norms. Just as with the column-regularity constraint, this row-regularity constraint—whether applied to $\mathring{\bm{G}}$ or to $\bm{G}$—mirrors the \gls{ldpc} design principle that requires a constant Hamming weight $d_c$ across all rows of the parity-check matrix $\bm{H}$.

To illustrate the concept of analog \gls{ldgm} \glspl{nsm} constructed with sparse analog generating matrices $\mathring{\bm{G}}$ having entries in $\mathbb{Z}$, we consider a representative family of cyclic \glspl{nsm}. These schemes are built upon regular cyclic shifts of a small number, say $\kappa \ge 2$, of sparse scaled filter impulse responses $\mathring{h}_i[p]$, $i = 0, 1, \ldots, \kappa - 1$, with integer-valued taps. The design inherently implements \emph{tail-biting}, and as a result, the achieved rate is an integer given by $\rho_G = \kappa$. Matrix $\mathring{\bm{G}}$ thus has $k = \kappa n$ rows, constructed from the cyclic shifts $\mathring{h}_i[p - j \bmod n]$, $j = 0, 1, \ldots, n - 1$, for each $i = 0, 1, \ldots, \kappa - 1$. This construction naturally satisfies the column-wise regularity constraint, since each column of the generating matrix $\mathring{\bm{G}}$ results from uniformly distributed cyclic shifts of the same set of filter responses.

To also satisfy the row-wise regularity constraint—ensuring that each row of $\mathring{\bm{G}}$ has the same \gls{sen}—it is necessary for the filter impulse responses $\mathring{h}_i[p]$, $i = 0, 1, \ldots, \kappa - 1$, to share a common energy. To simplify the design of such impulse responses, we consider $\kappa$ filter pattern vectors $\bm{\pi}_i$, $i = 0, 1, \ldots, \kappa - 1$, each of dimension $\lambda_i \ge 1$, with nonzero integer components and a common \gls{sen}, denoted by $\|\bm{\pi}\|^2 \triangleq \|\bm{\pi}_0\|^2 = \|\bm{\pi}_1\|^2 = \cdots = \|\bm{\pi}_{\kappa - 1}\|^2$. The $i$-th scaled impulse response $\mathring{h}_i[p]$ is then defined to have exactly $\lambda_i$ nonzero integer taps, whose magnitudes are determined by the components of $\bm{\pi}_i$.

Furthermore, the sparsity of the filter impulse responses $\mathring{h}_i[p]$—inherited from the small-sized pattern vectors $\bm{\pi}_i$—is crucial for reducing the complexity of \gls{nsm} detection when using message passing algorithms. Unlike trellis-based detection methods such as \gls{bcjr}, whose complexity grows exponentially with the filter length $L_0$, message passing complexity depends primarily on the number of nonzero entries in each filter, i.e., the dimension $\lambda_i$ of the pattern $\bm{\pi}_i$. As such, long filters with few nonzero taps (i.e., sparse and long) can be used without incurring significant complexity penalties, provided the filter pattern remains compact.

Beyond computational advantages, sparsity also plays a key structural role: it helps mitigate the presence of short cycles in the associated factor graph. This is analogous to the \emph{girth} criterion in \gls{ldpc} codes, where maximizing the length of the shortest cycle improves the effectiveness of message passing by reducing harmful correlation between messages. In the context of analog \gls{ldgm} \glspl{nsm}, sparse generating matrices lead to factor graphs with improved girth properties, thereby enhancing the reliability and convergence behavior of message passing detection.

Although tail-biting and filter sparsity may initially raise concerns about performance, both phenomena are well understood and effectively controlled in this setting. As discussed in Section~\ref{sssec:One-dimensional designed NSMs}, sparse filters derived from compact patterns such as $\bm{\pi}_0 = (3,3,3,2,2,1)$ have shown a strong trend toward achieving the asymptotic $2$-ASK performance as the filter length $L_0$ increases. Regarding confusion associated with tail-biting, as analyzed in Section~\ref{sec:One-dimensional NSMs and tail-biting}, even for the most unfavorable “duobinary” ($\mathring{h}_0[p] = \delta[p] + \delta[p-1]$) and “dicode” ($\mathring{h}_0[p] = \delta[p] - \delta[p-1]$) filters, the confusion probability decays exponentially with the modulated sequence length $K$ as $(\tfrac{3}{4})^K$, reaching values below $10^{-10}$ for typical packet sizes $K \ge 80$. For more favorable filters—such as those constructed from the aforementioned compact patterns—the decay is expected to be even faster. This exponential suppression of confusion confirms that tail-biting remains effective even when applied to long sparse filters in cyclic constructions, enabling practical, high-performance, and computationally efficient structured \gls{nsm} designs.

Altogether, sparsity and tail-biting—when carefully leveraged—lead to analog \gls{ldgm} \gls{nsm} constructions that are not only highly efficient in terms of spectral and computational resources but also robust in terms of performance. These structured designs remain compatible with message passing detection and seamlessly integrate into the compound \gls{ldgm}–\gls{ldpc} framework.

In Tables~\ref{table:Candidate Patterns Rate-2 Lambda_0 = Lambda_1 = 2}--\ref{table:Candidate Patterns Rate-2 Lambda_0 = Lambda_1 = 3}, we present several pairs of admissible pattern vectors $(\bm{\pi}_0, \bm{\pi}_1),$ with positive integer components and a common \gls{sen}, $\|\bm{\pi}\|^2,$ for rate-$2$ analog \gls{ldgm} \glspl{nsm}. The considered cases satisfy $2 \le \lambda_0 \le \lambda_1 \le 3.$ For each pair of pattern vectors, we compute the set of all possible bipolar combinations of their aggregate components and report the number of distinct values these combinations can take. To maximize this number, only vector pairs with distinct aggregate component values are retained. It should be noted, however, that relaxing this constraint would allow many more admissible vector pairs, which may still prove effective in practice.

Furthermore, for each pair, we evaluate the entropy, $\mathcal{H},$ of the resulting probability distribution, assuming that all $2^{\lambda_0 + \lambda_1}$ bipolar combinations are equiprobable. We also include the normalized entropy, $\mathcal{H} / \mathcal{H}_{\text{max}},$ where $\mathcal{H}_{\text{max}} = \lambda_0 + \lambda_1$ represents the maximum entropy achievable under this equiprobability assumption. The admissible vector pairs are listed in ascending order of their common \gls{sen}. Although a higher common norm frequently corresponds to larger values in the pattern vector components and is often associated with a higher entropy, $\mathcal{H},$ this trend does not hold universally. On one hand, since $\mathcal{H}$ is also the entropy per component of the scaled modulated vector $\mathring{\bm{s}}$, and given the partial statistical dependence among its components, the total entropy of $\mathring{\bm{s}}$ is upper bounded by $n \mathcal{H}.$ On the other hand, the total number of possible equiprobable input bipolar vectors $\bar{\bm{b}}$ is $2^k,$ leading to an input entropy of $k.$ Therefore, in order for the analog \gls{ldgm} modulator to be injective, the condition $k \le n \mathcal{H}$ must hold. In the case of the considered rate-$2$ construction, where $\rho_G = k/n = 2,$ this means that all admissible entropies must satisfy $\mathcal{H} \ge 2,$ which is indeed the case for all listed vector pairs in the tables.

The bipolar combinations of the aggregate pattern vectors $(\bm{\pi}_0, \bm{\pi}_1)$ directly determine the set of values that each component $\mathring{s}[j]$ of the scaled modulated vector can take, along with their associated probabilities. As a result, a higher normalized entropy $\mathcal{H} / \mathcal{H}_{\text{max}}$ reflects greater per-component discrimination power in the \gls{nsm}. This, in turn, enhances the performance of the message passing algorithm in the analog domain of the factor graph, particularly in computing the messages from the summation nodes to the variable bipolar nodes, by enabling more effective inference from the channel observations.

In Tables~\ref{table:Candidate Patterns Rate-3 Lambda_0 = Lambda_1 = Lambda_2 = 3}--\ref{table:Candidate Patterns Rate-3 Lambda_0 = Lambda_1 = Lambda_2 = 4}, we extend the analysis to several triplets of admissible pattern vectors $(\bm{\pi}_0, \bm{\pi}_1, \bm{\pi}_2),$ with common norm $\|\bm{\pi}\|^2,$ and satisfying $3 \le \lambda_0 \le \lambda_1 \le \lambda_2 \le 4,$ for rate-$3$ analog \gls{ldgm} \glspl{nsm}. For each configuration, we provide the same set of information and performance indicators as in the rate-$2$ case.The same considerations as those discussed for the rate-$2$ case apply here, and similar conclusions regarding injectivity, entropy, and discrimination power can be drawn.

Obviously, the number of distinct bipolar combinations, along with the entropy and normalized entropy, serve as subjective indicators of the analog \gls{ldgm} \gls{nsm}’s capability to discriminate signals received from the channel. In addition to these metrics, the minimum gap between different bipolar combinations—normalized by the common Euclidean norm of the pattern vectors, $\|\bm{\pi}\|$—provides a valuable measure of discrimination power at the detector level when processing noisy modulated symbols. From the data presented in Tables~\ref{table:Candidate Patterns Rate-2 Lambda_0 = Lambda_1 = 2}--\ref{table:Candidate Patterns Rate-3 Lambda_0 = Lambda_1 = Lambda_2 = 4}, it is observed that increasing the lengths of the pattern vectors, $\lambda_0, \lambda_1, \ldots, \lambda_{\kappa-1},$ tends to increase the entropy, $\mathcal{H},$ but simultaneously results in a decrease in the normalized entropy, $\mathcal{H}/\mathcal{H}_{\text{max}}.$ Conversely, expanding the integer range spanned by the components of the pattern vectors, $\bm{\pi}_0, \bm{\pi}_1, \ldots, \bm{\pi}_{\kappa-1},$ generally increases both the entropy and the normalized entropy. However, these enhancements come at the cost of a larger common Euclidean norm, $\|\bm{\pi}\|,$ which in turn reduces the minimum normalized gap between bipolar combinations. Therefore, all these indicators must be considered jointly to effectively evaluate the discrimination power of the detector when analyzing and processing the received symbols.

From another perspective, note that, consistent with the discussion in Section~\ref{Analog packing and digital packing}, for an analog \gls{ldgm} \gls{nsm} to achieve performance close to that of $2$-ASK, the lengths of the pattern vectors, $\lambda_0, \lambda_1, \ldots, \lambda_{\kappa-1},$ should be chosen as large as possible to introduce greater randomness and enhance the quasi-orthogonality among the rows of the generating matrix, $\mathring{\bm{G}}.$ This consideration is reminiscent of the behavior in regular \gls{ldpc} codes, where increasing the common Hamming weight, $d_c,$ of the rows of the parity-check matrix, $\bm{H},$ moves the code structure closer to that of random codes and improves performance~\cite{MacKay03}. However, since the relationship $m d_c = k d_v$ holds, increasing $d_c$ also increases the Hamming weight, $d_v,$ of the columns of $\bm{H}.$ This simultaneous increase in both $d_c$ and $d_v$ for regular \gls{ldpc} codes not only raises the complexity of the iterative decoding algorithm based on message passing, but also diminishes its effectiveness due to a reduction in the girth of the underlying Tanner graph, which corresponds to the length of its shortest cycle. These insights naturally extend to the analog \gls{ldgm} \gls{nsm}, where any increase in the total pattern length, $\lambda_0 + \lambda_1 + \cdots + \lambda_{\kappa-1},$ is expected not only to increase the computational complexity of the message passing algorithm during detection, but also to reduce its effectiveness.

For a representative example of the factor graph of the compounding of an \gls{ldpc} code with an analog \gls{ldgm} \gls{nsm}, we take $m = 4, k = 8,$ and $n = 4.$ As a coding scheme, we consider the $(8,4,4)$ extended Hamming code, which is equivalent to the Reed-Muller block code, $RM(1,3).$ This code, which results from the extension of the Hamming code $(7,4,3)$ by adding an overall even parity check, possesses the $4 \times 8$ binary parity check matrix
\begin{equation} \label{eq: Parity Check Matrix Rate 1/2 LDPC Code}
\bm{H} = \begin{bmatrix}
    1 & 0 & 1 & 0 & 1 & 0 & 1 & 0 \\
    0 & 1 & 1 & 0 & 0 & 1 & 1 & 0 \\
    0 & 0 & 0 & 1 & 1 & 1 & 1 & 0 \\
    1 & 1 & 1 & 0 & 0 & 0 & 0 & 1
\end{bmatrix}.
\end{equation}
For the specification of an analog \gls{ldgm} \gls{nsm}, we take $\kappa = 2,$ $\lambda_0 = 2$ and $\lambda_1 = 3,$ and consider the pair of pattern vectors $\bm{\pi}_0 = (4,5)$ and $\bm{\pi}_1 = (1,2,6),$ from Table~\ref{table:Candidate Patterns Rate-2 Lambda_0 = 2 Lambda_1 = 3}. We arbitrarily consider filters $\mathring{h}_0[p] = 4 \, \delta[p] - 5 \, \delta[p-2]$ and $\mathring{h}_1[p] = 2 \, \delta[p] - \delta[p-1] + 6 \, \delta[p-2]$ of common length $L_0 = L_1 = 3.$ The corresponding scaled generating matrix is
\begin{equation} \label{eq: Analog LDGM Rate 2 NSM}
\mathring{\bm{G}} = \begin{bmatrix}
    4 & 0 & -5 & 0  \\
    0 & 4 & 0 & -5  \\
    -5 & 0 & 4 & 0  \\
    0 & -5 & 0 & 4  \\
    2 & -1 & 6 & 0  \\
    0 & 2 & -1 & 6  \\
    6 & 0 & 2 & -1  \\
    -1 & 6 & 0 & 2    
\end{bmatrix}.
\end{equation}

\begin{figure}[!htbp]
    \centering
    \includegraphics[width=0.85\linewidth]{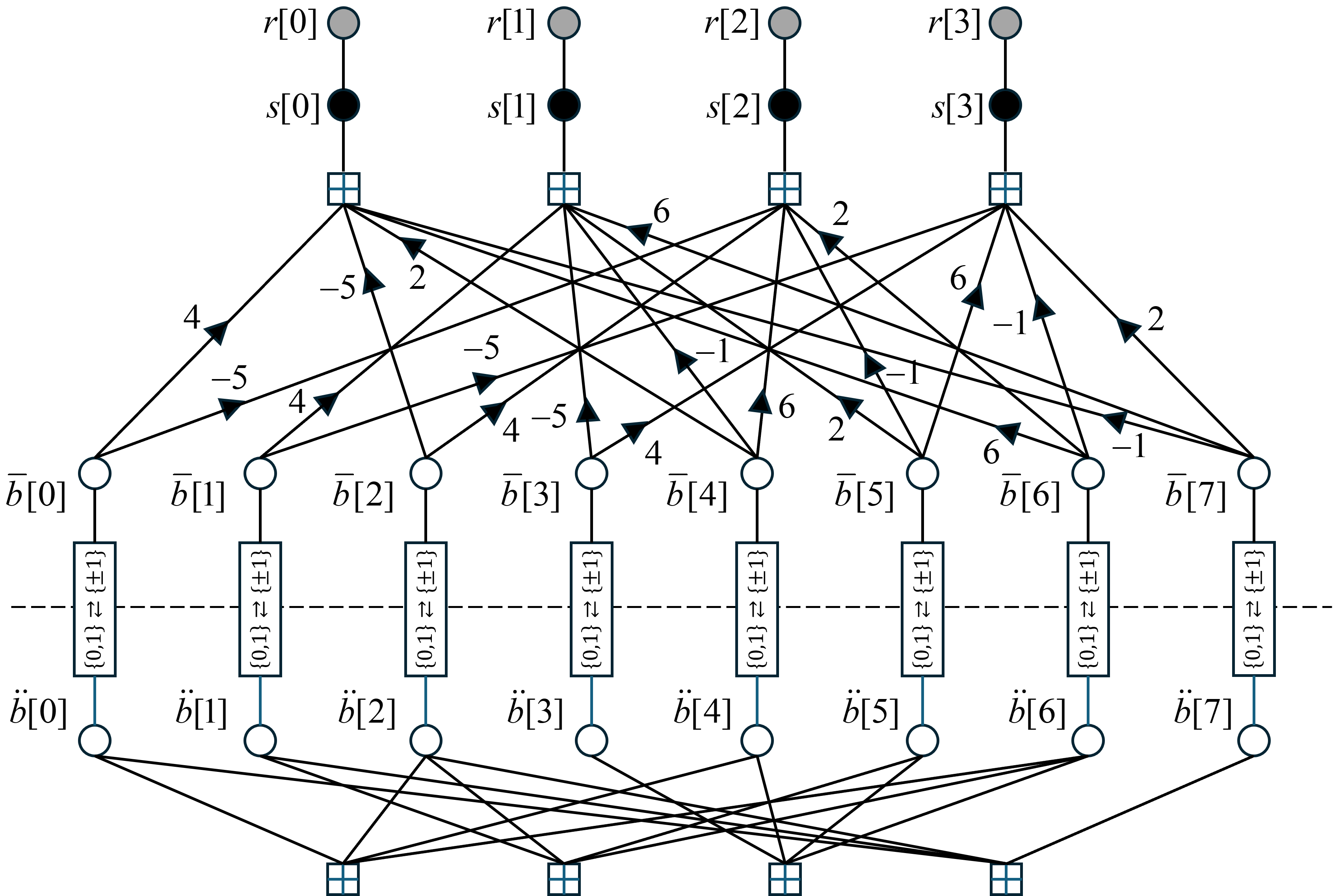} 
    \caption{Factor graph representation for the example compound coding scheme composed of a $(8,4,4)$ extended Hamming LDPC code and an analog LDGM NSM with parameters $\kappa = 2$, $\lambda_0 = 2$, and $\lambda_1 = 3$. The hybrid structure highlights the digital component defined by the parity-check matrix $\bm{H}$ and the analog modulation governed by the scaled generator matrix $\mathring{\bm{G}}$.}
    \label{fig:FactorGraphExampleLDPC_LDGM_NSM}
\end{figure}

Figure~\ref{fig:FactorGraphExampleLDPC_LDGM_NSM} shows, for this illustrative example, the underlying factor graph with its analog and digital parts. Table~\ref{table:Characteristics of LDPC & Analog LDGM NSM} presents the codewords of the extended Hamming code in both binary and bipolar forms, alongside the corresponding \gls{nsm} modulated codewords. The \gls{msed} for this compounded scheme is $d_{\text{min}}^2 = 208,$ while the average energy per information bit is given by $\|\mathring{\bm{G}}\|_F^2/(k-m) = 82,$ where $\|\mathring{\bm{G}}\|_F$ denotes the Frobenius norm of matrix $\mathring{\bm{G}}.$ Consequently, the normalized \gls{msed} offered by this compound scheme equals $208/82 = 104/41 \approx 2.5366,$ which is lower than $4,$ the \gls{msed} of uncoded $2$-ASK that achieves the same spectral efficiency. This outcome is expected due to the inherently poor performance of the analog \gls{ldgm} \gls{nsm} code stemming from its small length, $n = 4,$ and the pronounced adverse effect of tail-biting at such a short block length, which significantly weakens the overall scheme's robustness. It should be noted that the objective here is not to design the best compound scheme but to provide a straightforward illustrative example.

For completeness and illustrative purposes, we now consider the standard case of non-Gray coded $4$-ASK as a special and trivial instance of an analog \gls{ldgm} \gls{nsm}, characterized by the sparse generating matrix
\begin{equation} \label{eq: Analog LDGM Rate 2 Non Gray Coded 4-ASK}
\mathring{\bm{G}} = \begin{bmatrix}
    1 & 0 & 0 & 0  \\
    0 & 1 & 0 & 0  \\
    0 & 0 & 1 & 0  \\
    0 & 0 & 0 & 1  \\
    2 & 0 & 0 & 0  \\
    0 & 2 & 0 & 0  \\
    0 & 0 & 2 & 0  \\
    0 & 0 & 0 & 2    
\end{bmatrix}.
\end{equation}
As is well known, this matrix distributes the transmitted energy evenly across the modulated sequence components. However, the second half of the coded bits entering the \gls{ldgm} \gls{nsm} modulator receive four times the energy of the first four bits. This unbalanced energy distribution among coded bits constitutes a significant limitation, preventing the scheme from asymptotically achieving the performance of coded $2$-ASK.

The cyclic construction proposed above, which relies on defining $\kappa = k/n$ filters, $h_i[p],$ $i = 0, 1, \ldots, \kappa - 1,$ through the specification of $\kappa$ pattern vectors, $\bm{\pi}_i,$ and exemplified in (\ref{eq: Analog LDGM Rate 2 NSM}), imposes a significant amount of structure on the design of the scaled generating matrix, $\mathring{\bm{G}},$ of the analog \gls{ldgm} \gls{nsm}. This structural rigidity can severely limit the performance of the underlying \gls{nsm}, particularly for small dimensions, $\kappa,$ and short lengths, $n.$ As a first step towards introducing more randomness and reducing structural constraints, one may consider generating the rows of $\mathring{\bm{G}}$ in a randomized manner while still using a finite set of pattern vectors, $\bm{\pi}_i,$ $i=0,1,\ldots,\kappa-1.$ In this approach, rows of $\mathring{\bm{G}}$ associated with the same pattern vector $\bm{\pi}_i$ are not necessarily cyclic shifts of one another. The loss of the strict cyclic shift property may be relatively mild and achieved by applying random sign changes to the components of the pattern vector $\bm{\pi}_i,$ such that the magnitudes of the components remain cyclic shifts of each other. In a more extensive randomization approach, in addition to sign changes, the positions of the components of the pattern vector $\bm{\pi}_i$ are permuted randomly across rows. This permutation destroys the cyclic shift structure even in terms of the magnitude, resulting in rows that are no longer cyclic shifts of each other in any form.

One can simplify the design of sufficiently randomized scaled generating matrices by adopting a concept analogous to protograph-based \gls{ldpc} coding. In this approach, a protograph—a small bipartite graph—serves as a template for constructing larger codes through a process called lifting~\cite{Zaidi18}. Starting from a small base matrix or protomatrix, each nonzero entry is expanded into a permutation matrix to form a larger, structured parity-check matrix. Extending this concept to analog \gls{ldgm} \glspl{nsm} offers a highly structured yet flexible framework that facilitates control over short cycles and supports efficient detection implementations. Within this framework, \gls{qc} \gls{ldpc} codes—widely adopted in standards such as $5$G~\cite{Zaidi18,Morais22}—represent a specific subclass of protograph-based codes. In \gls{qc}-\gls{ldpc} codes, the lifting replaces each nonzero element of the base matrix with a circulant permutation matrix, which is a cyclic shift of the identity matrix. Analogously, in analog \gls{ldgm} \glspl{nsm}, these binary identity matrices can be replaced by diagonal matrices with nonzero integer entries, yielding a scaled generating matrix, $\mathring{\bm{G}},$ with a block-circulant structure. This block-circulant design preserves the benefits of protograph-based construction while additionally enabling low-complexity hardware implementation, fast parallel detection, and efficient storage and memory access.

For joint iterative analog \gls{ldgm} \gls{nsm} detection and \gls{ldpc} decoding, which can be viewed as a form of turbo-equalization, we naturally adopt and extend the two algorithmic strategies proposed in~\cite{Kurkoski02} for single-stream partial-response channels. These strategies, known as bit-based message passing and state-based message passing, are originally designed for sum-product inference over the analog part of the global factor graph. They remain applicable in the cyclic analog \gls{ldgm} \gls{nsm} setting considered here. However, their computational cost, implementation complexity, and practical feasibility scale very differently in this context. In particular, the state-based approach experiences exponential growth in complexity as the number of states increases rapidly with the sum of filters lengths in the cyclic analog \gls{ldgm} \gls{nsm} system. In contrast, the bit-based approach scales more favorably, benefiting from simpler, more localized message updates that can exploit the sparsity of the graph and parallel processing.

The bit-based message-passing approach operates directly at the level of transmitted bits. Messages are exchanged between variable nodes representing the binary codeword and factor nodes corresponding to the analog observations. These messages are iteratively updated following the sum-product rule, incorporating the contribution of each bit to the observed analog quantities via the analog generating matrix. This approach supports parallelism, scales well with the system size, and maintains manageable complexity, provided the graph has sparse connectivity and sufficient girth. While the multi-stream partial-response modulator inherently employs a cyclic generating matrix, which enables trellis-based state message passing, the bit-based method naturally extends to more general, non-cyclic generating matrices as discussed above. Such generalizations can increase the girth of the factor graph in the analog domain, potentially enhancing convergence and performance. At the same time, the absence of cyclic structure precludes the straightforward trellis construction required for state-based message passing, making bit-based message passing the more practical choice in these cases.

The state-based message-passing approach reformulates detection as inference over a Markov chain defined by the memory of the partial-response channel. This method relies critically on the presence of a well-defined trellis structure with a manageable number of states, which naturally arises from the cyclic generating matrix of the multi-stream partial response modulator. Within this cyclic framework, state-based message passing can converge rapidly and approach the performance of optimal \gls{bcjr} decoding. However, its complexity grows exponentially with the total filter memory. Specifically, for a cyclic analog \gls{ldgm} \gls{nsm} comprising $\kappa$ partial-response filters of lengths $L_0, L_1, \ldots, L_{\kappa-1},$ the number of states scales as $2^{L_0 + L_1 + \cdots + L_{\kappa-1} - \kappa}.$ This exponential increase quickly makes the state-based approach computationally prohibitive for large filter lengths. Moreover, in more general scenarios where the generating matrix is non-cyclic and lacks regular structure, the trellis representation ceases to exist, thus precluding the application of state-based message passing altogether.

In the bit-based message-passing framework, each received symbol is modeled as a noisy observation of a sparse linear combination of several bipolar input symbols. This sparsity stems from the structure of the analog generating matrix, which characterizes the modulation process and may be either cyclic—as in multi-stream partial response modulation—or non-cyclic in more general analog \gls{ldgm} \gls{nsm} constructions. Regardless of the cyclicity of the matrix, the inference procedure for recovering variable node information proceeds in an identical manner and depends critically on sparsity to ensure computational tractability. Efficient simplifications of this inference step have been proposed in the context of \gls{otfs} modulation, where similarly sparse observation models arise. In particular, the simplified message-passing algorithms developed in~\cite{Hong22, Raviteja18} facilitate low-complexity turbo equalization for systems with sparse delay-Doppler channels. These methods can be adapted to reduce the computational burden of message updates in the analog domain of \gls{ldgm} \gls{nsm} detection while maintaining convergence and performance. However, unlike the \gls{otfs} setting—where weak channel contributions may be approximated or neglected due to physical propagation characteristics—each nonzero entry in the sparse analog generating matrix of an \gls{ldgm} \gls{nsm} is deliberately designed and carries significant information. Therefore, no nonzero contribution can be omitted without adversely affecting detection performance.

As in~\cite{Kurkoski02}, the joint detection and decoding process unfolds over a total of $U$ \emph{global iterations}. Each global iteration consists of two successive stages: an \emph{analog-domain stage}, which handles modulation detection, and a \emph{digital-domain stage}, which performs \gls{ldpc} decoding. Inside each stage, a fixed number of \emph{local sub-iterations} is executed: $T$ sub-iterations in the analog domain and $S$ sub-iterations in the digital domain. These sub-iterations refine the exchanged messages within their respective domains before information is transferred between them. This schedule reflects the structure of the three-layer factor graph introduced earlier, in Figure~\ref{fig:Factor Graph Analog LDGM NSM-LDPC Code}, comprising a top analog layer associated with the channel and modulated symbols, a middle layer of bipolar variables that links analog and digital parts, and a bottom digital layer containing binary variable nodes and \gls{ldpc} parity-check constraints.

\begin{figure}[!htbp]
    \centering
    \includegraphics[width=0.45\linewidth]{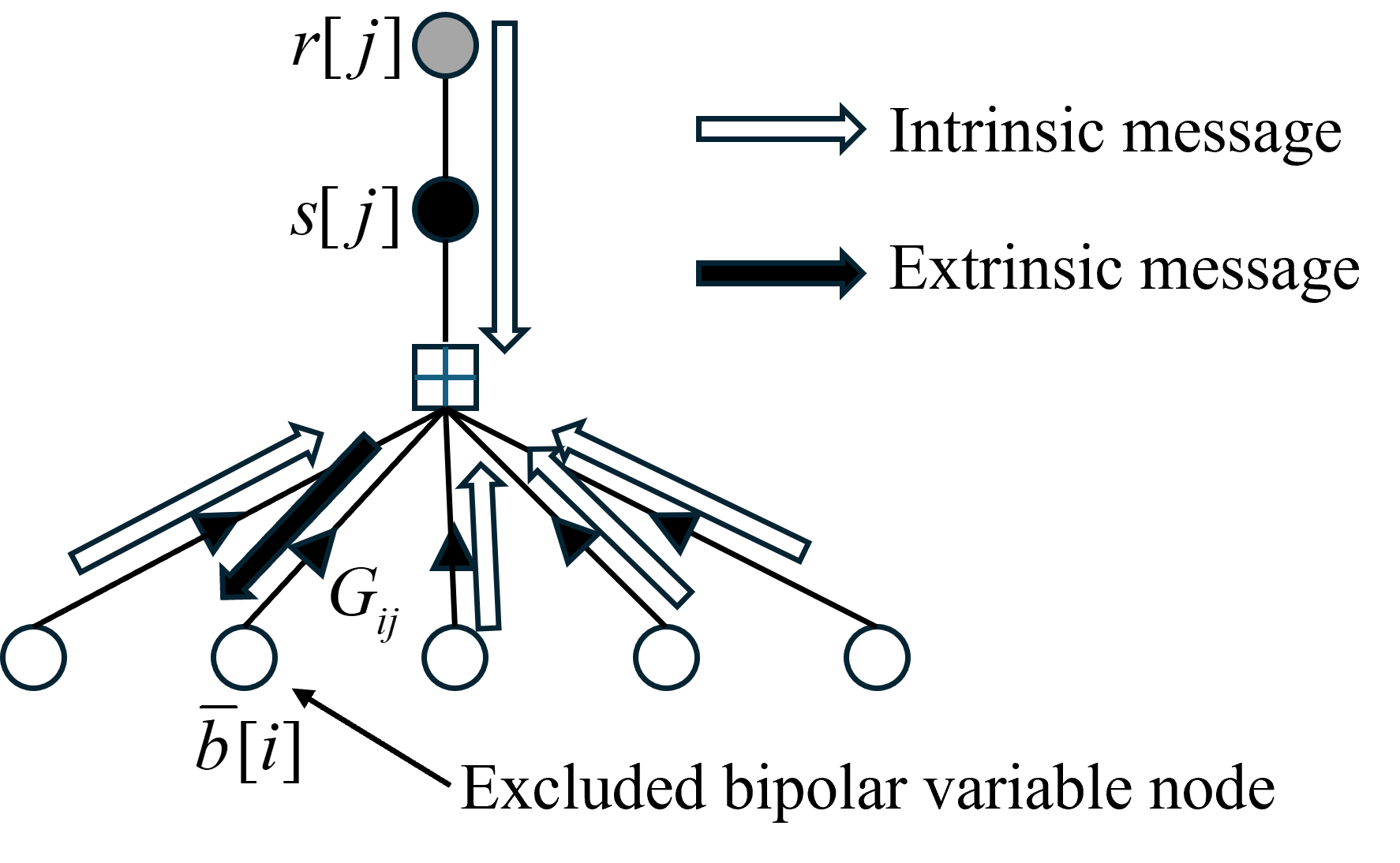}
    \caption{Extrinsic message computation at the summation node associated with modulated symbol $s[j]$. The summation node receives intrinsic messages from all connected bipolar variable nodes $\bar{b}[i]$, except the one currently being updated, and the noisy channel observation $r[j]$. It computes an extrinsic message to send back to the excluded bipolar variable node.}
    \label{fig:ExtrinsicMessageComputation}
\end{figure}

During the \emph{analog-domain stage}, iterative message passing is performed over $T$ local sub-iterations during each global iteration. In every sub-iteration, the summation node associated with modulated symbol $s[j]$ processes incoming intrinsic messages from all connected bipolar variable nodes $\bar{b}[i]$, except the one currently being updated. It also incorporates the corresponding noisy channel observation $r[j]$. Based on these inputs, it computes a new \emph{extrinsic message}, which is sent back to the excluded bipolar variable node. This refined message excludes the direct contribution of the target variable node and instead reflects the combined influence of the other connected variables and the observation $r[j]$. This step is illustrated in Figure~\ref{fig:ExtrinsicMessageComputation}, where all intrinsic messages to a summation node are shown, except the one under update, alongside the observation $r[j]$ and the outgoing extrinsic message.

\begin{figure}[!htbp]
    \centering
    \includegraphics[width=0.45\linewidth]{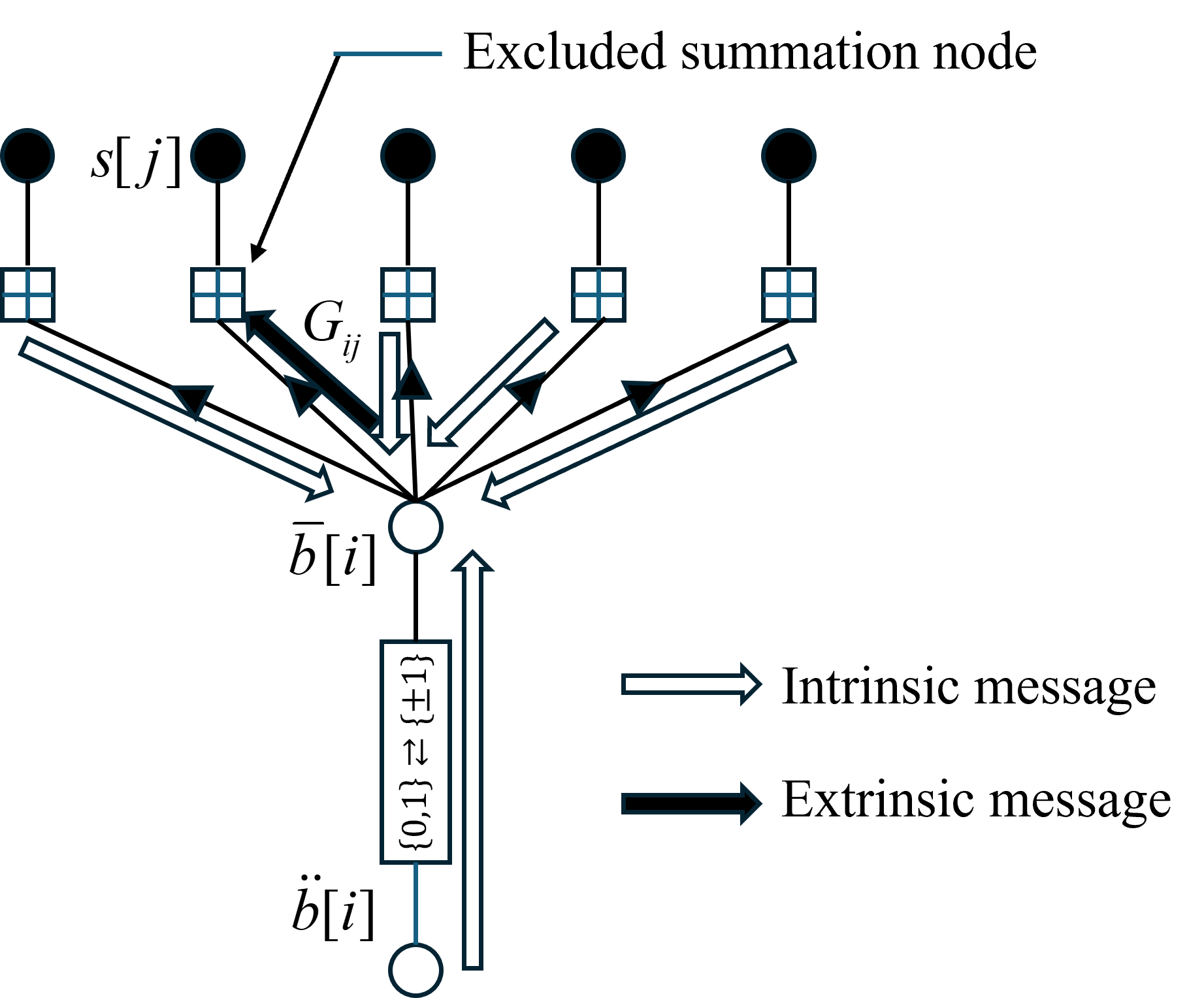}
    \caption{Equality constraint enforcement at a bipolar variable node during sub-iterations $1$ to $T-1$. The node combines all incoming extrinsic messages from summation nodes except the one it is replying to, together with intrinsic information from the digital domain, enforcing consistency and producing an enhanced extrinsic message sent back to the sidelined summation node.}
    \label{fig:EqualityConstraintLoopback}
\end{figure}

\begin{figure}[!htbp]
    \centering
    \includegraphics[width=0.45\linewidth]{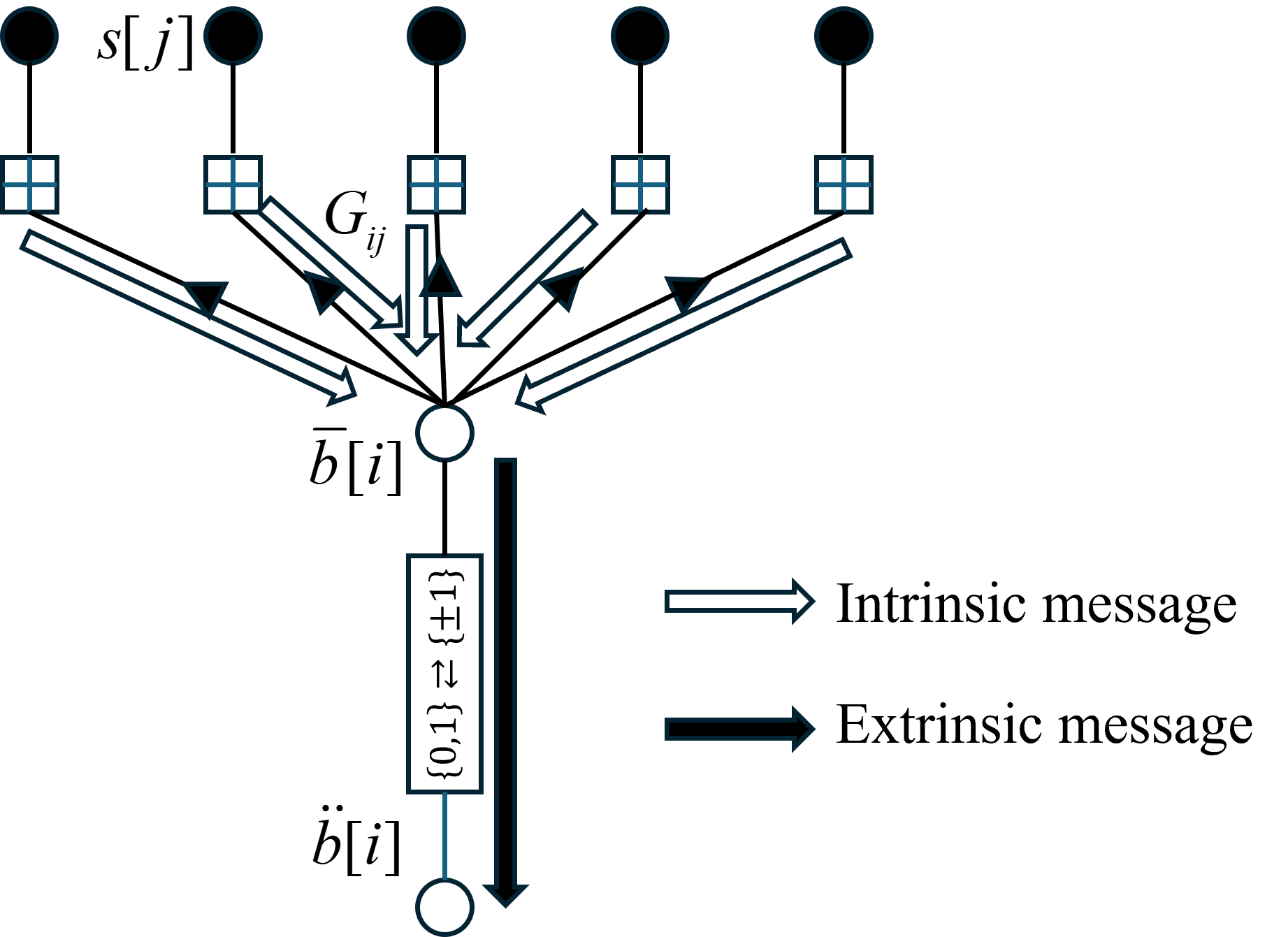}
    \caption{Equality constraint enforcement at the bipolar variable node during the final ($T$-th) sub-iteration of the analog domain. All incoming extrinsic messages from summation nodes are combined without exclusion to produce a final enhanced extrinsic message. This message is then passed transparently through the bipolar-to-binary conversion interface to the corresponding binary variable node in the digital domain.}
    \label{fig:BipolarToBinaryExtrinsicTransfer}
\end{figure}

After receiving these extrinsic messages from connected summation nodes, each bipolar variable node updates its outgoing messages differently depending on the sub-iteration:
\begin{itemize}

    \item In sub-iterations $1$ through $T{-}1$, each bipolar variable node combines all incoming intrinsic messages from connected summation nodes \emph{except} the one it is replying to, \emph{together with} the intrinsic message received from its corresponding binary variable node during the previous digital-domain stage. This set of messages is processed through an \emph{equality constraint}, resulting in an enhanced extrinsic message that is sent back to the excluded summation node. This procedure avoids self-reinforcement while integrating both analog and digital beliefs to iteratively refine the variable node’s estimate. This update mechanism is illustrated in Figure~\ref{fig:EqualityConstraintLoopback}, which shows how a bipolar variable node aggregates relevant intrinsic messages and produces a new extrinsic message directed to the sidelined summation node.

    \item In the final $T$-th sub-iteration, each bipolar variable node aggregates \emph{all} incoming messages from its connected summation nodes using the equality constraint, without excluding any. The result is a final, enforced extrinsic belief that is passed transparently downward through the bipolar-to-binary conversion interface to the corresponding binary variable node in the digital layer. This information transfer is depicted in Figure~\ref{fig:BipolarToBinaryExtrinsicTransfer}.

\end{itemize}

This processing schedule ensures that the analog domain refines its soft reliability information iteratively while avoiding self-reinforcement. Once the $T$ sub-iterations are completed, the most accurate belief for each bit—now expressed at the bipolar node level—is handed off to the \gls{ldpc} decoder in the digital domain to continue the global inference process.

During the \emph{digital-domain stage}, iterative message passing is carried out over $S$ local sub-iterations within each global iteration. In each sub-iteration, the process begins by using the extrinsic messages previously received from the analog domain—via the bipolar-to-binary interface—as intrinsic information at the binary variable nodes. These binary variable nodes forward messages to their connected parity-check nodes.

\begin{figure}[!htbp]
    \centering
    \includegraphics[width=0.4\linewidth]{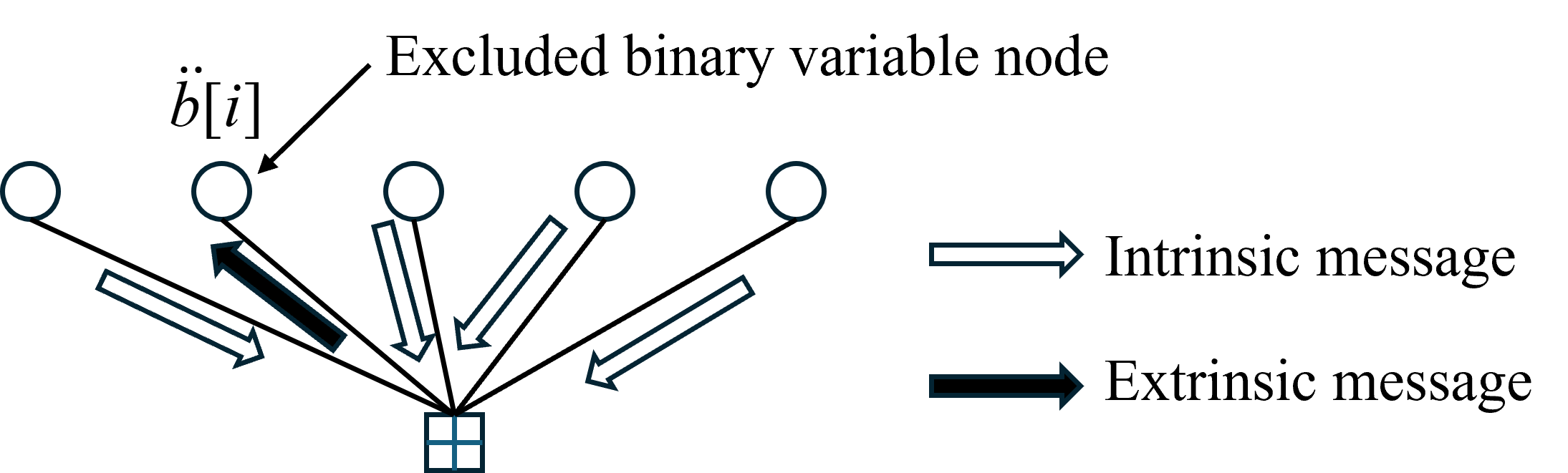}
    \caption{Extrinsic message computation at a parity-check node. Each parity-check node receives intrinsic messages from connected binary variable nodes except the one currently updated, enforces the parity constraint, and computes an extrinsic message directed to the excluded binary variable node.}
    \label{fig:ParityCheckExtrinsicComputation}
\end{figure}

Each parity-check node then updates its outgoing extrinsic messages by enforcing the parity constraint using the sum-product rule. Specifically, for every target binary variable node, the parity-check node aggregates incoming intrinsic messages from all connected binary variable nodes \emph{except} the target one. This exclusion avoids self-reinforcement, and the resulting extrinsic message is sent back to the sidelined binary variable node. This mechanism is illustrated in Figure~\ref{fig:ParityCheckExtrinsicComputation}, which shows the intrinsic messages to a parity-check node (excluding the one under update) and the resulting outgoing message toward the target binary variable node.

\begin{figure}[!htbp]
    \centering
    \includegraphics[width=0.45\linewidth]{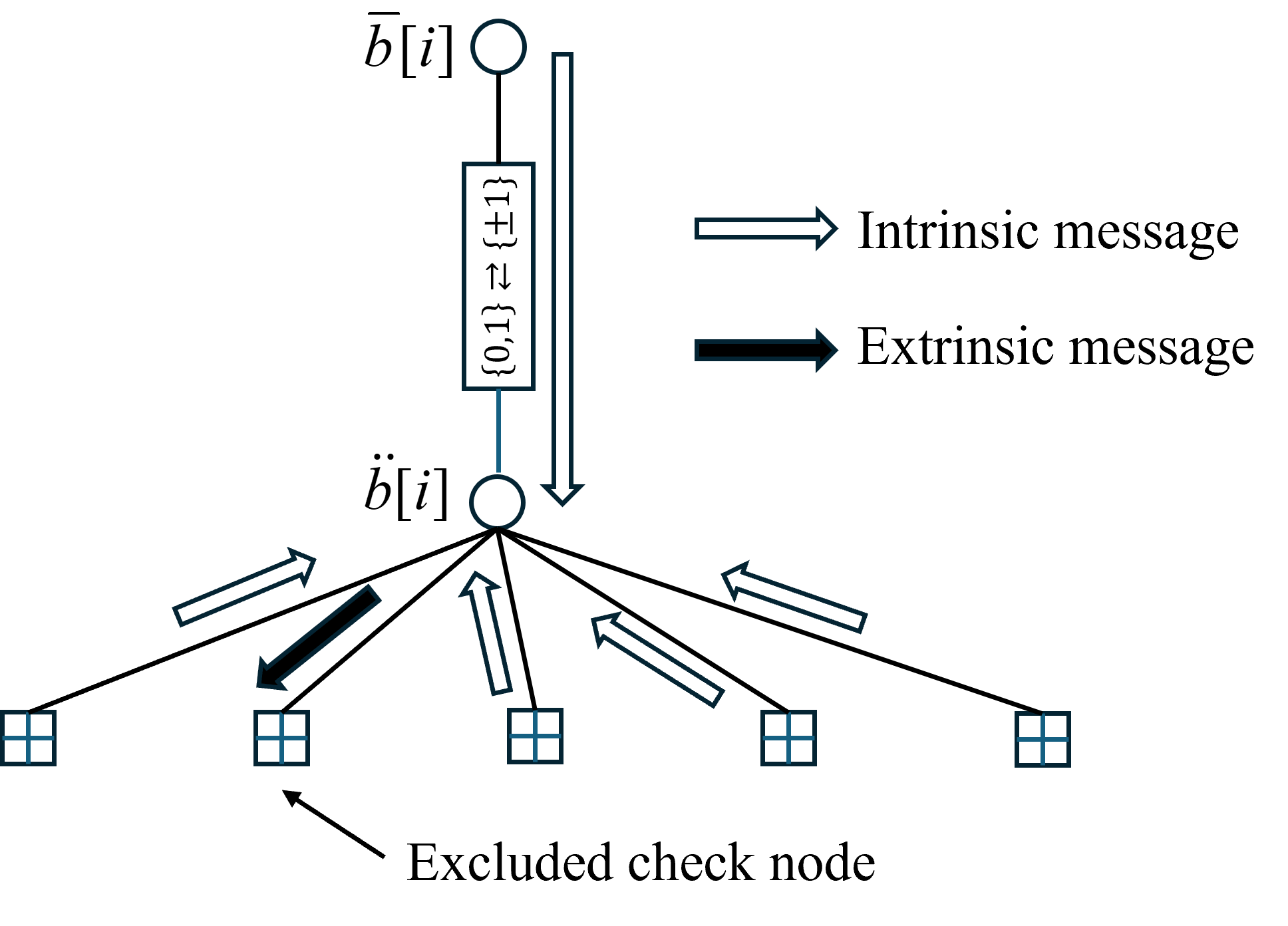}
    \caption{Equality constraint enforcement at a binary variable node during digital sub-iterations $1$ to $S-1$. The node combines all intrinsic messages from parity-check nodes except the one it is replying to, together with intrinsic information from the analog domain, enforcing consistency and producing an enhanced extrinsic message sent back to the sidelined parity-check node.}
    \label{fig:EqualityConstraintLoopbackDigital}
\end{figure}

\begin{figure}[!htbp]
    \centering
    \includegraphics[width=0.45\linewidth]{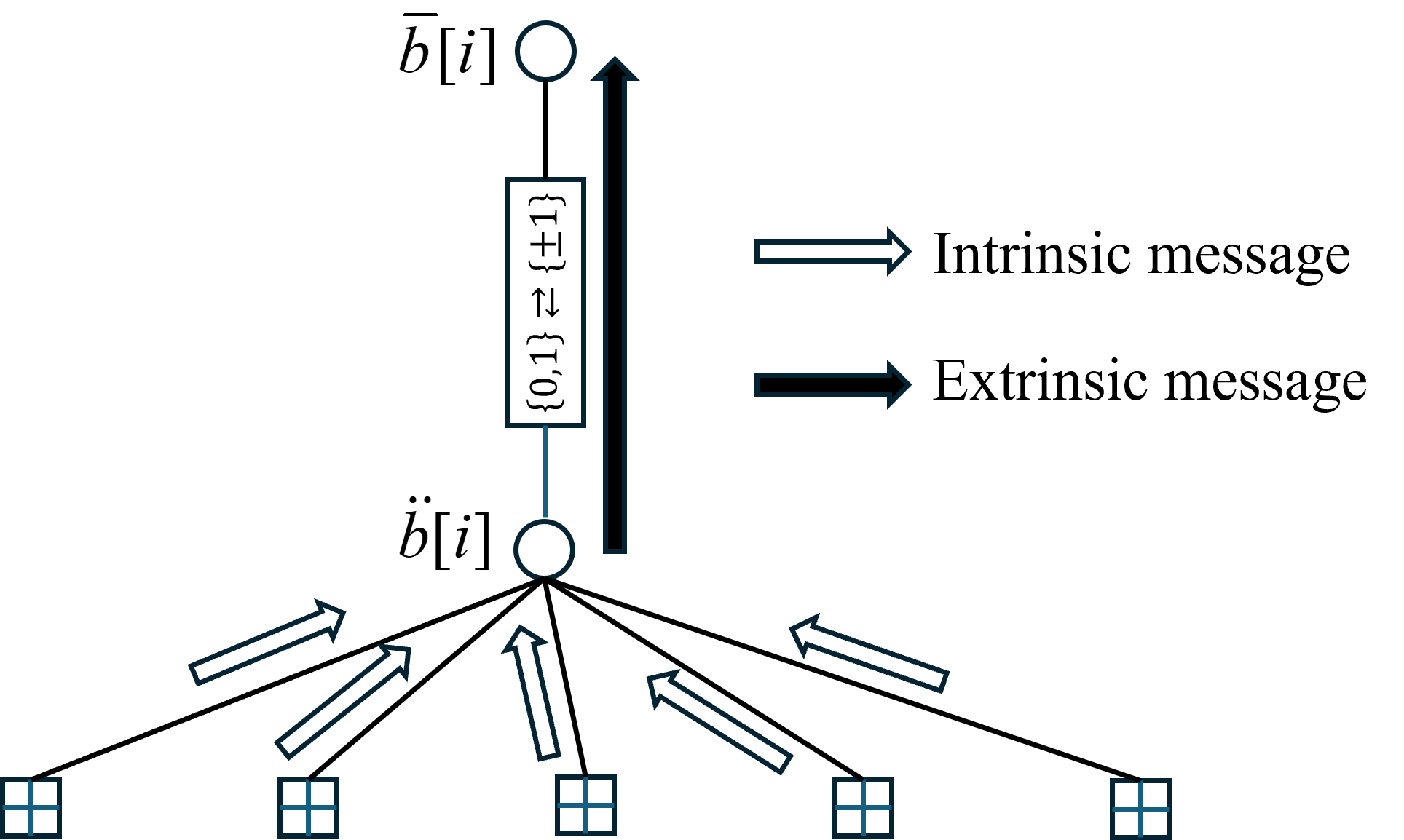}
    \caption{Equality constraint enforcement at the binary variable node during the final ($S$-th) digital sub-iteration. All intrinsic messages from parity-check nodes are combined without exclusion to produce a final enhanced extrinsic message. This message is then passed transparently through the binary-to-bipolar conversion interface back to the bipolar variable node in the analog domain.}
    \label{fig:BinaryToBipolarExtrinsicTransfer}
\end{figure}

Upon receiving extrinsic messages from connected parity-check nodes, each binary variable node updates its outgoing messages differently depending on the sub-iteration:
\begin{itemize}

    \item In sub-iterations $1$ through $S-1$, each binary variable node combines all incoming extrinsic messages from connected parity-check nodes \emph{except} the one it is replying to, \emph{together with} the intrinsic message received from its corresponding bipolar variable node during the previous analog-domain stage. This combination is enforced via an \emph{equality constraint}, yielding an enhanced extrinsic message that is sent back to the excluded parity-check node. This process avoids self-reinforcement while incorporating prior analog-domain beliefs to improve decoding convergence. The update mechanism is illustrated in Figure~\ref{fig:EqualityConstraintLoopbackDigital}, where incoming messages from parity-check nodes are combined at a binary variable node along with the analog-side intrinsic message, and the resulting extrinsic message is routed to the sidelined parity-check node.

    \item In the final $S$-th sub-iteration, each binary variable node aggregates \emph{all} incoming extrinsic messages from its connected parity-check nodes using the equality constraint. The resulting refined belief is then passed transparently to the corresponding bipolar variable node in the analog part through the binary-to-bipolar conversion interface. This final transfer is illustrated in Figure~\ref{fig:BinaryToBipolarExtrinsicTransfer}, which shows the full integration of incoming messages at a binary variable node and the direct transmission of the final enhanced extrinsic message to its bipolar counterpart.

\end{itemize}

This digital-domain update schedule allows the \gls{ldpc} decoder to iteratively refine the bit reliability estimates while avoiding self-reinforcement. Once the $S$ sub-iterations are completed, the most accurate belief for each bit—now expressed at the binary node level—is handed off to the analog domain to continue the global inference process. This alternating process between analog and digital stages embodies a turbo-like iterative structure that gradually improves bit estimates across the entire factor graph.

After completing the final global iteration $U$, hard decisions on the transmitted bits are made by computing, at each binary variable node, a \gls{llr} that fuses two sources of soft information: the extrinsic message passed from the analog domain at the end of the analog stage of iteration $U$, and the extrinsic message produced within the digital domain at the end of the corresponding digital stage. These two messages are combined through an \emph{equality constraint} applied at the variable node, yielding a final belief that reflects the total accumulated reliability across all stages of the iterative process. The signs of these \glspl{llr} determine the final bit decisions. This concludes the iterative procedure, which relies on the coordinated interaction between analog-domain detection and digital-domain \gls{ldpc} decoding to progressively refine the bit estimates and enhance overall decoding performance.

The joint message-passing decoding framework introduced in~\cite{Kurkoski02} uses a binary-domain precoding technique to mitigate error floors caused by ambiguous short patterns in partial-response channels. This precoding operates linearly over the binary field $\mathrm{GF}(2)$, but its behavior changes when translated to the real-valued bipolar domain through binary-to-bipolar mapping. In particular, the linear operations over $\mathrm{GF}(2)$ become nonlinear in $\mathbb{R}$ due to the nature of the conversion. While this precoding structurally resembles \gls{thp}, it cannot be considered a true instance of \gls{thp}. That is because \gls{thp}, although involving a nonlinear modulo operation, is defined over the field of real numbers $\mathbb{R}$ and remains linear in its filtering components within that domain. In contrast, the proposed precoding is fundamentally linear in the binary domain but induces nonlinear behavior when interpreted in $\mathbb{R}$. This distinction is crucial, as it highlights that the mechanism discussed here is not an analog-domain \gls{thp}, but rather a binary-domain analog that achieves similar interference cancellation goals through a different algebraic framework.

Unlike the approach taken in~\cite{Kurkoski02}, where admissible binary-domain precoding filters were selected from a predefined class of degree-$3$ polynomials and validated through simulation and ambiguity analysis, the methodology presented here offers a systematic and generalizable framework for deriving binary-domain precoders. Specifically, our approach identifies appropriate binary-domain precoding filters by analyzing how differences between bipolar input sequences propagate through the channel and impact the resulting real-valued output sequences. The key idea is to ensure that distinct binary input sequences, once mapped to their bipolar equivalents, produce distinguishable outputs at the channel output. This is done by studying the difference between two such real-valued output sequences and guaranteeing that this difference cannot vanish unless the original binary sequences are identical. In doing so, we establish a sufficient condition for avoiding ambiguity: if the modulo-$2$ reduction of a suitably normalized difference sequence at the output of the filter is nonzero, then the original real-valued difference sequence is also nonzero. This condition, in turn, can be enforced by applying a binary-domain precoder whose \glspl{tf} is the inverse of the normalized channel’s modulo-$2$ polynomial, where the channel taps have been first divided by their greatest common divisor to ensure the modulo-$2$ polynomial is non-null and therefore invertible over $\mathrm{GF}(2)$.

Importantly, the proposed analysis does not rely on empirical heuristics or exhaustive enumeration over small polynomial families. Instead, it formulates a sufficient condition based on modulo-$2$ arithmetic and filter inverses over $\mathrm{GF}(2)$ that guarantees injectivity of the composite precoding-plus-channel operation. This makes the technique not only principled but also extensible to a wide class of partial-response channels with arbitrary integer-valued impulse responses. Moreover, this framework recovers the precoding filters $1/(1 + D)$ for the “dicode” channel and $1/(1 + D + D^2 + D^3)$ for the \gls{epr4} channel, used in~\cite{Kurkoski02}, as special cases. It also provides a clearer explanation of why such precoding eliminates ambiguity in state message-passing algorithms and how it maintains distinguishability between different encoded sequences at the channel output.

Building on this foundation, the derivation we present next lays the groundwork for extending the concept of binary-domain precoding from simple scalar partial-response channels to more complex systems. In particular, the same principle will later be extended to the compounded \gls{ldpc} coding and analog \gls{ldgm} \gls{nsm} setting involving vector signal representations, where avoiding destructive collisions between distinct codewords becomes critical for the detection layer. The next steps develop this analysis through formal definitions and transformations, first focusing on the filtering case. The insights gained from this scalar-channel analysis are then used to address the more general vectorized and matrix transformation scenarios arising in the compounded \gls{ldpc} and analog \gls{ldgm} framework.

We begin by considering two distinct binary input sequences $\ddot{b}_0[l]$ and $\ddot{b}_1[l],$ each taking values in $\{0,1\}.$ These sequences are mapped into their bipolar counterparts via the transformation $\bar{b}[l] = 2\ddot{b}[l] - 1,$ which yields sequences $\bar{b}_0[l]$ and $\bar{b}_1[l]$ with values in $\{-1,+1\}.$ The difference between these bipolar sequences is then defined as $\Delta \bar{b}[l] \triangleq \bar{b}_1[l] - \bar{b}_0[l].$ Because each $\bar{b}[l]$ takes values in $\{ \pm 1 \},$ the difference $\Delta \bar{b}[l]$ can only assume values in $\{-2,0,2\},$ and is therefore always an even integer-valued sequence.

This bipolar difference sequence is passed through the channel, modeled as a finite impulse response filter with integer taps $\mathring{h}[l].$ The channel output sequences corresponding to the two bipolar inputs are $\mathring{s}_0[l] = \bar{b}_0[l] \circledast \mathring{h}[l]$ and $\mathring{s}_1[l] = \bar{b}_1[l] \circledast \mathring{h}[l].$ The difference between the output sequences is given by $\Delta \mathring{s}[l] \triangleq \mathring{s}_1[l] - \mathring{s}_0[l] = \Delta \bar{b}[l] \circledast \mathring{h}[l].$ Since both $\mathring{h}[l]$ and $\Delta \bar{b}[l]$ are integer-valued and $\Delta \bar{b}[l]$ is even-valued, it follows that $\Delta \mathring{s}[l]$ is also an even integer-valued sequence.

To simplify the analysis, we normalize the output difference by dividing by $2$:
\begin{equation}
\frac{\Delta \mathring{s}[l]}{2} = \left( \frac{\Delta \bar{b}[l]}{2} \right) \circledast \mathring{h}[l].
\end{equation}
Because $\Delta \bar{b}[l]/2$ takes values in $\{-1,0,1\}$ and $\mathring{h}[l]$ has integer taps, the normalized difference sequence $\Delta \mathring{s}[l]/2$ is integer-valued.

The central challenge is to ensure that $\Delta \mathring{s}[l] \neq 0$ whenever $\ddot{b}_0[l] \neq \ddot{b}_1[l]$. In other words, the channel must map distinct binary input sequences to distinct real-valued output sequences, avoiding ambiguity in the decoding process. To establish a sufficient condition for this injectivity, we consider the modulo-$2$ reduction of the normalized output difference:
\begin{equation}
\left( \frac{\Delta \mathring{s}[l]}{2} \bmod 2 \right).
\end{equation}
If this modulo-$2$ sequence is nonzero, then the original (unnormalized) output difference $\Delta \mathring{s}[l]$ cannot be zero, guaranteeing distinguishability.

Next, we express the modulo-$2$ normalized output difference in terms of the binary sequences at the input. Starting from the bipolar difference, observe that
\begin{equation}
\Delta \bar{b}[l] = \bar{b}_1[l] - \bar{b}_0[l] = 2(\ddot{b}_1[l] - \ddot{b}_0[l]),
\end{equation}
since $\bar{b}[l] = 2\ddot{b}[l] - 1.$ Note that the binary difference, $\ddot{b}_1[l] - \ddot{b}_0[l],$ can be expressed modulo $2$ as the binary sum $\ddot{b}_0[l] + \ddot{b}_1[l]$ over $\mathrm{GF}(2).$ Applying the modulo-$2$ operation to the normalized output difference, linearity over $\mathrm{GF}(2)$ implies
\begin{equation}
\left( \frac{\Delta \mathring{s}[l]}{2} \bmod 2 \right) = (\ddot{b}_0[l] + \ddot{b}_1[l]) \circledast \ddot{h}[l],
\end{equation}
where $\ddot{h}[l]$ denotes the modulo-$2$ version of the channel impulse response $\mathring{h}[l]$, obtained by reducing each of its integer taps modulo $2$. This relation reveals that the modulo-$2$ normalized output difference is the binary convolution of the modulo-$2$ channel with the binary sum of the two input sequences.

From this, the problem of ensuring non-vanishing output differences reduces to guaranteeing that $(\ddot{b}_0[l] + \ddot{b}_1[l]) \circledast \ddot{h}[l] \neq 0,$ for any distinct $\ddot{b}_0[l] \neq \ddot{b}_1[l].$ To enforce this, we apply a binary-domain precoding filter defined as the inverse of the modulo-$2$ channel polynomial $\ddot{h}(D) \triangleq \sum_l \ddot{h}[l] D^l,$ in the delay operator domain, where $D$ denotes the unit delay operator. That is, the precoder has \glspl{tf} $1/\ddot{h}(D).$

As previously mentioned in this section, before applying the modulo-$2$ reduction, the channel impulse response $\mathring{h}[l]$ is first normalized by dividing all its taps by their \gls{gcd}. This step ensures that the resulting impulse response contains at least one odd-valued tap, which guarantees that the modulo-$2$ reduction $\ddot{h}[l]$ is not identically zero. As a result, the corresponding polynomial $\ddot{h}(D)$ is invertible over $\mathrm{GF}(2)$, and the precoding filter $1/\ddot{h}(D)$ is well-defined and can be implemented in the binary domain.

By precoding the input sequences with $1/\ddot{h}(D)$, the convolution with $\ddot{h}(D)$ at the channel output effectively cancels, yielding
\begin{equation}
\left( \frac{\Delta \mathring{s}[l]}{2} \bmod 2 \right) = \ddot{b}_0[l] + \ddot{b}_1[l] \neq 0,
\end{equation}
for distinct inputs $\ddot{b}_0[l] \neq \ddot{b}_1[l].$ This establishes that the normalized modulo-$2$ output difference is nonzero and therefore the real-valued output difference $\Delta \mathring{s}[l]$ is nonzero as well. Hence, the binary precoding filter defined by $1/\ddot{h}(D)$ guarantees injectivity of the composite precoding-plus-channel operation.

Applying this framework to specific channels recovers known results: for the “dicode” channel with impulse response $\mathring{h}[l] = \delta[l]-\delta[l-1],$ the modulo-$2$ channel polynomial is $\ddot{h}(D) = 1 + D,$ yielding the classical precoder $1/(1 + D).$ For the \gls{epr4} channel with $\mathring{h}[l] = \delta[l]+\delta[l-1] -\delta[l-2]-\delta[l-3],$ the modulo-$2$ polynomial is $\ddot{h}(D) = 1 + D + D^2 + D^3,$ leading to the precoder $1/(1 + D + D^2 + D^3)$.

This derivation rigorously explains why the binary precoding filters eliminate ambiguity in the decoding process and maintain distinguishability between different encoded sequences at the channel output.

In some particular cases, however, full binary-domain precoding may not be necessary. For example, if the modulo-$2$ polynomial $\ddot{h}(D)$ associated with the channel has the form $\ddot{h}(D) = D^q$ for some $q \in \mathbb{N}$, then the precoding filter $1/\ddot{h}(D)$ reduces to a simple delay operation. Since delays do not affect injectivity in this context, no precoding is required. A concrete example is the channel $\mathring{h}[l] = 2\delta[l] + \delta[l-1],$ for which the modulo-$2$ version is $\ddot{h}(D) = D$.

More generally, when computing $\ddot{h}(D)$, the lowest power of $D$ with a nonzero coefficient may be strictly positive. Let $q$ denote the smallest exponent in $\ddot{h}(D)$ such that the corresponding coefficient is $1.$ Then, rather than inverting $\ddot{h}(D)$ directly, it suffices to precode using $1/(\ddot{h}(D)/D^q),$ which is a lower-degree polynomial over $\mathrm{GF}(2).$ This simplification reduces the number of memory elements needed in the precoder and hence its complexity.

This type of optimization appears explicitly in~\cite{Kurkoski02}, where the \gls{epr4} channel, having an integer impulse response $\mathring{h}(D) = (1 - D)(1 + D)^2,$ admits partial precoding. Instead of inverting the full modulo-$2$ polynomial $\ddot{h}(D),$ the authors precode only the $(1 - D)$ factor, which suffices to resolve ambiguities. This simplification is possible because $\mathring{h}(D)$ is reducible over the ring of integer polynomials, and its binary counterpart $\ddot{h}(D)$ inherits this factorization structure over $\mathrm{GF}(2).$ As a result, one can identify individual factors in $\ddot{h}(D)$ that contribute to the ambiguity and target them with partial precoding.

Furthermore, even if the integer-valued channel $\mathring{h}(D)$ is irreducible over the ring of polynomials with integer coefficients, it may still become reducible after modulo-$2$ reduction. In such cases, the polynomial $\ddot{h}(D)$ over $\mathrm{GF}(2)$ may admit nontrivial factorizations that enable partial precoding with lower-degree binary filters. If at least one of these factors is sufficient to prevent ambiguity between distinct binary inputs, then a significant reduction in precoding complexity can be achieved without sacrificing performance. To the best of our knowledge, this possibility has not been previously explored and constitutes a novel generalization of the binary-domain precoding approach initiated in~\cite{Kurkoski02}.

These observations motivate further exploration of partial precoding schemes based on factorization of $\ddot{h}(D)$ in $\mathrm{GF}(2),$ particularly in channels where full precoding would otherwise be complex or costly. Such an approach may be especially beneficial when targeting systems with strict hardware constraints or aiming to preserve other structural properties in the binary encoding layer.

In coded communication systems, the presence of error-correcting codes opens the possibility of integrating the binary-domain precoding operation into the broader binary encoding process. This integration is straightforward in the case of arbitrary block codes, where sparsity constraints are not critical: the precoding filter can be absorbed into the binary encoder without concern for structural overhead. However, in systems relying on \gls{ldpc} codes, preserving the sparsity of the parity-check matrix is essential for maintaining decoding efficiency. This constraint makes it problematic to incorporate a general binary filter directly into the encoding layer, as it risks introducing dense dependencies that hinder iterative decoding. Fortunately, as will be shown next, this challenge can be addressed within the framework of compounded \gls{ldpc} coding and analog \gls{ldgm} \gls{nsm}. There, we will introduce a mechanism that accommodates binary-domain precoding while preserving the necessary sparsity properties of the \gls{ldpc} structure.

Before turning to the compounded \gls{ldpc} and analog \gls{ldgm} framework, it is insightful to reinterpret the combined effect of the binary-domain precoding and the partial-response channel in a more abstract system-theoretic light. Specifically, this combination can be viewed as forming an \gls{arma} structure. In this interpretation, the precoder acts as an \gls{ar} component operating over $\mathrm{GF}(2)$, while the analog partial-response channel contributes a \gls{ma-mov} component over $\mathbb{R}$. This decomposition gives rise to a hybrid-domain direct form II \gls{arma} realization. However, unlike classical \gls{arma} systems, the \gls{ar} and \gls{ma-mov} parts here reside in distinct algebraic domains and therefore cannot be merged into a single reduced canonical form. Despite this separation, a one-to-one correspondence exists between the memory elements of the precoder and those of the channel, meaning that the overall number of trellis states remains unchanged. This structural insight explains why, in~\cite{Kurkoski02}, the inclusion of binary-domain precoding in the detection process does not increase state complexity.

Having established the scalar filtering case, the insights gained here form the basis for extending the analysis to vectorized and matrix-valued transformations encountered in compounded \gls{ldpc} coding and analog \gls{ldgm} \gls{nsm} frameworks, where vector-valued analog modulation replaces scalar filtering.

We begin by considering two distinct $k \times 1$ binary codewords, $\ddot{\bm{b}}_0$ and $\ddot{\bm{b}}_1,$ with components in $\{0,1\}.$ These vectors are valid codewords of the \gls{ldpc} code defined by a full-rank $m \times k$ parity-check matrix $\bm{H}$ of size, such that $\ddot{\bm{b}}_0 \bm{H}^T = \bm{0}$ and $\ddot{\bm{b}}_1 \bm{H}^T = \bm{0}.$ Their bipolar versions are defined as $\bar{\bm{b}}_0 = 2\ddot{\bm{b}}_0 - \bm{1}$ and $\bar{\bm{b}}_1 = 2\ddot{\bm{b}}_1 - \bm{1},$ where $\bm{1}$ denotes the $k \times 1$ all-ones vector.

These bipolar vectors are then passed through the analog \gls{ldgm} \gls{nsm} encoder, which is characterized by the scaled $k \times n$ generator matrix $\mathring{\bm{G}}.$ This matrix has integer entries and is generally sparse to enable low-complexity detection. The resulting modulated analog codewords are given by $\mathring{\bm{s}}_0 = \bar{\bm{b}}_0 \mathring{\bm{G}}$ and $\mathring{\bm{s}}_1 = \bar{\bm{b}}_1 \mathring{\bm{G}},$ where $\mathring{\bm{s}}_0$ and $\mathring{\bm{s}}_1$ are $1 \times n$ row vectors with components in $\mathbb{Z}.$

To ensure distinguishability at the detection layer, we require that $\mathring{\bm{s}}_0 \ne \mathring{\bm{s}}_1$ whenever $\ddot{\bm{b}}_0 \ne \ddot{\bm{b}}_1.$ Toward this end, we define the difference vector $\Delta \mathring{\bm{s}} \triangleq \mathring{\bm{s}}_1 - \mathring{\bm{s}}_0 = (\bar{\bm{b}}_1 - \bar{\bm{b}}_0) \mathring{\bm{G}} = \Delta \bar{\bm{b}} \mathring{\bm{G}}.$ The bipolar difference vector $\Delta \bar{\bm{b}} = \bar{\bm{b}}_1 - \bar{\bm{b}}_0$ can be expressed as $\Delta \bar{\bm{b}} = 2(\ddot{\bm{b}}_1 - \ddot{\bm{b}}_0),$
which implies that each component of $\Delta \bar{\bm{b}}$ belongs to the set $\{-2, 0, +2\}.$

As in the scalar case, we divide the output difference vector by $2$ to obtain
\begin{equation}
\frac{\Delta \mathring{\bm{s}}}{2} = (\ddot{\bm{b}}_1 - \ddot{\bm{b}}_0) \mathring{\bm{G}},
\end{equation}
and we examine the modulo-$2$ version of this expression to establish a sufficient condition for distinguishability. Letting $\ddot{\bm{G}}$ denote the binary matrix obtained by reducing each entry of $\mathring{\bm{G}}$ modulo $2,$ we compute
\begin{equation}
\left( \frac{\Delta \mathring{\bm{s}}}{2} \bmod 2 \right) = (\ddot{\bm{b}}_0 + \ddot{\bm{b}}_1) \ddot{\bm{G}},
\end{equation}
where the binary sum $\ddot{\bm{b}}_0 + \ddot{\bm{b}}_1$ is taken over $\mathrm{GF}(2).$

Since both $\ddot{\bm{b}}_0$ and $\ddot{\bm{b}}_1$ are valid \gls{ldpc} codewords, their sum $\ddot{\bm{b}}_0 + \ddot{\bm{b}}_1$ is itself a codeword belonging to the \gls{ldpc} code defined by the parity-check matrix $\bm{H}.$ Therefore, to guarantee that $\mathring{\bm{s}}_0 \neq \mathring{\bm{s}}_1$ for any pair of distinct codewords, it suffices to ensure that for every nonzero \gls{ldpc} codeword $\ddot{\bm{b}},$ the following condition holds: $\ddot{\bm{b}} \ddot{\bm{G}} \neq \bm{0}.$

Recall that $\bm{H}$ is an $m \times k$ full-rank parity-check matrix, which defines a dual code subspace of dimension $m$ in $\mathrm{GF}(2)^k.$ The \gls{ldpc} code itself forms a subspace of dimension $k - m.$ Consider a basis of the dual code formed by $m$ linearly independent row vectors of size $1 \times k.$ To satisfy the above condition, it is sufficient to construct the binary matrix $\ddot{\bm{G}}$ so that at least $k - m$ of its columns, when reduced modulo $2,$ complement this basis of the dual code to form a complete basis of $\mathrm{GF}(2)^k.$ In other words, aggregating the $m$ basis vectors of the dual code with the transposes of these $k - m$ columns of $\ddot{\bm{G}}$ yields a full basis of the ambient vector space. This condition ensures that any nonzero \gls{ldpc} codeword $\ddot{\bm{b}},$ which is orthogonal to all dual code basis vectors, must produce a nonzero scalar product with at least one of the selected columns of $\ddot{\bm{G}}.$ Consequently, the product $\ddot{\bm{b}} \ddot{\bm{G}}$ cannot be the zero vector, guaranteeing the distinguishability of modulated codewords. However, since $\ddot{\bm{G}}$ generally has $n$ columns, this requirement implies that $k - m \leq n,$ which restricts the overall spectral efficiency $\rho.$ Recall that the overall coded modulation rate is $\rho = \rho_H \rho_G = (k - m)/n.$ Thus, the sufficient condition framework inherently limits the rate to $\rho \leq 1.$

It is important to emphasize that this condition is a strong sufficient one, relying on the modulo-$2$ reduction of the generator matrix $\mathring{\bm{G}}.$ In practice, two distinct integer-valued vectors may differ by an even integer vector that reduces to zero modulo $2.$ Therefore, requiring the modulo-$2$ difference to be nonzero is quite restrictive and may not hold when aiming for spectral efficiencies greater than $1.$ For these higher-rate regimes, the sufficient condition based on modulo-$2$ reduction is no longer applicable, and more nuanced design and analysis are required. Still, the spanning condition provides valuable insight and guidance for the construction of $\mathring{\bm{G}}$ when $\rho \leq 1,$ ensuring fundamental distinguishability properties within this regime.

\begin{table}[H]
\centering
\caption{Candidate patterns for rate‑$2$ analog LDGM NSMs with $\kappa = 2$ and $\lambda_0 = \lambda_1 = 2$. The fourth column shows the number of distinct outcomes arising from all ${\pm1}$‑weighted sums of the combined components of $\bm{\pi}_0$ and $\bm{\pi}_1$. The fifth column gives the entropy, $\mathcal{H},$ of the resulting distribution, and the sixth column shows the normalized entropy, computed relative to $\mathcal{H}_{\text{max}} = \lambda_0 + \lambda_1 = 4$, which is the entropy of $2^{(\lambda_0 + \lambda_1)} = 16$ equally likely sign combinations.} \label{table:Candidate Patterns Rate-2 Lambda_0 = Lambda_1 = 2}
\begin{tabular}{|c|c|c|c|c|c|}
\hline
$\bm{\pi}_0$ & $\bm{\pi}_1$ & $\|\bm{\pi}\|^2$ & \# Diff. Vals & $\mathcal{H}$ & $\mathcal{H}/\mathcal{H}_{\text{max}}$ \\ \hline
(1,8) & (4,7) & 65 & 14 & 3.75 & 0.9375 \\ \hline
(2,9) & (6,7) & 85 & 14 & 3.75 & 0.9375 \\ \hline
(2,11) & (5,10) & 125 & 16 & 4 & 1 \\ \hline
(3,11) & (7,9) & 130 & 16 & 4 & 1 \\ \hline
(1,12) & (8,9) & 145 & 14 & 3.75 & 0.9375 \\ \hline
(1,13) & (7,11) & 170 & 16 & 4 & 1 \\ \hline
(4,13) & (8,11) & 185 & 16 & 4 & 1 \\ \hline
(3,14) & (6,13) & 205 & 16 & 4 & 1 \\ \hline
(5,14) & (10,11) & 221 & 16 & 4 & 1 \\ \hline
(5,15) & (9,13) & 250 & 16 & 4 & 1 \\ \hline
(3,16) & (11,12) & 265 & 16 & 4 & 1 \\ \hline
(1,17) & (11,13) & 290 & 16 & 4 & 1 \\ \hline
(4,17) & (7,16) & 305 & 16 & 4 & 1 \\ \hline
(1,18) & (6,17) & 325 & 14 & 3.75 & 0.9375 \\ \hline
(1,18) & (10,15) & 325 & 16 & 4 & 1 \\ \hline
(6,17) & (10,15) & 325 & 16 & 4 & 1 \\ \hline
(2,19) & (13,14) & 365 & 16 & 4 & 1 \\ \hline
(3,19) & (9,17) & 370 & 16 & 4 & 1 \\ \hline
(4,19) & (11,16) & 377 & 16 & 4 & 1 \\ \hline
(7,19) & (11,17) & 410 & 16 & 4 & 1 \\ \hline
(5,20) & (8,19) & 425 & 16 & 4 & 1 \\ \hline
(5,20) & (13,16) & 425 & 16 & 4 & 1 \\ \hline
(8,19) & (13,16) & 425 & 16 & 4 & 1 \\ \hline
(1,21) & (9,19) & 442 & 16 & 4 & 1 \\ \hline
(2,21) & (11,18) & 445 & 16 & 4 & 1 \\ \hline
\end{tabular}
\end{table}

\begin{table}[H]
\centering
\caption{Candidate patterns for rate‑$2$ analog LDGM NSMs with $\kappa = 2,$ $\lambda_0 = 2$ and $\lambda_1 = 3.$ The fourth column shows the number of distinct outcomes arising from all ${\pm1}$‑weighted sums of the combined components of $\bm{\pi}_0$ and $\bm{\pi}_1$. The fifth column gives the entropy, $\mathcal{H},$ of the resulting distribution, and the sixth column shows the normalized entropy, computed relative to $\mathcal{H}_{\text{max}} = \lambda_0 + \lambda_1 = 5$, which is the entropy of $2^{(\lambda_0 + \lambda_1)} = 32$ equally likely sign combinations.} \label{table:Candidate Patterns Rate-2 Lambda_0 = 2 Lambda_1 = 3}
\begin{tabular}{|c|c|c|c|c|c|}
\hline
$\bm{\pi}_0$ & $\bm{\pi}_1$ & $\|\bm{\pi}\|^2$ & \# Diff. Vals & $\mathcal{H}$ & $\mathcal{H}/\mathcal{H}_{\text{max}}$ \\ \hline
(4,5) & (1,2,6) & 41 & 19 & 4.0931 & 0.81863 \\ \hline
(3,6) & (2,4,5) & 45 & 19 & 4.1403 & 0.82806 \\ \hline
(1,7) & (3,4,5) & 50 & 19 & 4.1403 & 0.82806 \\ \hline
(2,7) & (1,4,6) & 53 & 21 & 4.2653 & 0.85306 \\ \hline
(1,8) & (2,5,6) & 65 & 21 & 4.2653 & 0.85306 \\ \hline
(4,7) & (2,5,6) & 65 & 21 & 4.2653 & 0.85306 \\ \hline
(5,7) & (1,3,8) & 74 & 23 & 4.3903 & 0.87806 \\ \hline
(5,8) & (2,6,7) & 89 & 23 & 4.3903 & 0.87806 \\ \hline
(3,9) & (1,5,8) & 90 & 23 & 4.3903 & 0.87806 \\ \hline
(3,9) & (4,5,7) & 90 & 23 & 4.3903 & 0.87806 \\ \hline
(1,10) & (2,4,9) & 101 & 25 & 4.5625 & 0.9125 \\ \hline
(1,10) & (4,6,7) & 101 & 23 & 4.3903 & 0.87806 \\ \hline
(7,8) & (2,3,10) & 113 & 23 & 4.3903 & 0.87806 \\ \hline
(6,9) & (1,4,10) & 117 & 23 & 4.3903 & 0.87806 \\ \hline
(6,9) & (2,7,8) & 117 & 23 & 4.3903 & 0.87806 \\ \hline
(1,11) & (3,7,8) & 122 & 23 & 4.3903 & 0.87806 \\ \hline
(1,11) & (4,5,9) & 122 & 23 & 4.4375 & 0.8875 \\ \hline
(2,11) & (3,4,10) & 125 & 27 & 4.6875 & 0.9375 \\ \hline
(2,11) & (5,6,8) & 125 & 23 & 4.3903 & 0.87806 \\ \hline
(4,11) & (1,6,10) & 137 & 23 & 4.3903 & 0.87806 \\ \hline
(1,12) & (3,6,10) & 145 & 27 & 4.6875 & 0.9375 \\ \hline
(8,9) & (3,6,10) & 145 & 27 & 4.6875 & 0.9375 \\ \hline
(5,11) & (1,8,9) & 146 & 27 & 4.6875 & 0.9375 \\ \hline
(5,11) & (4,7,9) & 146 & 23 & 4.3903 & 0.87806 \\ \hline
(7,10) & (1,2,12) & 149 & 26 & 4.625 & 0.925 \\ \hline
\end{tabular}
\end{table}

\begin{table}[H]
\centering
\caption{Candidate patterns for rate‑$2$ analog LDGM NSMs with $\kappa = 2$ and $\lambda_0 = \lambda_1 = 3$. The fourth column shows the number of distinct outcomes arising from all ${\pm1}$‑weighted sums of the combined components of $\bm{\pi}_0$ and $\bm{\pi}_1$. The fifth column gives the entropy, $\mathcal{H},$ of the resulting distribution, and the sixth column shows the normalized entropy, computed relative to $\mathcal{H}_{\text{max}} = \lambda_0 + \lambda_1 = 6$, which is the entropy of $2^{(\lambda_0 + \lambda_1)} = 64$ equally likely sign combinations.} \label{table:Candidate Patterns Rate-2 Lambda_0 = Lambda_1 = 3}
\begin{tabular}{|c|c|c|c|c|c|}
\hline
$\bm{\pi}_0$ & $\bm{\pi}_1$ & $\|\bm{\pi}\|^2$ & \# Diff. Vals & $\mathcal{H}$ & $\mathcal{H}/\mathcal{H}_{\text{max}}$ \\ \hline
(1,5,6) & (2,3,7) & 124 & 25 & 4.4528 & 0.74214 \\ \hline
(2,3,8) & (4,5,6) & 154 & 27 & 4.5931 & 0.76552 \\ \hline
(2,6,7) & (3,4,8) & 178 & 29 & 4.6945 & 0.78242 \\ \hline
(1,4,9) & (3,5,8) & 196 & 29 & 4.6213 & 0.77021 \\ \hline
(1,6,8) & (2,4,9) & 202 & 31 & 4.796 & 0.79933 \\ \hline
(1,2,10) & (4,5,8) & 210 & 31 & 4.8195 & 0.80326 \\ \hline
(1,2,4) & (5,8,10) & 210 & 31 & 4.8195 & 0.80326 \\ \hline
(1,3,10) & (2,5,9) & 220 & 31 & 4.8195 & 0.80326 \\ \hline
(1,3,10) & (5,6,7) & 220 & 31 & 4.8195 & 0.80326 \\ \hline
(1,4,10) & (2,7,8) & 234 & 33 & 4.8897 & 0.81495 \\ \hline
(3,7,8) & (4,5,9) & 244 & 31 & 4.7966 & 0.79943 \\ \hline
(3,4,10) & (5,6,8) & 250 & 33 & 4.9056 & 0.81761 \\ \hline
(2,5,10) & (4,7,8) & 258 & 33 & 4.898 & 0.81633 \\ \hline
(1,3,4) & (7,8,11) & 260 & 30 & 4.6525 & 0.77542 \\ \hline
(2,7,9) & (3,5,10) & 268 & 31 & 4.773 & 0.7955 \\ \hline
(1,4,11) & (5,7,8) & 276 & 33 & 4.8591 & 0.80984 \\ \hline
(1,8,14) & (2,3,4) & 290 & 33 & 4.9841 & 0.83068 \\ \hline
(1,8,9) & (3,4,11) & 292 & 33 & 4.898 & 0.81633 \\ \hline
(1,2,12) & (6,7,8) & 298 & 33 & 4.8744 & 0.8124 \\ \hline
(1,7,10) & (2,5,11) & 300 & 35 & 4.9917 & 0.83195 \\ \hline
(1,2,5) & (7,10,11) & 300 & 35 & 4.9917 & 0.83195 \\ \hline
(1,6,11) & (3,7,10) & 316 & 35 & 4.9369 & 0.82281 \\ \hline
(1,4,12) & (2,6,11) & 322 & 37 & 5.0542 & 0.84237 \\ \hline
(1,4,12) & (5,6,10) & 322 & 33 & 4.8744 & 0.8124 \\ \hline
(2,6,11) & (4,8,9) & 322 & 33 & 4.8827 & 0.81378 \\ \hline
\end{tabular}
\end{table}

\begin{table}[H]
\centering
\caption{Candidate patterns for rate‑$3$ analog LDGM NSMs with $\kappa = 3$ and $\lambda_0 = \lambda_1 = \lambda_2 = 3$. The fourth column shows the number of distinct outcomes arising from all ${\pm1}$‑weighted sums of the combined components of $\bm{\pi}_0,$ $\bm{\pi}_1$ and $\bm{\pi}_2$. The fifth column gives the entropy, $\mathcal{H},$ of the resulting distribution, and the sixth column shows the normalized entropy, computed relative to $\mathcal{H}_{\text{max}} = \lambda_0 + \lambda_1  + \lambda_2 = 9$, which is the entropy of $2^{(\lambda_0 + \lambda_1  + \lambda_2)} = 512$ equally likely sign combinations.} \label{table:Candidate Patterns Rate-3 Lambda_0 = Lambda_1 = Lambda_2 = 3}
\begin{tabular}{|c|c|c|c|c|c|c|}
\hline
$\bm{\pi}_0$ & $\bm{\pi}_1$ & $\bm{\pi}_2$ & $\|\bm{\pi}\|^2$ & \# Diff. Vals & $\mathcal{H}$ & $\mathcal{H}/\mathcal{H}_{\text{max}}$ \\ \hline
(2,5,12) & (3,8,10) & (4,6,11) & 173 & 60 & 5.53914 & 0.61546 \\ \hline
(1,9,10) & (2,3,13) & (5,6,11) & 182 & 61 & 5.57237 & 0.61915 \\ \hline
(2,4,13) & (3,6,12) & (5,8,10) & 189 & 62 & 5.60056 & 0.62228 \\ \hline
(1,3,14) & (5,9,10) & (6,7,11) & 206 & 65 & 5.66132 & 0.62904 \\ \hline
(1,2,15) & (3,5,14) & (7,9,10) & 230 & 67 & 5.73474 & 0.63719 \\ \hline
(1,2,15) & (3,10,11) & (5,6,13) & 230 & 67 & 5.73396 & 0.63711 \\ \hline
(1,2,15) & (5,6,13) & (7,9,10) & 230 & 67 & 5.73131 & 0.63681 \\ \hline
(1,2,16) & (4,7,14) & (6,9,12) & 261 & 72 & 5.83019 & 0.6478 \\ \hline
(1,3,16) & (4,5,15) & (8,9,11) & 266 & 71 & 5.81384 & 0.64598 \\ \hline
(2,3,16) & (5,10,12) & (6,8,13) & 269 & 72 & 5.82872 & 0.64764 \\ \hline
(1,10,13) & (3,6,15) & (5,7,14) & 270 & 73 & 5.85209 & 0.65023 \\ \hline
(1,9,14) & (2,7,15) & (3,10,13) & 278 & 75 & 5.86753 & 0.65195 \\ \hline
(1,6,16) & (2,8,15) & (4,9,14) & 293 & 76 & 5.91325 & 0.65703 \\ \hline
(1,6,16) & (2,8,15) & (7,10,12) & 293 & 74 & 5.89433 & 0.65493 \\ \hline
(1,6,16) & (4,9,14) & (7,10,12) & 293 & 76 & 5.9008 & 0.65564 \\ \hline
(2,8,15) & (4,9,14) & (7,10,12) & 293 & 76 & 5.9002 & 0.65558 \\ \hline
(1,12,13) & (3,4,17) & (5,8,15) & 314 & 77 & 5.92298 & 0.65811 \\ \hline
(1,12,13) & (3,7,16) & (5,8,15) & 314 & 79 & 5.93551 & 0.6595 \\ \hline
(3,4,17) & (5,8,15) & (7,11,12) & 314 & 77 & 5.92015 & 0.65779 \\ \hline
(3,4,17) & (7,11,12) & (8,9,13) & 314 & 77 & 5.92964 & 0.65885 \\ \hline
(1,11,14) & (2,5,17) & (7,10,13) & 318 & 79 & 5.96714 & 0.66302 \\ \hline
(1,10,15) & (6,11,13) & (7,9,14) & 326 & 79 & 5.9774 & 0.66416 \\ \hline
(1,2,18) & (3,8,16) & (4,12,13) & 329 & 78 & 5.98947 & 0.6655 \\ \hline
(2,6,17) & (3,8,16) & (4,12,13) & 329 & 80 & 5.99596 & 0.66622 \\ \hline
(2,10,15) & (3,8,16) & (4,12,13) & 329 & 82 & 5.98972 & 0.66552 \\ \hline
\end{tabular}
\end{table}

\begin{table}[H]
\centering
\caption{Candidate patterns for rate‑$3$ analog LDGM NSMs with $\kappa = 3,$ $\lambda_0 = \lambda_1 = 3$ and $\lambda_2 = 4.$ The fourth column shows the number of distinct outcomes arising from all ${\pm1}$‑weighted sums of the combined components of $\bm{\pi}_0,$ $\bm{\pi}_1$ and $\bm{\pi}_2$. The fifth column gives the entropy, $\mathcal{H},$ of the resulting distribution, and the sixth column shows the normalized entropy, computed relative to $\mathcal{H}_{\text{max}} = \lambda_0 + \lambda_1  + \lambda_2 = 10$, which is the entropy of $2^{(\lambda_0 + \lambda_1  + \lambda_2)} = 1024$ equally likely sign combinations.} \label{table:Candidate Patterns Rate-3 Lambda_0 = Lambda_1 = 3 Lambda_2 = 4}
\begin{tabular}{|c|c|c|c|c|c|c|}
\hline
$\bm{\pi}_0$ & $\bm{\pi}_1$ & $\bm{\pi}_2$ & $\|\bm{\pi}\|^2$ & \# Diff. Vals & $\mathcal{H}$ & $\mathcal{H}/\mathcal{H}_{\text{max}}$ \\ \hline
(1,6,13) & (5,9,10) & (2,3,7,12) & 206 & 69 & 5.67072 & 0.56707 \\ \hline
(3,8,12) & (6,9,10) & (1,2,4,14) & 217 & 70 & 5.70813 & 0.57081 \\ \hline
(4,9,11) & (5,7,12) & (2,3,6,13) & 218 & 71 & 5.7083 & 0.57083 \\ \hline
(1,2,15) & (3,5,14) & (6,7,8,9) & 230 & 71 & 5.74331 & 0.57433 \\ \hline
(1,2,15) & (3,10,11) & (6,7,8,9) & 230 & 73 & 5.74497 & 0.5745 \\ \hline
(1,2,15) & (7,9,10) & (3,4,6,13) & 230 & 71 & 5.74465 & 0.57446 \\ \hline
(1,2,15) & (7,9,10) & (3,6,8,11) & 230 & 73 & 5.74497 & 0.5745 \\ \hline
(3,10,11) & (5,6,13) & (1,2,9,12) & 230 & 73 & 5.75056 & 0.57506 \\ \hline
(3,9,12) & (7,8,11) & (2,5,6,13) & 234 & 73 & 5.7604 & 0.57604 \\ \hline
(2,3,15) & (6,9,11) & (1,4,5,14) & 238 & 71 & 5.76922 & 0.57692 \\ \hline
(2,3,15) & (6,9,11) & (5,7,8,10) & 238 & 73 & 5.77093 & 0.57709 \\ \hline
(1,4,15) & (3,8,13) & (2,6,9,11) & 242 & 73 & 5.78571 & 0.57857 \\ \hline
(1,4,15) & (3,8,13) & (5,6,9,10) & 242 & 73 & 5.77947 & 0.57795 \\ \hline
(1,7,14) & (5,10,11) & (2,3,8,13) & 246 & 75 & 5.79648 & 0.57965 \\ \hline
(3,4,15) & (5,9,12) & (1,2,7,14) & 250 & 73 & 5.80611 & 0.58061 \\ \hline
(2,5,15) & (3,7,14) & (4,6,9,11) & 254 & 75 & 5.81523 & 0.58152 \\ \hline
(2,5,15) & (6,7,13) & (1,3,10,12) & 254 & 75 & 5.8184 & 0.58184 \\ \hline
(2,5,15) & (6,7,13) & (3,8,9,10) & 254 & 75 & 5.81558 & 0.58156 \\ \hline
(1,3,16) & (4,5,15) & (6,7,9,10) & 266 & 75 & 5.84717 & 0.58472 \\ \hline
(1,11,12) & (4,5,15) & (6,7,9,10) & 266 & 77 & 5.84141 & 0.58414 \\ \hline
(1,11,12) & (4,9,13) & (3,5,6,14) & 266 & 77 & 5.85248 & 0.58525 \\ \hline
(1,10,13) & (3,6,15) & (2,8,9,11) & 270 & 79 & 5.86328 & 0.58633 \\ \hline
(1,10,13) & (5,7,14) & (2,8,9,11) & 270 & 79 & 5.86449 & 0.58645 \\ \hline
(1,10,13) & (5,7,14) & (3,6,9,12) & 270 & 79 & 5.8654 & 0.58654 \\ \hline
(3,6,15) & (5,7,14) & (1,2,11,12) & 270 & 77 & 5.8637 & 0.58637 \\ \hline
\end{tabular}
\end{table}

\begin{table}[H]
\centering
\caption{Candidate patterns for rate‑$3$ analog LDGM NSMs with $\kappa = 3,$ $\lambda_0 = 3$ and $\lambda_1 = \lambda_2 = 4.$ The fourth column shows the number of distinct outcomes arising from all ${\pm1}$‑weighted sums of the combined components of $\bm{\pi}_0,$ $\bm{\pi}_1$ and $\bm{\pi}_2$. The fifth column gives the entropy, $\mathcal{H},$ of the resulting distribution, and the sixth column shows the normalized entropy, computed relative to $\mathcal{H}_{\text{max}} = \lambda_0 + \lambda_1  + \lambda_2 = 11$, which is the entropy of $2^{(\lambda_0 + \lambda_1  + \lambda_2)} = 2048$ equally likely sign combinations.} \label{table:Candidate Patterns Rate-3 Lambda_0 = 3 Lambda_1 = Lambda_2 = 4}
\begin{tabular}{|c|c|c|c|c|c|c|}
\hline
$\bm{\pi}_0$ & $\bm{\pi}_1$ & $\bm{\pi}_2$ & $\|\bm{\pi}\|^2$ & \# Diff. Vals & $\mathcal{H}$ & $\mathcal{H}/\mathcal{H}_{\text{max}}$ \\ \hline
(1,5,14) & (2,4,9,11) & (3,7,8,10) & 222 & 75 & 5.72925 & 0.52084 \\ \hline
(2,7,13) & (1,6,8,11) & (4,5,9,10) & 222 & 77 & 5.73052 & 0.52096 \\ \hline
(1,8,13) & (2,3,10,11) & (4,5,7,12) & 234 & 77 & 5.76747 & 0.52432 \\ \hline
(4,7,13) & (1,5,8,12) & (2,3,10,11) & 234 & 77 & 5.76747 & 0.52432 \\ \hline
(1,4,15) & (2,6,9,11) & (3,5,8,12) & 242 & 77 & 5.78936 & 0.52631 \\ \hline
(2,5,15) & (1,3,10,12) & (4,6,9,11) & 254 & 79 & 5.82484 & 0.52953 \\ \hline
(5,8,13) & (1,6,10,11) & (2,3,7,14) & 258 & 81 & 5.83653 & 0.53059 \\ \hline
(4,9,13) & (1,2,6,15) & (3,7,8,12) & 266 & 81 & 5.85838 & 0.53258 \\ \hline
(1,10,13) & (2,4,5,15) & (3,6,9,12) & 270 & 81 & 5.86837 & 0.53349 \\ \hline
(1,10,13) & (2,4,5,15) & (6,7,8,11) & 270 & 83 & 5.87002 & 0.53364 \\ \hline
(1,10,13) & (2,8,9,11) & (3,4,7,14) & 270 & 83 & 5.86963 & 0.5336 \\ \hline
(3,6,15) & (1,2,11,12) & (5,8,9,10) & 270 & 83 & 5.86956 & 0.5336 \\ \hline
(3,6,15) & (1,5,10,12) & (2,4,9,13) & 270 & 81 & 5.86837 & 0.53349 \\ \hline
(3,6,15) & (1,5,10,12) & (2,8,9,11) & 270 & 83 & 5.86956 & 0.5336 \\ \hline
(7,10,11) & (1,3,8,14) & (2,4,5,15) & 270 & 81 & 5.86892 & 0.53354 \\ \hline
(7,10,11) & (1,3,8,14) & (2,4,9,13) & 270 & 83 & 5.86963 & 0.5336 \\ \hline
(7,10,11) & (1,6,8,13) & (2,4,5,15) & 270 & 83 & 5.87002 & 0.53364 \\ \hline
(7,10,11) & (2,4,5,15) & (3,6,9,12) & 270 & 83 & 5.86957 & 0.5336 \\ \hline
(7,9,12) & (2,3,6,15) & (4,5,8,13) & 274 & 83 & 5.87978 & 0.53453 \\ \hline
(3,10,13) & (1,4,6,15) & (2,7,9,12) & 278 & 83 & 5.88783 & 0.53526 \\ \hline
(1,5,16) & (2,3,10,13) & (4,8,9,11) & 282 & 83 & 5.90051 & 0.53641 \\ \hline
(7,8,13) & (1,2,9,14) & (5,6,10,11) & 282 & 85 & 5.90076 & 0.53643 \\ \hline
(7,8,13) & (1,3,4,16) & (5,6,10,11) & 282 & 83 & 5.90014 & 0.53638 \\ \hline
(3,9,14) & (1,2,5,16) & (4,7,10,11) & 286 & 83 & 5.90959 & 0.53724 \\ \hline
(6,9,13) & (1,2,5,16) & (4,7,10,11) & 286 & 85 & 5.91061 & 0.53733 \\ \hline
\end{tabular}
\end{table}

\begin{table}[H]
\centering
\caption{Candidate patterns for rate‑$3$ analog LDGM NSMs with $\kappa = 3$ and $\lambda_0 = \lambda_1 = \lambda_2 = 4$. The fourth column shows the number of distinct outcomes arising from all ${\pm1}$‑weighted sums of the combined components of $\bm{\pi}_0,$ $\bm{\pi}_1$ and $\bm{\pi}_2$. The fifth column gives the entropy, $\mathcal{H},$ of the resulting distribution, and the sixth column shows the normalized entropy, computed relative to $\mathcal{H}_{\text{max}} = \lambda_0 + \lambda_1  + \lambda_2 = 12$, which is the entropy of $2^{(\lambda_0 + \lambda_1  + \lambda_2)} = 4096$ equally likely sign combinations.} \label{table:Candidate Patterns Rate-3 Lambda_0 = Lambda_1 = Lambda_2 = 4}
\begin{tabular}{|c|c|c|c|c|c|c|}
\hline
$\bm{\pi}_0$ & $\bm{\pi}_1$ & $\bm{\pi}_2$ & $\|\bm{\pi}\|^2$ & \# Diff. Vals & $\mathcal{H}$ & $\mathcal{H}/\mathcal{H}_{\text{max}}$ \\ \hline
(1,2,3,16) & (4,6,7,13) & (5,8,9,10) & 270 & 85 & 5.87093 & 0.48924 \\ \hline
(1,2,11,12) & (3,4,7,14) & (5,8,9,10) & 270 & 87 & 5.87296 & 0.48941 \\ \hline
(1,2,11,12) & (4,6,7,13) & (5,8,9,10) & 270 & 89 & 5.87373 & 0.48948 \\ \hline
(1,5,10,12) & (2,4,9,13) & (6,7,8,11) & 270 & 89 & 5.87373 & 0.48948 \\ \hline
(1,5,10,12) & (2,8,9,11) & (3,4,7,14) & 270 & 87 & 5.87296 & 0.48941 \\ \hline
(1,5,10,12) & (2,8,9,11) & (4,6,7,13) & 270 & 89 & 5.87373 & 0.48948 \\ \hline
(1,6,7,14) & (2,3,10,13) & (4,8,9,11) & 282 & 89 & 5.90422 & 0.49202 \\ \hline
(1,4,8,15) & (2,5,9,14) & (6,7,10,11) & 306 & 93 & 5.96299 & 0.49692 \\ \hline
(1,4,8,15) & (2,9,10,11) & (5,6,7,14) & 306 & 93 & 5.96299 & 0.49692 \\ \hline
(1,2,4,17) & (5,8,10,11) & (6,7,9,12) & 310 & 93 & 5.97154 & 0.49763 \\ \hline
(1,2,7,16) & (3,6,11,12) & (4,5,10,13) & 310 & 91 & 5.97132 & 0.49761 \\ \hline
(1,7,8,14) & (3,6,11,12) & (4,5,10,13) & 310 & 93 & 5.97252 & 0.49771 \\ \hline
(1,2,12,13) & (3,7,8,14) & (4,9,10,11) & 318 & 95 & 5.99039 & 0.4992 \\ \hline
(1,5,6,16) & (3,7,8,14) & (4,9,10,11) & 318 & 93 & 5.98994 & 0.49916 \\ \hline
(1,2,6,17) & (3,5,10,14) & (4,7,11,12) & 330 & 93 & 6.01575 & 0.50131 \\ \hline
(1,2,6,17) & (3,5,10,14) & (4,8,9,13) & 330 & 93 & 6.01554 & 0.50129 \\ \hline
(1,3,8,16) & (2,7,9,14) & (5,6,10,13) & 330 & 95 & 6.01622 & 0.50135 \\ \hline
(1,3,8,16) & (4,7,11,12) & (5,6,10,13) & 330 & 95 & 6.01704 & 0.50142 \\ \hline
(1,8,11,12) & (2,7,9,14) & (5,6,10,13) & 330 & 97 & 6.01765 & 0.50147 \\ \hline
(1,8,11,12) & (3,4,7,16) & (5,6,10,13) & 330 & 95 & 6.01704 & 0.50142 \\ \hline
(1,4,11,14) & (2,5,7,16) & (3,6,8,15) & 334 & 93 & 6.02418 & 0.50201 \\ \hline
(1,4,11,14) & (2,5,7,16) & (3,9,10,12) & 334 & 95 & 6.02505 & 0.50209 \\ \hline
(1,7,11,13) & (2,4,8,16) & (3,5,9,15) & 340 & 95 & 6.03795 & 0.50316 \\ \hline
(1,6,7,16) & (2,5,12,13) & (3,4,11,14) & 342 & 95 & 6.04209 & 0.50351 \\ \hline
(1,3,4,18) & (2,9,11,12) & (5,6,8,15) & 350 & 95 & 6.05676 & 0.50473 \\ \hline
\end{tabular}
\end{table}

\begin{table}[H]
\centering
\caption{Characteristics of the compound of rate $\rho_H=1/2$ extended Hamming code, specified by the parity check matrix in (\ref{eq: Parity Check Matrix Rate 1/2 LDPC Code}), and the rate $\rho_G=2$ analog LDGM \gls{nsm}, specified by the analog LDGM generating matrix in (\ref{eq: Analog LDGM Rate 2 NSM}).} \label{table:Characteristics of LDPC & Analog LDGM NSM}
\begin{tabular}{|c|c|c|}
\hline
Binary Codeword & Bipolar Codeword & Modulated Codeword \\
\hline
$(0,0,0,0,0,0,0,0)$ & $(-1,-1,-1,-1,-1,-1,-1,-1)$ & $(-6,-6,-6,-6)$ \\ 
$(1,1,1,0,0,0,0,1)$ & $(1,1,1,-1,-1,-1,-1,1)$ & $(-10,14,-8,-12)$ \\ 
$(1,1,0,1,0,0,1,0)$ & $(1,1,-1,1,-1,-1,1,-1)$ & $(14,-8,-12,-10)$ \\ 
$(0,0,1,1,0,0,1,1)$ & $(-1,-1,1,1,-1,-1,1,1)$ & $(-6,-4,6,4)$ \\ 
$(1,0,1,1,0,1,0,0)$ & $(1,-1,1,1,-1,1,-1,-1)$ & $(-8,-12,-10,14)$ \\ 
$(0,1,0,1,0,1,0,1)$ & $(-1,1,-1,1,-1,1,-1,1)$ & $(-8,8,-8,8)$ \\ 
$(0,1,1,0,0,1,1,0)$ & $(-1,1,1,-1,-1,1,1,-1)$ & $(-4,6,4,-6)$ \\ 
$(1,0,0,0,0,1,1,1)$ & $(1,-1,-1,-1,-1,1,1,1)$ & $(12,10,-14,8)$ \\ 
$(0,1,1,1,1,0,0,0)$ & $(-1,1,1,1,1,-1,-1,-1)$ & $(-12,-10,14,-8)$ \\ 
$(1,0,0,1,1,0,0,1)$ & $(1,-1,-1,1,1,-1,-1,1)$ & $(4,-6,-4,6)$ \\ 
$(1,0,1,0,1,0,1,0)$ & $(1,-1,1,-1,1,-1,1,-1)$ & $(8,-8,8,-8)$ \\ 
$(0,1,0,0,1,0,1,1)$ & $(-1,1,-1,-1,1,-1,1,1)$ & $(8,12,10,-14)$ \\ 
$(1,1,0,0,1,1,0,0)$ & $(1,1,-1,-1,1,1,-1,-1)$ & $(6,4,-6,-4)$ \\ 
$(0,0,1,0,1,1,0,1)$ & $(-1,-1,1,-1,1,1,-1,1)$ & $(-14,8,12,10)$ \\ 
$(0,0,0,1,1,1,1,0)$ & $(-1,-1,-1,1,1,1,1,-1)$ & $(10,-14,8,12)$ \\ 
$(1,1,1,1,1,1,1,1)$ & $(1,1,1,1,1,1,1,1)$ & $(6,6,6,6)$ \\ 
\hline
\end{tabular}
\end{table}

\subsection{Integrating Source Coding, NSM Modulation, and LDPC Coding: A Unified Framework}

The design of reliable and efficient digital communication systems typically involves three fundamental processes: source coding, channel coding, and modulation. Traditionally, these operations are addressed separately, following the source–channel separation principle. However, in many practical scenarios—especially those constrained by delay, complexity, or finite block length—\gls{jscc} schemes offer superior performance. Among the various approaches to \gls{jscc}, lattice-based methods provide a promising framework due to their regular structure, geometric properties, and suitability for both multidimensional quantization and coded modulation.

In \gls{jscc}, the core objective is to integrate compression and error protection into a unified framework, removing the need for strict modularity between source coding, channel coding, and modulation. A real-valued source vector $\bm{x} \in \mathbb{R}^{1 \times q}$ is first quantized using a lattice-based source code, producing an approximation $\hat{\bm{x}} \in \mathbb{R}^{1 \times q}$ that minimizes distortion, typically measured using the \gls{sed}. The source coding lattice is designed to offer efficient covering in $\mathbb{R}^q$, thereby reducing the average quantization error. The quantization index is then mapped to a binary vector $\bar{\bm{b}} \in \{0,1\}^{1 \times k}$, serving as an intermediate representation between the source coding lattice and a lattice-based modulation scheme operating in dimension $n$. This binary vector is subsequently mapped to a channel input vector $\bm{s} \in \mathbb{R}^{1 \times n}$, chosen from a high packing-efficiency lattice constellation to enhance robustness against channel noise. After transmission through a noisy channel, the receiver observes a corrupted version of this signal, denoted by $\bm{r} \in \mathbb{R}^{1 \times n}$, from which a reconstruction of the original source is attempted.

The design of the source and modulation lattices typically targets opposing geometric objectives: efficient covering in \(\mathbb{R}^q\) to minimize quantization distortion, and efficient packing in \(\mathbb{R}^n\) to maximize noise immunity. These lattices operate in different dimensions and are connected via the intermediate bit vector \(\bar{\bm{b}}\), which defines a mapping ideally preserving geometric proximity across both domains. Ideally, two lattice points that are near in the source quantization lattice should be mapped to nearby points in the modulation lattice, and conversely, two nearby points in the modulation lattice should correspond to close points in the source domain—akin to a multidimensional Gray mapping. This bidirectional proximity preservation helps ensure that small perturbations, whether due to quantization or channel noise, lead to only minor deviations in the reconstructed signal. However, constructing such a distance-preserving mapping between lattices in different dimensions (\(q \neq n\)) is generally intractable. Even approximate solutions become difficult to implement, especially when shaping constraints—such as spherical or Voronoi shaping—are introduced to approach fundamental limits on distortion and power efficiency.

Early lattice-based \gls{jscc} efforts—such as those formalized in~\cite{RodriguezGuisantes97,RodriguezGuisantes00}—investigated joint optimization of source and channel lattices and proposed proximity-preserving mappings to mitigate geometric mismatch across the two coding domains. These studies demonstrated the limitations of conventional separation-based architectures and motivated the development of more integrated, structure-aware \gls{jscc} schemes. Building on these insights, several subsequent research directions have emerged over the past decades, differing in methodology and in the extent to which lattice structure is preserved or relaxed. These approaches can be broadly categorized into two main lines of work:

\begin{itemize}

  \item \textbf{Lattice-based extensions for structured joint coding:} More recent works have revisited the lattice approach in modern contexts such as distributed and federated learning. For example, \cite{AzimiAbarghouyi24} proposes using lattice codes for simultaneous quantization and channel coding in over-the-air model aggregation, effectively leveraging both packing and covering properties in a distributed \gls{jscc} setting.

  \item \textbf{Neural JSCC architectures:} Deep learning–based methods have emerged as a competitive alternative to analytical designs. In~\cite{Bourtsoulatze19}, convolutional autoencoders are trained end-to-end to perform both source compression and channel coding directly in the signal domain, without any explicit use of lattice structures. Similarly, \cite{Xu23} explores transformer-based architectures for semantic-aware \gls{jscc}. While these neural approaches offer strong empirical performance, they tend to obscure the geometric interpretability and analytical tractability that lattice-based methods naturally provide.

\end{itemize}

A third and structurally unified direction builds on the algebraic and graph-based synergy between source coding, channel coding, and modulation. This approach generalizes the two-part \gls{ldpc}–\gls{nsm} compounding described earlier, in Section~\ref{A natural and efficient synergy between NSM and LDPC coding}, into a broader three-part integration. The key idea is to treat the lossy source quantization step, the \gls{ldpc} error correction stage, and the analog \gls{nsm} modulation process as interdependent components governed by a shared latent binary variable. Rather than being connected through sequential blocks, these components are woven into a single factor-graph-driven structure that reveals a deep alignment between the roles of quantization, coding, and modulation.

Specifically, let $\bm{x}$ denote the real-valued source vector to be quantized. The system searches for a reconstruction $\hat{\bm{x}}$ that both minimizes the distortion with respect to $\bm{x}$ and satisfies binary parity-check constraints. The reconstruction takes the form $\hat{\bm{x}} = \bar{\bm{b}} \bm{G}_\text{F}$, where $\bm{G}_\text{F}$ is the source projection matrix, and $\bar{\bm{b}} \in \{\pm 1\}^k$ is a bipolar vector whose binary version $\ddot{\bm{b}} \in \text{GF}(2)^k$ satisfies the \gls{ldpc} constraint: $\ddot{\bm{b}} \bm{H}^T = \bm{0}.$

This same bipolar vector $\bar{\bm{b}}$ then produces the modulated sequence via $\bm{s} = \bar{\bm{b}} \bm{G}_\text{M}$, where $\bm{G}_\text{M}$ is the analog \gls{nsm} generating matrix. Both $\bm{G}_\text{F}$ and $\bm{G}_\text{M}$ may be chosen to have rational entries as a smart design choice to simplify processing complexity. In such cases, they can be scaled to their integer counterparts, denoted $\mathring{\bm{G}}_\text{F}$ and $\mathring{\bm{G}}_\text{M}$, for practical implementation, resulting in $\hat{\mathring{\bm{x}}} = \bar{\bm{b}} \mathring{\bm{G}}_\text{F}$ and $\mathring{\bm{s}} = \bar{\bm{b}} \mathring{\bm{G}}_\text{M}$.

To further reduce decoding complexity and maintain scalability, both $\mathring{\bm{G}}_\text{F}$ and $\mathring{\bm{G}}_\text{M}$ can be designed to be sparse. This sparsity ensures that each output symbol (in the case of $\mathring{\bm{G}}_\text{M}$) or each reconstructed component (in the case of $\mathring{\bm{G}}_\text{F}$) depends on only a small subset of the input bits, enabling efficient message passing during both quantization and detection. The resulting factor graph representation remains computationally tractable, even for large block lengths.

Both $\bm{G}_\text{F}$ and $\bm{G}_\text{M}$ may be chosen to have rational entries as a smart design choice to simplify processing complexity. In such cases, they can be scaled to their integer counterparts, denoted $\mathring{\bm{G}}_\text{F}$ and $\mathring{\bm{G}}_\text{M}$, for practical implementation, resulting in $\hat{\mathring{\bm{x}}} = \bar{\bm{b}} \mathring{\bm{G}}_\text{F}$ and $\mathring{\bm{s}} = \bar{\bm{b}} \mathring{\bm{G}}_\text{M}$.

As in Section~\ref{A natural and efficient synergy between NSM and LDPC coding}, regularization constraints can be applied to ensure that all columns (respectively, rows) of $\mathring{\bm{G}}_\text{M}$ have equal squared Euclidean norms, ensuring uniform energy distribution across output symbols (respectively, input bits). Identical constraints can be applied to $\mathring{\bm{G}}_\text{F}$ to promote balanced distortion and stability in quantization. These constraints mirror the degree-regularity principles of \gls{ldpc} parity-check matrices, which remain sparse by construction.

While it is natural to enforce uniform row norms independently within each matrix, an additional degree of flexibility becomes apparent when recognizing that $\mathring{\bm{G}}_\text{F}$ and $\mathring{\bm{G}}_\text{M}$ share the same number of rows, namely $k$, and that each row index $i$ in these matrices corresponds to the same entry $\bar{b}[i]$ of the bipolar vector. This alignment establishes a one-to-one correspondence between the rows of $\mathring{\bm{G}}_\text{F}$ and those of $\mathring{\bm{G}}_\text{M}$. As a result, it is not strictly necessary for all row norms within each matrix to be identical. Instead, the squared Euclidean norms of the $i$-th rows of $\mathring{\bm{G}}_\text{F}$ and $\mathring{\bm{G}}_\text{M}$ can vary individually, provided they satisfy a common proportionality constraint across the two matrices. That is, each pair of corresponding rows $(\mathring{\bm{h}}^\text{F}_i, \mathring{\bm{h}}^\text{M}_i)$ may have distinct norms as long as $\|\mathring{\bm{h}}^\text{M}_i\|^2 \propto \|\mathring{\bm{h}}^\text{F}_i\|^2$ for all $i \in \{0, 1, \ldots, k-1\}$. This proportionality preserves balance in the contributions of the underlying binary variables $\bar{b}[i]$ to both the quantization error and the transmission energy, enabling more nuanced trade-offs between source distortion and channel robustness.

These structural relationships are tightly connected to the dimensional rates already discussed. The quantization rate $\rho_\text{F} = q/k$ captures the compression ratio from the binary latent vector to the source space, and is typically smaller than $1$, reflecting that $\bar{\bm{b}}$ has more degrees of freedom than the source vector it reconstructs. On the other hand, the modulation rate $\rho_\text{M} = k/n$ is usually slightly greater than $1$, indicating that the analog \gls{nsm} stage maps the latent bits to a lower-dimensional analog signal than required by conventional $2$-ASK modulation. The overall system rate is therefore $\rho = q/n = \rho_\text{F} \rho_\text{M}$, which generally falls below $1$.

This asymmetry in rates—namely, $\rho_\text{F} < 1$ and $\rho_\text{M} > 1$—has direct implications for implementation. In particular, when scaled integer matrices $\mathring{\bm{G}}_\text{F}$ and $\mathring{\bm{G}}_\text{M}$ are used, the source synthesis matrix $\mathring{\bm{G}}_\text{F}$ typically requires a richer integer alphabet than the modulation matrix $\mathring{\bm{G}}_\text{M}$. This reflects the differing objectives: $\mathring{\bm{G}}_\text{F}$ must finely approximate continuous-valued sources, while $\mathring{\bm{G}}_\text{M}$ need only ensure robust transmission over the channel. Additionally, the \gls{ldpc} parity-check matrix $\bm{H}$, with $m$ rows, imposes $m$ parity constraints on the binary codeword $\ddot{\bm{b}}$. As $m$ increases, the solution space becomes more structured but also more constrained. If one wishes to increase $m$ while preserving the number of degrees of freedom (i.e., keeping $k - m$ constant), then $k$ must increase accordingly. This translates into a larger latent dimension $k$ and thus an increased number of rows in both $\bm{G}_\text{F}$ and $\bm{G}_\text{M}$, expanding the ambient space for both quantization and modulation. When the matrices are constrained to have scaled integer entries, this increase in dimensionality necessitates a corresponding enrichment of the integer alphabets used, so that the rows remain efficiently distinguishable in the higher-dimensional ambient space.

The analog \gls{nsm} generating matrix $\bm{G}_\text{M}$ can be constructed with pseudo-random rows that act as quasi-orthogonal vectors in $\mathbb{R}^n$, especially when their length approaches the full modulation dimension. In this setting, small subsets of rows behave locally as if drawn from an orthogonal basis, effectively creating a $2$-ASK-like environment. This enhances the local minimum-distance properties of the code and provides robustness against low-weight error events in the \gls{ldpc} code. A similar argument applies to $\bm{G}_\text{F}$: when constructed with random or pseudo-random synthesis vectors, it ensures that different $\bar{\bm{b}}$ vectors produce well-separated reconstructions $\hat{\bm{x}} = \bar{\bm{b}} \bm{G}_\text{F}$, reinforcing quantization stability and granularity.

Suppose now two distinct bipolar codewords $\bar{\bm{b}}_0$ and $\bar{\bm{b}}_1$ differ in $d_\text{H}$ positions. If the rows of $\bm{G}_\text{M}$ and $\bm{G}_\text{F}$ are pseudo-random and of large size, the vectors in the difference $\bar{\bm{b}}_0 - \bar{\bm{b}}_1$ act as selectors over quasi-orthogonal subsets of rows. When these rows have identical squared norms, the \gls{sed} between the two corresponding analog outputs—either modulated signals in $\mathbb{R}^n$ or reconstructed quantized vectors in $\mathbb{R}^q$—becomes approximately proportional to $d_\text{H}$. Therefore, the neighborhood structures in the source and modulation domains are naturally aligned through their shared dependence on $\bar{\bm{b}}$. Importantly, the closest quantized vectors tend to correspond to bipolar vectors differing in only a few positions, and these, in turn, yield modulated signals that are also closely spaced. Conversely, the nearest modulated signals—those lying close to each other in $\mathbb{R}^n$—typically correspond to bipolar vectors at small Hamming distance, which again implies that their associated reconstructions in $\mathbb{R}^q$ are also close. This bidirectional consistency implies that the system exhibits a local Gray-like behavior across both domains, without explicitly constructing a Gray mapping.

This property generalizes further: even when the matrices $\bm{G}_\text{F}$ and $\bm{G}_\text{M}$ do not have equal-norm rows, but rather their squared norms vary in a controlled way—such that the variation in $\bm{G}_\text{F}$ reflects that in $\bm{G}_\text{M}$ proportionally—the squared distances between analog points remain approximately preserved across domains. This insight stems from the generalized regularity condition mentioned earlier, in which the reconstruction and generating matrices are allowed to have rows with different norms, provided those norms are matched proportionally. As a result, the structure guarantees that even with this relaxed constraint, neighborhood relations are preserved across source and modulation domains. This observation highlights a major advantage of the proposed framework: it avoids the intractable search for explicit distance-preserving mappings between high-dimensional lattices of different dimension ($q \ne n$), while still achieving their key benefit.

It is important to emphasize that in this setting, the \gls{ldpc} code is not used to transmit a message in the conventional sense. There is no encoding of a separate information vector. Instead, the binary vector $\ddot{\bm{b}}$ itself plays the central role. Its consistency with the parity-check matrix $\bm{H}$, through the condition $\ddot{\bm{b}} \bm{H}^T = \bm{0}$, guarantees structural constraints that shape locally the geometry of both the source approximations and the modulated signals. Importantly, the \gls{ldpc} parity constraints limit $\bar{\bm{b}}$ to a subset of vectors that maintain these local geometric relationships. The most probable error events correspond to small Hamming distance deviations in $\bar{\bm{b}}$, which, due to the quasi-orthogonality and proportional row norm properties of $\bm{G}_\text{F}$ and $\bm{G}_\text{M}$, preserve the approximate Euclidean distance proportionality across source and modulation domains. Therefore, the presence of \gls{ldpc} coding enhances reliability without disrupting the natural local Gray-like behavior that arises from the joint matrix design. The \gls{ldpc} code thus acts as a discrete tool enforcing complementary local structuring in both the source and modulation spaces, improving quantization resolution and transmission robustness.

This complementary local structuring manifests differently across the two domains: in the modulation space, the \gls{ldpc} parity constraints impose a discrete packing on the bipolar vector $\bar{\bm{b}}$, organizing the signal points efficiently within the continuous lattice-like structure created by the analog \gls{ldgm} \gls{nsm}. Conversely, in the source space, the \gls{ldpc} constraints induce a form of digital covering that complements the analog covering properties of the overcomplete frame projection through $\bm{G}_\text{F}$. Together, this tandem of packing in the modulation domain and covering in the source domain establishes a robust geometric and probabilistic framework that underpins the joint source-channel-modulation coding scheme.

\begin{figure}[!htbp]
    \centering
    \includegraphics[width=0.85\textwidth]{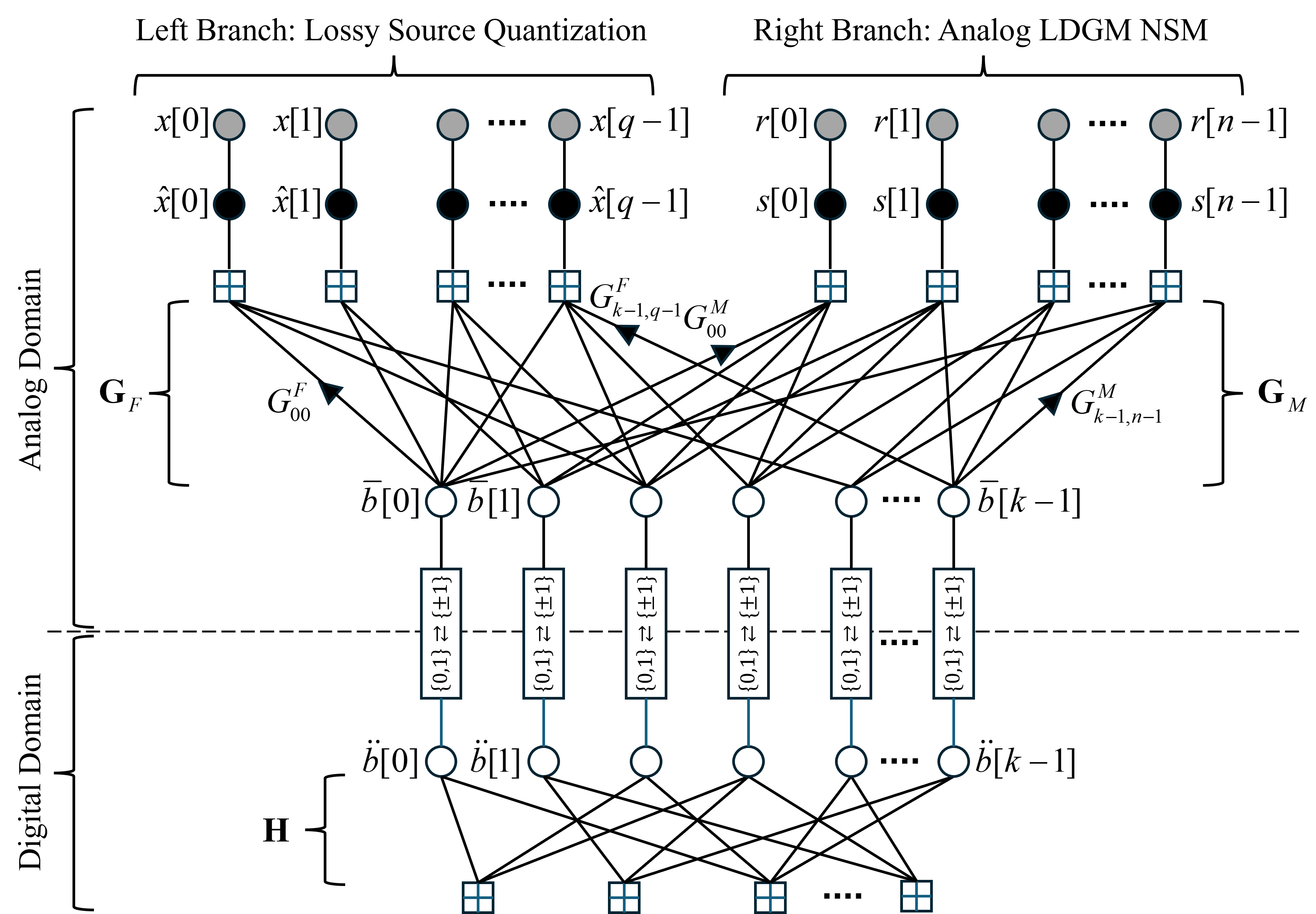}
    \caption{Y-shaped factor graph representing the joint integration of source coding, modulation, and LDPC parity constraints through the shared binary vector $\bar{\bm{b}}$. The left branch models lossy source quantization via the synthesis matrix $\bm{G}_\text{F}$, the right branch handles analog LDGM \gls{nsm} through $\bm{G}_\text{M}$, and the lower branch enforces LDPC code membership using the parity-check matrix $\bm{H}$.}
    \label{fig:Y_graph}
\end{figure}

This structural synergy among quantization, modulation, and parity constraints is most effectively captured by the factor-graph perspective. These interactions can be visualized through a Y-shaped factor graph, as illustrated in Figure~\ref{fig:Y_graph}, where the shared binary vector $\bar{\bm{b}}$ acts as a central latent variable coordinating the three computational branches. The left arm performs lossy source coding through projection onto an overcomplete frame using the synthesis matrix $\bm{G}_\text{F}$. The right arm carries out modulation via analog \gls{nsm} based on the generating matrix $\bm{G}_\text{M}$. The lower stem enforces digital parity constraints through the \gls{ldpc} parity-check matrix $\bm{H}$. This tripartite graphical model not only reveals the joint influence of all three components on the final reconstruction and modulation processes but also makes explicit the structural pathways by which reliability, fidelity, and efficiency are simultaneously achieved. The resulting architecture combines reconstruction fidelity, energy-efficient modulation, and error protection in a unified and structurally coherent design. 

Furthermore, this Y-shaped factor graph structure offers practical algorithmic advantages. The shared binary vector $\bar{\bm{b}}$ serves as the nexus of message passing, enabling seamless iterative exchange of soft information between source quantization, modulation, and parity-check constraint domains. This joint inference process improves convergence speed and decoding accuracy compared to separate or sequential processing blocks. The error correction coding provided by the \gls{ldpc} code not only structures the solution space but also strengthens the iterative message passing convergence and robustness at both the transmitter side—enabling iterative source quantization—and the receiver side—supporting turbo-like joint detection and decoding. Moreover, the sparse nature of the involved matrices $\bm{G}_\text{F}$, $\bm{G}_\text{M}$, and $\bm{H}$ keeps computational complexity manageable, even for large dimensions. These properties collectively enable scalable and robust implementations of the proposed joint source-channel-modulation coding system.

The proposed architecture exhibits a layered structure in how energy is distributed across target dimensions, a property we refer to as \emph{stratification}. Each bipolar symbol $\bar{b}[i]$ produces a localized signal in both the source and modulation domains through its corresponding $i$-th row in the reconstruction and generating matrices $\bm{G}_\text{F}$ and $\bm{G}_\text{M}$, respectively. The instantaneous power at coordinate index $j$ is simply the square of the corresponding matrix entry, written as $P_i[j] = (G_{\text{F},i}[j])^2$ or $P_i[j] = (G_{\text{M},i}[j])^2$, depending on the domain. These per-symbol power profiles reflect how energy from each component is distributed across the vector dimensions.

\begin{figure}[!htbp]
    \centering
    \includegraphics[width=0.9\linewidth]{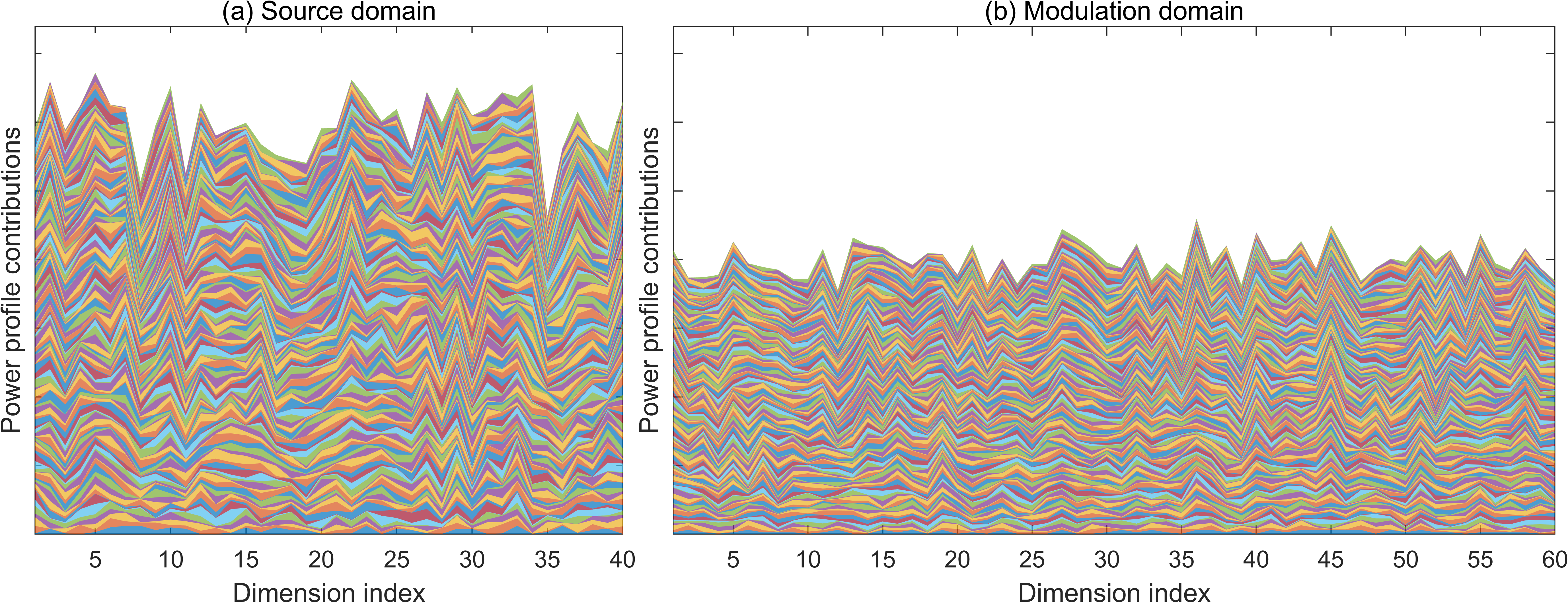}
    \caption{%
        Stratified instantaneous power profiles in the full-dimensional-span case for $q = 40$, $n = 60$, and $k = 120$. Each colored layer represents the squared values of a single bipolar symbol's row in the reconstruction matrix $\bm{G}_\text{F}$ (panel a) or the generating matrix $\bm{G}_\text{M}$ (panel b). Here, each row spans the full output dimension, resulting in globally distributed and relatively uniform profiles. Rows are normalized to unit energy for visualization and stacked vertically to reveal the structured energy distribution across dimensions. The vertical scaling between panels reflects the $n/q = 60/40 = 3/2$ ratio between modulation and source dimensions.
    }
    \label{fig:stratified_profiles_full}
\end{figure}

\begin{figure}[!htbp]
    \centering
    \includegraphics[width=0.9\linewidth]{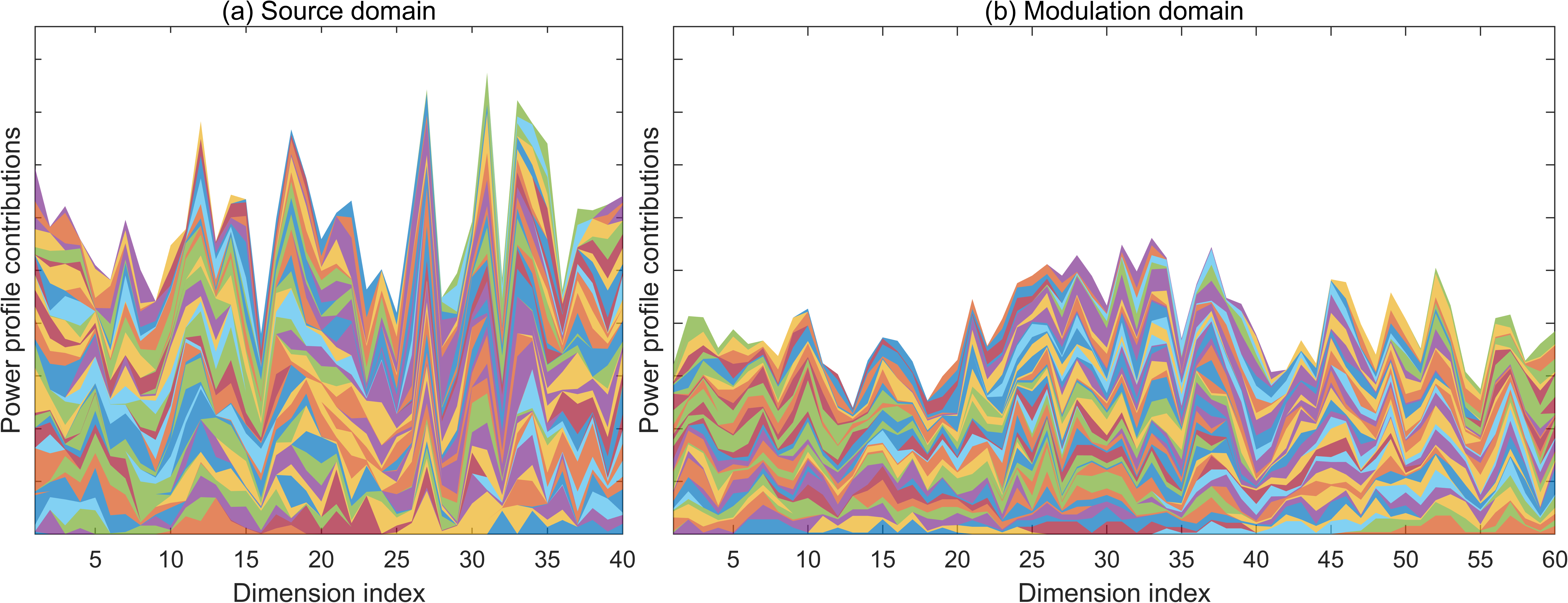}
    \caption{%
        Stratified instantaneous power profiles in the limited-dimensional-span case for parameters $q = 40$, $n = 60$, and $k = 120$. Each row in $\bm{G}_\text{F}$ and $\bm{G}_\text{M}$ contributes energy over a restricted support: $10$ dimensions in the source domain and $15$ dimensions in the modulation domain, respectively, simulating localized interactions. This localization leads to sharper and more concentrated energy patterns, particularly in the source domain. As in the full-span case, each row is normalized to unit energy for visualization, and the panels maintain the $3/2$ vertical scaling between modulation and source dimensions.
    }
    \label{fig:stratified_profiles_limited}
\end{figure}

Figures~\ref{fig:stratified_profiles_full} and~\ref{fig:stratified_profiles_limited} illustrate this concept for $q = 40$, $n = 60$, and $k = 120$. In both cases, each row is normalized to unit energy for visualization purposes. In Figure~\ref{fig:stratified_profiles_full}, the rows span the entire dimension range in each domain, producing smooth, globally distributed power patterns. In contrast, Figure~\ref{fig:stratified_profiles_limited} reflects a scenario where each bipolar symbol contributes only within a limited range of dimensions, mimicking local support or neighborhood constraints as required in practical message-passing algorithms. The resulting energy patterns are more concentrated and localized, particularly in the source domain. Despite these differences, the total energy contributed by all bipolar symbols remains the same. The modulation and source domains differ in size—in this example, $n/q = 60/40 = 3/2$—so the vertical scaling between panels reflects this ratio, with modulation dimensions spread across a wider base and accumulating energy more gradually. In both visualizations, rows are stacked vertically and color-coded to form a stratified view of energy flow. Note that the normalization here is performed on the original integer-valued matrices $\bm{G}_\text{F}$ and $\bm{G}_\text{M}$ without the scaling applied in the normalized versions $\mathring{\bm{G}}_\text{F}$ and $\mathring{\bm{G}}_\text{M}$. This allows for consistent visualization of the energy distribution per bipolar symbol while preserving the integer-structured nature of the matrices.

This Y-structured model generalizes the compound scheme described in Section~\ref{A natural and efficient synergy between NSM and LDPC coding}, where only the modulation and channel coding components were integrated. By introducing a third structural branch for source quantization, the proposed framework achieves full joint source-channel-modulation integration while preserving the sparsity and compatibility properties that made the \gls{ldpc}-\gls{nsm} synergy efficient in the first place. By analogy with the current Y-shaped architecture, the earlier two-branch structure involving only analog \gls{ldgm} modulation and \gls{ldpc} coding may be retrospectively referred to as an I-shaped structure, emphasizing its simpler vertical alignment and more limited scope. In the uncoded case, where the \gls{ldpc} branch is removed, the Y-shaped factor graph naturally degenerates into a v-shaped graph, with only source quantization and modulation branches linked by $\bar{\bm{b}}$. Likewise, the I-shaped structure reduces to a simpler i-shaped structure when the \gls{ldpc} coding component is omitted, reflecting progressively simpler system designs. These uncoded configurations can be interpreted as special cases of the more general coded architectures, obtained by deactivating the parity-check constraints. However, they are not recommended in practice due to their fragility: the absence of \gls{ldpc} coding makes the system more sensitive to channel noise and degrades the performance of iterative message-passing algorithms, both at the transmitter, during source quantization, and at the receiver, during turbo-like joint detection. Robustness and convergence are significantly enhanced when \gls{ldpc} coding is used to facilitate bidirectional message exchange—between source coding and parity constraints during quantization, and between modulation and parity constraints during detection—while maintaining structural consistency through the shared latent variable $\bar{\bm{b}}$.

At the transmitter, the system performs the following steps. First, it receives a source vector $\bm{x}$ to be quantized. It then searches for the best reconstruction $\hat{\bm{x}} = \bar{\bm{b}} \bm{G}_\text{F}$ minimizing distortion, where $\bar{\bm{b}}$ is the bipolar version of a binary codeword $\ddot{\bm{b}}$ satisfying the \gls{ldpc} constraint $\ddot{\bm{b}} \bm{H}^T = \bm{0}$. This constrained search can be handled through message passing on the Y-shaped factor graph, in a fashion similar to the turbo-decoding of the analog \gls{ldgm}-\gls{ldpc} compound scheme. Once the best $\bar{\bm{b}}$ (or equivalently, $\ddot{\bm{b}}$) is found, the corresponding analog signal $\bm{s} = \bar{\bm{b}} \bm{G}_\text{M}$ is computed and transmitted through the physical channel.

At the receiver, the process reverses. The system observes a received vector $\bm{r}$ resulting from the channel transformation of $\bm{s}$. It seeks a binary vector $\ddot{\bm{b}}$ such that its bipolar version $\bar{\bm{b}}$ yields a modulated signal $\bm{s} = \bar{\bm{b}} \bm{G}_\text{M}$ that is as close as possible to $\bm{r}$ in \gls{sed}, and such that $\ddot{\bm{b}}$ satisfies the \gls{ldpc} code constraint $\ddot{\bm{b}} \bm{H}^T = \bm{0}$. This again amounts to a joint decoding task, that can be performed via belief propagation or sum-product algorithms on the Y-structured factor graph. Once the best $\bar{\bm{b}}$ is identified, the original source is approximately reconstructed as $\hat{\bm{x}} = \bar{\bm{b}} \bm{G}_\text{F}$.

This Y-structured model achieves tight coupling among the source, channel, and modulation layers, through aligned sparse operators and a unified binary representation. In contrast to traditional modular pipelines or deep black-box solutions, it offers a tractable and interpretable \gls{jscc} system grounded in linear algebra, factor graphs, and sparse coding principles.

To illustrate the joint source–modulation design, we consider the following parameter values: the source dimension $q = 4$, the latent binary dimension $k = 12$, the modulation dimension $n = 6$, and the \gls{ldpc} code includes $m = 6$ parity-check equations.

The \gls{ldpc} parity-check matrix $\bm{H}$ is chosen to be a regular binary matrix of size $m \times k = 6 \times 12$ with column weight $d_v = 2$ and row weight $d_c = 4$. Here, $d_v$ denotes the number of ones per column (variable node degree), and $d_c$ denotes the number of ones per row (check node degree). Such a matrix enforces balanced structural constraints on the binary codewords $\ddot{\bm{b}}$. We consider the following parity-check matrix for our \gls{ldpc} code example:
\begin{equation}
\bm{H} =
\left[
\begin{array}{cccccccccccc}
1 & 1 & 0 & 0 & 1 & 0 & 0 & 0 & 1 & 0 & 0 & 0 \\
0 & 0 & 1 & 1 & 0 & 1 & 0 & 0 & 0 & 1 & 0 & 0 \\
0 & 0 & 0 & 0 & 0 & 0 & 1 & 1 & 0 & 0 & 1 & 1 \\
1 & 0 & 1 & 0 & 0 & 0 & 1 & 0 & 0 & 0 & 0 & 1 \\
0 & 1 & 0 & 1 & 0 & 1 & 0 & 1 & 0 & 0 & 0 & 0 \\
0 & 0 & 0 & 0 & 1 & 0 & 0 & 0 & 1 & 1 & 1 & 0
\end{array}
\right]
\label{eq:H_regular}
\end{equation}

The scaled source synthesis matrix $\mathring{\bm{G}}_\text{F} \in \mathbb{Z}^{12 \times 4}$ and the scaled \gls{nsm} generating matrix $\mathring{\bm{G}}_\text{M} \in \mathbb{Z}^{12 \times 6}$ are both constructed using the tail-biting, cyclic shift structure previously described in Section~\ref{A natural and efficient synergy between NSM and LDPC coding}. Each matrix is generated by cyclically shifting a small number of sparse integer-valued filters derived from pattern vectors.

For $\mathring{\bm{G}}_\text{F}$, we use three filters: $\mathring{h}_0^\text{F}[p] = \delta[p] - 8\,\delta[p - 2]$, $\mathring{h}_1^\text{F}[p] = 7\,\delta[p] + 4\,\delta[p - 1]$, and $\mathring{h}_2^\text{F}[p] = 6\,\delta[p] + 5\,\delta[p - 1] - 2\,\delta[p - 2]$, based on the pattern vectors $(1,8)$, $(4,7)$, and $(2,5,6)$, respectively, from Table~\ref{table:Candidate Patterns Rate-2 Lambda_0 = 2 Lambda_1 = 3}. Each of these filters is cyclically shifted four times to populate the $12 \times 4$ scaled source synthesis matrix $\mathring{\bm{G}}_\text{F}$:
\begin{equation}
\mathring{\bm{G}}_\text{F} =
\begin{bmatrix}
1 & 0 & -8 & 0 \\
0 & 1 & 0 & -8 \\
-8 & 0 & 1 & 0 \\
0 & -8 & 0 & 1 \\
7 & 4 & 0 & 0 \\
0 & 7 & 4 & 0 \\
0 & 0 & 7 & 4 \\
4 & 0 & 0 & 7 \\
6 & 5 & -2 & 0 \\
0 & 6 & 5 & -2 \\
-2 & 0 & 6 & 5 \\
5 & -2 & 0 & 6
\end{bmatrix}
\label{eq:G_F_example}
\end{equation}

For $\mathring{\bm{G}}_\text{M}$, we use two filters: $\mathring{h}_0^\text{M}[p] = \delta[p] + 8\,\delta[p - 3]$ and $\mathring{h}_1^\text{M}[p] = 4\,\delta[p] - 7\,\delta[p - 1]$, derived from the pattern vectors $(1,8)$ and $(4,7)$ in Table~\ref{table:Candidate Patterns Rate-2 Lambda_0 = Lambda_1 = 2}. Each filter is cyclically shifted six times to generate the $12 \times 6$ scaled modulation generating matrix $\mathring{\bm{G}}_\text{M}$:
\begin{equation}
\mathring{\bm{G}}_\text{M} =
\begin{bmatrix}
1 & 0 & 0 & 8 & 0 & 0 \\
0 & 1 & 0 & 0 & 8 & 0 \\
0 & 0 & 1 & 0 & 0 & 8 \\
8 & 0 & 0 & 1 & 0 & 0 \\
0 & 8 & 0 & 0 & 1 & 0 \\
0 & 0 & 8 & 0 & 0 & 1 \\
4 & -7 & 0 & 0 & 0 & 0 \\
0 & 4 & -7 & 0 & 0 & 0 \\
0 & 0 & 4 & -7 & 0 & 0 \\
0 & 0 & 0 & 4 & -7 & 0 \\
0 & 0 & 0 & 0 & 4 & -7 \\
-7 & 0 & 0 & 0 & 0 & 4
\end{bmatrix}
\label{eq:G_M_example}
\end{equation}

The cyclic construction ensures column-wise regularity due to uniform shifting of each filter across the matrix columns. Row-wise norm regularity is enforced by using pattern vectors with common squared Euclidean norms. This ensures that the energy contribution from each latent bit remains consistent across both quantization and modulation branches. The selected pattern vectors also offer high discrimination and entropy characteristics, aiding in the robustness and convergence of belief propagation on the Y-shaped factor graph.

\begin{figure}[!htbp]
    \centering
    \includegraphics[width=0.85\textwidth]{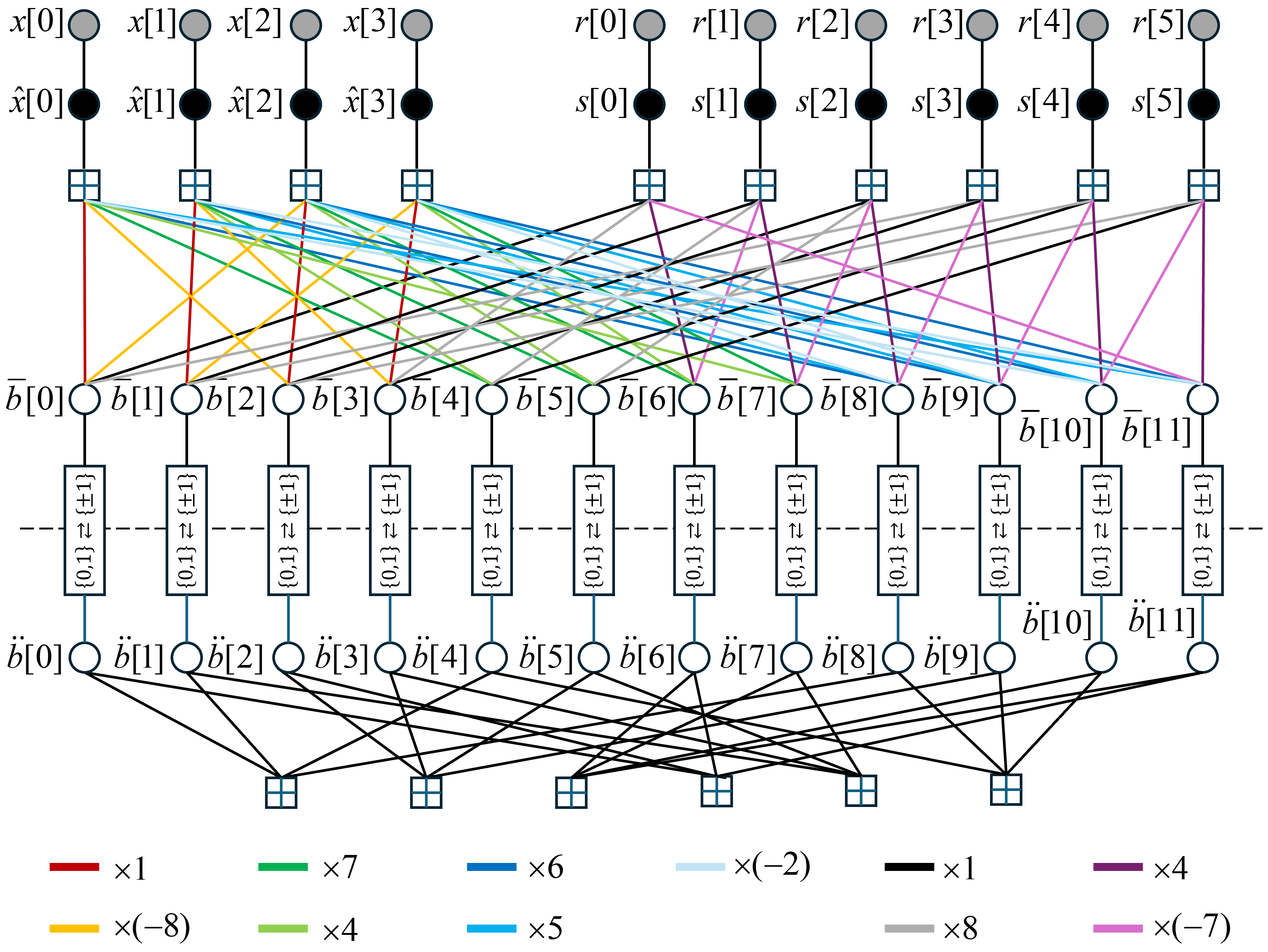}  
    \caption{Y-shaped factor graph corresponding to the joint source-modulation-coding example with $q = 4$, $k = 12$, and $n = 6$. The central binary vector $\bar{\bm{b}}$ is shared across three branches: the lower branch enforces LDPC code membership via the $6 \times 12$ parity-check matrix $\bm{H}$ with $m = 6$ constraints; the right branch applies analog LDGM NSM through the $12 \times 6$ matrix $\mathring{\bm{G}}_\text{M}$; and the left branch performs lossy source quantization using the $12 \times 4$ matrix $\mathring{\bm{G}}_\text{F}$.}
    \label{fig:Y_graph_example}
\end{figure}

A visual representation of the corresponding Y-shaped factor graph for this specific example is shown in Figure~\ref{fig:Y_graph_example}. It reflects the structural interplay between the \gls{ldpc} parity constraints, the analog modulation defined by $\mathring{\bm{G}}_\text{M}$, and the source reconstruction via $\mathring{\bm{G}}_\text{F}$. The latent binary vector $\bar{\bm{b}}$ connects all three branches, providing a unified interface for quantization, modulation, and error correction.

To demonstrate the generality of the proposed framework, we now consider a special case that mimics classical uniform scalar quantization and non-Gray coded $4$-ASK modulation using simplified integer-valued filters. We retain the same matrix sizes ($q = 4$, $k = 12$, $n = 6$). For the source quantization matrix $\mathring{\bm{G}}_\text{F}$, we use the filters $h_0^\text{F}[p] = \delta[p]$, $h_1^\text{F}[p] = 2\,\delta[p]$, and $h_2^\text{F}[p] = 4\,\delta[p]$, each cyclically shifted 4 times. For the modulation matrix $\mathring{\bm{G}}_\text{M}$, we use $h_0^\text{M}[p] = \delta[p]$ and $h_1^\text{M}[p] = 2\,\delta[p]$, each cyclically shifted 6 times. These choices generate matrices with bipolar entries drawn from the sets $\{1, 2, 4\}$ for source coding and $\{1, 2\}$ for modulation. This alphabet asymmetry reflects the greater shaping flexibility typically required in quantization, where fine control over signal representation is important, compared to modulation, which often targets compact signal constellations like $4$-ASK for energy efficiency and simplicity.

Despite reflecting familiar classical setups, this configuration exposes an important design issue: partial energy imbalance across the latent components. The bipolar components $\bar{b}[0]$ to $\bar{b}[3]$ are energy-matched, with $\|\mathring{h}_0^\text{F}\|^2 = \|\mathring{h}_0^\text{M}\|^2 = 1$, as are $\bar{b}[6]$ and $\bar{b}[7]$, with $\|\mathring{h}_1^\text{F}\|^2 = \|\mathring{h}_1^\text{M}\|^2 = 4$. However, mismatches occur at $\bar{b}[4]$ and $\bar{b}[5]$, where the energy is $4$ in the source domain but only $1$ in the modulation domain. An even larger discrepancy exists for $\bar{b}[8]$ to $\bar{b}[11]$, which have energy $16$ in the source domain versus $4$ in the modulation domain.

While these imbalances do not necessarily prevent the operation of iterative message-passing algorithms—and could even have mixed effects on convergence—they are suboptimal from a purely information-theoretic and geometric perspective. Balanced or proportionally matched energy allocation across the branches ensures that each latent component contributes coherently and consistently to the global inference task. This alignment is particularly important when aiming for high intrinsic efficiency in joint quantization and modulation.

Thus, although classical schemes such as uniform quantization with $8$ levels and non-Gray coded $4$-ASK modulation can be modeled within the proposed Y-shaped \gls{jscc} framework, they are not ideal design points under energy consistency criteria. The general framework not only captures these legacy configurations but also reveals their limitations—highlighting the importance of deliberate energy structuring as a means to unlock improved performance and better geometric alignment across the source and modulation spaces.

\subsection{Structural Similarities Between NSMs and Sparse/Structured Multiple Access}

Several \gls{ma-mac} techniques developed in the context of 5G and beyond incorporate features such as sparse signal representations, structured mappings, graph-based processing, or power-domain differentiation. Among these, \gls{noma}~\cite{Dai15}, \gls{scma}~\cite{Nikopour13}, \gls{pdma}~\cite{Dai18}, \gls{musa}~\cite{Yuan16}, \gls{lds-cdma}~\cite{Hoshyar08,Hoshyar10}, and \gls{idma}~\cite{Ping06} are particularly relevant. These techniques exploit non-orthogonal transmission and structured user-resource mappings—whether via power levels, sparse codebooks, or signature sequences—enabling overloaded transmission and efficient multi-user detection.

Compounded analog \gls{ldgm} \gls{nsm} and \gls{ldpc} coding, introduced in Section~\ref{A natural and efficient synergy between NSM and LDPC coding}, shares conceptual ground with several of these techniques. Although originally developed for single-user transmission, the proposed framework is inherently well-suited for multi-user scenarios due to its reliance on sparse generating matrices—composed of real-valued or appropriately scaled integer entries—and its compatibility with message-passing–based inference. Different users can be associated with distinct submatrices or analog basis functions drawn from the \gls{nsm} generator, effectively creating user-specific analog signatures. This idea parallels the separation achieved by user-dedicated codebooks in \gls{scma} or the signature sequences in \gls{lds-cdma}. The key distinction lies in the joint modulation and coding structure of the proposed system, where the analog \gls{ldgm} layer is not simply a mapping mechanism but also contributes to redundancy, shaping, and structural diversity across users. This enables the system to simultaneously realize multi-user access and facilitate low-complexity joint decoding through belief propagation.

The integration of \gls{ldpc} coding into the \gls{nsm} graph introduces additional robustness through extrinsic information exchange, which can significantly improve detection and decoding performance under interference or overloading conditions. Importantly, the joint factor graph remains sparse and structured, allowing efficient scalability with respect to the number of users and the complexity of the generating matrix.

Further extensions of this approach are natural. For instance, the analog signatures generated through \gls{nsm} can be combined with power-domain separation mechanisms characteristic of \gls{pd-noma}, enhancing user separation in highly overloaded regimes without altering the base graph structure~\cite{Dai15}. Likewise, by designing \gls{nsm} modulation patterns across spatial, temporal, and frequency dimensions, the framework can be adapted to exploit spatial diversity and beamforming in multi-antenna systems. This extension aligns well with existing \gls{mimo} architectures and preserves the graph-based iterative detection benefits of the original single-user design~\cite{Gesbert07,Rusek14}. These capabilities collectively suggest that the \gls{nsm}–\gls{ldpc} paradigm, beyond its original scope, offers a rich and flexible design space to address the multiple access challenges of future wireless systems.

\subsection{Structural Similarities Between NSMs and Sparse Superposition Coding}

\gls{ssc}~\cite{Rush17} is a modulation and coding strategy designed to approach the Shannon capacity of the \gls{awgn} channel using structured sparse linear combinations. In a typical \gls{ssc} framework, the input information bits are first encoded using an outer error-correcting code such as an \gls{ldpc} code, producing $p \log_2(q)$ coded bits. These coded bits are then mapped onto a sparse superposition of vectors as follows: the bit sequence is divided into $p$ sections, each representing $\log_2(q)$ bits. For each section, a group of $q$ real-valued vectors of length $n$ is defined, where $n$ denotes the dimension of the transmitted modulated vector. These vectors form a local dictionary from which exactly one vector is selected based on the $\log_2(q)$ bits assigned to that section.

When $q \ll n$, the vectors within each group tend to be approximately orthogonal with high probability, allowing each group to function like a local orthogonal modulation system. However, this approximate orthogonality holds only when $q$ is sufficiently smaller than $n$, as increasing $q$ reduces the quality of this orthogonality. The overall transmitted signal is formed by summing one vector from each of the $p$ groups, yielding a sparse linear superposition of $p$ vectors chosen from a total pool of $pq$ candidates. Due to the law of large numbers and high dimensionality, the entries of the resulting modulated vector tend to approximate Gaussian random variables.

Decoding \gls{ssc} typically relies on \gls{amp}, a low-complexity iterative algorithm that leverages the structured sparsity and known dictionary for efficient inference. \gls{amp} produces soft estimates of the sparse codeword components and exchanges soft information iteratively with the outer \gls{ldpc} decoder, forming a turbo-like decoding loop. Notably, this iterative decoder is often designed heuristically, with approximations in the message updates that can degrade performance relative to ideal joint decoding.

While \gls{ssc} and the analog \gls{ldgm} \gls{nsm} framework both use superpositions of analog signals and iterative message-passing decoding, they differ fundamentally in structure and operation. \gls{ssc} selects exactly one vector per group without further modulation, resulting in a sparse sum of vectors. In contrast, the analog \gls{ldgm} \gls{nsm} uses a full set of $k$ generating vectors (the rows of the analog \gls{ldgm} matrix), each modulated by bipolar latent symbols ($\pm 1$), yielding a dense linear combination. Thus, \gls{ssc}'s signal is sparse in terms of dictionary vectors, whereas \gls{nsm}'s signal is dense but structured via the sparse generating matrix and latent modulation.

Furthermore, \gls{ssc} often introduces group-wise power imbalances to enhance the convergence of iterative decoding algorithms. By contrast, the \gls{nsm} approach can maintain uniform power across components when the generating matrix rows satisfy regularity constraints. In such cases, the joint structure of the analog modulation and \gls{ldpc} coding layers supports decoder convergence without relying on power shaping.

Overall, while \gls{ssc} and analog \gls{ldgm} \gls{nsm} share the core idea of structured superposition and iterative decoding, \gls{ssc} relies on sparse selection from approximately orthogonal dictionaries per group, and \gls{nsm} exploits globally distributed bipolar modulation over a sparse generating matrix. This results in differing trade-offs in decoding complexity, performance, and modulation structure, with \gls{nsm}–\gls{ldpc} offering a fully joint factor graph framework for modulation and coding that avoids heuristic approximations inherent in \gls{ssc} iterative decoding.



\section{Conclusion and Perspectives}

In this work, a broad theoretical and numerical study of \glspl{nsm} and their various subclasses and generalizations was developed. Within this framework, \gls{ms-prs} represented a key subclass that was investigated in depth. Other extensions and generalizations of the \gls{nsm} paradigm were also discussed toward the end of the report, although without the same level of analytical and numerical detail.

The study was motivated by the observation that deliberately introducing controlled \gls{isi} in \gls{ftn} signaling, under certain compression factors, could enhance spectral efficiency while preserving the \gls{msed} of the reference modulation, $2$-ASK. Building on this insight, the constructive principle of controlled ISI was transposed to the \gls{nsm} framework, particularly to \gls{ms-prs}, where the objective was likewise to preserve the $2$-ASK \gls{msed} while achieving higher symbol rates. In this perspective, the deliberate introduction of interference became a design strategy rather than a limitation, providing a guiding principle for the structural optimization and performance enhancement of \glspl{nsm}.
The present approach was developed from the start in a sampled discrete-time framework at the Nyquist rate, which defined the class of \glspl{nsm}. It drew conceptual inspiration from Mazo’s original \gls{ftn} paradigm, where “sinc” pulses were used as shaping filters. In that formulation, the transmitted signal is constructed from time-shifted “sinc” pulses that deliberately introduce controlled \gls{isi}. By transposing this continuous-time formulation into the discrete-time Nyquist domain, the study identified a new path that led naturally to the conception of \gls{ms-prs} as a subclass within the broader \gls{nsm} family.

The first exploratory experiments of this work were carried out using the rate-$2$ \gls{ms-prs} configuration, which laid the foundation for the subsequent development of \glspl{nsm}. In these experiments, the main data stream was demultiplexed into two substreams, each filtered by one of two filters of equal length. Common filter lengths of $2$, $3$, $4$, and $5$ taps were considered to observe how increasing the filter length affected interference phenomena. The primary focus was on destructive interference between the two substreams, although a secondary effect related to \gls{sisi} within each substream was also observed. Thousands of filter pairs were generated and evaluated under varying energy-balancing conditions.

These experiments revealed several key insights:
\begin{itemize}
    \item \textbf{Risk and promise in balanced-energy scenarios:} The riskiest situations occurred when the two filters had comparable energies, either balanced or nearly balanced. These energy-balanced configurations proved to be the most promising for achieving the target \gls{msed} of $2$-ASK. This conclusion was supported by the theoretical upper bound that had been derived for the maximum achievable \gls{msed} as a function of energy balancing between the two filters.
    \item \textbf{Effect of filter length:} Increasing the common filter length in these energy-balanced scenarios mitigated the risk of poor performance, which arose from unlucky choices of filter pairs that failed to achieve the target \gls{msed}. Longer filters also reduced sensitivity to the specific selection of filters and provided additional protection against both destructive interference between substreams and \gls{sisi} within each substream, thereby enhancing overall robustness and achievable performance.
    \item \textbf{Energy balancing and performance optimization:} The experiments indicated that the most promising energy-balancing configurations were those around equal filter energies. However, these configurations also proved to be the most sensitive to the specific choice of filters, revealing a trade-off between risk and potential performance gain. This situation echoed the adages “Nothing ventured, nothing gained” and “With great power comes great responsibility.”
\end{itemize}

Overall, these early observations provided a crucial foundation for the later development of the \gls{nsm} and \gls{ms-prs} paradigms, and guided subsequent theoretical, structural, and numerical investigations. Building on the empirical findings from the rate-$2$ \gls{ms-prs} experiments, the \gls{nsm} framework demonstrated several structural and conceptual advantages that extended beyond the specific configurations tested. These properties enabled greater flexibility in filter design, optimization, and multidimensional generalization, providing a solid theoretical foundation for high-performance signaling.

A central result of this study was that the \gls{nsm} framework proved far stronger and more flexible than the \gls{ftn} paradigm from which it was originally inspired. Several structural and conceptual reasons explained this superiority.

First, the choice of the shaping filter was decoupled from that of the discrete filters that, in \gls{ftn} systems, had played the role of equivalent polyphase components of Mazo’s “sinc” scheme. This separation enabled the independent design of all filters involved in an \gls{nsm}, with complete freedom in selecting their lengths and tap values. In particular, discrete alphabets, especially rational ones (equivalent to integer alphabets after renormalization), were found to simplify the optimization process for the best \glspl{nsm} and to considerably reduce the computational burden of modulation and detection, since operations could be performed partly in the integer domain.

Second, the optimization of \glspl{nsm} was performed independently of the shaping filter, which ultimately serves only as an information carrier. This independence enables the modulated signal to be transmitted through a wide variety of physical mechanisms such as \gls{sc}, \gls{mc}, spread spectrum, or spatial multiplexing, without any loss of performance and with the ability to adapt these transmission modes dynamically. Whereas \gls{ftn} relies on an equivalent discrete-time filter of infinite impulse response that must be truncated, \glspl{nsm} are inherently finite impulse response systems, optimized directly under a chosen complexity constraint. This eliminates the trade-off in \gls{ftn} between severe truncation, which degrades performance, and mild truncation, which renders optimal detection algorithms such as the Viterbi algorithm for \gls{ml} decoding~\cite{Forney73} or the \gls{bcjr} algorithm for \gls{map} detection~\cite{Bahl74} computationally intractable.

Finally, the \gls{nsm} framework is fundamentally extensible, reaching beyond the one- and two-dimensional cases corresponding to single- and multi-carrier \gls{ftn}. It enables optimizations in three, four, five, or even higher dimensions, as well as mixed-dimensional configurations that balance performance and complexity. This multidimensional flexibility provides substantial advantages over classical \gls{ftn} systems, which are inherently limited to one- or two-dimensional structures. The same formalism can also be generalized toward analog \gls{ldgm} \glspl{nsm}, which integrate efficiently with \gls{ldpc} error-correction coding and analog \gls{ldgm}-based source coding, offering a unified and scalable foundation for future physical-layer architectures.

A clear illustration of the structural and performance advantages of \glspl{nsm} can be found in a simple block \gls{nsm} of rate $5/4$. This example was shown to outperform Mazo’s “sinc”-based \gls{ftn} system under practical truncation conditions. Only when the “sinc” filter is truncated extremely mildly, requiring an unrealistically large receiver trellis, does \gls{ftn} achieve comparable performance. This demonstrates the superior balance achieved by \glspl{nsm} between complexity and performance: they can maintain the same \gls{msed} as $2$-ASK while operating at nearly the same rate as Mazo’s limit. The rate-$5/4$ example thus provides a tangible benchmark of how \glspl{nsm} combine high spectral efficiency with implementation simplicity.

A second major outcome of this study lies in the distinct behavior of \glspl{nsm} in the analog domain compared with classical error-correction codes in the digital domain. In convolutional codes, the symbolic parameter $N$, which reflects the weight of input error sequences in transfer-function analyses, is instantiated as $N=1$. This implies that the multiplicity, and hence the probability, of any error event is independent of its input Hamming weight. In contrast, in the analog \gls{nsm} domain, our analysis revealed that $N=1/2$ is an intrinsic property that naturally emerges from the underlying analog structure. This finding profoundly changes the statistical nature of error events: their multiplicities, and therefore their probabilities of occurrence, decrease exponentially with the Hamming weight of the difference between bipolar input sequences. Consequently, only error events with small Hamming weights were shown to play a significant role at moderate \gls{snr} and within practical \gls{bep} ranges. This discovery marks a fundamental conceptual difference between the analog and digital domains and establishes a theoretical basis for their complementary roles in communication system design. Beyond its theoretical significance, this property also underlies how \gls{nsm} structures can be optimized, by focusing on low-Hamming-weight events and exploiting the flexibility of the filter design space to mitigate their impact.

These findings clarify how \gls{nsm} filters should be conceived. To ensure that an \gls{nsm} achieves the \gls{msed} of $2$-ASK, its discrete filters must be designed so that the \glspl{sed} associated with low-Hamming-weight input difference sequences are at least equal to that of $2$-ASK. Preserving these distances for low-Hamming-weight error events is therefore essential to strictly and perfectly attain the $2$-ASK benchmark. Although infinite-impulse-response designs are theoretically admissible, practical realizations of \glspl{nsm} must rely on finite-impulse-response filters to ensure feasibility and controllable complexity. Because the shaping filter is conceptually decoupled from the discrete filters that define the \gls{nsm} itself, one can freely select their respective lengths, coefficients, and even dimensional structures. This independence enables optimization in one, two, or several dimensions, extending beyond the classical single- and multi-carrier \gls{ftn} configurations, and makes it possible to balance performance and complexity through mixed-dimensional designs. Theoretical analysis further suggests that error events whose non-null differences remain confined within a finite portion of the input sequence are the most likely to produce squared distances below that benchmark.

Building on these insights, the potential combination of \glspl{nsm} with digital-domain error-correction schemes offers promising perspectives for unified physical-layer architectures. The analog and digital domains can be viewed as governed by an exclusion principle: error events in the digital coding domain and those in the analog modulation domain are structurally distinct and thus intrinsically unlikely to coincide in a harmful way. This separation is further reinforced when interleaving is applied between coded bits and \gls{nsm} inputs, randomizing the alignment of error events across domains. More broadly, this conceptual framework extends naturally to joint constructions combining \gls{ldpc} coding and analog \gls{ldgm} \glspl{nsm}, providing a flexible foundation for next-generation communication architectures that integrate the complementary advantages of analog and digital domains.

By focusing on the “sinc” pulse, our analysis remained fully compatible with the Nyquist sampling theorem: sampling the continuous-time \gls{ftn} signal at the Nyquist rate preserved all information without aliasing or distortion. This exact information preservation under Nyquist-rate sampling, ensured by the “sinc” pulse, is not maintained with other shaping filters such as raised-cosine or root-raised-cosine pulses with nonzero roll-off, whose spectra exceed the Nyquist band. Nevertheless, the “sinc”-based formulation proved theoretically sufficient and conceptually fertile, providing a rigorous foundation for the development of the broader \gls{nsm} paradigm.

Within this perspective, the discrete-time \gls{nsm} framework can be viewed as a natural evolution of Mazo’s formulation, expressing the principle of controlled interference directly in the discrete domain, without the need for oversampling or time compression. This reformulation enables a fully digital treatment of the problem, in which filter coefficients, symbol mappings, and interference patterns can be optimized under explicit complexity constraints. As a result, \glspl{nsm} offer a more versatile and practically tractable formalism.

In summary, while the \gls{ftn} paradigm originally emerged from a continuous-time, “sinc”-based construction, \glspl{nsm} extend it into a general discrete-time framework that encompasses a much wider range of modulation structures. This generalization preserves Mazo’s core principle of controlled ISI yet opens the way to systematic optimization, scalable implementation, and eventual integration within unified analog–digital transmission architectures.

Building on the one-dimensional \gls{nsm} framework derived from \gls{sftn}, the analysis was extended to \gls{2d} rate-$2$ \glspl{nsm} using rational-tap filters, equivalent to integer taps after appropriate scaling. This extension introduces new optimization opportunities, particularly by exploiting the increased degrees of freedom provided by the additional dimension for filter design and for preserving the \gls{msed}. The multidimensional formulation was then expanded to rate-$3$ \glspl{nsm}, conceptually constructed from combinations of one-, two-, and four-dimensional filter structures, highlighting the potential for several dimensions to coexist within a single modulation scheme.

These findings established that multidimensional \glspl{nsm} can serve as a powerful tool for structural optimization, offering richer design spaces while maintaining control over the \gls{msed}. Although practical challenges remain, such as designing tractable detection strategies for higher-dimensional \glspl{nsm} and managing the multidimensional filter span, the conceptual groundwork demonstrates the viability and flexibility of multidimensional \gls{nsm} structures for future high-performance communication systems.

In the study of rate-$2$ \glspl{nsm}, we focused on filters with rational taps, which can be equivalently scaled to integer values. This choice simplified the optimization process and highlighted key performance patterns. Most configurations led to \glspl{msed} below that of $2$-ASK. Among the configurations we examined, only one achieved the $2$-ASK benchmark, confirming that \glspl{nsm} with rational filter taps are indeed capable of attaining the $2$-ASK \gls{msed}. Such configurations exist but are rare and identifying them can be time-consuming and challenging.

Extending the lengths of multiple filters is expected to increase the potential for improved performance. Configurations where more than one filter has length greater than one offer greater design flexibility and a higher likelihood of preserving \glspl{msed} compared to setups with only a single long filter.

Rational taps also made the optimization process tractable, even for very large filter lengths. They are expected to support extensions to higher-rate \glspl{nsm}, such as rate-$3$ (providing the same spectral efficiency as $64$-QAM) or rate-$4$ (providing the same spectral efficiency as $256$-QAM) designs. In these higher-rate scenarios, candidate structures may either inherit optimized elements from lower-rate \glspl{nsm} to further reduce search complexity or be fully optimized from scratch to maximize performance within the constraints of the chosen rational taps.

A key tool in this optimization is the use of equivalence relations, which reduce the number of distinct filters that must be examined while preserving the possibility of identifying optimal configurations. Together, rational taps, equivalence relations, and potential inheritance strategies provide a practical and systematic framework for optimizing \glspl{nsm} across rates.

Building upon the initial observations presented earlier, additional experiments were conducted to examine how the distribution of energy between filters influences performance across various \gls{nsm} configurations. These experiments confirm that the balanced-energy mode, where the filters have strictly equal energies, remains the most promising in terms of achievable performance, yet also the most sensitive to specific filter selections. This dual nature makes balanced configurations simultaneously the riskiest and the most rewarding in terms of potential performance gain, an outcome consistent with the earlier conceptual analysis.

The empirical results further showed that increasing the filter length effectively mitigates the vulnerability associated with balanced-energy configurations. Longer filters provide additional robustness, reducing the likelihood of performance degradation caused by unfavorable filter combinations. This trend persists consistently across all tested setups, confirming that the interplay between energy balancing and filter length is a key factor governing \gls{nsm} performance.

Overall, these experiments provided practical validation of the theoretical trends identified earlier: while balanced-energy designs carry inherent sensitivity, their risks can be controlled through appropriate filter-length selection. This balance between risk and robustness forms a cornerstone of the \gls{nsm} design philosophy and reinforces the constructive role of controlled interference within the broader framework of multi-stream signaling.

Experiments with one-dimensional rate-$2$ \glspl{nsm} using real-tapped filters of modest span revealed a simple, high-value design rule. At similar detection complexity, configurations in which one filter has unit length and the other has a slightly larger span consistently achieve superior \gls{msed} compared to designs where both filters have lengths greater than one. This trend was observed for the tractable search region explored, notably for configurations where the longer filter had a length between $3$ and $7$ and the alternative designs had both filters with lengths between $2$ and $4$. Extending the search to longer filters was not tractable because, with real-tapped filters, the associated optimization space expanded rapidly and could not be explored efficiently within a reasonable time, preventing objective comparison under equal detection-complexity constraints. Consequently, for the range of small filter lengths that can be explored exhaustively and reliably, focusing on structures with a single unit-length filter provides the best compromise between achieved \gls{msed} and detection complexity.

In this work, every \gls{nsm} under investigation was constructed with only one of the two filters having a length greater than one, a choice consistently applied across all examined configurations. This design decision, made for simplicity and methodological coherence, continues to provide a robust baseline and is expected to remain near-optimal even for filter lengths that could not be exhaustively explored. However, future investigations addressing larger filter lengths should not overlook configurations where more than one filter has a length greater than one, as they might reveal additional performance gains.

During the early phase of this work, we focused on one-dimensional rate-$2$ \glspl{nsm} with real-tapped filters. For these systems, we developed a three-step analytical procedure to derive closed-form expressions for the filters and their associated \gls{msed}. The method was applied under both unconstrained and constrained configurations. It succeeded up to filter lengths $8$ in the unconstrained case and $6$ in the constrained one. Beyond these limits, closed-form expressions could no longer be obtained despite extensive effort. The precise reason for this failure remains uncertain. It is likely rooted in the second or third step of the procedure, either in the identification of the error events that determine the \gls{msed} or in the inability of symbolic computation to cope with the rapidly increasing algebraic complexity.

When the filter whose length exceeds one reaches ten taps, the problem was expected to be intractable from the outset, mainly because of the existence of a multitude, and possibly an infinite number, of distinct optimal \glspl{nsm} achieving the same \gls{msed} as $2$-ASK. This multiplicity adds to the earlier computational difficulties and jointly prevents symbolic derivation. To overcome this barrier in the specific case where the longer filter has length $10$, a promising direction is to extend the optimization criterion beyond the first minimum distance and also include the second \gls{msed}. This extension could lead to a unique optimum solution up to the established equivalence relations, thereby most likely enabling the closed-form determination of the filters for the best \gls{nsm} with a longer filter of length $10$.

For rate-$2$ \glspl{nsm} with rational taps, we began our investigation with the simple family of \glspl{nsm} whose filters have bipolar non-null taps. By exploiting modulo-two calculus and algebraic tools, including m-sequences, primitive polynomials, and polynomial factorization combined with rational fraction decomposition over the binary Galois field, we were able not only to maximize the minimum length of error events associated with the first \gls{msed}, but also to determine the Hamming weight of the corresponding input difference sequences. The first \gls{msed} of these \glspl{nsm} is half that of $2$-ASK, while the second \gls{msed} precisely coincides with that of $2$-ASK. Fortunately, the Hamming weight of the input-difference sequences associated with the first \gls{msed} grows rapidly with the length of the longer filter, becoming significant once the filter length exceeds $8$ taps. This increased Hamming weight, together with the analog-domain instantiation $N=1/2$ in the \gls{tf}, causes the multiplicity of these low-distance error events to vanish extremely rapidly. As a result, these events have negligible impact on system performance for moderate \gls{snr} values and practical ranges of \gls{bep}, leaving the second \gls{msed} as the dominant factor in determining performance. Consequently, despite the first \gls{msed} being only half of $2$-ASK, these \glspl{nsm} achieve performance tightly aligned with $2$-ASK in practical operating conditions. This observation also illustrates that the first \gls{msed} alone is not always a sufficient predictor of practical performance, in a way similar to how the minimum Hamming distance does not fully determine performance in classical digital coding, as evidenced by the behavior of turbo codes.

Building on this initial investigation, we expanded the framework to a richer rational alphabet and examined numerous candidate \gls{nsm} families, each specified by an integer pattern designed to balance energy between the filters. Many of these patterns failed to achieve the \gls{msed} of $2$-ASK, but one particular pattern was identified through systematic optimization of its underlying family and successfully reached the $2$-ASK benchmark. This promising result motivates further exploration of alternative filter patterns and larger lengths, extending beyond those already analyzed, to uncover additional high-performance \gls{nsm} configurations.

The optimization of rate-$3$ \gls{ms-prs} \glspl{nsm} with real taps was carried out using two complementary methodologies. The first relied on inheritance from optimized rate-$2$ \gls{ms-prs} structures. In this approach, the filter of length greater than one that achieved the best \gls{msed} performance for the rate-$2$ case was reused together with two additional unit-length filters. Candidate values for the longer filter ranged from $2$ to $8$ taps. The energy distribution among the three filters was determined according to the tightness principle, ensuring consistent energy balancing while maintaining the \gls{msed} properties of the inherited design.

The second methodology performed a global optimization directly on the rate-$3$ configuration, again under the tightness principle. Although this direct optimization is more involved, its additional complexity arises mainly from the slightly more intricate branch metrics used in the trellis for \gls{msed} determination. The rate-$3$ trellis includes two substreams in addition to the one corresponding to the longest filter, whereas only one such additional substream is involved in the rate-$2$ case. Nevertheless, since two of the filters have length one and thus introduce no memory, the overall trellis structure remains manageable, and the computational burden continues to be dominated by the length of the single filter whose span exceeds one symbol.

Across the explored design space, the inheritance procedure often produced the same optimal configurations as the global optimization, confirming that good rate-$3$ designs can be efficiently constructed from high-quality rate-$2$ building blocks. In a number of instances, however, the global procedure yielded slightly better performance. Detailed analysis showed that these improvements stem from subtle adjustments in energy sharing and interference balance among the three substreams that are only visible when all three filters are co-optimized. These observations provide practical insight: inheritance is an effective and lower-complexity design shortcut for many cases, while full joint optimization can capture marginal gains when they exist.

The study of the rate-$3$ \gls{ms-prs} case also illustrates how much performance improvement can be achieved through a configuration in which only one filter among three has length greater than one. In particular, for a longest-filter length of eight taps, an asymptotic gain of about $5.2$ dB was observed with respect to $8$-ASK, or equivalently to its \gls{2d} $64$-QAM counterpart. Despite this notable gain, the detection trellis remains of reasonable complexity thanks to the two unit-length filters, which introduce no additional memory. This configuration therefore demonstrates that significant \gls{snr} improvements can be obtained without a prohibitive increase in detection complexity, suggesting a practical and scalable pathway for extending the same design philosophy to higher-rate \gls{ms-prs} \glspl{nsm}.

For rate-$(Q+1)/Q$ \glspl{nsm} with rational-tap filters, it has been shown that the \gls{msed} of $2$-ASK can be achieved very efficiently. In these \gls{ms-prs} configurations, exactly one filter has a length greater than one, while all other filters have unit length. Rational taps, combined with equivalence-class representatives, allow a systematic and tractable search over candidate filters, ensuring energy balance among the $Q+1$ substreams, which is a necessary condition to achieve the $2$-ASK \gls{msed}. The memory introduced by the longer filter is minimal, spanning at most three previous substream symbols, corresponding to a detection trellis with at most eight states. By comparison, rate-$2$ \glspl{nsm} achieving the same \gls{msed} require a trellis with $512$ states for real-tap filters and $1024$ states for rational-tap filters. This demonstrates that rate-$(Q+1)/Q$ \gls{ms-prs} dramatically reduce detection complexity while maintaining optimal performance.

Building on these results, for the highest achievable rate of $3/2$ ($Q=2$), a very simple \gls{ms-prs} with a two-state trellis reached the $2$-ASK \gls{msed} and provided a \gls{bep} performance nearly identical to that of $2$-ASK. More generally, rate-$(Q+1)/Q$ \gls{ms-prs} not only dramatically reduced detection complexity compared to both higher-rate \glspl{nsm} and classical \gls{sftn} schemes, but also preserved the flexibility regarding the shaping filter that had been previously emphasized. In contrast, \gls{ftn} schemes operating in the \gls{sc} framework typically relied on specific pulse shapes, such as \gls{rc}, \gls{rrc}, or \gls{pswf}. These shaping filters often led to more complex detection structures and, in many cases, lower spectral efficiency, while constraining the transmitted pulse to a predefined form. As a reminder, the \gls{ms-prs} paradigm retains the freedom to adopt any shaping filter, further reinforcing its practical simplicity and design flexibility.

Drawing inspiration from \gls{mftn}, we extended the one-dimensional \gls{nsm} framework, originally derived from \gls{sftn} systems, to \gls{2d} rate-$2$ \glspl{nsm} with rational-tap filters. Building on this development, we introduced the concept of \gls{3d} rate-$2$ \glspl{nsm}, demonstrating that additional dimensions provided further degrees of freedom for filter optimization and the preservation of \gls{msed}. The multidimensional formulation also allowed us to conceptually outline rate-$3$ \glspl{nsm} combining one-, two-, and four-dimensional filter structures, enabling the coexistence of multiple dimensions within a single modulation scheme. Importantly, the study showed that the framework is not fundamentally limited: higher-dimensional \glspl{nsm} and mixed-dimensional configurations remain feasible in principle, offering virtually unlimited flexibility.

These multidimensional \glspl{nsm} preserved the \gls{msed} across the sub-streams by exploiting the additional degrees of freedom offered by higher dimensions. Working in two or three dimensions generally allowed intuitive reasoning about the placement and span of filters, facilitating the design of configurations that guaranteed the $2$-ASK \gls{msed}. The conceptual framework enabled sub-streams to be optimally arranged in two-, three-, or even mixed-dimensional configurations, maintaining performance while increasing modulation flexibility.

Detection complexity varies substantially across multidimensional \glspl{nsm}. For two- and three-dimensional configurations, no trellis representation exists, and complexity grows exponentially with packet size. Small packets suffered from disproportionately high overhead due to the span of multidimensional filters, reducing the effective transmission rate and spectral efficiency. In contrast, for larger packets, the relative overhead decreases, but the overall detection complexity increases drastically. This trade-off represents a current bottleneck in the design of multidimensional \glspl{nsm} and highlights the importance of developing advanced detection methods capable of managing high complexity, at the possible price of suboptimality.

A conceptual extension to rate-$3$ \gls{ms-prs} \glspl{nsm} was proposed, combining one-, two-, and four-dimensional filters within a single modulation scheme. In this configuration, the second and third filters both had lengths greater than one. The design aimed to illustrate the theoretical potential of mixed-dimensional \glspl{nsm} while preserving the \gls{msed}.

Even without numerical optimization or simulations, visualizing the system simultaneously in one-, two-, and four dimensions made it straightforward to understand how the three substreams interacted. Geometrical insight showed that each substream individually met the $2$-ASK \gls{msed} requirement, and when combined, their interactions did not reduce the overall minimum distance. This was ensured by the tightness constraints applied to all three filters, despite their operation in different dimensions, which makes it conceptually clear that the target \gls{msed} is preserved.

This conceptual configuration demonstrates that multidimensional \glspl{nsm} can accommodate multiple substreams without destructive interference and that there is no fundamental limit to the number or combination of dimensions that can be used in \gls{nsm} design.

A simplified two-stage \gls{ml} detection algorithm was investigated to address the complexity of multidimensional detection. The algorithm first detects the substream associated with the filter having a span greater than one, then proceeds with the detection of the remaining substreams, all filtered by unit-length, memoryless filters. This simplification was initially developed for the rate-$5/4$ block \gls{nsm} and later applied successfully to multidimensional rate-$2$ \glspl{nsm} in both two- and three-dimensional configurations. It provided exact \gls{ml} detection while drastically reducing the overall computational burden. Nevertheless, detection in higher-dimensional \glspl{nsm} remains demanding, highlighting the need for further complexity reduction through suboptimal approaches.

As a possible future direction, sphere decoding~\cite{Pohst81,Fincke85,Guo20,Bedeer17} is a promising alternative for multidimensional \glspl{nsm} where trellis-based detection becomes infeasible. Sphere decoding can confine the search region in the multidimensional signal space, thereby reducing computational load. Moreover, its soft-output variants can support iterative detection structures, making it attractive for practical receiver implementations in higher-dimensional \glspl{nsm}.

Another perspective is provided by the iterative detection scheme used in multi-stream \gls{ftn} systems, as discussed by Anderson \emph{et al.}~\cite{Anderson13}. In that framework, which is structurally comparable to \gls{2d} \glspl{nsm}, detection relied on soft information exchanged both within each substream and across substreams during iterations, enabling the mitigation of self-interference within substreams as well as inter-substream interference. In practical \gls{mc} implementations, \gls{mmse}-type detection can be applied to jointly process all substreams, providing a low-complexity iterative approach to interference suppression. While Anderson \emph{et al.} did not explicitly call this turbo-equalization in the \gls{mc} case, the iterative refinement of soft symbol estimates is conceptually equivalent and can naturally be extended to multidimensional \glspl{nsm} in the analog domain, offering a flexible framework for managing interference and enhancing performance.

Finally, the concept of tail-biting is discussed as a means to mitigate the loss in spectral efficiency due to filter span in finite packet transmissions. Extending the principle of tail-biting from convolutional coding to multidimensional \glspl{nsm}, this approach is expected to preserve the \gls{nsm}’s \gls{msed} close to the target $2$-ASK \gls{msed} for sufficiently large packets. However, as packet sizes decrease, there is an increasing risk that the \gls{nsm}’s \gls{msed} falls below the $2$-ASK target. Despite this limitation, tail-biting provides a valuable conceptual strategy for maintaining spectral efficiency in finite-length multidimensional signaling, while helping the \gls{nsm} preserve \gls{msed} close to the $2$-ASK target.

To accelerate \gls{nsm} optimization, particularly for rational-tap filters of length greater than one, we identified and applied a family of discrete transformations that preserve the \gls{msed}. These transformations play a key role in reducing the effective search space by grouping filters into equivalence classes, allowing only one representative per class to be evaluated without loss of generality.

In one dimension, the applied transformations included global sign inversion, which multiplies all taps by $-1$, time translation, which shifts the tap sequence, and time inversion or reversal, which reverses the tap order. Additional transformations such as alternate-sign change, which multiplies the tap at position $n$ by $(-1)^n$, and scrambling by periodic bipolar sequences were also used. The latter applies to \glspl{nsm} of rate $(Q+1)/Q$ with $Q \ge 2$, where the scrambling sequence has a period equal to $Q$. All these operations preserve \gls{msed} by maintaining the relative \glspl{sed} between the output sequences corresponding to different input symbol combinations.

The notion of time inversion naturally generalized to higher dimensions through geometrical symmetries. In two dimensions, horizontal, vertical, and diagonal symmetries were considered, while in three dimensions, horizontal, vertical, depth, face-to-face, and edge-to-edge symmetries were applied. Rotations in two and three dimensions further extended these symmetries, representing higher-order generalizations of the one-dimensional time reversal. Applying this complete set of \gls{msed}-preserving transformations enables the efficient exploration of high-dimensional \gls{nsm} design spaces while guaranteeing the inclusion of all distinct filter configurations.

Rational-tap filters played a central role in making large-scale \gls{nsm} optimization tractable. After appropriate normalization, they were represented by integer-valued taps, which allowed the search domain to be rigorously partitioned into a finite set of equivalence classes. By restricting exhaustive searches to a single representative from each class, every optimal filter configuration remained available for discovery while redundant evaluations were avoided. This representative-based scanning was therefore both exhaustive and computationally efficient.

The use of equivalence-class representatives had a dramatic impact on the number of candidate filters that needed to be evaluated individually. In rate $(Q+1)/Q$ one-dimensional \glspl{nsm} and in multidimensional \glspl{nsm} (two-, three-dimensional, and beyond), the richness of the \gls{msed}-preserving equivalence transformations led to a strong reduction in the number of distinct equivalence classes to consider, making the optimization process far more tractable.

In the configurations studied, only one filter had a length greater than one, while all other filters had unit length. Moving to configurations where more than one filter has length greater than one would dramatically increase the complexity of the optimization process, and such cases have not yet been systematically investigated. Equivalence relations therefore become crucial in these extreme, high-complexity scenarios. Exploring these promising configurations represents one of the most interesting and impactful directions for future work, offering the potential to uncover new high-performance \gls{nsm} structures while keeping the search computationally feasible.

A key distinction between analog-domain \glspl{nsm} and conventional digital error-correction codes lies in the behavior of error-event multiplicities. In classical convolutional codes, it is well established that, in transfer-function analyses, the symbolic parameter $N$ is instantiated as $N=1$. This implies that the multiplicity (and therefore the probability of occurrence) of any error event remains constant, regardless of the Hamming weight of the input difference sequence.

In contrast, our detailed analysis of \glspl{nsm} in the analog domain showed that $N$ must be instantiated as $N=1/2$. This reveals a fundamental difference between the analog and digital domains: in the analog case, the multiplicity of error events decreases exponentially with the Hamming weight of the input difference. As a result, events associated with large Hamming-weight input differences become exponentially unlikely.

Moreover, the probability of an error event decreases exponentially with its output-difference \gls{sed}. These two exponential mechanisms (one acting on the combinatorial multiplicity of input differences and the other on their Euclidean separation) work together to suppress the overall significance of most error events. Consequently, only events that simultaneously exhibit low input-difference Hamming weight and low output-difference \gls{sed} have a meaningful influence on system performance at moderate \gls{snr} and practical \gls{bep} ranges.

This insight provides a clear guideline for \gls{nsm} filter optimization. In practice, filters should be designed so that, for every relevant input difference, either the input Hamming weight or the output \gls{sed} is sufficiently large to minimize the impact of the corresponding error event on overall performance.

When \glspl{nsm} are combined with digital error-correction coding, an intrinsic structural independence emerges between the analog and digital domains. Coded binary sequences are converted into bipolar sequences that serve as inputs to the \gls{nsm} modulator. Due to the inherent differences in how error events manifest in these two domains, input-difference patterns in the analog \gls{nsm} that produce low \glspl{sed} almost never correspond exactly to output differences at the digital decoding level. This exclusion principle ensures that harmful low-distance analog \gls{nsm} events rarely align with decoder-level error events, substantially reducing their impact on overall performance. These mechanisms also extend naturally to multidimensional \glspl{nsm}, where the exponential suppression of error-event probability for input differences with large Hamming weight continues to play a central role.

Interleaving further strengthens the exclusion principle, which arises from the combination of \glspl{nsm} with error-correction coding. By randomizing the association between coded bits and \gls{nsm} inputs, interleaving reduces the probability that low-Euclidean-distance analog \gls{nsm} error events align in their input differences with digital-codeword differences.

Two complementary mechanisms are responsible for drastically reducing the impact of low-distance \gls{nsm} error events on overall system performance. The first is the exponential suppression of error-event probability for input differences with large Hamming weight in the \gls{nsm} domain. The second is the exclusion principle resulting from error-correction coding. This dual effect provides a clear advantage over classical modulations such as $16$-QAM or $64$-QAM, where neither mechanism is present.

The transfer-function framework was employed to derive upper bounds on the \gls{bep} of \glspl{nsm}. Both full and \glspl{rtf} were used, in both complete and truncated representations, depending on the feasibility of symbolic computation. The full \gls{tf}, expressed as a bivariate function of the symbolic variables $N$ and $D$, provided the most general analytical representation of the error-event structure, but its determination in closed form was found to be intractable in several important cases.

For \glspl{nsm} characterized by real-valued tapped filters, the exponents of the distance variable $D$ were non-integer, which prevented symbolic calculus and precluded a closed-form expression. Even when the taps were rational and could be renormalized to integer values, the symbolic expansion rapidly became unmanageable for long filters because of the combinatorial growth in the number of terms. As a result, full analytical expressions could only be obtained for short or low-dimensional \glspl{nsm} with limited filter length.

To overcome these limitations, a \gls{rtf} was introduced. This function was derived from the complete one by performing a partial differentiation with respect to the multiplicity variable $N$, multiplying by $N$, and then instantiating $N$ to the value $1/2$. This operation effectively eliminated one symbolic variable and yielded a univariate function that depended only on $D$. The \gls{rtf} was designed to preserve all information necessary to establish \gls{bep} upper bounds while significantly simplifying computation. Its reduced algebraic complexity also allowed its use in degenerate \gls{nsm} cases where the full \gls{tf} could not be determined.

When neither the full nor the \gls{rtf} could be obtained symbolically, iterative algorithms were employed to produce truncated versions of the \gls{rtf} after a small number of iterations. For \glspl{nsm} with real tapped filters, the algorithms operated numerically, progressively accumulating the contributions of dominant error events in increasing order of their Euclidean distance. For \glspl{nsm} with rational tapped filters that could be renormalized to integer values, the algorithms operated symbolically, using successive polynomial additions and multiplications in the variable $D$, followed by truncations. In all cases, these approaches were found to be efficient and rapidly convergent. After a few iterations, the truncated functions stabilized, yielding accurate \gls{bep} upper bounds that closely matched the simulated \glspl{ber}.

Although higher-dimensional \glspl{nsm} did not admit either complete or \glspl{rtf}, it was shown that optimized one-dimensional extracts of these multidimensional structures could still be effectively analyzed using the reduced transfer-function methodology. These one-dimensional counterparts retained the essential local characteristics of the multidimensional modulation, enabling partial optimization through analytical evaluation of their \glspl{rtf}. This approach provided valuable insight into multidimensional designs and served as a practical bridge between analytical tractability and higher-dimensional performance optimization.

Overall, this methodological contribution confirmed that the \gls{rtf}, whether it was determined in closed form or obtained through the iterative procedures, was sufficient to capture the essential characteristics of the modulation’s error structure. Its computational simplicity and accuracy made it a practical and reliable analytical tool for the study of \glspl{nsm}, even in the presence of real-valued filters, long filter spans, or extrapolated higher-dimensional configurations for which complete transfer functions were unattainable.

The comparative analysis between analytical and simulated results confirmed the accuracy and robustness of the transfer-function approach developed in this work. The \gls{bep} upper bounds derived from both the complete and the \glspl{rtf} were found to coincide remarkably well with the simulated \gls{ber} curves across all configurations that were examined. This close agreement demonstrated that the analytical framework captured the essential statistical behavior of error events in \glspl{nsm}, thereby validating both theoretical modeling and numerical implementation.

In particular, the bounds obtained from the truncated \glspl{rtf}, whether determined symbolically or produced through the iterative algorithm, consistently reproduced the shape and slope of the simulated \gls{ber} curves over the full signal-to-noise ratio range of interest. This confirmed that the dominant contributions to the \gls{bep} were effectively captured within the first few terms of the truncated expansion and that higher-order terms had negligible influence on the predicted performance. The truncated analytical forms therefore provided an accurate and computationally efficient substitute for exhaustive Monte-Carlo simulations.

The cross-validation between analytical and simulated results played a decisive role in establishing the reliability of the proposed methodology. The observed coincidence between theory and simulation was not limited to a few isolated cases but was observed systematically across one-dimensional \glspl{nsm}, including rate-$2$ configurations with real taps. This systematic agreement confirmed that the proposed analytical tools remained valid across the different modulation configurations studied and that the assumptions underlying the reduced transfer-function derivation were sound.

Overall, the cross-validation between simulation and theory confirmed that the transfer-function framework, and in particular its reduced and truncated forms, provided a dependable analytical foundation for performance evaluation. The demonstrated accuracy of the \gls{bep} upper bounds validated the entire \gls{nsm} design and analysis methodology, showing that reliable predictions of error performance could be obtained long before any implementation or large-scale simulation was carried out.

The generalization of \glspl{nsm} through analog \gls{ldgm} constructions introduced a new perspective on modulation design, emphasizing the structured organization of transmitted signals. In this framework, the modulation process was represented explicitly through a generating matrix, where each row corresponded to an input bipolar data symbol and each column corresponded to a transmitted modulated symbol. Regularity conditions were established across both rows and columns. Row regularity ensures that each input symbol contributes equally to the transmitted energy, while column regularity guarantees that the average power of transmitted symbols remains uniform across the modulation sequence.

These regularity constraints in the analog domain find a clear conceptual parallel with the design principles of \gls{ldpc} codes in the digital domain. In \gls{ldpc} codes, regularity is defined both over the variable nodes, ensuring that each coded bit participates in a fixed number of parity-check equations, and over the check nodes, ensuring that each parity-check equation involves a fixed number of coded bits. Analog \gls{ldgm} structures impose an analogous form of energy regularity, operating directly in the analog domain on symbol amplitudes rather than bit weights. Irregular analog \gls{ldgm} matrices remain a viable option, offering additional flexibility at the potential cost of greater design complexity, just as irregular \gls{ldpc} codes trade uniformity for enhanced error-correction performance in digital systems.

Using structured generating matrices in combination with regularity constraints, analog \gls{ldgm} \glspl{nsm} extend the \gls{nsm} paradigm into a second, broader design dimension. They provide a principled approach to balancing energy across substreams and modulated symbols, echoing the success of structured code designs in the digital domain. This perspective paves the way for high-performance, analytically tractable, and computationally efficient modulation strategies that could profoundly influence the design of $6$G physical layers and beyond.

In analog \gls{ldgm} \glspl{nsm}, randomness may be incorporated into the generating matrices, which, when present, suppresses harmful alignments of error events and helps preserve \glspl{msed}, thereby ensuring robust performance. This principle mirrors the benefits of randomness in \gls{ldpc} codes, where pseudo-random connections between variable nodes and check nodes avoid short cycles in the Tanner graph, distribute parity-check constraints effectively, and improve iterative decoding performance. By controlling the structural organization of the analog \gls{ldgm} matrices, the modulation framework can exploit randomness to maintain minimum-distance properties and suppress detrimental interactions between substreams. These properties are absent in classical modulations such as $16$-QAM or $64$-QAM, highlighting the transformative potential of analog \gls{ldgm} \glspl{nsm} for future high-performance communication systems.

The analog \gls{ldgm} \gls{nsm} framework forms a powerful bridge between modulation and coding theory. It extends to the modulation domain the same conceptual principles that underly \gls{ldpc} codes: sparse and structured connectivity, iterative processing, and robustness achieved through controlled randomness. This unified viewpoint allows modulation design to adopt the language and analytical tools of modern coding theory while preserving the geometrical intuition inherent to signal-space representation. It thus provides a coherent foundation for the joint optimization of modulation and coding, paving the way toward unified and efficient $6$G and beyond physical-layer architectures.

When the generating matrices of analog \gls{ldgm} \glspl{nsm} or the parity-check matrices of \gls{ldpc} codes are deliberately designed with a regular structure, embedded randomness may be limited. In such configurations, interleaving can be selectively introduced to restore statistical independence between symbol and codeword components. This additional randomness decorrelates potential error events while preserving the benefits of regular geometry, ensuring that iterative detection and decoding remain efficient and easily parallelizable.

Within this general framework, \gls{2d} and higher-dimensional \glspl{nsm} can be interpreted naturally as structured analog \glspl{ldgm}, whose geometrical organization of modulated symbols explicitly reveals which components contribute most significantly to interference. This spatial disposition guides the equalization or turbo-equalization process in \gls{nsm} detection, allowing the receiver to focus on the most influential neighboring symbols in the multidimensional space. The resulting organization provides both theoretical clarity and implementation efficiency, reinforcing the relevance of analog \gls{ldgm} \glspl{nsm} as a powerful and intuitive basis for next-generation physical-layer design.

The extension of the analog \gls{ldgm} concept to source coding introduces the third paradigm shift in communication system design. Following the transition from \gls{ftn} signaling to \glspl{nsm} (first shift) and from \glspl{nsm} to analog \gls{ldgm} \glspl{nsm} (second shift), this step unifies source coding, error-correction coding, and modulation within a single analog \gls{ldgm} framework, creating a structured analog-domain representation throughout the processing chain.

In this framework, analog \gls{ldgm}-based source coding generates bipolar sequences that serve directly as inputs to the analog \gls{ldgm} \gls{nsm} at the transmitter. Error-correction coding, typically realized via \gls{ldpc} codes due to their structural similarity to analog \glspl{ldgm}, acts as an intermediate structurer. At the transmitter, it organizes and “cleans” the bipolar sequences produced by the source encoder, thereby reducing the covering radius at the source coding stage. Error-correction coding operates exclusively on codewords, bypassing the use of information bits and avoiding any associated mapping or demapping steps. At the receiver, the detected bipolar symbols undergo a similar structured cleaning process. Here, the \gls{ldpc} decoder serves to increase the packing radius at the analog \gls{ldgm} \gls{nsm} stage, which improves detection performance. As at the transmitter, this operation focuses on codewords rather than information bits.

This unified approach eliminates the need for Gray mappings or explicit bit-to-symbol assignments. Source coding, error correction, and modulation are tightly coupled within the analog \gls{ldgm} paradigm, while each maintains a distinct functional role: the source encoder produces structured bipolar symbols, the \gls{ldpc} code imposes local structure and enables iterative cleaning, and the analog \gls{ldgm} \gls{nsm} maps these sequences into transmitted signals with predictable interference and analytically tractable properties.

The resulting framework supports parallelized, message-passing-based implementations while retaining the analytical transparency and structural regularity characteristic of the analog \gls{ldgm} principle. By treating source coding, modulation, and error-correction coding as tightly integrated but functionally distinct components, this third paradigm shift offers a unified, highly flexible, and potentially near-optimal architecture for next-generation systems, where the traditional separation of source coding, error-correction coding, and modulation is replaced by a strongly joint analog-domain processing approach.

For decades, the research community devoted tremendous effort to error-correction coding, leading to major advances such as \gls{ldpc}, turbo, and polar codes. In contrast, modulation design, although equally fundamental to the communication chain, remained comparatively underexplored. The prevailing assumption was that, once an adequate constellation and a powerful error-correcting code were selected, system performance was essentially optimized. This view, historically justified by the remarkable progress achieved in coding theory, nonetheless restricted the exploration of the wide design space available in the modulation domain.

This work highlighted that modulation design itself represents a major and independent dimension of performance optimization. The introduction of \glspl{nsm} marked a conceptual shift in this direction. \glspl{nsm} demonstrated that modulation could actively regulate error-event multiplicities and intentionally shape interference through carefully designed filter structures. This controlled interference preserves the \gls{msed} of $2$-ASK, ensuring robustness without compromising spectral efficiency. Moreover, the multiplicity of low-distance error events decreases exponentially with the Hamming weight of the input difference, directly limiting their contribution to the overall error rate. As a result, modulation becomes not merely a passive transmission mechanism but an active contributor to robustness and efficiency.

Building on this foundation, analog \gls{ldgm} \glspl{nsm} were proposed as a natural and promising generalization of the \gls{nsm} concept. These architectures, inspired by the sparse matrix organization of \gls{ldgm} and \gls{ldpc} codes, are expected to promote deeper synergies between modulation and coding, potentially bridging long-standing gaps between these traditionally distinct domains. The analog-domain formulation of such \glspl{nsm} may open the way to novel design principles and analytical tools that enable unified physical-layer representations. In the perspective of analog \gls{ldgm} \glspl{nsm}, additional embedded irregularity or structured randomness can suppress harmful alignment of error events and further enhance robustness.

When coupled with \gls{ldpc} error-correction coding alone, analog \gls{ldgm} \glspl{nsm} form a powerful coded-modulation architecture in which information words remain meaningful: mappings between information words and codewords are defined, and demapping from decoded codewords enables information recovery. When combined with analog \gls{ldgm}-based source coding, on the other hand, the system may achieve an even more integrated form, eliminating the need for discrete information bits altogether. Finally, in a fully joint source–error-correction–modulation framework, the error-correction code acts as a structuring mechanism, ensuring local consistency between the source and modulation domains while supporting message-passing interactions that preserve low complexity.

These prospective directions aim to stimulate renewed research interest in modulation as a central element of system design. Modulation, long regarded as a fixed and secondary component, can evolve into a core enabler of physical-layer innovation. The perspectives presented in this technical report, particularly the analog \gls{ldgm} \gls{nsm} framework and its potential coupling with source and error-correction coding, are intended to encourage future theoretical, algorithmic, and experimental investigations. They outline a promising research path toward more unified and flexible architectures for $6$G and beyond, where modulation may once again become a key driver of progress.

Looking forward, several promising research directions arise from the \gls{nsm} and analog \gls{ldgm} framework:
\begin{itemize}

    \item \textbf{Digital-–analog packing optimization in error-correction–aided \glspl{nsm}:} In systems combining \glspl{nsm} with error-correction coding, the length of the coded sequence—corresponding to the number of \gls{nsm} substreams or, equivalently, the number of rows in the analog \gls{ldgm} generating matrix—can be flexibly adjusted for given information and modulated sequence sizes. Increasing this length introduces additional degrees of freedom that enhance packing efficiency and improve performance, but at the expense of higher computational complexity. Identifying the optimal balance between these opposing effects remains a central challenge for practical system design.

    \item \textbf{Joint source coding, error-correction coding, and analog \gls{ldgm} \glspl{nsm}:} In the perspective of a fully joint design, the length of the codewords exchanged between the analog \gls{ldgm} source encoder, the error-correction code, and the analog \gls{ldgm} \gls{nsm} modulator determines the effective modulation resolution. Longer codewords increase the granularity of the modulated sequences and improve source representation, potentially reducing distortion. However, if the modulation granularity exceeds the precision needed relative to the noise variance at the detector, further increases yield negligible improvement while substantially increasing computational complexity. Consequently, for a given channel \gls{snr}, an optimal trade-off must be found to balance modulation granularity with processing cost. Properly tuned, this joint design has the potential to approach information-theoretic limits, achieve near-theoretical efficiency, and minimize end-to-end distortion, highlighting the power of integrated source-channel-modulation optimization.

    \item \textbf{Extension to non-binary alphabets:} As in FTN~\cite{Rusek06b,Dasalukunte14b}, the \gls{nsm} framework, and its generalization to analog \gls{ldgm} \glspl{nsm}, is naturally generalizable to alphabets beyond the conventional binary/bipolar case. For alphabets whose size is a power of two, such as $4$-ASK or $8$-ASK, non-Gray mappings effectively correspond to reusing the same underlying binary/bipolar filter with different scaling factors. Consequently, these configurations do not constitute a genuine extension of the binary/bipolar paradigm and introduce strong energy imbalances among the underlying substreams. In contrast, alphabets that are not powers of two—such as a ternary digital alphabet $\{0,1,2\}$ and the corresponding ternary analog alphabet $\{-1,0,+1\}$—cannot be mapped onto any equivalent binary/bipolar system. Restricting attention to \glspl{nsm} and analog \gls{ldgm} \glspl{nsm}, such non-binary alphabets offer new opportunities for modulation design and filter optimization. This perspective naturally extends to more complex setups, including joint channel coding and modulation, or fully joint source coding, channel coding, and modulation. In these scenarios, the analog \gls{ldgm} paradigm at the source encoder/decoder and modulator, combined with digital \gls{ldpc} channel coding, could potentially exploit the richer structure of non-conventional alphabets to improve performance, enhance design flexibility, and provide novel error-protection properties.

    \item \textbf{Exchange of soft decisions in the analog \gls{ldgm} framework:} In the proposed joint source coding, channel coding, and modulation setup, iterative message passing occurs between the analog \gls{ldgm} source encoder and the digital channel decoder at the transmitter, and between the modulator/demodulator and the channel decoder at the receiver. The \gls{ldpc} code primarily provides local structural constraints and assists in cleaning the exchanged information, but its presence is not directly felt in the exchange between source coding and modulation. In the discussions presented in the report, the analysis was restricted to the exchange of hard-decided binary or bipolar symbols between the source coding and modulation stages at both the transmitter and receiver sides. While this simplifying assumption helped clarify the underlying principles, it limits performance by discarding reliability information available during message passing. A natural extension is to replace these hard decisions with soft decisions, derived from the latest intrinsic and extrinsic information in the respective iterative processes. Soft exchanges are more informative: when a bipolar symbol is unreliable, exchanging its soft-decision value is preferable to using a hard decision in $\{-1,+1\}$. This soft-information exchange represents a promising extension of the analog \gls{ldgm} framework for source coding and modulation.

    \item \textbf{Reduced-complexity detection algorithms for higher-dimensional \glspl{nsm}:} As \gls{nsm} dimensionality increases—whether in \gls{2d}, \gls{3d}, or mixed-dimensional configurations—the computational burden of optimum \gls{ml} detection grows rapidly. While proposed two-stage detection strategies offer substantial complexity reduction compared to direct one-stage ML detection, the complexity remains prohibitive for larger packet sizes. This forces the use of smaller packets to maintain tractable complexity, which in turn makes the fixed filter spans relatively more costly, significantly reducing spectral efficiency. For small packet sizes, multidimensional tail-biting is risky, as it could compromise the achievable \gls{msed}. To enable practical use of higher-dimensional \glspl{nsm} and mixed-dimensional designs, alternative suboptimal detection algorithms must be explored. Prospective approaches include sphere decoding, possibly augmented with soft outputs when error correction coding is employed, as well as multi-stream-inspired methods akin to those proposed by Anderson \emph{et al.}~\cite{Anderson13}. Systematic assessment of these algorithms’ performance, complexity, and trade-offs will be crucial to realizing the benefits of multidimensional \glspl{nsm} in practical transmission scenarios.

    \item \textbf{Joint matrix optimization and controlled interleaving assessment:} In combined error-correction–modulation systems, the generating matrix of the analog \gls{ldgm} \gls{nsm} should be jointly optimized with the parity-check matrix of the \gls{ldpc} code to maintain low densities while ensuring strong minimum coded \glspl{sed}. In the fully joint source–error-correction–modulation configuration, the generating matrices of the analog \gls{ldgm} source and modulation components, together with the parity-check matrix of the error-correction code, must be co-optimized to achieve the lowest possible densities compatible with satisfactory distortion levels over various channel conditions and \glspl{snr}. The error-correction code may be configured with either shallow or strong structuring, depending on the desired redundancy level and the targeted resolution in both the source and modulation spaces. Furthermore, the role of interleaving must be systematically assessed: determining whether performance gains are achieved, what interleaver structures are most beneficial, and how they interact with the analog \gls{ldgm} framework. A single interleaver may be placed between the error-correction code and the modulator, or two interleavers may be used in the fully joint configuration—one between the source code and the error-correction code, and another between the error-correction code and the modulator. In highly ordered multidimensional \glspl{nsm}, such interleaving may prove essential to enhance robustness while preserving the underlying structural coherence.

    \item \textbf{Extension of quasi-cyclic and protograph structures to analog \gls{ldgm} systems:} In the digital domain, protograph-based and quasi-cyclic (QC) \gls{ldpc} codes \cite{Liva06,Li17,Song23,Smarandache22,Dehghan18} have proven instrumental in simplifying both the specification and implementation of sparse parity-check matrices. Their compact base-matrix representations and cyclically shifted identity submatrices provide structural regularity that supports efficient iterative message-passing decoding and scalable hardware realization. A similar line of reasoning can be extended to the analog \gls{ldgm} paradigm, where generating matrices operate over integer or rational alphabets rather than binary fields. In this analog framework, two base matrices can be introduced to fully characterize the structure: a primary base matrix specifying the circular shifts of identity submatrices, and a secondary base matrix specifying the corresponding integer or real scaling (homothety) factors. Each submatrix of the analog \gls{ldgm} generating matrix is then obtained as a scaled and circularly shifted identity matrix, with a null submatrix produced when the scaling factor equals zero. This dual-base representation generalizes the QC and protograph concepts to the analog domain, providing a compact and systematic way to describe structured analog \gls{ldgm} systems. Two main application scenarios can be distinguished:
	\begin{enumerate}
		\item \textbf{LDPC coding combined with analog \gls{ldgm} modulation:} The modulator has its own pair of base matrices, defining the circular shifts and scaling factors of the generating matrix. The LDPC code itself can be represented using a classical protograph structure. Joint optimization of the modulator's base matrices and the LDPC protograph can enhance performance by aligning energy distributions and structural irregularities. Interleaving between the LDPC code and the modulator may further improve robustness against low-Euclidean-distance error events.
		\item \textbf{Fully joint source coding, channel coding, and modulation:} In this scenario, both the analog \gls{ldgm} source encoder and the modulator have their respective pairs of base matrices, which can be jointly optimized to improve end-to-end performance. The LDPC code retains a classical protograph representation and may also be optimized jointly with the source and modulation base matrices. Interleaving between the LDPC code and each of the source code and the modulator can help scatter low-Euclidean-distance error events, supporting better iterative convergence and source reconstruction quality.
	\end{enumerate}
	Structured analog constructions of this type could substantially reduce the complexity of specification and iterative processing in analog \gls{ldgm}-based modulation and source-coding systems, while providing a unified framework for joint optimization in both two-component (LDPC + modulator) and fully joint three-component (source + LDPC + modulator) scenarios, paving the way toward modular and hardware-efficient implementations.
	
	\item \textbf{Block-diagonal analog \gls{ldgm} and \gls{ldpc} structures with interleaving for scalable joint coding and modulation:} A promising approach for simplifying the implementation of analog \gls{ldgm} systems and \gls{ldpc} channel coding is the use of block-diagonal generating matrices for the analog components and block-diagonal parity-check matrices for the \gls{ldpc} code. In this paradigm, the mother analog \gls{ldgm} or \gls{ldpc} matrix can be used for modulation, source coding, or channel coding, and is decomposed into smaller diagonal blocks, each corresponding to an independent sub-component:
	\begin{itemize}
		\item For modulation, each diagonal block defines a separate analog \gls{ldgm} modulator, reducing the overall modulation operation into multiple parallel modulators.  
		\item For source coding, each diagonal block defines a separate analog \gls{ldgm} source encoder, reducing the overall source encoding operation into multiple independent sub-blocks.  
		\item For channel coding, the block-diagonal principle can be applied to the \gls{ldpc} parity-check matrix, implementing multiple smaller \gls{ldpc} encoders in parallel.  
	\end{itemize}
	Each elementary sub-block, corresponding to a portion of a source encoder, modulator, or \gls{ldpc} encoder, must already provide sufficiently good performance so that interleaving and message passing are effective. Full performance is restored by the iterations of the message-passing algorithm combined with carefully designed interleaving:
	\begin{itemize}
		\item Source–\gls{ldpc} interleaving ensures bidirectional spreading: each source encoder sub-block is spread across all \gls{ldpc} decoder sub-blocks, and each \gls{ldpc} decoder sub-block is also spread across all source encoder sub-blocks at the transmitter. This allows iterative message passing between the source encoder and the \gls{ldpc} decoder to propagate reliability information effectively across all blocks.  
		\item \gls{ldpc}–modulator interleaving ensures bidirectional belief spreading: each demodulator sub-block spreads information over all \gls{ldpc} decoder sub-blocks, and each \gls{ldpc} decoder sub-block spreads information back over all demodulator sub-blocks at the receiver. This iterative exchange reinforces message propagation across the entire set of sub-blocks.  
		\item Row-column block interleaving provides additional spreading within each analog \gls{ldgm} or \gls{ldpc} sub-block. While more sophisticated interleavers could be designed, row-column block interleaving is simple to implement and expected to achieve efficient propagation of reliability information in these configurations.  
	\end{itemize}
	This methodology applies to both channel coding plus modulation alone (no source coding) and joint source coding–channel coding–modulation systems. In both cases, iterative message passing combined with interleaving allows the system to approximate the performance of the original full-size mother matrices, while maintaining the benefits of hardware simplicity, modularity, and parallel processing. By atomizing the analog \gls{ldgm} and \gls{ldpc} matrices, designers gain a flexible, scalable, and implementation-friendly architecture, where each elementary block can be independently optimized, reused across system configurations, and connected via interleaving to achieve high-performance joint coding and modulation.

    \item \textbf{Message-passing strategies and algorithmic optimization:} A key research direction concerns the design and evaluation of message-passing algorithms governing the interaction between analog \gls{ldgm} components and the \gls{ldpc} error-correction structure. It remains to determine whether identical or distinct message-passing schemes should be employed between the analog \gls{ldgm} source encoder and the \gls{ldpc} decoder at the transmitter, and between the analog \gls{ldgm} \gls{nsm} detector and the \gls{ldpc} decoder at the receiver. These algorithms must be assessed for their ability to approach optimal or near-optimal bipolar sequences exchanged between the source and modulation domains, while leveraging the error-cleaning capability provided by the error-correction code. Alternative formulations—including sigma–delta-inspired quantization and other iterative source-coding algorithms with integrated \gls{ldpc}-based cleaning—should also be explored. In parallel, the introduction of structural regularity, enabled for instance by multidimensional \glspl{nsm} or coordinated matrix designs, should be investigated as a means to simplify processing, support parallelism, and reduce complexity while preserving performance.

    \item \textbf{Tail-biting extensions to structured \glspl{nsm}:} Originally developed in the digital domain for convolutional codes, the tail-biting principle naturally extends to structured \glspl{nsm} such as \gls{ms-prs} modulations, which share a comparable sequential organization. This concept can also be generalized to \gls{2d}, \gls{3d}, and mixed-dimensional \glspl{nsm}. However, its use raises key research questions: tail biting may degrade performance for short packets by lowering the achievable \gls{msed} and increasing the multiplicity of low-distance error events, and it may prove unsuitable for certain degenerate \glspl{nsm} such as rate-$2$ “biphase” or “dicode” configurations. A detailed investigation is therefore required to evaluate under which conditions tail biting can preserve the \gls{msed} while maintaining a limited increase in error-event multiplicity across multidimensional and mixed-dimensional \glspl{nsm}.

    \item \textbf{Exploration of \gls{ms-prs} with richer rational filter structures:} Thus far, only \gls{ms-prs} configurations with a single filter of length greater than one have been investigated. Extending this study to include rational \gls{ms-prs} structures where multiple filters have lengths exceeding one could provide greater design flexibility, helping to maintain the target \gls{msed} of $2$-ASK while reducing the multiplicities of low-distance error events. Equivalence relations may be leveraged to keep the search space tractable even for long filters or high \gls{nsm} rates. Initial results for rate-$2$ \glspl{ms-prs} showed that most filter configurations (except one) failed to achieve the $2$-ASK \gls{msed}, indicating that richer filter structures could facilitate this goal. Moreover, allowing longer filters across multiple branches could prove advantageous in channels with selectivity in time, frequency, or space, by introducing intrinsic modulation-level diversity—complementing the diversity typically recovered through interleaving and error-correction coding, and enabling combined analog–digital diversity gains.

    \item \textbf{Extending \glspl{nsm} with non-linearity for enhanced performance, diversity, and \gls{papr} reduction:} All studies and perspectives presented so far have remained within the linear framework, where contributions are linear combinations of filtered bipolar sequences (\gls{ms-prs}) or linear combinations of generating-matrix rows weighted by bipolar symbols (analog \gls{ldgm} \gls{nsm}). A natural extension is to allow non-linear transformations within the \gls{nsm} paradigm, inspired by research on non-linear \gls{scma} variants~\cite{Huang22}. Such non-linear \glspl{nsm} could exploit additional degrees of freedom in the mapping from bipolar sequences or generating-matrix rows to transmitted signals, potentially achieving performance beyond the $2$-ASK benchmark, without relying on channel coding. Non-linearity could also introduce additional diversity at the modulation level when transmitting over selective channels, analogous to the diversity benefits observed in linear \glspl{nsm} when combining multi-branch filter structures with interleaving. In the one-dimensional case, non-linear \glspl{nsm} may admit a trellis-like representation, reminiscent of Ungerboeck’s trellis-coded modulation, which could further enhance performance or facilitate structured detection. Moreover, non-linear \glspl{nsm} present opportunities for reducing the \gls{papr}, a critical consideration in single-carrier transmission systems for efficient power amplifier operation. While complex tapped filters were suggested in the linear case to reduce \gls{papr}, incorporating non-linear operations could achieve even greater \gls{papr} reduction, offering a practical advantage for power-limited transmitters. These extensions suggest a rich design space where \glspl{nsm} can be optimized for enhanced performance, increased diversity, and reduced \gls{papr}.

    \item \textbf{Discrete-time \gls{cpm}-inspired multistream modulation for constant-envelope generation:} Pushing the \gls{papr} reduction philosophy to its conceptual extreme, a discrete-time paradigm inspired by \gls{cpm} \cite{Aulin81A,Aulin81B,Anderson86} can be envisioned. In classical \gls{cpm}, the modulated signal is generated directly in continuous time with constant modulus. The primitive of the continuous-time shaping filter is used to filter the stream of data symbols, thereby constructing the phase signal that governs the transmitted signal. In the discrete multistream framework considered here, an analogous principle can be adopted: the modulated samples are generated with perfectly constant modulus in discrete time, while a subsequent shaping filter, such as a square-\gls{rrc} filter, is applied in a single-carrier context to produce the final complex envelope of the transmitted signal. This final shaping stage inevitably induces a small \gls{papr} increase with respect to the ideal discrete-time constant-modulus sequence, but such an effect is inherent to the discrete framework and remains limited when the discrete generation enforces a constant-envelope sequence. In continuous-time \gls{cpm}, the shaping filter is specified and its primitive is then used to filter the data symbols and generate the phase of the transmitted signal. In the discrete-time framework proposed here, the notion of a continuous-time primitive is not directly applicable. This difficulty is overcome by defining, directly in discrete form, the equivalents of the primitive shaping filters of classical \gls{cpm} for each substream. Hence, there is no need to define discrete shaping filters, since all relevant information is carried by their discrete primitive equivalents. Each of these discrete primitive functions starts from zero and asymptotically reaches unity after a finite transition interval. The finite transition region corresponds to what would have been the support of the underlying discrete shaping filter associated with that primitive. The optimization process therefore focuses on this transition interval, as it governs the discrete phase evolution and, consequently, the spectral properties of the transmitted signal. Distinct modulation indices can be associated with the different discrete primitive filters, forming a flexible multi-index configuration analogous to the multi-$h$ structure of advanced \gls{cpm} schemes \cite{Itoh90,Sasase91}. As in multi-$h$ \gls{cpm}, the design objectives are to maximize the \gls{msed} and to minimize the occupied spectral bandwidth. In the proposed discrete framework, the overall shaping effect results from two combined sources: the choice of the discrete primitive functions defining the multiple substreams, and the final continuous-time shaping filter (e.g., \gls{rrc}) used to produce the single-carrier transmitted signal. Furthermore, the discrete multistream formulation with a bipolar alphabet naturally accommodates richer modulation alphabets. For illustration, a quaternary alphabet can be obtained by using two bipolar substreams filtered by two shaping filters that are scaled versions of each other by a factor of two. Such a construction yields a $4$-ASK equivalent signaling scheme but does not fully exploit the flexibility of the discrete multistream structure, since both substreams use filters of identical shape. Allowing the discrete primitive functions to differ in form and associating distinct modulation indices with them provides additional design degrees of freedom. This enhanced flexibility goes beyond what is achievable in classical \gls{cpm}, in multi-$h$ \gls{cpm}, and in richer-alphabet \gls{cpm} variants, enabling the exploration of optimized trade-offs between \gls{msed} and spectral occupancy within a unified discrete-time constant-envelope framework.

    \item \textbf{Energy balancing and turbo-equalization:} Energy imbalances between \gls{nsm} filters can facilitate the bootstrap phase of turbo-equalization when \glspl{nsm} are combined with error-correction coding. Once the iterative process has converged, however, balanced-energy configurations yield the best performance and are required to achieve the \gls{ber} performance of $2$-ASK. While our previous studies of rational-tap \glspl{nsm} focused exclusively on balanced energies, future work should extend the analysis to unbalanced configurations, which could better support the bootstrap phase of turbo-equalization, especially at low \glspl{snr}.

    \item \textbf{Multiplicity-aware design criterion for \glspl{nsm}:} Till now, studies in \gls{ftn}, starting from the seminal work of Mazo, focused on the \gls{msed} of the \gls{ftn} system as a design criterion. We followed this same approach in our studies on \glspl{nsm}. However, using the \gls{msed} alone is not sufficient to faithfully characterize performance, particularly at low to moderate \glspl{snr}. This limitation was evidenced in the rate-$2$ \gls{ms-prs} modulations with bipolar non-null taps considered in our work. Even though their \gls{msed} was only half that of $2$-ASK, these \glspl{nsm} were able to closely approach the \gls{bep} of $2$-ASK. The rationale behind this behavior is that the multiplicity of error events corresponding to this sub-$2$-ASK \gls{msed} rapidly diminishes—exponentially—when the length of the filter exceeding one grows beyond $8$. This phenomenon highlights that good \gls{nsm} design should not only ensure that the \gls{msed} is acceptable, but also explicitly consider the multiplicities of error events with \glspl{sed} below the $2$-ASK \gls{msed}. These multiplicities should vanish as filter lengths increase. By doing so, their impact on performance is effectively suppressed, particularly at low to moderate \glspl{snr}. We believe this behavior is general and can be naturally achieved in analog \gls{ldgm} \glspl{nsm} thanks to the randomness intrinsically offered by their generating matrices. This randomness helps prevent destructive contributions to the output (modulated) sequence differences from the nonzero components of error-event input sequences.

    \item \textbf{Tightness of filters and its impact on \gls{nsm} performance:} We observed that tightness was consistently verified for rate-$2$ \gls{ms-prs} modulations with real tapped filters as long as the longest filter’s length remained below $9$. Once the longest filter reached a length of $10$, this tightness disappeared. The tightness characteristic enhances packing in the modulated signal space, which contributes to good performance for short filter lengths of the longest filter. However, it also dramatically increases the multiplicity of error events corresponding to the \gls{msed}, which we found to be the origin of degraded performance compared to $2$-ASK, particularly at low \gls{snr} in terms of \gls{ber}. Owing to its efficient packing, tightness was adopted as a guiding principle in the optimization of all other \glspl{nsm} using rational tapped filters. We believe that the negative impact of tightness at low to moderate \glspl{snr} can be mitigated in coded scenarios through the natural mismatch between error events of the modulation and those of the error-correction code, a phenomenon we previously referred to as the “exclusion principle.” Consequently, even if tightness significantly affects performance in the uncoded \gls{nsm} case, leveraging the exclusion principle can ensure good performance at low to moderate \glspl{snr} for coded \glspl{nsm} in terms of coded \gls{ber}. The precise effect of tightness under coding, particularly the extent to which the exclusion principle can counteract the multiplicity-induced degradation, requires further investigation. Since coded \gls{ber} represents the ultimate target for system performance, understanding and quantifying this interaction is crucial for the design of high-performance \glspl{nsm} in practical, error-corrected scenarios.

    \item \textbf{Interleaving and low-Hamming-weight error event mitigation in coded \glspl{nsm}:} To approach the coded \gls{bep} performance of $2$-ASK as closely as desired, \glspl{nsm} must ensure that critical low-Hamming-weight error events of the error-correction code lead to output modulated sequence differences with Euclidean distances identical to those in $2$-ASK. Two key principles support this goal: (i) balancing the energies of all \gls{nsm} filters, analogous to the uniform symbol energy in $2$-ASK, and (ii) preventing destructive overlap after filtering in the output modulated sequence differences of the non-null components of the input sequence differences corresponding to a critical code error event. One simple way to achieve this is by scattering these non-null components: thanks to the bounded lengths of the \gls{nsm} filters and the low Hamming weight of the critical code error events, the contributions of these components to the output sequence differences can be easily distributed in a non-overlapping fashion, preserving the target Euclidean distances. The efficiency of this scattering increases with modulated packet size, allowing progressively more critical code error events to be mitigated. This mechanism exhibits two parallels with turbo codes: the interleaver in \glspl{nsm} disperses low-Hamming-weight sequences, while in \glspl{pccc}, interleaving distributes confined low-weight binary sequences to allow recursive processing to exploit dispersed information sequences of low Hamming weight efficiently.

    \item \textbf{Understanding and overcoming the general intractability of closed-form derivations for optimal filters in rate-$2$ \glspl{nsm}:} The derivation of closed-form expressions for the optimal filters of one-dimensional rate-$2$ \glspl{nsm} with real taps becomes intractable when the filter length exceeds moderate values, specifically for lengths greater than $7$ in the balanced-energy framework and greater than $8$ in the unbalanced-energy framework. This difficulty may arise from any of the three stages of the analytical procedure: numerical optimization, identification of the relevant error events, or symbolic determination of the closed forms. As a first remedy, the numerical search for the best filters can be substantially improved. The optimization algorithm initially employed was a rudimentary simulated-annealing-like method, chosen primarily to enable rapid progress and obtain interesting results quickly, rather than to guarantee with great confidence the optimality of the obtained solutions. However, the failure to derive closed-form expressions for longer filters may partly stem from the suboptimal convergence properties of this basic algorithm. A significant enhancement would therefore be to replace it with more advanced metaheuristic optimization techniques, such as \gls{pso}~\cite{Kennedy95}, the \gls{gwo}~\cite{Mirjalili14}, \gls{cs}~\cite{Yang09}, \gls{sso}~\cite{Abualigah24}, \gls{kh}~\cite{Gandomi12}, or the \gls{woa}~\cite{ Mirjalili16}. These methods offer a much higher capability to escape local extrema and to converge toward the true global optima, potentially yielding more accurate filter candidates for symbolic derivation. A second improvement concerns the identification of the error events corresponding to the \gls{msed}. The current algorithm explores all potential error events below a given length threshold, which leads to exponential complexity and prevents exhaustive identification for long filters. In our work, the trellis structure of all error events was already used to determine the \gls{msed} of a given \gls{nsm} based on its underlying filter coefficients. We believe that this trellis-based algorithm can be augmented to provide, with a computational cost that grows linearly with the error-event length, the set of all error events corresponding to the \gls{msed}. This enhancement could be achieved by performing a multiple aggregated trace-back on the existing trellis structure. Incorporating such an improvement would significantly strengthen the reliability and scalability of the second stage of the derivation, thereby enhancing the overall feasibility of obtaining closed-form expressions for longer optimal filters.

    \item \textbf{Addressing the multiplicity and structural indeterminacy of optimal solutions for 10-tap filters in rate-$2$ \glspl{nsm}:} A distinct and particularly challenging situation arises when the longest filter in a rate-$2$ \gls{nsm} reaches a length of $10$ taps. In this configuration, both the balanced- and unbalanced-energy frameworks converge and achieve, for the first time, the same \gls{msed} as $2$-ASK. However, this convergence reveals a deeper difficulty: the existence of an extremely large, and possibly infinite, set of non-equivalent optimal filters achieving the same \gls{msed}. For real-tap filters, this multiplicity translates into a continuum of optimal solutions, causing the corresponding filters to lose any refined structural regularity or algebraic organization. This structural indeterminacy is believed to be the primary reason why closed-form expressions could not be obtained for this case. To overcome this limitation, a promising research direction is to extend the optimization criterion beyond the first \gls{msed}. By also considering the second \gls{msed}, it may become possible to discriminate among the numerous \gls{msed}-equivalent solutions and isolate a much smaller subset of filters—potentially a unique class up to known equivalence relations. By being uniquely defined in this refined sense, the optimal filters could recover a structured organization in their associated \gls{msed} error events, which would, in turn, facilitate the analytical derivation of their closed-form expressions.

    \item \textbf{Adapting the analog \gls{ldgm} framework to source and channel characteristics:} The analog \gls{ldgm} paradigm offers a natural ability to adjust to the characteristics of both the source and the modulation processes. In the case of standalone analog \gls{ldgm} \glspl{nsm}, where the modulation operates independently from any source-coding stage, regularity principles were established in close analogy with those of \gls{ldpc} codes, yet applied to the generating matrix in the analog domain instead of the parity-check matrix in the digital domain. Two complementary forms of regularity were identified: regularity with respect to the rows and regularity with respect to the columns of the generating matrix. Row regularity ensures that the corresponding bipolar information symbols are balanced in energy, a crucial property for achieving the performance of coded $2$-ASK. Column regularity, on the other hand, guarantees that the generated signal samples exhibit uniform average energy or power, which is a desirable characteristic for transmission over channels with uniform quality. However, in static frequency-selective channels, such as those encountered in \gls{adsl}~\cite{Starr03,Yu07} systems, subcarriers experience different \gls{snr} levels, and the well-known water-filling principle~\cite{Gallager72,Cover06} dictates that subcarriers with higher \glspl{snr} be allocated more power and larger constellation sizes. The analog \gls{ldgm} \gls{nsm} framework can inherently follow this principle through the deliberate introduction of irregularity in the generating matrix. By adapting the average energy per column to the quality of the subcarrier to which the corresponding signal component is mapped, the system can allocate more energy and involve a greater number of active rows for columns associated with high-quality subcarriers. This results in an effective analog-domain realization of water-filling that statically aligns the modulation structure with channel conditions. The same principle extends naturally to the joint source-coding, channel-coding, and modulation configuration when the analog \gls{ldgm} paradigm is also employed for source coding. Classical transforms such as the \gls{dct}~\cite{Ahmed74,Rao90} or the \gls{klt}~\cite{Fukunaga90,Rao01,Papoulis02}, also known as \gls{pca}~\cite{Jolliffe02,Jolliffe16}, produce output samples with inherently unequal average energies. To adapt to this property, irregularity can again be introduced among the columns of the analog \gls{ldgm} generating matrix at the source encoder: samples with higher average energy are associated with columns of higher energy and denser row participation. This tailored irregularity enhances source reconstruction quality while minimizing distortion. These structural adaptation mechanisms highlight the intrinsic flexibility of the analog \gls{ldgm} paradigm, positioning it as a promising candidate for context-aware physical-layer design capable of matching its internal organization to both the source and the transmission environment.

    \item \textbf{Joint source, channel, and modulation optimization with irregular generating-matrix rows and no performance loss:} When row regularity is not enforced in the generating matrices of the analog \gls{ldgm} \gls{nsm} framework, flexibility is introduced in the allocation of energy among the rows. In the case of standalone \gls{ldpc}-coded modulation (i.e., without joint source coding), this situation resembles \gls{ssc}~\cite{Rush17}: energy imbalances among rows may improve iterative decoding convergence, but they are not theoretically optimal from an information-theoretic perspective. Therefore, while the framework allows such irregularities, their use is not encouraged in this isolated modulation scenario. However, in the fully joint source–channel–modulation configuration, where both the source code and the modulator adopt the analog \gls{ldgm} paradigm, unbalanced energy allocation among the rows of both generating matrices can be applied without any theoretical loss of performance, assuming an ideal decoder exists. Each exchanged bipolar symbol corresponds to a row in the source-code generating matrix and a row in the modulator generating matrix. By maintaining proportional energy allocation between the corresponding rows across source and modulator (i.e., the ratio of energies for each symbol is constant), the system ensures that more influential symbols in the source reconstruction naturally benefit from higher transmitted energy, reducing their error probability. Consequently, the potentially greater impact of these symbols on reconstruction is offset by their enhanced protection, preserving optimality. This row-wise energy irregularity allows for hierarchical reconstruction of the source and modulator signals, analogous to non-Gray 4-ASK mappings, where certain bipolar symbols carry more significance. The approach can reduce the number of bipolar symbols exchanged between the source and modulator through the \gls{ldpc} code, leading to lower computational complexity for the same reconstruction quality. Further, \gls{ldpc} code irregularity (through unequal parity-check participation) can complement this design by reinforcing protection for lower-energy, more vulnerable bipolar symbols. Individually, these low-energy symbols have limited influence on reconstruction, but collectively their errors can accumulate and degrade source reconstruction quality. By involving these symbols in a larger number of parity checks, the \gls{ldpc} code mitigates the cumulative effect of such errors, ensuring reliable overall performance. Finally, this framework maintains complete flexibility regarding column irregularity in both generating matrices, which can still be leveraged independently to adapt to non-uniform source statistics or static channel SNR variations. The combination of row- and column-level flexibility positions the analog \gls{ldgm} paradigm as a highly adaptable tool for context-aware joint source–channel–modulation optimization.

    \item \textbf{Extending the \gls{nsm} paradigm to \gls{noma}~\cite{ Islam17, Ding17, Xia20, Liu21,Hu22} systems:} in both cases, several independent information streams are intentionally superposed on the same transmission resources, and reliable separation must be achieved at the receiver. In rate-$2$ \glspl{nsm}, the most critical configuration corresponds to energy balance between the two filters, which maximizes the risk of destructive interference while also defining the condition required to achieve the maximum theoretical \gls{msed}. In \gls{noma}, an equivalent situation arises when the simultaneous transmissions of two users reach the base station with comparable received powers, making user separation and interference cancellation particularly challenging. Extending the \gls{nsm} paradigm to \gls{noma} would consist in assigning to each user a dedicated partial-response filter and optimizing these filters jointly over a controlled range of energy imbalance, thereby enabling robust separation even in near-balanced power conditions. Moreover, adopting the analog \gls{ldgm} \gls{nsm} framework for each user introduces beneficial structured randomness through distinct, nearly uncorrelated generating matrices, which enhances user separability and mitigates destructive interference. This perspective opens the way to a unified design of joint modulation–multiple-access schemes grounded in \gls{nsm} principles.

    \item \textbf{Adaptive matrix-based NSM framework:}  A central long-term perspective for joint source coding, channel coding, and analog LDGM NSM modulation is the development of a fully adaptive transmitter--receiver chain, where adaptation is enabled through pre-optimized finite batteries of generating matrices for the source code and modulator, alongside a battery of parity-check matrices for the LDPC channel code. These matrices allow the system to flexibly respond to changing channel conditions and source statistics by selectively adapting one or more components while keeping the others unchanged. Three elementary adaptation scenarios illustrate the versatility of this approach:  
    \begin{enumerate}

        \item \textbf{Channel-code-only adaptation:} When the channel degrades, the source code and modulator remain unchanged. Adaptation occurs by increasing the number of check nodes in the LDPC code, effectively reducing the granularity of the source and modulator spaces. This reinforces the channel code’s ability to clean residual errors induced by increased noise variance, thereby preserving source reconstruction quality.  

        \item \textbf{Source-code-only adaptation:} When the channel degrades, the LDPC code and modulator remain unchanged, while the source generating matrix adapts by reducing the number of columns. This decreases the number of transmitted source samples for the same modulated signal size, improving per-sample resolution and reducing source distortion. Energies of the rows of the source generating matrix are decreased, column density is increased, and a richer symbol alphabet is deployed to preserve discriminability at the source decoder.  

        \item \textbf{Modulator-only adaptation:} When the channel degrades, the source code and LDPC code remain unchanged. Adaptation occurs exclusively at the modulator by increasing the number of columns in its generating matrix, effectively lengthening the transmitted signal. Column sparsity is increased and the average energy per row is raised, improving the quality of bipolar symbols delivered to the source decoder through the LDPC code, thereby reducing distortion.

    \end{enumerate}
    This perspective highlights that single-component adaptations can be combined for joint adaptation, allowing the system to dynamically select the most appropriate configuration. By maintaining batteries of pre-optimized generating and parity-check matrices, the system can flexibly trade spectral efficiency, source distortion, and processing/decoding effort in response to instantaneous channel and source conditions. This framework positions the analog LDGM NSM paradigm as a flexible, implementation-ready foundation for adaptive physical-layer design in next-generation communication systems.

\end{itemize}

These perspectives collectively define an ambitious research agenda for the coming years. By integrating modulation design with coding theory, multidimensional \glspl{nsm}, and analog \gls{ldgm} generalizations, this work establishes a foundation for $6$G and beyond, enabling communication systems that are flexible, high-performance, and computationally efficient.


\newpage

\bibliographystyle{IEEEtran}  
\bibliography{Bibliography}     
\newpage

\appendix


\section{Tight Estimate of the BEP for the Rate-5/4 Block NSM} \label{app:Tight Estimate BEP Rate 5/4}

Before enumerating all error events leading to a \gls{med} for the proposed block modulation of rate $5/4,$ we need to understand how each error event contributes to the overall \gls{bep} estimate, using union bounding. To this end, it is important to underline the difference between the present bipolar signaling, with alphabet $\{ \pm 1 \}$, and the more conventional binary signaling, with alphabet in the Galois field $\{ 0, 1 \},$ which uses modulo $2$ addition and multiplication. In binary signaling, $100 \, \%$ of input blocks are capable of leading to a $0$ or a $1$ in each component of the input block difference. This is also the case for bipolar signaling, when a component of the input block difference is equal to $0.$ However, only  $50 \, \%$ of the input blocks can lead to either $-2$ or $+2$ in one component of the input block difference, in bipolar signaling. Indeed, only those input blocks, $\bm{b}$ with corresponding component equal to $+1$ (respectively, $-1$) can potentially lead to a difference of $-2$ (respectively, $+2$), in $\Delta \bm{b}$. As a consequence, when evaluating the contribution of each error event, we need to apply a weighting factor equal to $(1/2)^w,$ where $w$ is the number of non-null components of $\Delta \bm{b}.$

This is being said, to facilitate the discussion, we split the error events of interest into two categories, those for which $\Delta b[4] = 0$ and those for which $\Delta b[4] = \pm 2.$ For those error events with $\Delta b[4] = 0$, and leading to a \gls{msed} of $4,$ all other components of $\Delta \bm{b}$ should be null, except one, which should take a non-null value in $\{\pm 2\}.$ There are $8$ such input difference blocks, with a single non-null component. These blocks, which cause only one error among the $5$ components of the input block, $\bm{b},$ need to be weighted by a factor of $1/2$ according to the above discussion. Hence, the overall contribution of these input difference blocks, to the evaluation of an estimate of the \gls{bep}, is a weighted multiplicity factor of $4.$

Now, for those error events with $\Delta b[4] = a,$ $a \in \{ \pm 2\}$, and leading to a \gls{msed} of $4,$ all other components of $\Delta \bm{b}$ must take values in the set $\{ -a, 0\}.$ In this way, all components of the corresponding modulated block difference, $\Delta \bm{s},$ take always their values in the bipolar alphabet $\{ \pm 1\},$ guaranteeing the sought \gls{msed} of $4.$

Let $w,$ $0 \le w \le 4,$ denotes number of non-null components among the first $4$ components of such input block difference, $\Delta \bm{b}.$ Then, the corresponding error event causes $w+1$ bipolar (or binary) errors, one of them being the fifth component $\Delta b[4].$ Since $\Delta b[4] = \pm 2$ with probability one, such input block difference should be weighted by a factor of $(1/2)^w.$ Now, if we put together all input block differences having a common value of $w,$ the number of which is equal to $\binom{4}{w},$ then the aggregate contribution to the weighted multiplicity factor is $(w+1) (1/2)^w \binom{4}{w}.$

In summary, the global weighted multiplicity factor of the binary probability error estimate is equal to $4+\sum_{w=0}^4 (w+1) (1/2)^w \binom{4}{w} = \tfrac{253}{16}.$ Remembering that all accounted for binary errors affect all $5$ bipolar inputs of each input block, $\bm{b},$ the final sought multiplicative factor is $\tfrac{253}{80}.$ Since the proposed modulation preserves the \gls{med} of $2$-ASK, this multiplicative factor should be put in front of the corresponding \gls{bep}, leading to the \gls{bep} estimate in (\ref{eq:BEP Estimate Rate 5/4}).



\section{Tight Estimate of the BEP for the Rate-2 NSM} \label{app:Tight Estimate BEP Rate 2}

The derivation of a tight estimate of the \gls{bep} for the rate-$2$ \gls{nsm} is more involving than that of the “duobinary” channel, given in (\ref{eq:BEP Estimate Duobinary Channel}). It requires the computation of an underlying \gls{tf},
\begin{equation} \label{eq:Transfer Function Definition}
T(N,D) = \sum_{l > 0,m \ge 0} \mu(l,m) N^l D^m,    
\end{equation}
similar to convolutional codes, where $\mu(l,m)$ is the number of error events exhibiting an input difference sequence with $l$ non-null components, in $\{ \pm 2\},$ and an output difference sequence with \gls{sen} $m$.

\begin{figure}[!htbp]
    \centering
    \includegraphics[width=1\textwidth]{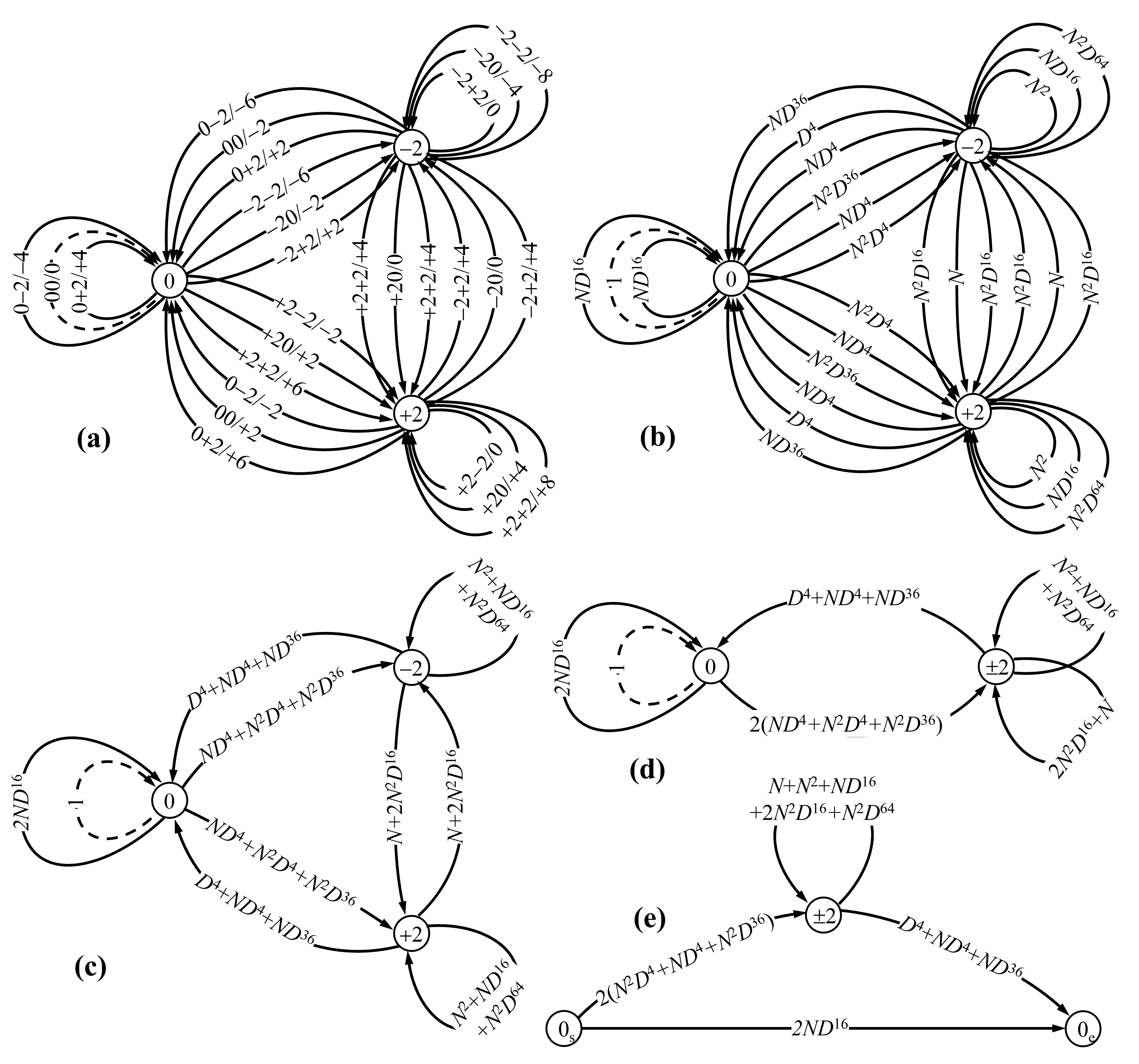}
    \caption{State diagrams of the input/output sequences differences of the rate-$2$ Nyquist signaling modulation: (a) State diagram with Input/Output labels, (b) State diagram with $N^iD^j$ labels, (c) Simplified state diagram, aggregating $N^i D^i$ labels with identical starting and ending states, (d) Simplified state diagram with state merging, (e) Modified state diagram for TF computation. Branch with zero Input/Output difference shown with dashed line.}
    \label{fig:State Diagram Input/output Difference Rate-2 Modulation}
\end{figure}

To derive the aforementioned \gls{tf}, $T(N,D),$ we begin by depicting, in Figure~\ref{fig:State Diagram Input/output Difference Rate-2 Modulation}(a), the state diagram of the input/output sequence differences resulting from the trellis section shown in Figure~\ref{fig:Trellis Input Difference Duobinary Channel Rate-2 Modulation}(b). Each branch in this state diagram has an input/output label of the form $\Delta \bar{b}_0[k] \, \Delta \bar{b}_1[k] / \Delta \mathring{s}[k],$ just like the trellis section in Figure~\ref{fig:Trellis Input Difference Duobinary Channel Rate-2 Modulation}(b). Because we are interested in listing all error events that can occur and lead to decision errors at the demodulator, we have shown the “null” branch with input/output label $00/0,$ in Figure~\ref{fig:State Diagram Input/output Difference Rate-2 Modulation}(a), in a dashed line. This branch, as well as the corresponding branches in Figures~\ref{fig:Trellis Input Difference Duobinary Channel Rate-2 Modulation}(b)--\ref{fig:Trellis Input Difference Duobinary Channel Rate-2 Modulation}(d), should not be included in any error event and, as a result, should not be considered when computing the \gls{tf}.

Given the desired role played by the \gls{tf}, in exhibiting all error events according to their input/output characteristics, we relabel, in Figure~\ref{fig:State Diagram Input/output Difference Rate-2 Modulation}(b), the state diagram of Figure~\ref{fig:State Diagram Input/output Difference Rate-2 Modulation}(a), by assigning labels of the form $N^i D^j,$ to all state diagram branches. For each branch label, $N^i D^j,$ in the state diagram of Figure~\ref{fig:State Diagram Input/output Difference Rate-2 Modulation}(b), $i$ stands for the number of non-null components in $\Delta \bar{b}_0[k] \, \Delta \bar{b}_1[k]$ and $j$ stands for the square of $\Delta \mathring{s}[k],$ $\Delta \bar{b}_0[k] \, \Delta \bar{b}_1[k] / \Delta \mathring{s}[k]$ being the label of the same branch in the state diagram of Figure~\ref{fig:State Diagram Input/output Difference Rate-2 Modulation}(a).

In order to simplify the derivation of the \gls{tf}, $T(N,D),$ we start by aggregating, in Figure~\ref{fig:State Diagram Input/output Difference Rate-2 Modulation}(c), the labels in Figure~\ref{fig:State Diagram Input/output Difference Rate-2 Modulation}(b), for branches starting at the same state and ending in the same state of the state diagram. Clearly, the “null” branch, with label $N^0D^0=1,$ which should be discarded for the expression of the \gls{tf}, is not merged with the other branches departing from and arriving at the same $0$ state.

To further simplify the derivation of the said \gls{tf}, we notice that states $-2$ and $+2$ play identical roles in term of aggregate branch labels. These states are therefore merged, in Figure~\ref{fig:State Diagram Input/output Difference Rate-2 Modulation}(d), into a single state, named $\pm 2.$ As a result, in Figure~\ref{fig:State Diagram Input/output Difference Rate-2 Modulation}(c), the labels of branches linking state $0$ to states $-2$ and $+2$ are counted twice, and labels of branches leaving state $-2$ to state $+2$ or state $+2$ to state $-2$ are replaced by a single branch with the same label, beginning and ending at the same aggregate state, $\pm 2.$

As shown in Figure~\ref{fig:State Diagram Input/output Difference Rate-2 Modulation}(e), the final step, before determining the \gls{tf}, is to discard the “null” branch, with label $1,$ and split the $0$ state into a starting state, designated as $0_\text{s},$ and an ending state, designated as $0_\text{e}.$ This split of state $0$ reflects the fact that each error event must leave and return to state $0$ once.

We arrive at the desired \gls{tf}
\begin{equation}
T(N,D) = 2ND^{16} + \frac{2N \left( (1+N)D^4 + ND^{36}\right)^2}{1 - \left( N(1+N) + N(1+2N)D^{16} + N^2D^{64} \right)},
\end{equation}
using Figure~\ref{fig:State Diagram Input/output Difference Rate-2 Modulation}(e), as our starting point, and the same procedure that is typically used to derive the \glspl{tf} of convolutional codes. The expansion of this \gls{tf}, limited to the two terms of least powers of $D,$ leads to
\begin{equation} \label{eq:Expansion Transfer Function Rate-2 Modulation}
T(N,D) = \frac{2N (1+N)^2}{1 - N(1+N)} D^8 + 2ND^{16} + \cdots.
\end{equation}
This expansion corroborates the fact, seen in Sub-section \ref{ssec:Modulation of Rate 2}, that the \gls{msed} of the rate-$2$ modulation with scaled filters, $\mathring{h}_m[k],$ $m=0,1,$ is indeed equal to $8.$

Each error event, with the label $N^l D^m,$ contains $l$ non-null components, in $\{ \pm 2\},$ resulting in $l$ errors in the input sequence. Furthermore, it occurs with probability $(\tfrac{1}{2})^l,$ in conformity with the discussion that took place at the beginning of Section~\ref{app:Tight Estimate BEP Rate 5/4}. As a result, the elementary contribution of each error event to the \gls{bep} is calculated by partially deriving the label $N^l D^m$  with respect to $N,$ multiplying by $N,$ and finally evaluating at $N=1/2.$ Applying the exact same steps to the expansion of the \gls{tf} in (\ref{eq:Expansion Transfer Function Rate-2 Modulation}) results in
\begin{equation} \label{eq:Multiplicity and Distance Rate-2 Modulation}
\dot{T}(D) \triangleq \left. N \cfrac{\partial T(N,D)}{\partial N} \right|_{N=1/2} = 51 D^8 + \cdots,
\end{equation}
which will next be referred to as the \gls{rtf}, in order to distinguish it from the original \gls{tf}, $T(N,D),$ and the \gls{atf}, $T(J, N,D),$ which will be introduced in Appendix~\ref{app:Iterative Determination Transfer Function Rate-2 NSM}.

The tight estimate of the \gls{bep} of the proposed rate-$2$ \gls{nsm}, as shown in (\ref{eq:Tight Estimate BEP Rate 2}), results from keeping only error events at the \gls{msed}. With $2$-ASK as reference, it requires a normalization of the multiplicity factor $51$ in (\ref{eq:Multiplicity and Distance Rate-2 Modulation}) by $2,$ since two bipolar (or binary) symbols are transmitted per transmission opportunity. It also requires the normalization of the square-root argument within the $\operatorname{erfc}(\cdot)$ function by $(1/6) \times 2 \times (1/4) = 1/12.$ The first term, $1/6,$ in the normalization results from the fact that the average energy per transmitted symbol in the scaled version of the proposed modulation is $6.$ The second normalization term, namely $2,$ results from the fact that two bipolar symbols are comprised within each transmitted symbol. Finally, the third and last normalization term, $1/4,$ results from, $4,$ the reference \gls{msed} of $2$-ASK.

To bring this appendix to a close, it is worth noting the explosive increase to $51/2$ in the multiplicity of the \gls{bep} of the rate-$2$ modulation, in (\ref{eq:Tight Estimate BEP Rate 2}), as opposed to the multiplicity of $4,$ embedded in the \gls{bep} approximation of the “duobinary” channel, in (\ref{eq:BEP Estimate Duobinary Channel}). This expected result is the result of two contributions. First, there is the degeneracy that the “duobinary” channel left behind, which is illustrated in Figure~\ref{fig:State Diagram Input/output Difference Rate-2 Modulation}(b) by the branches labeled $N$ that connect states $-2$ and $+2$ to each other. On the other hand, there is a brand-new degeneracy that is stronger (two times stronger) and coming from looping branches with label $N^2,$ connecting each of states $-2$ and $+2$ to itself.

\section{Determination of the MSEDs of Rate-2 NSMs} \label{app:d_min^2 Rate-2 NSM}

The determination of the \gls{msed} of a rate-$2$ \gls{nsm} requires the use of the underlying trellis of input/output sequences differences. Remember that a sample of such a trellis for the rate-$2$ \gls{nsm}, with filter lengths $L_0=2$ and $L_1=1,$ was already provided in Figure~\ref{fig:Trellis Input Difference Duobinary Channel Rate-2 Modulation}, and was thoroughly discussed in Subsection~\ref{ssec:Modulation of Rate 2}. To proceed didactically, we distinguish between the cases where filter lengths, $L_0$ and $L_1,$ are both greater than one and the cases where $L_0$ is greater than one, but $L_1$ is equal to one. As exemplified in Figure~\ref{fig:Trellis Input Difference Duobinary Channel Rate-2 Modulation}(b), for $L_0=2$ and $L_1=1,$ the latter case leads to parallel branches between consecutive trellis states. These parallel branches require special attention and treatment that the former case, where both $L_0$ and $L_1$ are greater than one, does not.

\subsection{Rate-2 NSMs with Both Filters Longer Than One} \label{sapp:d_min^2 Rate-2 NSM L_0>1, L_1>1}

When both $L_0$ and $L_1$ are greater than one, connected consecutive states, delimiting the $k$-th section of the trellis of input/output sequences differences, are of the form $\bm{\sigma}[k-1] \triangleq \Delta \bar{b}_0[k-(L_0-1)] \ldots \Delta \bar{b}_0[k-1], \Delta \bar{b}_1[k-(L_1-1)] \ldots \Delta \bar{b}_1[k-1]$ and $\bm{\sigma}[k] \triangleq \Delta \bar{b}_0[k-(L_0-2)] \ldots \Delta \bar{b}_0[k], \Delta \bar{b}_1[k-(L_1-2)] \ldots \Delta \bar{b}_1[k],$ respectively. As a matter of fact, each trellis section is delimited to the right by $Q$ states and to the left by $Q$ other states, where $Q \triangleq 3^{L_0-1} \times 3^{L_1-1} = 3^{L_0+L_1-2}$ is always greater than or equal to $9$.

Starting from an arbitrary state $\bm{\sigma}[k-1],$ to the left of the $k$-th section, we can reach $9$ states $\bm{\sigma}[k],$ to the right of the same section, with associated $\Delta \bar{b}_0[k]$ and $\Delta \bar{b}_1[k]$ freely taking their values in the set $\{0, \pm 2\}.$ Alternatively, each fixed state $\bm{\sigma}[k],$ to the right of the $k$-th section, can be reached by $9$ states, $\bm{\sigma}[k-1],$ to the left of the same section, with associated $\Delta \bar{b}_0[k-(L_0-1)]$ and $\Delta \bar{b}_1[k-(L_1-1)]$ freely taking their values in the set $\{0, \pm 2\}.$

As in the trellis of the rate-$2$ \gls{nsm}, with $L_0=2$ and $L_1=1,$ shown in Figure~\ref{fig:Trellis Input Difference Duobinary Channel Rate-2 Modulation}(b), to each branch connecting state $\bm{\sigma}[k-1]$ to state $\bm{\sigma}[k],$ corresponds an input pair, $\Delta \bar{b}_0[k]\,\Delta \bar{b}_1[k],$ with values in $\{0, \pm 2\}^2 = \{00, 0\pm2, \pm20, \pm2\pm2\},$ and an output 
$\Delta s[k] = \Delta s(\bm{\sigma}[k-1], \bm{\sigma}[k]) \triangleq \sum_{l=0}^{L_0-1} h_0[l] \Delta \bar{b}_0[k-l] + \sum_{l=0}^{L_1-1} h_1[l] \Delta \bar{b}_1[k-l].$ 

However, because the emphasis here is on determining the \gls{nsm}'s \gls{msed}, $d_{\text{min}}^2,$ $m(\bm{\sigma}[k-1], \bm{\sigma}[k]) \triangleq (\Delta s(\bm{\sigma}[k-1], \bm{\sigma}[k]))^2$ is the appropriate metric for each branch in the trellis connecting states $\bm{\sigma}[k-1]$ and $\bm{\sigma}[k],$ as long as both states are not simultaneously equal to the “zero” state. The zero state, denoted by $\bm{0},$ is such that, if $\bm{\sigma}[k] = \bm{0},$ for some $k,$ then $\Delta \bar{b}_0[k-l]=0,$ for $0 \le l < L_0-1,$ and $\Delta \bar{b}_1[k-l]=0,$ for $0 \le l < L_1-1.$

Remember that among all admissible error events, the least cumulated metric is precisely the sought-after \gls{msed}. However, every error event is, by definition, a path in the trellis that must leave and return to the zero state, $\bm{0},$ once. Therefore, to prevent trivial or degenerate error paths that remain at the zero state, and to ensure a meaningful search for the \gls{msed}, the infinite metric, $m(\bm{0}, \bm{0}) = \infty,$ should be assigned to the branches in the trellis that connect states $\bm{\sigma}[k\!-\!1] = \bm{0}$ and $\bm{\sigma}[k] = \bm{0}.$ This assignment effectively forbids direct transitions from the null state $\bm{\sigma}[k\!-\!1] = \bm{0}$ to the null state $\bm{\sigma}[k] = \bm{0}$ during the search process in the trellis of input/output sequence differences.

For an efficient description of the algorithm for calculating the \gls{msed}, $d_{\text{min}}^2,$ we introduce the following complementary and helpful notations. First, we denote by $\Sigma$ the set of $Q$ states that delimit any section of the trellis, either from the left or from the right, and by $\Sigma^*$ the set $\Sigma$ deprived of the zero state $\bm{0}.$ We also denote by $\pi(\bm{\sigma}[k])$ the set of all $9$ states located to the left of the $k$-th section that are capable of reaching state $\bm{\sigma}[k]$ on the right of the same section. Finally, we denote by $M(\bm{\sigma}[k], k)$ the cumulated metric at any state $\bm{\sigma}[k]$ on the right of the $k$-th section of the trellis.

Algorithm~\ref{alg:d_min^2 Rate-2 NSM L_0>1, L_1>1} details the different steps needed to determine the \gls{msed}, $d_{\text{min}}^2,$ and the degeneracy of a rate-$2$ \gls{nsm}, specified by its filters $h_0[k],$ $h_1[k],$ when $L_0 > 1$ and $L_1 > 1.$ If, at a certain section of the trellis of index $k \le K,$ where $K$ is the maximum authorized number of processed sections in the trellis, the metrics $M(\bm{\sigma}[k], k)$ of all states $\bm{\sigma}[k] \in \Sigma,$ other than the zero state $\bm{0},$ are equal to or larger than the current value of $d_{\text{min}}^2,$ the algorithm stops and the \gls{nsm} is declared non-degenerate. Otherwise, the algorithm continues iterating through the maximum permitted number $K$ of trellis sections, and the \gls{nsm} is declared degenerate.

\begin{algorithm}[H]
\caption{Determination of the MSED, $d_{\text{min}}^2,$ of a rate-$2$ NSM with $L_0 > 1$ and $L_1 > 1$}
\label{alg:d_min^2 Rate-2 NSM L_0>1, L_1>1}
\begin{algorithmic}[1]
\Require $h_0[k],$ $h_1[k],$ $L_0 > 1,$ $L_1 > 1$
\Ensure $d_{\text{min}}^2,$ $\text{Degeneracy}$

\State $\infty \gets 10^6$ \Comment{Numerical substitute for infinity}
\State $k \gets 0$
\For{$\Delta \bar{b}_m[k\!-\!l] \in \{0, \pm 2\}, \; l = 0, \dots, L_m\!-\!1,\; m = 0,1,$}
    \State $\bm{\sigma}[k\!-\!1] \gets \Delta \bar{b}_0[k\!-\!(L_0\!-\!1)] \ldots \Delta \bar{b}_0[k\!-\!1], \Delta \bar{b}_1[k\!-\!(L_1\!-\!1)] \ldots \Delta \bar{b}_1[k\!-\!1]$
    \State $\bm{\sigma}[k] \gets \Delta \bar{b}_0[k\!-\!(L_0\!-\!2)] \ldots \Delta \bar{b}_0[k], \Delta \bar{b}_1[k\!-\!(L_1\!-\!2)] \ldots \Delta \bar{b}_1[k]$
    \State $\Delta s(\bm{\sigma}[k\!-\!1], \bm{\sigma}[k]) \gets \sum_{l=0}^{L_0-1} h_0[l] \Delta \bar{b}_0[k\!-\!l] + \sum_{l=0}^{L_1-1} h_1[l] \Delta \bar{b}_1[k\!-\!l]$
    \State $m(\bm{\sigma}[k\!-\!1], \bm{\sigma}[k]) \gets \left(\Delta s(\bm{\sigma}[k\!-\!1], \bm{\sigma}[k])\right)^2$
\EndFor

\State $m(\bm{0}, \bm{0}) \gets \infty$ \label{step:start determination minimum squared Euclidean distance algorithm common step 2}

\For{$\bm{\sigma}[k\!-\!1] \in \Sigma$} \label{step:start determination minimum squared Euclidean distance algorithm common step}
    \State $M(\bm{\sigma}[k\!-\!1], k\!-\!1) \gets \infty$
\EndFor
\State $M(\bm{0}, k\!-\!1) \gets 0$

\State $K \gets 10^4$ \Comment{Maximum trellis depth before declaring the NSM degenerate}
\State $d_{\text{min}}^2 \gets \infty$
\State $M_{\text{min}} \gets 0$

\While{$(k \le K) \lor (M_{\text{min}} < d_{\text{min}}^2)$}
    \For{$\bm{\sigma}[k] \in \Sigma$}
        \State $M(\bm{\sigma}[k], k) \gets \min_{\bm{\sigma}[k\!-\!1] \in \pi(\bm{\sigma}[k])} \left( M(\bm{\sigma}[k\!-\!1], k\!-\!1) + m(\bm{\sigma}[k\!-\!1], \bm{\sigma}[k]) \right)$
    \EndFor
    \State $d_{\text{min}}^2 \gets \min(d_{\text{min}}^2, M(\bm{0}, k))$
    \State $M_{\text{min}} \gets \min_{\bm{\sigma}[k] \in \Sigma^*} M(\bm{\sigma}[k], k)$ \Comment{Minimum over all non-zero states}
    \State $k \gets k + 1$
\EndWhile

\State $\text{Degeneracy} \gets 0$
\If{$k = K + 1$}
    \State $\text{Degeneracy} \gets 1$
\EndIf
\end{algorithmic}
\end{algorithm}

\subsection{Rate-2 NSMs with Only One Filter Longer Than One} \label{ssec:Determination Minimum Squared Euclidean Distance L_0>1 L_1=1}

When $L_0 > 1$ and $L_1 = 1,$ the second input data sequence difference, $\Delta \bar{b}_1[k],$ no longer intervenes in the specification of the states in the trellis of input/output sequence differences. As a matter of fact, the connected consecutive states that delimit the $k$-th section of the trellis have the form $\bm{\sigma}[k\!-\!1] \triangleq \Delta \bar{b}_0[k\!-\!(L_0\!-\!1)] \ldots \Delta \bar{b}_0[k\!-\!1]$ and $\bm{\sigma}[k] \triangleq \Delta \bar{b}_0[k\!-\!(L_0\!-\!2)] \ldots \Delta \bar{b}_0[k],$ respectively. Each trellis section is thus delimited on the left and right by $Q \triangleq 3^{L_0-1}$ states, with $Q$ equal to or greater than $3.$

Each state $\bm{\sigma}[k\!-\!1]$ on the left of the $k$-th section is connected to three distinct states $\bm{\sigma}[k]$ on the right of the same section. Moreover, each such state-to-state connection is realized through three parallel branches. As a result, from each state $\bm{\sigma}[k\!-\!1],$ a total of nine branches depart, targeting three possible successor states $\bm{\sigma}[k],$ with three branches leading to each. Symmetrically, each state $\bm{\sigma}[k]$ on the right receives nine incoming branches from three possible predecessor states $\bm{\sigma}[k\!-\!1],$ with three branches arriving from each.

While $\bar{b}_1[k] \in \{0, \pm 2\}$ specifies which of the three parallel branches connects a given pair of states $\bm{\sigma}[k\!-\!1]$ and $\bm{\sigma}[k],$ the values $\Delta \bar{b}_0[k]$ and $\Delta \bar{b}_0[k\!-\!(L_0\!-\!1)]$ determine the subsequent and preceding states of $\bm{\sigma}[k\!-\!1]$ and $\bm{\sigma}[k],$ respectively. As such, and similarly to the trellis of the rate-$2$ \gls{nsm} with $L_0 = 2$ and $L_1 = 1,$ shown in Figure~\ref{fig:Trellis Input Difference Duobinary Channel Rate-2 Modulation}~(b), each branch connecting state $\bm{\sigma}[k\!-\!1]$ to state $\bm{\sigma}[k]$ corresponds to an input pair $\Delta \bar{b}_0[k]\,\Delta \bar{b}_1[k],$ with values in $\{0, \pm 2\}^2,$ and an output 
\begin{equation}
\Delta s[k] = \Delta s(\bm{\sigma}[k\!-\!1], \bm{\sigma}[k], \Delta \bar{b}_1[k]) \triangleq \sum_{l=0}^{L_0-1} h_0[l] \Delta \bar{b}_0[k\!-\!l] + h_1[0] \Delta \bar{b}_1[k].
\end{equation}

The goal here, as in Subsection~\ref{sapp:d_min^2 Rate-2 NSM L_0>1, L_1>1}, is to determine the \gls{nsm}'s \gls{msed}, $d_{\text{min}}^2.$ Owing to the fact that three parallel branches connect any pair of consecutive connected states in the trellis, these branches can be logically merged into a single effective branch. This merged branch is then associated with a single metric, $m(\bm{\sigma}[k\!-\!1], \bm{\sigma}[k]),$ which no longer explicitly depends on the second input sequence difference, $\Delta \bar{b}_1[k].$

As long as the two connected states $\bm{\sigma}[k\!-\!1]$ and $\bm{\sigma}[k]$ are not both equal to the zero state, the corresponding branch metric is defined as the minimum of the three individual metrics associated with the three possible values of $\Delta \bar{b}_1[k].$ Formally, this metric is given by
\begin{equation}
    m(\bm{\sigma}[k\!-\!1], \bm{\sigma}[k]) \triangleq \min_{\Delta \bar{b}_1[k] \in \{0, \pm 2\}} \left( \Delta s(\bm{\sigma}[k\!-\!1], \bm{\sigma}[k], \Delta \bar{b}_1[k]) \right)^2.
\end{equation}
Here, the zero state $\bm{0}$ is defined such that, if $\bm{\sigma}[k] = \bm{0}$ for some $k,$ then $\Delta \bar{b}_0[k\!-\!l] = 0$ for all $0 \le l < L_0 \!-\! 1.$

Recall that, for all admissible error events used in computing the minimum metric, at least one of the input sequence differences $\Delta \bar{b}_0[k]$ or $\Delta \bar{b}_1[k]$ must be non-zero. As a matter of fact, the branch connecting $\bm{\sigma}[k\!-\!1] = \bm{0}$ to $\bm{\sigma}[k] = \bm{0},$ with $\Delta \bar{b}_1[k] = 0,$ is forbidden and should be assigned an infinite individual metric prior to the merging of branches. In other words, when $\bm{\sigma}[k\!-\!1] = \bm{\sigma}[k] = \bm{0},$ the appropriate merged metric is
\begin{equation}
    m(\bm{0}, \bm{0}) \triangleq \min_{\Delta \bar{b}_1[k] \in \{\pm 2\}} \left( \Delta s(\bm{0}, \bm{0}, \Delta \bar{b}_1[k]) \right)^2.
\end{equation}

Using the same notations and methodological guidelines as in Subsection~\ref{sapp:d_min^2 Rate-2 NSM L_0>1, L_1>1}, Algorithm~\ref{alg:d_min^2 Rate-2 NSM L_0>1, L_1=1} outlines the various steps required to compute the \gls{msed}, $d_{\text{min}}^2,$ and to assess the degeneracy of a rate-$2$ \gls{nsm}, in the case where $L_0 > 1$ and $L_1 = 1.$

\begin{algorithm}
\caption{Determination of the MSED, $d_{\text{min}}^2,$ of a rate-$2$ NSM with $L_0 > 1$ and $L_1 = 1$}
\label{alg:d_min^2 Rate-2 NSM L_0>1, L_1=1}
\begin{algorithmic}[1]
\Require $h_0[k],$ $h_1[0],$ $L_0 > 1$
\Ensure $d_{\text{min}}^2,$ $\text{Degeneracy}$

\State $\infty \gets 10^6$ \Comment{Numerical substitute for infinity}
\State $k \gets 0$

\For{$\Delta \bar{b}_0[k\!-\!l] \in \{0, \pm 2\}, \; l = 0, \dots, L_0\!-\!1$}
    \State $\bm{\sigma}[k\!-\!1] \gets \Delta \bar{b}_0[k\!-\!(L_0\!-\!1)] \ldots \Delta \bar{b}_0[k\!-\!1]$
    \State $\bm{\sigma}[k] \gets \Delta \bar{b}_0[k\!-\!(L_0\!-\!2)] \ldots \Delta \bar{b}_0[k]$

    \For{$\Delta \bar{b}_1[k] \in \{0, \pm 2\}$}
        \State $\Delta s(\bm{\sigma}[k\!-\!1], \bm{\sigma}[k], \Delta \bar{b}_1[k]) \gets \sum_{l=0}^{L_0-1} h_0[l] \Delta \bar{b}_0[k\!-\!l] + h_1[0] \Delta \bar{b}_1[k]$
    \EndFor

    \State $m(\bm{\sigma}[k\!-\!1], \bm{\sigma}[k]) \gets \min_{\Delta \bar{b}_1[k] \in \{0, \pm 2\}} \left( \Delta s(\bm{\sigma}[k\!-\!1], \bm{\sigma}[k], \Delta \bar{b}_1[k]) \right)^2$
\EndFor

\State $m(\bm{0}, \bm{0}) \gets \min_{\Delta \bar{b}_1[k] \in \{\pm 2\}} \left( \Delta s(\bm{0}, \bm{0}, \Delta \bar{b}_1[k]) \right)^2$

\State \textbf{Continue with Algorithm~\ref{alg:d_min^2 Rate-2 NSM L_0>1, L_1>1} from Step~\ref{step:start determination minimum squared Euclidean distance algorithm common step}}

\end{algorithmic}
\end{algorithm}



\section{Iterative Determination of the TFs of Rate-2 NSMs} \label{app:Iterative Determination Transfer Function Rate-2 NSM}

The \gls{tf} of a rate-$2$ \gls{nsm}, as defined in (\ref{eq:Transfer Function Definition}), plays a crucial role in identifying the error events with \gls{msed} and in providing either an approximation or an upper bound of the \gls{bep}. Its computation typically relies on a modified version of the underlying state diagram of input/output sequence differences, labeled with $N^i D^j,$ where the zero state $\bm{0}$ is split into a starting state, $\bm{0}_s,$ and an ending state, $\bm{0}_e.$ This construction is illustrated progressively in Figures~\ref{fig:State Diagram Input/output Difference Rate-2 Modulation}(b)--\ref{fig:State Diagram Input/output Difference Rate-2 Modulation}(e), for the case $L_0 = 2$ and $L_1 = 1.$

The \gls{tf} of a rate-$2$ Nyquist signaling modulation can be computed manually by establishing symbolic relationships between the \glspl{tf} of intermediate states—based on the modified state diagram—and solving for the \gls{tf} of the final zero state, $\bm{0}_e.$ Alternatively, it can be derived more directly, though less intuitively, by applying Mason's gain formula~\cite{Wicker95, Mason56}, a well-known method for computing the \gls{tf} of a linear signal-flow graph, to the same modified diagram.

The aforementioned approaches—commonly used to characterize convolutional codes in terms of distance properties and error probability performance—are not well suited for determining the \glspl{tf} of \glspl{nsm}. First, convolutional codes operate in the digital domain, where all distances are integers, whereas \glspl{nsm} function in the analog domain, where distances are real-valued. Consequently, due to the inherent imprecision in expressing \gls{nsm} filters with real coefficients, many terms in the \gls{tf} that should correspond to the same real-valued distance are mistakenly treated as distinct. Second, the number of states in the state diagram grows exponentially as $3^{L_0 + L_1 - 2}.$ For practical filter lengths considered here—whose sum may easily exceed $10$—this number can surpass $6561.$ Clearly, such a large state space renders the manual application of either symbolic methods or Mason’s formula infeasible, especially when compounded by computationally intensive operations such as symbolic manipulation and Taylor series expansion.

To overcome the limitations discussed above, we propose an iterative computation of a truncated version of the \gls{tf}, $T(N,D),$ for the rate-$2$ \gls{nsm}. Recall that the occurrence probability of an arbitrary error event, associated with a term $N^l D^m$ in $T(N,D),$ decreases exponentially with exponent $m.$ Consequently, the terms that most strongly influence the \gls{bep} at high \gls{snr} are those with the smallest values of $m.$ To reflect this, we first introduce a lexicographic ordering of the $N^l D^m$ terms in the \gls{tf} defined in (\ref{eq:Transfer Function Definition}). Specifically, a term $N^l D^m$ is said to precede another term $N^{l^\prime} D^{m^\prime}$ in $T(N,D)$ if either $m < m^\prime,$ or $l < l^\prime$ when $m = m^\prime.$ Finally, we retain only a finite number, $P,$ of the earliest and most influential terms from the \gls{tf} $T(N,D),$ and we denote the resulting truncated \gls{tf} by $T(N,D;P).$

We now focus exclusively on the case where $L_0 > 1$ and $L_1 > 1,$ in order to keep the presentation as clear as possible. The alternative case, where $L_0 > 1$ and $L_1 = 1,$ is not addressed here, as it is expected to be more intricate due to the aggregate labels that result from merging parallel branches between connected states in the state diagram, as illustrated in Figures~\ref{fig:State Diagram Input/output Difference Rate-2 Modulation}(c)--\ref{fig:State Diagram Input/output Difference Rate-2 Modulation}(e). We also distinguish between degenerate and non-degenerate \glspl{nsm}. For degenerate \glspl{nsm}, an infinite number of terms $N^l D^m$ with exponent $m$ equal to the \gls{msed} appear in the \gls{tf} $T(N,D).$ Consequently, the terms corresponding to the \gls{med} are only partially represented in the truncated \gls{tf}, regardless of the chosen truncation length $P$. Furthermore, higher-order squared Euclidean distances, which naturally appear in $T(N,D),$ are entirely absent from the truncated version $T(N,D;P).$ To address this issue, recall that the \gls{rtf} $\dot{T}(D),$ defined in (\ref{eq:Multiplicity and Distance Rate-2 Modulation}), is what truly matters when estimating the \gls{bep} at moderate to high \glspl{snr}. Therefore, in the degenerate case, we systematically estimate the truncated version of $\dot{T}(D),$ rather than the truncated version of the more detailed original \gls{tf} $T(N,D).$ This simplified approach can also be applied in the non-degenerate case, provided the focus remains on obtaining a reliable estimate of the \gls{bep} at high \gls{snr}.

For non-degenerate \glspl{nsm}, the number of terms $N^l D^m$ corresponding to a given exponent $m$ of the symbol $D$ is typically finite. As a result, for a moderate truncation length $P,$ the truncated \gls{tf} $T(N,D;P)$ can capture the contributions of all error events that are critical to determining the \gls{bep}. This favorable property makes it possible to define an augmented version,
\begin{equation} \label{eq:Augmented Transfer Function Definition}
T(J,N,D) = \sum_{k > 0, l > 0, m \ge 0} \mu(k,l,m) J^k N^l D^m,    
\end{equation}
of the original \gls{tf} $T(N,D),$ where each term $J^k N^l D^m$ now includes, in addition to the standard parameters $l$ and $m,$ the length $k$ of the associated error event, measured as the number of trellis sections it spans in the input/output difference diagram. To obtain this augmented representation, every branch label in the state diagram originally expressed as $N^i D^j$ is replaced by $J N^i D^j,$ where $J$ is a new symbolic variable whose exponent is incremented by $1$ for each additional trellis section traversed by the error event.

The \gls{atf} $T(J,N,D)$ enables a more precise characterization of non-degenerate \glspl{nsm} by revealing the lengths of all error events associated with the \gls{msed}. This information is required in Appendix~\ref{app:Closed-Form Expressions Optimum Filters Non-Degenerate Rate-2 NSMs} for the exact identification of the input sequence differences corresponding to these minimum-distance error events. These input sequence differences must produce output sequence differences whose norms equal the \gls{med} after filtering through the \gls{nsm} filters $h_m[k],$ $m = 0,1.$ As shown in Appendix~\ref{app:Closed-Form Expressions Optimum Filters Non-Degenerate Rate-2 NSMs}, for a non-degenerate optimized \gls{nsm} with numerically computed filters, these identities yield as many independent equations as there are unknowns in $h_m[k],$ $m=0,1,$ thereby enabling the determination of closed-form expressions for the filters.

The iterative algorithms used to compute the truncated versions of the three \gls{tf} variants discussed earlier are based on a modified state diagram, such as the one shown in Figure~\ref{fig:State Diagram Input/output Difference Rate-2 Modulation}(e). This diagram is derived by removing the unique self-loop branch at the zero state $\bm{0},$ and by splitting this state into a starting state $\bm{0}_s$ and an ending state $\bm{0}_e.$ As in Section~\ref{sapp:d_min^2 Rate-2 NSM L_0>1, L_1>1} of Appendix~\ref{app:d_min^2 Rate-2 NSM}, the set $\Sigma^*$ denotes all intermediate states of the modified state diagram, excluding the split zero states $\bm{0}_s$ and $\bm{0}_e.$ The set $\Sigma,$ in contrast, includes all states in the modified diagram, including $\bm{0}_s$ and $\bm{0}_e.$ Furthermore, $\pi(\bm{\sigma})$ continues to denote the set of states capable of reaching a given state $\bm{\sigma}$ in the modified diagram, as previously defined. It is important to note that $\pi(\bm{\sigma})$ remains unchanged for all $\bm{\sigma} \in \Sigma^*.$ However, since the connection between $\bm{0}_s$ and $\bm{0}_e$ has been removed, we now have $\pi(\bm{0}_s) = \emptyset,$ and $\pi(\bm{0}_e)$ includes only eight states.

\subsection{Determination of a Truncated Version of the Original TF} \label{ssec:Determination Truncated Original Transfer Function}

For the iterative computation of the truncated version $T(N,D;P)$ of the original \gls{tf} $T(N,D),$ we must first evaluate truncated versions $T_{\bm{\sigma}}(N,D;P)$ of the intermediate \glspl{tf} $T_{\bm{\sigma}}(N,D),$ corresponding to each intermediate state $\bm{\sigma} \in \Sigma^*$ in the modified state diagram. Like $T(N,D;P),$ each $T_{\bm{\sigma}}(N,D;P)$ retains only the first $P$ most significant terms from its full counterpart $T_{\bm{\sigma}}(N,D).$ Recall that the truncated \gls{tf} $T_{\bm{0}_e}(N,D;P),$ associated with the final zero state $\bm{0}_e,$ is exactly the truncated global \gls{tf} $T(N,D;P)$ that we aim to compute. Furthermore, since the \gls{tf} of the initial zero state $\bm{0}_s$ is defined by $T_{\bm{0}_s}(N,D) = 1,$ its truncated version remains unchanged and is always given by $T_{\bm{0}_s}(N,D;P) = 1.$

Under the assumption that $L_0 > 1$ and $L_1 > 1,$ every pair of connected states $\bm{\sigma}^\prime$ and $\bm{\sigma}$ in the modified state diagram is linked by a single branch. The label of this branch, which takes the form $N^i D^j,$ is denoted by $L_{\bm{\sigma}^\prime \bm{\sigma}}(N,D).$

The computation of the \gls{tf} $T(N,D)$ can be expressed as a symbolic system of equations, where each equation corresponds to an unknown \gls{tf} $T_{\bm{\sigma}}(N,D)$ $\bm{\sigma} \in \Sigma^* \cup \{\bm{0}_e\}$. These equations are constructed based on the labeled branches of the modified state diagram, relating each state's \gls{tf} to those of its predecessor states. The only exception is the initial zero state $\bm{0}_s$, whose \gls{tf} is known and fixed as $T_{\bm{0}_s}(N,D) = 1$. For each of the remaining states $\bm{\sigma} \in \Sigma^* \cup \{\bm{0}_e\}$, the corresponding symbolic equation takes the form
\begin{equation} \label{eq:Equations Relating Transfer Functions}
T_{\bm{\sigma}}(N,D) = \sum_{\bm{\sigma}^\prime \in \pi(\bm{\sigma})} T_{\bm{\sigma^\prime}}(N,D) L_{\bm{\sigma}^\prime \bm{\sigma}}(N,D),
\end{equation}
which expresses the \gls{tf} at state $\bm{\sigma}$ in terms of the \glspl{tf} of all predecessor states $\bm{\sigma}^\prime$ that can reach it via a labeled branch.

In its symbolic form, the iterative algorithm for computing the truncated \gls{tf} $T(N,D;P)$ is based on a reformulation of the set of equations in (\ref{eq:Equations Relating Transfer Functions}), defined for each state $\bm{\sigma} \in \Sigma^* \cup \{\bm{0}_e\}$. Although these original equations are not recursive, they are adapted into recursive update rules that enable iterative computation. Let $T_{\bm{\sigma}}^{(q)}(N,D;P)$ denote the truncated \gls{tf} associated with state $\bm{\sigma}$ at iteration $q$. At the initial iteration $q=0,$ the truncated \glspl{tf} are initialized as $T_{\bm{0}_s}^{(0)}(N,D;P) = 1$ and $T_{\bm{\sigma}}^{(0)}(N,D;P) = 0$ for all $\bm{\sigma} \neq \bm{0}_s$. The truncated \gls{tf} at state $\bm{0}_s$ remains fixed throughout the iterative process, such that $T_{\bm{0}_s}^{(q)}(N,D;P) = 1$ for all $q > 0$. For every other state $\bm{\sigma} \in \Sigma^* \cup \{\bm{0}_e\},$ the truncated \gls{tf} is updated at each iteration $q > 0$ through two successive steps, which will be detailed in the following paragraphs. These values evolve from one iteration to the next until convergence is achieved.

Under the influence of (\ref{eq:Equations Relating Transfer Functions}), temporary truncated \glspl{tf} are computed in the first step of iteration $q > 0$ as
\begin{equation} \label{eq:Temporary Truncated Transfer Functions}
\tilde{T}_{\bm{\sigma}}^{(q)}(N,D;P) \triangleq \sum_{\bm{\sigma}^\prime \in \pi(\bm{\sigma})} T_{\bm{\sigma^\prime}}^{(q-1)}(N,D;P) L_{\bm{\sigma}^\prime \bm{\sigma}}(N,D).
\end{equation}
Since the set $\pi(\bm{\sigma})$ contains at most $9$ elements, the temporary \glspl{tf} include at most $9P$ terms.

In the second step of iteration $q > 0$, terms in each temporary truncated \gls{tf} that share common exponents in the symbols $N$ and $D$ are combined by summing their multiplicities. The resulting terms are then arranged in the lexicographic order introduced at the beginning of Section~\ref{app:Iterative Determination Transfer Function Rate-2 NSM}. By retaining only the first $P$ terms after grouping and ordering, the truncated \gls{tf} $T_{\bm{\sigma}}^{(q)}(N,D;P)$ is obtained for each state $\bm{\sigma} \neq \bm{0}_s$. If $\tilde{T}_{\bm{\sigma}}^{(q)}(N,D;P)$ contains $P$ or fewer terms, it is taken as is.

The iterative algorithm terminates after a finite number of iterations, denoted by $Q,$ as soon as
\begin{equation}
T_{\bm{\sigma}}^{(Q)}(N,D;P) = T_{\bm{\sigma}}^{(Q-1)}(N,D;P),
\end{equation}
for each $\bm{\sigma} \in \Sigma.$ We conjecture that $T_{\bm{0}_e}^{(Q)}(N,D;P)$ is exactly the sought truncated \gls{tf} $T(N,D;P).$ This fact is intuitively understandable, because the multiplicities of $T_{\bm{\sigma}}^{(q)}(N,D;P)$ are always integer numbers, dominated by the integer multiplicities of $T_{\bm{\sigma}}(N,D;P),$ and because the multiplications in (\ref{eq:Temporary Truncated Transfer Functions}), with terms having positive exponents in the symbols $N$ and $D,$ can be interpreted as some kind of causal filtering. The latter property is necessary due to the truncation operation applied to $\tilde{T}_{\bm{\sigma}}^{(q)}(N,D;P)$ at the end of the second step of each iteration $q > 0$ of the algorithm.

For a pseudo-code description of the iterative algorithm, it is necessary to adopt a more suitable vector representation for the functions in (\ref{eq:Temporary Truncated Transfer Functions}). At the state level, for each state $\bm{\sigma} \in \Sigma$ and iteration $q,$ any truncated \gls{tf} $T_{\bm{\sigma}}^{(q)}(N,D;P),$ comprising $P_{\bm{\sigma}}^{(q)}$ terms with $P_{\bm{\sigma}}^{(q)} \le P,$ is represented by three vectors $\bm{l}_{\bm{\sigma}}^{(q)},$ $\bm{m}_{\bm{\sigma}}^{(q)},$ and $\bm{\mu}_{\bm{\sigma}}^{(q)},$ each of size $P_{\bm{\sigma}}^{(q)}.$ The vectors $\bm{l}_{\bm{\sigma}}^{(q)}$ and $\bm{m}_{\bm{\sigma}}^{(q)}$ specify, respectively, the exponents of $N$ and $D$ in the terms $N^l D^m,$ which appear in lexicographic order in $T_{\bm{\sigma}}^{(q)}(N,D;P),$ while $\bm{\mu}_{\bm{\sigma}}^{(q)}$ contains the corresponding multiplicities. Similarly, the temporary truncated \gls{tf} $\tilde{T}_{\bm{\sigma}}^{(q)}(N,D;P)$ is associated with vectors $\tilde{\bm{l}}_{\bm{\sigma}}^{(q)},$ $\tilde{\bm{m}}_{\bm{\sigma}}^{(q)},$ and $\tilde{\bm{\mu}}_{\bm{\sigma}}^{(q)}$ having a common size $\tilde{P}_{\bm{\sigma}}^{(q)}$ that can be up to $9$ times larger than $P.$ Furthermore, the sought truncated \gls{tf} $T(N,D;P)$ is described by three vectors $\bm{l},$ $\bm{m},$ and $\bm{\mu},$ each of size $P.$ At the branch level, each symbolic label $L_{\bm{\sigma}^\prime \bm{\sigma}}(N,D) = N^i D^j,$ with an integer exponent $i$ for $N$ and a real-valued exponent $j$ for $D,$ is represented by two scalar labels $i_{\bm{\sigma}^\prime \bm{\sigma}} \triangleq i$ and $j_{\bm{\sigma}^\prime \bm{\sigma}} \triangleq j.$ The real values of the exponents in $D$ arise from the computation of squared Euclidean norms of error events, which involve real-valued sequences. This distinction—integer exponents for $N$ and real-valued exponents for $D$—motivates the “exploded” vector representation of \glspl{tf}, as it avoids the complexities of directly manipulating non-integer exponents in $D$ within the iterative computations.

The pseudo-code for the iterative determination of the truncated \gls{tf} $T(N,D;P)$ of $T(N,D)$ is presented in Algorithm~\ref{alg:Truncated Transfer Function T(N,D;P) Rate-2 NSM}.

\begin{algorithm}
\caption{Determination of the truncated transfer function $T(N,D;P)$ for a rate-$2$ NSM}
\label{alg:Truncated Transfer Function T(N,D;P) Rate-2 NSM}
\begin{algorithmic}[1]
\Require $P$, integer exponents $i_{\bm{\sigma}^\prime \bm{\sigma}}$, real exponents $j_{\bm{\sigma}^\prime \bm{\sigma}}$, for all $\bm{\sigma}^\prime \in \pi(\bm{\sigma})$, $\bm{\sigma} \in \Sigma^* \cup \{\bm{0}_e\}$
\Ensure Vectors $\bm{l}$, $\bm{m}$, and $\bm{\mu}$ representing $T(N,D;P)$

\Statex \Comment{Step 1: Initialize transfer function at starting state $\bm{0}_s$}
\State $\bm{l}_{\bm{0}_s}^{(0)} \gets [0]$
\State $\bm{m}_{\bm{0}_s}^{(0)} \gets [0]$
\State $\bm{\mu}_{\bm{0}_s}^{(0)} \gets [1]$

\Statex \Comment{Step 2: Initialize transfer functions at other states}
\For{$\bm{\sigma} \in \Sigma, \bm{\sigma} \neq \bm{0}_s$}
    \State $\bm{l}_{\bm{\sigma}}^{(0)} \gets [\;]$
    \State $\bm{m}_{\bm{\sigma}}^{(0)} \gets [\;]$
    \State $\bm{\mu}_{\bm{\sigma}}^{(0)} \gets [\;]$
\EndFor

\State $q \gets 0$
\Repeat
    \State $q \gets q + 1$
    
    \Statex \Comment{Step 3: Reset starting state transfer function at iteration $q$}
    \State $\bm{l}_{\bm{0}_s}^{(q)} \gets [0]$
    \State $\bm{m}_{\bm{0}_s}^{(q)} \gets [0]$
    \State $\bm{\mu}_{\bm{0}_s}^{(q)} \gets [1]$
    
    \Statex \Comment{Step 4: Update transfer functions at other states}
    \For{$\bm{\sigma} \in \Sigma^* \cup \{\bm{0}_e\}$}
        \State $\tilde{\bm{l}}_{\bm{\sigma}}^{(q)} \gets [\;]$
        \State $\tilde{\bm{m}}_{\bm{\sigma}}^{(q)} \gets [\;]$
        \State $\tilde{\bm{\mu}}_{\bm{\sigma}}^{(q)} \gets [\;]$
        
        \For{$\bm{\sigma}^\prime \in \pi(\bm{\sigma})$}
            \State $\tilde{\bm{l}}_{\bm{\sigma}}^{(q)} \gets [\tilde{\bm{l}}_{\bm{\sigma}}^{(q)}, \bm{l}_{\bm{\sigma}^\prime}^{(q-1)} + i_{\bm{\sigma}^\prime \bm{\sigma}}]$
            \State $\tilde{\bm{m}}_{\bm{\sigma}}^{(q)} \gets [\tilde{\bm{m}}_{\bm{\sigma}}^{(q)}, \bm{m}_{\bm{\sigma}^\prime}^{(q-1)} + j_{\bm{\sigma}^\prime \bm{\sigma}}]$
            \State $\tilde{\bm{\mu}}_{\bm{\sigma}}^{(q)} \gets [\tilde{\bm{\mu}}_{\bm{\sigma}}^{(q)}, \bm{\mu}_{\bm{\sigma}^\prime}^{(q-1)}]$
        \EndFor
        
        \State $(\tilde{\bm{l}}_{\bm{\sigma}}^{(q)}, \tilde{\bm{m}}_{\bm{\sigma}}^{(q)}, \tilde{\bm{\mu}}_{\bm{\sigma}}^{(q)}) \gets \text{Group}(\tilde{\bm{l}}_{\bm{\sigma}}^{(q)}, \tilde{\bm{m}}_{\bm{\sigma}}^{(q)}, \tilde{\bm{\mu}}_{\bm{\sigma}}^{(q)})$
        \State $(\bm{l}_{\bm{\sigma}}^{(q)}, \bm{m}_{\bm{\sigma}}^{(q)}, \bm{\mu}_{\bm{\sigma}}^{(q)}) \gets \text{Truncate}(\tilde{\bm{l}}_{\bm{\sigma}}^{(q)}, \tilde{\bm{m}}_{\bm{\sigma}}^{(q)}, \tilde{\bm{\mu}}_{\bm{\sigma}}^{(q)})$
    \EndFor
\Until{$\forall \bm{\sigma} \in \Sigma^* \cup \{\bm{0}_e\}$,
    $\bm{l}_{\bm{\sigma}}^{(q)} = \bm{l}_{\bm{\sigma}}^{(q-1)}$,
    $\bm{m}_{\bm{\sigma}}^{(q)} = \bm{m}_{\bm{\sigma}}^{(q-1)}$,
    $\bm{\mu}_{\bm{\sigma}}^{(q)} = \bm{\mu}_{\bm{\sigma}}^{(q-1)}$}

\Statex \Comment{Step 5: Output the converged transfer function at ending zero state}
\State $\bm{l} \gets \bm{l}_{\bm{0}_e}^{(q)}$
\State $\bm{m} \gets \bm{m}_{\bm{0}_e}^{(q)}$
\State $\bm{\mu} \gets \bm{\mu}_{\bm{0}_e}^{(q)}$
\end{algorithmic}
\end{algorithm}

\subsection{Determination of a Truncated Version of the RTF} \label{ssec:Determination Truncated Reduced Transfer Function}

We assume that the terms of the \gls{rtf}, $\dot{T}(D),$ are arranged in ascending order according to the exponents of $D.$ The truncated version, $\dot{T}(D;P),$ of $\dot{T}(D)$ is obtained by keeping only the first $P$ terms in $\dot{T}(D).$ To enable an iterative estimation of $\dot{T}(D;P),$ we introduce the \gls{rtf} associated with each state $\bm{\sigma} \in \Sigma^* \cup \{\bm{0}_s, \bm{0}_e\},$ given by
\begin{equation} \label{eq:Reduced Transfer Functions}
    \dot{T}_{\bm{\sigma}}(D) \triangleq \left. N \frac{\partial T_{\bm{\sigma}}(N,D)}{\partial N} \right|_{N=1/2},
\end{equation}
along with its corresponding truncated version, $\dot{T}_{\bm{\sigma}}(D;P).$ Similar to $\dot{T}(D;P),$ the truncated function $\dot{T}_{\bm{\sigma}}(D;P)$ retains only the first $P$ terms of $\dot{T}_{\bm{\sigma}}(D),$ arranged in increasing order of the exponents of $D.$

Deriving both sides of (\ref{eq:Equations Relating Transfer Functions}) with respect to $N$, multiplying the result by $N$, and evaluating at $N = 1/2$ yields
\begin{equation} \label{eq:Equations Relating Reduced Transfer Functions}
\dot{T}_{\bm{\sigma}}(D) = \sum_{\bm{\sigma}^\prime \in \pi(\bm{\sigma})} \dot{T}_{\bm{\sigma^\prime}}(D) L_{\bm{\sigma}^\prime \bm{\sigma}}(D)
+ \sum_{\bm{\sigma}^\prime \in \pi(\bm{\sigma})} T_{\bm{\sigma^\prime}}(D) \dot{L}_{\bm{\sigma}^\prime \bm{\sigma}}(D),
\end{equation}
where
\begin{equation} \label{eq:Projected Branch Label}
L_{\bm{\sigma}^\prime \bm{\sigma}}(D) \triangleq \left. L_{\bm{\sigma}^\prime \bm{\sigma}}(N,D) \right|_{N=1/2},
\end{equation}
\begin{equation} \label{eq:Projected Transfer Function}
T_{\bm{\sigma}}(D) \triangleq \left. T_{\bm{\sigma}}(N,D) \right|_{N=1/2},
\end{equation}
and
\begin{equation} \label{eq:Projected Differentiated Branch Label}
\dot{L}_{\bm{\sigma}^\prime \bm{\sigma}}(D) \triangleq \left. N \frac{\partial L_{\bm{\sigma}^\prime \bm{\sigma}}(N,D)}{\partial N} \right|_{N=1/2}.
\end{equation}

For a branch connecting states $\bm{\sigma}^\prime$ and $\bm{\sigma}$ with label $L_{\bm{\sigma}^\prime \bm{\sigma}}(N,D) = N^i D^j,$ for some exponents $i$ and $j,$ the projected quantities become $L_{\bm{\sigma}^\prime \bm{\sigma}}(D) = (\tfrac{1}{2})^i D^j$ and $\dot{L}_{\bm{\sigma}^\prime \bm{\sigma}}(D) = i (\tfrac{1}{2})^i D^j = i L_{\bm{\sigma}^\prime \bm{\sigma}}(D).$ In parallel, for any state $\bm{\sigma} \ne \bm{0}_s,$ the instantiated \glspl{tf} $T_{\bm{\sigma}}(D)$ can be evaluated using
\begin{equation} \label{eq:Equations Relating Instatiated Transfer Functions}
T_{\bm{\sigma}}(D) = \sum_{\bm{\sigma}^\prime \in \pi(\bm{\sigma})} T_{\bm{\sigma^\prime}}(D) L_{\bm{\sigma}^\prime \bm{\sigma}}(D),
\end{equation}
which directly follow from (\ref{eq:Equations Relating Transfer Functions}) by setting $N = 1/2.$

All the equations above are to be interpreted with the initial conditions at the starting state $\bm{0}_s$, where the \gls{rtf} satisfies $\dot{T}_{\bm{0}_s}(D) = 0$ and the instantiated \gls{tf} satisfies $T_{\bm{0}_s}(D) = 1.$

In symbolic form, the iterative algorithm for determining the truncated \gls{rtf} $\dot{T}(D;P)$ is based on a combination of equations~(\ref{eq:Equations Relating Reduced Transfer Functions}) and~(\ref{eq:Equations Relating Instatiated Transfer Functions}). For each state $\bm{\sigma}$ and each iteration $q$ of the algorithm, this approach involves the introduction of the truncated \gls{rtf} $\dot{T}_{\bm{\sigma}}^{(q)}(D;P)$ and the truncated instantiated \gls{tf} $T_{\bm{\sigma}}^{(q)}(D;P)$. The initialization is given by $\dot{T}_{\bm{\sigma}}^{(0)}(D;P) = 0$ for all $\bm{\sigma} \in \Sigma^* \cup \{\bm{0}_s,\bm{0}_e\}$, $T_{\bm{0}_s}^{(0)}(D;P) = 1$ and $T_{\bm{\sigma}}^{(0)}(D;P) = 0$ for all $\bm{\sigma} \in \Sigma^* \cup \{\bm{0}_e\}$. For all subsequent iterations $q > 0$ the values $\dot{T}_{\bm{\sigma}}^{(q)}(D;P)$ and $T_{\bm{\sigma}}^{(q)}(D;P)$ remain fixed for $\bm{\sigma} = \bm{0}_s$ which implies $\dot{T}_{\bm{0}_s}^{(q)}(D;P) = 0$ and $T_{\bm{0}_s}^{(q)}(D;P) = 1$ for all $q > 0$. As in the case of the original \gls{tf} described in Subsection~\ref{ssec:Determination Truncated Original Transfer Function}, the truncated instantiated and \gls{rtf}s are updated in two steps at each iteration $q > 0$.

The temporary instantiated and reduced \glspl{tf}, motivated by equations~(\ref{eq:Equations Relating Reduced Transfer Functions}) and~(\ref{eq:Equations Relating Instatiated Transfer Functions}), are computed in the first step of iteration $q > 0$ as
\begin{equation} \label{eq:Temporary Instantiated Truncated Transfer Functions}
\tilde{T}_{\bm{\sigma}}^{(q)}(D;P) \triangleq \sum_{\bm{\sigma}^\prime \in \pi(\bm{\sigma})} T_{\bm{\sigma^\prime}}^{(q-1)}(D;P) L_{\bm{\sigma}^\prime \bm{\sigma}}(D),
\end{equation}
and
\begin{equation} \label{eq:Temporary Reduced Truncated Transfer Functions}
\tilde{\dot{T}}_{\bm{\sigma}}^{(q)}(D;P) \triangleq \sum_{\bm{\sigma}^\prime \in \pi(\bm{\sigma})} \dot{T}_{\bm{\sigma^\prime}}^{(q-1)}(D;P) L_{\bm{\sigma}^\prime \bm{\sigma}}(D)
+ \sum_{\bm{\sigma}^\prime \in \pi(\bm{\sigma})} T_{\bm{\sigma^\prime}}^{(q-1)}(D;P) \dot{L}_{\bm{\sigma}^\prime \bm{\sigma}}(D),
\end{equation}
respectively. As observed in Subsection~\ref{ssec:Determination Truncated Original Transfer Function}, equation~(\ref{eq:Temporary Instantiated Truncated Transfer Functions}) implies that $\tilde{T}_{\bm{\sigma}}^{(q)}(D;P)$ contains at most $9P$ terms. However, due to the double summation on the right-hand side of equation~(\ref{eq:Temporary Reduced Truncated Transfer Functions}), the function $\tilde{\dot{T}}_{\bm{\sigma}}^{(q)}(D;P)$ may contain up to $18P$ terms.

In the second step of iteration $q > 0$ the terms in $\tilde{T}_{\bm{\sigma}}^{(q)}(D;P)$ and $\tilde{\dot{T}}_{\bm{\sigma}}^{(q)}(D;P)$ that share the same exponent in the symbol $D$ are merged by summing their multiplicities. Once the merging is complete, the resulting terms from both functions are sorted in ascending order according to the exponents of $D$. The truncated \glspl{tf} $T_{\bm{\sigma}}^{(q)}(D;P)$ and $\dot{T}_{\bm{\sigma}}^{(q)}(D;P)$ are then obtained by retaining all terms of $\tilde{T}_{\bm{\sigma}}^{(q)}(D;P)$ and $\tilde{\dot{T}}_{\bm{\sigma}}^{(q)}(D;P)$ if they contain $P$ or fewer terms, or by keeping only the first $P$ terms when the number of terms exceeds $P$.

For degenerate \glspl{nsm} convergence is not guaranteed after a finite number of iterations even though the multiplicities that define these functions are rational. As a result, the algorithm stops after $Q$ iterations once the multiplicities of $\dot{T}_{\bm{\sigma}}^{(Q-1)}(D;P)$ and $\dot{T}_{\bm{\sigma}}^{(Q)}(D;P)$ differ in absolute value by less than a precision value $\varepsilon$. When the same stopping condition is applied to non degenerate \glspl{nsm} it ensures perfect convergence.

The pseudo-code corresponding to the iterative determination of the truncated version $\dot{T}_{\bm{\sigma}}(D;P)$ of the \gls{rtf} $\dot{T}_{\bm{\sigma}}(D)$ is provided in Algorithm~\ref{alg:Truncated Reduced Transfer Function T(D;P) Rate-2 NSM}. At the level of the modified state diagram, the algorithm employs vector notations to represent the desired truncated \gls{rtf} $\dot{T}_{\bm{\sigma}}(D;P)$ along with the truncated \glspl{rtf} $\dot{T}_{\bm{\sigma}}^{(q)}(D;P)$ and the truncated instantiated \glspl{tf} $T_{\bm{\sigma}}^{(q)}(D;P)$ at iteration $q$.

\begin{algorithm}
\caption{Determination of the truncated RTF $\dot{T}(D;P)$ for a rate-$2$ NSM}
\label{alg:Truncated Reduced Transfer Function T(D;P) Rate-2 NSM}
\begin{algorithmic}[1]
\Require $P$, $\mu_{\bm{\sigma}^\prime \bm{\sigma}}$, $j_{\bm{\sigma}^\prime \bm{\sigma}}$, $\dot{\mu}_{\bm{\sigma}^\prime \bm{\sigma}}$, and $\dot{j}_{\bm{\sigma}^\prime \bm{\sigma}}$, for all $\bm{\sigma}^\prime \in \pi(\bm{\sigma})$, $\bm{\sigma} \in \Sigma^* \cup \{\bm{0}_e\}$
\Ensure $\dot{\bm{m}}$ and $\dot{\bm{\mu}}$
\Statex \Comment{Step 1: Initialize transfer functions at $\bm{0}_s$}
\State $\bm{m}_{\bm{0}_s}^{(0)} \gets [0]$
\State $\bm{\mu}_{\bm{0}_s}^{(0)} \gets [1]$
\State $\dot{\bm{m}}_{\bm{0}_s}^{(0)} \gets [0]$
\State $\dot{\bm{\mu}}_{\bm{0}_s}^{(0)} \gets [0]$
\Statex \Comment{Step 2: Initialize transfer functions at other states}
\For{$\bm{\sigma} \in \Sigma^* \cup \{\bm{0}_e\}$}
    \State $\bm{m}_{\bm{\sigma}}^{(0)} \gets [\;]$
    \State $\bm{\mu}_{\bm{\sigma}}^{(0)} \gets [\;]$
    \State $\dot{\bm{m}}_{\bm{\sigma}}^{(0)} \gets [\;]$
    \State $\dot{\bm{\mu}}_{\bm{\sigma}}^{(0)} \gets [\;]$
\EndFor
\State $q \gets 0$
\Repeat
    \State $q \gets q + 1$
    \Statex \Comment{Step 3: Preserve fixed values at $\bm{0}_s$}
    \State $\bm{m}_{\bm{0}_s}^{(q)} \gets [0]$
    \State $\bm{\mu}_{\bm{0}_s}^{(q)} \gets [1]$
    \State $\dot{\bm{m}}_{\bm{0}_s}^{(q)} \gets [0]$
    \State $\dot{\bm{\mu}}_{\bm{0}_s}^{(q)} \gets [0]$
    \Statex \Comment{Step 4: Iteratively compute updated transfer functions}
    \For{$\bm{\sigma} \in \Sigma^* \cup \{\bm{0}_e\}$}
        \State $\tilde{\bm{m}}_{\bm{\sigma}}^{(q)} \gets [\;]$
        \State $\tilde{\bm{\mu}}_{\bm{\sigma}}^{(q)} \gets [\;]$
        \State $\tilde{\dot{\bm{m}}}_{\bm{\sigma}}^{(q)} \gets [\;]$
        \State $\tilde{\dot{\bm{\mu}}}_{\bm{\sigma}}^{(q)} \gets [\;]$
        \For{$\bm{\sigma}^\prime \in \pi(\bm{\sigma})$}
            \State $\tilde{\bm{m}}_{\bm{\sigma}}^{(q)} \gets [\tilde{\bm{m}}_{\bm{\sigma}}^{(q)}, \bm{m}_{\bm{\sigma^\prime}}^{(q-1)} + j_{\bm{\sigma}^\prime \bm{\sigma}}]$
            \State $\tilde{\bm{\mu}}_{\bm{\sigma}}^{(q)} \gets [\tilde{\bm{\mu}}_{\bm{\sigma}}^{(q)}, \bm{\mu}_{\bm{\sigma^\prime}}^{(q-1)} \times \mu_{\bm{\sigma}^\prime \bm{\sigma}}]$
            \State $\tilde{\dot{\bm{m}}}_{\bm{\sigma}}^{(q)} \gets [\tilde{\dot{\bm{m}}}_{\bm{\sigma}}^{(q)}, \dot{\bm{m}}_{\bm{\sigma^\prime}}^{(q-1)} + j_{\bm{\sigma}^\prime \bm{\sigma}}, \bm{m}_{\bm{\sigma^\prime}}^{(q-1)} + \dot{j}_{\bm{\sigma}^\prime \bm{\sigma}}]$
            \State $\tilde{\dot{\bm{\mu}}}_{\bm{\sigma}}^{(q)} \gets [\tilde{\dot{\bm{\mu}}}_{\bm{\sigma}}^{(q)}, \dot{\bm{\mu}}_{\bm{\sigma^\prime}}^{(q-1)} \times \mu_{\bm{\sigma}^\prime \bm{\sigma}}, \bm{\mu}_{\bm{\sigma^\prime}}^{(q-1)} \times \dot{\mu}_{\bm{\sigma}^\prime \bm{\sigma}}]$
        \EndFor
        \State $(\tilde{\bm{m}}_{\bm{\sigma}}^{(q)}, \tilde{\bm{\mu}}_{\bm{\sigma}}^{(q)}) \gets \text{Group}(\tilde{\bm{m}}_{\bm{\sigma}}^{(q)}, \tilde{\bm{\mu}}_{\bm{\sigma}}^{(q)})$ \Comment{Group terms with identical exponents}
        \State $(\bm{m}_{\bm{\sigma}}^{(q)}, \bm{\mu}_{\bm{\sigma}}^{(q)}) \gets \text{Truncate}(\tilde{\bm{m}}_{\bm{\sigma}}^{(q)}, \tilde{\bm{\mu}}_{\bm{\sigma}}^{(q)})$ \Comment{Retain only first $P$ terms}
        \State $(\tilde{\dot{\bm{m}}}_{\bm{\sigma}}^{(q)}, \tilde{\dot{\bm{\mu}}}_{\bm{\sigma}}^{(q)}) \gets \text{Group}(\tilde{\dot{\bm{m}}}_{\bm{\sigma}}^{(q)}, \tilde{\dot{\bm{\mu}}}_{\bm{\sigma}}^{(q)})$
        \State $(\dot{\bm{m}}_{\bm{\sigma}}^{(q)}, \dot{\bm{\mu}}_{\bm{\sigma}}^{(q)}) \gets \text{Truncate}(\tilde{\dot{\bm{m}}}_{\bm{\sigma}}^{(q)}, \tilde{\dot{\bm{\mu}}}_{\bm{\sigma}}^{(q)})$
    \EndFor
\Until{$\bm{m}_{\bm{\sigma}}^{(q)} = \bm{m}_{\bm{\sigma}}^{(q-1)}$ and $\| \bm{\mu}_{\bm{\sigma}}^{(q)} - \bm{\mu}_{\bm{\sigma}}^{(q-1)} \|_\infty < \varepsilon$ for all $\bm{\sigma} \in \Sigma^* \cup \{\bm{0}_e\}$} \Comment{$\| \cdot \|_\infty$ is the supremum norm}
\Statex \Comment{Step 5: Output result at end state}
\State $\dot{\bm{m}} \gets \dot{\bm{m}}_{\bm{0}_e}^{(q)}$
\State $\dot{\bm{\mu}} \gets \dot{\bm{\mu}}_{\bm{0}_e}^{(q)}$
\end{algorithmic}
\end{algorithm}

The exponents and multiplicities corresponding to terms of the form $D^m$ in the target \gls{tf} $\dot{T}_{\bm{\sigma}}(D;P)$, ordered by increasing exponents, are denoted by $\bm{m}_{\bm{\sigma}}$ and $\bm{\mu}_{\bm{\sigma}}$ respectively. In the same manner, $\bm{m}_{\bm{\sigma}}^{(q)}$ and $\bm{\mu}_{\bm{\sigma}}^{(q)}$ represent the exponents and multiplicities associated with $T_{\bm{\sigma}}^{(q)}(D;P)$, while $\dot{\bm{m}}_{\bm{\sigma}}^{(q)}$ and $\dot{\bm{\mu}}_{\bm{\sigma}}^{(q)}$ represent those associated with $\dot{T}_{\bm{\sigma}}^{(q)}(D;P)$.

We recall that at the branch level, each branch with symbolic label $L_{\bm{\sigma}^\prime \bm{\sigma}}(N,D) = N^i D^j$ for some exponents $i$ and $j$ corresponds to an instantiated symbolic label $L_{\bm{\sigma}^\prime \bm{\sigma}}(D) = (\tfrac{1}{2})^i D^j$ and a reduced symbolic label $\dot{L}_{\bm{\sigma}^\prime \bm{\sigma}}(D) = i (\tfrac{1}{2})^i D^j$. The instantiated symbolic label is represented by the two scalar labels $\mu_{\bm{\sigma}^\prime \bm{\sigma}} \triangleq (\tfrac{1}{2})^i$ and $j_{\bm{\sigma}^\prime \bm{\sigma}} \triangleq j$. The reduced symbolic label is associated with the two scalar labels $\dot{\mu}_{\bm{\sigma}^\prime \bm{\sigma}} \triangleq i (\tfrac{1}{2})^i$ and $\dot{j}_{\bm{\sigma}^\prime \bm{\sigma}} \triangleq j$.

\subsection{Determination of a Truncated Version of the ATF} \label{ssec:Determination Truncated Augmented Transfer Function}

We present an iterative algorithm that determines a truncated version $T(J, N, D;P)$ of the \gls{atf} $T(J, N, D)$ in which only the first $P$ terms are retained. As in Subsection~\ref{ssec:Determination Truncated Original Transfer Function}, for each state $\bm{\sigma} \in \Sigma^*$, we introduce an \gls{atf} $T_{\bm{\sigma}}(J, N, D)$ and its truncated version $T_{\bm{\sigma}}(J, N, D;P)$ in which only the first $P$ terms in $T_{\bm{\sigma}}(J, N, D)$ are kept. To accomplish this, we extend the previous lexicographic order by incorporating the exponent of the symbol $J$ in each term of $T(J, N, D)$ or $T_{\bm{\sigma}}(J, N, D)$. Terms with identical exponents in symbols $N$ and $D$ are ordered according to this lexicographic rule in ascending order of their exponents in symbol $J$.

As with the original \gls{tf}, $T(N,D),$ we assume $T_{\bm{0}_s}(J, N,D)=1.$ Additionally, all \glspl{atf} for states, $\bm{\sigma},$ other than $\bm{0}_s,$ are expressed as
\begin{equation} \label{eq:Equations Relating Augmented Transfer Functions}
T_{\bm{\sigma}}(J,N,D) = \sum_{\bm{\sigma}^\prime \in \pi(\bm{\sigma})} T_{\bm{\sigma^\prime}}(J,N,D) L_{\bm{\sigma}^\prime \bm{\sigma}}(J,N,D),
\end{equation}
where $L_{\bm{\sigma}^\prime \bm{\sigma}}(J,N,D)$ is the new branch label, of the form $J N^i D^j,$ created by multiplying the original branch label, $L_{\bm{\sigma}^\prime \bm{\sigma}}(N,D)$ by symbol $J.$

The iterative algorithm initializes the truncated \glspl{atf} as $T_{\bm{0}_s}^{(0)}(J, N, D;P) = 1$ and $T_{\bm{\sigma}}^{(0)}(J, N, D;P) = 0$ for $\bm{\sigma} \ne \bm{0}_s$. Furthermore, the \gls{tf} associated with state $\bm{0}_s$ remains unchanged and is given by $T_{\bm{0}_s}^{(q)}(J, N, D;P) = 1$ in all subsequent iterations $q > 0$. At each iteration $q > 0$, the truncated \glspl{atf} corresponding to all other states in $\Sigma$ are updated in two steps and are therefore affected by the iteration process.

As in (\ref{eq:Temporary Truncated Transfer Functions}), at the first step of iteration $q > 0$, temporary truncated \glspl{atf} are computed as
\begin{equation} \label{eq:Temporary Truncated Augmented Transfer Functions}
\tilde{T}_{\bm{\sigma}}^{(q)}(J,N,D;P) \triangleq \sum_{\bm{\sigma}^\prime \in \pi(\bm{\sigma})} T_{\bm{\sigma^\prime}}^{(q-1)}(J,N,D;P) L_{\bm{\sigma}^\prime \bm{\sigma}}(J,N,D).
\end{equation}
Then, in the second step of iteration $q$, terms in each temporary function that share the same exponents in $J$, $N$ and $D$ are merged by adding their multiplicities. The resulting terms are then arranged in the lexicographic order defined earlier. For $\bm{\sigma} \ne \bm{0}_s$, the truncated \gls{atf} $T_{\bm{\sigma}}^{(q)}(J,N,D;P)$ is obtained by keeping $\tilde{T}_{\bm{\sigma}}^{(q)}(J,N,D;P)$ unchanged if it contains $P$ or fewer terms, or by retaining only the first $P$ terms otherwise.

Because all term multiplicities in $T_{\bm{\sigma}}^{(q)}(J, N, D; P)$ are integers, the iterative algorithm terminates after a finite number of iterations, $Q$, as soon as $T_{\bm{\sigma}}^{(Q)}(J, N, D; P) = T_{\bm{\sigma}}^{(Q-1)}(J, N, D; P)$ holds for all relevant states.

The pseudo-code corresponding to the iterative algorithm for determining $T_{\bm{\sigma}}(J, N, D; P)$ is presented in Algorithm~\ref{alg:Truncated Augmented Transfer Function T(N,D;P) Rate-2 NSM}. In addition to the vector notations introduced in Subsection~\ref{ssec:Determination Truncated Original Transfer Function}, this pseudo-code employs three additional vectors $\bm{k}_{\bm{\sigma}}^{(q)}$, $\tilde{\bm{k}}_{\bm{\sigma}}^{(q)}$, and $\bm{k}$, which represent the exponents of $J$ in the terms $J^k N^l D^m$ arranged in lexicographic order in $T_{\bm{\sigma}}^{(q)}(J, N, D; P)$, $\tilde{T}_{\bm{\sigma}}^{(q)}(J, N, D; P)$, and $T(J, N, D; P)$, respectively. Since the exponent of $J$ in each branch label $L_{\bm{\sigma}^\prime \bm{\sigma}}(J, N, D)$ is always $1$, there is no need to explicitly represent this exponent in the branch labels.

\begin{algorithm}
\caption{Determination of the truncated \gls{atf} $T(J,N,D;P)$ for a rate-$2$ NSM}
\label{alg:Truncated Augmented Transfer Function T(N,D;P) Rate-2 NSM}
\begin{algorithmic}[1]
\Require $P$, $i_{\bm{\sigma}^\prime \bm{\sigma}}$, $j_{\bm{\sigma}^\prime \bm{\sigma}}$, for all $\bm{\sigma}^\prime \in \pi(\bm{\sigma})$, $\bm{\sigma} \in \Sigma^* \cup \{\bm{0}_e\}$
\Ensure $\bm{k}$, $\bm{l}$, $\bm{m}$ and $\bm{\mu}$

\Statex \Comment{Step 1: Initialize transfer functions at $\bm{0}_s$}
\State $\bm{k}_{\bm{0}_s}^{(0)} \gets [0]$
\State $\bm{l}_{\bm{0}_s}^{(0)} \gets [0]$
\State $\bm{m}_{\bm{0}_s}^{(0)} \gets [0]$
\State $\bm{\mu}_{\bm{0}_s}^{(0)} \gets [1]$

\Statex \Comment{Step 2: Initialize transfer functions at other states}
\For{$\bm{\sigma} \in \Sigma^* \cup \{\bm{0}_e\}$}
    \State $\bm{k}_{\bm{\sigma}}^{(0)} \gets [\;]$
    \State $\bm{l}_{\bm{\sigma}}^{(0)} \gets [\;]$
    \State $\bm{m}_{\bm{\sigma}}^{(0)} \gets [\;]$
    \State $\bm{\mu}_{\bm{\sigma}}^{(0)} \gets [\;]$
\EndFor

\State $q \gets 0$
\Repeat
    \State $q \gets q + 1$
    
    \Statex \Comment{Step 3: Fix values at $\bm{0}_s$}
    \State $\bm{k}_{\bm{0}_s}^{(q)} \gets [0]$
    \State $\bm{l}_{\bm{0}_s}^{(q)} \gets [0]$
    \State $\bm{m}_{\bm{0}_s}^{(q)} \gets [0]$
    \State $\bm{\mu}_{\bm{0}_s}^{(q)} \gets [1]$
    
    \Statex \Comment{Step 4: Update transfer functions at other states}
    \For{$\bm{\sigma} \in \Sigma^* \cup \{\bm{0}_e\}$}
        \State $\tilde{\bm{k}}_{\bm{\sigma}}^{(q)} \gets [\;]$
        \State $\tilde{\bm{l}}_{\bm{\sigma}}^{(q)} \gets [\;]$
        \State $\tilde{\bm{m}}_{\bm{\sigma}}^{(q)} \gets [\;]$
        \State $\tilde{\bm{\mu}}_{\bm{\sigma}}^{(q)} \gets [\;]$
        \For{$\bm{\sigma}^\prime \in \pi(\bm{\sigma})$}
            \State $\tilde{\bm{k}}_{\bm{\sigma}}^{(q)} \gets [\tilde{\bm{k}}_{\bm{\sigma}}^{(q)}, \bm{k}_{\bm{\sigma^\prime}}^{(q-1)} + 1]$
            \State $\tilde{\bm{l}}_{\bm{\sigma}}^{(q)} \gets [\tilde{\bm{l}}_{\bm{\sigma}}^{(q)}, \bm{l}_{\bm{\sigma^\prime}}^{(q-1)} + i_{\bm{\sigma}^\prime \bm{\sigma}}]$
            \State $\tilde{\bm{m}}_{\bm{\sigma}}^{(q)} \gets [\tilde{\bm{m}}_{\bm{\sigma}}^{(q)}, \bm{m}_{\bm{\sigma^\prime}}^{(q-1)} + j_{\bm{\sigma}^\prime \bm{\sigma}}]$
            \State $\tilde{\bm{\mu}}_{\bm{\sigma}}^{(q)} \gets [\tilde{\bm{\mu}}_{\bm{\sigma}}^{(q)}, \bm{\mu}_{\bm{\sigma^\prime}}^{(q-1)}]$
        \EndFor
        \State $(\tilde{\bm{k}}_{\bm{\sigma}}^{(q)}, \tilde{\bm{l}}_{\bm{\sigma}}^{(q)}, \tilde{\bm{m}}_{\bm{\sigma}}^{(q)}, \tilde{\bm{\mu}}_{\bm{\sigma}}^{(q)}) \gets \text{Group}(\tilde{\bm{k}}_{\bm{\sigma}}^{(q)}, \tilde{\bm{l}}_{\bm{\sigma}}^{(q)}, \tilde{\bm{m}}_{\bm{\sigma}}^{(q)}, \tilde{\bm{\mu}}_{\bm{\sigma}}^{(q)})$
        \State $(\bm{k}_{\bm{\sigma}}^{(q)}, \bm{l}_{\bm{\sigma}}^{(q)}, \bm{m}_{\bm{\sigma}}^{(q)}, \bm{\mu}_{\bm{\sigma}}^{(q)}) \gets \text{Truncate}(\tilde{\bm{k}}_{\bm{\sigma}}^{(q)}, \tilde{\bm{l}}_{\bm{\sigma}}^{(q)}, \tilde{\bm{m}}_{\bm{\sigma}}^{(q)}, \tilde{\bm{\mu}}_{\bm{\sigma}}^{(q)})$
    \EndFor
\Until {$(\bm{k}_{\bm{\sigma}}^{(q)} = \bm{k}_{\bm{\sigma}}^{(q-1)}) \land (\bm{l}_{\bm{\sigma}}^{(q)} = \bm{l}_{\bm{\sigma}}^{(q-1)}) \land (\bm{m}_{\bm{\sigma}}^{(q)} = \bm{m}_{\bm{\sigma}}^{(q-1)}) \land (\bm{\mu}_{\bm{\sigma}}^{(q)} = \bm{\mu}_{\bm{\sigma}}^{(q-1)}), \forall \bm{\sigma} \in \Sigma^* \cup \{\bm{0}_e\}$}

\Statex \Comment{Step 5: Output transfer function at final state}
\State $\bm{k} \gets \bm{k}_{\bm{0}_e}^{(q)}$
\State $\bm{l} \gets \bm{l}_{\bm{0}_e}^{(q)}$
\State $\bm{m} \gets \bm{m}_{\bm{0}_e}^{(q)}$
\State $\bm{\mu} \gets \bm{\mu}_{\bm{0}_e}^{(q)}$
\end{algorithmic}
\end{algorithm}



\section{Closed-Form Expressions of Optimum Filters for Rate-2 NSMs under Unconstrained Optimization} \label{app:Closed-Form Expressions Optimum Filters Non-Degenerate Rate-2 NSMs}

This appendix presents the symbolic derivation of the optimal filter coefficients and the energy allocation parameter~$\eta$ for rate-$2$ \glspl{nsm} optimized under the unconstrained design framework. The numerical optimization procedure, the extraction of dominant minimum-distance error events, and the symbolic validation steps are detailed in Section~\ref{Rate-2 guaranteeing, minimum Euclidean distance approaching NSMs with real filters' coefficients}.

The symbolic derivations focus on filter lengths $L_0$ ranging from $3$ to $8$, with $L_1 = 1$, where consistent structural patterns and tightness conditions have been observed in the dominant error events that yield the \gls{msed}, $d_{\text{min}}^2$. These structural properties enable the derivation of closed-form expressions for both the normalized filter coefficients and the energy allocation parameter $\eta$, leading to a complete symbolic characterization of the corresponding non-normalized filters $h_0[k]$ and $h_1[k]$. For the simplest cases, such as $L_0 = 3$ and $4$, the derivations can be obtained manually. For more intricate scenarios such as $L_0 = 5$, $7$, and $8$, symbolic expressions were obtained using \textsc{Matlab}'s symbolic computation engine. The case $L_0 = 6$ follows a relatively direct derivation.

The cases $L_0 = 9$ and $L_0 = 10$ are excluded from this symbolic treatment, albeit for different reasons. For $L_0 = 9$, the dominant error events responsible for $d_{\text{min}}^2$ have been identified (as shown in Table~\ref{table:Input Differences Vectors L_0=9 L_1=1}), but closed-form expressions for the filters $h_m[k]$ could not be derived. This may be due to an incomplete identification of all necessary error patterns, or to the inherent complexity of the symbolic resolution process, which exceeded the capabilities of \textsc{Matlab}'s symbolic engine. For $L_0 = 10$, the situation is notably different: the \gls{msed} exactly equals that of conventional 2-ASK transmission, and in this case, the energy allocation parameter satisfies $\eta = 5/2$, leading to energy balance between the two filters. Moreover, this optimal performance is achieved by an infinite family of filter pairs, up to equivalence transformations, and the associated tightness conditions are no longer active—an indication of the structural flexibility permitted by this configuration. The multiplicity of optimal solutions, however, likely prevents the derivation of a unique symbolic representation. One possible path to overcoming this issue would be to isolate, among the many optimal filters, the one that maximizes the second \gls{msed}, which might yield a unique solution amenable to symbolic analysis. However, this approach has not been explored, and the symbolic derivation for $L_0 = 10$ remains an open challenge. As a result, closed-form expressions for both $L_0 = 9$ and $L_0 = 10$ are not included in this appendix.

\subsection{Closed-form Expressions for \texorpdfstring{$\bm{L_0=3}$ and $\bm{L_1=1}$}{L0=3 and L1=1}}

For the case where $L_0=3$ and $L_1=1,$ we start by determining the four unknowns in filters' vectors $\bm{h}_0 \triangleq (h_0[0], h_0[1], h_0[2])$ and $\bm{h}_1 \triangleq (h_1[0]).$ We begin by taking into account the three shortest-length error events with $\Delta \bar{\hat{\bm{b}}}_0^n=(2)$ and $\Delta \bar{\hat{\bm{b}}}_1^n=(-2,0,0), (0,0,0)$ and $(0,0,-2).$ Using \textsc{Matlab} notations for appending row vectors, the corresponding squared Euclidean distances, which must all be equal to $d_{\text{min}}^2,$ are given respectively by
\begin{equation}
    a_s \triangleq \| 2 \bm{h}_0 + h_1[0] (-2,0,0) \|^2,
\end{equation}
\begin{equation}
    a_0 \triangleq \| 2 \bm{h}_0 \|^2
\end{equation}
and
\begin{equation}
    a_e \triangleq \|2 \bm{h}_0 + h_1[0] (0,0,-2) \|^2.
\end{equation}
Equating $a_s$ and $a_0$ leads to the first tightness constraint $h_1[0]=2h_0[0],$ while equating $a_e$ and $a_0$ leads to the second tightness constraint $h_1[0]=2h_0[2].$

We now require two more equations because there are four unknowns and we already have two equations that link them. To establish the third equation, we consider the error event with $\Delta \bar{\hat{\bm{b}}}_0^n=(2,-2)$ and $\Delta \bar{\hat{\bm{b}}}_1^n=(0,0,0,0),$ for which the \gls{sed} is given by
\begin{equation}
    b \triangleq \|2(\bm{h}_0,0)-2(0,\bm{h}_0) \|^2.
\end{equation}
Equating $a_0$ and $b$ leads to $h_0[1]^2 = 2 (h_0[1]-h_0[0])^2,$ which gives two solutions, namely $h_0[1] = \pm \sqrt{2} h_0[0] /(1 \pm \sqrt{2}).$

The fourth and last equation, which is referred to as the energy equation, translates the fact that the average symbol energy,
\begin{equation}
    c \triangleq \| \bm{h}_0 \|^2 + \| \bm{h}_1 \|^2,
\end{equation}
is equal to that of conventional $4$-ASK. Letting $c=5,$ where $5$ is the average symbol energy of $4$-ASK, results in
\begin{equation}
h_0[0]^2 = \frac{15 \pm 10\sqrt{2}}{20 \pm 12\sqrt{2}}.   
\end{equation}
Since,
\begin{equation}
\eta_1 = h_1[0]^2 = 4 h_1[0]^2 = 4 \frac{15 \pm 10\sqrt{2}}{20 \pm 12\sqrt{2}}.   
\end{equation}
must be greater than half the average symbol energy, $5/2,$ to get the best achievable \gls{med}, we only preserve the solution
\begin{equation}
    h_1[0]^2 =  4 \frac{15 + 10\sqrt{2}}{20 + 12\sqrt{2}}.
\end{equation}
Now, by virtue of the relationships established in Subsection~\ref{ssec:Modulation of Rate 2} between the filters of equivalent \glspl{nsm}, we can retain the positive value of $h_1[0].$ After simplification, we have
\begin{equation}
    h_1[0] = 2 h_0[0] = 2 h_0[2] = \sqrt{\frac{15+5\sqrt{2}}{7}}
\end{equation}
and
\begin{equation}
    h_0[1] = \sqrt{\frac{25-15\sqrt{2}}{14}}.
\end{equation}
From these equations, we can deduce that
\begin{equation}
    \eta_1 = h_1[0]^2 = \frac{15+5\sqrt{2}}{7},
\end{equation}
\begin{equation}
    \eta_0 = 5-\eta_1 = \frac{20-5\sqrt{2}}{7},
\end{equation}
and
\begin{equation}
    d_{\text{min}}^2 = 4 \eta_0 = \frac{20(4-\sqrt{2})}{7}.
\end{equation}
The corresponding asymptotic gain in dB is approximated as
\begin{equation}
    \text{Gain [dB]} = 10 \log_{10}(d_{\text{min}}^2/4) = 10 \log_{10} \left( \frac{20-5\sqrt{2}}{7} \right) \approx 2.664646175845514.
\end{equation}

\subsection{Closed-Form Expressions for \texorpdfstring{$\bm{L_0=4}$ and $\bm{L_1=1}$}{L0=4 and L1=1}}

For the case where $L_0=4$ and $L_1=1,$ we begin by taking into account the three shortest-length error events with $\Delta \bar{\hat{\bm{b}}}_0^n=(2)$ and $\Delta \bar{\hat{\bm{b}}}_1^n=(-2,0,0,0), (0,0,0,0)$ and $(0,0,0,-2).$ The corresponding squared Euclidean distances, which must equate to $d_{\text{min}}^2,$ are given respectively by
\begin{equation}
    a_s \triangleq \| 2 \bm{h}_0 + h_1[0] (-2,0,0,0) \|^2,
\end{equation}
\begin{equation}
    a_0 \triangleq \| 2 \bm{h}_0 \|^2
\end{equation}
and
\begin{equation}
    a_e \triangleq \|2 \bm{h}_0 + h_1[0] (0,0,0,-2) \|^2,
\end{equation}
where $\bm{h}_0 \triangleq (h_0[0], h_0[1], h_0[2], h_0[3])$ and $\bm{h}_1 \triangleq (h_1[0]).$ Equating $a_s$ and $a_0$ leads to the first tightness constraint $h_1[0]=2h_0[0],$ while equating $a_e$ and $a_0$ leads to the second tightness constraint $h_1[0]=2h_0[3].$

We need three more equations in addition to this pair of equations, because we have five unknowns. The third equation is obtained by considering the error event with $\Delta \bar{\hat{\bm{b}}}_0^n=(2,2,2)$ and $\Delta \bar{\hat{\bm{b}}}_1^n=(0,-2,-2,-2,-2,0),$ for which the \gls{sed} is given by
\begin{equation}
    b \triangleq \|2(\bm{h}_0,0,0)+2(0,\bm{h}_0,0)+2(0,0,\bm{h}_0) + h_1[0] (0,-2,-2,-2,-2,0) \|^2.
\end{equation}
The error event with $\Delta \bar{\hat{\bm{b}}}_0^n=(2,2,2,0,-2)$ and $\Delta \bar{\hat{\bm{b}}}_1^n=(0,-2,-2,-2,0,0,0,0)$ can be used to support the fourth equation. The corresponding \gls{sed} is expressed as
\begin{equation}
    c \triangleq \|2(\bm{h}_0,0,0,0,0)+2(0,\bm{h}_0,0,0,0)+2(0,0,\bm{h}_0,0,0)-2(0,0,0,0,\bm{h}_0) + h_1[0] (0,-2,-2,-2,0,0,0,0) \|^2.
\end{equation}

The fifth and last equation is the energy equation,
\begin{equation} \label{eq:Energy Equation L_0=4 L_1=1}
    d \triangleq \| \bm{h}_0 \|^2 + \| \bm{h}_1 \|^2,
\end{equation}
which allows us to account for the fact that the average symbol energy is equal to $5.$

On the one hand, developing $b-a_0$ and equating it to zero leads to 
\begin{equation}
    0= b-a_0 = 2(h_0[0]-h_0[1]-h_0[2])(2h_0[0]-h_0[1]-h_0[2]).
\end{equation}
As a result, we have either $h_0[0]=h_0[1]+h_0[2]$ or $h_0[0]=(h_0[1]+h_0[2])/2.$  Since it can be demonstrated that the latter equation results in a smaller effective \gls{med}, the former equation corresponds to the correct solution.

On the other hand, equating $b-c$ to zero leads to
\begin{equation}
    h_0^2[1] + h_0^2[2] + 2h_0[0](h_0[2]-h_0[1])=0,
\end{equation}
which, when $h_0[0]$ is replaced by it expression as a function of $h_0[1]$ and $h_0[2],$ gives $h_0^2[1]=3h_0^2[2].$ Consequently, we have either $h_0[1]=\sqrt{3}h_0[2]$ or $h_0[1]=-\sqrt{3}h_0[2].$ It can be verified that the final solution results in a smaller effective \gls{med} and should therefore be rejected. The right solution that should be kept is $h_0[1]=\sqrt{3}h_0[2].$

Injecting the tightness constraints $h_1[0]=2h_0[0]=2h_0[3]$ and the equalities $h_0[0]=h_0[1]+h_0[2]$ and $h_0[1]=\sqrt{3}h_0[2]$ in the energy equation in (\ref{eq:Energy Equation L_0=4 L_1=1}) results in
\begin{equation} \label{eq:h_1[0] L_0=4 L_1=1}
    h_1[0] = 2 h_0[0] = 2 h_0[3] = \sqrt{\frac{5}{11} (5 + \sqrt{3})}
\end{equation}
and
\begin{equation}
    h_0[1] = \sqrt{3} h_0[2] = \sqrt{\frac{15}{88} (7-3 \sqrt{3})}.
\end{equation}

From (\ref{eq:h_1[0] L_0=4 L_1=1}), we deduce that
\begin{equation}
    \eta_1 = h_1[0]^2 = \frac{25+5 \sqrt{3}}{11},
\end{equation}
\begin{equation}
    \eta_0 = 5 - \eta_1= \frac{30-5 \sqrt{3}}{11}. 
\end{equation}
and
\begin{equation}
    d_{\text{min}}^2 = 4 \eta_0 = \frac{20(6-\sqrt{3})}{11}.
\end{equation}
The asymptotic gain in dB is therefore approximated as
\begin{equation}
    \text{Gain [dB]} = 10 \log_{10}(d_{\text{min}}^2/4) = 10 \log_{10} \left( \frac{30-5 \sqrt{3}}{11} \right) \approx 2.877965599259757.
\end{equation}

\subsection{Closed-Form Expressions for \texorpdfstring{$\bm{L_0=5}$ and $\bm{L_1=1}$}{L0=5 and L1=1}}

In this situation, we have six unknowns: $h_0[k],$ $k=0,1,\ldots,4$ and $h_1[0].$ As a result, using the brute force method, we require six equations. However, if we look at Table~\ref{table:Best Filters Numerical Form Rate-2 NSM}, we can see that $h_0[2]$ should be zero. Taking this into consideration, we may minimize the number of equations required to five.

Using the three shortest-length error events with $\Delta \bar{\hat{\bm{b}}}_0^n=(2)$ and $\Delta \bar{\hat{\bm{b}}}_1^n=(-2,0,0,0,0),$ $(0,0,0,0,0)$ and $(0,0,0,0,-2),$ we get the tightness relationships $h_1[0] = 2 h_0[0] = 2 h_0[4],$ which are equivalent to two equations. Using the energy equation, we have one more equation. To complete the set of equations that lead to the closed-for expressions of $h_m[k],$ $m=0,1,$ we consider from Table~\ref{table:Input Differences Vectors L_0=5 L_1=1} the three \gls{med} error events specified by the input differences vectors:
\begin{itemize}
    \item $\Delta \bar{\hat{\bm{b}}}_0^n=(2)$ and $\Delta \bar{\hat{\bm{b}}}_1^n=(0,0,0,0,0),$
    \item $\Delta \bar{\hat{\bm{b}}}_0^n=(2,2,2,2)$ and $\Delta \bar{\hat{\bm{b}}}_1^n=(0,-2,-2,-2,-2,-2,-2,0),$ and
    \item $\Delta \bar{\hat{\bm{b}}}_0^n=(2,-2,2,-2,0,0,-2)$ and $\Delta \bar{\hat{\bm{b}}}_1^n=(0,0,0,0,0,0,0,2,0,0,0).$
\end{itemize}

After solving all of the gathered equations with \textsc{Matlab}'s symbolic processing, we get the following closed-form expressions of $h_m[k],$ $m=0,1,$
\begin{itemize}
    \item $h_1[0] = 2 h_0[0] = 2 h_0[4] = z,$
    \item $h_0[1] = \frac{5439}{16000} z^7 - \frac{3577}{1600} z^5 + \frac{631}{160} z^3 - \frac{1}{2} z,$
    \item $h_0[2] = 0,$ and
    \item $h_0[3] = \frac{2331}{3200} z^7 - \frac{8553}{1600} z^5 + \frac{2063}{160} z^3 - 10 z,$
\end{itemize}
where $z$ is the solution to equation $777 z^8 - 6220 z^6 + 17900 z^4 - 22000 z^2 + 10000 = 0,$ which is numerically closest to $1.704938908106564.$

From the above, we deduce that $\eta_1 = h_1[0]^2 = z^2 \approx 2.906816680375603,$ $\eta_0 = 5 - \eta_1 = 5-z^2 \approx 2.093183319624397$ and $d_{\text{min}}^2 = 4 \eta_0 = 20 - 4 z^2 \approx 8.372733278497586.$ The asymptotic gain in dB is consequently approximated as
\begin{equation}
    \text{Gain [dB]} = 10 \log_{10}(d_{\text{min}}^2/4) \approx 3.208072652306580.
\end{equation}

\subsection{Closed-Form Expressions for \texorpdfstring{$\bm{L_0=6}$ and $\bm{L_1=1}$}{L0=6 and L1=1}}

We can see from Table~\ref{table:Best Filters Numerical Form Rate-2 NSM} that this scenario is the simplest of all those studied previously. Aside from the tightness criteria that result in $h_1[0] = 2 h_0[0] = 2 h_5[0],$ we observe that $h_0[1] = h_0[4] = 0$ and, more crucially, $h_0[0] = 2 h_0[2] = h_0[3].$ These observations avoid using any of the error events indicated in Table~\ref{table:Input Differences Vectors L_0=6 L_1=1}. We get the filters' vectors %
\begin{equation}
    \bm{h}_0 \triangleq (h_0[0], h_0[1], h_0[2], h_0[3], h_0[4], h_0[5]) = \sqrt{\frac{5}{29}} (2, 0, 1, 2, 0, 2)
\end{equation}
and
\begin{equation}
\bm{h}_1 \triangleq (h_1[0]) = \sqrt{\frac{5}{29}} (4)
\end{equation}
by simply evaluating the energy equation and injecting the prior equalities.

These results lead us to the conclusions that $\eta_1 = h_1[0]^2 = 80/29,$ $\eta_0 = 5 - \eta_1 = 65/29$ and $d_{\text{min}}^2 = 4 \eta_0 = 260/29.$ As a result,
\begin{equation}
    \text{Gain [dB]} = 10 \log_{10}(d_{\text{min}}^2/4) = 10 \log_{10}(65/29) \approx 3.505153587438994.
\end{equation}
is a good approximation for the asymptotic gain in dB.

Before concluding this section, keep in mind that there is a very simple scaled version of the optimum filters for $L_0=6$ and $L_1=1$ with integer coefficients. This scaled version is specified by specified by $\mathring{h}_0[k] = 2 \delta[k] + \delta[k-2] + 2 \delta[k-3] + 2 \delta[k-5]$ and $\mathring{h}_1[k] = 4 \delta[k].$ As a result, only the \gls{nsm} explored exhaustively in Subsection~\ref{ssec:Modulation of Rate 2}, where $L_0=2,$ and the current \gls{nsm}, where $L_0=6,$ allow simplified expressions of the optimum filters with integer coefficients.

\subsection{Closed-Form Expressions for \texorpdfstring{$\bm{L_0=7}$ and $\bm{L_1=1}$}{L0=7 and L1=1}}

For the scenario where $L_0=7$ and $L_1=1,$ we have eight unknowns: $h_0[k],$ $k=0,1,\ldots,6$ and $h_1[0].$ As above, we avoid the brute force approach by noticing, from Table~\ref{table:Best Filters Numerical Form Rate-2 NSM}, that $h_0[2] = h_0[4] = 0$. Taking this into consideration, we may reduce the number of equations required to $6$.

Using the four shortest-length error events with $\Delta \bar{\hat{\bm{b}}}_0^n=(2)$ and $\Delta \bar{\hat{\bm{b}}}_1^n = (0, 0, 0, 0, 0, 0, 0)$, $(-2, 0, 0, 0, 0, 0, 0),$ $(0, -2, 0, 0, 0, 0, 0),$ and $(0, 0, 0, 0, 0, 0, -2),$ we get the tightness relationships $h_1[0] = 2 h_0[0] = 2 h_0[1] = 2 h_0[6],$ which are equivalent to three equations. Using the energy equation, we have one more equation. To complete the set of equations that lead to the closed-for expressions of $h_m[k],$ $m=0,1,$ we consider from Table~\ref{table:Input Differences Vectors L_0=7 L_1=1} the three \gls{med} error events specified by the input differences vectors:
\begin{itemize}

    \item $\Delta \bar{\hat{\bm{b}}}_0^n=(2)$ and $\Delta \bar{\hat{\bm{b}}}_1^n=(0,0,0,0,0,0,0),$ and

    \item $\Delta \bar{\hat{\bm{b}}}_0^n=(2,-2,-2)$ and $\Delta \bar{\hat{\bm{b}}}_1^n=(0,0,2,0,0,0,0,2,0),$

    \item $\Delta \bar{\hat{\bm{b}}}_0^n=(2,2,-2,0,-2,2)$ and $\Delta \bar{\hat{\bm{b}}}_1^n=(0,-2,0,0,0,0,-2,0,0,0,0,0).$

\end{itemize}

Solving all the resulting equations with \textsc{Matlab}'s symbolic processing, we get the following closed-form expressions
\begin{itemize}

    \item $h_1[0] = 2 h_0[0] = 2 h_0[1] = 2 h_0[6] = z,$
    
    \item $h_0[2] = h_0[4] = 0,$
    
    \item $h_0[3] = - \frac{175711}{1776000} z^7 + \frac{48179}{44400} z^5 - \frac{3791}{888} z^3 + \frac{647}{111} z,$ and
    
    \item $h_0[5] = \frac{16849}{355200} z^7 - \frac{49859}{88800} z^5 + \frac{1477}{888} z^3 - \frac{257}{222} z,$

\end{itemize}
where $z$ is the solution to equation $2407 z^8 - 25740 z^6 + 110000 z^4 - 216000 z^2 + 160000 = 0,$ which is numerically closest to $1.644657696635838.$

From the above, we have $\eta_1 = h_1[0]^2 = z^2 \approx 2.704898939103502,$ $\eta_0 = 5 - \eta_1 = 5-z^2 \approx 2.295101060896498$ and $d_{\text{min}}^2 = 4 \eta_0 = 20 - 4 z^2 \approx 9.180404243585992$ and the asymptotic gain in dB is approximated as
\begin{equation}
    \text{Gain [dB]} = 10 \log_{10}(d_{\text{min}}^2/4) \approx 3.608018137178796.
\end{equation}

\subsection{Closed-Form Expressions for \texorpdfstring{$\bm{L_0=8}$ and $\bm{L_1=1}$}{L0=8 and L1=1}}
\label{Closed-form expressions for L0 = 8 and L1 = 1}

There are nine unknowns in this scenario: $h_0[k],$ $k=0,1,\ldots,7,$ and $h_1[0].$ As before, by noting from Table~\ref{table:Best Filters Numerical Form Rate-2 NSM} that $h_0[3] = h_0[4] = h_0[5] = 0,$ we may simplify the symbolic determination of the closed-form expressions of the unknowns. We obtain the tightness relationships $h_1[0] = 2 h_0[0] = 2 h_0[2] = 2 h_0[7]$ based on the shortest-length error events with $\Delta \bar{\hat{\bm{b}}}_0^n=(2)$ and $\Delta \bar{\hat{\bm{b}}}_1^n = (0,0,0,0,0,0,0,0)$, $(-2,0,0,0,0,0,0,0),$ $(0,0,-2,0,0,0,0,0),$ and $(0,0,0,0,0,0,0,-2).$ These equalities are equivalent to seven equations when combined with the energy equation. We consider the three \gls{med} error events using input differences 
\begin{itemize}

    \item $\Delta \bar{\hat{\bm{b}}}_0^n=(2)$ and $\Delta \bar{\hat{\bm{b}}}_1^n=(0,0,0,0,0,0,0,0,0),$ and
    
    \item $\Delta \bar{\hat{\bm{b}}}_0^n=(2,-2,-2,2,-2)$ and $\Delta \bar{\hat{\bm{b}}}_1^n=(0,0,0,0,2,0,0,0,2,0,0,0),$
    
    \item $\Delta \bar{\hat{\bm{b}}}_0^n=(2,-2,0,2,-2,0,0,0,2,2)$ and $\Delta \bar{\hat{\bm{b}}}_1^n=(0,0,0,0,0,0,0,0,0,-2,-2,0,0,0,0,-2,0).$

\end{itemize}
from Table~\ref{table:Input Differences Vectors L_0=8 L_1=1} to complete this set of equations.

The following closed-form expressions
\begin{itemize}

    \item $h_1[0] = 2 h_0[0] = 2 h_0[2] = 2 h_0[7] = \sqrt{8570} \sqrt{17 \sqrt{29} + 793}/1714,$
    
    \item $h_0[1] = \sqrt{17 \sqrt{29} + 793} \; (143 \sqrt{8570} - 9 \sqrt{248530})/620468,$
    
    \item $h_0[3] = h_0[4] = h_0[5] = 0,$ and
    
    \item $h_0[6] = \sqrt{17 \sqrt{29} + 793} \; (179 \sqrt{8570} - 10 \sqrt{248530})/620468,$

\end{itemize}
of $h_m[k],$ $m=0,1,$ are obtained by solving all of the above equations using \textsc{Matlab}'s symbolic processing.

From the facts that $\eta_1 = h_1[0]^2 \approx 2.580361148545177,$ $\eta_0 = 5 - \eta_1 \approx 2.419638851454823$ and $d_{\text{min}}^2 = 4 \eta_0 \approx 9.678555405819292,$ the asymptotic gain in dB can be approximated as
\begin{equation}
    \text{Gain [dB]} = 10 \log_{10}(d_{\text{min}}^2/4) \approx 3.837505492346273.
\end{equation}



\section{Closed-Form Expressions of Optimum Filters for Rate-2 NSMs under Constrained Optimization} \label{app:Closed-Form Expressions Optimum Filters Rate-2 NSMs Constrained Optimization}

This appendix presents the symbolic derivation of the optimal filter coefficients for rate-$2$ \glspl{nsm} obtained under \emph{constrained optimization}, where both filters share equal energy contributions. Specifically, the energy constraint $\eta_0 = \eta_1 = \eta$ is imposed, with $\eta = 5/2$ chosen such that the total average symbol energy equals that of a conventional $4$-ASK modulation. The numerical optimization procedures and resulting \gls{nsm} designs are discussed in detail in Section~\ref{Rate-2 guaranteeing, minimum Euclidean distance approaching NSMs with real filters' coefficients}.

The symbolic derivations focus on filter lengths $L_0$ ranging from $3$ to $7$, with $L_1 = 1$, for which all error events leading to the \gls{msed} $d_{\text{min}}^2$ have been identified. These dominant error structures enable the derivation of closed-form expressions for the optimal normalized filter $h_0[k]$. The filter $h_1[k]$ is fixed to $h_1[k] = \sqrt{5/2} \, \delta[k]$ as a direct consequence of the energy constraint $\| h_1[k] \|^2 = \eta_1 = 5/2$ imposed in the constrained optimization setting. The sign-reversed version $h_1[k] = -\sqrt{5/2} \, \delta[k]$ is equally valid, but the positive version is adopted without loss of generality. For $L_0 = 3$, $4$, $5$, and $6$, the complete sets of minimum-distance error vectors are reported in Tables~\ref{table:Input Differences Vectors L_0=3 L_1=1 Constrained Optimization}–\ref{table:Input Differences Vectors L_0=6 L_1=1 Constrained Optimization}, and symbolic solutions are derived based on these patterns. The case $L_0 = 7$ also admits a symbolic solution, although only a reduced set of error events is presented, as the organized structure observed in the numerically optimized $h_0[k]$—such as nearly equal component values and components close to zero—guides and simplifies the symbolic derivation.

A special case where $L_0 = L_1 = 2$ is also included in this appendix, though it does not originate from numerical optimization. Instead, it is constructed analytically by designing two filters—such as $\bm{h}_0 = (2,1)$ and $\bm{h}_1 = (1,2)$, up to sign changes—to achieve exact energy balancing while preserving both the average symbol energy and the \gls{msed} $d_{\text{min}}^2$ of a conventional $4$-ASK modulation after normalization by $\sqrt{2}$.This configuration introduces memory and, despite requiring trellis-based detection—which is more complex than the memoryless detection of $4$-ASK—offers benefits in systems employing error correction and iterative decoding such as turbo-equalization, due to the balanced energy distribution between the two filters and their corresponding bipolar input streams. Although not derived via formal optimization, this case illustrates how guided filter design can yield high-quality constrained \glspl{nsm} and is included here for completeness.

\subsection{Closed-Form Expressions for \texorpdfstring{$\bm{L_0=L_1=2}$ }{L0=L1=2}}
\label{app:Closed-form expressions for L0 = L1 = 2}

As noted earlier, the numerical optimization did not yield a noteworthy \gls{nsm} with $L_0=2$ and $L_1=1$ that outperforms $4$-ASK. However, with $L_0 = L_1 = 2,$ we obtain an \gls{nsm} defined by the filters $\bm{h}_0 = (2,1)/\sqrt{2}$ and $\bm{h}_1 = (1,2)/\sqrt{2},$ which achieves the same \gls{msed} as $4$-ASK, while also ensuring balanced energy contributions from the input sequences $\bar{b}_0[k]$ and $\bar{b}_1[k].$ This balance is guaranteed by the equal filter norms, $\|\bm{h}_0\|^2 = \|\bm{h}_1\|^2 = 5/2.$ In contrast, $4$-ASK assigns unequal energy contributions: $\|h_0[k]\|^2 = \|\delta[k]\|^2 =1,$ from $\bar{b}_0[k],$ and $\|h_1[k]\|^2 = \|2\,\delta[k]\|^2 = 4,$ from $\bar{b}_1[k].$

To verify that the proposed \gls{nsm} matches the asymptotic performance of $4$-ASK, it suffices to show that its \gls{msed} is also $4.$ For any pair of input sequences differences $(\Delta \bar{b}_0[k],\Delta \bar{b}_1[k]),$ the resulting modulated sequence difference $\Delta s[k]$ has a squared norm lower-bounded by $4.$ This bound is achieved, for example, by the input sequence pairs $(\bar{b}_0^0[k] = \delta[k], \bar{b}_1^0[k] = -\delta[k])$ and $(\bar{b}_0^1[k] = -\delta[k], \bar{b}_1^1[k] = \delta[k]),$ producing the modulated sequences $s^0[k] = \bar{b}_0^0[k] \circledast h_0[k] + \bar{b}_1^0[k] \circledast h_1[k] = (\delta[k] - \delta[k-1])/\sqrt{2}$ and $s^1[k] = \bar{b}_0^1[k] \circledast h_0[k] + \bar{b}_1^1[k] \circledast h_1[k] = (- \delta[k] + \delta[k-1])/\sqrt{2}.$ The corresponding difference is $\Delta s[k] = s^1[k] - s^0[k] = - \sqrt{2}\,\delta[k] + \sqrt{2}\,\delta[k-1]),$ with \gls{sen} $\| \Delta s[k] \|^2 = 4.$ This confirms that the \gls{nsm} achieves the same minimum distance as $4$-ASK.

Before examining other $(L_0,L_1)$ configurations, with $L_0 \ge 3$ and $L_1=1,$ observe that scaling the filters $\bm{h}_0$ and $\bm{h}_1$ by $\sqrt{2}$ yields scaled filters $\mathring{\bm{h}}_0 \triangleq \sqrt{2}\,\bm{h}_0 = (2,1)$ and $\mathring{\bm{h}}_1 \triangleq \sqrt{2}\,\bm{h}_1 = (1,2),$ both with integer-valued taps. This property offers practical advantages, including simplified modulation at the transmitter, more efficient metric computation at the receiver, and easier analysis of the \gls{nsm}’s characteristics—such as its transfer function and distance spectrum.

Interestingly, the scaled filters $\mathring{\bm{h}}_0 = (2,1)$ and $\mathring{\bm{h}}_1 = (1,2)$ can be interpreted as time-domain combinations of the basic $4$-ASK filters $\bm{h}_0 = (1)$ and $\bm{h}_1 = (2),$ formed by concatenating them in opposite orders. Specifically, $\mathring{\bm{h}}_0$ corresponds to the sequence $(\bm{h}_1, \bm{h}_0)$ and $\mathring{\bm{h}}_1$ to $(\bm{h}_0, \bm{h}_1).$ This construction results in filters that are time-reverses of each other, offering equal contributions to the average symbol energy of the input sequences $\bar{b}_0[k]$ and $\bar{b}_1[k].$

\subsection{Closed-Form Expressions for \texorpdfstring{$\bm{L_0=3}$ and $\bm{L_1=1}$}{L0=3 and L1=1}}

For the case where $L_0=3$ and $L_1=1,$ we start by determining the three unknowns in filter vector $\bm{h}_0 \triangleq (h_0[0], h_0[1], h_0[2]).$ We begin by taking into account the three shortest-length error events with $\Delta \bar{\hat{\bm{b}}}_0^n=(2)$ and $\Delta \bar{\hat{\bm{b}}}_1^n=(-2,-2,0), (0,-2,0)$ and $(0,-2,-2).$ The corresponding \glspl{sed}, which must all be equal to $d_{\text{min}}^2,$ are given respectively by
\begin{equation}
    a_s \triangleq \| 2 \bm{h}_0 + h_1[0] (-2,-2,0) \|^2,
\end{equation}
\begin{equation}
    a_0 \triangleq \| 2 \bm{h}_0 + h_1[0] (0,-2,0) \|^2
\end{equation}
and
\begin{equation}
    a_e \triangleq \|2 \bm{h}_0 + h_1[0] (0,-2,-2) \|^2.
\end{equation}
Equating $a_s$ and $a_0$ leads to the first tightness constraint $h_1[0]=2h_0[0],$ while equating $a_e$ and $a_0$ leads to the second tightness constraint $h_1[0]=2h_0[2].$ Given that $h_1[0] = \sqrt{\eta_1} = \sqrt{5/2},$ we conclude that $h_0[0] = h_0[2] = \sqrt{5/8}.$

We now require one more equation to determine the remaining unknown, $h_0[2]$. To establish this equation, we consider the error event with $\Delta \bar{\hat{\bm{b}}}_0^n=(2,-2)$ and $\Delta \bar{\hat{\bm{b}}}_1^n=(0,0,0,0),$ for which the \gls{sed} is given by
\begin{equation}
    b \triangleq \|2(\bm{h}_0,0)-2(0,\bm{h}_0) \|^2.
\end{equation}
Equating $a_0$ and $b$ leads to $h_0[1]^2 = 2 h_0[0]^2,$ which gives two solutions, namely $h_0[1] = \pm \sqrt{2} \, h_0[0].$

Substituting the solution $\bm{h}_0 = \sqrt{\tfrac{5}{8}}(1,-\sqrt{2},1),$ into the expression of $a_0$ yields a \gls{msed}, $d_{\text{min}}^2=20(1+1/\sqrt{2}),$ which exceeds the upper bound of $10$ achieved by $2$-ASK with bipolar alphabet $\{ \pm \sqrt{5/2}\},$ consistent with the average symbol energy of $5$ for $4$-ASK. Therefore, this solution must be discarded. In contrast, the \gls{msed} associated with the first solution,
\begin{equation}
    \bm{h}_0 = \sqrt{\frac{5}{8}} \,(1,\sqrt{2},1) = \left( \frac{\sqrt{10}}{4}, \frac{\sqrt{5}}{2}, \frac{\sqrt{10}}{4}   \right),
\end{equation}
is 
\begin{equation}
    d_{\text{min}}^2 = 20 \, (1-\tfrac{1}{\sqrt{2}}),
\end{equation}
which is admissible. The corresponding asymptotic gain is approximately
\begin{equation}
    \text{Gain [dB]} = 10 \log_{10}(d_{\text{min}}^2/4) = 10 \log_{10} \left( 5 \, (1-\tfrac{1}{\sqrt{2}}) \right) \approx 1.656793211661651.
\end{equation}

Observe that the detection trellis of the current \gls{nsm}, with parameters $L_0 = 3$ and $L_1 = 1$, comprises $4$ states—the same number as the trellis of the previously discussed \gls{nsm}, with $L_0 = L_1 = 2.$ As a result, both \glspl{nsm} involve the same detection complexity at the receiver. However, while the configuration with $L_0 = L_1 = 2$ offers no gain over $4$-ASK, the present configuration provides a nonzero performance gain. This clearly justifies favoring the current \gls{nsm} over the previously considered one.

Recall that for the unconstrained optimized \gls{nsm} with $L_0 = 3$ and $L_1 = 1,$ the \gls{msed} was $d_{\text{min}}^2 = \tfrac{20}{7} \, (4-\sqrt{2}),$ corresponding to an asymptotic gain of approximately $\text{Gain [dB]} = 10 \log_{10} (\tfrac{5}{7} \, (4-\sqrt{2})) \approx 2.664646175845514.$ Imposing the constraint that the input sequences, $\bar{b}_0[k]$ and $\bar{b}_1[k],$ contribute equally to the average symbol energy results in a degradation of this gain by approximately $1.00785296418$ dB. In light of the discussion at the beginning of the Appendix, this reduction is naturally detrimental to the initial stage of the iterative turbo-detection algorithm.

Even before examining other values of $L_0,$ we can already observe a trend in the degradation caused by imposing energy constraints in the optimization. Specifically, as $L_0$ increases from $3$ to $10,$ the gain achievable through unconstrained optimization—given by $10 \, \log_{10} (\eta_0)$ in dB—increases. This occurs because the upper bound on the \gls{msed}, in (\ref{Upper-Bound Minimum Squared Euclidean Distance}), is always attained in the unconstrained case, with $\eta = \eta_0 \le \eta_1,$ and the gain improves as both $\eta_0$ and $\eta_1$ converge toward $5/2,$ with $\eta_0$ increasing and $\eta_1$ decreasing, under the constraint $\eta_0 + \eta_1 = 5.$ Thus, unconstrained optimization naturally evolves toward the balanced energy allocation of the constrained case, where $\eta_0 = \eta_1 = 5/2.$ As a result, the performance gap between the unconstrained and constrained optimized \glspl{nsm} is expected to diminish as $L_0$ increases.

\subsection{Closed-Form Expressions for \texorpdfstring{$\bm{L_0=4}$ and $\bm{L_1=1}$}{L0=4 and L1=1}}

For the case where $L_0=4$ and $L_1=1,$ we begin by taking into account the three shortest-length error events with $\Delta \bar{\hat{\bm{b}}}_0^n=(2)$ and $\Delta \bar{\hat{\bm{b}}}_1^n=(-2,-2,0,0), (0,-2,0,0)$ and $(0,-2,0,2).$ The corresponding \glspl{sed}, which must equate to $d_{\text{min}}^2,$ are given respectively by
\begin{equation}
    a_s \triangleq \| 2 \bm{h}_0 + h_1[0] (-2,-2,0,0) \|^2,
\end{equation}
\begin{equation}
    a_0 \triangleq \| 2 \bm{h}_0  + h_1[0] (0,-2,0,0) \|^2
\end{equation}
and
\begin{equation}
    a_e \triangleq \|2 \bm{h}_0 + h_1[0] (0,-2,0,-2) \|^2,
\end{equation}
where $\bm{h}_0 \triangleq (h_0[0], h_0[1], h_0[2], h_0[3]).$ Equating $a_s$ and $a_0$ leads to the first tightness constraint $h_1[0]=2h_0[0],$ while equating $a_e$ and $a_0$ leads to the second tightness constraint $h_1[0]=-2h_0[3].$

To derive a closed form for the best filter, $\bm{h}_0,$ we proceed differently from the unconstrained optimization framework and use directly the symbolic math toolbox of Matlab. In addition to the tightness constraints, we consider three linked equations that originate from equating the norms of the error events of \gls{msed}, in Table \ref{table:Input Differences Vectors L_0=4 L_1=1 Constrained Optimization}.

The first equation results from equating $b-a_0$ to zero, where
\begin{equation}
    b \triangleq \|2(\bm{h}_0,0,0)+2(0,\bm{h}_0,0)+2(0,0,\bm{h}_0) + h_1[0] (0,-2,-2,0,2,0) \|^2.
\end{equation}
is the \gls{sed} corresponding to the error event with $\Delta \bar{\hat{\bm{b}}}_0^n=(2,2,2)$ and $\Delta \bar{\hat{\bm{b}}}_1^n=(0,-2,-2,0,2,0).$ The second equation results from setting $c-a_0$ to zero, where
\begin{equation}
    c \triangleq \|2(\bm{h}_0,0,0,0,0)+2(0,\bm{h}_0,0,0,0)+2(0,0,\bm{h}_0,0,0)-2(0,0,0,0,\bm{h}_0) + h_1[0] (0,-2,-2,0,2,2,0,0) \|^2
\end{equation}
is the \gls{sed} associated to the error event with $\Delta \bar{\hat{\bm{b}}}_0^n=(2,2,2,0,-2)$ and $\Delta \bar{\hat{\bm{b}}}_1^n=(0,-2,-2,0,2,2,0,0).$ The third equation is obtained by equating $d$ and $a_0,$ where
\begin{equation}
    d \triangleq \|2(\bm{h}_0,0,0,0,0)+2(0,\bm{h}_0,0,0,0)+2(0,0,\bm{h}_0,0,0)+2(0,0,0,0,\bm{h}_0) + h_1[0] (0,-2,-2,0,0,0,0,0) \|^2
\end{equation}
is the \gls{sed} associated to the error event with $\Delta \bar{\hat{\bm{b}}}_0^n=(2,2,2,0,2)$ and $\Delta \bar{\hat{\bm{b}}}_1^n=(0,-2,-2,0,0,0,0,0).$

The last used equation,
\begin{equation}
    e \triangleq \| \bm{h}_0 \|^2 = 5/2,
\end{equation}
accounts for the fact that the contribution of the first input sequence $\bar{b}_0[k],$ to the average symbol energy is equal to $5/2.$

Solving with Matlab symbolic math toolbox, we obtain, up to a sign change, two solutions. The first solution, $\bm{b}_0 = \sqrt{\tfrac{5}{8}} \, (1, \tfrac{1}{2}\,(1-\sqrt{3}), \tfrac{1}{2}\,(1+\sqrt{3}), -1),$ when injected in the expression of $a_0,$ $b,$ $c$ or $d,$ leads to a \gls{msed}, $d_{\text{min}}^2 = 15+5\, \sqrt{3},$ which is larger than that of $2$-ASK, when its average symbol energy is aligned with that of $4$-ASK. Hence, this solution must be rejected. The other solution,
\begin{equation}
    \bm{h}_0 = \sqrt{\frac{5}{8}} \, \left(1,\frac{1+\sqrt{3}}{2},\frac{1-\sqrt{3}}{2},-1 \right) = \left( \frac{\sqrt{10}}{4}, \frac{\sqrt{10}+\sqrt{30}}{8}, \frac{\sqrt{10}-\sqrt{30}}{8}, -\frac{\sqrt{10}}{4}   \right),
\end{equation}
is admissible, since it leads to acceptable \gls{msed}, $d_{\text{min}}^2 = 15-5 \, \sqrt{3}.$
The corresponding asymptotic gain, in dB, with respect to $4$-ASK, is approximated as
\begin{equation}
    \text{Gain [dB]} = 10 \log_{10}(d_{\text{min}}^2/4) = 10 \log_{10} \left( \frac{15-5 \sqrt{3}}{4} \right) \approx 2.000118644331985.
\end{equation}

Recall that the unconstrained optimized \gls{nsm}, with $L_0 = 4$ and $L_1 = 1,$ achieves a \gls{msed} of $d_{\text{min}}^2 = \tfrac{20}{11}(6-\sqrt{3}),$ which corresponds to an asymptotic gain (in dB) over $4$-ASK given by $\text{Gain [dB]} = 10 \log_{10} \left( \frac{30-5 \sqrt{3}}{11} \right) \approx 2.877965599259757.$ Consequently, the constrained optimized \gls{nsm} experiences a reduction in asymptotic gain relative to the unconstrained case of approximately $0.87784695492$ dB. As anticipated, this loss is slightly smaller than the corresponding reduction observed for $L_0 = 3$ and $L_1 = 1.$

\subsection{Closed-Form Expressions for \texorpdfstring{$\bm{L_0=5}$ and $\bm{L_1=1}$}{L0=5 and L1=1}}

For the case where $L_0=5$ and $L_1=1,$ we begin by taking into account the three shortest-length error events with $\Delta \bar{\hat{\bm{b}}}_0^n=(2)$ and $\Delta \bar{\hat{\bm{b}}}_1^n=(-2,0,0,-2,0), (0,0,0,-2,0)$ and $(0,0,0,-2,-2).$ The corresponding \glspl{sed}, which must equate to $d_{\text{min}}^2,$ are given respectively by
\begin{equation}
    a_s \triangleq \| 2 \bm{h}_0 + h_1[0] (-2,0,0,-2,0) \|^2,
\end{equation}
\begin{equation}
    a_0 \triangleq \| 2 \bm{h}_0  + h_1[0] (0,0,0,-2,0) \|^2
\end{equation}
and
\begin{equation}
    a_e \triangleq \|2 \bm{h}_0 + h_1[0] (0,0,0,-2,-2) \|^2,
\end{equation}
where $\bm{h}_0 \triangleq (h_0[0], h_0[1], h_0[2], h_0[3], h_0[4]).$ Equating $a_s$ and $a_0$ leads to the first tightness constraint $h_1[0]=2h_0[0],$ while equating $a_e$ and $a_0$ leads to the second tightness constraint $h_1[0]=-2h_0[4].$

As in the case where $L_0=4$ and $L_1=1,$ for the current case with $L_0=5$ and $L_1=1,$ we consider not only the tightness constraints but also three additional equations derived by equating the norms of the error events corresponding to the \gls{msed}, as shown in Table \ref{table:Input Differences Vectors L_0=5 L_1=1 Constrained Optimization}.

We obtain the closed form expressions of the components of $\bm{h}_0$ as
\begin{itemize}
    \item $h_0[0] = -h_0[4] = h_1[0]/2 = \sqrt{10}/4,$
    \item $h_0[1] = \tfrac{92968}{63442625}\,z^7 - \tfrac{436432\,\sqrt{10}}{63442625}\,z^6 + \tfrac{359106}{2537705}\,z^5 - \tfrac{1837744\,\sqrt{10}}{12688525}\,z^4 + \tfrac{2111027}{2537705}\,z^3 - \tfrac{268297\,\sqrt{10}}{2537705}\,z^2 + \tfrac{543585}{1015082}\,z - \tfrac{5055\,\sqrt{10}}{44134},$
    \item $h_0[2] = \tfrac{157528}{63442625}\,z^7 - \tfrac{147272\,\sqrt{10}}{12688525}\,z^6 + \tfrac{3081362}{12688525}\,z^5 - \tfrac{3249324\,\sqrt{10}}{12688525}\,z^4 + \tfrac{3944164}{2537705}\,z^3 - \tfrac{862709\,\sqrt{10}}{5075410}\,z^2 + \tfrac{991157}{507541}\,z - \tfrac{18700\,\sqrt{10}}{22067},$ and
    \item $h_0[3] = z,$
\end{itemize}
where $z$ is the solution to equation $\tfrac{92968}{63442625}\,z^7 - \tfrac{436432\,\sqrt{10}}{63442625}\,z^6 + \tfrac{359106}{2537705}\,z^5 -\tfrac{1837744\,\sqrt{10}}{12688525}\,z^4 + \tfrac{2111027}{2537705}\,z^3 - \tfrac{268297\,\sqrt{10}}{2537705}\,z^2 + \tfrac{543585}{1015082}\,z - \tfrac{5055\,\sqrt{10}}{44134}= 0,$ which is numerically closest to $1.035860604572035.$ In numerical form, filter $h_0[k]$ is approximated as $\bm{h}_0 = (0.790569415042095, 0.374822289162865, -0.191052504412710,$ $1.035860604572035,  0.790569415042095).$

Based on the above, the \gls{msed} is approximately $d_{\text{min}}^2 \approx 6.897284604453374,$ which yields an asymptotic gain relative to $4$-ASK of $\text{Gain [dB]} = 10 \log_{10}(d_{\text{min}}^2/4) \approx 2.366181554371827.$ This represents a loss of approximately $0.84189109793$ dB, compared to the unconstrained case with the same parameters, $L_0=5$ and $L_1=1.$

\subsection{Closed-Form Expressions for \texorpdfstring{$\bm{L_0=6}$ and $\bm{L_1=1}$}{L0=6 and L1=1}}

Using the same methodology applied in the case $L_0=5$ and $L_1=1,$ and the \gls{msed} error events listed in Table \ref{table:Input Differences Vectors L_0=6 L_1=1 Constrained Optimization}, we derive the filter $h_0[k].$ This filter has coefficients $h_0[0] = h_0[5] = \sqrt{10}/4,$ while $h_0[1]$ through $h_0[4]$ are degree-$15$ polynomials in a parameter $z,$ which itself is a root of a degree-$16$ equation. Due to the complexity of these expressions, we provide the filter in numerical form: $\bm{h}_0 = (0.790569415042095, 0.381490134032264, -0.684574552553092, -0.689502871008588,$ $\\-0.400510612223563, 0.790569415042095).$ The resulting \gls{msed} is approximately $d_{\text{min}}^2 \approx 7.673335660277092,$ corresponding to an asymptotic gain of $\text{Gain [dB]} = 10 \log_{10}(d_{\text{min}}^2/4) \approx 2.829242049462369.$ This reflects a loss of about $0.67591153797$ dB compared to the unconstrained case with the same parameters, $L_0=6$ and $L_1=1.$

\subsection{Closed-Form Expressions for \texorpdfstring{$\bm{L_0=7}$ and $\bm{L_1=1}$}{L0=7 and L1=1}}

Following the same approach used for the cases with $L_0 \le 6,$ and relying on the tightness property, we have $h_0[0] = -h_0[6] = h_1[0]/2 = \sqrt{10}/4.$ In addition, by analyzing the numerical values of the optimized filter $h_0[k],$ we were able to propose exact closed-form expressions for all components, except $h_0[5],$ which was instead determined using the norm constraint. Specifically, we observe that $h_0[1] = h_0[2] = 0,$ $h_0[3] = -h_1[0]/4 = -\sqrt{10}/8$ and $h_0[4] = h_1[0]/2 = \sqrt{10}/4.$ Taking into account the norm constraint $\| \bm{h}_0 \|^2 = \eta_0 = 5/2,$ we then deduce $h_0[5] = \sqrt{30}/8.$ In summary, we obtain the closed filter expression
\begin{equation}
    \bm{h}_0 = \sqrt{\frac{5}{8}} \, \left(1,0,0,-\frac{1}{2},1,\frac{\sqrt{3}}{2},-1 \right) = \left( \frac{\sqrt{10}}{4}, 0, 0, -\frac{\sqrt{10}}{8}, \frac{\sqrt{10}}{4}, \frac{\sqrt{30}}{8}, -\frac{\sqrt{10}}{4}   \right).
\end{equation}
Using one of the error events from Table \ref{table:Input Differences Vectors L_0=7 L_1=1 Constrained Optimization}, the \gls{msed} is computed as $d_{\text{min}}^2 = (130-55 \, \sqrt{3})/4 \approx 8.684301395927937.$ The corresponding asymptotic gain, relative to $4$-ASK, is
\begin{equation}
    \text{Gain [dB]} = 10 \log_{10}(d_{\text{min}}^2/4) = 10 \log_{10} \left( \frac{130-55 \sqrt{3}}{16} \right) \approx 3.366748962684391,
\end{equation}
which reflects a loss of approximately $0.24126917449$ dB compared to the unconstrained case with the same parameters 
($L_0=7,$ $L_1=1$).



\section{Determination of the Multiplicities of Error Events with MSED for Rate-2 NSMs with Simple Rational Filter Coefficients} \label{app:Multiplicities Minimum Euclidean Distance Rate-2 NSMs Simple Coefficients}

As we have shown in Subsection~\ref{sssec:One-dimensional designed NSMs}, the lengths of all error events with \gls{msed}, for rate-$2$ \glspl{nsm} with simple rational coefficients are of the form $\kappa(2^{L_0-2}-1)+1,$ $\kappa$ being a positive integer, when $q^n(x)$ is of the form $q^n(x) = (x + 1)p^n(x),$ where $p^n(x)$ is a primitive polynomial of degree $L_0-2.$ This is precisely the case for the filter coefficients that maximize the shortest error event with \gls{msed}, when $L_0=5$ and $L_1=1,$ as shown in particular in Table~\ref{table:Characteristics First Minimum Euclidean Distance Simple Filters Coefficients L_0=5 L_1=1}. Indeed, in this case, we have precisely $q^n(x) = x^4 + x^2 + x + 1 = (x + 1)(x^3 + x^2 + 1),$ where $p^n(x)=x^3 + x^2 + 1$ is the primitive polynomial.

For each of the $4$ non-equivalent filters $\mathring{\bm{h}}_0 = \mathring{\bm{h}}^l_0, l = 0,1,2$ and $3,$ characterized globally in Table~\ref{table:Characteristics First Minimum Euclidean Distance Simple Filters Coefficients L_0=5 L_1=1}, for $L_0=5$ and $L_1=1,$ any error event of length $\kappa(2^{L_0-2}-1)+1, \kappa \geq 1$ can be built as the succession, with partial overlapping, of $\kappa$ building blocks, among those specified in Tables~\ref{table:Building Bricks Simple Filters Coefficients L_0=5 L_1=1 h_0=h^0_0}--\ref{table:Building Bricks Simple Filters Coefficients L_0=5 L_1=1 h_0=h^3_0}. The construction of any error event must adhere to the following rules:

\begin{enumerate}

    \item \textbf{Rule 1:} Any two consecutive building blocks must only have one component in common. This means that the last component of the preceding building block must be positioned in front of the first component of the subsequent building block.

    \item \textbf{Rule 2:} Each of the $\kappa-1$ first building blocks must belong to the primary category. It is imperative that the last component of $\Delta \mathring{\bm{b}}_1$ in these building blocks be null. Failure to do so may result in one of two unpleasant scenarios when the first component of $\Delta \mathring{\bm{b}}_1$ in the next block is added to the last component of $\Delta \mathring{\bm{b}}_1$ in one of these blocks. In the first scenario, $\Delta \mathring{b}_1[k]$ belongs to the set $\{\pm 2\},$ which is banned. Remember that $\Delta \mathring{b}_1[k]$ must exclusively be in the set $\{0,\pm 1\}.$ In the second scenario, $\Delta \mathring{b}_1[k]=0,$ which is clearly an acceptable value. The problem is that, when it comes to weight counting for multiplicity determination, this component's contribution to $\Delta \mathring{b}_1[k]$ is equal to $2,$ rather than null, as it should be.    
    
    \item \textbf{Rule 3:} When $\kappa \geq 2,$ subsequent building blocks should have overlapping components in $\Delta \bar{\bm{s}},$ with opposite signs. This requirement is necessary because the final difference sequence, $\Delta \bar{s}[k],$ must be precisely null for inner (non-boundary) components, to ensure a \gls{msed} of $5.$ Accordingly, building blocks with $\Delta \bar{\bm{s}}=(\pm 1, 0,0,0, 0,0,0, 1)$ can only be followed by building blocks with $\Delta \bar{\bm{s}}=(- 1, 0,0,0, 0,0,0, \pm 1),$ while building blocks with $\Delta \bar{\bm{s}}=(\pm 1, 0,0,0, 0,0,0, -1)$ may only be followed by building blocks with $\Delta \bar{\bm{s}}=(1, 0,0,0, 0,0,0, \pm 1).$ Hence, at each subsequent block, only half of the primary building blocks are allowed.

    \item \textbf{Rule 4:} The previous restriction, in Rule 3, does not apply to the first building block, as the first component in $\Delta \bar{\bm{s}},$ can freely take values in the set $\{\pm 1\}$.
    
    \item \textbf{Rule 5:} It is possible to use any primary or secondary building block as the last, $kappa$-th, block in the error event, as long as the constraint of opposing signs in $\Delta \bar{\bm{s}}$ for consecutive building blocks, in Rule 3, is observed.

\end{enumerate}

Given the previous explanation, and focusing on the first filter, $\mathring{\bm{h}}_0 = \mathring{\bm{h}}^0_0,$ which is completely specified in Table~\ref{table:Building Bricks Simple Filters Coefficients L_0=5 L_1=1 h_0=h^0_0}, we can see that the error event with length $\kappa(2^{L_0-2}-1)+1,$ should make the following contributions to the partial \gls{tf}, $T^0(N),$ which characterizes \gls{msed} error events:
\begin{enumerate}

    \item Contribution of the first building block (all primary building blocks can be used to start the error event): $2N^5(1+N)^2 + 2N^5(1+N)^2 = 4N^5(1+N)^2$ ($2N^5(1+N)^2$ for primary blocks with first component in $\Delta \bar{\bm{s}}$ equal to $1$ and another $2N^5(1+N)^2$ for primary blocks with first component in $\Delta \bar{\bm{s}}$ equal to $-1$).

    \item Contribution of the $\kappa-2$ intermediate (non-boundary) building blocks (only half of the primary building blocks are allowed, either with first component in $\Delta \bar{\bm{s}}$ equal to $1$ or with first component in $\Delta \bar{\bm{s}}$ equal to $-1,$ depending on the previous building block last component in $\Delta \bar{\bm{s}}$): $(2N^5(1+N)^2)^{\kappa-2}$.
    
    \item Contribution of the last building block in the error event (secondary blocks can be utilized here, however only half of the primary and secondary blocks are allowed as in the previous situation for intermediate building blocks): $2N^5(1+N)^2 + 2N^6(1+N)^2 = N^5(1+N)^3$ ($2N^5(1+N)^2$ for half of the primary blocks and another $2N^6(1+N)^2$ for half of the secondary blocks)

\end{enumerate}
To put it briefly, for the first filter, $\mathring{\bm{h}}_0 = \mathring{\bm{h}}^0_0,$ we may say that the aggregate contribution of error events of length $\kappa(2^{L_0-2}-1)+1$ to the partial \gls{tf} $T^0(N)$ is equal to $(4N^5(1+N)^2) \times (2N^5(1+N)^2)^{\kappa-2} \times (N^5(1+N)^3) = 2(1+N)(2N^5(1+N)^2)^\kappa.$ Hence, the partial \gls{tf} is given by
\begin{equation} \label{eq:Partial Transfer Function h^0_0}
T^0(N) = \sum_{\kappa \geq 1} 2(1+N)(2N^5(1+N)^2)^\kappa = \frac{4N^5(1+N)^3}{1-2N^5(1+N)^2}.
\end{equation}
Given that the \gls{nsm} at hand has rate $R=2$, the multiplicity of error events with \gls{msed}, $d_{\text{min}}^2=5,$ in the \gls{bep} union bound is equal to
\begin{equation} \label{eq:Multiplicity Minimum Squared Distance h^0_0}
\mu^0 = \frac{1}{R} \left. N \cfrac{\partial T^0(N)}{\partial N} \right|_{N=1/2} = \frac{10287}{6050} \approx 1.7.
\end{equation}

As a result, for the first filter, $\mathring{\bm{h}}_0 = \mathring{\bm{h}}^0_0,$ the \gls{bep} can be approximated by 
\begin{equation} \label{eq:Tight Estimate BEP Filter h^0_0}
        \text{BEP}^0 \approx \frac{\mu^0}{2} \operatorname{erfc} \left( \sqrt{ \frac{1}{2} \frac{E_b}{N_0}} \right) = \frac{10287}{12100} \operatorname{erfc} \left( \sqrt{ \frac{1}{2} \frac{E_b}{N_0}} \right),
\end{equation}
at high \gls{snr}.

For the second filter, $\mathring{\bm{h}}_0 = \mathring{\bm{h}}^1_0$ and third filter, $\mathring{\bm{h}}_0 = \mathring{\bm{h}}^2_0,$ the contributions of the building blocks to an error event are the same, as shown in Tables~\ref{table:Building Bricks Simple Filters Coefficients L_0=5 L_1=1 h_0=h^1_0} and~\ref{table:Building Bricks Simple Filters Coefficients L_0=5 L_1=1 h_0=h^2_0}. Hence, both filters have the same partial \gls{tf}, denoted here as $T^{12}(N)$. Following the same steps as for the first filter, $\mathring{\bm{h}}_0 = \mathring{\bm{h}}^0_0,$ we may say that the aggregate contribution of error events of length $\kappa(2^{L_0-2}-1)+1$ to the partial \gls{tf} $T^12(N)$ is equal to $(2N^4(1+N)^2(1+N^2)) \times (N^4(1+N)^2(1+N^2))^{\kappa-2} \times (N^5(1+N)^2(1+N^2) = 2(1+N)(N^4(1+N)^2(1+N^2))^\kappa.$ Thus, the partial \gls{tf} is given in this case by
\begin{equation} \label{eq:Partial Transfer Function h^12_0}
T^{12}(N) = \sum_{\kappa \geq 1} 2(1+N)(N^4(1+N)^2(1+N^2))^\kappa = \frac{2N^4(1+N)^3(1+N^2)}{1-N^4(1+N)^2(1+N^2)}.
\end{equation}
The corresponding multiplicity of the \gls{msed}, $d_{\text{min}}^2=5,$ in the \gls{bep} union bound is
\begin{equation} \label{eq:Multiplicity Minimum Squared Distance h^12_0}
\mu^{12} = \frac{1}{R} \left. N \cfrac{\partial T^{12}(N)}{\partial N} \right|_{N=1/2} = \frac{184599}{89042} \approx 2.07
\end{equation}
for both filters, $\mathring{\bm{h}}^1_0$ and $\mathring{\bm{h}}^2_0.$

This leads to the \gls{bep} approximation 
\begin{equation} \label{eq:Tight Estimate BEP Filter h^12_0}
        \text{BEP}^{12} \approx \frac{\mu^{12}}{2} \operatorname{erfc} \left( \sqrt{ \frac{1}{2} \frac{E_b}{N_0}} \right) = \frac{184599}{178084} \operatorname{erfc} \left( \sqrt{ \frac{1}{2} \frac{E_b}{N_0}} \right),
\end{equation}
for both filters, $\mathring{\bm{h}}^1_0$ and $\mathring{\bm{h}}^2_0,$ at high \gls{snr}.

For the fourth and last filter, $\mathring{\bm{h}}_0 = \mathring{\bm{h}}^3_0,$ the contributions of the building blocks to an error event are shown in Table~\ref{table:Building Bricks Simple Filters Coefficients L_0=5 L_1=1 h_0=h^3_0}. Following the same steps as for the first filter, we may say that the aggregate contribution of error events of length $\kappa(2^{L_0-2}-1)+1$ to the partial \gls{tf} $T^3(N)$ is equal to $(2N^4(1+2N+N^4)(1+N)) \times (N^4(1+2N+N^4)(1+N))^{\kappa-2} \times (N^4(1+2N+N^4)(1+N)^2) = 2(1+N)(N^4(1+2N+N^4)(1+N))^\kappa.$ Thus, the partial \gls{tf} is given in this case by
\begin{equation} \label{eq:Partial Transfer Function h^3_0}
T^3(N) = \sum_{\kappa \geq 1} 2(1+N)(N^4(1+2N+N^4)(1+N))^\kappa = \frac{2N^4(1+2N+N^4)(1+N)^2}{1-N^4(1+2N+N^4)(1+N)}.
\end{equation}
The corresponding multiplicity of the \gls{msed}, $d_{\text{min}}^2=5,$ in the \gls{bep} union bound is
\begin{equation} \label{eq:Multiplicity Minimum Squared Distance h^3_0}
\mu^3 = \frac{1}{R} \left. N \cfrac{\partial T^3(N)}{\partial N} \right|_{N=1/2} = \frac{791991}{341138} \approx 2.32.
\end{equation}

Based on this multiplicity, the \gls{bep} can be approximated at high \gls{snr} by
\begin{equation} \label{eq:Tight Estimate BEP Filter h^3_0}
        \text{BEP}^{3} \approx \frac{\mu^3}{2} \operatorname{erfc} \left( \sqrt{ \frac{1}{2} \frac{E_b}{N_0}} \right) = \frac{791991}{682276} \operatorname{erfc} \left( \sqrt{ \frac{1}{2} \frac{E_b}{N_0}} \right),
\end{equation}
for the fourth and last filter, $\mathring{\bm{h}}^3_0.$

In Figure~\ref{fig:BER-BEP-NSM-2-FilterSimpleCoefficients-FilterLength_5}, all three \gls{bep} approximations in (\ref{eq:Tight Estimate BEP Filter h^0_0}), (\ref{eq:Tight Estimate BEP Filter h^12_0}) and (\ref{eq:Tight Estimate BEP Filter h^3_0}), are displayed. At moderate to high \glspl{snr}, it is readily apparent they are in perfect agreement with the simulated \gls{ber} for the four filters, $\mathring{\bm{h}}^i_0, i = 0,1,2$ and $3.$ At low \glspl{snr}, the error events with second \gls{msed} play a noticeable role because the corresponding term, $0.5 \operatorname{erfc} (\sqrt{E_b/N_0}),$ starts to be as important as the term, $0.5 \operatorname{erfc} (\sqrt{0.5 E_b/N_0}),$ corresponding to the error events with \gls{msed}. This is why the approximate \gls{bep} curves start to separate and detach from the \gls{ber} curves, passing below them at low \gls{snr}.



\section{Characterization of the Multiplicities of Error Events with Second MSED for Rate-2 NSMs with Simple Rational Filter Coefficients}
\label{app:Characterization Second Minimum Euclidean Distance Rate-2 NSMs Simple Coefficients}

In Appendix~\ref{app:Multiplicities Minimum Euclidean Distance Rate-2 NSMs Simple Coefficients}, we have demonstrated in detail that, for the least examined initial filter length $L_0 = 5,$ error events with \gls{msed} $d_{\text{min}}^2 = 5,$ do not have marginal probability error multiplicities. However, as we demonstrated in Subsection~\ref{sssec:One-dimensional designed NSMs}, when the first filter length $L_0$ increases, the probability error multiplicities of these error events with \gls{msed} vanish exponentially to $0.$ It is clear that these error events don't matter much for moderate to large values of $L_0$ in the practical working ranges of binary error probability. Next, we show that what actually matters for performed binary error probability are error events with second \gls{msed} $2d_{\text{min}}^2 = 10.$ These error events have a global error probability multiplicity that never goes to zero, regardless of whether $L_0$ is large or not. Before proceeding, it is useful to note that this observation is applicable most of the time in other \gls{nsm} system configurations.

A first category of error events with second \gls{msed} $2d_{\text{min}}^2 = 10,$ can be inspired from error events with \gls{msed} $d_{\text{min}}^2 = 5.$ As stated in Subsection~\ref{sssec:One-dimensional designed NSMs}, error events with \gls{msed} $d_{\text{min}}^2 = 5$ are of lengths $K = \kappa(K_{\text{min}}-1)+1, \kappa \geq 1,$ where $K_{\text{min}}$ is the length of the shortest error event with \gls{msed}. Remember that, for the \gls{nsm} systems studied in Subsection~\ref{sssec:One-dimensional designed NSMs}, $K_{\text{min}}=2^{L_0-2},$ for $L_0 \neq 14$ and $K_{\text{min}} = 2^{L_0-2}-1,$ for $L_0 = 14.$ Let $\Delta \mathring{b}_0[l]$ and $\Delta \mathring{b}_1[k] = \pm \delta[k] \pm \delta[k-(K-1)],$ be two finite-length input sequence differences that result, in compliance to (\ref{eq:Scaled Modulated Sequence Difference One-Dimensional NSM}), to a modulated sequence difference $\Delta \bar{s}[k],$ corresponding to an error event with length $K$ and \gls{msed} of $d_{\text{min}}^2 = 5.$ Then, according to (\ref{eq:Galois Field One-Dimensional NSM}), $\Delta \Ddot{s}[k]$ and $\Delta \Ddot{b}_0[k],$ the binary equivalents of $\Delta \bar{s}[k]$ and $\Delta \mathring{b}_0[k],$ are such that $\Delta \Ddot{s}[k] = \Delta \Ddot{b}_0[k] \circledast \Ddot{h}_0[k] = \delta[k] + \delta[k-(K-1)].$ When expressed in polynomial form, this leads to $\Delta \Ddot{s}(x) = \Delta \Ddot{b}_0(x) \Ddot{h}_0(x) = x^{K-1} + 1,$ as seen in (\ref{eq:Galois Field One-Dimensional NSM Polynomial Form}).

Now, if we consider all error events with input difference sequences $\Delta \mathring{\beta}_0[k]$ and $\Delta \mathring{\beta}_1[k]$ such that, $\Delta \Ddot{\beta}_0(x),$ the polynomial equivalent of $\Delta\Ddot{\beta}_0[k]$ is determined by $\Delta \Ddot{\beta}_0(x) = (x^L+1) \Delta \Ddot{b}_0(x),$ for some integer $L,$ $1 \leq L < K-1.$ Then, the corresponding binary counterpart, $\Delta \Ddot{\sigma}[k],$ of the associated output difference sequences, $\Delta \bar{\sigma}[k],$ lead, in polynomial form, to $\Delta \Ddot{\sigma}(x) = \Delta \Ddot{\beta}_0(x) \Ddot{h}_0(x) = (x^L+1) \Delta \Ddot{b}_0(x) \Ddot{h}_0(x) = (x^L+1)(x^{K-1} + 1) = x^{K+L-1} + x^{K-1} + x^L + 1.$ This identity says that it could be possible, in some cases, to find eligible error events with input difference sequences $\Delta \mathring{\beta}_0[k]$ and $\Delta \mathring{\beta}_1[k],$ such that the associated output difference $\Delta \bar{\sigma}[k],$ is specified as $\Delta \bar{\sigma}(x) = \pm x^{K+L-1} \pm x^{K-1} \pm x^L \pm 1,$ in polynomial form. If such eligible sequences exist, then they correspond precisely to error events with second \gls{msed} $2d_{\text{min}}^2 = 10.$ For illustration purpose, Tables~\ref{table:Characteristics Second Minimum Euclidean Distance Simple Filters Coefficients L_0=5 L_1=1 (x + 1)}, \ref{table:Characteristics Second Minimum Euclidean Distance Simple Filters Coefficients L_0=5 L_1=1 (x^2 + 1)} and~\ref{table:Characteristics Second Minimum Euclidean Distance Simple Filters Coefficients L_0=5 L_1=1 (x^3 + 1)} show that such error events, that are inherited from \gls{msed} error events, indeed exist, for the simple case where $L_0=5,$ $L_1=1,$ $K=8$ and $L=1,2$ and $3.$

The set of first category of error events, if not empty, is inherited from the set of error events with \gls{msed}. As shown in Subsection~\ref{sssec:One-dimensional designed NSMs}, since the latter set has a vanishing error probability multiplicity, as $L_0$ increases, it is also expected that the multiplicity of the set of first category of error events vanishes too, as $L_0$ increases. Indeed, it is anticipated that the binary equivalents, $\Delta\Ddot{\beta}_0[k]$ and $\Delta\Ddot{\beta}_1[k],$ of the associated input difference sequences, $\Delta \mathring{\beta}_0[k]$ and $\Delta \mathring{\beta}_1[k],$ have large Hamming weights, which together determine the exponent of the $\tfrac{1}{2}$ term reflecting their contribution to error probability multiplicity.

The second category of error events, with second \gls{msed}, $2d_{\text{min}}^2 = 10,$ encompasses all error events with least Hamming weights in the binary equivalents, $\Delta\Ddot{\beta}_0[k]$ and $\Delta\Ddot{\beta}_1[k],$ of the input difference sequences, $\Delta \mathring{\beta}_0[k]$ and $\Delta \mathring{\beta}_1[k].$ This category is therefore expected to contribute significantly to the error probability multiplicity o of error events with second \gls{msed}. As such, it should play an important role in the binary error probability lower bound at moderate values of the \glspl{snr}.

The determination of the multiplicities of error events under this second category follow a similar analysis as that followed in Appendix~\ref{app:Tight Estimate BEP Rate 5/4} for the rate $5/4$ \gls{nsm} presented in Subsection~\ref{ssec:Minimum Euclidean Distance Guaranteeing 5/4-NSM}. 

The first type of error events of maximum multiplicity contributions are those where $\Delta \mathring{b}_0[k] = 0$ and $\Delta \mathring{b}_1[k] = \pm \delta[k].$ It corresponds to output difference sequences $\Delta \bar{s}[k] = \pm 2 \delta[k].$ Its contribution to the transfer function, with regards to the second \gls{msed}, is given by $2N.$

The second type of error events of maximum multiplicity contributions are those where $\Delta \mathring{b}_0[k] = \alpha \delta[k],$ with $\alpha \in \{\pm 1\}.$ In this case, there are $16$ authorized input difference sequences, $\Delta \mathring{b}_1[k],$ that intimately depend on the chosen first filter, $\mathring{h}_0[k].$ Let us assume that this filter is of the form $\mathring{h}_0[k] = h_0[0] \delta[k] + h_0[k_1] \delta[k-k_1] + h_0[k_2] \delta[k-k_2] + h_0[L_0-1] \delta[k-(L_0-1)],$ where $0<k_1<k_2<L_0-1.$ Then, the $16$ allowed input difference sequences, $\Delta \mathring{b}_1[k],$ are of the form $\Delta \mathring{b}_1[k] = \Delta \mathring{b}_1[0] \delta[k] + \Delta \mathring{b}_1[k_1] \delta[k-k_1] + \Delta \mathring{b}_1[k_2] \delta[k-k_2] + \Delta \mathring{b}_1[L_0-1] \delta[k-(L_0-1)],$ where $\Delta \mathring{b}_1[k] \in \{0, - \alpha h_0[k] \},$ for $k \in \{ 0, k_1, k_2, L_0-1 \}.$ The $16$ corresponding output difference sequences, $\Delta \bar{s}[k],$ are of the form $\Delta \bar{s}[k] = \pm \delta[k] \pm \delta[k-k_1] \pm \delta[k-k_2] \pm \delta[k-(L_0-1)].$ These $16$ error events in the second type contribute significantly to the multiplicity of the second \gls{msed} in the binary error probability. Their significative number stems from the tightness property inherited from the peculiar choice of the filters coefficients. Their contribution to the transfer function, with regards to the second \gls{msed}, is given by $2N(1+N)^4.$

To sum up, the total contribution of the second type of error events to the transfer function, with regards to the second \gls{msed}, is equal to
\begin{equation} \label{eq:Partial Transfer Function Second Minimum Squared Distance}
T^s(N) = 2N+2N(1+N)^4.
\end{equation}
The resulting contribution to the multiplicity
of the second \gls{msed} is therefore given by
\begin{equation} \label{eq:Multiplicity Second Minimum Squared Distance}
\mu^s = \frac{1}{R} \left. N \cfrac{\partial T^s(N)}{\partial N} \right|_{N=1/2} = \frac{205}{32} \approx 6.4,
\end{equation}
where $R=2$ is the rate of the \gls{nsm} at hand. As a result, for the second type of error events, the \gls{bep} contribution can be approximated as 
\begin{equation} \label{eq:Contribution Second Minimum Squared Distance BEP}
\text{BEP}^s \approx \frac{\mu^s}{2} \operatorname{erfc} \left( \sqrt{\frac{E_b}{N_0}} \right) = \frac{205}{64} \operatorname{erfc} \left( \sqrt{\frac{E_b}{N_0}} \right),
\end{equation}
at high \glspl{snr}. Notice here that the error probability multiplicity corresponding to the second type of errors doesn't depend on $L_0,$ and therefore doesn't vanish to zero when $L_0$ increases. The \gls{bep} curve corresponding to $\text{BEP}^s$ in (\ref{eq:Contribution Second Minimum Squared Distance BEP}) is shown in Figure~\ref{fig:BER-BEP-NSM-2-FilterSimpleCoefficients-FilterLength_14}, for $L_0=14.$ As can be noticed from this figure, there is an almost perfect match between the \gls{ber} curve obtained by simulation and the \gls{bep} curve corresponding to $\text{BEP}^s,$ at moderate \glspl{snr} values. This fact goes in the direction that, if there are other categories of error events with second \gls{msed}, their multiplicities should vanish in the same way as the first category.

To conclude this Appendix, it is of utmost importance to suggest, as future work, a way to reduce the relatively huge multiplicity of $\tfrac{205}{32},$ obtained in (\ref{eq:Multiplicity Second Minimum Squared Distance}). Remember that this huge multiplicity arises from the tightness property underlying filters $\mathring{h}_0[k] = \delta[k] \pm \delta[k-k_0] \pm \delta[k-k_1] \pm \delta[k-L_0+1]$ and $\mathring{h}_1[k] = 2\delta[k].$ As a consequence, to reduce it, the best way is to chose filters that do not show tightness. Keeping in mind that filters $\mathring{h}_0[k]$ and $\mathring{h}_1[k]$ must present the same energy (same \gls{sen}) and integer coefficients, immediate efficient choices can be obtained by taking $\mathring{h}_1[k] = 3\delta[k]$ or $5\delta[k].$

To avoid tightness conditions for the former choice, where $\mathring{h}_1[k] = 3\delta[k],$ the coefficients of filter $\mathring{h}_0[k]$ should be exclusively in the set $\{0, \pm 1 \}.$ Being of the same \gls{sen} as $\mathring{h}_1[k],$ which is equal to $9,$ this filter must show $9$ non-null coefficients in the set $\{\pm 1 \}.$ As a consequence, $L_0$ mus be at least equal to $9.$ \glspl{nsm} with these filters are expected to present strictly lower \glspl{msed} than $2$-ASK. However, the cumulative of the multiplicities of all error events, with lower \gls{sed} than $2$-ASK, can be made very small with respect to practical working values of the $BEP,$ above $10^{-10},$ as is the case for the filters treated in this appendix and in Section~\ref{Rate 2 Approaching NSM Simple Rational Coefficients}. The way to accomplish this is to maximize the minimum length of these error events, through an adequate choice of filter $\mathring{h}_0[k].$ This aspect can be studied in detail by casting all involved differences sequences into equivalent ones in the Galois field $\text{GF}(3)=\{0, 1, 2\},$ using modular $3$ arithmetic.

For the latter choice, where $\mathring{h}_1[k] = 5\delta[k],$ the coefficients of filter $\mathring{h}_0[k]$ should be exclusively in the set $\{0, \pm 1, \pm 2 \}.$ The equality constraint for the squared Euclidean norms of $\mathring{h}_0[k]$ and $\mathring{h}_1[k]$ requires that $\mathring{h}_0[k]$ has $25$ as \gls{sen}. On the one hand, the candidate filters $\mathring{h}_0[k],$ with the smallest number of non-null coefficients, should have exactly $6$ non-null coefficients in the set $\{ \pm 2 \}$ and one non-null coefficient in the set $\{ \pm 1 \}.$ The corresponding minimum filter length is therefore lower bounded by $7.$ On the other hand, the candidate filters $\mathring{h}_0[k],$ with the largest number of non-null coefficients, should have $25$ non-null coefficients in the set $\{ \pm 1 \}.$ Obviously, the minimum filter length for $\mathring{h}_0[k]$ is $L_0=25.$ Intermediate candidate filters can be obtained by replacing any non-null coefficient in $\mathring{h}_0[k],$ in the set $\{ \pm 2 \}$ by four non-null coefficients in the set $\{ \pm 1 \}.$ The discussions above, for the case where $\mathring{h}_1[k] = 3\delta[k],$ apply here. We can use modular arithmetic in the Galois field $\text{GF}(5)=\{0, 1, 2, 3, 4\}.$



\section{Tight Estimate of the BEP for the Rate-3/2 NSM} \label{app:Tight Estimate BEP Rate 3/2}

\begin{figure}[!htbp]
    \centering
    \includegraphics[width=0.9\textwidth]{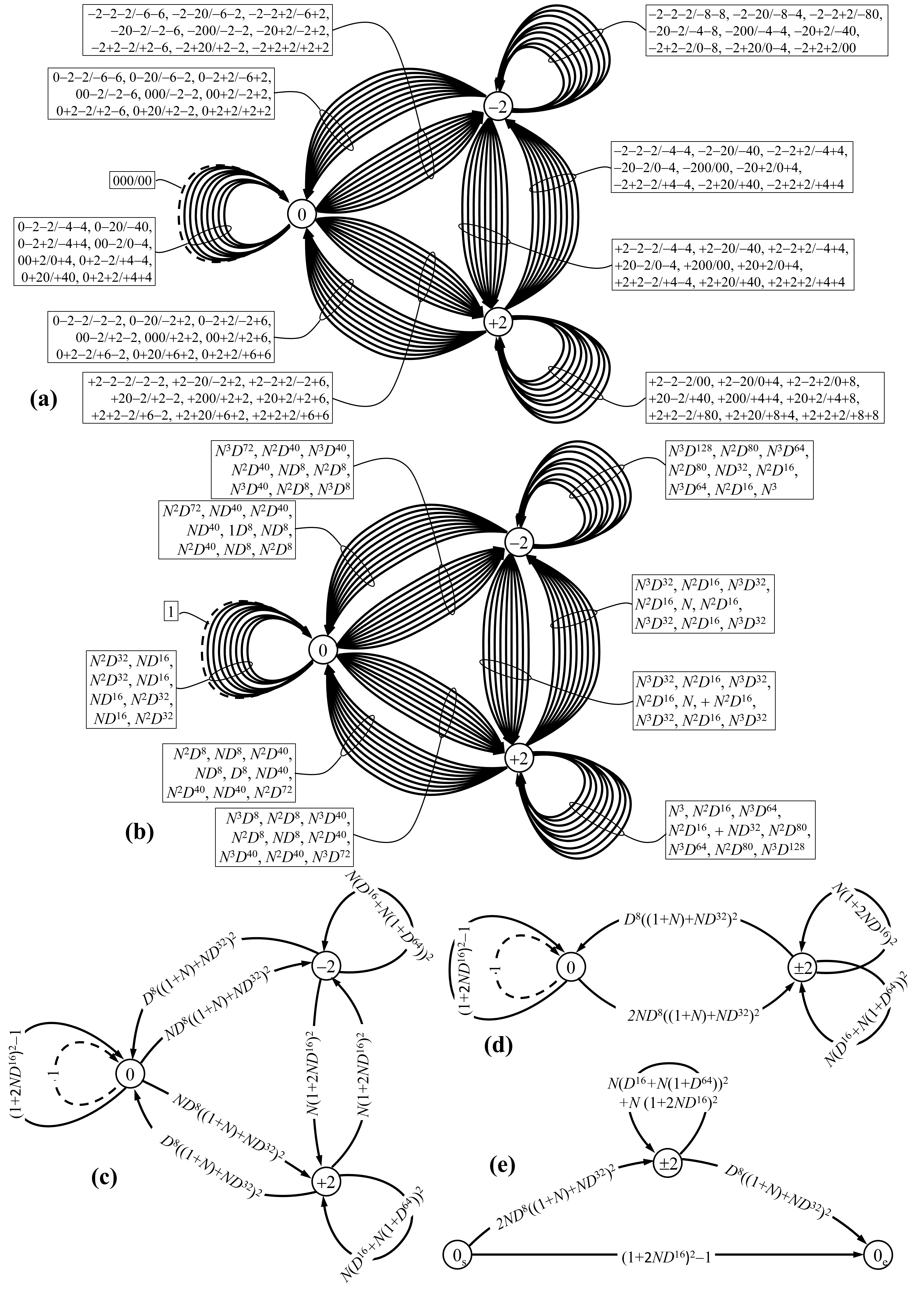}
    \caption{State diagrams of the input/output sequences differences of the rate-$3/2$ Nyquist signaling modulation: (a) State diagram with Input/Output labels, (b) State diagram with $N^iD^j$ labels, (c) Simplified state diagram, aggregating $N^i D^i$ labels with identical starting and ending states, (d) Simplified state diagram with state merging, (e) Modified state diagram for TF computation. Branch with zero Input/Output difference shown with dashed line.}
    \label{fig:State Diagram Input/output Difference Rate-3/2 Modulation}
\end{figure}

The procedure for deriving a close estimate of the \gls{bep} for the rate-$3/2$ \gls{nsm} is identical to that in Appendix~\ref{app:Tight Estimate BEP Rate 2}, for the rate-$2$ \gls{nsm} described in Subsection~\ref{ssec:Modulation of Rate 2}. To begin, we determine the \gls{nsm} \gls{tf}, $T(N,D),$ using the state diagram of the input/output sequence differences given in Figure~\ref{fig:State Diagram Input/output Difference Rate-3/2 Modulation}(a), arising from the trellis section shown in Figure~\ref{fig:Trellis Input Difference Filter h_0 Rate-3/2 Modulation}(b). Similar to the trellis section in Figure~\ref{fig:Trellis Input Difference Filter h_0 Rate-3/2 Modulation}(b), each branch in this state diagram has an input/output label of the form $\Delta \bar{b}_0[k] \, \Delta \bar{b}_1[k] \, \Delta \bar{b}_2[k] / \Delta \mathring{s}[2k] \, \Delta \mathring{s}[2k+1].$ Because we want to identify every error event that could lead to demodulator decision errors, we've shown the “null” branch, with labels $000/00,$ as a dashed line in Figure~\ref{fig:State Diagram Input/output Difference Rate-3/2 Modulation}(a). This branch, which is also shown as a dashed line in Figures~\ref{fig:State Diagram Input/output Difference Rate-3/2 Modulation}(b)--\ref{fig:State Diagram Input/output Difference Rate-3/2 Modulation}(d), should not be a part of any error event and, as a result, should not be taken into account when calculating the \gls{tf}.

In order to move forward with the derivation of the rate-$3/2$ \gls{nsm} \gls{tf}, we rename, in Figure~\ref{fig:State Diagram Input/output Difference Rate-3/2 Modulation}(b), the branches of the state diagram of Figure~\ref{fig:State Diagram Input/output Difference Rate-3/2 Modulation}(a), by assigning labels of the form $N^i D^j.$ In each branch in the state diagram of Figure~\ref{fig:State Diagram Input/output Difference Rate-3/2 Modulation}(b), with label $N^i D^j,$ $i$ represents the number of non-null components in $\Delta \bar{b}_0[k] \, \Delta \bar{b}_1[k] \, \Delta \bar{b}_2[k],$ while $j$ represents the squared norm of $(\Delta \mathring{s}[2k], \Delta \mathring{s}[2k+1]).$

To simplify the calculation of the \gls{tf}, $T(N,D),$ we begin by aggregating, in Figure~\ref{fig:State Diagram Input/output Difference Rate-3/2 Modulation}(c), the labels of the branches in Figure~\ref{fig:State Diagram Input/output Difference Rate-3/2 Modulation}(b) which begin in a common state and end in a common state. Evidently, the “null” branch, labeled $N^0 D^0=1,$ which should be ignored for the \gls{tf} expression, is not merged with the other branches departing from and arriving at the same $0$ state.

To further simplify the derivation of the said \gls{tf}, states $-2$ and $+2,$ which play identical roles in terms of aggregate branch labels, are merged, in Figure~\ref{fig:State Diagram Input/output Difference Rate-3/2 Modulation}(d), into a single state, named $\pm 2.$ Because of this, in Figure~\ref{fig:State Diagram Input/output Difference Rate-3/2 Modulation}(c), the labels of branches connecting state $0$ to states $-2$ and $+2$ are aggregated, and labels of branches leaving state $-2$ to state $+2$ or state $+2$ to state $-2$ are replaced by a single branch with the same label that starts and ends at the same aggregate state, $\pm 2.$

The final step before finding the \gls{tf}, as illustrated in Figure~\ref{fig:State Diagram Input/output Difference Rate-3/2 Modulation}(e), is to remove the “null” branch, labeled $1,$ and split the $0$ state into a starting state, labeled $0_\text{s},$ and an ending state, labeled $0_\text{e}.$

Using Figure~\ref{fig:State Diagram Input/output Difference Rate-3/2 Modulation}(e) and a process similar to that used to derive the \gls{tf} of convolutional codes, we arrive at the desired \gls{tf},
\begin{equation} \label{eq:Transfer Function Rate-3/2 Modulation}
T(N,D) = (1+2ND^{16})^2-1 + \frac{2N D^{16} \left( (1+N) + ND^{32}\right)^4}{1 - N \left( \left(D^{16}+N(1+D^{64}) \right)^2 + \left( 1+2N D^{16} \right)^2 \right)}.
\end{equation}
The expansion of this \gls{tf}, which is limited to the main terms of least power of $D,$ which correspond to the \gls{nsm}'s minimal squared Euclidean distance, yields
%
%
\begin{equation} \label{eq:Expansion Transfer Function Rate-3/2 Modulation}
T(N,D) = \left(4\,N-\frac{2\,N\,{\left(1+N\right)}^4}{N\,\left(1+N^2\right)-1}\right)\,D^{16}+\left(4\,N^2+\frac{12\,N^3\,{\left(1+N\right)}^4}{{\left(N\,\left(1+N^2\right)-1\right)}^2}\right)\,D^{32} + \cdots.
\end{equation}
Since the smallest squared Euclidean distance with scaled filters, $\mathring{h}_m[k],$ $m=0,1,2,$ by a scaling factor equal to $2,$ is $16,$ this expansion supports the hypothesis that the rate-$3/2$ modulation's minimum squared Euclidean distance with normalized filters is equal to $4.$ As a result, the proposed \gls{nsm} with rate $3/2$ has the extremely pleasant property of achieving the minimum Euclidean distance of $2$-ASK perfectly.

In order to determine an approximation of the \gls{bep} for the suggested rate-$3/2$ \gls{nsm}, we apply the same rules as in Appendix~\ref{app:Tight Estimate BEP Rate 2} and derive an expansion of the reduced \gls{tf},
%
\begin{equation} \label{eq:Multiplicity and Distance Rate-3/2 Modulation}
\dot{T}(D) = \left. N \cfrac{\partial T(N,D)}{\partial N} \right|_{N=1/2} = 65\,D^{16}+488\,D^{32}+3300\,D^{48}+19800\,D^{64}+112480\,D^{80} + \cdots,
\end{equation}
limited to its major terms.

The suggested rate-$3/2$ \gls{nsm}'s tight estimate of the \gls{bep}, provided in (\ref{eq:Tight Estimate BEP Rate 3-2}), is obtained by preserving only error events with minimum squared Euclidean distance. With $2$-ASK as a reference, the multiplicity factor, $65,$ in (\ref{eq:Multiplicity and Distance Rate-3/2 Modulation}), must be divided by three because three bipolar symbols are sent at each section of the \gls{nsm}'s trellis, shown in Figure~\ref{fig:Trellis Filter h_0 Rate-3/2 Modulation}(b). However, unlike the rate-$2$ \gls{nsm} discussed in Subsection~\ref{ssec:Modulation of Rate 2}, it does not require any normalization within the square-root argument of the error function complementary, $\operatorname{erfc}(\cdot),$ because all three bipolar symbols use filters with identical energies and are thus transmitted with the same average energy.

To conclude this appendix, it is important to note that the multiplicity of the \gls{bep} of the rate-$3/2$ modulation, in (\ref{eq:Tight Estimate BEP Rate 2}), has increased explosively to $65/3$ as contrasted to the multiplicity of $1,$ for $2$-ASK. This expected result is due to two different sorts of degeneration, which are more clearly visible when looking at the transitions in the input/output sequence trellis in Figure~\ref{fig:Trellis Input Difference Filter h_0 Rate-3/2 Modulation}(b). On the one hand, with a non-null input difference of $-2+2+2$ (respectively, $+2-2-2$), but a null output difference of $00,$ the branch connecting state $-2$ (respectively, $+2$) to itself is the cause of the first type of degeneration. Due to the fact that all three input differences are simultaneously non-null, in the set $\{ \pm 2 \},$ transiting over such a branches happens with a relatively low probability of $(\tfrac{1}{2})^3.$ On the other hand, with a non-null input difference of $+200$ (respectively, $-200$), but a null output difference of $00,$ the branch connecting state $-2$ (respectively, $+2$) to state $+2$ (respectively, $-2$) is the cause of the second type of degeneration. This kind of degeneration is closely related to the quincunx alternating-sign error event that has been pointed out in Subsection~\ref{ssec:Two-Dimensional Rate-2 NSMs}. Because it happens with a larger probability of $1/2,$ its contribution to the multiplicity factor in the approximate expression of the \gls{bep} is greater than that of the first type of degeneration.



\section{Symbolic Determination of the RTFs of Rate-(Q+1)/Q NSMs} \label{app:Symbolic Determination Reduced Transfer Function Rate-(Q+1)/Q NSM}

The iterative determination of the truncated \gls{tf}, $T(N,D;P),$ of rate-$2$ \glspl{nsm}, described in Subsection~\ref{ssec:Determination Truncated Reduced Transfer Function} of Appendix~\ref{app:Iterative Determination Transfer Function Rate-2 NSM}, and achieved in Algorithm~\ref{alg:Truncated Transfer Function T(N,D;P) Rate-2 NSM}, was affordable numerically. The requirement to have $L_1 > 1,$ in addition to $L_0 > 1,$ was indeed a perfect configuration since it prevented parallel branches between any two states in the state diagram.

When $L_1 = 1,$ which is a pretty common arrangement in this study, parallel branches between any two connected states in the state diagram, cannot be avoided anymore. For any rate-$(Q+1)/Q$ \gls{nsm}, $Q \ge 1,$ state diagram admits two possible configurations. In the first configuration, any two connected states, $\bm{\sigma}^\prime$ and $\bm{\sigma},$ are bound by $3^Q$ parallel branches, corresponding to the $3^Q$ possible values taken by $Q$ consecutive input sequence difference, $\Delta \bar{b}_1[k],$ with components in the ternary set $\{0, \pm 2 \}.$ An example of such a state diagram configuration is shown in Figures~\ref{fig:State Diagram Input/output Difference Rate-3/2 Modulation}(a) and~\ref{fig:State Diagram Input/output Difference Rate-3/2 Modulation}(b), for the rate-$3/2$ \gls{nsm} with $Q=2$ and $3^Q=9$ parallel branches between any two connected states, studied in Section~\ref{Rate-3/2 Approaching NSMs}. This first state diagram configuration is adopted in Algorithm~\ref{alg:One-Shot Reduced Transfer Function T(D) Rate-(Q+1)/Q NSM} for the symbolic determination of the \gls{rtf}, $\dot{T}(D),$ of an arbitrary rate-$(Q+1)/Q$ \gls{nsm}.

\begin{algorithm}[H]
\caption{One-shot, \emph{one-step}, symbolic determination of the RTF, $\dot{T}(D),$ of a rate-$(Q+1)/Q$ NSM and its truncated version $\dot{T}(D;P),$ \emph{without} prior computation of TF $T(N,D)$}\label{alg:One-Shot Reduced Transfer Function T(D) Rate-(Q+1)/Q NSM}
\begin{algorithmic}
\Require $\mathring{\bm{h}}_0 = (\mathring{h}_0[0], \mathring{h}_0[1], \ldots, \mathring{h}_0[L_0-1]),$ $\mathring{\bm{h}}_1 = (\mathring{h}_1[0]),$ $Q \ge 1$ and $Q \mid L_0$, $P$ \Comment{$L_0$ must be a multiple of $Q$, $P$ is the Taylor series truncation order}
\Ensure $\dot{T}(D),$ $\dot{T}(D;P)$

\State Follows Algorithm~\ref{alg:One-Shot Two-Step Reduced Transfer Function T(D) Rate-(Q+1)/Q NSM} up to Step \ref{step:last two-step algorithm common step}

\State $L_{\bm{\sigma}^\prime \bm{\sigma}}(D) \gets \left.L_{\bm{\sigma}^\prime \bm{\sigma}}(N,D)\right|_{N=1/2}, (\bm{\sigma}^\prime, \bm{\sigma}) \in \Sigma \times \Sigma$
\State $\dot{L}_{\bm{\sigma}^\prime \bm{\sigma}}(D) \gets \left. N \tfrac{\partial }{\partial N} L_{\bm{\sigma}^\prime \bm{\sigma}}(N,D)\right|_{N=1/2}, (\bm{\sigma}^\prime, \bm{\sigma}) \in \Sigma \times \Sigma$
\State $\bm{L}(D) \gets (L_{\bm{\sigma}^\prime \bm{\sigma}}(D))_{(\bm{\sigma}^\prime, \bm{\sigma}) \in \Sigma \times \Sigma}$ \Comment{$\bm{L}(D)$ is a $|\Sigma| \times |\Sigma|$ square matrix, with $|\Sigma| = 3^{M-1}$}
\State $\dot{\bm{L}}(D) \gets (\dot{L}_{\bm{\sigma}^\prime \bm{\sigma}}(D))_{(\bm{\sigma}^\prime, \bm{\sigma}) \in \Sigma \times \Sigma}$ \Comment{$\dot{\bm{L}}(D)$ is a $|\Sigma| \times |\Sigma|$ square matrix}
\State Let $\bm{T}^e(D) = (T_{\bm{\sigma}}^e(D))_{\bm{\sigma} \in \Sigma}$ and $\dot{\bm{T}}^e(D) = (\dot{T}^e_{\bm{\sigma}}(D))_{\bm{\sigma} \in \Sigma} $ be two symbolic $1 \times|\Sigma|$ row vectors \Comment{$\bm{T}^e(D)$ is a version of $\bm{T}(D),$ which focuses on the ending state $\bm{\sigma}=\bm{0}_e$}
\State $\bm{T}^s(D) \gets \bm{T}^e(D)$ \Comment{$\bm{T}^s(D)$ is a version of $\bm{T}(D),$ which focuses on the starting state $\bm{\sigma}=\bm{0}_s$}
\State $T^s_{\bm{0}}(D) \gets 1$ \Comment{For the assessment of all error events, we set $T_{\bm{0}}(D) = 1$ when the zero state is $\bm{\sigma}=\bm{0}_s$}
\State $\dot{\bm{T}}^s(D) \gets \dot{\bm{T}}^e(D)$ \Comment{$\dot{\bm{T}}^s(D)$ is a version of $\dot{\bm{T}}(D),$ which focuses on the starting state $\bm{\sigma}=\bm{0}_s$}
\State $\dot{T}^s_{\bm{0}}(D) \gets 0$ \Comment{For the assessment of all error events, we set $\dot{T}_{\bm{0}}(D) = 0$ when the zero state is $\bm{\sigma}=\bm{0}_s$}
\State $\left( \begin{smallmatrix} \bm{T}^e(D) & \dot{\bm{T}}^e(D) \end{smallmatrix} \right) \gets  \text{Solve} \left( \left( \begin{smallmatrix} \bm{T}^e(D) & \dot{\bm{T}}^e(D) \end{smallmatrix} \right) = \left( \begin{smallmatrix} \bm{T}^s(D) & \dot{\bm{T}}^s(D) \end{smallmatrix} \right) \left( \begin{smallmatrix} \bm{L}(D) &  \dot{\bm{L}}(D) \\ \bm{0}_{|\Sigma| \times |\Sigma|} & \bm{L}(D) \end{smallmatrix} \right)  \right)$ \Comment{$\bm{0}_{|\Sigma| \times |\Sigma|}$ is the $|\Sigma| \times |\Sigma|$ null square matrix}
\State $\dot{T}(D) \gets \dot{T}^e_{\bm{0}}(D)$ \Comment{The symbolic expression of the reduced transfer function, $\dot{T}(D),$ when the zero state is seen as the to ending state $\bm{\sigma} = \bm{0}_e$}
\State $\dot{T}(D;P) \gets \text{Taylor} \left(\dot{T}(D), P\right)$ \Comment{Truncated Taylor series of the reduced transfer function $\dot{T}(D)$}

\end{algorithmic}
\end{algorithm}

In the second configuration, we use the fact that $L_1 = 1,$ and therefore that there is no memory with respect to the input sequence difference, $\Delta \bar{b}_1[k].$ To any pair of connected states, $\bm{\sigma}^\prime$ and $\bm{\sigma},$ corresponds $Q$ consecutive auxiliary state diagram sections, interconnected by $Q-1$ auxiliary intermediate states. Each introduced auxiliary section, among $Q,$ corresponds to a single input sequence difference, $\Delta \bar{b}_1[k],$ and hence showcases only three parallel branches in each auxiliary section between states $\bm{\sigma}^\prime$ and $\bm{\sigma}.$ This configuration is illustrated in Figure~\ref{fig:Trellis Input Difference Filter h_0 Rate-4/3 Modulation}, for the rate-$4/3$ \gls{nsm} with $Q=3,$ studied in Subsection~\ref{A basic rate-4/3 NSM inspired by a previous basic rate-3/2 NSM}, through its trellis. From the trellis point of view, it has the merit of factorizing and considerably simplifying the computation of branches and states metrics, when maximum-likelihood detection with the Viterbi algorithm is used. From the state diagram point of view, which is our main focus here, it greatly simplifies the derivation of the branch labels, needed in \gls{tf} derivation. This second state diagram configuration is adopted in Algorithms~\ref{alg:Simplified One-Shot Reduced Transfer Function T(D) Rate-(Q+1)/Q NSM} and~\ref{alg:Simplified Iterative Reduced Transfer Function T(D) Rate-(Q+1)/Q NSM} for the symbolic determination of the \gls{rtf}, $\dot{T}(D),$ of an arbitrary rate-$(Q+1)/Q$ \gls{nsm}.

\begin{algorithm}[H]
\caption{Simplified one-shot, \emph{one-step}, symbolic determination of the RTF, $\dot{T}(D),$ of a rate-$(Q+1)/Q$ NSM and its truncated version $\dot{T}(D;P),$ \emph{without} prior computation of TF $T(N,D)$}\label{alg:Simplified One-Shot Reduced Transfer Function T(D) Rate-(Q+1)/Q NSM}
\begin{algorithmic}[1]
\Require $\mathring{\bm{h}}_0 = (\mathring{h}_0[0], \mathring{h}_0[1], \ldots, \mathring{h}_0[L_0-1]),$ $\mathring{\bm{h}}_1 = (\mathring{h}_1[0]),$ $Q \ge 1$ and $Q \mid L_0,$ $P$
\Ensure $\dot{T}(D),$ $\dot{T}(D;P)$

\State Follows Algorithm~\ref{alg:Simplified One-Shot Two-Step Reduced Transfer Function T(D) Rate-(Q+1)/Q NSM} up to Step \ref{step:last simplified two-step algorithm common step}

\State $L_{\bm{\sigma}^\prime \bm{\sigma}}^+(D) \gets \left.L_{\bm{\sigma}^\prime \bm{\sigma}}(N,D)\right|_{N=1/2}+\left.L_{-\bm{\sigma}^\prime \bm{\sigma}}(N,D)\right|_{N=1/2}, \bm{\sigma}^\prime \in \Sigma^+,  \bm{\sigma} \in \Sigma^+ \cup \{ \bm{0} \}$
\State $L_{\bm{\sigma}^\prime \bm{\sigma}}^+(D) \gets \left.L_{\bm{\sigma}^\prime \bm{\sigma}}(N,D)\right|_{N=1/2}, \bm{\sigma}^\prime = \bm{0}, \bm{\sigma} \in \Sigma^+ \cup \{ \bm{0} \}$
\State $\dot{L}_{\bm{\sigma}^\prime \bm{\sigma}}^+(D) \gets \left. N \tfrac{\partial }{\partial N} L_{\bm{\sigma}^\prime \bm{\sigma}}(N,D)\right|_{N=1/2} + \left. N \tfrac{\partial }{\partial N} L_{\bm{-\sigma}^\prime \bm{\sigma}}(N,D)\right|_{N=1/2}, \bm{\sigma}^\prime \in \Sigma^+,  \bm{\sigma} \in \Sigma^+ \cup \{ \bm{0} \}$
\State $\dot{L}_{\bm{\sigma}^\prime \bm{\sigma}}^+(D) \gets \left. N \tfrac{\partial }{\partial N} L_{\bm{\sigma}^\prime \bm{\sigma}}(N,D)\right|_{N=1/2}, \bm{\sigma}^\prime = \bm{0},  \bm{\sigma} \in \Sigma^+ \cup \{ \bm{0} \}$

\State $\bm{L}^+(D) \gets (L_{\bm{\sigma}^\prime \bm{\sigma}}^+(D))_{(\bm{\sigma}^\prime, \bm{\sigma}) \in (\Sigma^+ \cup \{ \bm{0} \}) \times (\Sigma^+ \cup \{ \bm{0} \})}$
\State $\dot{\bm{L}}^+(D) \gets (\dot{L}_{\bm{\sigma}^\prime \bm{\sigma}}^+(D))_{(\bm{\sigma}^\prime, \bm{\sigma}) \in (\Sigma^+ \cup \{ \bm{0} \}) \times (\Sigma^+ \cup \{\bm{0} \})}$ \label{step:last algorithm common step}

\State Let $\bm{T}_+^e(D) = (T_{\bm{\sigma}}^e(D))_{\bm{\sigma} \in \Sigma^+ \cup \{ \bm{0} \}}$ and $ \dot{\bm{T}}_+^e(D) = (\dot{T}^e_{\bm{\sigma}}(D))_{\bm{\sigma} \in \Sigma^+ \cup \{ \bm{0} \}} $ be two symbolic $1 \times (|\Sigma^+|+1)$ row vectors

\State $\bm{T}_+^s(D) \gets \bm{T}_+^e(D),$ $T^s_{\bm{0}}(D) \gets 1$
\State $\dot{\bm{T}}_+^s(D) \gets \dot{\bm{T}}_+^e(D),$ $\dot{T}^s_{\bm{0}}(D) \gets 0$

\State $\left( \begin{smallmatrix} \bm{T}_+^e(D) & \dot{\bm{T}}_+^e(D) \end{smallmatrix} \right) \gets  \text{Solve} \left( \left( \begin{smallmatrix} \bm{T}_+^e(D) & \dot{\bm{T}}_+^e(D) \end{smallmatrix} \right) = \left( \begin{smallmatrix} \bm{T}_+^s(D) & \dot{\bm{T}}_+^s(D) \end{smallmatrix} \right) \left( \begin{smallmatrix} \bm{L}^+(D) &  \dot{\bm{L}}^+(D) \\ \bm{0}_{(|\Sigma^+|+1) \times (|\Sigma^+|+1)} & \bm{L}^+(D) \end{smallmatrix} \right)  \right)$
\State $\dot{T}(D) \gets \dot{T}^e_{\bm{0}}(D)$
\State $\dot{T}(D;P) \gets \text{Taylor} \left(\dot{T}(D), P\right)$

\end{algorithmic}
\end{algorithm}

\begin{algorithm}[H]
\caption{Simplified iterative, \emph{one-step} symbolic determination of the truncated version $\dot{T}(D;P)$ of the RTF, $\dot{T}(D),$ of a rate-$(Q+1)/Q$ NSM, \emph{without} intermediate TF $T(N,D)$}\label{alg:Simplified Iterative Reduced Transfer Function T(D) Rate-(Q+1)/Q NSM}
\begin{algorithmic}
\Require $\mathring{\bm{h}}_0 = (\mathring{h}_0[0], \mathring{h}_0[1], \ldots, \mathring{h}_0[L_0-1]),$ $\mathring{\bm{h}}_1 = (\mathring{h}_1[0]),$ $Q \ge 1$ and $Q \mid L_0,$ $P$
\Ensure $\dot{T}(D;P)$

\State Follows Algorithm~\ref{alg:Simplified One-Shot Reduced Transfer Function T(D) Rate-(Q+1)/Q NSM} up to Step \ref{step:last algorithm common step}

\State Let $\bm{T}_+(D) = (T_{\bm{\sigma}}(D))_{\bm{\sigma} \in \Sigma^+ \cup \{ \bm{0} \}}$ and $\dot{\bm{T}}_+(D) = (\dot{T}_{\bm{\sigma}}(D))_{\bm{\sigma} \in \Sigma^+ \cup \{ \bm{0} \}} $ be two symbolic $1 \times (|\Sigma^+|+1)$ row vectors

\State $\bm{T}_+(D) \gets \bm{0}_{1 \times (|\Sigma^+|+1)}$ \Comment{$\bm{0}_{1 \times (|\Sigma^+|+1)}$ is the $1 \times (|\Sigma^+|+1)$ null row vector}
\State $\dot{\bm{T}}_+(D) \gets \bm{0}_{1 \times (|\Sigma^+|+1)}$ \Comment{$\bm{0}_{1 \times (|\Sigma^+|+1)}$ is the $1 \times (|\Sigma^+|+1)$ null row vector}
\State $q \gets 0$ \Comment{Iteration number initialized to $0$}

\Repeat
 \State $\bm{\xi}(D) \gets \dot{\bm{T}}_+(D)$ \Comment{$\bm{\xi}(D)$ stores the old value of $\dot{\bm{T}}_+(D),$ obtained at the previous iteration}
 \State $T_{\bm{0}}(D) \gets 1$
 \State $\dot{T}_{\bm{0}}(D) \gets 0$
 \State $\bm{\zeta}(D) \gets \bm{T}_+(D) \bm{L}^+(D)$ \Comment{$\bm{\zeta}(D)$ stores the new value of $\bm{T}_+(D),$ obtained at the current iteration}
 \State $\dot{\bm{T}}_+(D) \gets \bm{T}_+(D) \dot{\bm{L}}^+(D) + \dot{\bm{T}}_+(D) \bm{L}^+(D)$ \Comment{Notice that we use here the old value of $\bm{T}_+(D),$ obtained at the previous iteration}
 \State $\bm{T}_+(D) \gets \bm{\zeta}(D)$
 \State $\bm{T}_+(D) \gets \text{Truncate}(\bm{T}_+(D), P)$ \Comment{Truncation to at most $P$ terms, with only terms up to $D^{P-1}$ retained}
 \State $\dot{\bm{T}}_+(D) \gets \text{Truncate}(\dot{\bm{T}}_+(D), P)$ 
\Until {$\dot{\bm{T}}_+(D) = \bm{\xi}(D)$}

\State $\dot{T}(D;P) \gets \dot{T}_{\bm{0}}(D)$

\end{algorithmic}
\end{algorithm}

For the symbolic determination of the \gls{rtf}, $\dot{T}(D),$ of any rate-$(Q+1)/Q$ \gls{nsm}, we rely on a modified state diagram, similar to the one introduced in Appendix~\ref{app:Iterative Determination Transfer Function Rate-2 NSM}. This modified state diagram is obtained by removing the unique branch looping into the zero state, $\bm{0},$ and splitting this zero state into a starting zero state, $\bm{0}_s,$ and an ending state, $\bm{0}_e.$ Due to the two roles that the zero state, $\bm{0},$ plays in the modified state diagram, we introduce the two new sets of states, $\Sigma_s \triangleq \Sigma^* \cup \{ \bm{0}_s \}$ and $\Sigma_e \triangleq \Sigma^* \cup \{ \bm{0}_e \},$ where, as in Appendices~\ref{app:d_min^2 Rate-2 NSM} and~\ref{app:Iterative Determination Transfer Function Rate-2 NSM}, $\Sigma^*$ denotes the set of all intermediate states, excluding the split zero states $\bm{0}_s$ and $\bm{0}_e.$ For completeness, we also need the set of all sets, $\Sigma = \Sigma^* \cup \{ \bm{0} \},$ in the original state diagram. Notice that $\Sigma_s,$ $\Sigma_e$ and $\Sigma$ have a common size $|\Sigma_s| = |\Sigma_e| = |\Sigma| = 3^{M-1},$ where $M \triangleq L_0/Q.$

Moreover, we use (\ref{eq:Equations Relating Instatiated Transfer Functions}) and (\ref{eq:Equations Relating Reduced Transfer Functions}), and rewrite them, in an altered aggregate form, suitable for matrix writing, as 
\begin{equation}
\label{eq:Aggregate Equations Relating Reduced Transfer Functions}
\left\{
\begin{aligned}
T_{\bm{\sigma}}(D) & = \sum_{\bm{\sigma}^\prime \in \Sigma_s} T_{\bm{\sigma}^\prime}(D) L_{\bm{\sigma}^\prime \bm{\sigma}}(D), \quad \bm{\sigma} \in \Sigma_e, \\
\dot{T}_{\bm{\sigma}}(D) & = \sum_{\bm{\sigma}^\prime \in \Sigma_s} \dot{T}_{\bm{\sigma}^\prime}(D) L_{\bm{\sigma}^\prime \bm{\sigma}}(D)
+ \sum_{\bm{\sigma}^\prime \in \Sigma_s} T_{\bm{\sigma}^\prime}(D) \dot{L}_{\bm{\sigma}^\prime \bm{\sigma}}(D), \quad \bm{\sigma} \in \Sigma_e,
\end{aligned}
\right.
\end{equation}

To prepare for the algorithmic description of the symbolic determination of the \gls{rtf}, $\dot{T}(D),$ in Algorithm~\ref{alg:One-Shot Reduced Transfer Function T(D) Rate-(Q+1)/Q NSM}, we introduce the $1 \times |\Sigma|$ row vectors $\bm{T}^s = (T_{\bm{\sigma}}(D))_{\bm{\sigma} \in \Sigma_s}$ and $\dot{\bm{T}}^s = (\dot{T}_{\bm{\sigma}}(D))_{\bm{\sigma} \in \Sigma_s},$ on the starting zero state side, and $\bm{T}^e \triangleq (T_{\bm{\sigma}}(D))_{\bm{\sigma} \in \Sigma_e}$ and $\dot{\bm{T}}^e \triangleq (\dot{T}_{\bm{\sigma}}(D))_{\bm{\sigma} \in \Sigma_e},$ on the ending zero state side. We also introduce the $|\Sigma| \times |\Sigma|$ matrices $\bm{L} \triangleq (L_{\bm{\sigma}^\prime \bm{\sigma}}(D))_{(\bm{\sigma}^\prime, \bm{\sigma}) \in \Sigma_s \times \Sigma_e}$ and $\dot{\bm{L}} \triangleq (\dot{L}_{\bm{\sigma}^\prime \bm{\sigma}}(D))_{(\bm{\sigma}^\prime, \bm{\sigma}) \in \Sigma_s \times \Sigma_e}.$ Then, the aggregated $2 |\Sigma|$ equations in (\ref{eq:Aggregate Equations Relating Reduced Transfer Functions}) can be compacted in matrix form as
\begin{equation} \label{eq: Original Symbolic Matrix System Reduced Transfer Function}
\left( \begin{matrix} \bm{T}^e(D) & \dot{\bm{T}}^e(D) \end{matrix} \right) = \left( \begin{matrix} \bm{T}^s(D) & \dot{\bm{T}}^s(D) \end{matrix} \right) \left( \begin{matrix} \bm{L}(D) &  \dot{\bm{L}}(D) \\ \bm{0}_{|\Sigma| \times |\Sigma|} & \bm{L}(D) \end{matrix} \right),
\end{equation}
where $\bm{0}_{|\Sigma| \times |\Sigma|}$ is the $|\Sigma| \times |\Sigma|$ null matrix. After solving (\ref{eq: Original Symbolic Matrix System Reduced Transfer Function}), the desired \gls{rtf}, $\dot{T}(D),$ is determined as $\dot{T}(D)=\dot{T}_{\bm{0}_e}(D),$ $\dot{T}_{\bm{0}_e}(D)$ being the component of $\bm{T}^e,$ associated to ending state, $\bm{0}_e.$

As demonstrated in Algorithm~\ref{alg:One-Shot Reduced Transfer Function T(D) Rate-(Q+1)/Q NSM}, this linear system of $2 |\Sigma|$ unknowns implies the processing of a $2 |\Sigma| \times 2 |\Sigma|$ square matrix. The size of this matrix explodes exponentially as $M=L_0/Q$ increases, leading to an explosion of symbolic computation complexity, when $L_0$ increases, for a given value of $Q$, and therefore a given \gls{nsm} rate $\rho = (Q+1)/Q.$

One way to simplify the resolution, of symbolic matrix system (\ref{eq: Original Symbolic Matrix System Reduced Transfer Function}), is to proceed in two steps, by solving for $\bm{T}^e(D)$ first, using equation $\bm{T}^e(D) = \bm{T}^s(D) \bm{L}(D),$ and $\dot{\bm{T}}^e(D)$ second, using equation $\dot{\bm{T}}^e(D) = \bm{T}^s(D) \dot{\bm{L}}(D) + \dot{\bm{T}}^s(D) \bm{L}(D).$ This two-step approach reduces the common size of involved matrices by half, with respect to the original approach, leading to a significant reduction in complexity, even though it requires the processing to be repeated twice.

Another different way to reduce the complexity of the symbolic derivation of the \gls{rtf}, is to notice that $L_{\bm{\sigma}^\prime \bm{\sigma}}(D) = L_{-\bm{\sigma}^\prime,-\bm{\sigma}}(D)$ for any pair of states, $(\bm{\sigma}^\prime, \bm{\sigma})$ in $\bar{\Sigma} \triangleq \Sigma^* \cup \{\bm{0}_s, \bm{0}_e\}.$ Based on this observation, we can affirm that the solutions, $T_{\bm{\sigma}}(D)$ and $\dot{T}_{\bm{\sigma}}(D),$ $\bm{\sigma} \in \Sigma^*,$ to the symbolic equations in (\ref{eq:Aggregate Equations Relating Reduced Transfer Functions}) and symbolic matrix equation in (\ref{eq: Original Symbolic Matrix System Reduced Transfer Function}), is such that $T_{\bm{\sigma}}(D) = T_{-\bm{\sigma}}(D)$ and $\dot{T}_{\bm{\sigma}}(D) = \dot{T}_{-\bm{\sigma}}(D),$ $\bm{\sigma} \in \bar{\Sigma}.$ In light of this, the set of states $\Sigma^*$ must be split into two complementary subsets, $\Sigma^{+*}$ and $\Sigma^{-*} = -\Sigma^{+*} \triangleq \{-\bm{\sigma} \mid \bm{\sigma} \in \Sigma^{+*} \},$ such that  $\Sigma^{+*} \cup \Sigma^{-*} = \Sigma^*$ and $\Sigma^{+*} \cap \Sigma^{-*} = \varnothing.$ Without loss of generality, we can define $\Sigma^{+*}$ as the set of non-null states with positive first non-null component. In this case, $\Sigma^{-*}$ will be the set of non-null states with negative first non-null component. Clearly, subsets $\Sigma^{+*}$ and $\Sigma^{-*}$ have a common size, $|\Sigma^{+*}| = |\Sigma^{-*}| = |\Sigma^*|/2 = (3^{M-1}-1)/2.$

Based on the previous considerations, we can reduce the $2 |\Sigma_e| = 2\times3^{M-1}$ equations in (\ref{eq:Aggregate Equations Relating Reduced Transfer Functions}), corresponding to the $|\Sigma_e| = 3^{M-1}$ states in $\Sigma_e,$ to only  $2( |\Sigma^{+*}| +1) = 3^{M-1}+1$ states in the set $\Sigma_e^+ \triangleq \Sigma^{+*} \cup \{ \bm{0}_e \}.$ By grouping terms in $T_{\bm{\sigma}^\prime}(D)$ and $T_{-\bm{\sigma}^\prime}(D),$ on the one hand, and terms in $\dot{T}_{\bm{\sigma}^\prime}(D)$ and $\dot{T}_{-\bm{\sigma}^\prime}(D),$ on the other hand, for $\bm{\sigma} \in \Sigma^{+*},$ we end up with the new reduced set of equations
\begin{equation}
\label{eq:Simplified Aggregate Equations Relating Reduced Transfer Functions}
\left\{
\begin{aligned}
T_{\bm{\sigma}}(D) & = T_{\bm{0}_s}(D) L_{\bm{0}_s \bm{\sigma}}(D) + \sum_{\bm{\sigma}^\prime \in \Sigma^+} T_{\bm{\sigma}^\prime}(D) (L_{\bm{\sigma}^\prime \bm{\sigma}}(D) + L_{-\bm{\sigma}^\prime, \bm{\sigma}}(D)), \quad \bm{\sigma} \in \Sigma_e^+, \\
\dot{T}_{\bm{\sigma}}(D) & = \dot{T}_{\bm{0}_s}(D) L_{\bm{0}_s \bm{\sigma}}(D)
\sum_{\bm{\sigma}^\prime \in \Sigma^+} \dot{T}_{\bm{\sigma}^\prime}(D) (L_{\bm{\sigma}^\prime \bm{\sigma}}(D) + L_{-\bm{\sigma}^\prime, \bm{\sigma}}(D)) \\
& + T_{\bm{0}_s}(D) \dot{L}_{\bm{0}_s \bm{\sigma}}(D)
+ \sum_{\bm{\sigma}^\prime \in \Sigma^+} T_{\bm{\sigma}^\prime}(D) (\dot{L}_{\bm{\sigma}^\prime \bm{\sigma}}(D) + \dot{L}_{-\bm{\sigma}^\prime, \bm{\sigma}}(D)), \quad \bm{\sigma} \in \Sigma_e^+.
\end{aligned}
\right.
\end{equation}

In order to write (\ref{eq:Simplified Aggregate Equations Relating Reduced Transfer Functions}) in compact matrix form, we introduce
\begin{equation}
L_{\bm{\sigma}^\prime \bm{\sigma}}^+(D) \triangleq 
    \begin{cases}
      L_{\bm{\sigma}^\prime \bm{\sigma}}(D) + L_{-\bm{\sigma}^\prime, \bm{\sigma}}(D), & \text{if $\bm{\sigma}^\prime \in \Sigma^+$ and $\bm{\sigma} \in \Sigma_e^+$}, \\
      L_{\bm{\sigma}^\prime \bm{\sigma}}(D), & \text{if $\bm{\sigma}^\prime = \bm{0}_s$ and $\bm{\sigma} \in \Sigma_e^+$}.
    \end{cases}
\end{equation}
We also introduce the $1 \times |\Sigma_s^+|$ row vectors $\bm{T}_+^s = (T_{\bm{\sigma}}(D))_{\bm{\sigma} \in \Sigma_s^+}$ and $\dot{\bm{T}}_+^s = (\dot{T}_{\bm{\sigma}}(D))_{\bm{\sigma} \in \Sigma_s^+},$ on the starting zero state side, and $\bm{T}_+^e \triangleq (T_{\bm{\sigma}}(D))_{\bm{\sigma} \in \Sigma_e^+}$ and $\dot{\bm{T}}_+^e \triangleq (\dot{T}_{\bm{\sigma}}(D))_{\bm{\sigma} \in \Sigma_e^+},$ on the ending zero state side. We finally introduce the $|\Sigma_s^+| \times |\Sigma_e^+|$ matrices $\bm{L}^+ \triangleq (L_{\bm{\sigma}^\prime \bm{\sigma}}^+(D))_{(\bm{\sigma}^\prime, \bm{\sigma}) \in \Sigma_s^+ \times \Sigma_e^+}$ and $\dot{\bm{L}}^+ \triangleq (\dot{L}_{\bm{\sigma}^\prime \bm{\sigma}}^+(D))_{(\bm{\sigma}^\prime, \bm{\sigma}) \in \Sigma_s^+ \times \Sigma_e^+}.$ Then, the aggregated $2 |\Sigma_e^+|$ equations in (\ref{eq:Simplified Aggregate Equations Relating Reduced Transfer Functions}) can be compacted in matrix form as
\begin{equation} \label{eq: Simplified Symbolic Matrix System Reduced Transfer Function}
\left( \begin{matrix} \bm{T}_+^e(D) & \dot{\bm{T}}_+^e(D) \end{matrix} \right) = \left( \begin{matrix} \bm{T}_+^s(D) & \dot{\bm{T}}_+^s(D) \end{matrix} \right) \left( \begin{matrix} \bm{L}^+(D) &  \dot{\bm{L}}^+(D) \\ \bm{0}_{|\Sigma_s^+| \times |\Sigma_e^+|} & \bm{L}^+(D) \end{matrix} \right),
\end{equation}
where $\bm{0}_{|\Sigma_s^+| \times |\Sigma_e^+|}$ is the $|\Sigma_s^+| \times |\Sigma_e^+|$ null matrix (notice here that $|\Sigma_s^+| = |\Sigma_e^+| = (3^{M-1}+1)/2$). After solving (\ref{eq: Simplified Symbolic Matrix System Reduced Transfer Function}), the desired \gls{rtf}, $\dot{T}(D),$ is determined as $\dot{T}(D)=\dot{T}_{\bm{0}_e}(D),$ where $\dot{T}_{\bm{0}_e}(D)$ is now the component of $\bm{T}_+^e,$ associated to ending state, $\bm{0}_e.$

As shown in Algorithm~\ref{alg:Simplified One-Shot Reduced Transfer Function T(D) Rate-(Q+1)/Q NSM}, this new linear system has exactly $2 |\Sigma_e^+|$ unknowns and therefore requires the processing of a $2 |\Sigma_s^+| \times 2 |\Sigma_e^+|$ square matrix. When comparing the original system in (\ref{eq: Original Symbolic Matrix System Reduced Transfer Function}) with the simplified one in (\ref{eq: Simplified Symbolic Matrix System Reduced Transfer Function}), we observe a progression from the resolution of a $2 \cdot 3^{M-1} \times 2 \cdot 3^{M-1}$ symbolic matrix problem to a more straightforward $(3^{M-1}+1) \times (3^{M-1}+1)$ symbolic matrix problem. This reduction of complexity can be pushed further by by solving symbolic matrix system (\ref{eq: Simplified Symbolic Matrix System Reduced Transfer Function}) in two steps. We use equation $\bm{T}_+^e(D) = \bm{T}_+^s(D) \bm{L}^+(D),$ to solve for $\bm{T}_+^e(D)$ first, and then equation $\dot{\bm{T}}_+^e(D) = \bm{T}_+^s(D) \dot{\bm{L}}^+(D) + \dot{\bm{T}}_+^s(D) \bm{L}^+(D),$ to solve for $\dot{\bm{T}}^e(D).$ Based on this additional simplification, we move from the resolution of a unique symbolic equation, involving a $2 |\Sigma_s^+| \times 2 |\Sigma_e^+|$ square matrix, to the successive resolution of two symbolic equations, involving two $|\Sigma_s^+| \times |\Sigma_e^+|$ square matrices. Hence, we observe a progression from the resolution of a $2 \cdot 3^{M-1} \times 2 \cdot 3^{M-1}$ matrix problem to the resolution of two $\tfrac{1}{2}(3^{M-1}+1) \times \tfrac{1}{2}(3^{M-1}+1)$ matrix problems.

Algorithm~\ref{alg:Simplified One-Shot Reduced Transfer Function T(D) Rate-(Q+1)/Q NSM} has been successfully utilized to derive the truncated version, $\dot{T}(D;P),$ to order $P,$ of the \gls{rtf}, $\dot{T}(D),$ of the majority of the \glspl{nsm} listed in Tables~\ref{table:NSMs Rate-3/2 Filter Pattern (1, 1, 1, 1)}–-\ref{table:NSMs Rate-5/4 Filter Pattern (2, 1, 1, 1, 1, 1, 1, 1, 1, 1, 1, 1, 1, 0, 0, 0)}. However, for some rate-$(Q+1)/Q$ \glspl{nsm}, this algorithm, which was implemented in \textsc{Matlab}, failed to yield a solution, even after a significant length of time. We therefore turned to Algorithm~\ref{alg:Simplified Iterative Reduced Transfer Function T(D) Rate-(Q+1)/Q NSM}, which has the benefit of never failing, in such circumstances.

The \textsc{Matlab} implementations of Algorithms~\ref{alg:Simplified One-Shot Reduced Transfer Function T(D) Rate-(Q+1)/Q NSM} and~\ref{alg:Simplified Iterative Reduced Transfer Function T(D) Rate-(Q+1)/Q NSM} have been validated by comparing the first terms of the \gls{rtf} $\dot{T}(D),$ calculated experimentally, in Table~\ref{table:NSMs Rate-3/2 Filter Pattern (1, 1, 1, 1)}, and theoretically, in (\ref{eq:Multiplicity and Distance Rate-3/2 Modulation}), for filter $\mathring{h}_0[k]=\delta[k]+\delta[k-1]+\delta[k-2]+\delta[k-3].$


\newpage

\section*{Biographies}

\vspace{0.25cm}

\begin{flushleft}


\begin{biography}
{\includegraphics[width=2.54cm,height=3.18cm,clip,keepaspectratio]{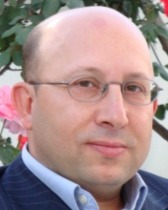}}{}
\textbf{Mohamed Siala} received the engineering degree from École Polytechnique, Palaiseau, France, in 1988, the telecommunications engineering degree from Télécom Paris, Paris, France, in 1990, and the Ph.D. degree in digital communications from Télécom Paris in 1995, with a focus on coding for the magnetic recording channel. From 1990 to 2001, he worked in France at Alcatel Radio-Téléphones, Wavecom, and Orange Labs on GSM physical-layer design, channel estimation techniques for the INMARSAT ICO mobile satellite system, and the development and standardization of 3G radio interfaces. Since 2001, he has been with the Higher School of Communications of Tunis (SUP’COM), Tunisia, where he is currently a Professor in the Department of Applied Mathematics, Signals, and Communications. His research interests include multicarrier modulation, waveform design, channel estimation and synchronization, MIMO systems, adaptive transmission, and cooperative and cognitive radio networks.
\end{biography}

\vspace{0.25cm}

\begin{biography}{\includegraphics[width=1in,height=1.25in,clip,keepaspectratio]{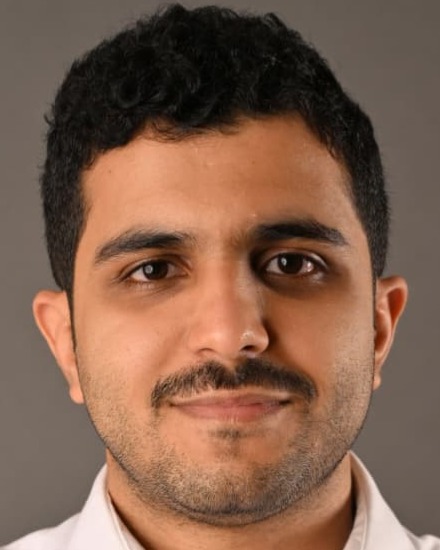}}{}
\textbf{Abdullah Al-Nafisah} holds an M.S. degree in Electrical and Computer Engineering from King Abdullah University of Science and Technology (KAUST). He earned a double major with highest honors in Electrical Engineering and Physics from King Fahd University of Petroleum and Minerals (KFUPM). His research interests include RF electronics and signal processing for telecommunications. He has worked on chaotic circuits with an emphasis on practical, low-cost designs for synchronous and secure communications. Also, he developed a novel CFOA-based filter circuit with a tunable bandwidth. He is currently working on real-time FPGA systems for different research projects.

\end{biography}

\vspace{0.25cm}

\begin{biography}{\includegraphics[width=1in,height=1.25in,clip,keepaspectratio]{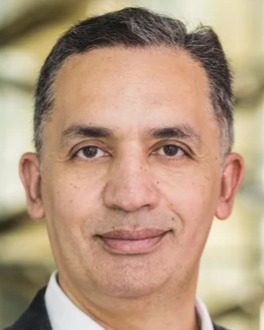}}{}
\textbf{Tareq Al-Naffouri} received the B.S. (Honors) degree in Mathematics and Electrical Engineering from King Fahd University of Petroleum and Minerals (KFUPM), Saudi Arabia. He obtained the M.S. degree in Electrical Engineering from the Georgia Institute of Technology, USA, in 1998, and the Ph.D. degree in Electrical Engineering from Stanford University, USA, in 2004. He has held visiting scholar appointments at the California Institute of Technology (2005–2006) and served as a Fulbright Scholar at the University of Southern California (2008). He is currently a Professor in the Electrical and Computer Engineering Program at King Abdullah University of Science and Technology (KAUST), Saudi Arabia, and Principal Investigator of the Information Science Lab. His research interests include sparse, adaptive, and statistical signal processing, with applications to wireless communications, localization, inference and learning, smart cities, and IoT-enabled networks.
\end{biography}

\end{flushleft}

\end{document}